%% file: EA-HSIIfin_v12_sent.tex
\documentclass[11pt,a4paper]{JHEP3} 
\pdfoutput=1
\usepackage{amsmath,amssymb}
\usepackage{euscript}
\usepackage{slashed}
\usepackage{amsfonts}
\usepackage{soul}
\def\be{\begin{eqnarray}}
\def\ee{\end{eqnarray}}
\newcommand{\al}[1]{\begin{align}#1\end{align}}
\newcommand{\refb}[1]{(\ref{#1})}
\def\0{\nonumber}

\def\EF{\EuScript F}
\def\EG{\EuScript G}\def\ER{\EuScript R}

\usepackage[utf8]{inputenc}
\usepackage{amsmath}
\usepackage{amsthm}
\usepackage{amssymb}
\usepackage{esint}

\allowdisplaybreaks
 
\preprint{SISSA/42/2017/FISI\\ZTF-EP-17-06\\{\tt hep-th/170m.xxxxx } }

\title{One-loop effective actions and higher spins. II}

\author{ L.~Bonora$^{a}$, M.~Cvitan$^{b}$, P.~Dominis
Prester$^{c}$, S.Giaccari$^{b}$,T.~Štemberga$^{b}$
\\\textit{${}^{a}$ International School for Advanced Studies (SISSA),\\Via
Bonomea 265, 34136 Trieste, Italy, and INFN, Sezione di
Trieste\\}%
\textit{${}^{b}$ Department of Physics, Faculty of Science, University of Zagreb, \\
Bijenička cesta 32, 10000 Zagreb, Croatia\\}%
\textit{${}^{c}$ Department of Physics, University of Rijeka,\\
Radmile Matejčić 2, 51000 Rijeka, Croatia\\}%
E-mail: \email{bonora@sissa.it}, \email{mcvitan@phy.hr}, 
\email{pprester@phy.uniri.hr},
\email{sgiaccari@phy.hr}, \email{tstember@phy.hr}}

\abstract{In this paper we continue and improve the analysis of the effective actions obtained by integrating out a scalar and a fermion field coupled to external symmetric sources, started in the previous paper. The first subject we study is the geometrization of the results obtained there, that is we express them in terms of covariant Jacobi tensors. The second subject concerns the treatment of tadpoles and seagull terms in order to implement off-shell covariance in the initial model. The last and by far largest part of the paper is a repository of results concerning all two point correlators (including mixed ones) of symmetric currents of any spin  up to 5 and in any dimensions between 3 and 6. In the massless case we also provide formulas for any spin in any dimension.}


\begin{document}

\section{Introduction}
\label{sec:intro}

This paper is a follow-up of \cite{BCDGLS}. In that paper we analyzed the 
two-point functions of conserved currents of two models (a free scalar and a free Dirac fermion model
coupled to diverse backgrounds) in various dimensions. For a background, represented by a 
completely symmetric field, the two-point function of the current minimally coupled to it is the basic 
ingredient of its (quadratic) effective action (EA). We found in  \cite{BCDGLS} that the effective
action for any background field obtained in this way  is based on the corresponding linearized Fronsdal kinetic operator, \cite{Fronsdal}, in the nonlocal form introduced by Francia and Sagnotti, \cite{FS}. In view of constructing a covariant action for a completely symmetric tensor field, this
result is promising. It suggests that integrating out scalar or fermion fields (or any other field by
which one can form conserved currents) may be a useful way to analyze the dynamics of higher spin fields. But of course
what we have done in \cite{BCDGLS} is only the beginning. The crucial next step is the calculation of the three-point functions of conserved currents, the analysis of the (lowest order) interaction terms in
the effective actions and their consistency with covariance. 
Before arriving at the three-point functions, it is however necessary to
improve our analysis of the quadratic EA. In fact in the course of our research 
we realized that it inevitably branches out in different directions. At the same time, in \cite{BCDGLS},
several aspects and questions were left behind . In this and a subsequent paper
we would like to cover as thoroughly as possible any aspect of the quadratic EA's.

The first issue is the geometrization (at the linear level) of our results in \cite{BCDGLS}. They were
expressed there mostly in terms of a projection operator, which is very convenient in that context because it
automatically ensures conservation. But, in this way, the 
geometrical content of the resulting equations of motion or the EA remains implicit.
Now the formulation of our results in terms of geometrical objects is essential, 
if our target is to arrive at covariant EA's. One first aim of this paper will be to geometrize the results of \cite{BCDGLS}. We will do it in terms of Jacobi tensors.

A second related important point is related to local subtractions. In \cite{BCDGLS} we found several violations  of the Ward Identities induced by the conservation of the initial current (which induces a gauge invariance of the
relevant minimal coupling). Such violations consist of local terms, so that it is rather elementary to recover conservation by
subtracting local counterterms from the EA. There is nothing special in this, it is a very ordinary procedure. The
interesting point is that it is in general not necessary to do it, because the perturbative field theory formalism already automatically takes care of covariance, provided one takes into account not only the two-point bubble diagrams but also other diagrams such as tadpole and seagull ones. Now, from a practical point of view it is much easier to subtract easily identifiable local counterterms, than calculating additional diagrams to guarantee conservation. 
The latter could appear as an academic exercise for spin 1 and 2, where we already know the covariant form of the minimal coupling. But, it is
important to show that dimensional regularization,
which we use, is giving manifestly covariant expressions (without
subtractions by hand). For spin 3 and higher it may be a very useful and even necessary calculation. The reason is that
seagull diagrams are related to terms in the initial action that do not belong to the minimal model we start with (a scalar or fermion field minimally coupled to a background field). Conservation (without subtractions) requires the presence of such additional terms and constraints not only their form but also their coefficients. It is clear, that when
we consider higher spin backgrounds, this remark may be used in order to determine additional action terms, as well as conditions for their coefficients. This goes in the direction of constructing an initial off shell covariant model, an important target in itself and a necessary step in the construction of a covariant EA.

The third important issue is represented by mixed two-point correlators. In \cite{BCDGLS} we have considered only two-point functions of each current with itself. Of course this provides basic information about the relevant EA. However
higher spin theories are known to be consistent only if they encompass an 
infinite number of fields (although in 3d consistent theories may exist with a finite number of fields). It is obvious that this requires not only the knowledge of the two-point correlator of each higher spin current with itself, but also of
any two currents (mixed correlators) coupled to fields that may enter the action. This part has the structure of a repository of results about the two point correlators of symmetric currents of spin up to 5 in dimension $3\leq d\leq 6$ for both the massive scalar and fermion theory.
In 3d we also
consider the odd parity sector which emerges from the parity-breaking fermion mass term, and we find a nice generalization of 
Pope and Townsend's Chern-Simons-like action in the case when different higher-spin fields are taken into consideration.
  
In this paper we will deal with these three issues. Other topics, such as  the discussion of the ambiguities inherent in the choice of 
the conserved currents in the initial matter model, will be included in a subsequent article. This is a good point to mention
that our research is indebted to several preexisting works, in particular with \cite{Sakharov, Maldacena} as far as the inspiration is concerned, with \cite{Babu,Dunne,Maldacena-Zhiboedov,bekaert,Gama,Closset,giombi,GMPTWY} as far as the methods are concerned and with \cite{FS,campoleoni,solvay2004,sorokin,Francia1,Francia2,Francia2012} for HS theories. Other papers of ours, related to the present one are, beside \cite{BCDGLS}, \cite{BCLPS,BCDGLS1}. 

The paper is organized as follows. In the next section we show how to geometrize the results of \cite{BCDGLS}
and of this paper, that is how to express them in terms of Jacobi tensors. In section 3 we discuss the issue of 
tadpole and seagull terms and how they guarantee covariance without subtractions in the case of spin 1 and 2.
Section 4 forms the bulk of the paper. After an explanatory introduction we list all possible conserved two-point correlators
for currents up to spin 5, including the mixed ones. This part of the paper is intended as a source book. It contains 
the complete correlators as well as their UV and IR expansions. Several results were already contained in 
\cite{BCDGLS}. We have left them here for completeness. Finally, section 5 is devoted to some conclusions.

\section{Geometry in effective actions}
\label{s:geometry}

The construction of interacting quantum field theories with massless higher spin
($s>2$) fields still poses an interesting theoretical problem. On the one hand,
there are different "no-go" theorems putting serious constraints on such
theories, especially in flat space-time. On the other hand, we have significant higher spin results: free fields can be
constructed in the same manner as in lower spin cases (see, e.g.
\cite{Weinberg1}); a few cubic interaction terms have been constructed in the literature (see \cite{solvay2004,sorokin});
 most notably, a full consistent covariant HS theory in AdS background has been constructed by
Vasilev and collaborators \cite{Vasiliev}. In our previous paper, \cite{BCDGLS}, we remarked that free 
lower spin field theories possess conserved higher spin currents which simply "beg" to be coupled to higher spin fields. 
Therefore such simple models seem to be a useful tool to study higher spin theories. A basic ingredient of the approach in \cite{BCDGLS} is the connection between the on shell conservation of the initial free field theory current and the gauge invariance of the minimal coupling term with the higher spin field, which induces a gauge invariance of the linearized higher spin EA (or covariance of the corresponding equation of motion).  In \cite{BCDGLS} this invariance was left somewhat implicit. There is, however, a way to make it explicit, by expressing the results in terms of covariant `geometric' tensors constructed out of the symmetric higher spin fields. In this section we would like to make connection with such a geometrization program.

In the sequel we first introduce well-known definitions and properties about higher 
spin tensors, their linearized eom's and their possible geometrical formulations. Then we show how to use this  material to 
express the results obtained in \cite{BCDGLS,BCLPS} and in this paper in a geometric language. 

Differences between lower spin ($s\le2$) and higher spin theories emerge already
at the level of classical free field theories. The simplest way to construct a
free theory of higher spin field is provided by the Fronsdal equation, \cite{Fronsdal,Curtright,Singh}:
\be \label{froeq}
\mathcal{F} \equiv \Box \varphi - \partial\, \partial \cdot \varphi + \partial^2
\varphi' = 0
\ee
where the spin-$s$ field is described by the completely symmetric rank-$s$ tensor
field $\varphi \equiv \varphi_{\mu_1\cdots\mu_s}$. In this expression standard HS
conventions from \cite{FS,Francia1,Francia2}  are
assumed.\footnote{Conventions assume symmetrization over free indices with
minimal number of terms and without any symmetry factors. Also, a prime denotes
contraction of a pair of indices, so, e.g., $\varphi' \equiv
\varphi_{\mu_1\cdots\mu_{s-2}} = \eta^{\mu_{s-1}\mu_s}
\varphi_{\mu_1\cdots\mu_s}$ is a completely symmetric rank-$(s-2)$ tensor
field.} The Fronsdal equation (\ref{froeq}) is invariant under local
transformations parametrised by traceless completely symmetric rank-$(s-1)$
tensor fields $\Lambda \equiv \Lambda_{\mu_1\cdots\mu_{s-1}}$
\be \label{gdiff}
\delta \varphi = \partial \Lambda
\ee
with $\Lambda' = 0$. While this gauge symmetry guarantees that the field
propagates only free spin-$s$ excitations, we see that for $s\ge3$ the gauge
symmetry is constrained to trace-free parameters $\Lambda$. One can rewrite the Fronsdal equation in an
unconstrained form by introducing a rank-$(s-3)$ compensator field $\alpha$
transforming on (unconstrained) gauge transformations (\ref{gdiff}) as $\delta
\alpha = \Lambda'$, in the following way
\be \label{froequn}
\mathcal{F} = \partial^3 \alpha
\ee
This equation is invariant under the unconstrained gauge transformations
(\ref{gdiff}) because the variation of $\alpha$ exactly cancels the variation of the
Fronsdal tensor. Most important, the system $\varphi,\alpha$ can be cast in a (local) Lagrangian form. 
By the partial gauge fixing condition $\alpha=0$ one obtains the
original Fronsdal's equation (\ref{froeq}).

The generalization $\EF^{(n)}$ of the 
Fronsdal differential operator, which is gauge invariant for $n$ large enough, is given in terms of the
recursive equation
\be
\EF^{(n+1)}=\EF^{(n)}+\frac 1{(n+1)(2n+1)} \frac {\partial^2}{\square}
{\EF^{(n)}}' -\frac 1{n+1} \frac {\partial}{\square} \partial\cdot
\EF^{(n)}\label{Fronsdalrecur}
\ee
with $\EF^{(0)}= \square\varphi$. So, in particular,
\be
\EF^{(1)} \equiv \EF= \square \varphi -\partial\partial \cdot
\varphi+\partial^2\varphi'\label{Fronsdal}
\ee
is the original Fronsdal operator. However, the connection with our results cannot be 
in terms of the tensor $\EF^{(n)}$, because the latter does not satisfy a conservation law, 
while our results are conserved two-point functions.  To make the connection one constructs the
Einstein-like tensor
\be
\EG^{(n)} = \sum_{p=0}^n (-1)^p \frac {(n-p)!} {2^p n!} \,\eta^p
\,\EF^{(n)[p]}\label{Einstein}
\ee
where the superscript in square bracket denotes the number of time $\EF^{(n)}$
has been traced, and $\eta$ is the Minkowski metric. The association of $\varphi$ with the spin $s$ is as follows:
\be
\left\{ \begin{matrix} s= 2n &\quad s\quad {\rm even}\\
s= 2n-1 &\quad s \quad{\rm odd}\end{matrix}\right.\0
\ee
The $\EG^{(n)}$ tensor is divergenceless
\be
\partial\cdot\EG^{(n)}=0 \label{divergence0}
\ee
The free (unconstrained) linearized equations of motion for $\varphi$ are
\be
\EG^{(n)}=0\label{varphieom}
\ee
Once again, it can be shown that such an equation can be cast 
in local Lagrangian form, provided one introduces auxiliary fields (compensators).
$\EG^{(n)}$ are the objects that can be directly connected with the
LHS of \eqref{Ephi} below.

\subsection{Geometrization in terms of Jacobi tensors}

In \cite{BCDGLS} all the two-point correlators and corresponding effective
actions are presented in momentum space and expressed in terms of the projector
\be
\pi^{(k)}_{\mu\nu}= \eta_{\mu\nu} - \frac {k_\mu k_\nu}{k^2}\label{proj}
\ee
Applied to $k^\nu$ gives 0, so any two-point function expressed in terms of it
alone is conserved.
We showed that any conserved correlator for spin $s$ can be written in terms of
the following structures:
\be
\tilde A^{(s)}_0(k,n_1,n_2)&=&  (n_1 \!\cdot\!
\pi^{(k)}\!\cdot
\!n_2)^s\label{A0}\\
\tilde A^{(s)}_1(k,n_1,n_2)&=& (n_1 \!\cdot\!
\pi^{(k)}\!\cdot
\!n_2)^{s-2} (n_1 \!\cdot\! \pi^{(k)}\!\cdot \!n_1)(n_2
\!\cdot\! \pi^{(k)}\!\cdot \!n_2)\label{A1}\\
\ldots && \ldots\ldots\0\\
\tilde A^{(s)}_l(k,n_1,n_2) &=&  (n_1 \!\cdot\!
\pi^{(k)}\!\cdot
\!n_2)^{s-2l} (n_1 \!\cdot\! \pi^{(k)}
\!\cdot \!n_1)^l(n_2 \!\cdot\! \pi^{(k)}\!\cdot \!n_2)^l\label{Al}\\
\ldots && \ldots\ldots\0
\ee
where $n_1,n_2$ are generic polarization vectors, and 
$n_1\!\cdot\! \pi^{(k)}\!\cdot \!n_2= n_1^\mu \pi^{(k)}_{\mu\nu}n_2^\nu$. 
There are $\lfloor s/2 \rfloor$ independent such terms.
The generic term in the final formulas are combinations of $\tilde
A^{(s)}_l(k,n_1,n_2)$ with numerical coefficients $a_l$, say
\be 
\tilde E^{(s)}(k,n_1,n_2)= \sum_{l=0}^{\lfloor s/2 \rfloor} a_l \tilde
A^{(s)}_l(k,n_1,n_2)\label{tildeE}
\ee
preceded by a function of $f(|k|,m)$ and the mass $m$\footnote{This function 
can be expanded in series of $m/|k|$ or $|k|/m$ near the IR and UV, respectively, 
which gives the tomographic expansions considered in  \cite{BCDGLS}. The latter 
clearly show that the structures of the two-point functions (and corresponding 
linearized EA's) are determined by the unique Fronsdal operator appropriate for 
the given source, although, generally, the operator appears in a nonlocal form 
and in different gauges. In this paper we consider only these operators and 
disregard the function $f$.}.   Eq.(\ref{tildeE}) can be
easily translated into a corresponding differential operator by Fourier
anti-transforming
\be 
 E^{(s)}(\partial,n_1,n_2)= \sum_{l=0}^{\lfloor s/2 \rfloor} a_l  
A^{(s)}_l(\partial,n_1,n_2)\label{E}
\ee
These are the types of differential operators that appear in the EA's acting on
the spin $s$ field $\varphi_{\mu_1\ldots\mu_s}$. The corresponding eom will take
the following form.
Set $(m^s \cdot \varphi)=\frac 1{s!} m_{\mu_1\ldots\mu_s}\varphi^{\mu_1\ldots
\mu_s}$ and
$n_1=n\equiv \{ n_\mu\}$, $n_2= {\partial}_m \equiv \left\{\frac
{\partial}{\partial m_\mu}\right\}$. The
eom are 
\be
\frac 1{s!} (\partial_n)^s E^{(s)}(\partial,n,\partial_m)(m^s \cdot
\varphi)=0\label{Ephi}
\ee
multiplied by a function of $k^2$ and $m$. 

The purpose of this section is to rewrite the equations such as (\ref{Ephi}) in
the geometrical form of \cite{FS}. 

To this end let us introduce the {\it symbol} of $\EG^{(n)}$, $\tilde{\cal
G}^{(n)}(k,n_1,n_2)$, as follows. First we saturate all its $s$
naked indices of $\EG^{(n)}$ with $n_1$ polarizations, then we Fourier transform it and replace
the Fourier transform of $\varphi$, $\tilde \varphi$, with a symmetric tensor made out of the product of $s$ polarizations $n_2$. Finally we define
\be 
{\cal G}^{(n)}\equiv\frac 1{s!} \,(\partial_n)^s {\cal
G}^{(n)}(\partial,n,\partial_m)\, (m^s \cdot \varphi)\label{symbol}
\ee
Then the connection between \eqref{varphieom} and \eqref{Ephi} is given by
\be
\frac 1{k^2}\, \tilde{\cal G}^{(n)}(k,n_1,n_2)= \sum_{l=0}^{\lfloor s/2 \rfloor}
(-1)^l \left( \begin{matrix} \lfloor s/2 \rfloor\\ l\end{matrix} \right)
\tilde A_l^{(s)}(k, n_1,n_2), \label{GnEn}
\ee
which corresponds to a particular choice of the coefficients $a_l$ in \eqref{tildeE}.

Of course we are interested not only in the relation \eqref{GnEn}, but in
expressing all the $\tilde A_l^{(s)}(k, n_1,n_2)$ in terms 
of the $\tilde{\cal G}^{(n)}(k,n_1,n_2)$. To do so we have to take the
successive traces of \eqref{GnEn}. We have, for instance
\be
{\tilde{\cal G}}^{(n)}{}' = -2\lfloor s/2 \rfloor (2\lfloor s/2 \rfloor+D-4)  {\tilde{\cal
G}^{(n-1)}}\,(n_2 \!\cdot\! \pi^{(k)}\!\cdot \!n_2)\label{1sttrace}
\ee 
In general
\be 
{\tilde{\cal G}^{(n)[p]}} = (-2)^p \frac {(2\lfloor s/2 \rfloor+D-4)!!(\lfloor s/2 \rfloor)!}{(2\lfloor s/2 \rfloor +D-2p-4)!!(\lfloor s/2
\rfloor-p)!}
{\tilde{\cal G}^{(n-p)}}\,(n_2 \!\cdot\! \pi^{(k)}\!\cdot
\!n_2)^p\label{pthtrace}
\ee 
and
\be 
{\tilde{\cal G}^{(n)[n]}} = (-2)^n \frac {(2\lfloor s/2 \rfloor+D-4)!!(\lfloor s/2 \rfloor)!}{(D-4)!!}
{\tilde{\cal G}^{(0)}}\,(n_2 \!\cdot\! \pi^{(k)}\!\cdot \!n_2)^n\label{nthtrace}
\ee 
for $s$ even, with ${\tilde{\cal G}^{(0)}}={k^2}$, and
\be 
{\tilde{\cal G}^{(n)[n-1]}} = (-2)^{n-1} \frac {(2\lfloor s/2 \rfloor+D-4)!!(\lfloor s/2 \rfloor)!}{(D-4)!!}
{\tilde{\cal G}^{(1)}}\,(n_2 \!\cdot\! \pi^{(k)}\!\cdot
\!n_2)^{n-1}\label{n-1thtrace}
\ee 
for $s$ odd, with  ${\tilde{\cal G}^{(1)}}={k^2}(n_1 \!\cdot\!
\pi^{(k)}\!\cdot \!n_2) $.

Now, using \eqref{GnEn}, one can write
\be
(n_1 \!\cdot\! \pi^{(k)}\!\cdot \!n_2)^s &\equiv& \tilde A_0^{(s)}(k,
n_1,n_2)=\frac 1{k^2}
\tilde{\cal G}^{(n)}(k, n_1,n_2) \label{A0EG}\\
&+& \sum_{l=0}^{\lfloor s/2 \rfloor-1} (-1)^l \left( \begin{matrix} \lfloor s/2 \rfloor \\ l+1\end{matrix} \right)
(n_1 \!\cdot\! \pi^{(k)}\!\cdot \!n_1)^{l+1}
(n_1 \!\cdot\! \pi^{(k)}\!\cdot \!n_2)^{s-2l-2} (n_2 \!\cdot\! \pi^{(k)}\!\cdot
\!n_2)^{l+1} \0
\ee
for even $s$, and a similar expression for odd $s$. Now the strategy consists in
repeating the same step for the 
second line in \eqref{A0EG}, by using \eqref{1sttrace} and successively 
\eqref{nthtrace}. The end result is
\be
{k^2} (n_1 \!\cdot\! \pi^{(k)}\!\cdot \!n_2)^s = \sum_{p=0}^{\lfloor s/2 \rfloor}\left( -\frac 12 \right)^p 
\frac
{(2\lfloor s/2 \rfloor+D-2p-4)!!}{p!(2\lfloor s/2 \rfloor+D-4)!!} 
 (n_1 \!\cdot\! \pi^{(k)}\!\cdot \!n_1)^p\, \tilde {\cal G}^{(n)[p]}(k,
n_1,n_2)\0\\ \label{n1n2s}
\ee
In a similar way one can obtain
\be 
&&{k^2} (n_1 \!\cdot\!\pi^{(k)}\!\cdot\!n_2)^{s-2l} (n_1 \!\cdot\!
\pi^{(k)}\!\cdot \!n_1)^l(n_2 \!\cdot\! \pi^{(k)}\!\cdot \!n_2)^l\label{n1n2l}\\
&&=\frac{1}{\left(\begin{matrix} \lfloor s/2 \rfloor\\ l \end{matrix} \right)} \sum_{p=l}^{\lfloor s/2 \rfloor}\left( -\frac 12 \right)^p 
\left( \begin{matrix} p\\ l \end{matrix} \right)\frac
{(2\lfloor s/2 \rfloor+D-2p-4)!!}{p!(2\lfloor s/2 \rfloor+D-4)!!}
 (n_1 \!\cdot\! \pi^{(k)}\!\cdot \!n_1)^p\, \tilde {\cal G}^{(n)[p]}(k,
n_1,n_2)\0
\ee
In conclusion any expression of the type \eqref{tildeE}, i.e. any conserved
structure, can be expressed in terms of the generalized Einstein 
symbols $\tilde {\cal G}^{(n)}(k, n_1,n_2)$ and its traces. Thus any EA (or any
eom) we obtain from our models, by integrating out matter, 
can be expressed in terms of the generalized Einstein tensor $\EG^{(n)}$ and its
traces preceded by a function of $\square$ 
and the mass $m$ of the model, with suitable multiples of the operator
\be
\eta^{\mu\nu}- \frac {\partial^\mu \partial^\nu}{\square} \0
\ee
acting on the traces. Using \eqref{Einstein} one can replace the dependence on
${\EG}^{(n)}$ of such expressions with the dependence on
${\EF}^{(n)}$. The geometrization program can be completed by introducing  the
Jacobi tensors $\ER_{\mu_1,\ldots\mu_s\nu_1\ldots \nu_s}$ (one of the possible generalizations of the 4d Riemann tensor, \cite{deWitFreedman,Damour}) by means of
\be
\frac 1{(s!)^2} (m^s\!\cdot\! \ER^{(s)} \!\cdot\!n^s) = \sum_{l=0}^s \frac
{(-1)^l} {s!(s-l)!l!} (m\!\cdot\!\partial)^{s-l} (n\!\cdot\!\partial)^{l}
(m^l\!\cdot\! \varphi \!\cdot\!n^{s-l})\label{ER}
\ee
The tensors $\ER^{(s)}$ are connected to the $\EF^{(n)}$ as follows:
\be
\EF^{(n)}= \begin{cases} \frac 1{\square^{n-1}} \ER^{(s)[n]} &
\quad s=2n\\
                            \frac 1{\square^{n-1}}\partial\!\cdot\! \ER^{(s)[n-1]} & \quad s=2n-1
                           \end{cases} \label{EFER}
\ee
where the traces in square brackets refer to the first set of indices. 
In this way we can express any EA or any eom in terms of $\ER^{(s)}$ and traces
(in the second set of indices) thereof.

Since above we have referred to \cite{FS}, we feel that, to end this section, it
is opportune for us to clarify the
context in which our results are derived and point out the differences with the
spirit of \cite{FS,Francia1,Francia2}.
In these papers the initial purpose was to write down a generalization of the
Fronsdal equations for higher spin in 
such a way as to avoid the constraints needed in the original formulation of
\cite{Fronsdal}. The authors of \cite{FS}
chose to sacrifice locality in favor of an unconstrained gauge symmetry. The
typical (linearized) non-local 
equation of motion one obtains in this way is \eqref{varphieom}. As we have already pointed out, it can be shown
that such an equation can be cast 
in Lagrangian form, provided one introduces auxiliary fields (compensators).
Therefore one can say that the nonlocality
of \eqref{varphieom} is a gauge artifact, with no physical implication. However equations of motion invariant under unrestricted gauge symmetry
are far from unique. There actually exist several families of them
depending on arbitrary parameters 
(by the way, this is evident by reversing the argument above and starting from
the generic operator \eqref{E}, instead of
the completely fixed one \eqref{GnEn}). These are all equally valid as long as
the field $\varphi$ is considered in isolation 
and the linearized eom is the free one, \eqref{varphieom}. However, if the spin
$s$ system is minimally coupled to a conserved current
the question arises as to whether the propagating degrees of freedom are the
truly physical ones, i.e. those corresponding to the
appropriate little group representation for massless fields. The authors of
\cite{Francia1,Francia2} were able to prove that 
there exist only one choice for the Einstein-like tensor which is Lagrangian and
satisfies such a physicality condition.

Such `physical' Einstein tensors do not correspond, in general, to the kinetic
operators we find in our effective action in section \ref{s:alltypes} below. This is not
surprising, as our main goal is covariance: 
our purpose is to arrive at a covariant EA with respect to a completely unfolded
gauge symmetry. In a logical development
the next step will be to introduce auxiliary fields to eliminate nonlocalities.
Following this we would need to gauge-fix the action and introduce appropriate ghosts to produce the physical propagators. At that point
would the problem handled by \cite{Francia1,Francia2} come to the surface. However, we would like to recall that
our immediate prospect is to construct the linearized covariant EA in preparation for the analysis of the three-point function.

\section{Tadpoles, seagulls and conservation}
\label{s:seagull}

In this section we wish to illustrate the role of tadpole and seagull diagrams
in implementing conservation in two-point correlators.
In \cite{BCDGLS}, in order to evaluate the two point correlators of conserved
currents we computed only the bubble diagrams formed by 
two internal scalar or fermion lines and two vertices. In this way we found
several violations  of the relevant Ward identities.
Such violations consist of local terms, so that it was rather elementary to
recover conservation by
subtracting local counterterms from the EA. However it is in general not
necessary to do this,
because the perturbative field theory formalism already automatically takes care
of it provided one takes into account not only the 
two-point bubble diagrams but also other diagrams such as tadpole and seagull
ones, \cite{Jackiw,Parker}. 
Although this is a rather well-known fact, we would like to show it in detail
here for spin 1 and 2 as a guide for the more challenging
higher spin cases. The reason is that seagull diagrams reflect the presence  in the initial action of
additional terms, additional with respect to the minimal couplings (symbolically $\int j\varphi$), which are on-shell covariant, 
but off-shell non-covariant.  One of the crucial steps in our program is clearly
implementing off-shell gauge covariance of the initial models, that is adding to the minimal couplings in the relevant actions the terms that render them off-shell covariant, at least to the lowest order in a perturbative approach to the gauge symmetry. We know such additional terms exactly in the case of spin 1 and spin 2 (because we know the full covariant action), but not yet for higher spins. In the latter cases, however, we can implement off-shell current conservation by satisfying the corresponding two-point function Ward identity. In turn this requires considering tadpoles and seagull terms. The latter, in particular, originate from the above additional terms, which in this way may hopefully be identified\footnote{An approach related to ours is outlined in \cite{bekaert}. It is based on Weyl quantization. Its main advantage is that it provides a full quantum action and quantum symmetry for the initial scalar model. It will be interesting to compare the two approaches.}. 

Hereafter in this section we work out the cases of spin 1 and spin 2 in any dimension (in 3d for the odd parity part) in detail, showing the role of tadpole and seagull terms in the Ward identities for two-point functions of spin 1 and 2 respectively, and their origin in the various terms of the initial actions. We keep the derivation at a pedagogical level and, for completeness, we analyze the full structure of the relevant two-point functions and, in particular, their IR and UV expansions, as well as their contributions to the EA's.

Starting from the generating function 
\be
Z[a]=e^{i W[a]}=\int D\psi D \bar{\psi} e^{i (S_0 + S_{int}[a])}
\ee
where $a$ is the external higher spin field, we will compute the effective
action for the external source fields up to the quadratic order:
\be 
i\,W[a]&=&i\,W[0] +\int d^dx\,
a_{\mu_{1}\ldots \mu_{s}}(x)\Theta^{\mu_{1} \ldots \mu_{s}}(x)\0\\
&&+\frac {1}{2!} \int d^dx d^dy\,
a_{\mu_{1}\ldots \mu_{s}}(x) a_{\nu_{1}\ldots \nu_{s}}(y)T^{\mu_{1} \ldots
\mu_{s}\nu_{1}\ldots\nu_{s}}(x,y)+\ldots  \label{Was}
\ee
 where
 \be 
\Theta^{\mu_{1} \ldots \mu_{s}}(x)=\frac{\delta \left(i\,W[a]\right)}{\delta
a_{\mu_{1}\ldots \mu_{s}}(x)}\Big{|}_{a=0}
\ee
is a tadpole (1-point function) and 
\be 
T^{\mu_{1} \ldots \mu_{s} \nu_{1}\ldots\nu_{s}}(x, y)=\frac{\delta^2
\left(i\,W[a]\right)}{\delta a_{\mu_{1}\ldots \mu_{s}}(x) \delta
a_{\nu_{1}\ldots \nu_{s}}(y)}\Big{|}_{a=0}
\ee
is a 2-point function. Using Feynman diagrams we wish to compute the 2-point function 
including not only the bubble diagram (as in \cite{BCDGLS}) but also tadpoles and seagulls.  

The one-loop 1-pt correlator for the external field is (up to the linear order):
 \be 
\langle\!\langle J^{\mu_{1} \ldots  \mu_{s}}(x)\rangle\!\rangle &=& \frac{\delta
W[a] }{\delta a_{\mu_{1}\ldots \mu_{s}}(x)}\0\\
&=&-i\left(\Theta^{\mu_{1} \ldots \mu_{s}}(x) + \int d^dy \, a_{\nu_{1}\ldots
\nu_{s}}(y)T^{\mu_{1} \ldots \mu_{s}\nu_{1}\ldots\nu_{s}}(x,y)+\ldots
\right)\label{EOM}
\ee
External spin $s$ fields $a_{\mu_1\ldots \mu_s}$ are in particular 
\be  
\text{Spin 1}\quad a_\mu & = & A_\mu \quad\text{ gauge field}\\
\text{Spin 2}\quad a_{\mu\nu}& = & h_{\mu\nu} \quad\text{ graviton field} \0
\ee

We will need one-loop conservation which for spin 1 reads
\be 
\partial_\mu\langle\!\langle J^{\mu}(x) \rangle \! \rangle=0 \label{conss1} 
\ee
Ward identity for the two-point function in momentum space can be written as
\be
k_\mu  \tilde T^{\mu\nu}(k)=0
\ee

Furthermore, for spin 2, the energy-momentum tensor is defined with
$\langle\!\langle T^{\mu\nu}(x) \rangle \! \rangle =\frac 2{\sqrt{g}}\frac
{\delta W}{\delta h_{\mu\nu}(x)}$. The full conservation law of the
energy-momentum tensor is 
\be
\nabla_\mu \langle\!\langle T^{\mu\mu}(x) \rangle \! \rangle=0 \label{conss2}
\ee
Hence, the Ward identity for one-point function is
\be
\partial_\mu \Theta^{\mu\mu}(x) =0
\ee 
while for two-point correlator we have
\be
 \partial_{\mu} T^{\mu\mu\nu\nu}(x,y)&=&
\frac 12 \eta^{\nu\nu}\delta (x-y)\partial_\mu \Theta^{\mu\mu}(x)+\frac 12
\Theta^{\nu\nu}(x)\partial^\mu \delta (x-y)\0\\
&&- \partial_\mu\left(\delta
\left(x-y\right)\Theta^{\mu\nu}\left(x\right)\right)\eta^{\mu\nu}
\ee
As we will see, the tadpole contribution is $\tilde \Theta^{\mu\mu}(k)=\tilde
\Theta\, \eta^{\mu\mu}$ where $\tilde \Theta$ is a constant. The Ward identity
in momentum space is now
\be
 k_\mu \tilde T^{\mu\mu\nu\nu} (k) =\left[ -   k^{\nu} \eta^{\mu\nu}
+\frac 12 k^\mu \eta^{\nu\nu}\right]\tilde \Theta\label{WI}
\ee

\subsection{Fermions - spin 1}

The action for the  theory of fermions interacting with gauge field can be
written as
\be
S=\int\text{d}x \left[\bar{\psi}\left(i\gamma^\mu D_\mu-m\right)\psi\right] 
\ee
where $D_\mu=\partial_\mu-i\,A_\mu$. There is one fermion-fermion-photon vertex
 \be
V^\mu_{ffp}&:& i\gamma^\mu	\label{Vs1ffA}
\ee

In the case of fermions coupled to gauge field the tadpole diagram vanishes,
while the seagull is zero because the theory is linear in the gauge field. The
only contribution we get from the 2-pt correlator ((11.7) from \cite{BCDGLS})
which in the momentum space  reads
\be
\tilde T^{\mu\nu}(k) 
&=&\frac{2^{-d+\lfloor\frac{d}{2}\rfloor} \, i\, \pi ^{-\frac{d}{2}} m^{d-2}}{
4m^2-k^2}\Gamma \left(1-\frac{d}{2}\right)\0\\
&&\times\left(-4m^2 +\, _2F_1\left[1,1-\frac{d}{2};\frac{3}{2};\frac{k^2}{4
m^2}\right](4m^2+(d-2)k^2)\right)\pi^{\mu\nu}\label{2pts1}
\ee
where $\pi^{\mu\nu}=\eta^{\mu\nu}-\frac{k^\mu k^\nu}{k^2}$ is the projector.
Since the 2-point correlator can be expressed in terms of the projector, it
satisfies Ward identity (\ref{conss1}) We can expand the two-point correlator in
the IR region
\be 
\tilde T^{\mu\nu}(k) &=&-2^{1-d+\lfloor \frac d2\rfloor} i\, m^{d-2} \pi^{-\frac
d2}\sum_{n=1}^\infty \frac{ n\, m^{-2n} \Gamma\left(2n-\frac d2\right)}{2^{n}
(2n+1)!!}k^{2n}\pi^{\mu\nu}\label{2pts1IR}
\ee
Using the Fourier transform of  (\ref{2pts1IR}) in the one-loop $1$-point  function
(\ref{EOM}) we get
\be 
\langle\!\langle J^\mu(x)\rangle\!\rangle=2^{1-d+\lfloor \frac d2\rfloor} 
m^{d-2} \pi^{-\frac d2}\sum_{n=1}^\infty \frac{(-1)^n n\, m^{-2n}
\Gamma\left(2n-\frac d2\right)}{2^n (2n+1)!!}\Box^{n-1}\partial_\nu F^{\mu\nu}
\ee
The one-loop 1-point correlator satisfies (\ref{conss1}) Using the same
expansion in the IR (\ref{2pts1IR}) for the  effective action (\ref{Was}) we
obtain
\be 
W&=&2^{-1-d+\lfloor \frac d2\rfloor}  m^{d-2} \pi^{-\frac d2}\sum_{n=1}^\infty
\frac{(-1)^n n\, m^{-2n} \Gamma\left(2n-\frac d2\right)}{2^n (2n+1)!!}\int
\text{d}^dxF_{\mu\nu}\Box^{n-1}F^{\mu\nu}\0\\
&\stackrel{\text{IR}}{=}&-\frac{2^{-2-d+\lfloor \frac d2\rfloor}}{3}\,  m^{d-4}
 \pi^{-\frac d2}\Gamma\left(2-\frac d2\right)\int \text{d}^dx 
F_{\mu\nu}F^{\mu\nu}
\ee
So, in the IR region (large m) we get the Maxwell action.

Furthermore, the dominating term in the UV $\left(O(m^0)\right)$ of
(\ref{2pts1}) corresponds to the massless case (B.2) from \cite{BCDGLS}
\be
\tilde T^{\mu\nu}(k) 
\stackrel{UV}{=}-\frac{2^{2-2d+\lfloor \frac d2\rfloor} \, \pi^{\frac
32-\frac d2}(d-2)}{\left(-1+e^{i\pi d}\right)\Gamma\left(\frac
{d+1}2\right)}(k^2)^{\frac d2-1}\pi^{\mu\nu}
\ee
The effective action in the UV is then
\be
W\stackrel{UV}{=} \frac{(-1)^{\frac d2}2^{1-2d+\lfloor \frac d2\rfloor} \,
\pi^{\frac 32-\frac d2}(d-2)}{\left(-1+e^{i\pi d}\right)\Gamma\left(\frac
{d+1}2\right)}F^{\mu\nu}\Box^{\frac d2-2}F_{\mu\nu}
\ee

\subsubsection{Odd parity part}
\label{odd-spin1}
For the analysis of the odd parity correlators 
we will restrict ourselves to $d=3$. The odd part of the two-point correlator is
non-vanishing only in $3d$ and it is given by
\be
\tilde T^{\mu\nu}_{o}(k)=\frac {m}{2\pi k}
\text{ArcCoth}\left(\frac{2m}{k}\right) \epsilon^{\mu\nu\lambda}k_\lambda
\label{2pts1co}
\ee
The expansion of (\ref{2pts1co}) in the IR reads
\be
\tilde T^{\mu\nu}_{o}(k)
=\frac{1}{\pi }\sum_{n=0}^\infty\frac{k^{2n}m^{-2n}}{2^{2(n+1)}(2n+1)}
\epsilon^{\mu\nu\lambda}k_\lambda
\ee
Using the IR expansion in (\ref{EOM}), the odd part of the one-loop 1-point correlator is now
\be
\langle\!\langle J^\mu(x)\rangle\!\rangle
=\frac{1}{\pi }\sum_{n=0}^\infty\frac{(-1)^n m^{-2n}}{2^{2n+3}(2n+1)}
\epsilon^{\mu\nu\lambda}\Box^nF_{\lambda\nu}
\ee
and just like the even parity part satisfies (\ref{conss1}). The effective
action in the IR (the dominating term)
 \be 
W&\stackrel{IR}{=}&\frac {1}{8\pi}\epsilon^{\mu\nu\lambda}\int
d^3x\,A_\mu\partial_\nu A_\lambda+\ldots
\ee
corresponds to Chern-Simons term in 3d
\be 
S_{CS}=\frac {1}{8\pi}\int d^3x\,Tr\left(A \wedge d A+\frac 23 A\wedge A\wedge
A\right)
\ee

\subsection{Scalars - spin 1}

The action in the scalar QED model is

\be
S=\int \text{d}^dx\left[D_\mu\varphi^\dagger
D^\mu\varphi-m^2\varphi^\dagger\varphi\right]
\ee
where $D_\mu=\partial_\mu-i \,A_\mu$. The full covariant action is
\be
S=\int \text{d}x\left[\partial_\mu\varphi^\dagger
\partial^\mu\varphi+i\,
A_\mu\left(\varphi^\dagger\partial^\mu\varphi-\partial^\mu\varphi^\dagger\varphi
\right)
+A_\mu A^\mu\varphi^\dagger\varphi-m^2\varphi^\dagger\varphi\right]
\ee
In the scalar model the scalar-scalar-photon vertex is
\be
V^\mu_{ssp}(p,p')&:& -i(p+p')^\mu
\ee
and we also have scalar-scalar-photon-photon vertex (coming from $\int d^d x
A^\mu A_\mu\varphi^\dagger\varphi$ term in Lagrangian)
\be
V^{\mu\nu}_{sspp}(p,p') &:& 2i\eta^{\mu\nu}\label{Vsspp}
\ee

The two-point function for the massive scalar in any dimension $d$ for spin
$s=1$ is
\be
\tilde T^{\mu\nu}(k)=- 2^{1-d}\, i\,  \pi ^{-d/2}
   m^{d-2} \Gamma \left(1-\frac{d}{2}\right)
   \left( \,\,
   {_2F_1}\left[1,1-\frac{d}{2};\frac{3}{2};\frac{k^2}{4
   m^2}\right]\pi^{\mu\nu}+\frac{k^\mu k^\nu}{k^2}\right)\label{JmnS}
\ee
which has a non-conserved part. However, since the theory is quadratic in the external photon field $A$ we also have a
seagull diagram (which is obtained by joining with a unique a fermion line the two fermion legs of the vertex \eqref{Vsspp}) for which we obtain
\be
\tilde T_{(s)}^{\mu\nu}(k) = 2^{1-d}i \pi^{-\frac d2}m^{d-2}\Gamma \left(1-\frac
d2\right)\eta^{\mu\nu} \label{JmunuS2}
\ee
After combining  (\ref{JmnS}) and (\ref{JmunuS2}) we can write down the
full 2-point function
\be 
\tilde T^{\mu\nu}(k) = 2^{1-d}i\pi^{-\frac d2}m^{d-2}\Gamma \left(1-\frac
d2\right)\left(1-{ _2F_1} \left[1,1-\frac d2;\frac 32;\frac{k^2}{4 m^2}\right]
\right)\pi^{\mu\nu},\label{JmnSC}
\ee 
which is conserved.

Expanding the two-point function (\ref{JmnSC})  in the IR gives
\be 
\tilde T^{\mu\nu}(k)& =&-2^{-d} i\,  m^{d-4} \pi^{-\frac d2}\sum_{n=0}^\infty
\frac{ \, m^{-2n} \Gamma\left(2+n-\frac d2\right)}{2^{n}
(2n+3)!!}k^{2n+2}\pi^{\mu\nu}\label{JmnScIR}
\ee 
Using the IR expansion together with (\ref{EOM}), the one-loop $1$-point function
(\ref{EOM}) now reads 
\be 
\langle\!\langle J^\mu\rangle\!\rangle=-2^{-d}  m^{d-4} \pi^{-\frac
d2}\sum_{n=0}^\infty \frac{(-1)^{n} \, m^{-2n} \Gamma\left(2+n-\frac
d2\right)}{2^n (2n+3)!!}\Box^{n}\partial_\nu F^{\mu\nu}
\ee
On the other hand, the dominating term of the effective action in the IR region
is
\be 
W\stackrel{\text{IR}}{=}-\frac{2^{-d}}{3} m^{d-4}  \pi^{-\frac
d2}\Gamma\left(2-\frac d2\right)\int \text{d}^dx  F_{\mu\nu}F^{\mu\nu}
\ee
In the IR (for large mass m) we get the Maxwell action.

The leading order term of the expansion in the UV (term $m^0$ corresponds to
(B.13) from \cite{BCDGLS})
\be 
\tilde T^{\mu\nu}(k) \stackrel{UV}{=} -\frac{2^{3-2d} \, \pi^{\frac 32-\frac
d2}(k^2)^{\frac d2-1}}{\left(-1+e^{i\pi d}\right)\Gamma\left(\frac
{d+1}2\right)}\pi^{\mu\nu}
\ee 
Hence, the effective action in the UV is
\be
W\stackrel{UV}{=} -i\frac{(-1)^{\frac d2}2^{3-2d} \, \pi^{\frac 32-\frac
d2}}{\left(-1+e^{i\pi d}\right)\Gamma\left(\frac {d+1}2\right)}\int d^dx
F^{\mu\nu}\Box^{\frac d2-2}F_{\mu\nu}\label{s1scalUV}
\ee

\subsection{Fermions - spin 2}

Let us consider the free fermion theory in a generic dimension $d$
\be
S&=& \int d^d x \, \sqrt{|g|} \left[\, i\overline {\psi} E_a^m
\gamma^a\left(\partial_m +\frac 12 \Omega_m \right)\psi
-m\bar{\psi}\psi\right]\label{fermionaction} 
\ee
where $E^m_a$ is the inverse vierbein.  From now on we will  set  $g_{\mu\nu}=
\eta_{\mu\nu}+ h_{\mu\nu}$. Using the following expansions
\be
g^{\mu\nu}= \eta^{\mu\nu}-h^{\mu\nu} +(h^2)^{\mu\nu}+\ldots, &\qquad &
\sqrt{|g|}=  1+\frac 12  h+\frac 18   h^2 -\frac 14   h^{\mu\nu}h_{\mu\nu}
+\ldots ,\0\\
e_a^\mu= \delta_a^\mu -\frac 12 h_a^\mu +\frac 38 (h^2)_a^\mu +\ldots, &\qquad &
e^a_\mu= \delta^a_\mu +\frac 12 h^a_\mu -\frac 18 (h^2)^a_\mu +\ldots
\ee
we can expand the parity even part of the action (\ref{fermionaction}) in powers
of $h$:
\be 
S_e&=& \int d^dx \, \Big{[} \frac{i}{2} \overline {\psi}\gamma^m
{\stackrel{\leftrightarrow}{\partial}}_m  \psi -m\bar\psi\psi+\frac 12 h_\mu^\mu
\left( \frac{i}{2} \overline {\psi}\gamma^m
{\stackrel{\leftrightarrow}{\partial}}_m  \psi -m\bar\psi\psi\right) -\frac i4 
 \overline {\psi}h^m_a
\gamma^a {\stackrel{\leftrightarrow}{\partial}}_m  \psi \0\\
&&\qquad\quad+ \frac 1{ 8} \left( (h_\mu^\mu)^2-2 h_\mu^\nu h_\nu^\mu\right)
\left(\frac{i}{2} \overline {\psi}\gamma^m
{\stackrel{\leftrightarrow}{\partial}}_m  \psi -m\bar\psi\psi\right)\0\\
&&\qquad\quad-\frac i8 h^\mu_\mu \overline {\psi}h^m_a
\gamma^a {\stackrel{\leftrightarrow}{\partial}}_m  \psi +\frac {3i}{16}
 \overline {\psi} (h^2)^m_a \gamma^a {\stackrel{\leftrightarrow}{\partial}}_m 
\psi+\ldots \Big{]} \label{expandaction}
\ee
There is one fermion-fermion-graviton vertex\footnote{We use the convention according to which two repeated identical indices represent a symmetrized couple of indices, and so on.}:
\be
V^{\mu\mu}_{ffh}(p,p')&:&  -\frac i4 (p+p')^\mu \gamma^\mu +\frac i4
\eta^{\mu\mu} (\slashed{p}+\slashed{p'}-2m)	\label{V1ffh}
\ee
and one vertex with two fermions and two gravitons:
\be
V^{\mu\mu\nu\nu}_{ffhh}(p,p')&:&\frac{ 3i}{16} \left( (p+p')^\mu \gamma^{\nu}
\eta^{\mu\nu} +  (p+p')^{\nu} \gamma^{\mu} \eta^{\mu\nu}
\right)\0\\
&&+\frac i8  (\slashed{p}+\slashed{p'}-2m) \left( \eta^{\mu\mu} \eta^{\nu\nu}-
 2\eta^{\mu\nu} \eta^{\mu\nu}\right)\0\\
&&-\frac i{8}\left((p+p')^\mu \gamma^\mu \eta^{\nu\nu}+(p+p')^{\nu}
\gamma^{\nu}\eta^{\mu\mu}\right)\label{V3ffhh}
\ee
We can also expand the odd parity part of the action (the latter contains a part proportional to the completely antisymmetric symbol). We will restrict ourselves to 3d because only in this case can we get a non-vanishing contribution to the effective action and 1-point correlator. 
\be 
S_o&=&  \frac 1{16}\int d^3x \, 
 \epsilon^{abc} \partial_a h_{b\sigma}h_c^\sigma \overline {\psi} \psi
\ee
The relevant  vertex with two fermions and two gravitons is
\be
V^{\mu\mu\nu\nu}_{\epsilon, ffhh}&:&\frac 1{16} \, \eta^{\mu\nu}
\epsilon^{\mu\nu\lambda} \,(k-k')_\lambda \label{Veffhh3d}
\ee 
\subsubsection{Even parity part}

The tadpole contribution is now
\be 
\tilde \Theta^{\mu\mu}(k)= \label{tad1D}
 -2^{-2-d+\lfloor \frac d2\rfloor}\,i\,m^d \pi^{\frac d2}\Gamma\left(-\frac
d2\right) \eta^{\mu\mu}
=\tilde \Theta\, \eta^{\mu\mu}
\ee
where $\tilde \Theta$ is a constant. Since the theory of gravity is non-linear
we have a contribution from the seagull term, which can be written as
\be 
\tilde T^{\mu\mu\nu\nu}_{(s)}(k) \label{tad2D}
= 2^{-3-d+\lfloor \frac d2\rfloor}\, i\,m^d \pi^{\frac d2}\Gamma\left(-\frac
d2\right) \left(3\eta^{\mu\nu}\eta^{\mu\nu}-2 \eta^{\mu\mu}\eta^{\nu\nu}\right)
\ee
The bubble diagram contributes two parts, the transverse (conserved) part, 
\be
\tilde T^{\mu\mu\nu\nu}_t(k)&=&-\frac1{d(d+1)k^2} 2^{-2-d+\lfloor \frac
d2\rfloor}\, i\,m^d\pi^{\frac d2}\Gamma\left(1-\frac d2\right)\0\\ \label{2ptc}
&&\left[\left(-8m^2+(d+1)k^2+{_2F_1}\left[1,-\frac d2,\frac 12,\frac
{k^2}{4m^2}\right](8m^2+(d-1)k^2)\right)\pi^{\mu\nu}\pi^{\mu\nu} \right. \0\\
&& \left. +\left(-4m^2+(d+1)k^2+{_2F_1}\left[1,-\frac d2,\frac 12,\frac
{k^2}{4m^2}\right](4m^2-k^2)\right)\pi^{\mu\mu}\pi^{\nu\nu}\right] 
\ee
whose expansion in the IR is
\be
\tilde T^{\mu\mu\nu\nu}_t(k)=-2^{-3-d+\lfloor \frac d2\rfloor} i\, m^d \pi^{-\frac d2}\sum_{n=1}^\infty
\frac{  m^{-2n} \Gamma\left(n-\frac d2\right)}{2^{n}
(2n+1)!!}k^{2n}\left((2n-1)\pi^{\mu\nu}\pi^{\mu\nu}-\pi^{\mu\mu}\pi^{\nu\nu}
\right),
\ee
and the non-transverse (non-conserved) part  
\be
{\tilde T^{\mu\mu\nu\nu}_{nt}}(k)&=& -2^{-3-d+\lfloor \frac d2\rfloor}\,
i\,m^d\pi^{\frac d2}\Gamma\left(-\frac d2\right)
\left(\eta^{\mu\nu}\eta^{\mu\nu}- \eta^{\mu\mu}\eta^{\nu\nu}\right).\label{2ptnc}
\ee
Taking formulas (\ref{tad1D}), (\ref{tad2D}), (\ref{2ptc}) and (\ref{2ptnc}) and
substituting them in (\ref{WI}) we can see that the Ward identity is satisfied
for any dimension $d$.

The one-loop 1-point function (energy-momentum tensor) defined as
$\langle\!\langle T^{\mu\nu}(x) \rangle \! \rangle =\frac 2{\sqrt{g}}\frac
{\delta W}{\delta h_{\mu\nu}(x)}$ now becomes
\be 
\langle\!\langle T^{\mu\mu}(x) \rangle\! \rangle
&=&-2^{-1-d+\lfloor \frac d2\rfloor}\,m^d \pi^{-\frac
d2}\left[\Gamma\left(-\frac d2\right)g^{\mu\mu}
+\sum_{n=1}^\infty \frac{(-1)^n\, m^{-2n} \Gamma\left(n-\frac d2\right)}{2^{n+1}
(2n+1)!!}\right.\0\\
&&\left. \times\left((2n-1)\Box^{n-1}
G^{\mu\mu}+(n-1)\Box^{n-2}(\eta^{\mu\mu}
\Box-\partial^\mu\partial^\mu)R\right)\right]+O(h^2)
\ee
where $G_{\mu\mu}=R_{\mu\mu}-\frac 12 \eta_{\mu\mu}R$ is the Einstein tensor. The
energy-momentum tensor is clearly divergence free (\ref{conss2}). For the
effective action in the IR we obtain (in the even parity sector)
\be
W &\stackrel{\text{IR}}{=}&- 2^{-1-d+\lfloor \frac d2\rfloor}m^d\pi^{-\frac
d2} \int \text{d}^dx \sqrt{g}\times\left[\Gamma\left(-\frac d2\right)- \frac
{\Gamma\left(1-\frac d2\right)}{24m^2} R \right.\0\\
&&\left. - \frac {\Gamma\left(2-\frac d2\right)}{80m^4}
\left(R_{\mu\nu\lambda\rho}R^{\mu\nu\lambda\rho}-2R_{\mu\nu}R^{\mu\nu}+\frac 13
R^2\right)+\ldots\right]+O(h^3)
\ee
The divergent part of the effective action for $d=4$ (i.e.\ $d=4+\varepsilon$) is 
\be \label{Wfs2d4ir}
W &\stackrel{\text{IR}}{=}& \frac{1}{8\pi^2\varepsilon} \int \text{d}^4x \sqrt{g} \left(m^4+ \frac 1{12}
m^{2}R- \frac 1{40}\mathcal{W}^2+\ldots\right)+O(h^3)
\ee
The first term  is a cosmological constant term and the second is the linearized
Einstein-Hilbert action. The third term ($m^0$ term) is the Weyl density
$\mathcal{W}^2=R_{\mu\nu\lambda\rho}R^{\mu\nu\lambda\rho}-2R_{\mu\nu}R^{\mu\nu}
+\frac 13 R^2$ (conformal invariant in 4d).

The dominating term in the UV  ($O(m^0)$ term corresponds to (B.3) from
\cite{BCDGLS}) of the transverse part $\tilde T_{t}^{\mu\mu\nu\nu}(k)$ is
\be
\tilde T_t^{\mu\mu\nu\nu}(k)
\stackrel{UV}{=} \frac{2^{-3-2d+\lfloor \frac d2\rfloor} \, \pi^{\frac
32-\frac d2}(k^2)^{\frac d2}}{\left(-1+e^{i\pi d}\right)\Gamma\left(\frac
{d+3}2\right)}((d-1)\pi^{\mu\nu}\pi^{\mu\nu}-\pi^{\mu\mu}\pi^{\nu\nu})
\ee
The effective action in the UV is then
\be \label{Wfs2d4uv}
W &\stackrel{UV}{=}&(-1)^{\frac d2} \frac{2^{-4-2d+\lfloor \frac
d2\rfloor}\pi^{\frac 32-\frac d2}}{(-1+e^{i\pi d})\Gamma\left(\frac
{d+3}2\right)} \int \text{d}^dx \sqrt{g}
\left[(d-4)R_{\mu\nu\lambda\rho}\Box^{\frac d2
-2}R^{\mu\nu\lambda\rho}\right.\0\\
&&\left. +6 \left(R_{\mu\nu\lambda\rho}\Box^{\frac d2
-2}R^{\mu\nu\lambda\rho}-2R_{\mu\nu}\Box^{\frac d2 -2}R^{\mu\nu}+\frac 13
R\Box^{\frac d2 -2}R\right)+\ldots\right]+O(h^3)
\ee

\subsubsection{Odd parity part}
\label{odd-spin2}

In 3d the contribution from the seagull diagram with vertex (\ref{Veffhh3d}) becomes
\be 
\tilde T^{\mu\mu\nu\nu}_{(s,o)}(k)=
-\frac{m^2}{16\pi}\eta^{\mu\nu}\epsilon^{\mu\nu\lambda}k_\lambda \label{tad3do}
\ee
The odd part of the two-point correlator is non-vanishing only in $3d$ (the vertex
is (\ref{V1ffh})). The transverse part can be written as 
\be
\tilde T^{\mu\mu\nu\nu}_{t,o}(k)&=&-\frac {m}{64\pi k}
\left((k^2-4m^2)\text{ArcCoth}\left(\frac{2m}{k}\right)+2mk\right)\pi^{\mu\nu}
\epsilon^{\mu\nu\lambda}k_\lambda  \label{2ptco}
\ee
and the expansion of $\tilde T^{\mu\mu\nu\nu}_{t,o}(k)$ in the IR is
\be
\tilde T^{\mu\mu\nu\nu}_{t,o}(k)=-\frac{1}{64\pi}\sum_{n=0}^\infty\frac{k^{2(n+1)}m^{-2n}}{4^{2n}(4(n+1)^2-1)}
\pi^{\mu\nu} \epsilon^{\mu\nu\lambda}k_\lambda
\ee
The odd non-transverse part reads\footnote{In the notation from the previous
paper $\pi^{\mu\nu} \epsilon^{\mu\nu\lambda}k_\lambda$ corresponds to $(n_1\cdot \pi\cdot
n_2)\epsilon(k\cdot n_1\cdot n_2)$}
\be
\tilde T^{\mu\mu\nu\nu}_{nt,o}(k)&=& 
\frac{m^2}{16\pi}\eta^{\mu\nu}\epsilon^{\mu\nu\lambda}k_\lambda\label{2ptnco}
\ee
and can be canceled by the seagull contribution (\ref{tad3do}). So, only the
transverse odd part remains.
The odd part of the one-loop 1-pt function (energy-momentum tensor) 
\be 
\langle\!\langle T^{\mu\mu}(x) \rangle\! \rangle
=\frac{1}{32\pi}\sum_{n=0}^\infty \frac{(-1)^n\, m^{-2n} }{4^{2n}
(4(n+1)^2-1)}\Box^{n}C^{\mu\mu}
\ee
where $C^{\mu\mu}$ is linearized the Cotton tensor (\ref{cotton}).
The effective action in the IR (the dominating term)
\be 
W\stackrel{IR}{=}-\frac {1}{384\pi}\epsilon^{\mu\nu\lambda}\int
d^3x\,h_{\nu\nu}\left(\partial_\lambda\partial^\mu\partial^\nu
h_{\mu\mu}-\partial_\lambda\Box h_\mu^\nu\right)
+O(h^3)\ee
corresponds to gravitational Chern-Simons term in 3d
 \be 
S_{gCS}=\frac {1}{192\pi}\epsilon^{\mu\nu\lambda}\int
d^3x\,\left(\partial_\mu\omega_\nu{}^{ab}\omega_\lambda^{ b a}+\frac 23
\omega_{\mu a}{}^b\omega_{\nu b}{}^c\omega_{\lambda c}{}^a \right)
\ee

\subsection{Scalars - spin 2}

Let us consider the action of a scalar field $\varphi$ in a curved space
($g_{\mu\nu}=\eta_{\mu\nu}+h_{\mu\nu}$) with a scalar curvature coupling
\be
S=\int d^dx\sqrt{g}\left(g^{\mu\nu}\partial_\mu \varphi^{\dagger}
\partial_\nu\varphi-m^2\varphi^{\dagger}\varphi +\xi R \varphi^{\dagger}\varphi 
\right)
\ee
Let us redefine $\phi=g^{\frac 14}\varphi$. The expansion of the action in the
external field h is
\be
S&=&\int d^dx\left[\eta^{\mu\nu}\partial_\mu \phi^{\dagger}
\partial_\nu\phi-m^2\phi^{\dagger}\phi  +  h^{\mu\nu}\left(\frac 14
\phi^{\dagger}\stackrel{\leftrightarrow}{\partial}_\mu\stackrel{\leftrightarrow}
{\partial}_\nu\phi+\left(\xi-\frac
14\right)(\partial_\mu\partial_\nu-\Box\eta_{\mu\nu})\phi^{\dagger}
\phi\right)\right.\0\\
&&\left. +h^{\mu\sigma}h_\sigma^\nu\partial_\mu\phi^{\dagger}\partial_\nu\phi
+\frac 1{16}h\Box h\phi^{\dagger}\phi +\left(-\frac{\xi}4+\frac
18\right)\partial_\mu h \partial^\mu h\phi^{\dagger}\phi-2\xi
h^{\mu\nu}\partial_\nu\partial_\lambda
h_\mu^\lambda\phi^{\dagger}\phi\right.\0\\
&&\left.+ \xi h^{\mu\nu}\Box h_{\mu\nu}\phi^{\dagger}\phi -\xi \partial_\nu
h^{\mu\nu}\partial_\lambda h_\mu^\lambda \phi^{\dagger}\phi +\frac
34\xi\partial_\lambda h_{\mu\nu}\partial^\lambda
h^{\mu\nu}\phi^{\dagger}\phi-\frac 12\xi\partial_\lambda h^{\mu\nu}\partial_\nu
h_\mu^\lambda\phi^{\dagger}\phi\right.\0\\
&&\left.\left(\xi-\frac 14\right) h^{\mu\nu}\partial_\mu\partial_\nu
h\phi^{\dagger}\phi+ \left(\xi-\frac 14\right)\partial_\mu h\partial_\nu
h^{\mu\nu}\phi^{\dagger}\phi\right]
\ee
The scalar-scalar-graviton vertex is:
\be
V^{\mu\mu}_{ssh}(p,p')&:&  -\frac i4 (p^\mu+p'^\mu)^2  -i\left(\xi-\frac
14\right)\left((p'^\mu-p^\mu)^2 - \eta^{\mu\mu} (p'-p)^2\right)	\label{Vssh}
\ee
and there is a vertex with two scalars and two gravitons:
\be
V^{\mu\mu\nu\nu}_{sshh}(p,p',k,k')&:&i\eta^{\mu\nu}\left( p'^\mu p^\nu +p^\mu
p'^\nu \right)-i\Big{[}\left(\xi-\frac 14\right)\left(\eta^{\mu\mu}k^\nu
k^\nu+\eta^{\nu\nu}k^\mu k^\mu \right)\0\\
&&+2\left(\xi\eta^{\mu\nu}\eta^{\mu\nu}+\frac
1{16}\eta^{\mu\mu}\eta^{\nu\nu}\right)k^2-4\xi\eta^{\mu\nu}k^{\mu}
k^{\nu}\Big{]}\0\\
&&- i\Big{[}\left(\left(\frac
14-\frac{\xi}2\right)\eta^{\mu\mu}\eta^{\nu\nu}+\frac
32\xi\eta^{\mu\nu}\eta^{\mu\nu}\right)k\cdot k' \\
&&+\left(\xi-\frac 14\right)\left(\eta^{\mu\mu}k^\nu k'^\nu+\eta^{\nu\nu}k^\mu
k'^\mu\right)-2\xi\eta^{\mu\nu}k^\mu k'^\nu-\xi\eta^{\mu\nu} k^\nu k'^\mu
\Big{]}\0 \label{Vsshh}
\ee

The result for the tadpole diagram is
\be 
\tilde \Theta^{\mu\mu}= 2^{-d-1} \,i\, \pi ^{-d/2} m^d \Gamma
   \left(-\frac{d}{2}\right) \,\eta^{\mu\mu}\label{tads}
\ee
while the contribution from the seagull term is
\be
\tilde T_{(s)}^{\mu\mu\nu\nu}(k)&=&-2^{-4-d} \,i  \pi ^{-d/2} m^{d-2} \Gamma
   \left(-\frac{d}{2}\right) \0\\
   &&\times\left(d k^2 (1-4 \xi ) \eta ^{\mu \mu } \eta ^{\nu \nu }+4 \eta^{\mu
\nu }\eta^{\mu \nu } \left(4 m^2-d k^2 \xi \right)+8 d \xi  \eta ^{\mu \nu }
k^{\mu } k^{\nu}\right)\label{seag}
\ee
Furthermore, the transverse part of the bubble diagram reads
\be
\tilde T_{t}^{\mu\mu\nu\nu}(k)&=&-\frac{1}{3 d \left(d^2-1\right) k^4}i
2^{-d-2} e^{-\frac{1}{2} i \pi  d} \pi ^{-d/2} (-m^2)^{d/2} m^{-2} \Gamma
   \left(1-\frac{d}{2}\right)\0\\
   && \Big{[} \Big{(}12 \left(d^2-1\right) k^4 m^2 \left(8 \xi ^2-8 \xi
+1\right)+d
   \left(d^2-1\right) k^6 \left(24 \xi ^2-1\right)\0\\
   &&+24 d k^2 m^4 (3-8 \xi ) -192 k^2 m^4 \xi +96
   m^6 \0\\
   &&+ \left(-6 k^4 m^2 \left(d^2
   (1-4 \xi )^2+d (8 \xi -2)-2 \left(8 \xi ^2-8 \xi +1\right)\right)\right.\0\\
   &&\left.+24 k^2 m^4 (d (8 \xi -2)+8
   \xi )-96 m^6\right)\, _2F_1\left[1,-\frac{d}{2};-\frac{1}{2};\frac{k^2}{4
m^2}\right]\Big{)}\pi ^{\mu \mu } \pi ^{\nu \nu }\0\\
   &&  +\left(-12 d^2 k^4 m^2+d \left(d^2-1\right) k^6+48 d k^2 m^4-96 k^2
m^4+12
   k^4 m^2+192 m^6\right.\0\\
   &&\left.-12 m^2 \left(k^2-4 m^2\right)^2 \,
   _2F_1\left[1,-\frac{d}{2};-\frac{1}{2};\frac{k^2}{4 m^2}\right]
\right)\pi^{\mu \nu }\pi^{\mu\nu}\Big{]}\label{2ptsc}
\ee
The expansion of the transverse part $\tilde T_{t}^{\mu\mu\nu\nu}(k)$ in the IR is
\be
\tilde T_{t}^{\mu\mu\nu\nu}(k)&=& 2^{-3-d} i\, m^{d-4} \pi^{-\frac d2}
k^4\sum_{n=0}^\infty \frac{ \, m^{-2n}\Gamma\left(2+n-\frac
d2\right)}{2^{n}\,(2n+5)!!}k^{2n}\0\\
&&\times\left(\pi^{\mu\nu}\pi^{\mu\nu}+\frac{a(n,\xi)}2\pi ^{\mu \mu } \pi ^{\nu
\nu }\right)
\ee
where $a(n,\xi)$ is a constant
\be
a(n,\xi)=(2n+5)(2n+3)(4\xi-1)^2+2(2n+5)(4\xi-1)+1
\ee
The non-transverse part of the bubble diagram is
\be
\tilde T_{nt}^{\mu\mu\nu\nu}(k)&=&\frac{2^{-4-d}}{3}   \,i\,\pi ^{-d/2} 
m^{d-2} \Gamma
   \left(-\frac{d}{2}\right) \0\\
   &&\left(\eta^{\mu \nu }\eta^{\mu \nu } \left(24 m^2-2 d k^2\right)+4 \,d\,
   \eta ^{\mu \nu } k^{\mu } k^{\nu }+ 2 d\, (6 \xi -1) \eta ^{\mu \mu } k^{\nu
}k^\nu\right.\0\\
   &&\left.+\eta ^{\nu \nu }
   \left(\eta ^{\mu \mu } \left(d k^2 (5-24 \xi )+12 m^2\right)+2 d (6 \xi -1)
k^{\mu }k^\mu\right) \right)\label{2ptsnc}
\ee
The seagull diagram and  the non-transverse part of 2-pt function together give
\be
\tilde T_{(s)}^{\mu\mu\nu\nu}(k)+\tilde T_{nt}^{\mu\mu\nu\nu}(k)&=&- 2^{-d-2}
i \pi ^{-d/2} m^d \Gamma
   \left(-\frac{d}{2}\right) \left(2 \eta ^{\mu \nu }\eta ^{\mu \nu }-\eta ^{\mu
\mu } \eta ^{\nu \nu }\right)\\
   &&+ 2^{-d-1} i \pi ^{-d/2}  m^{d-2}  \left(\xi -\frac{1}{6}\right) 
 \Gamma \left(1-\frac{d}{2}\right)  k^2\left(\pi^{\mu \nu }\pi^{\mu \nu }-\pi
^{\mu \mu } \pi
   ^{\nu \nu }\right)\0
\ee
Taking formulas (\ref{tads}), (\ref{seag}), (\ref{2ptsc}) and (\ref{2ptsnc}) and
substituting them in (\ref{WI}) we can see that the Ward identity is satisfied
for any dimension $d$.

The one-loop 1-point correlator 
\be 
\langle\!\langle T^{\mu\mu}(x) \rangle\! \rangle
&=&-2^{-d}\,m^d\pi^{-\frac d2}\left[\Gamma\left(-\frac
d2\right)g^{\mu\mu}-\frac{2\Gamma\left(1-\frac d2\right)}{m^2}\left(\xi-\frac
16\right)G^{\mu\mu} \right.\0\\
&&\left.+\sum_{n=2}^\infty \frac{(-1)^n\, m^{-2n} \Gamma\left(n-\frac
d2\right)}{2^{n} (2n+1)!!}\Box^{n-2}\right.\0\\
&&\left.\times\left(-2\Box
G^{\mu\mu}+\left(1-\frac{a(n,\xi)}2\right)(\eta^{\mu\mu}
\Box-\partial^\mu\partial^\mu)R\right)\right]
+O(h^2)\ee
satisfies (\ref{conss2}).
For the effective action in the IR we obtain 
\be
W [h] &\stackrel{\text{IR}}{=}&2^{-d}m^d\pi^{-\frac d2}\int d^dx
\sqrt{g}\left[\Gamma\left(-\frac d2\right)-\frac{\Gamma\left(1-\frac
d2\right)}{2m^2}
\left(\xi-\frac 16\right)R\right.\0\\
&&\left. +\frac {\Gamma\left(2-\frac d2\right)}{120m^4}
\left(R_{\mu\nu\lambda\rho}R^{\mu\nu\lambda\rho}+\frac {a(0,\xi)}2
R^2\right)+\ldots\right]+O(h^3)
\ee
For $\xi=\frac 16$ (the conformal value) the third term in the expansion is proportional
to
\be
&\propto& m^{d-4} \int
d^dx\sqrt{g}\left(R_{\mu\nu\lambda\rho}R^{\mu\nu\lambda\rho}-\frac 13 R^2\right)
\ee
We can use the Gauss-Bonnet theorem 
\be 
R_{\mu\nu\lambda\rho}R^{\mu\nu\lambda\rho}-4R_{\mu\nu}R^{\mu\nu}+R^2=\text{total
derivative}\label{GB}
\ee
to write the divergent part of the effective action in $d=4$ as a Weyl square density
\be \label{Wss2d4ir}
W &\stackrel{\text{IR}}{=}& -\frac1{16\pi^2\varepsilon} \int d^4x\sqrt{g} \left(m^4+\frac
1{30}\mathcal{W}^2\right)+O(h^3)
\ee

In the massless case ($m^0$ is the dominating term in the UV) we have
\be
\tilde T^{\mu\mu\nu\nu}(k)&\stackrel{\text{UV}}{=}&\frac{2^{-1-2d} \, \pi^{\frac 32-\frac
d2}(k^2)^{\frac d2}}{\left(-1+e^{i\pi d}\right)\Gamma\left(\frac
{d+3}2\right)}\left(\pi^{\mu\nu}\pi^{\mu\nu}+\frac{b(d,\xi)}2
\pi^{\mu\mu}\pi^{\nu\nu}\right)
\ee
where
\be
b(d,\xi)=(d^2-1)(4\xi-1)^2+2(d+1)(4\xi-1)+1
\ee
The effective action in the UV now becomes
\be \label{Wss2d4uv}
W &\stackrel{UV}{=}&(-1)^{\frac d2} \frac{2^{-2-2d+\lfloor \frac
d2\rfloor}\pi^{\frac 32-\frac d2}}{(-1+e^{i\pi d})\Gamma\left(\frac
{d+3}2\right)} \int \text{d}^dx \,  \left(R^{\mu\nu\lambda\rho}\Box^{\frac d2
-2}R_{\mu\nu\lambda\rho}+\frac {b(d,\xi)}2 R\Box^{\frac d2 -2}R\right)
\ee
After we use (\ref{GB}) and put $\xi=\frac 16$ in 4d we will again get the Weyl square
density
\be \label{Wss2d4x16uv}
W &\stackrel{UV}{=}& \int \text{d}^dx \, \mathcal{W}^2
\ee

\section{All types of correlator: a repository} 
\label{s:alltypes}

This last section of the paper is a systematic collection of results concerning all types of two-point correlators, including the mixed ones, for symmetric currents of of spin up to 5 and in dimension $3\leq d\leq 6$. It also contains results concerning the correlators of currents of any spin and in any dimensions, in the case of massless models, for which it is possible to write down very compact formulas.

The two point amplitudes in question for fermion and of scalar currents for spins up to 5, are schematically denoted as follows:
\al{\tilde{T}_{{\mu_1\ldots \mu_{s_1}}{\nu_1\ldots \nu_{s_2}}}(k) \equiv
\langle {\tilde J}_{\mu_1\ldots \mu_{s_1}}(-k) {\tilde J}_{\nu_1\ldots \nu_{s_2}}(k) \rangle\,,}
Scalar and fermion currents are given by
\be
{\tilde{T}}^\text{s}_{\mu_1\ldots\mu_s}=i^s \varphi^\dagger 
\left({\stackrel{\leftrightarrow}{\partial}}_\mu \right)^{s}\varphi
\,,\quad
{\tilde{T}}^\text{f}_{\mu_1\ldots\mu_s}=i^{s-1} \bar{\psi}\gamma_\mu\left({\stackrel{\leftrightarrow}{\partial}}_\mu \right)^{s-1}\psi\label{Jsimple}
\ee
(For fermions in case $s=0$ we use ${\tilde{T}}^\text{f}_{s=0}=\bar{\psi}\psi$.)
These currents will be henceforth referred to as {\em simple} currents.
In the fermionic case the two point correlator is 
 \be 
{\tilde{T}}^\textrm{f}_{\mu_1 \ldots \mu_{s_1}\nu_1\ldots\nu_{s_2}}(k)& =& 
-\int\frac
{d^dp}{(2\pi)^d}\, {\rm Tr} 
\left( \frac {i} {\slashed{p}-m} \gamma_\sigma 
\frac {{ i}} {\slashed{p}-\slashed{k}-m} \gamma_\tau \right)
V^\sigma_{\mu_1\ldots\mu_{s_1}}V^\tau_{\nu_1\ldots\nu_{s_2}}\label{Jmunu0}
\ee
whereas in the scalar case it is
 \be 
{\tilde{T}}^\textrm{s}_{\mu_1 \ldots \mu_{s_1}\nu_1\ldots\nu_{s_2}}(k)& =& 
\int\frac
{d^dp}{(2\pi)^d}\, 
\frac{1}{(p^2-m^2)((p-k)^2-m^2)}
V_{\mu_1\ldots\mu_{s_1}}V_{\nu_1\ldots\nu_{s_2}} \label{Jfs}
\ee
with the Feynman vertices for fermions and scalars respectively 
\be
V^\sigma_{\mu_1\ldots\mu_s} = i \, \delta^\sigma_\mu \, (2 p_\mu- k_\mu)^{s-1}
\,,\quad
V_{\mu_1\ldots\mu_s} = i \, (2 p_\mu- k_\mu)^{s} \label{Jss}
\ee

To label the correlators we  often suppress writing indices and add the number of space-time dimensions in the subscript on the left hand side. Additionally, when $s_1, s_2 \neq 0$, we split the amplitudes in the transverse and the non-transverse part, so for the correlator of e.g.\ fermionic spin-$s_1$ and spin-$s_2$ currents in $d$ dimensions we write:
\be \label{defsplitcnc}
{\tilde{T}}^\textrm{f}_{s_1,s_2,d} = 
{\tilde{T}}^\textrm{f,t}_{s_1,s_2,d} + 
{\tilde{T}}^\textrm{f,nt}_{s_1,s_2,d} 
\ee
There is no preferred way to do the splitting in \refb{defsplitcnc} because one can always add some transverse quantity to ${\tilde{T}}^\textrm{t}$ and subtract the same quantity from ${\tilde{T}}^\textrm{nt}$. However, it always happens that the non-transverse part can be chosen to be a polynomial in $k$ and $m$ (i.e.\ local). Here, we always make this choice so that  the non-transverse part is local. After this choice is made there is still some remaining freedom in the splitting into the transverse and the non-transverse part in \refb{defsplitcnc}, nevertheless the quantities we define below do not depend on this remaining freedom. 

One such quantity is ${\tilde{T}}^\textrm{f,UV-IR}_{s_1,s_2,d}$, the difference between the UV and the IR expansions in the shortly explained sense. Since, as explained above, the non-transverse part is always local the non-transverse parts of UV and IR are the same and therefore cancel so that only transverse parts remain in the expression for ${\tilde{T}}^\textrm{f,UV-IR}_{s_1,s_2,d}$
\be \label{defUVIR}
{\tilde{T}}^\textrm{f,UV-IR}_{s_1,s_2,d} = 
{\tilde{T}}^\textrm{f,UV}_{s_1,s_2,d} -
{\tilde{T}}^\textrm{f,IR}_{(0)s_1,s_2,d} 
\ee
where the UV and IR expansions are denoted by ${\tilde{T}}^\textrm{f,t,UV}_{s_1,s_2,d}$ and ${\tilde{T}}^\textrm{f,t,IR}_{s_1,s_2,d}$ respectively, and $ {\tilde{T}}^\textrm{f,IR}_{(0)s_1,s_2,d}$ is the part of the IR expansion of order ${\cal O}(m^n)$ with $n\geq 0$.

Another such quantities are the divergences of the correlators:
\be \label{defdiv}
\left(k \cdot {\tilde{T}}^\textrm{f}_{s_1,s_2,d}\right)_{\mu_2 \ldots \mu_{s_1}\nu_1\ldots\nu_{s_2}} = k^{\mu_1}
\left({\tilde{T}}^\textrm{f,nt}_{s_1,s_2,d}\right)_{\mu_1 \ldots \mu_{s_1}\nu_1\ldots\nu_{s_2}} \0\\
\left( {\tilde{T}}^\textrm{f}_{s_1,s_2,d} \cdot k \right)_{\mu_1 \ldots \mu_{s_1}\nu_2\ldots\nu_{s_2}}
=  
k^{\nu_1}
\left({\tilde{T}}^\textrm{f,nt}_{s_1,s_2,d}\right)_{\mu_1 \ldots \mu_{s_1}\nu_1\ldots\nu_{s_2}}
\ee
The definitions \refb{defsplitcnc}, \refb{defUVIR}, \refb{defdiv} are analogous in the scalar case.

Before listing the results for the massive models, it is worth to show some general formulas (for any spin and any dimension)  that it was possible to obtain for the massless case. (We recall that the results for the massless cases correspond to the dominant term in the UV expansion of the massive case.) In addition some general formulas are easy to write in terms of particular linear combination of the previous currents which become traceless in the massless case. These ``traceless'' versions of the currents can be defined in the following way:
\be
{\tilde{T}}^{\textrm{st}}_{\mu_1\ldots\mu_s}=
\sum_{l=0}^{\lfloor \frac{s}{2}\rfloor} 
a^{{\textrm{s}}}_{s,l}\left(\Box\pi_{\mu\mu}\right)^l
 {\tilde{T}}^{\textrm{s}}_{\mu_1\ldots\mu_{s-2l}}
\,,  \quad
{\tilde{T}}^{\textrm{ft}}_{\mu_1\ldots\mu_s}=
\sum_{l=0}^{\lfloor \frac{s-1}{2}\rfloor} 
{a^{\textrm{f}}_{s,l}}\left(\Box\pi_{\mu\mu}\right)^l
 {\tilde{T}}^{\textrm{f}}_{\mu_1\ldots\mu_{s-2l}}  
\ee
where
\be
 a^{{\textrm{s}}}_{s,l}=\frac{{(-1)^l}s!\,\Gamma\left(s+\frac{d-3}{2}-l\right)}{2^{2l}{l!}(s-2l)!\,\Gamma\left(s+\frac{d-3}{2}\right)}\, ,\quad
{ a^{\textrm{f}}_{s,l}=\frac{(-1)^l(s-1)!\,\Gamma\left(s+\frac{d-3}{2}-l\right)}{2^{2l}l!(s-2l-1)!\,\Gamma\left(s+\frac{d-3}{2}\right)}}\label{tralS} 
\ee
It is easy to see that amplitudes for two general spins $s_1$ and $s_2$ for the ``traceless'' currents can be written as linear combinations of the amplitudes \refb{Jfs} and \refb{Jss} of the ``simple'' currents \refb{Jsimple}
\be
{\tilde{T}}^\text{st}_{\mu_1\ldots\mu_{s_1}\nu_1\ldots\nu_{s_2}} =
\sum_{l=0}^{\lfloor \frac{s_1}{2}\rfloor}
\sum_{k=0}^{\lfloor \frac{s_2}{2}\rfloor}
a^{{\textrm{s}}}_{s_1,l} a^{{\textrm{s}}}_{s_2,k}  \left(k^2\eta_{\mu\mu}-k^2_\mu\right)^l \left(k^2\eta_{\nu\nu}-k^2_\nu\right)^k {\tilde{T}}^\text{s}_{\mu_1\ldots\mu_{s_1-2l}\nu_1\ldots\nu_{s_2-2k}}
\0
\ee

\be
{\tilde{T}}^\text{ft}_{\mu_1\ldots\mu_{s_1}\nu_1\ldots\nu_{s_2}} =
\sum_{l=0}^{\lfloor \frac{s_1-1}{2}\rfloor}
\sum_{k=0}^{\lfloor \frac{s_2-1}{2}\rfloor}
a^{{\textrm{f}}}_{s_1,l} a^{{\textrm{f}}}_{s_2,k} 
\left(k^2\eta_{\mu\mu}-k^2_\mu\right)^l \left(k^2\eta_{\nu\nu}-k^2_\nu\right)^k 
{\tilde{T}}^\text{f}_{\mu_1\ldots\mu_{s_1-2l}\nu_1\ldots\nu_{s_2-2k}}
\0
\ee

The result for the traceless currents in the massless limit is
\be
{\tilde{T}}^\text{st,massless}_{\mu_1\ldots\mu_{s}\nu_1\ldots\nu_{s}} &=&(-1)^s \frac{ 2^{4-2d-s}  \pi ^{\frac 32-\frac{d}{2}}s! \left(k^2\right)^{\frac d2+s-2}}{ \left(-1+e^{i \pi  d}\right) \Gamma \left(\frac{d+2s-1}2\right)}\sum_{l=0}^{\lfloor \frac{s}{2}\rfloor} {a_{s,l}^{\textrm{s}}}\pi^l_{\mu\mu}\pi^l_{\nu\nu}\pi_{\mu\nu}^{s-2l}\\
&=&(-1)^s \frac{ 2^{4-2d-s}  \pi ^{\frac 32-\frac{d}{2}} s!\left(k^2\right)^{\frac d2+s-2}}{ \left(-1+e^{i \pi  d}\right) \Gamma \left(\frac{d+2s-1}2\right)}
 \pi^s_{\mu\nu}\,\,{_2 F_1}\left(\frac{1-s}2,-\frac s2,\frac{5-d-2s}2,\frac{\pi_{\mu\mu}\pi_{\nu\nu}}{\pi_{\mu\nu}^2}\right)\0
\ee
We note that for traceless currents mixed spin terms are zero i.e.\ the result vanishes for spin ${s_1} \neq s_2$. For simple currents this is not the case
and the general expression for spin $s_1\times s_2$, ${s_2 \geqslant s_1}$ is
\be
{\tilde{T}}^\text{s,massless}_{\mu_1\ldots\mu_{s_1}\nu_1\ldots\nu_{s_2}} &=&(-1)^{\frac{s_1+s_2}2} \frac{\left(2\lfloor\frac{s_2+1}{2}\rfloor-1\right)!!\left(2\lfloor\frac{s_2+1}{2}\rfloor\right)!! 2^{4-2d-\frac{s_1+s_2}2}  \pi ^{\frac 32-\frac{d}{2}} \left(k^2\right)^{\frac{d+s_1+s_2}2-2}}{\left(2\lfloor\frac{s_2}{2}\rfloor-2\lfloor\frac{s_1}{2}\rfloor\right)!! \left(-1+e^{i \pi  d}\right) \Gamma \left(\frac{d+s_1+s_2-1}2\right)}\0\\
&&\times\pi^{\frac{s_2-s_1}{2}}_{\nu\nu}\sum_{l=0}^{\lfloor \frac{{s_1}}{2}\rfloor}\frac{s_1!(s_2-s_1)!!}{2^{\frac{l(l+1)}2}(s_1-2l)!(s_2-s_1+2l)!!}\pi^l_{\mu\mu}\pi^l_{\nu\nu}\pi_{\mu\nu}^{{s_1}-2l}
\ee
For fermions in the massless limit it also happens that only the diagonal ({$s_1=s_2\equiv s$ and $s>0$}) amplitudes survive for the traceless currents
\al{
{\tilde{T}}^\text{ft,massless,even}_{\mu_1\ldots\mu_{s}\nu_1\ldots\nu_{s}} & = 
(-1)^s \frac{ 2^{3-2d-s+\lfloor\frac d2 \rfloor}  \pi ^{\frac 32-\frac{d}{2}} (s-1)!(d-3+s)\left(k^2\right)^{\frac d2+s-2}}{ \left(-1+e^{i \pi  d}\right) \Gamma \left(\frac{d+2s-1}2\right)}\sum_{l=0}^{\lfloor \frac{s}{2}\rfloor} {a_{s,l}^{\textrm{s}}}\pi^l_{\mu\mu}\pi^l_{\nu\nu}\pi_{\mu\nu}^{s-2l}
\0\\
 & = (-1)^s \frac{ 2^{3-2d-s+\lfloor\frac d2 \rfloor}  \pi ^{\frac 32-\frac{d}{2}}(s-1)!(d-3+s) \left(k^2\right)^{\frac d2+s-2}}{ \left(-1+e^{i \pi  d}\right) \Gamma \left(\frac{d+2s-1}2\right)}
 \0\\
 & \quad \quad
\times \pi^s_{\mu\nu}\,\,{_2 F_1}\left(\frac{1-s}2,-\frac s2,\frac{5-d-2s}2,\frac{\pi_{\mu\mu}\pi_{\nu\nu}}{\pi_{\mu\nu}^2}\right)
}
The formula above is valid for $d\geq 4$ and for the even part in $d=3$. For the odd part in $d=3$ we obtain for traceless currents, for the dominant term in the UV,  a
general expression for spin $s_1\times s_2$, ${s_2 \geqslant s_1}$,  ${s_1>0,\, s_2>0}$
\be
{\tilde{T}}^{\text{ft,UV dominant,odd}}_{\mu_1\ldots\mu_{s_1}\nu_1\ldots\nu_{s_2};3D}&=&(-1)^{\frac{s_1+s_2}2} \frac{i m k^{s_1+s_2-3}}{2^{s_2+1}}\pi_{\nu\nu}^{\frac{s_2-s_1}2}\sum_{l=0}^{{\lfloor \frac{s_1-1}{2}\rfloor}}\frac{(-1)^l\Gamma\left(s_1-l\right)}{2^{2l} l!\Gamma\left(s_1-2l\right)}\pi^l_{\mu\mu}\pi^l_{\nu\nu}\pi_{\mu\nu}^{s_1-2l-1}\epsilon_{\sigma\mu\nu}k^\sigma\0\\
 &=&(-1)^{\frac{s_1+s_2}2} \frac{i m k^{s_1+s_2-3}}{2^{s_2+1}}\pi_{\nu\nu}^{\frac{s_2-s_1}2}
 \pi^{s_1-1}_{\mu\nu}
\0\\&& \quad\times \,\,{_2 F_1}\left(\frac{1-s_1}2,{1}-\frac {s_1}2,1-s_1,\frac{\pi_{\mu\mu}\pi_{\nu\nu}}{\pi_{\mu\nu}^2}\right)\epsilon_{\sigma\mu\nu}k^\sigma
 \label{eq:PT}
\ee
In Appendix \ref{sec:3dtraceless} we show that this formula is a straightforward generalization of  the linearized action proposed long ago by Pope and Townsend, \cite{pope}, for conformal higher spin fields.
In the case of simple currents we instead get
\be
{\tilde{T}}^{\text{f,UV dominant,odd}}_{\mu_1\ldots\mu_{s_1}\nu_1\ldots\nu_{s_2};3D} &=&(-1)^{\frac{s_1+s_2}2} \frac{\left(2\lfloor\frac{s_2-1}{2}\rfloor\right)!!\left(s_1+s_2-2\lfloor\frac{s_1-1}{2}\rfloor-3\right)!!  m k^{s_1+s_2-3}}{2^2(s_1+s_2-2)!!{\left(2\lfloor\frac{s_2}{2}\rfloor-2\lfloor\frac{s_1}{2}\rfloor\right)!!} }\\
&&\times\pi^{\frac{s_2-s_1}{2}}_{\nu\nu}\epsilon_{\sigma\mu\nu}k^\sigma\sum_{l=0}^{{\lfloor \frac{s_1-1}{2}\rfloor}}\frac{(s_1-1)!(s_2-s_1)!!}{2^{\frac{l(l+1)}2}(s_1-2l-1)!(s_2-s_1+2l)!!}\pi^l_{\mu\mu}\pi^l_{\nu\nu}\pi_{\mu\nu}^{{s_1}-2l-1}\0
\ee
In the case of 
simple currents it is possible to write the formula for the IR expansion of the transverse part:
\be
{\tilde{T}}^{\text{f,t,IR,odd}}_{\mu_1\ldots\mu_{s_1}\nu_1\ldots\nu_{s_2};3D} &=&(-1)^{\frac{s_1+s_2}2-1} \frac{\left(2\lfloor\frac{s_2-1}{2}\rfloor\right)!!\left(s_1+s_2-2\lfloor\frac{s_1-1}{2}\rfloor-3\right)!!  k^{s_1+s_2-2}}{2^2\pi(s_1+s_2-1)!!{\left(2\lfloor\frac{s_2}{2}\rfloor-2\lfloor\frac{s_1}{2}\rfloor\right)!!} }\\
&&\times\pi^{\frac{s_2-s_1}{2}}_{\nu\nu}\epsilon_{\sigma\mu\nu}k^\sigma\sum_{l=0}^{{\lfloor \frac{s_1-1}{2}\rfloor}}\frac{(s_1-1)!(s_2-s_1)!!}{2^{\frac{l(l+1)}2}(s_1-2l-1)!(s_2-s_1+2l)!!}\pi^l_{\mu\mu}\pi^l_{\nu\nu}\pi_{\mu\nu}^{{s_1}-2l-1}\0
\ee

\vskip 2cm
In the rest of the section we list the results for the massive case. The results are given for $d=3,4,5,6$ and spin $s\leq 5$. For even $d$, we use $d \rightarrow d + \varepsilon$ and expand around $\varepsilon$. For odd $d$ this is not necessary. It is convenient to use the following shorthand notation \al{
L_n &= \frac{2}{\varepsilon } + \log \left(\frac{m^2}{4 \pi }\right)+\gamma-\sum_{k=1}^{n}\frac{1}{k}
}
as well as
\al{
K &= \log \left( -\frac{k^2}{m^2} \right)
\0\\
P &= \frac{2}{\varepsilon } + \log \left(-\frac{k^2}{4 \pi }\right)+\gamma
}
We see that there is a relationship 
\be
P = K + L_0
\ee
Furthermore we define
\al{
T &= -\frac{2 i \coth ^{-1}\left(\frac{2 m}{k}\right)}{\pi } \0\\
S &= \sqrt{4 m^2-k^2} \csc ^{-1}\left(\frac{2 m}{k}\right)
}
It turns out that $T$ is useful in even dimensions $d$ and $S$ is useful in odd. The branches of the functions $T$ and $S$ are chosen such that the IR and UV expansions are
\al{
T& \quad \stackrel{IR}{=}\quad
-\frac{i k}{\pi  m}-\frac{i k^3}{12 \pi  m^3}-\frac{i k^5}{80 \pi  m^5} + \ldots
\0\\
S& \quad \stackrel{IR}{=}\quad
k   - \frac{k^3}{ 12 m^2}  - \frac{k^5}{120 m^4}  + \ldots
}
and
\al{
T& \quad \stackrel{UV}{=}\quad
1-\frac{4 i m}{\pi  k}-\frac{16 i m^3}{3 \pi  k^3}-\frac{64 i m^5}{5 \pi  k^5} + \ldots
\0\\
S& \quad \stackrel{UV}{=}\quad
 \frac{k K}{2}
 -\frac{m^2 \left(1+K\right)}{k}
 +\frac{m^4 \left(1-2 K\right)}{2 k^3}
 +\frac{m^6 \left(5-6 K\right)}{3 k^5} + \ldots
}
In the results for UV-IR which follow, the difference is shown for the terms containing the powers of $m$ and $k$ that ``overlap'' in UV and IR in sense that those powers appear both in UV and in IR expansions. The rest, i.e. the UV expansion that does not overlap with the IR, is denoted by ellipses.
\vskip 1cm 
The following results are organized as follows: first come the ones for the scalar model, sec.4.1-4.3, then those of the fermion models, sec. 4.4-4.6. Sections 4.1 and 4.4 contain the full transverse analytic expressions of the correlators. Sections 4.2 and 4.5 contain the UV and IR expansions of the latter, as well as the above-mentioned UV-IR expressions. Sections 4.3. and 4.6 are devoted to the non-transverse local parts of the correlators. These are the non-transverse expressions which have not been already discussed in the previous section and should be eliminated by the use of tadpole and seagull terms. 

\vskip 1cm
The method to obtain the results below has been explained in \cite{BCDGLS} and is largely based on the approach of Davydychev and collaborators, \cite{BoosDavy}, see \cite{Blima}.

\input{ss12t.tex}
\input{fs12t.tex}

\section{Conclusion}
\label{s:Conclusion}
We finally sum up our major results. In this paper we have pursued further the program started in \cite{BCDGLS}, considering in particular the quadratic part of the effective action.

First of all, we have discussed the relevant issue of the geometric interpretation of the obtained effective actions. The basic outcome of \cite{BCDGLS} was that, upon considering on-shell conserved higher spin currents, the corresponding effective actions inherit an off-shell gauge invariance once a finite number of local counterterms are subtracted. In particular our (linearized) gauge invariance involves unconstrained fully symmetric parameters and is the same as the one considered  in \cite{FS,Francia1,Francia2,Francia2012}. We are therefore naturally led to the problem of expressing our results in the geometric language of \cite{deWitFreedman,Damour}. This is done in full generality in section \ref{s:geometry}.

Another relevant issue is whether it is possible to construct a gauge invariant effective action without the subtraction of ad hoc non-invariant counterterms. The answer to this question of course requires the choice of a specific regularization scheme. We decided to work in dimensional regularization, which turns out to be particularly convenient for the lower spin cases. In fact, in section \ref{s:seagull} we have explictly shown that for spin 1 and spin 2 gauge fields it is generally possible to introduce additional local terms allowed by covariance (involving the spin 0 current) such that the  effective action is gauge invariant with no non-covariant subtractions needed thanks to the tadpole and seagull diagrams entering the Ward Identities. This is no surprise, as in these cases we already know fully off-shell covariant versions of QED and gravity coupled to ordinary scalar and spin $1/2$ matter. Nevertheless our explict computations work as a promising test for higher spin gauge fields. Related questions are whether this whole procedure depends on the choice of the on-shell conserved current that is minimally coupled in the first place and whether it can be restricted to a finite number of higher spin currents. In the cases of spin 1 and 2 we checked there is no need to introduce higher spin currents and and off-shell invariant couplings between the matter and gauge sectors can be obtained for any choice of on-shell conserved Noether currents. 
A particularly interesting choice for the currents and the couplings for the spin 2 case in $d=4$  is the one for which the dominating term in the UV expansion is the Weyl density ((\ref{Wfs2d4uv}), (\ref{Wss2d4x16uv})), corresponding to the emerging conformal symmetry in the massless case. This term is also found in the corresponding IR expansion ((\ref{Wfs2d4ir}), (\ref{Wss2d4ir})). The case of higher spin gauge fields, in particular as far as the choice of different currents is concerned, will be treated in details in a subsequent work.

The major task of this paper was the completion of the construction of the quadratic part of the higher spin effective action started in \cite{BCDGLS,BCLPS}. We did it in section \ref{s:alltypes} presenting all two point correlators of symmetric currents of any spin up to 5 and in any dimension between 3 and 6. We also spelled out UV and IR expansions, finding compact formulae for dominating terms in the two limits. In particular, we also included the mixed ones which were not considered in our previous work. The results of section \ref{s:alltypes} show that these terms turn out to have the usual structure found for the diagonal ones, i.e.\ the sum of a nonlocal transverse part and a local longitudinal one. We expect that their presence is crucial when one tries to test the invariance of the effective action beyond the lowest order in the gauge fields as well as when one tries to introduce tadpoles and seagulls in the Ward Identities. In $d=3$  odd-parity kinetic terms are present when spin $1/2$ matter is integrated out. We find that for the traceless currents considered in section \ref{s:alltypes} the UV limit coincides with an interesting generalization of the conformal higher spin action found in \cite{pope}. Although made up of the same invariant building blocks, the kinetic terms appearing in this section do not coincide with the ones found in \cite{FS, Francia1} by coupling the gauge theory to matter and considering the analysis of the propagating degrees of freedom. The fundamental reason is that our effective actions are obtained after a subtraction procedure that is required to give invariant terms, but still allows for a wide class of possible choices. A proper discussion of the propagating degrees of freedom in our case would require the removal of nonlocalities by introducing compensating fields and the consideration of physical propagators after proper gauge fixing. However, the next logical step is the computation of three-point functions which would provide an insight in gauge invariance beyond the lowest order and therefore prepare the ground for the construction of fully covariant effective actions.

\vskip 5cm

{\bf Acknowledgements.}
This research has been supported by the Croatian
Science Foundation under the project No.~8946 and by the University of
Rijeka under the research support No.~13.12.1.4.05. 

\newpage
\vskip 1cm

\noindent{\bf \large Appendices}
\appendix
\section{Spin 2 - expansions}

Here, we list several useful expansions of geometrical quantities in terms of spin 2 field $h_{\mu\nu}$. 

\noindent
Riemann tensor
\be
R_{\mu\nu\lambda\rho}=-\frac 12 \partial_\mu\partial_\lambda h_{\nu\rho}+\frac
12 \partial_\mu\partial_\rho h_{\nu\lambda}+\frac 12
\partial_\nu\partial_\lambda h_{\mu\rho}-\frac 12 \partial_\nu\partial_\rho
h_{\mu\lambda}
\ee
Ricci tensor
\be
R_{\mu\nu}=-\frac 12 \partial_\mu\partial_\nu h+\frac 12 \partial_\mu
\partial_\lambda h_\nu^\lambda+\frac 12 \partial_\nu \partial_\lambda
h_\mu^\lambda-\frac 12\Box h_{\mu\nu}
\ee
Ricci scalar
\be
R&=&-\Box h+\partial_\mu \partial_\nu h^{\mu\nu}\0\\
&&+h^{\mu\nu}\partial_\mu\partial_\nu h-2h^{\mu\nu}\partial_\nu\partial_\lambda
h_\mu^\lambda-\frac 14\partial_\nu h\partial^\nu h- \partial_\nu
h^{\mu\nu}\partial_\lambda h_\mu^\lambda+ \partial_\mu h\partial_\nu
h^{\mu\nu}+h^{\mu\nu}\Box h_{\mu\nu}\0\\
&&-\frac 12\partial_\lambda h^{\mu\nu}\partial_\nu h_\mu^\lambda+\frac
34\partial_\lambda h_{\mu\nu}\partial^\lambda h^{\mu\nu}
\ee
Einstein-Hilbert 
\be
\sqrt{g}R=-\Box h+\partial_\mu \partial_\nu h^{\mu\nu}+\frac
12\left(h^{\mu\nu}\Box h_{\mu\nu}+2h^{\mu\nu}\partial_\mu\partial_\nu h-h\Box
h-2h^{\mu\lambda}\partial_\mu\partial_\lambda h_\nu^\lambda\right)
\ee
\al{\sqrt{g} R & \quad\leftrightarrow\quad 
k^2 \pi_{\mu\mu} h^{\mu\mu} 
+
h^{\mu\mu} \left(
-\frac{1}{4} k^2 \left(\pi_{\mu\nu}^2-\pi_{\mu\mu}\pi_{\nu\nu}\right)
\right)h^{\nu\nu}
}
Riemann squared
\be
R^2_{\mu\nu\lambda\rho}=h^{\mu\nu}
\partial_\mu\partial_\nu\partial_\lambda\partial_\rho
h^{\lambda\rho}-2h^{\mu\nu}\partial_\mu\partial_\lambda\Box
h_\nu^\lambda+h^{\mu\nu}\Box^2 h_{\mu\nu}
\ee
\al{
\sqrt{g} R_{\kappa\lambda\rho\sigma}R^{\kappa\lambda\rho\sigma} & \quad\leftrightarrow\quad
h^{\mu\mu}\left( 
k^4 \pi_{\mu\nu}^2
\right)h^{\nu\nu}
}
Ricci tensor squared
\be
R^2_{\mu\nu}=\frac 14 h^{\mu\nu}\Box^2 h_{\mu\nu}-\frac 12
h^{\mu\nu}\partial_\mu\partial_\nu\Box h- \frac 12
h^{\mu\nu}\partial_\mu\partial_\lambda\Box h_\nu^\lambda +\frac 12
h^{\mu\nu}\partial_\mu\partial_\nu\partial_\lambda\partial_\rho
h^{\lambda\rho}+\frac 14 h\Box^2 h
\ee
\al{
\sqrt{g} R_{\kappa\lambda}R^{\kappa\lambda} & \quad\leftrightarrow\quad
h^{\mu\mu}\left( 
\frac{1}{4}k^4 \left(\pi_{\mu\nu}^2+\pi_{\mu\mu}\pi_{\nu\nu}
\right)
\right)h^{\nu\nu}
}
Ricci scalar squared
\be
R^2= h^{\mu\nu}\partial_\mu\partial_\nu\partial_\lambda\partial_\rho
h^{\lambda\rho}-2h^{\mu\nu}\partial_\mu\partial_\nu\Box h+h\Box^2 h
\ee
\al{
\sqrt{g} R^2 & \quad\leftrightarrow\quad
h^{\mu\mu}\left( 
k^4  \pi_{\mu\mu}\pi_{\nu\nu} 
\right)h^{\nu\nu}
}
Weyl density 
\al{
W_{\mu\nu\rho\sigma} &= R_{\mu\nu\rho\sigma} -
\frac{1}{d-2} \left(
R_{\mu\rho}g_{\nu\sigma}
-R_{\mu\sigma}g_{\nu\rho} 
-R_{\nu\rho}g_{\mu\sigma}
+R_{\nu\sigma}g_{\mu\rho} 
\right)
+\\\0 & \quad +
\frac{R}{(d-1)(d-2)} \left(
g_{\mu\rho}g_{\nu\sigma}
-g_{\mu\sigma}g_{\nu\rho} \right) 
}
\al{
W_{\mu\nu\rho\sigma}W^{\mu\nu\rho\sigma} &= 
R_{\mu\nu\rho\sigma}R^{\mu\nu\rho\sigma} -
\frac{4}{d-2} R_{\mu\nu}R^{\mu\nu}
+\frac{2}{d^2-3d+2} R^2
}
\al{
\sqrt{g} W_{\kappa\lambda\rho\sigma}W^{\kappa\lambda\rho\sigma} & \quad\leftrightarrow\quad
h^{\mu\mu}\left( 
\frac{d-3}{d-2}k^4 \pi_{\mu\nu}^2-
\frac{d-3}{(d-1)(d-2)}k^4 \pi_{\mu\mu} \pi_{\nu\nu}
\right)h^{\nu\nu}
}
Weyl density for $d=4$
\be
\mathcal{W}^2&=&R_{\mu\nu\lambda\rho}^2-2R_{\mu\nu}^2+\frac 13 R^2\\
&=&\frac 12 h^{\mu\nu}\Box^2 h_{\mu\nu}- 
h^{\mu\nu}\partial_\mu\partial_\lambda\Box h_\nu^\lambda+\frac 13
h^{\mu\nu}\partial_\mu\partial_\nu\Box h +\frac 13
h^{\mu\nu}\partial_\mu\partial_\nu\partial_\lambda\partial_\rho
h^{\lambda\rho}-\frac 16 h\Box^2 h\0
\ee
Cotton tensor for $d=3$
\be 
\label{cotton}
C_{\mu\nu}=\epsilon_\mu{}^{\tau\rho}\partial_\tau\left(R_{\rho\nu}-\frac{1}{d-1}
g_{\nu\rho}R\right)
=\frac 12 \epsilon_\mu{}^{\rho\tau}\partial_\tau\left(\Box
h_{\nu\rho}-\partial_\lambda\partial_\nu h_\rho^\lambda\right)
\ee

\section{Higher spin traceless actions}
\label{sec:3dtraceless}

In this appendix we will review the parity-odd actions that are expected in 3D for conformal higher spin in Minkowski background, showing that they coincide with the UV limits of  the amplitudes considered in sections \ref{odd-spin1}, \ref{odd-spin2}, 
\ref{FA}\footnote{It is understood the UV limit should be carried out as described in \cite{BCDGLS1}, introducing an ad hoc flavor index.}. 
The action considered in \cite{pope} can be easily generalized to the case when the quadratic terms involve fields of different spins, namely
\begin{eqnarray}
I_{s_1,s_2} & = & \frac{1}{2}\int d^{3}x\sum_{r=0}^{s_1-1}\binom{2s_1}{2r+1}\overset{(s_1)}{h}^{\alpha_{1}\ldots\alpha_{2s_1}}\left(\square\right)^{r}\partial_{\alpha_{1}}\phantom{}^{\beta_{1}}\ldots\partial_{\alpha_{2s_1-2r-1}}\phantom{}^{\beta_{2s_1-2r-1}}\nonumber\\
 &  & \partial^{\beta_{2s_1+1}\beta_{2s_1+2}}\ldots\partial^{\beta_{2s_2-1}\beta_{2s_2}}\overset{(s_2)}{h}_{\beta_{1}\ldots\beta_{2s_1-2r-1}\alpha_{2s_1-2r}\ldots\alpha_{2s_1}\beta_{2s_1+1}\ldots\beta_{2s_2}},
\end{eqnarray}
where we assume $s_1\leq s_2$ and $\partial_{\alpha}\phantom{}^{\beta}=\left(\gamma^{\mu}\right)_{\alpha}\phantom{}^{\beta}\partial_{\mu}$. We define $\left(\gamma_{\mu}\right)_{\alpha\beta}=\varepsilon_{\beta\gamma}\left(\gamma_{\mu}\right)_{\alpha}\phantom{}^{\gamma}$ and $\left(\gamma_{\mu}\right)^{\alpha\beta}=\varepsilon^{\alpha\gamma}\left(\gamma_{\mu}\right)_{\gamma}\phantom{}^{\beta}$, in agreement with the conventions of Wess and Bagger. 
Going from the spinor notation $h^{\alpha_{1}\ldots\alpha_{2s}}=h^{\mu_{1}\ldots\mu_{s}}\left(\gamma_{\mu_{1}}\right)^{\alpha_{1}\alpha_{2}}\ldots\left(\gamma_{\mu_{s}}\right)^{\alpha_{2s-1}\alpha_{2s}}$ to the standard tensor one, we get
\begin{eqnarray*}
I_{s_1,s_2} & = & \frac{1}{2}\int d^{3}x\sum_{r=0}^{s_1-1}\binom{2s_1}{2r+1}\overset{(s_1)}{h}^{\mu_{1}\ldots\mu_{s_1}}\left(\gamma_{\mu_{1}}\right)^{\alpha_{1}\alpha_{2}}\ldots\left(\gamma_{\mu_{s_1}}\right)^{\alpha_{2s_1-1}\alpha_{2s_1}}\left(\square\right)^{r}\\
 &  & \partial_{\alpha_{1}}\phantom{}^{\beta_{1}}\ldots\partial_{\alpha_{2s_1-2r-1}}\phantom{}^{\beta_{2s_1-2r-1}}\delta_{\alpha_{2s_1-2r}}^{\beta_{2s_1-2r}}\ldots\delta_{\alpha_{2s_1}}^{\beta_{2s_1}} \partial^{\beta_{2s_1+1}\beta_{2s_1+2}}\ldots\partial^{\beta_{2s_2-1}\beta_{2s_2}}\\
 & & \left(\gamma_{\nu_{1}}\right)_{\beta_{1}\beta_{2}}\ldots\left(\gamma_{\nu_{s_2}}\right)_{\beta_{2s_2-1}\beta_{2s_2}}\overset{(s_2)}{h}^{\nu_{1}\ldots\nu_{s_2}}\\
 & = & \frac{1}{2}\int d^{3}x\sum_{r=0}^{s_1-1}\binom{2s_1}{2r+1}\overset{(s_1)}{h}^{\mu_{1}\ldots\mu_{s_1}}\left(\square\right)^{r}\left[\left(\gamma_{\mu_{1}}\right)^{\alpha_{1}\alpha_{2}}\partial_{\alpha_{1}}\phantom{}^{\beta_{1}}\left(\gamma_{\nu_{1}}\right)_{\beta_{1}\beta_{2}}\partial_{\alpha_{2}}\phantom{}^{\beta_{2}}\right]\ldots\\
 &  & \ldots\left[\left(\gamma_{\mu_{s_1-r-1}}\right)^{\alpha_{2s_1-2r-3}\alpha_{2s_1-2r-2}}\partial_{\alpha_{2s_1-2r-3}}\phantom{}^{\beta_{2s_1-2r-3}}
\right.\0\\&& \qquad\left.
 \left(\gamma_{\nu_{s_1-r-1}}\right)_{\beta_{2s_1-2r-3}\beta_{2s_1-2r-2}}\partial_{\alpha_{2s_1-2r-2}}\phantom{}^{\beta_{2s_1-2r-2}}\right]\\
 &  & \left[\left(\gamma_{\mu_{s_1-r}}\right)^{\alpha_{2s_1-2r-1}\alpha_{2s_1-2r}}\partial_{\alpha_{2s_1-2r-1}}\phantom{}^{\beta_{2s_1-2r-1}}
 \left(\gamma_{\nu_{s_1-r}}\right)_{\beta_{2s_1-2r-1}\beta_{2s_1-2r}}\delta_{\alpha_{2s_1-2r}}^{\beta_{2s_1-2r}}\right]\\
 &  & \left[\left(\gamma_{\mu_{s_1-r+1}}\right)^{\alpha_{2s_1-2r+1}\alpha_{2s_1-2r+2}}\left(\gamma_{\nu_{s-r+1}}\right)_{\beta_{2s-2r+1}\beta_{2s-2r+2}}\delta_{\alpha_{2s-2r+1}}^{\beta_{2s_1-2r+1}}\delta_{\alpha_{2s_1-2r+2}}^{\beta_{2s_1-2r+2}}\right]\ldots\\
 &  & \ldots\left[\left(\gamma_{\mu_{s_1}}\right)^{\alpha_{2s_1-1}\alpha_{2s_1}}\left(\gamma_{\nu_{s_1}}\right)_{\beta_{2s_1-1}\beta_{2s_1}}\delta_{\alpha_{2s_1-1}}^{\beta_{2s_1-1}}\delta_{\alpha_{2s_1}}^{\beta_{2s_1}}\right]\\
 & &  \left[\partial^{\beta_{2s_1+1}\beta_{2s_1+2}}\left(\gamma_{\nu_{s_1+1}}\right)_{\beta_{2s_1+1}\beta_{2s_1+2}}\right]\ldots\left[\partial^{\beta_{2s_2-1}\beta_{2s_2}}\left(\gamma_{\nu_{s_2}}\right)_{\beta_{2s_2-1}\beta_{2s_2}}\right]\overset{(s_2)}{h}^{\nu_{1}\ldots\nu_{s}}\,,
\end{eqnarray*}
where, keeping in mind that in $D=3$ $\left(\gamma_\mu\right)_{\alpha\beta}=\left(\gamma_\mu\right)_{\beta\alpha}$, we can easily recognize the traces
\begin{eqnarray*}
Tr\left[\gamma_{\mu_{i}}\left(\gamma\centerdot\partial\right)\gamma_{\nu_{i}}\left(\gamma\centerdot\partial\right)\right]
& = & 
\left[\left(\gamma_{\mu_{i}}\right)^{\alpha_{2i-1}\alpha_{2i}}\partial_{\alpha_{2i-1}}\phantom{}^{\beta_{2i-1}}\left(\gamma_{\nu_{i}}\right)_{\beta_{2i-1}\beta_{2i}}\partial_{\alpha_{2i}}\phantom{}^{\beta_{2i}}\right] ,\\
-Tr\left[\gamma_{\mu_{s_1-r}}\left(\gamma\centerdot\partial\right)\gamma_{\nu_{s_1-r}}\right]
& = & 
\left[\left(\gamma_{\mu_{s_1-r}}\right)^{\alpha_{2s_1-2r-1}\alpha_{2s_1-2r}}\partial_{\alpha_{2s_1-2r-1}}\phantom{}^{\beta_{2s_1-2r-1}}
\right. \0\\ && \quad \left.
\left(\gamma_{\nu_{s_1-r}}\right)_{\beta_{2s_1-2r-1}\beta_{2s_1-2r}}\delta_{\alpha_{2s_1-2r}}^{\beta_{2s_1-2r}}\right] 
,\\
 -Tr\left[\gamma_{\mu_{j}}\gamma_{\nu_{j}}\right]
  & = &
\left[\left(\gamma_{\mu_{j}}\right)^{\alpha_{2j-1}\alpha_{2j}}\left(\gamma_{\nu_{j}}\right)_{\beta_{2j-1}\beta_{2j}}\delta_{\alpha_{2j-1}}^{\beta_{2j-1}}\delta_{\alpha_{2j}}^{\beta_{2j}}\right]  ,\\
 -Tr\left[\left(\gamma\centerdot\partial\right)\gamma_{\nu_{k}}\right] & = & \left[\partial^{\beta_{2k-1}\beta_{2k}}\left(\gamma_{\nu_{k}}\right)_{\beta_{2k-1}\beta_{2k}}\right]\,.
\end{eqnarray*}
In order to evaluate these traces, we need the
following rules
\begin{eqnarray*}
Tr\left[\gamma_{\mu}\gamma_{\nu}\right] & = & 2\eta_{\mu\nu}\,,\\
Tr\left[\gamma_{\mu}\gamma_{\nu}\gamma_{\rho}\right] & = & -2
\varepsilon_{\mu\nu\rho}\,,\\
Tr\left[\gamma_{\mu}\gamma_{\nu}\gamma_{\rho}\gamma_{\sigma}\right] & = & 2\left(\eta_{\mu\nu}\eta_{\rho\sigma}-\eta_{\mu\rho}\eta_{\nu\sigma}+\eta_{\mu\sigma}\eta_{\nu\rho}\right)\,,
\end{eqnarray*}
which imply
\be
Tr\left[\left(\gamma\centerdot\partial\right)\gamma_{\nu_{k}}\right] & = & 2\partial_{\nu_{k}}\,,\0\\
Tr\left[\gamma_{\mu_{s-r}}\left(\gamma\centerdot\partial\right)\gamma_{\nu_{s-r}}\right] & = & 2
\varepsilon_{\mu_{s-r}\nu_{s-r}\rho}\partial^{\rho}\,,\0\\
Tr\left[\gamma_{\mu_{i}}\left(\gamma\centerdot\partial\right)\gamma_{\nu_{i}}\left(\gamma\centerdot\partial\right)\right] & = & 2\left(2\partial_{\mu_{i}}\partial_{\nu_{i}}-\eta_{\mu_{i}\nu_{i}}\square\right)\,.\0
\ee
We can therefore rewrite the action $I_{s_1,s_2}$ as
\be
I_{s_1,s_2} & = & \frac{1}{2}\left(-2\right)^{s_2}
\int d^{3}x\sum_{r=0}^{s_1-1}\binom{2s_1}{2r+1}\overset{(s_1)}{h}^{\mu_{1}\ldots\mu_{s_1}}\left(\square\right)^{r}\left(\eta_{\mu_{1}\nu_{1}}\square-
2\partial_{\mu_{1}}\partial_{\nu_{1}}\right)\ldots\0\\
&  & \ldots\left(\eta_{\mu_{s_1-r-1}\nu_{s_1-r-1}}\square-2\partial_{\mu_{s_1-r-1}}
\partial_{\nu_{s_1-r-1}}\right)\varepsilon_{\mu_{s_1-r}\nu_{s_1-r}\rho}\partial^{\rho}\0\\
&  & \eta_{\mu_{s_1-r+1}\nu_{s_1-r+1}}\ldots\eta_{\mu_{s_1}\nu_{s_1}}\partial_{\nu_{s_1+1}}\ldots\partial_{\nu_{s_2}}\overset{(s_2)}{h}^{\nu_{1}\ldots\nu_{s_2}}\0\\
& = & -\left(-2\right)^{s_2-1}
\int d^{3}x\sum_{r=0}^{s_1-1}\binom{2s_1}{2r+1}\overset{(s_1)}{h}^{\mu_{1}\ldots\mu_{s_1}}\left(\square\right)^{r}\varepsilon_{\mu_{1}\nu_{1}\rho}
\partial^{\rho}\eta_{\mu_{2}\nu_{2}}\ldots\eta_{\mu_{r+1}\nu_{r+1}}\0\\
&  & \left(\eta_{\mu_{r+2}\nu_{r+2}}\square-2\partial_{\mu_{r+2}}\partial_{\nu_{r+2}}\right)\ldots
\left(\eta_{\mu_{s_1}\nu_{s_1}}\square-2\partial_{\mu_{s_1}}\partial_{\nu_{s_1}}\right)\0\\
& &\partial_{\nu_{s_1+1}}\ldots\partial_{\nu_{s_2}} \overset{(s_2)}{h}^{\nu_{1}\ldots\nu_{s_2}}\,. \label{Is}
\ee
By elementary manipulations, one can write the action \eqref{Is} as a triple summation
\be
I_{s_1,s_2}
& = & -\left(-2\right)^{s_2-1}
\int d^{3}x \overset{(s_1)}{h}^{\mu_{1}\ldots\mu_{s_1}}\varepsilon_{\mu_{1}\nu_{1}\rho}\partial^{\rho}\sum_{r=0}^{s_1-1}
\sum_{j=0}^{s_1-r-1}\sum_{k=0}^{s_1-j-1}\0\\
&  & \left(-2\right)^{j}\binom{2s_1}{2r+1}\binom{s_1-r-1}{j}\binom{s_1-j-1}{k}\left(\eta_{\mu_{2}\nu_{2}}\square-\partial_{\mu_{2}}\partial_{\nu_{2}}\right)\ldots
\left(\eta_{\mu_{k+1}\nu_{k+1}}\square-\partial_{\mu_{k+1}}\partial_{\nu_{k+1}}\right)\0\\
&  &\left(\partial_{\mu_{k+2}}\partial_{\nu_{k+2}}\right)\ldots\left(\partial_{\mu_{s_1}}\partial_{\nu_{s_1}}
\right)\partial_{\nu_{s_1+1}}\ldots\partial_{\nu_{s_2}} \overset{(s_2)}{h}^{\nu_{1}\ldots\nu_{s_2}}\,.
\ee
We can now switch the order of summation
\be
I_{s_1,s_2}& = & -\left(-2\right)^{s_2-1}
\int d^{3}x \overset{(s_1)}{h}^{\mu_{1}\ldots\mu_{s_1}}\varepsilon_{\mu_{1}\nu_{1}\rho}\partial^{\rho}
\sum_{k=0}^{s_1-1}\left(-\right)^{s_1-k-1}\sum_{j=0}^{s_1-k-1}\sum_{r=0}^{s_1-j-1}\0\\
&  & \left(-2\right)^{j}\binom{2s_1}{2r+1}\binom{s_1-r-1}{j}\binom{s_1-j-1}{k}\left(\eta_{\mu_{2}\nu_{2}}\square-\partial_{\mu_{2}}\partial_{\nu_{2}}\right)\ldots
\left(\eta_{\mu_{k+1}\nu_{k+1}}\square-\partial_{\mu_{k+1}}\partial_{\nu_{k+1}}\right)\0\\
&  &\left(-\partial_{\mu_{k+2}}\partial_{\nu_{k+2}}\right)\ldots\left(-\partial_{\mu_{s_1}}\partial_{\nu_{s_1}}
\right)\partial_{\nu_{s_1+1}}\ldots\partial_{\nu_{s_2}} \overset{(s_2)}{h}^{\nu_{1}\ldots\nu_{s_2}}\,,
\ee
and perform two summations using the tables or Mathematica
\begin{eqnarray*}
&&\sum_{j=0}^{s_1-k-1}\sum_{r=0}^{s_1-j-1}\left(-2\right)^{j}\binom{2s_1}{2r+1}\binom{s_1-r-1}{j}\binom{s_1-j-1}{k} 
\0\\&& \qquad=  2^{-1+2s_1}\binom{s_1-1}{k}\phantom{}_{2}F_{1}\left(\frac{1}{2}-s_1,1+k-s_1,1-2s_1;2\right)\,,
\end{eqnarray*}
thus obtaining
\be
I_{s_1,s_2}& = & -\left(-2\right)^{s_2-1}2^{-1+2s_1}
\int d^{3}x \overset{(s_1)}{h}^{\mu_{1}\ldots\mu_{s_1}}\varepsilon_{\mu_{1}\nu_{1}\rho}\partial^{\rho}
\sum_{k=0}^{s_1-1}\left(-\right)^{s_1-k-1}\binom{s_1-1}{k}\0\\
&  &\phantom{}_{2}F_{1}\left(\frac{1}{2}-s_1,1+k-s_1,1-2s_1;2\right) \left(\eta_{\mu_{2}\nu_{2}}\square-\partial_{\mu_{2}}\partial_{\nu_{2}}\right)\ldots
\left(\eta_{\mu_{k+1}\nu_{k+1}}\square-\partial_{\mu_{k+1}}\partial_{\nu_{k+1}}\right)\0\\
&  &\left(-\partial_{\mu_{k+2}}\partial_{\nu_{k+2}}\right)\ldots\left(-\partial_{\mu_{s_1}}\partial_{\nu_{s_1}}
\right)\partial_{\nu_{s_1+1}}\ldots\partial_{\nu_{s_2}} \overset{(s_2)}{h}^{\nu_{1}\ldots\nu_{s_2}}\,,
\ee
or  equivalently
\be
I_{s_1,s_2}& = & -\left(-2\right)^{s_2-1}2^{-1+2s_1}
\int d^{3}x \overset{(s_1)}{h}^{\mu_{1}\ldots\mu_{s_1}}\varepsilon_{\mu_{1}\nu_{1}\rho}\partial^{\rho}
\sum_{k=0}^{s_1-1}\left(-\right)^{k}\binom{s_1-1}{s_1-1-k}\0\\
&  &\phantom{}_{2}F_{1}\left(-k,\frac{1}{2}-s_1,1-2s_1;2\right)\left(-\partial_{\mu_{2}}\partial_{\nu_{2}}\right)\ldots\left(-\partial_{\mu_{k+1}}\partial_{\nu_{k+1}}
\right)\0\\
&  & \left(\eta_{\mu_{k+2}\nu_{k+2}}\square-\partial_{\mu_{k+2}}\partial_{\nu_{k+2}}\right)\ldots
\left(\eta_{\mu_{s_1}\nu_{s_1}}\square-\partial_{\mu_{s_1}}\partial_{\nu_{s_1}}\right)\partial_{\nu_{s_1+1}}\ldots\partial_{\nu_{s_2}} \overset{(s_2)}{h}^{\nu_{1}\ldots\nu_{s_2}}\,.
\ee
Using the recursion relation $\phantom{}_{2}F_{1}\left(-k,\frac{1}{2}-s,1-2s;2\right)=\frac{2-2s+k}{k+1}\phantom{}_{2}F_{1}\left(-k-2,\frac{1}{2}-s,1-2s;2\right)$
and the starting values $\phantom{}_{2}F_{1}\left(0,\frac{1}{2}-s,1-2s;2\right)=1$
and $\phantom{}_{2}F_{1}\left(-1,\frac{1}{2}-s,1-2s;2\right)=0$, one finds
\begin{eqnarray*}
\phantom{}_{2}F_{1}\left(-2j-1,\frac{1}{2}-s,1-2s;2\right) & = & 0\,,\\
\phantom{}_{2}F_{1}\left(-2j,\frac{1}{2}-s,1-2s;2\right) & = & \frac{\Gamma\left(2j+1\right)\Gamma\left(1-2s\right)}{\Gamma\left(2j-2s+1\right)}P_{k}^{\left(-2s,s-2j-\frac{1}{2}\right)}\\
 & = & \frac{\Gamma\left(2j+1\right)\Gamma\left(1-2s\right)}{\Gamma\left(2j-2s+1\right)}\binom{j-s-\frac{1}{2}}{j}\\
 & = & \frac{\left(-1\right)^{j}}{2^{2j}}\frac{\Gamma\left(2j\right)\Gamma\left(s-j\right)}{\Gamma\left(s\right)\Gamma(j)}\,.
\end{eqnarray*}
The summation is therefore only over even $k=2j$ and, using the fact that $h$'s are traceless and the convenient notation of projectors $\pi_{\mu\nu}=\eta_{\mu\nu}-\frac{\partial_\mu \partial_\nu}{\square}$. we can write down
\be
I_{s_1,s_2}& = & -\left(-2\right)^{s_2-1}2^{-1+2s_1}
\int d^{3}x \overset{(s_1)}{h}^{\mu_{1}\ldots\mu_{s_1}}\varepsilon_{\mu_{1}\nu_{1}\rho}\partial^{\rho}\square^{s_1-1}
\sum_{j=0}^{\left[\frac{s_1-1}2\right]}\left(-\right)^{j}\frac1{2^{2j}}\binom{s_1-1-j}{j}\0\\
&  & \pi_{\mu_2 \mu_3}\ldots\pi_{\mu_{2j} \mu_{2j+1}}\pi_{\nu_2 \nu_3}\ldots\pi_{\nu_{2j} \nu_{2j+1}}\pi_{\mu_{2j+2} \nu_{2j+2}}\ldots\pi_{\mu_{s_1} \nu_{s_1}}\partial_{\nu_{s_1+1}}\ldots\partial_{\nu_{s_2}} \overset{(s_2)}{h}^{\nu_{1}\ldots\nu_{s_2}}\,.
\ee
One can go to momentum representation by the substitutions $\overset{(s)}{h}(x)=\int \frac{d^{3}k}{(2\pi)^3} e^{-i k x}\overset{(s)}{h}(k) $,
\be
I_{s_1,s_2}& = & -\left(-2\right)^{s_2-1}2^{-1+2s_1}(-i)^{s_2-s_1+1}
\\&&\quad
\int d^{3}k \overset{(s_1)}{h}^{\mu_{1}\ldots\mu_{s_1}}\varepsilon_{\mu_{1}\nu_{1}\rho}k^{\rho}(-k^2)^{s_1-1}
\sum_{j=0}^{\left[\frac{s_1-1}2\right]}\left(-\right)^{j}\frac1{2^{2j}}\binom{s_1-1-j}{j}\0\\
&&\quad\quad \pi_{\mu_2 \mu_3}\ldots\pi_{\mu_{2j} \mu_{2j+1}}\pi_{\nu_2 \nu_3}\ldots\pi_{\nu_{2j} \nu_{2j+1}}\pi_{\mu_{2j+2} \nu_{2j+2}}\ldots\pi_{\mu_{s_1} \nu_{s_1}}k_{\nu_{s_1+1}}\ldots k_{\nu_{s_2}} \overset{(s_2)}{h}^{\nu_{1}\ldots\nu_{s_2}}\,,
\0\ee
which corresponds to the amplitude \eqref{eq:PT} up to an overall constant.  This follows from the fact that $h$ is traceless and $s_2-s_1$ is even so we can substitute $k_{\nu_{s_1+1}}\ldots k_{\nu_{s_2}}$ with $\pi$'s.

 \end{document}

%% file: ss12t.tex
\subsection{Scalar amplitudes}
Scalars, spin 0 x 0, dimension 3:
\al{\label{eq:sc:0:0:3:}\tilde{T}_{0,0;3\text{D}}^{\text{s}} & = -\frac{ T}{8}\frac{1}{k} }
Scalars, spin 0 x 0, dimension 4:
\al{\label{eq:sc:0:0:4:}\tilde{T}_{0,0;4\text{D}}^{\text{s}} & = \frac{i  }{8 \pi ^2}\left(1-\frac{L_0}{2}\right)-\frac{i  S}{8 \pi ^2}\frac{1}{k} }
Scalars, spin 0 x 0, dimension 5:
\al{\label{eq:sc:0:0:5:}\tilde{T}_{0,0;5\text{D}}^{\text{s}} & = -\frac{i }{32 \pi ^2} m+\frac{ T}{32 \pi } \left(-\frac{1}{4} k+\frac{m^2}{k}\right) }
Scalars, spin 0 x 0, dimension 6:
\al{\label{eq:sc:0:0:6:}\tilde{T}_{0,0;6\text{D}}^{\text{s}} & = \frac{i }{16 \pi ^3} \left( \left(\frac{1}{9}-\frac{L_0}{24}\right) k^2+ \left(-\frac{7}{12}+\frac{L_0}{4}\right) m^2\right)+\frac{i  S}{48 \pi ^3} \left(-\frac{1}{4} k+\frac{m^2}{k}\right) }
Scalars, spin 0 x 2, dimension 3:
\al{\label{eq:sc:0:2:3:}\tilde{T}_{0,2;3\text{D}}^{\text{s,t}} & = k^2 \pi _{\nu\nu} \left(-\frac{i }{4 \pi }\frac{ m}{k^2}+ T \left(\frac{1}{16}\frac{1}{k}-\frac{1}{4}\frac{ m^2}{k^3}\right)\right) }
\al{\label{eq:snc:0:2:3:}\tilde{T}_{0,2;3\text{D}}^{\text{s,nt}} & = \eta_{\nu\nu} \left(\frac{i }{2 \pi } m\right) }
Scalars, spin 0 x 2, dimension 4:
\al{\label{eq:sc:0:2:4:}\tilde{T}_{0,2;4\text{D}}^{\text{s,t}} & = k^2 \pi _{\nu\nu} \left(\frac{i }{6 \pi ^2} \left( \left(-\frac{1}{3}+\frac{L_0}{8}\right)+\frac{m^2}{k^2}\right)+\frac{i  S}{6 \pi ^2} \left(\frac{1}{4}\frac{1}{k}-\frac{ m^2}{k^3}\right)\right) }
\al{\label{eq:snc:0:2:4:}\tilde{T}_{0,2;4\text{D}}^{\text{s,nt}} & = \eta_{\nu\nu} \left(-\frac{i  L_1}{8 \pi ^2} m^2\right) }
Scalars, spin 0 x 2, dimension 5:
\al{\label{eq:sc:0:2:5:}\tilde{T}_{0,2;5\text{D}}^{\text{s,t}} & = k^2 \pi _{\nu\nu} \left(\frac{i }{32 \pi ^2} \left(\frac{1}{4} m+\frac{m^3}{k^2}\right)+\frac{ T}{32 \pi } \left(\frac{1}{16} k-\frac{1}{2}\frac{ m^2}{k}+\frac{m^4}{k^3}\right)\right) }
\al{\label{eq:snc:0:2:5:}\tilde{T}_{0,2;5\text{D}}^{\text{s,nt}} & = \eta_{\nu\nu} \left(-\frac{i }{12 \pi ^2} m^3\right) }
Scalars, spin 0 x 2, dimension 6:
\al{\label{eq:sc:0:2:6:}\tilde{T}_{0,2;6\text{D}}^{\text{s,t}} & = k^2 \pi _{\nu\nu} \left(\frac{i }{12 \pi ^3} \left( \left(-\frac{23}{1200}+\frac{L_0}{160}\right) k^2+ \left(\frac{43}{240}-\frac{L_0}{16}\right) m^2-\frac{1}{5}\frac{ m^4}{k^2}\right)+
\right.\0\\ & \quad \quad \left.
 + \frac{i  S}{60 \pi ^3} \left(\frac{1}{16} k-\frac{1}{2}\frac{ m^2}{k}+\frac{m^4}{k^3}\right)\right) }
\al{\label{eq:snc:0:2:6:}\tilde{T}_{0,2;6\text{D}}^{\text{s,nt}} & = \eta_{\nu\nu} \left(\frac{i  L_2}{64 \pi ^3} m^4\right) }
Scalars, spin 0 x 4, dimension 3:
\al{\label{eq:sc:0:4:3:}\tilde{T}_{0,4;3\text{D}}^{\text{s,t}} & = k^4 \pi _{\nu\nu}^2 \left(\frac{i }{4 \pi } \left(\frac{5 }{4}\frac{ m}{k^2}-3 \frac{ m^3}{k^4}\right)+ T \left(-\frac{3 }{64}\frac{1}{k}+\frac{3 }{8}\frac{ m^2}{k^3}-\frac{3 }{4}\frac{ m^4}{k^5}\right)\right) }
\al{\label{eq:snc:0:4:3:}\tilde{T}_{0,4;3\text{D}}^{\text{s,nt}} & = k_{\nu }^2 \eta_{\nu\nu} \left(\frac{i }{\pi } m\right)+\eta_{\nu\nu}^2 \left(\frac{i }{2 \pi } \left(- k^2 m+4  m^3\right)\right) }
Scalars, spin 0 x 4, dimension 4:
\al{\label{eq:sc:0:4:4:}\tilde{T}_{0,4;4\text{D}}^{\text{s,t}} & = k^4 \pi _{\nu\nu}^2 \left(\frac{i }{5 \pi ^2} \left( \left(\frac{23}{120}-\frac{L_0}{16}\right)-\frac{7 }{6}\frac{ m^2}{k^2}+2 \frac{ m^4}{k^4}\right)+\frac{i  S}{5 \pi ^2} \left(-\frac{1}{8}\frac{1}{k}+\frac{m^2}{k^3}-2 \frac{ m^4}{k^5}\right)\right) }
\al{\label{eq:snc:0:4:4:}\tilde{T}_{0,4;4\text{D}}^{\text{s,nt}} & = k_{\nu }^2 \eta_{\nu\nu} \left(-\frac{i  L_1}{4 \pi ^2} m^2\right)+\eta_{\nu\nu}^2 \left(\frac{i }{8 \pi ^2} \left( L_1 k^2 m^2-3  L_2 m^4\right)\right) }
Scalars, spin 0 x 4, dimension 5:
\al{\label{eq:sc:0:4:5:}\tilde{T}_{0,4;5\text{D}}^{\text{s,t}} & = k^4 \pi _{\nu\nu}^2 \left(\frac{i }{8 \pi ^2} \left(-\frac{1}{32} m-\frac{1}{3}\frac{ m^3}{k^2}+\frac{1}{2}\frac{ m^5}{k^4}\right)+
\right.\0\\ & \quad \quad \left.
 + \frac{ T}{16 \pi } \left(-\frac{1}{64} k+\frac{3 }{16}\frac{ m^2}{k}-\frac{3 }{4}\frac{ m^4}{k^3}+\frac{m^6}{k^5}\right)\right) }
\al{\label{eq:snc:0:4:5:}\tilde{T}_{0,4;5\text{D}}^{\text{s,nt}} & = k_{\nu }^2 \eta_{\nu\nu} \left(-\frac{i }{6 \pi ^2} m^3\right)+\eta_{\nu\nu}^2 \left(\frac{i }{\pi ^2} \left(\frac{1}{12} k^2 m^3-\frac{1}{5} m^5\right)\right) }
Scalars, spin 0 x 4, dimension 6:
\al{\label{eq:sc:0:4:6:}\tilde{T}_{0,4;6\text{D}}^{\text{s,t}} & = k^4 \pi _{\nu\nu}^2 \left(\frac{i }{\pi ^3} \left( \left(\frac{11}{14700}-\frac{L_0}{4480}\right) k^2+ \left(-\frac{337}{33600}+\frac{L_0}{320}\right) m^2+\frac{1}{42}\frac{ m^4}{k^2}-
\right.\right.\0\\ & \quad \quad \quad \left.\left.
 - \frac{1}{35}\frac{ m^6}{k^4}\right)+\frac{i  S}{35 \pi ^3} \left(-\frac{1}{64} k+\frac{3 }{16}\frac{ m^2}{k}-\frac{3 }{4}\frac{ m^4}{k^3}+\frac{m^6}{k^5}\right)\right) }
\al{\label{eq:snc:0:4:6:}\tilde{T}_{0,4;6\text{D}}^{\text{s,nt}} & = k_{\nu }^2 \eta_{\nu\nu} \left(\frac{i  L_2}{32 \pi ^3} m^4\right)+\eta_{\nu\nu}^2 \left(\frac{i }{32 \pi ^3} \left(-\frac{ L_2}{2} k^2 m^4+ L_3 m^6\right)\right) }
Scalars, spin 1 x 1, dimension 3:
\al{\label{eq:sc:1:1:3:}\tilde{T}_{1,1;3\text{D}}^{\text{s,t}} & = k^2 \pi _{\mu\nu} \left(-\frac{i }{4 \pi }\frac{ m}{k^2}+ T \left(\frac{1}{16}\frac{1}{k}-\frac{1}{4}\frac{ m^2}{k^3}\right)\right) }
\al{\label{eq:snc:1:1:3:}\tilde{T}_{1,1;3\text{D}}^{\text{s,nt}} & = \eta_{\mu\nu} \left(\frac{i }{2 \pi } m\right) }
Scalars, spin 1 x 1, dimension 4:
\al{\label{eq:sc:1:1:4:}\tilde{T}_{1,1;4\text{D}}^{\text{s,t}} & = k^2 \pi _{\mu\nu} \left(\frac{i }{6 \pi ^2} \left( \left(-\frac{1}{3}+\frac{L_0}{8}\right)+\frac{m^2}{k^2}\right)+\frac{i  S}{6 \pi ^2} \left(\frac{1}{4}\frac{1}{k}-\frac{ m^2}{k^3}\right)\right) }
\al{\label{eq:snc:1:1:4:}\tilde{T}_{1,1;4\text{D}}^{\text{s,nt}} & = \eta_{\mu\nu} \left(-\frac{i  L_1}{8 \pi ^2} m^2\right) }
Scalars, spin 1 x 1, dimension 5:
\al{\label{eq:sc:1:1:5:}\tilde{T}_{1,1;5\text{D}}^{\text{s,t}} & = k^2 \pi _{\mu\nu} \left(\frac{i }{32 \pi ^2} \left(\frac{1}{4} m+\frac{m^3}{k^2}\right)+\frac{ T}{32 \pi } \left(\frac{1}{16} k-\frac{1}{2}\frac{ m^2}{k}+\frac{m^4}{k^3}\right)\right) }
\al{\label{eq:snc:1:1:5:}\tilde{T}_{1,1;5\text{D}}^{\text{s,nt}} & = \eta_{\mu\nu} \left(-\frac{i }{12 \pi ^2} m^3\right) }
Scalars, spin 1 x 1, dimension 6:
\al{\label{eq:sc:1:1:6:}\tilde{T}_{1,1;6\text{D}}^{\text{s,t}} & = k^2 \pi _{\mu\nu} \left(\frac{i }{12 \pi ^3} \left( \left(-\frac{23}{1200}+\frac{L_0}{160}\right) k^2+ \left(\frac{43}{240}-\frac{L_0}{16}\right) m^2-\frac{1}{5}\frac{ m^4}{k^2}\right)+
\right.\0\\ & \quad \quad \left.
 + \frac{i  S}{60 \pi ^3} \left(\frac{1}{16} k-\frac{1}{2}\frac{ m^2}{k}+\frac{m^4}{k^3}\right)\right) }
\al{\label{eq:snc:1:1:6:}\tilde{T}_{1,1;6\text{D}}^{\text{s,nt}} & = \eta_{\mu\nu} \left(\frac{i  L_2}{64 \pi ^3} m^4\right) }
Scalars, spin 1 x 3, dimension 3:
\al{\label{eq:sc:1:3:3:}\tilde{T}_{1,3;3\text{D}}^{\text{s,t}} & = k^4 \pi _{\nu\nu} \pi _{\mu\nu} \left(\frac{i }{4 \pi } \left(\frac{5 }{4}\frac{ m}{k^2}-3 \frac{ m^3}{k^4}\right)+ T \left(-\frac{3 }{64}\frac{1}{k}+\frac{3 }{8}\frac{ m^2}{k^3}-\frac{3 }{4}\frac{ m^4}{k^5}\right)\right) }
\al{\label{eq:snc:1:3:3:}\tilde{T}_{1,3;3\text{D}}^{\text{s,nt}} & =  \left(k_{\nu }^2 \eta_{\mu\nu}+k_{\mu } k_{\nu } \eta_{\nu\nu}\right) \left(\frac{i }{2 \pi } m\right)+\eta_{\mu\nu} \eta_{\nu\nu} \left(\frac{i }{2 \pi } \left(- k^2 m+4  m^3\right)\right) }
Scalars, spin 1 x 3, dimension 4:
\al{\label{eq:sc:1:3:4:}\tilde{T}_{1,3;4\text{D}}^{\text{s,t}} & = k^4 \pi _{\nu\nu} \pi _{\mu\nu} \left(\frac{i }{5 \pi ^2} \left( \left(\frac{23}{120}-\frac{L_0}{16}\right)-\frac{7 }{6}\frac{ m^2}{k^2}+2 \frac{ m^4}{k^4}\right)+\frac{i  S}{5 \pi ^2} \left(-\frac{1}{8}\frac{1}{k}+\frac{m^2}{k^3}-2 \frac{ m^4}{k^5}\right)\right) }
\al{\label{eq:snc:1:3:4:}\tilde{T}_{1,3;4\text{D}}^{\text{s,nt}} & =  \left(k_{\nu }^2 \eta_{\mu\nu}+k_{\mu } k_{\nu } \eta_{\nu\nu}\right) \left(-\frac{i  L_1}{8 \pi ^2} m^2\right)+\eta_{\mu\nu} \eta_{\nu\nu} \left(\frac{i }{8 \pi ^2} \left( L_1 k^2 m^2-3  L_2 m^4\right)\right) }
Scalars, spin 1 x 3, dimension 5:
\al{\label{eq:sc:1:3:5:}\tilde{T}_{1,3;5\text{D}}^{\text{s,t}} & = k^4 \pi _{\nu\nu} \pi _{\mu\nu} \left(\frac{i }{8 \pi ^2} \left(-\frac{1}{32} m-\frac{1}{3}\frac{ m^3}{k^2}+\frac{1}{2}\frac{ m^5}{k^4}\right)+
\right.\0\\ & \quad \quad \left.
 + \frac{ T}{16 \pi } \left(-\frac{1}{64} k+\frac{3 }{16}\frac{ m^2}{k}-\frac{3 }{4}\frac{ m^4}{k^3}+\frac{m^6}{k^5}\right)\right) }
\al{\label{eq:snc:1:3:5:}\tilde{T}_{1,3;5\text{D}}^{\text{s,nt}} & =  \left(k_{\nu }^2 \eta_{\mu\nu}+k_{\mu } k_{\nu } \eta_{\nu\nu}\right) \left(-\frac{i }{12 \pi ^2} m^3\right)+\eta_{\mu\nu} \eta_{\nu\nu} \left(\frac{i }{\pi ^2} \left(\frac{1}{12} k^2 m^3-\frac{1}{5} m^5\right)\right) }
Scalars, spin 1 x 3, dimension 6:
\al{\label{eq:sc:1:3:6:}\tilde{T}_{1,3;6\text{D}}^{\text{s,t}} & = k^4 \pi _{\nu\nu} \pi _{\mu\nu} \left(\frac{i }{\pi ^3} \left( \left(\frac{11}{14700}-\frac{L_0}{4480}\right) k^2+ \left(-\frac{337}{33600}+\frac{L_0}{320}\right) m^2+\frac{1}{42}\frac{ m^4}{k^2}-
\right.\right.\0\\ & \quad \quad \quad \left.\left.
 - \frac{1}{35}\frac{ m^6}{k^4}\right)+\frac{i  S}{35 \pi ^3} \left(-\frac{1}{64} k+\frac{3 }{16}\frac{ m^2}{k}-\frac{3 }{4}\frac{ m^4}{k^3}+\frac{m^6}{k^5}\right)\right) }
\al{\label{eq:snc:1:3:6:}\tilde{T}_{1,3;6\text{D}}^{\text{s,nt}} & =  \left(k_{\nu }^2 \eta_{\mu\nu}+k_{\mu } k_{\nu } \eta_{\nu\nu}\right) \left(\frac{i  L_2}{64 \pi ^3} m^4\right)+\eta_{\mu\nu} \eta_{\nu\nu} \left(\frac{i }{32 \pi ^3} \left(-\frac{ L_2}{2} k^2 m^4+ L_3 m^6\right)\right) }
Scalars, spin 1 x 5, dimension 3:
\al{\label{eq:sc:1:5:3:}\tilde{T}_{1,5;3\text{D}}^{\text{s,t}} & = k^6 \pi _{\nu\nu}^2 \pi _{\mu\nu} \left(\frac{i }{\pi } \left(-\frac{11 }{32}\frac{ m}{k^2}+\frac{5 }{3}\frac{ m^3}{k^4}-\frac{5 }{2}\frac{ m^5}{k^6}\right)+
\right.\0\\ & \quad \quad \left.
 +  T \left(\frac{5 }{128}\frac{1}{k}-\frac{15 }{32}\frac{ m^2}{k^3}+\frac{15 }{8}\frac{ m^4}{k^5}-\frac{5 }{2}\frac{ m^6}{k^7}\right)\right) }
\al{\label{eq:snc:1:5:3:}\tilde{T}_{1,5;3\text{D}}^{\text{s,nt}} & = k_{\mu } k_{\nu }^3 \eta_{\nu\nu} \left(\frac{i }{\pi } m\right)+k_{\nu }^4 \eta_{\mu\nu} \left(\frac{i }{2 \pi } m\right)+k_{\mu } k_{\nu } \eta_{\nu\nu}^2 \left(\frac{i }{\pi } \left(-\frac{1}{2} k^2 m+\frac{10 }{3} m^3\right)\right)+
\0\\ & \quad 
 + k_{\nu }^2 \eta_{\mu\nu} \eta_{\nu\nu} \left(\frac{i }{3 \pi } \left(-3  k^2 m+20  m^3\right)\right)+
\0\\ & \quad 
 + \eta_{\mu\nu} \eta_{\nu\nu}^2 \left(\frac{i }{\pi } \left(\frac{1}{2} k^4 m-\frac{10 }{3} k^2 m^3+8  m^5\right)\right) }
Scalars, spin 1 x 5, dimension 4:
\al{\label{eq:sc:1:5:4:}\tilde{T}_{1,5;4\text{D}}^{\text{s,t}} & = k^6 \pi _{\nu\nu}^2 \pi _{\mu\nu} \left(\frac{i }{7 \pi ^2} \left( \left(-\frac{22}{105}+\frac{L_0}{16}\right)+\frac{29 }{15}\frac{ m^2}{k^2}-\frac{20 }{3}\frac{ m^4}{k^4}+8 \frac{ m^6}{k^6}\right)+
\right.\0\\ & \quad \quad \left.
 + \frac{i  S}{7 \pi ^2} \left(\frac{1}{8}\frac{1}{k}-\frac{3 }{2}\frac{ m^2}{k^3}+6 \frac{ m^4}{k^5}-8 \frac{ m^6}{k^7}\right)\right) }
\al{\label{eq:snc:1:5:4:}\tilde{T}_{1,5;4\text{D}}^{\text{s,nt}} & = k_{\mu } k_{\nu }^3 \eta_{\nu\nu} \left(-\frac{i  L_1}{4 \pi ^2} m^2\right)+k_{\nu }^4 \eta_{\mu\nu} \left(-\frac{i  L_1}{8 \pi ^2} m^2\right)+
\0\\ & \quad 
 + k_{\mu } k_{\nu } \eta_{\nu\nu}^2 \left(\frac{i }{8 \pi ^2} \left( L_1 k^2 m^2-5  L_2 m^4\right)\right)+
\0\\ & \quad 
 + k_{\nu }^2 \eta_{\mu\nu} \eta_{\nu\nu} \left(\frac{i }{4 \pi ^2} \left( L_1 k^2 m^2-5  L_2 m^4\right)\right)+
\0\\ & \quad 
 + \eta_{\mu\nu} \eta_{\nu\nu}^2 \left(\frac{i }{4 \pi ^2} \left(-\frac{ L_1}{2} k^4 m^2+\frac{5  L_2}{2} k^2 m^4-5  L_3 m^6\right)\right) }
Scalars, spin 1 x 5, dimension 5:
\al{\label{eq:sc:1:5:5:}\tilde{T}_{1,5;5\text{D}}^{\text{s,t}} & = k^6 \pi _{\nu\nu}^2 \pi _{\mu\nu} \left(\frac{i }{32 \pi ^2} \left(\frac{5 }{64} m+\frac{73 }{48}\frac{ m^3}{k^2}-\frac{55 }{12}\frac{ m^5}{k^4}+5 \frac{ m^7}{k^6}\right)+
\right.\0\\ & \quad \quad \left.
 + \frac{ T}{32 \pi } \left(\frac{5 }{256} k-\frac{5 }{16}\frac{ m^2}{k}+\frac{15 }{8}\frac{ m^4}{k^3}-5 \frac{ m^6}{k^5}+5 \frac{ m^8}{k^7}\right)\right) }
\al{\label{eq:snc:1:5:5:}\tilde{T}_{1,5;5\text{D}}^{\text{s,nt}} & = k_{\mu } k_{\nu }^3 \eta_{\nu\nu} \left(-\frac{i }{6 \pi ^2} m^3\right)+k_{\nu }^4 \eta_{\mu\nu} \left(-\frac{i }{12 \pi ^2} m^3\right)+
\0\\ & \quad 
 + k_{\mu } k_{\nu } \eta_{\nu\nu}^2 \left(\frac{i }{3 \pi ^2} \left(\frac{1}{4} k^2 m^3- m^5\right)\right)+k_{\nu }^2 \eta_{\mu\nu} \eta_{\nu\nu} \left(\frac{i }{3 \pi ^2} \left(\frac{1}{2} k^2 m^3-2  m^5\right)\right)+
\0\\ & \quad 
 + \eta_{\mu\nu} \eta_{\nu\nu}^2 \left(\frac{i }{\pi ^2} \left(-\frac{1}{12} k^4 m^3+\frac{1}{3} k^2 m^5-\frac{4 }{7} m^7\right)\right) }
Scalars, spin 1 x 5, dimension 6:
\al{\label{eq:sc:1:5:6:}\tilde{T}_{1,5;6\text{D}}^{\text{s,t}} & = k^6 \pi _{\nu\nu}^2 \pi _{\mu\nu} \left(\frac{i }{7 \pi ^3} \left( \left(-\frac{563}{181440}+\frac{L_0}{1152}\right) k^2+ \left(\frac{1091}{20160}-\frac{L_0}{64}\right) m^2-\frac{1}{5}\frac{ m^4}{k^2}+
\right.\right.\0\\ & \quad \quad \quad \left.\left.
 + \frac{13 }{27}\frac{ m^6}{k^4}-\frac{4 }{9}\frac{ m^8}{k^6}\right)+\frac{i  S}{21 \pi ^3} \left(\frac{1}{192} k-\frac{1}{12}\frac{ m^2}{k}+\frac{1}{2}\frac{ m^4}{k^3}-\frac{4 }{3}\frac{ m^6}{k^5}+\frac{4 }{3}\frac{ m^8}{k^7}\right)\right) }
\al{\label{eq:snc:1:5:6:}\tilde{T}_{1,5;6\text{D}}^{\text{s,nt}} & = k_{\mu } k_{\nu }^3 \eta_{\nu\nu} \left(\frac{i  L_2}{32 \pi ^3} m^4\right)+k_{\nu }^4 \eta_{\mu\nu} \left(\frac{i  L_2}{64 \pi ^3} m^4\right)+
\0\\ & \quad 
 + k_{\mu } k_{\nu } \eta_{\nu\nu}^2 \left(\frac{i }{32 \pi ^3} \left(-\frac{ L_2}{2} k^2 m^4+\frac{5  L_3}{3} m^6\right)\right)+
\0\\ & \quad 
 + k_{\nu }^2 \eta_{\mu\nu} \eta_{\nu\nu} \left(\frac{i }{16 \pi ^3} \left(-\frac{ L_2}{2} k^2 m^4+\frac{5  L_3}{3} m^6\right)\right)+
\0\\ & \quad 
 + \eta_{\mu\nu} \eta_{\nu\nu}^2 \left(\frac{i }{32 \pi ^3} \left(\frac{ L_2}{2} k^4 m^4-\frac{5  L_3}{3} k^2 m^6+\frac{5  L_4}{2} m^8\right)\right) }
Scalars, spin 2 x 2, dimension 3:
\al{\label{eq:sc:2:2:3:}\tilde{T}_{2,2;3\text{D}}^{\text{s,t}} & = k^4 \pi _{\mu\nu}^2 \left(\frac{i }{2 \pi } \left(\frac{3 }{4}\frac{ m}{k^2}-\frac{ m^3}{k^4}\right)+ T \left(-\frac{1}{32}\frac{1}{k}+\frac{1}{4}\frac{ m^2}{k^3}-\frac{1}{2}\frac{ m^4}{k^5}\right)\right)+
\0\\ & \quad 
 + k^4 \pi _{\mu\mu} \pi _{\nu\nu} \left(\frac{i }{4 \pi } \left(-\frac{1}{4}\frac{ m}{k^2}-\frac{ m^3}{k^4}\right)+ T \left(-\frac{1}{64}\frac{1}{k}+\frac{1}{8}\frac{ m^2}{k^3}-\frac{1}{4}\frac{ m^4}{k^5}\right)\right) }
\al{\label{eq:snc:2:2:3:}\tilde{T}_{2,2;3\text{D}}^{\text{s,nt}} & = k_{\mu } k_{\nu } \eta_{\mu\nu} \left(\frac{i }{\pi } m\right)+\eta_{\mu\mu} \eta_{\nu\nu} \left(\frac{2 i }{3 \pi } m^3\right)+\eta_{\mu\nu}^2 \left(\frac{i }{\pi } \left(-\frac{1}{2} k^2 m+\frac{4 }{3} m^3\right)\right) }
Scalars, spin 2 x 2, dimension 4:
\al{\label{eq:sc:2:2:4:}\tilde{T}_{2,2;4\text{D}}^{\text{s,t}} & = k^4 \pi _{\mu\nu}^2 \left(\frac{i }{3 \pi ^2} \left( \left(\frac{23}{300}-\frac{L_0}{40}\right)- \left(\frac{41}{120}+\frac{L_0}{8}\right)\frac{ m^2}{k^2}+\frac{4 }{5}\frac{ m^4}{k^4}\right)+
\right.\0\\ & \quad \quad \left.
 + \frac{i  S}{15 \pi ^2} \left(-\frac{1}{4}\frac{1}{k}+2 \frac{ m^2}{k^3}-4 \frac{ m^4}{k^5}\right)\right)+
\0\\ & \quad 
 + k^4 \pi _{\mu\mu} \pi _{\nu\nu} \left(\frac{i }{3 \pi ^2} \left( \left(\frac{23}{600}-\frac{L_0}{80}\right)+ \left(-\frac{43}{120}+\frac{L_0}{8}\right)\frac{ m^2}{k^2}+\frac{2 }{5}\frac{ m^4}{k^4}\right)+
\right.\0\\ & \quad \quad \left.
 + \frac{i  S}{15 \pi ^2} \left(-\frac{1}{8}\frac{1}{k}+\frac{m^2}{k^3}-2 \frac{ m^4}{k^5}\right)\right) }
\al{\label{eq:snc:2:2:4:}\tilde{T}_{2,2;4\text{D}}^{\text{s,nt}} & = k_{\mu } k_{\nu } \eta_{\mu\nu} \left(-\frac{i  L_1}{4 \pi ^2} m^2\right)+\eta_{\mu\mu} \eta_{\nu\nu} \left(-\frac{i  L_2}{8 \pi ^2} m^4\right)+\eta_{\mu\nu}^2 \left(\frac{i }{4 \pi ^2} \left(\frac{ L_1}{2} k^2 m^2- L_2 m^4\right)\right) }
Scalars, spin 2 x 2, dimension 5:
\al{\label{eq:sc:2:2:5:}\tilde{T}_{2,2;5\text{D}}^{\text{s,t}} & = k^4 \pi _{\mu\nu}^2 \left(\frac{i }{6 \pi ^2} \left(-\frac{1}{64} m-\frac{1}{3}\frac{ m^3}{k^2}+\frac{1}{4}\frac{ m^5}{k^4}\right)+
\right.\0\\ & \quad \quad \left.
 + \frac{ T}{8 \pi } \left(-\frac{1}{192} k+\frac{1}{16}\frac{ m^2}{k}-\frac{1}{4}\frac{ m^4}{k^3}+\frac{1}{3}\frac{ m^6}{k^5}\right)\right)+
\0\\ & \quad 
 + k^4 \pi _{\mu\mu} \pi _{\nu\nu} \left(\frac{i }{24 \pi ^2} \left(-\frac{1}{32} m+\frac{1}{3}\frac{ m^3}{k^2}+\frac{1}{2}\frac{ m^5}{k^4}\right)+
\right.\0\\ & \quad \quad \left.
 + \frac{ T}{16 \pi } \left(-\frac{1}{192} k+\frac{1}{16}\frac{ m^2}{k}-\frac{1}{4}\frac{ m^4}{k^3}+\frac{1}{3}\frac{ m^6}{k^5}\right)\right) }
\al{\label{eq:snc:2:2:5:}\tilde{T}_{2,2;5\text{D}}^{\text{s,nt}} & = k_{\mu } k_{\nu } \eta_{\mu\nu} \left(-\frac{i }{6 \pi ^2} m^3\right)+\eta_{\mu\mu} \eta_{\nu\nu} \left(-\frac{i }{15 \pi ^2} m^5\right)+\eta_{\mu\nu}^2 \left(\frac{i }{3 \pi ^2} \left(\frac{1}{4} k^2 m^3-\frac{2 }{5} m^5\right)\right) }
Scalars, spin 2 x 2, dimension 6:
\al{\label{eq:sc:2:2:6:}\tilde{T}_{2,2;6\text{D}}^{\text{s,t}} & = k^4 \pi _{\mu\nu}^2 \left(\frac{i }{3 \pi ^3} \left( \left(\frac{11}{7350}-\frac{L_0}{2240}\right) k^2+ \left(-\frac{337}{16800}+\frac{L_0}{160}\right) m^2+
\right.\right.\0\\ & \quad \quad \quad \left.\left.
 +  \left(\frac{65}{2688}+\frac{L_0}{64}\right)\frac{ m^4}{k^2}-\frac{2 }{35}\frac{ m^6}{k^4}\right)+
\right.\0\\ & \quad \quad \left.
 + \frac{i  S}{35 \pi ^3} \left(-\frac{1}{96} k+\frac{1}{8}\frac{ m^2}{k}-\frac{1}{2}\frac{ m^4}{k^3}+\frac{2 }{3}\frac{ m^6}{k^5}\right)\right)+
\0\\ & \quad 
 + k^4 \pi _{\mu\mu} \pi _{\nu\nu} \left(\frac{i }{3 \pi ^3} \left( \left(\frac{11}{14700}-\frac{L_0}{4480}\right) k^2+ \left(-\frac{337}{33600}+\frac{L_0}{320}\right) m^2+
\right.\right.\0\\ & \quad \quad \quad \left.\left.
 +  \left(\frac{127}{2688}-\frac{L_0}{64}\right)\frac{ m^4}{k^2}-\frac{1}{35}\frac{ m^6}{k^4}\right)+
\right.\0\\ & \quad \quad \left.
 + \frac{i  S}{35 \pi ^3} \left(-\frac{1}{192} k+\frac{1}{16}\frac{ m^2}{k}-\frac{1}{4}\frac{ m^4}{k^3}+\frac{1}{3}\frac{ m^6}{k^5}\right)\right) }
\al{\label{eq:snc:2:2:6:}\tilde{T}_{2,2;6\text{D}}^{\text{s,nt}} & = k_{\mu } k_{\nu } \eta_{\mu\nu} \left(\frac{i  L_2}{32 \pi ^3} m^4\right)+\eta_{\mu\mu} \eta_{\nu\nu} \left(\frac{i  L_3}{96 \pi ^3} m^6\right)+\eta_{\mu\nu}^2 \left(\frac{i }{16 \pi ^3} \left(-\frac{ L_2}{4} k^2 m^4+\frac{ L_3}{3} m^6\right)\right) }
Scalars, spin 2 x 4, dimension 3:
\al{\label{eq:sc:2:4:3:}\tilde{T}_{2,4;3\text{D}}^{\text{s,t}} & = k^6 \pi _{\mu\nu}^2 \pi _{\nu\nu} \left(\frac{i }{\pi } \left(-\frac{3 }{8}\frac{ m}{k^2}+\frac{4 }{3}\frac{ m^3}{k^4}-2 \frac{ m^5}{k^6}\right)+ T \left(\frac{1}{32}\frac{1}{k}-\frac{3 }{8}\frac{ m^2}{k^3}+\frac{3 }{2}\frac{ m^4}{k^5}-2 \frac{ m^6}{k^7}\right)\right)+
\0\\ & \quad 
 + k^6 \pi _{\mu\mu} \pi _{\nu\nu}^2 \left(\frac{i }{\pi } \left(\frac{1}{32}\frac{ m}{k^2}+\frac{1}{3}\frac{ m^3}{k^4}-\frac{1}{2}\frac{ m^5}{k^6}\right)+
\right.\0\\ & \quad \quad \left.
 +  T \left(\frac{1}{128}\frac{1}{k}-\frac{3 }{32}\frac{ m^2}{k^3}+\frac{3 }{8}\frac{ m^4}{k^5}-\frac{1}{2}\frac{ m^6}{k^7}\right)\right) }
\al{\label{eq:snc:2:4:3:}\tilde{T}_{2,4;3\text{D}}^{\text{s,nt}} & = k_{\mu }^2 k_{\nu }^2 \eta_{\nu\nu} \left(\frac{i }{2 \pi } m\right)+k_{\mu } k_{\nu }^3 \eta_{\mu\nu} \left(\frac{i }{\pi } m\right)+k_{\nu }^2 \eta_{\mu\mu} \eta_{\nu\nu} \left(\frac{4 i }{3 \pi } m^3\right)+k_{\mu }^2 \eta_{\nu\nu}^2 \left(\frac{2 i }{3 \pi } m^3\right)+
\0\\ & \quad 
 + k_{\mu } k_{\nu } \eta_{\mu\nu} \eta_{\nu\nu} \left(\frac{i }{3 \pi } \left(-3  k^2 m+16  m^3\right)\right)+k_{\nu }^2 \eta_{\mu\nu}^2 \left(\frac{i }{\pi } \left(-\frac{1}{2} k^2 m+\frac{8 }{3} m^3\right)\right)+
\0\\ & \quad 
 + \eta_{\mu\mu} \eta_{\nu\nu}^2 \left(\frac{i }{\pi } \left(-\frac{2 }{3} k^2 m^3+\frac{8 }{5} m^5\right)\right)+
\0\\ & \quad 
 + \eta_{\mu\nu}^2 \eta_{\nu\nu} \left(\frac{i }{\pi } \left(\frac{1}{2} k^4 m-\frac{8 }{3} k^2 m^3+\frac{32 }{5} m^5\right)\right) }
Scalars, spin 2 x 4, dimension 4:
\al{\label{eq:sc:2:4:4:}\tilde{T}_{2,4;4\text{D}}^{\text{s,t}} & = k^6 \pi _{\mu\nu}^2 \pi _{\nu\nu} \left(\frac{i }{\pi ^2} \left( \left(-\frac{88}{3675}+\frac{L_0}{140}\right)+ \left(\frac{823}{4200}+\frac{L_0}{40}\right)\frac{ m^2}{k^2}-\frac{16 }{21}\frac{ m^4}{k^4}+\frac{32 }{35}\frac{ m^6}{k^6}\right)+
\right.\0\\ & \quad \quad \left.
 + \frac{i  S}{35 \pi ^2} \left(\frac{1}{2}\frac{1}{k}-6 \frac{ m^2}{k^3}+24 \frac{ m^4}{k^5}-32 \frac{ m^6}{k^7}\right)\right)+
\0\\ & \quad 
 + k^6 \pi _{\mu\mu} \pi _{\nu\nu}^2 \left(\frac{i }{\pi ^2} \left( \left(-\frac{22}{3675}+\frac{L_0}{560}\right)+ \left(\frac{337}{4200}-\frac{L_0}{40}\right)\frac{ m^2}{k^2}-\frac{4 }{21}\frac{ m^4}{k^4}+
\right.\right.\0\\ & \quad \quad \quad \left.\left.
 + \frac{8 }{35}\frac{ m^6}{k^6}\right)+\frac{i  S}{35 \pi ^2} \left(\frac{1}{8}\frac{1}{k}-\frac{3 }{2}\frac{ m^2}{k^3}+6 \frac{ m^4}{k^5}-8 \frac{ m^6}{k^7}\right)\right) }
\al{\label{eq:snc:2:4:4:}\tilde{T}_{2,4;4\text{D}}^{\text{s,nt}} & = k_{\mu }^2 k_{\nu }^2 \eta_{\nu\nu} \left(-\frac{i  L_1}{8 \pi ^2} m^2\right)+k_{\mu } k_{\nu }^3 \eta_{\mu\nu} \left(-\frac{i  L_1}{4 \pi ^2} m^2\right)+k_{\nu }^2 \eta_{\mu\mu} \eta_{\nu\nu} \left(-\frac{i  L_2}{4 \pi ^2} m^4\right)+
\0\\ & \quad 
 + k_{\mu }^2 \eta_{\nu\nu}^2 \left(-\frac{i  L_2}{8 \pi ^2} m^4\right)+k_{\mu } k_{\nu } \eta_{\mu\nu} \eta_{\nu\nu} \left(\frac{i }{4 \pi ^2} \left( L_1 k^2 m^2-4  L_2 m^4\right)\right)+
\0\\ & \quad 
 + k_{\nu }^2 \eta_{\mu\nu}^2 \left(\frac{i }{2 \pi ^2} \left(\frac{ L_1}{4} k^2 m^2- L_2 m^4\right)\right)+\eta_{\mu\mu} \eta_{\nu\nu}^2 \left(\frac{i }{4 \pi ^2} \left(\frac{ L_2}{2} k^2 m^4- L_3 m^6\right)\right)+
\0\\ & \quad 
 + \eta_{\mu\nu}^2 \eta_{\nu\nu} \left(\frac{i }{2 \pi ^2} \left(-\frac{ L_1}{4} k^4 m^2+ L_2 k^2 m^4-2  L_3 m^6\right)\right) }
Scalars, spin 2 x 4, dimension 5:
\al{\label{eq:sc:2:4:5:}\tilde{T}_{2,4;5\text{D}}^{\text{s,t}} & = k^6 \pi _{\mu\nu}^2 \pi _{\nu\nu} \left(\frac{i }{8 \pi ^2} \left(\frac{1}{64} m+\frac{7 }{16}\frac{ m^3}{k^2}-\frac{11 }{12}\frac{ m^5}{k^4}+\frac{m^7}{k^6}\right)+
\right.\0\\ & \quad \quad \left.
 + \frac{ T}{8 \pi } \left(\frac{1}{256} k-\frac{1}{16}\frac{ m^2}{k}+\frac{3 }{8}\frac{ m^4}{k^3}-\frac{ m^6}{k^5}+\frac{m^8}{k^7}\right)\right)+
\0\\ & \quad 
 + k^6 \pi _{\mu\mu} \pi _{\nu\nu}^2 \left(\frac{i }{32 \pi ^2} \left(\frac{1}{64} m-\frac{11 }{48}\frac{ m^3}{k^2}-\frac{11 }{12}\frac{ m^5}{k^4}+\frac{m^7}{k^6}\right)+
\right.\0\\ & \quad \quad \left.
 + \frac{ T}{32 \pi } \left(\frac{1}{256} k-\frac{1}{16}\frac{ m^2}{k}+\frac{3 }{8}\frac{ m^4}{k^3}-\frac{ m^6}{k^5}+\frac{m^8}{k^7}\right)\right) }
\al{\label{eq:snc:2:4:5:}\tilde{T}_{2,4;5\text{D}}^{\text{s,nt}} & = k_{\mu }^2 k_{\nu }^2 \eta_{\nu\nu} \left(-\frac{i }{12 \pi ^2} m^3\right)+k_{\mu } k_{\nu }^3 \eta_{\mu\nu} \left(-\frac{i }{6 \pi ^2} m^3\right)+k_{\nu }^2 \eta_{\mu\mu} \eta_{\nu\nu} \left(-\frac{2 i }{15 \pi ^2} m^5\right)+
\0\\ & \quad 
 + k_{\mu }^2 \eta_{\nu\nu}^2 \left(-\frac{i }{15 \pi ^2} m^5\right)+k_{\mu } k_{\nu } \eta_{\mu\nu} \eta_{\nu\nu} \left(\frac{i }{3 \pi ^2} \left(\frac{1}{2} k^2 m^3-\frac{8 }{5} m^5\right)\right)+
\0\\ & \quad 
 + k_{\nu }^2 \eta_{\mu\nu}^2 \left(\frac{i }{3 \pi ^2} \left(\frac{1}{4} k^2 m^3-\frac{4 }{5} m^5\right)\right)+\eta_{\mu\mu} \eta_{\nu\nu}^2 \left(\frac{i }{5 \pi ^2} \left(\frac{1}{3} k^2 m^5-\frac{4 }{7} m^7\right)\right)+
\0\\ & \quad 
 + \eta_{\mu\nu}^2 \eta_{\nu\nu} \left(\frac{i }{\pi ^2} \left(-\frac{1}{12} k^4 m^3+\frac{4 }{15} k^2 m^5-\frac{16 }{35} m^7\right)\right) }
Scalars, spin 2 x 4, dimension 6:
\al{\label{eq:sc:2:4:6:}\tilde{T}_{2,4;6\text{D}}^{\text{s,t}} & = k^6 \pi _{\mu\nu}^2 \pi _{\nu\nu} \left(\frac{i }{5 \pi ^3} \left( \left(-\frac{563}{317520}+\frac{L_0}{2016}\right) k^2+ \left(\frac{1091}{35280}-\frac{L_0}{112}\right) m^2-
\right.\right.\0\\ & \quad \quad \quad \left.\left.
 -  \left(\frac{407}{4480}+\frac{L_0}{64}\right)\frac{ m^4}{k^2}+\frac{52 }{189}\frac{ m^6}{k^4}-\frac{16 }{63}\frac{ m^8}{k^6}\right)+
\right.\0\\ & \quad \quad \left.
 + \frac{i  S}{105 \pi ^3} \left(\frac{1}{48} k-\frac{1}{3}\frac{ m^2}{k}+2 \frac{ m^4}{k^3}-\frac{16 }{3}\frac{ m^6}{k^5}+\frac{16 }{3}\frac{ m^8}{k^7}\right)\right)+
\0\\ & \quad 
 + k^6 \pi _{\mu\mu} \pi _{\nu\nu}^2 \left(\frac{i }{5 \pi ^3} \left( \left(-\frac{563}{1270080}+\frac{L_0}{8064}\right) k^2+ \left(\frac{1091}{141120}-\frac{L_0}{448}\right) m^2+
\right.\right.\0\\ & \quad \quad \quad \left.\left.
 +  \left(-\frac{233}{4480}+\frac{L_0}{64}\right)\frac{ m^4}{k^2}+\frac{13 }{189}\frac{ m^6}{k^4}-\frac{4 }{63}\frac{ m^8}{k^6}\right)+
\right.\0\\ & \quad \quad \left.
 + \frac{i  S}{105 \pi ^3} \left(\frac{1}{192} k-\frac{1}{12}\frac{ m^2}{k}+\frac{1}{2}\frac{ m^4}{k^3}-\frac{4 }{3}\frac{ m^6}{k^5}+\frac{4 }{3}\frac{ m^8}{k^7}\right)\right) }
\al{\label{eq:snc:2:4:6:}\tilde{T}_{2,4;6\text{D}}^{\text{s,nt}} & = k_{\mu }^2 k_{\nu }^2 \eta_{\nu\nu} \left(\frac{i  L_2}{64 \pi ^3} m^4\right)+k_{\mu } k_{\nu }^3 \eta_{\mu\nu} \left(\frac{i  L_2}{32 \pi ^3} m^4\right)+k_{\nu }^2 \eta_{\mu\mu} \eta_{\nu\nu} \left(\frac{i  L_3}{48 \pi ^3} m^6\right)+
\0\\ & \quad 
 + k_{\mu }^2 \eta_{\nu\nu}^2 \left(\frac{i  L_3}{96 \pi ^3} m^6\right)+k_{\mu } k_{\nu } \eta_{\mu\nu} \eta_{\nu\nu} \left(\frac{i }{4 \pi ^3} \left(-\frac{ L_2}{8} k^2 m^4+\frac{ L_3}{3} m^6\right)\right)+
\0\\ & \quad 
 + k_{\nu }^2 \eta_{\mu\nu}^2 \left(\frac{i }{8 \pi ^3} \left(-\frac{ L_2}{8} k^2 m^4+\frac{ L_3}{3} m^6\right)\right)+
\0\\ & \quad 
 + \eta_{\mu\mu} \eta_{\nu\nu}^2 \left(\frac{i }{32 \pi ^3} \left(-\frac{ L_3}{3} k^2 m^6+\frac{ L_4}{2} m^8\right)\right)+
\0\\ & \quad 
 + \eta_{\mu\nu}^2 \eta_{\nu\nu} \left(\frac{i }{8 \pi ^3} \left(\frac{ L_2}{8} k^4 m^4-\frac{ L_3}{3} k^2 m^6+\frac{ L_4}{2} m^8\right)\right) }
Scalars, spin 3 x 3, dimension 3:
\al{\label{eq:sc:3:3:3:}\tilde{T}_{3,3;3\text{D}}^{\text{s,t}} & = k^6 \pi _{\mu\nu}^3 \left(\frac{i }{\pi } \left(-\frac{7 }{16}\frac{ m}{k^2}+\frac{2 }{3}\frac{ m^3}{k^4}-\frac{ m^5}{k^6}\right)+ T \left(\frac{1}{64}\frac{1}{k}-\frac{3 }{16}\frac{ m^2}{k^3}+\frac{3 }{4}\frac{ m^4}{k^5}-\frac{ m^6}{k^7}\right)\right)+
\0\\ & \quad 
 + k^6 \pi _{\mu\mu} \pi _{\mu\nu} \pi _{\nu\nu} \left(\frac{i }{2 \pi } \left(\frac{3 }{16}\frac{ m}{k^2}+2 \frac{ m^3}{k^4}-3 \frac{ m^5}{k^6}\right)+
\right.\0\\ & \quad \quad \left.
 +  T \left(\frac{3 }{128}\frac{1}{k}-\frac{9 }{32}\frac{ m^2}{k^3}+\frac{9 }{8}\frac{ m^4}{k^5}-\frac{3 }{2}\frac{ m^6}{k^7}\right)\right) }
\al{\label{eq:snc:3:3:3:}\tilde{T}_{3,3;3\text{D}}^{\text{s,nt}} & = k_{\mu }^2 k_{\nu }^2 \eta_{\mu\nu} \left(\frac{3 i }{2 \pi } m\right)+ \left(k_{\nu }^2 \eta_{\mu\mu} \eta_{\mu\nu}+k_{\mu } k_{\nu } \eta_{\mu\mu} \eta_{\nu\nu}+k_{\mu }^2 \eta_{\mu\nu} \eta_{\nu\nu}\right) \left(\frac{2 i }{\pi } m^3\right)+
\0\\ & \quad 
 + k_{\mu } k_{\nu } \eta_{\mu\nu}^2 \left(\frac{i }{2 \pi } \left(-3  k^2 m+8  m^3\right)\right)+\eta_{\mu\mu} \eta_{\mu\nu} \eta_{\nu\nu} \left(\frac{i }{5 \pi } \left(-10  k^2 m^3+24  m^5\right)\right)+
\0\\ & \quad 
 + \eta_{\mu\nu}^3 \left(\frac{i }{\pi } \left(\frac{1}{2} k^4 m-\frac{4 }{3} k^2 m^3+\frac{16 }{5} m^5\right)\right) }
Scalars, spin 3 x 3, dimension 4:
\al{\label{eq:sc:3:3:4:}\tilde{T}_{3,3;4\text{D}}^{\text{s,t}} & = k^6 \pi _{\mu\nu}^3 \left(\frac{i }{\pi ^2} \left( \left(-\frac{44}{3675}+\frac{L_0}{280}\right)+ \left(\frac{149}{4200}+\frac{3 L_0}{40}\right)\frac{ m^2}{k^2}-\frac{8 }{21}\frac{ m^4}{k^4}+\frac{16 }{35}\frac{ m^6}{k^6}\right)+
\right.\0\\ & \quad \quad \left.
 + \frac{i  S}{35 \pi ^2} \left(\frac{1}{4}\frac{1}{k}-3 \frac{ m^2}{k^3}+12 \frac{ m^4}{k^5}-16 \frac{ m^6}{k^7}\right)\right)+
\0\\ & \quad 
 + k^6 \pi _{\mu\mu} \pi _{\mu\nu} \pi _{\nu\nu} \left(\frac{i }{\pi ^2} \left( \left(-\frac{22}{1225}+\frac{3 L_0}{560}\right)+ \left(\frac{337}{1400}-\frac{3 L_0}{40}\right)\frac{ m^2}{k^2}-\frac{4 }{7}\frac{ m^4}{k^4}+
\right.\right.\0\\ & \quad \quad \quad \left.\left.
 + \frac{24 }{35}\frac{ m^6}{k^6}\right)+\frac{i  S}{35 \pi ^2} \left(\frac{3 }{8}\frac{1}{k}-\frac{9 }{2}\frac{ m^2}{k^3}+18 \frac{ m^4}{k^5}-24 \frac{ m^6}{k^7}\right)\right) }
\al{\label{eq:snc:3:3:4:}\tilde{T}_{3,3;4\text{D}}^{\text{s,nt}} & = k_{\mu }^2 k_{\nu }^2 \eta_{\mu\nu} \left(-\frac{3 i  L_1}{8 \pi ^2} m^2\right)+ \left(k_{\nu }^2 \eta_{\mu\mu} \eta_{\mu\nu}+k_{\mu } k_{\nu } \eta_{\mu\mu} \eta_{\nu\nu}+k_{\mu }^2 \eta_{\mu\nu} \eta_{\nu\nu}\right) \left(-\frac{3 i  L_2}{8 \pi ^2} m^4\right)+
\0\\ & \quad 
 + k_{\mu } k_{\nu } \eta_{\mu\nu}^2 \left(\frac{i }{4 \pi ^2} \left(\frac{3  L_1}{2} k^2 m^2-3  L_2 m^4\right)\right)+
\0\\ & \quad 
 + \eta_{\mu\mu} \eta_{\mu\nu} \eta_{\nu\nu} \left(\frac{i }{4 \pi ^2} \left(\frac{3  L_2}{2} k^2 m^4-3  L_3 m^6\right)\right)+
\0\\ & \quad 
 + \eta_{\mu\nu}^3 \left(\frac{i }{2 \pi ^2} \left(-\frac{ L_1}{4} k^4 m^2+\frac{ L_2}{2} k^2 m^4- L_3 m^6\right)\right) }
Scalars, spin 3 x 3, dimension 5:
\al{\label{eq:sc:3:3:5:}\tilde{T}_{3,3;5\text{D}}^{\text{s,t}} & = k^6 \pi _{\mu\nu}^3 \left(\frac{i }{16 \pi ^2} \left(\frac{1}{64} m+\frac{53 }{48}\frac{ m^3}{k^2}-\frac{11 }{12}\frac{ m^5}{k^4}+\frac{m^7}{k^6}\right)+
\right.\0\\ & \quad \quad \left.
 + \frac{ T}{16 \pi } \left(\frac{1}{256} k-\frac{1}{16}\frac{ m^2}{k}+\frac{3 }{8}\frac{ m^4}{k^3}-\frac{ m^6}{k^5}+\frac{m^8}{k^7}\right)\right)+
\0\\ & \quad 
 + k^6 \pi _{\mu\mu} \pi _{\mu\nu} \pi _{\nu\nu} \left(\frac{i }{32 \pi ^2} \left(\frac{3 }{64} m-\frac{11 }{16}\frac{ m^3}{k^2}-\frac{11 }{4}\frac{ m^5}{k^4}+3 \frac{ m^7}{k^6}\right)+
\right.\0\\ & \quad \quad \left.
 + \frac{ T}{32 \pi } \left(\frac{3 }{256} k-\frac{3 }{16}\frac{ m^2}{k}+\frac{9 }{8}\frac{ m^4}{k^3}-3 \frac{ m^6}{k^5}+3 \frac{ m^8}{k^7}\right)\right) }
\al{\label{eq:snc:3:3:5:}\tilde{T}_{3,3;5\text{D}}^{\text{s,nt}} & = k_{\mu }^2 k_{\nu }^2 \eta_{\mu\nu} \left(-\frac{i }{4 \pi ^2} m^3\right)+ \left(k_{\nu }^2 \eta_{\mu\mu} \eta_{\mu\nu}+k_{\mu } k_{\nu } \eta_{\mu\mu} \eta_{\nu\nu}+k_{\mu }^2 \eta_{\mu\nu} \eta_{\nu\nu}\right) \left(-\frac{i }{5 \pi ^2} m^5\right)+
\0\\ & \quad 
 + k_{\mu } k_{\nu } \eta_{\mu\nu}^2 \left(\frac{i }{\pi ^2} \left(\frac{1}{4} k^2 m^3-\frac{2 }{5} m^5\right)\right)+\eta_{\mu\mu} \eta_{\mu\nu} \eta_{\nu\nu} \left(\frac{i }{5 \pi ^2} \left(k^2 m^5-\frac{12 }{7} m^7\right)\right)+
\0\\ & \quad 
 + \eta_{\mu\nu}^3 \left(\frac{i }{\pi ^2} \left(-\frac{1}{12} k^4 m^3+\frac{2 }{15} k^2 m^5-\frac{8 }{35} m^7\right)\right) }
Scalars, spin 3 x 3, dimension 6:
\al{\label{eq:sc:3:3:6:}\tilde{T}_{3,3;6\text{D}}^{\text{s,t}} & = k^6 \pi _{\mu\nu}^3 \left(\frac{i }{5 \pi ^3} \left( \left(-\frac{563}{635040}+\frac{L_0}{4032}\right) k^2+ \left(\frac{1091}{70560}-\frac{L_0}{224}\right) m^2+
\right.\right.\0\\ & \quad \quad \quad \left.\left.
 +  \left(\frac{59}{4480}-\frac{3 L_0}{64}\right)\frac{ m^4}{k^2}+\frac{26 }{189}\frac{ m^6}{k^4}-\frac{8 }{63}\frac{ m^8}{k^6}\right)+
\right.\0\\ & \quad \quad \left.
 + \frac{i  S}{105 \pi ^3} \left(\frac{1}{96} k-\frac{1}{6}\frac{ m^2}{k}+\frac{m^4}{k^3}-\frac{8 }{3}\frac{ m^6}{k^5}+\frac{8 }{3}\frac{ m^8}{k^7}\right)\right)+
\0\\ & \quad 
 + k^6 \pi _{\mu\mu} \pi _{\mu\nu} \pi _{\nu\nu} \left(\frac{i }{5 \pi ^3} \left( \left(-\frac{563}{423360}+\frac{L_0}{2688}\right) k^2+ \left(\frac{1091}{47040}-\frac{3 L_0}{448}\right) m^2+
\right.\right.\0\\ & \quad \quad \quad \left.\left.
 +  \left(-\frac{699}{4480}+\frac{3 L_0}{64}\right)\frac{ m^4}{k^2}+\frac{13 }{63}\frac{ m^6}{k^4}-\frac{4 }{21}\frac{ m^8}{k^6}\right)+
\right.\0\\ & \quad \quad \left.
 + \frac{i  S}{35 \pi ^3} \left(\frac{1}{192} k-\frac{1}{12}\frac{ m^2}{k}+\frac{1}{2}\frac{ m^4}{k^3}-\frac{4 }{3}\frac{ m^6}{k^5}+\frac{4 }{3}\frac{ m^8}{k^7}\right)\right) }
\al{\label{eq:snc:3:3:6:}\tilde{T}_{3,3;6\text{D}}^{\text{s,nt}} & = k_{\mu }^2 k_{\nu }^2 \eta_{\mu\nu} \left(\frac{3 i  L_2}{64 \pi ^3} m^4\right)+ \left(k_{\nu }^2 \eta_{\mu\mu} \eta_{\mu\nu}+k_{\mu } k_{\nu } \eta_{\mu\mu} \eta_{\nu\nu}+k_{\mu }^2 \eta_{\mu\nu} \eta_{\nu\nu}\right) \left(\frac{i  L_3}{32 \pi ^3} m^6\right)+
\0\\ & \quad 
 + k_{\mu } k_{\nu } \eta_{\mu\nu}^2 \left(\frac{i }{16 \pi ^3} \left(-\frac{3  L_2}{4} k^2 m^4+ L_3 m^6\right)\right)+
\0\\ & \quad 
 + \eta_{\mu\mu} \eta_{\mu\nu} \eta_{\nu\nu} \left(\frac{i }{32 \pi ^3} \left(- L_3 k^2 m^6+\frac{3  L_4}{2} m^8\right)\right)+
\0\\ & \quad 
 + \eta_{\mu\nu}^3 \left(\frac{i }{16 \pi ^3} \left(\frac{ L_2}{4} k^4 m^4-\frac{ L_3}{3} k^2 m^6+\frac{ L_4}{2} m^8\right)\right) }
Scalars, spin 3 x 5, dimension 3:
\al{\label{eq:sc:3:5:3:}\tilde{T}_{3,5;3\text{D}}^{\text{s,t}} & = k^8 \pi _{\mu\nu}^3 \pi _{\nu\nu} \left(\frac{i }{4 \pi } \left(\frac{27 }{16}\frac{ m}{k^2}-\frac{73 }{12}\frac{ m^3}{k^4}+\frac{55 }{3}\frac{ m^5}{k^6}-20 \frac{ m^7}{k^8}\right)+
\right.\0\\ & \quad \quad \left.
 +  T \left(-\frac{5 }{256}\frac{1}{k}+\frac{5 }{16}\frac{ m^2}{k^3}-\frac{15 }{8}\frac{ m^4}{k^5}+5 \frac{ m^6}{k^7}-5 \frac{ m^8}{k^9}\right)\right)+
\0\\ & \quad 
 + k^8 \pi _{\mu\mu} \pi _{\mu\nu} \pi _{\nu\nu}^2 \left(\frac{i }{4 \pi } \left(-\frac{15 }{64}\frac{ m}{k^2}-\frac{73 }{16}\frac{ m^3}{k^4}+\frac{55 }{4}\frac{ m^5}{k^6}-15 \frac{ m^7}{k^8}\right)+
\right.\0\\ & \quad \quad \left.
 +  T \left(-\frac{15 }{1024}\frac{1}{k}+\frac{15 }{64}\frac{ m^2}{k^3}-\frac{45 }{32}\frac{ m^4}{k^5}+\frac{15 }{4}\frac{ m^6}{k^7}-\frac{15 }{4}\frac{ m^8}{k^9}\right)\right) }
\al{\label{eq:snc:3:5:3:}\tilde{T}_{3,5;3\text{D}}^{\text{s,nt}} & = k_{\mu }^3 k_{\nu }^3 \eta_{\nu\nu} \left(\frac{i }{2 \pi } m\right)+k_{\mu }^2 k_{\nu }^4 \eta_{\mu\nu} \left(\frac{3 i }{2 \pi } m\right)+k_{\mu } k_{\nu }^3 \eta_{\mu\mu} \eta_{\nu\nu} \left(\frac{4 i }{\pi } m^3\right)+
\0\\ & \quad 
 +  \left(k_{\nu }^4 \eta_{\mu\mu} \eta_{\mu\nu}+k_{\mu }^3 k_{\nu } \eta_{\nu\nu}^2\right) \left(\frac{2 i }{\pi } m^3\right)+k_{\mu }^2 k_{\nu }^2 \eta_{\mu\nu} \eta_{\nu\nu} \left(\frac{i }{2 \pi } \left(-3  k^2 m+24  m^3\right)\right)+
\0\\ & \quad 
 + k_{\mu } k_{\nu }^3 \eta_{\mu\nu}^2 \left(\frac{i }{2 \pi } \left(-3  k^2 m+16  m^3\right)\right)+
\0\\ & \quad 
 +  \left(k_{\mu } k_{\nu } \eta_{\mu\mu} \eta_{\nu\nu}^2+k_{\mu }^2 \eta_{\mu\nu} \eta_{\nu\nu}^2\right) \left(\frac{i }{\pi } \left(-2  k^2 m^3+8  m^5\right)\right)+
\0\\ & \quad 
 + k_{\nu }^2 \eta_{\mu\mu} \eta_{\mu\nu} \eta_{\nu\nu} \left(\frac{i }{\pi } \left(-4  k^2 m^3+16  m^5\right)\right)+
\0\\ & \quad 
 + k_{\mu } k_{\nu } \eta_{\mu\nu}^2 \eta_{\nu\nu} \left(\frac{i }{2 \pi } \left(3  k^4 m-16  k^2 m^3+64  m^5\right)\right)+
\0\\ & \quad 
 + k_{\nu }^2 \eta_{\mu\nu}^3 \left(\frac{i }{\pi } \left(\frac{1}{2} k^4 m-\frac{8 }{3} k^2 m^3+\frac{32 }{3} m^5\right)\right)+
\0\\ & \quad 
 + \eta_{\mu\mu} \eta_{\mu\nu} \eta_{\nu\nu}^2 \left(\frac{i }{7 \pi } \left(14  k^4 m^3-56  k^2 m^5+96  m^7\right)\right)+
\0\\ & \quad 
 + \eta_{\mu\nu}^3 \eta_{\nu\nu} \left(\frac{i }{\pi } \left(-\frac{1}{2} k^6 m+\frac{8 }{3} k^4 m^3-\frac{32 }{3} k^2 m^5+\frac{128 }{7} m^7\right)\right) }
Scalars, spin 3 x 5, dimension 4:
\al{\label{eq:sc:3:5:4:}\tilde{T}_{3,5;4\text{D}}^{\text{s,t}} & = k^8 \pi _{\mu\nu}^3 \pi _{\nu\nu} \left(\frac{i }{7 \pi ^2} \left( \left(\frac{563}{5670}-\frac{L_0}{36}\right)- \left(\frac{2159}{2520}+\frac{3 L_0}{8}\right)\frac{ m^2}{k^2}+\frac{32 }{5}\frac{ m^4}{k^4}-\frac{416 }{27}\frac{ m^6}{k^6}+
\right.\right.\0\\ & \quad \quad \quad \left.\left.
 + \frac{128 }{9}\frac{ m^8}{k^8}\right)+\frac{i  S}{21 \pi ^2} \left(-\frac{1}{6}\frac{1}{k}+\frac{8 }{3}\frac{ m^2}{k^3}-16 \frac{ m^4}{k^5}+\frac{128 }{3}\frac{ m^6}{k^7}-\frac{128 }{3}\frac{ m^8}{k^9}\right)\right)+
\0\\ & \quad 
 + k^8 \pi _{\mu\mu} \pi _{\mu\nu} \pi _{\nu\nu}^2 \left(\frac{i }{7 \pi ^2} \left( \left(\frac{563}{7560}-\frac{L_0}{48}\right)+ \left(-\frac{1091}{840}+\frac{3 L_0}{8}\right)\frac{ m^2}{k^2}+\frac{24 }{5}\frac{ m^4}{k^4}-
\right.\right.\0\\ & \quad \quad \quad \left.\left.
 - \frac{104 }{9}\frac{ m^6}{k^6}+\frac{32 }{3}\frac{ m^8}{k^8}\right)+
\right.\0\\ & \quad \quad \left.
 + \frac{i  S}{7 \pi ^2} \left(-\frac{1}{24}\frac{1}{k}+\frac{2 }{3}\frac{ m^2}{k^3}-4 \frac{ m^4}{k^5}+\frac{32 }{3}\frac{ m^6}{k^7}-\frac{32 }{3}\frac{ m^8}{k^9}\right)\right) }
\al{\label{eq:snc:3:5:4:}\tilde{T}_{3,5;4\text{D}}^{\text{s,nt}} & = k_{\mu }^3 k_{\nu }^3 \eta_{\nu\nu} \left(-\frac{i  L_1}{8 \pi ^2} m^2\right)+k_{\mu }^2 k_{\nu }^4 \eta_{\mu\nu} \left(-\frac{3 i  L_1}{8 \pi ^2} m^2\right)+k_{\mu } k_{\nu }^3 \eta_{\mu\mu} \eta_{\nu\nu} \left(-\frac{3 i  L_2}{4 \pi ^2} m^4\right)+
\0\\ & \quad 
 +  \left(k_{\nu }^4 \eta_{\mu\mu} \eta_{\mu\nu}+k_{\mu }^3 k_{\nu } \eta_{\nu\nu}^2\right) \left(-\frac{3 i  L_2}{8 \pi ^2} m^4\right)+
\0\\ & \quad 
 + k_{\mu }^2 k_{\nu }^2 \eta_{\mu\nu} \eta_{\nu\nu} \left(\frac{i }{4 \pi ^2} \left(\frac{3  L_1}{2} k^2 m^2-9  L_2 m^4\right)\right)+
\0\\ & \quad 
 + k_{\mu } k_{\nu }^3 \eta_{\mu\nu}^2 \left(\frac{i }{2 \pi ^2} \left(\frac{3  L_1}{4} k^2 m^2-3  L_2 m^4\right)\right)+
\0\\ & \quad 
 +  \left(k_{\mu } k_{\nu } \eta_{\mu\mu} \eta_{\nu\nu}^2+k_{\mu }^2 \eta_{\mu\nu} \eta_{\nu\nu}^2\right) \left(\frac{i }{4 \pi ^2} \left(\frac{3  L_2}{2} k^2 m^4-5  L_3 m^6\right)\right)+
\0\\ & \quad 
 + k_{\nu }^2 \eta_{\mu\mu} \eta_{\mu\nu} \eta_{\nu\nu} \left(\frac{i }{2 \pi ^2} \left(\frac{3  L_2}{2} k^2 m^4-5  L_3 m^6\right)\right)+
\0\\ & \quad 
 + k_{\mu } k_{\nu } \eta_{\mu\nu}^2 \eta_{\nu\nu} \left(\frac{i }{2 \pi ^2} \left(-\frac{3  L_1}{4} k^4 m^2+3  L_2 k^2 m^4-10  L_3 m^6\right)\right)+
\0\\ & \quad 
 + k_{\nu }^2 \eta_{\mu\nu}^3 \left(\frac{i }{\pi ^2} \left(-\frac{ L_1}{8} k^4 m^2+\frac{ L_2}{2} k^2 m^4-\frac{5  L_3}{3} m^6\right)\right)+
\0\\ & \quad 
 + \eta_{\mu\mu} \eta_{\mu\nu} \eta_{\nu\nu}^2 \left(\frac{i }{4 \pi ^2} \left(-\frac{3  L_2}{2} k^4 m^4+5  L_3 k^2 m^6-\frac{15  L_4}{2} m^8\right)\right)+
\0\\ & \quad 
 + \eta_{\mu\nu}^3 \eta_{\nu\nu} \left(\frac{i }{\pi ^2} \left(\frac{ L_1}{8} k^6 m^2-\frac{ L_2}{2} k^4 m^4+\frac{5  L_3}{3} k^2 m^6-\frac{5  L_4}{2} m^8\right)\right) }
Scalars, spin 3 x 5, dimension 5:
\al{\label{eq:sc:3:5:5:}\tilde{T}_{3,5;5\text{D}}^{\text{s,t}} & = k^8 \pi _{\mu\nu}^3 \pi _{\nu\nu} \left(\frac{i }{\pi ^2} \left(-\frac{1}{1024} m-\frac{25 }{384}\frac{ m^3}{k^2}+\frac{2 }{15}\frac{ m^5}{k^4}-\frac{7 }{24}\frac{ m^7}{k^6}+\frac{1}{4}\frac{ m^9}{k^8}\right)+
\right.\0\\ & \quad \quad \left.
 + \frac{ T}{4 \pi } \left(-\frac{1}{1024} k+\frac{5 }{256}\frac{ m^2}{k}-\frac{5 }{32}\frac{ m^4}{k^3}+\frac{5 }{8}\frac{ m^6}{k^5}-\frac{5 }{4}\frac{ m^8}{k^7}+\frac{m^{10}}{k^9}\right)\right)+
\0\\ & \quad 
 + k^8 \pi _{\mu\mu} \pi _{\mu\nu} \pi _{\nu\nu}^2 \left(\frac{i }{2 \pi ^2} \left(-\frac{3 }{2048} m+\frac{7 }{256}\frac{ m^3}{k^2}+\frac{1}{5}\frac{ m^5}{k^4}-\frac{7 }{16}\frac{ m^7}{k^6}+\frac{3 }{8}\frac{ m^9}{k^8}\right)+
\right.\0\\ & \quad \quad \left.
 + \frac{ T}{16 \pi } \left(-\frac{3 }{1024} k+\frac{15 }{256}\frac{ m^2}{k}-\frac{15 }{32}\frac{ m^4}{k^3}+\frac{15 }{8}\frac{ m^6}{k^5}-\frac{15 }{4}\frac{ m^8}{k^7}+3 \frac{ m^{10}}{k^9}\right)\right) }
\al{\label{eq:snc:3:5:5:}\tilde{T}_{3,5;5\text{D}}^{\text{s,nt}} & = k_{\mu }^3 k_{\nu }^3 \eta_{\nu\nu} \left(-\frac{i }{12 \pi ^2} m^3\right)+k_{\mu }^2 k_{\nu }^4 \eta_{\mu\nu} \left(-\frac{i }{4 \pi ^2} m^3\right)+k_{\mu } k_{\nu }^3 \eta_{\mu\mu} \eta_{\nu\nu} \left(-\frac{2 i }{5 \pi ^2} m^5\right)+
\0\\ & \quad 
 +  \left(k_{\nu }^4 \eta_{\mu\mu} \eta_{\mu\nu}+k_{\mu }^3 k_{\nu } \eta_{\nu\nu}^2\right) \left(-\frac{i }{5 \pi ^2} m^5\right)+
\0\\ & \quad 
 + k_{\mu }^2 k_{\nu }^2 \eta_{\mu\nu} \eta_{\nu\nu} \left(\frac{i }{\pi ^2} \left(\frac{1}{4} k^2 m^3-\frac{6 }{5} m^5\right)\right)+k_{\mu } k_{\nu }^3 \eta_{\mu\nu}^2 \left(\frac{i }{\pi ^2} \left(\frac{1}{4} k^2 m^3-\frac{4 }{5} m^5\right)\right)+
\0\\ & \quad 
 +  \left(k_{\mu } k_{\nu } \eta_{\mu\mu} \eta_{\nu\nu}^2+k_{\mu }^2 \eta_{\mu\nu} \eta_{\nu\nu}^2\right) \left(\frac{i }{\pi ^2} \left(\frac{1}{5} k^2 m^5-\frac{4 }{7} m^7\right)\right)+
\0\\ & \quad 
 + k_{\nu }^2 \eta_{\mu\mu} \eta_{\mu\nu} \eta_{\nu\nu} \left(\frac{i }{\pi ^2} \left(\frac{2 }{5} k^2 m^5-\frac{8 }{7} m^7\right)\right)+
\0\\ & \quad 
 + k_{\mu } k_{\nu } \eta_{\mu\nu}^2 \eta_{\nu\nu} \left(\frac{i }{\pi ^2} \left(-\frac{1}{4} k^4 m^3+\frac{4 }{5} k^2 m^5-\frac{16 }{7} m^7\right)\right)+
\0\\ & \quad 
 + k_{\nu }^2 \eta_{\mu\nu}^3 \left(\frac{i }{3 \pi ^2} \left(-\frac{1}{4} k^4 m^3+\frac{4 }{5} k^2 m^5-\frac{16 }{7} m^7\right)\right)+
\0\\ & \quad 
 + \eta_{\mu\mu} \eta_{\mu\nu} \eta_{\nu\nu}^2 \left(\frac{i }{\pi ^2} \left(-\frac{1}{5} k^4 m^5+\frac{4 }{7} k^2 m^7-\frac{16 }{21} m^9\right)\right)+
\0\\ & \quad 
 + \eta_{\mu\nu}^3 \eta_{\nu\nu} \left(\frac{i }{3 \pi ^2} \left(\frac{1}{4} k^6 m^3-\frac{4 }{5} k^4 m^5+\frac{16 }{7} k^2 m^7-\frac{64 }{21} m^9\right)\right) }
Scalars, spin 3 x 5, dimension 6:
\al{\label{eq:sc:3:5:6:}\tilde{T}_{3,5;6\text{D}}^{\text{s,t}} & = k^8 \pi _{\mu\nu}^3 \pi _{\nu\nu} \left(\frac{i }{7 \pi ^3} \left( \left(\frac{1627}{1372140}-\frac{L_0}{3168}\right) k^2+ \left(-\frac{12701}{498960}+\frac{L_0}{144}\right) m^2+
\right.\right.\0\\ & \quad \quad \quad \left.\left.
 +  \left(\frac{25903}{443520}+\frac{3 L_0}{64}\right)\frac{ m^4}{k^2}-\frac{692 }{1485}\frac{ m^6}{k^4}+\frac{256 }{297}\frac{ m^8}{k^6}-\frac{64 }{99}\frac{ m^{10}}{k^8}\right)+
\right.\0\\ & \quad \quad \left.
 + \frac{i  S}{693 \pi ^3} \left(-\frac{1}{16} k+\frac{5 }{4}\frac{ m^2}{k}-10 \frac{ m^4}{k^3}+40 \frac{ m^6}{k^5}-80 \frac{ m^8}{k^7}+64 \frac{ m^{10}}{k^9}\right)\right)+
\0\\ & \quad 
 + k^8 \pi _{\mu\mu} \pi _{\mu\nu} \pi _{\nu\nu}^2 \left(\frac{i }{7 \pi ^3} \left( \left(\frac{1627}{1829520}-\frac{L_0}{4224}\right) k^2+ \left(-\frac{12701}{665280}+\frac{L_0}{192}\right) m^2+
\right.\right.\0\\ & \quad \quad \quad \left.\left.
 +  \left(\frac{24667}{147840}-\frac{3 L_0}{64}\right)\frac{ m^4}{k^2}-\frac{173 }{495}\frac{ m^6}{k^4}+\frac{64 }{99}\frac{ m^8}{k^6}-\frac{16 }{33}\frac{ m^{10}}{k^8}\right)+
\right.\0\\ & \quad \quad \left.
 + \frac{i  S}{231 \pi ^3} \left(-\frac{1}{64} k+\frac{5 }{16}\frac{ m^2}{k}-\frac{5 }{2}\frac{ m^4}{k^3}+10 \frac{ m^6}{k^5}-20 \frac{ m^8}{k^7}+16 \frac{ m^{10}}{k^9}\right)\right) }
\al{\label{eq:snc:3:5:6:}\tilde{T}_{3,5;6\text{D}}^{\text{s,nt}} & = k_{\mu }^3 k_{\nu }^3 \eta_{\nu\nu} \left(\frac{i  L_2}{64 \pi ^3} m^4\right)+k_{\mu }^2 k_{\nu }^4 \eta_{\mu\nu} \left(\frac{3 i  L_2}{64 \pi ^3} m^4\right)+k_{\mu } k_{\nu }^3 \eta_{\mu\mu} \eta_{\nu\nu} \left(\frac{i  L_3}{16 \pi ^3} m^6\right)+
\0\\ & \quad 
 +  \left(k_{\nu }^4 \eta_{\mu\mu} \eta_{\mu\nu}+k_{\mu }^3 k_{\nu } \eta_{\nu\nu}^2\right) \left(\frac{i  L_3}{32 \pi ^3} m^6\right)+
\0\\ & \quad 
 + k_{\mu }^2 k_{\nu }^2 \eta_{\mu\nu} \eta_{\nu\nu} \left(\frac{i }{16 \pi ^3} \left(-\frac{3  L_2}{4} k^2 m^4+3  L_3 m^6\right)\right)+
\0\\ & \quad 
 + k_{\mu } k_{\nu }^3 \eta_{\mu\nu}^2 \left(\frac{i }{8 \pi ^3} \left(-\frac{3  L_2}{8} k^2 m^4+ L_3 m^6\right)\right)+
\0\\ & \quad 
 +  \left(k_{\mu } k_{\nu } \eta_{\mu\mu} \eta_{\nu\nu}^2+k_{\mu }^2 \eta_{\mu\nu} \eta_{\nu\nu}^2\right) \left(\frac{i }{32 \pi ^3} \left(- L_3 k^2 m^6+\frac{5  L_4}{2} m^8\right)\right)+
\0\\ & \quad 
 + k_{\nu }^2 \eta_{\mu\mu} \eta_{\mu\nu} \eta_{\nu\nu} \left(\frac{i }{16 \pi ^3} \left(- L_3 k^2 m^6+\frac{5  L_4}{2} m^8\right)\right)+
\0\\ & \quad 
 + k_{\mu } k_{\nu } \eta_{\mu\nu}^2 \eta_{\nu\nu} \left(\frac{i }{8 \pi ^3} \left(\frac{3  L_2}{8} k^4 m^4- L_3 k^2 m^6+\frac{5  L_4}{2} m^8\right)\right)+
\0\\ & \quad 
 + k_{\nu }^2 \eta_{\mu\nu}^3 \left(\frac{i }{8 \pi ^3} \left(\frac{ L_2}{8} k^4 m^4-\frac{ L_3}{3} k^2 m^6+\frac{5  L_4}{6} m^8\right)\right)+
\0\\ & \quad 
 + \eta_{\mu\mu} \eta_{\mu\nu} \eta_{\nu\nu}^2 \left(\frac{i }{32 \pi ^3} \left( L_3 k^4 m^6-\frac{5  L_4}{2} k^2 m^8+3  L_5 m^{10}\right)\right)+
\0\\ & \quad 
 + \eta_{\mu\nu}^3 \eta_{\nu\nu} \left(\frac{i }{8 \pi ^3} \left(-\frac{ L_2}{8} k^6 m^4+\frac{ L_3}{3} k^4 m^6-\frac{5  L_4}{6} k^2 m^8+ L_5 m^{10}\right)\right) }
Scalars, spin 4 x 4, dimension 3:
\al{\label{eq:sc:4:4:3:}\tilde{T}_{4,4;3\text{D}}^{\text{s,t}} & = k^8 \pi _{\mu\nu}^4 \left(\frac{i }{2 \pi } \left(\frac{15 }{16}\frac{ m}{k^2}-\frac{5 }{12}\frac{ m^3}{k^4}+\frac{11 }{3}\frac{ m^5}{k^6}-4 \frac{ m^7}{k^8}\right)+
\right.\0\\ & \quad \quad \left.
 +  T \left(-\frac{1}{128}\frac{1}{k}+\frac{1}{8}\frac{ m^2}{k^3}-\frac{3 }{4}\frac{ m^4}{k^5}+2 \frac{ m^6}{k^7}-2 \frac{ m^8}{k^9}\right)\right)+
\0\\ & \quad 
 + k^8 \pi _{\mu\mu} \pi _{\mu\nu}^2 \pi _{\nu\nu} \left(\frac{i }{2 \pi } \left(-\frac{3 }{16}\frac{ m}{k^2}-\frac{21 }{4}\frac{ m^3}{k^4}+11 \frac{ m^5}{k^6}-12 \frac{ m^7}{k^8}\right)+
\right.\0\\ & \quad \quad \left.
 +  T \left(-\frac{3 }{128}\frac{1}{k}+\frac{3 }{8}\frac{ m^2}{k^3}-\frac{9 }{4}\frac{ m^4}{k^5}+6 \frac{ m^6}{k^7}-6 \frac{ m^8}{k^9}\right)\right)+
\0\\ & \quad 
 + k^8 \pi _{\mu\mu}^2 \pi _{\nu\nu}^2 \left(\frac{i }{4 \pi } \left(-\frac{3 }{64}\frac{ m}{k^2}+\frac{11 }{16}\frac{ m^3}{k^4}+\frac{11 }{4}\frac{ m^5}{k^6}-3 \frac{ m^7}{k^8}\right)+
\right.\0\\ & \quad \quad \left.
 +  T \left(-\frac{3 }{1024}\frac{1}{k}+\frac{3 }{64}\frac{ m^2}{k^3}-\frac{9 }{32}\frac{ m^4}{k^5}+\frac{3 }{4}\frac{ m^6}{k^7}-\frac{3 }{4}\frac{ m^8}{k^9}\right)\right) }
\al{\label{eq:snc:4:4:3:}\tilde{T}_{4,4;3\text{D}}^{\text{s,nt}} & = k_{\mu }^3 k_{\nu }^3 \eta_{\mu\nu} \left(\frac{2 i }{\pi } m\right)+k_{\mu }^2 k_{\nu }^2 \eta_{\mu\mu} \eta_{\nu\nu} \left(\frac{4 i }{\pi } m^3\right)+
\0\\ & \quad 
 +  \left(k_{\mu } k_{\nu }^3 \eta_{\mu\mu} \eta_{\mu\nu}+k_{\mu }^3 k_{\nu } \eta_{\mu\nu} \eta_{\nu\nu}\right) \left(\frac{8 i }{\pi } m^3\right)+k_{\mu }^2 k_{\nu }^2 \eta_{\mu\nu}^2 \left(\frac{i }{\pi } \left(-3  k^2 m+8  m^3\right)\right)+
\0\\ & \quad 
 +  \left(k_{\nu }^2 \eta_{\mu\mu}^2 \eta_{\nu\nu}+k_{\mu }^2 \eta_{\mu\mu} \eta_{\nu\nu}^2\right) \left(\frac{16 i }{5 \pi } m^5\right)+
\0\\ & \quad 
 + k_{\mu } k_{\nu } \eta_{\mu\mu} \eta_{\mu\nu} \eta_{\nu\nu} \left(\frac{i }{5 \pi } \left(-40  k^2 m^3+128  m^5\right)\right)+
\0\\ & \quad 
 +  \left(k_{\nu }^2 \eta_{\mu\mu} \eta_{\mu\nu}^2+k_{\mu }^2 \eta_{\mu\nu}^2 \eta_{\nu\nu}\right) \left(\frac{i }{5 \pi } \left(-20  k^2 m^3+64  m^5\right)\right)+
\0\\ & \quad 
 + k_{\mu } k_{\nu } \eta_{\mu\nu}^3 \left(\frac{i }{3 \pi } \left(6  k^4 m-8  k^2 m^3+\frac{256 }{5} m^5\right)\right)+
\0\\ & \quad 
 + \eta_{\mu\mu}^2 \eta_{\nu\nu}^2 \left(\frac{i }{5 \pi } \left(-8  k^2 m^5+\frac{96 }{7} m^7\right)\right)+
\0\\ & \quad 
 + \eta_{\mu\mu} \eta_{\mu\nu}^2 \eta_{\nu\nu} \left(\frac{i }{5 \pi } \left(20  k^4 m^3-64  k^2 m^5+\frac{768 }{7} m^7\right)\right)+
\0\\ & \quad 
 + \eta_{\mu\nu}^4 \left(\frac{i }{\pi } \left(-\frac{1}{2} k^6 m+\frac{2 }{3} k^4 m^3-\frac{64 }{15} k^2 m^5+\frac{256 }{35} m^7\right)\right) }
Scalars, spin 4 x 4, dimension 4:
\al{\label{eq:sc:4:4:4:}\tilde{T}_{4,4;4\text{D}}^{\text{s,t}} & = k^8 \pi _{\mu\nu}^4 \left(\frac{i }{5 \pi ^2} \left( \left(\frac{563}{19845}-\frac{L_0}{126}\right)+ \left(\frac{2297}{17640}-\frac{27 L_0}{56}\right)\frac{ m^2}{k^2}+
\right.\right.\0\\ & \quad \quad \quad \left.\left.
 +  \left(\frac{1339}{560}-\frac{3 L_0}{8}\right)\frac{ m^4}{k^4}-\frac{832 }{189}\frac{ m^6}{k^6}+\frac{256 }{63}\frac{ m^8}{k^8}\right)+
\right.\0\\ & \quad \quad \left.
 + \frac{i  S}{105 \pi ^2} \left(-\frac{1}{3}\frac{1}{k}+\frac{16 }{3}\frac{ m^2}{k^3}-32 \frac{ m^4}{k^5}+\frac{256 }{3}\frac{ m^6}{k^7}-\frac{256 }{3}\frac{ m^8}{k^9}\right)\right)+
\0\\ & \quad 
 + k^8 \pi _{\mu\mu} \pi _{\mu\nu}^2 \pi _{\nu\nu} \left(\frac{i }{5 \pi ^2} \left( \left(\frac{563}{6615}-\frac{L_0}{42}\right)+ \left(-\frac{1091}{735}+\frac{3 L_0}{7}\right)\frac{ m^2}{k^2}+
\right.\right.\0\\ & \quad \quad \quad \left.\left.
 +  \left(\frac{1221}{280}+\frac{3 L_0}{4}\right)\frac{ m^4}{k^4}-\frac{832 }{63}\frac{ m^6}{k^6}+\frac{256 }{21}\frac{ m^8}{k^8}\right)+
\right.\0\\ & \quad \quad \left.
 + \frac{i  S}{35 \pi ^2} \left(-\frac{1}{3}\frac{1}{k}+\frac{16 }{3}\frac{ m^2}{k^3}-32 \frac{ m^4}{k^5}+\frac{256 }{3}\frac{ m^6}{k^7}-\frac{256 }{3}\frac{ m^8}{k^9}\right)\right)+
\0\\ & \quad 
 + k^8 \pi _{\mu\mu}^2 \pi _{\nu\nu}^2 \left(\frac{i }{5 \pi ^2} \left( \left(\frac{563}{52920}-\frac{L_0}{336}\right)+ \left(-\frac{1091}{5880}+\frac{3 L_0}{56}\right)\frac{ m^2}{k^2}+
\right.\right.\0\\ & \quad \quad \quad \left.\left.
 +  \left(\frac{699}{560}-\frac{3 L_0}{8}\right)\frac{ m^4}{k^4}-\frac{104 }{63}\frac{ m^6}{k^6}+\frac{32 }{21}\frac{ m^8}{k^8}\right)+
\right.\0\\ & \quad \quad \left.
 + \frac{i  S}{35 \pi ^2} \left(-\frac{1}{24}\frac{1}{k}+\frac{2 }{3}\frac{ m^2}{k^3}-4 \frac{ m^4}{k^5}+\frac{32 }{3}\frac{ m^6}{k^7}-\frac{32 }{3}\frac{ m^8}{k^9}\right)\right) }
\al{\label{eq:snc:4:4:4:}\tilde{T}_{4,4;4\text{D}}^{\text{s,nt}} & = k_{\mu }^3 k_{\nu }^3 \eta_{\mu\nu} \left(-\frac{i  L_1}{2 \pi ^2} m^2\right)+k_{\mu }^2 k_{\nu }^2 \eta_{\mu\mu} \eta_{\nu\nu} \left(-\frac{3 i  L_2}{4 \pi ^2} m^4\right)+
\0\\ & \quad 
 +  \left(k_{\mu } k_{\nu }^3 \eta_{\mu\mu} \eta_{\mu\nu}+k_{\mu }^3 k_{\nu } \eta_{\mu\nu} \eta_{\nu\nu}\right) \left(-\frac{3 i  L_2}{2 \pi ^2} m^4\right)+
\0\\ & \quad 
 + k_{\mu }^2 k_{\nu }^2 \eta_{\mu\nu}^2 \left(\frac{i }{2 \pi ^2} \left(\frac{3  L_1}{2} k^2 m^2-3  L_2 m^4\right)\right)+
\0\\ & \quad 
 +  \left(k_{\nu }^2 \eta_{\mu\mu}^2 \eta_{\nu\nu}+k_{\mu }^2 \eta_{\mu\mu} \eta_{\nu\nu}^2\right) \left(-\frac{i  L_3}{2 \pi ^2} m^6\right)+
\0\\ & \quad 
 + k_{\mu } k_{\nu } \eta_{\mu\mu} \eta_{\mu\nu} \eta_{\nu\nu} \left(\frac{i }{2 \pi ^2} \left(3  L_2 k^2 m^4-8  L_3 m^6\right)\right)+
\0\\ & \quad 
 +  \left(k_{\nu }^2 \eta_{\mu\mu} \eta_{\mu\nu}^2+k_{\mu }^2 \eta_{\mu\nu}^2 \eta_{\nu\nu}\right) \left(\frac{i }{4 \pi ^2} \left(3  L_2 k^2 m^4-8  L_3 m^6\right)\right)+
\0\\ & \quad 
 + k_{\mu } k_{\nu } \eta_{\mu\nu}^3 \left(\frac{i }{\pi ^2} \left(-\frac{ L_1}{2} k^4 m^2+\frac{ L_2}{2} k^2 m^4-\frac{8  L_3}{3} m^6\right)\right)+
\0\\ & \quad 
 + \eta_{\mu\mu}^2 \eta_{\nu\nu}^2 \left(\frac{i }{4 \pi ^2} \left( L_3 k^2 m^6-\frac{3  L_4}{2} m^8\right)\right)+
\0\\ & \quad 
 + \eta_{\mu\mu} \eta_{\mu\nu}^2 \eta_{\nu\nu} \left(\frac{i }{4 \pi ^2} \left(-3  L_2 k^4 m^4+8  L_3 k^2 m^6-12  L_4 m^8\right)\right)+
\0\\ & \quad 
 + \eta_{\mu\nu}^4 \left(\frac{i }{\pi ^2} \left(\frac{ L_1}{8} k^6 m^2-\frac{ L_2}{8} k^4 m^4+\frac{2  L_3}{3} k^2 m^6- L_4 m^8\right)\right) }
Scalars, spin 4 x 4, dimension 5:
\al{\label{eq:sc:4:4:5:}\tilde{T}_{4,4;5\text{D}}^{\text{s,t}} & = k^8 \pi _{\mu\nu}^4 \left(\frac{i }{5 \pi ^2} \left(-\frac{1}{512} m-\frac{73 }{192}\frac{ m^3}{k^2}+\frac{1}{15}\frac{ m^5}{k^4}-\frac{7 }{12}\frac{ m^7}{k^6}+\frac{1}{2}\frac{ m^9}{k^8}\right)+
\right.\0\\ & \quad \quad \left.
 + \frac{ T}{2 \pi } \left(-\frac{1}{5120} k+\frac{1}{256}\frac{ m^2}{k}-\frac{1}{32}\frac{ m^4}{k^3}+\frac{1}{8}\frac{ m^6}{k^5}-\frac{1}{4}\frac{ m^8}{k^7}+\frac{1}{5}\frac{ m^{10}}{k^9}\right)\right)+
\0\\ & \quad 
 + k^8 \pi _{\mu\mu} \pi _{\mu\nu}^2 \pi _{\nu\nu} \left(\frac{i }{5 \pi ^2} \left(-\frac{3 }{512} m+\frac{7 }{64}\frac{ m^3}{k^2}+\frac{6 }{5}\frac{ m^5}{k^4}-\frac{7 }{4}\frac{ m^7}{k^6}+\frac{3 }{2}\frac{ m^9}{k^8}\right)+
\right.\0\\ & \quad \quad \left.
 + \frac{ T}{2 \pi } \left(-\frac{3 }{5120} k+\frac{3 }{256}\frac{ m^2}{k}-\frac{3 }{32}\frac{ m^4}{k^3}+\frac{3 }{8}\frac{ m^6}{k^5}-\frac{3 }{4}\frac{ m^8}{k^7}+\frac{3 }{5}\frac{ m^{10}}{k^9}\right)\right)+
\0\\ & \quad 
 + k^8 \pi _{\mu\mu}^2 \pi _{\nu\nu}^2 \left(\frac{i }{10 \pi ^2} \left(-\frac{3 }{2048} m+\frac{7 }{256}\frac{ m^3}{k^2}-\frac{1}{5}\frac{ m^5}{k^4}-\frac{7 }{16}\frac{ m^7}{k^6}+\frac{3 }{8}\frac{ m^9}{k^8}\right)+
\right.\0\\ & \quad \quad \left.
 + \frac{ T}{16 \pi } \left(-\frac{3 }{5120} k+\frac{3 }{256}\frac{ m^2}{k}-\frac{3 }{32}\frac{ m^4}{k^3}+\frac{3 }{8}\frac{ m^6}{k^5}-\frac{3 }{4}\frac{ m^8}{k^7}+\frac{3 }{5}\frac{ m^{10}}{k^9}\right)\right) }
\al{\label{eq:snc:4:4:5:}\tilde{T}_{4,4;5\text{D}}^{\text{s,nt}} & = k_{\mu }^3 k_{\nu }^3 \eta_{\mu\nu} \left(-\frac{i }{3 \pi ^2} m^3\right)+k_{\mu }^2 k_{\nu }^2 \eta_{\mu\mu} \eta_{\nu\nu} \left(-\frac{2 i }{5 \pi ^2} m^5\right)+
\0\\ & \quad 
 +  \left(k_{\mu } k_{\nu }^3 \eta_{\mu\mu} \eta_{\mu\nu}+k_{\mu }^3 k_{\nu } \eta_{\mu\nu} \eta_{\nu\nu}\right) \left(-\frac{4 i }{5 \pi ^2} m^5\right)+
\0\\ & \quad 
 + k_{\mu }^2 k_{\nu }^2 \eta_{\mu\nu}^2 \left(\frac{i }{\pi ^2} \left(\frac{1}{2} k^2 m^3-\frac{4 }{5} m^5\right)\right)+ \left(k_{\nu }^2 \eta_{\mu\mu}^2 \eta_{\nu\nu}+k_{\mu }^2 \eta_{\mu\mu} \eta_{\nu\nu}^2\right) \left(-\frac{8 i }{35 \pi ^2} m^7\right)+
\0\\ & \quad 
 + k_{\mu } k_{\nu } \eta_{\mu\mu} \eta_{\mu\nu} \eta_{\nu\nu} \left(\frac{i }{5 \pi ^2} \left(4  k^2 m^5-\frac{64 }{7} m^7\right)\right)+
\0\\ & \quad 
 +  \left(k_{\nu }^2 \eta_{\mu\mu} \eta_{\mu\nu}^2+k_{\mu }^2 \eta_{\mu\nu}^2 \eta_{\nu\nu}\right) \left(\frac{i }{5 \pi ^2} \left(2  k^2 m^5-\frac{32 }{7} m^7\right)\right)+
\0\\ & \quad 
 + k_{\mu } k_{\nu } \eta_{\mu\nu}^3 \left(\frac{i }{3 \pi ^2} \left(- k^4 m^3+\frac{4 }{5} k^2 m^5-\frac{128 }{35} m^7\right)\right)+
\0\\ & \quad 
 + \eta_{\mu\mu}^2 \eta_{\nu\nu}^2 \left(\frac{i }{35 \pi ^2} \left(4  k^2 m^7-\frac{16 }{3} m^9\right)\right)+
\0\\ & \quad 
 + \eta_{\mu\mu} \eta_{\mu\nu}^2 \eta_{\nu\nu} \left(\frac{i }{5 \pi ^2} \left(-2  k^4 m^5+\frac{32 }{7} k^2 m^7-\frac{128 }{21} m^9\right)\right)+
\0\\ & \quad 
 + \eta_{\mu\nu}^4 \left(\frac{i }{3 \pi ^2} \left(\frac{1}{4} k^6 m^3-\frac{1}{5} k^4 m^5+\frac{32 }{35} k^2 m^7-\frac{128 }{105} m^9\right)\right) }
Scalars, spin 4 x 4, dimension 6:
\al{\label{eq:sc:4:4:6:}\tilde{T}_{4,4;6\text{D}}^{\text{s,t}} & = k^8 \pi _{\mu\nu}^4 \left(\frac{i }{5 \pi ^3} \left( \left(\frac{1627}{4802490}-\frac{L_0}{11088}\right) k^2+ \left(-\frac{12701}{1746360}+\frac{L_0}{504}\right) m^2+
\right.\right.\0\\ & \quad \quad \quad \left.\left.
 +  \left(-\frac{166489}{3104640}+\frac{27 L_0}{448}\right)\frac{ m^4}{k^2}+ \left(-\frac{126691}{665280}+\frac{L_0}{32}\right)\frac{ m^6}{k^4}+\frac{512 }{2079}\frac{ m^8}{k^6}-
\right.\right.\0\\ & \quad \quad \quad \left.\left.
 - \frac{128 }{693}\frac{ m^{10}}{k^8}\right)+
\right.\0\\ & \quad \quad \left.
 + \frac{i  S}{693 \pi ^3} \left(-\frac{1}{40} k+\frac{1}{2}\frac{ m^2}{k}-4 \frac{ m^4}{k^3}+16 \frac{ m^6}{k^5}-32 \frac{ m^8}{k^7}+\frac{128 }{5}\frac{ m^{10}}{k^9}\right)\right)+
\0\\ & \quad 
 + k^8 \pi _{\mu\mu} \pi _{\mu\nu}^2 \pi _{\nu\nu} \left(\frac{i }{5 \pi ^3} \left( \left(\frac{1627}{1600830}-\frac{L_0}{3696}\right) k^2+ \left(-\frac{12701}{582120}+\frac{L_0}{168}\right) m^2+
\right.\right.\0\\ & \quad \quad \quad \left.\left.
 +  \left(\frac{24667}{129360}-\frac{3 L_0}{56}\right)\frac{ m^4}{k^2}- \left(\frac{31583}{110880}+\frac{L_0}{16}\right)\frac{ m^6}{k^4}+\frac{512 }{693}\frac{ m^8}{k^6}-
\right.\right.\0\\ & \quad \quad \quad \left.\left.
 - \frac{128 }{231}\frac{ m^{10}}{k^8}\right)+
\right.\0\\ & \quad \quad \left.
 + \frac{i  S}{231 \pi ^3} \left(-\frac{1}{40} k+\frac{1}{2}\frac{ m^2}{k}-4 \frac{ m^4}{k^3}+16 \frac{ m^6}{k^5}-32 \frac{ m^8}{k^7}+\frac{128 }{5}\frac{ m^{10}}{k^9}\right)\right)+
\0\\ & \quad 
 + k^8 \pi _{\mu\mu}^2 \pi _{\nu\nu}^2 \left(\frac{i }{5 \pi ^3} \left( \left(\frac{1627}{12806640}-\frac{L_0}{29568}\right) k^2+ \left(-\frac{12701}{4656960}+\frac{L_0}{1344}\right) m^2+
\right.\right.\0\\ & \quad \quad \quad \left.\left.
 +  \left(\frac{24667}{1034880}-\frac{3 L_0}{448}\right)\frac{ m^4}{k^2}+ \left(-\frac{23777}{221760}+\frac{L_0}{32}\right)\frac{ m^6}{k^4}+\frac{64 }{693}\frac{ m^8}{k^6}-
\right.\right.\0\\ & \quad \quad \quad \left.\left.
 - \frac{16 }{231}\frac{ m^{10}}{k^8}\right)+
\right.\0\\ & \quad \quad \left.
 + \frac{i  S}{231 \pi ^3} \left(-\frac{1}{320} k+\frac{1}{16}\frac{ m^2}{k}-\frac{1}{2}\frac{ m^4}{k^3}+2 \frac{ m^6}{k^5}-4 \frac{ m^8}{k^7}+\frac{16 }{5}\frac{ m^{10}}{k^9}\right)\right) }
\al{\label{eq:snc:4:4:6:}\tilde{T}_{4,4;6\text{D}}^{\text{s,nt}} & = k_{\mu }^3 k_{\nu }^3 \eta_{\mu\nu} \left(\frac{i  L_2}{16 \pi ^3} m^4\right)+k_{\mu }^2 k_{\nu }^2 \eta_{\mu\mu} \eta_{\nu\nu} \left(\frac{i  L_3}{16 \pi ^3} m^6\right)+
\0\\ & \quad 
 +  \left(k_{\mu } k_{\nu }^3 \eta_{\mu\mu} \eta_{\mu\nu}+k_{\mu }^3 k_{\nu } \eta_{\mu\nu} \eta_{\nu\nu}\right) \left(\frac{i  L_3}{8 \pi ^3} m^6\right)+
\0\\ & \quad 
 + k_{\mu }^2 k_{\nu }^2 \eta_{\mu\nu}^2 \left(\frac{i }{8 \pi ^3} \left(-\frac{3  L_2}{4} k^2 m^4+ L_3 m^6\right)\right)+
\0\\ & \quad 
 +  \left(k_{\nu }^2 \eta_{\mu\mu}^2 \eta_{\nu\nu}+k_{\mu }^2 \eta_{\mu\mu} \eta_{\nu\nu}^2\right) \left(\frac{i  L_4}{32 \pi ^3} m^8\right)+
\0\\ & \quad 
 + k_{\mu } k_{\nu } \eta_{\mu\mu} \eta_{\mu\nu} \eta_{\nu\nu} \left(\frac{i }{4 \pi ^3} \left(-\frac{ L_3}{2} k^2 m^6+ L_4 m^8\right)\right)+
\0\\ & \quad 
 +  \left(k_{\nu }^2 \eta_{\mu\mu} \eta_{\mu\nu}^2+k_{\mu }^2 \eta_{\mu\nu}^2 \eta_{\nu\nu}\right) \left(\frac{i }{8 \pi ^3} \left(-\frac{ L_3}{2} k^2 m^6+ L_4 m^8\right)\right)+
\0\\ & \quad 
 + k_{\mu } k_{\nu } \eta_{\mu\nu}^3 \left(\frac{i }{2 \pi ^3} \left(\frac{ L_2}{8} k^4 m^4-\frac{ L_3}{12} k^2 m^6+\frac{ L_4}{3} m^8\right)\right)+
\0\\ & \quad 
 + \eta_{\mu\mu}^2 \eta_{\nu\nu}^2 \left(\frac{i }{32 \pi ^3} \left(-\frac{ L_4}{2} k^2 m^8+\frac{3  L_5}{5} m^{10}\right)\right)+
\0\\ & \quad 
 + \eta_{\mu\mu} \eta_{\mu\nu}^2 \eta_{\nu\nu} \left(\frac{i }{4 \pi ^3} \left(\frac{ L_3}{4} k^4 m^6-\frac{ L_4}{2} k^2 m^8+\frac{3  L_5}{5} m^{10}\right)\right)+
\0\\ & \quad 
 + \eta_{\mu\nu}^4 \left(\frac{i }{4 \pi ^3} \left(-\frac{ L_2}{16} k^6 m^4+\frac{ L_3}{24} k^4 m^6-\frac{ L_4}{6} k^2 m^8+\frac{ L_5}{5} m^{10}\right)\right) }
Scalars, spin 5 x 5, dimension 3:
\al{\label{eq:sc:5:5:3:}\tilde{T}_{5,5;3\text{D}}^{\text{s,t}} & = k^{10} \pi _{\mu\nu}^5 \left(\frac{i }{\pi } \left(-\frac{31 }{64}\frac{ m}{k^2}-\frac{23 }{24}\frac{ m^3}{k^4}-\frac{32 }{15}\frac{ m^5}{k^6}+\frac{14 }{3}\frac{ m^7}{k^8}-4 \frac{ m^9}{k^{10}}\right)+
\right.\0\\ & \quad \quad \left.
 +  T \left(\frac{1}{256}\frac{1}{k}-\frac{5 }{64}\frac{ m^2}{k^3}+\frac{5 }{8}\frac{ m^4}{k^5}-\frac{5 }{2}\frac{ m^6}{k^7}+5 \frac{ m^8}{k^9}-4 \frac{ m^{10}}{k^{11}}\right)\right)+
\0\\ & \quad 
 + k^{10} \pi _{\mu\mu} \pi _{\mu\nu}^3 \pi _{\nu\nu} \left(\frac{i }{\pi } \left(\frac{5 }{64}\frac{ m}{k^2}+\frac{125 }{24}\frac{ m^3}{k^4}-\frac{32 }{3}\frac{ m^5}{k^6}+\frac{70 }{3}\frac{ m^7}{k^8}-20 \frac{ m^9}{k^{10}}\right)+
\right.\0\\ & \quad \quad \left.
 +  T \left(\frac{5 }{256}\frac{1}{k}-\frac{25 }{64}\frac{ m^2}{k^3}+\frac{25 }{8}\frac{ m^4}{k^5}-\frac{25 }{2}\frac{ m^6}{k^7}+25 \frac{ m^8}{k^9}-20 \frac{ m^{10}}{k^{11}}\right)\right)+
\0\\ & \quad 
 + k^{10} \pi _{\mu\mu}^2 \pi _{\mu\nu} \pi _{\nu\nu}^2 \left(\frac{i }{2 \pi } \left(\frac{15 }{256}\frac{ m}{k^2}-\frac{35 }{32}\frac{ m^3}{k^4}-8 \frac{ m^5}{k^6}+\frac{35 }{2}\frac{ m^7}{k^8}-15 \frac{ m^9}{k^{10}}\right)+
\right.\0\\ & \quad \quad \left.
 +  T \left(\frac{15 }{2048}\frac{1}{k}-\frac{75 }{512}\frac{ m^2}{k^3}+\frac{75 }{64}\frac{ m^4}{k^5}-\frac{75 }{16}\frac{ m^6}{k^7}+\frac{75 }{8}\frac{ m^8}{k^9}-\frac{15 }{2}\frac{ m^{10}}{k^{11}}\right)\right) }
\al{\label{eq:snc:5:5:3:}\tilde{T}_{5,5;3\text{D}}^{\text{s,nt}} & = k_{\mu }^4 k_{\nu }^4 \eta_{\mu\nu} \left(\frac{5 i }{2 \pi } m\right)+k_{\mu }^3 k_{\nu }^3 \eta_{\mu\mu} \eta_{\nu\nu} \left(\frac{20 i }{3 \pi } m^3\right)+
\0\\ & \quad 
 +  \left(k_{\mu }^2 k_{\nu }^4 \eta_{\mu\mu} \eta_{\mu\nu}+k_{\mu }^4 k_{\nu }^2 \eta_{\mu\nu} \eta_{\nu\nu}\right) \left(\frac{20 i }{\pi } m^3\right)+
\0\\ & \quad 
 + k_{\mu }^3 k_{\nu }^3 \eta_{\mu\nu}^2 \left(\frac{i }{3 \pi } \left(-15  k^2 m+40  m^3\right)\right)+
\0\\ & \quad 
 +  \left(k_{\mu } k_{\nu }^3 \eta_{\mu\mu}^2 \eta_{\nu\nu}+k_{\mu }^3 k_{\nu } \eta_{\mu\mu} \eta_{\nu\nu}^2\right) \left(\frac{16 i }{\pi } m^5\right)+
\0\\ & \quad 
 +  \left(k_{\nu }^4 \eta_{\mu\mu}^2 \eta_{\mu\nu}+k_{\mu }^4 \eta_{\mu\nu} \eta_{\nu\nu}^2\right) \left(\frac{8 i }{\pi } m^5\right)+
\0\\ & \quad 
 + k_{\mu }^2 k_{\nu }^2 \eta_{\mu\mu} \eta_{\mu\nu} \eta_{\nu\nu} \left(\frac{i }{\pi } \left(-20  k^2 m^3+96  m^5\right)\right)+
\0\\ & \quad 
 +  \left(k_{\mu } k_{\nu }^3 \eta_{\mu\mu} \eta_{\mu\nu}^2+k_{\mu }^3 k_{\nu } \eta_{\mu\nu}^2 \eta_{\nu\nu}\right) \left(\frac{i }{\pi } \left(-20  k^2 m^3+64  m^5\right)\right)+
\0\\ & \quad 
 + k_{\mu }^2 k_{\nu }^2 \eta_{\mu\nu}^3 \left(\frac{i }{\pi } \left(5  k^4 m+64  m^5\right)\right)+k_{\mu } k_{\nu } \eta_{\mu\mu}^2 \eta_{\nu\nu}^2 \left(\frac{i }{7 \pi } \left(-56  k^2 m^5+160  m^7\right)\right)+
\0\\ & \quad 
 +  \left(k_{\nu }^2 \eta_{\mu\mu}^2 \eta_{\mu\nu} \eta_{\nu\nu}+k_{\mu }^2 \eta_{\mu\mu} \eta_{\mu\nu} \eta_{\nu\nu}^2\right) \left(\frac{i }{7 \pi } \left(-112  k^2 m^5+320  m^7\right)\right)+
\0\\ & \quad 
 + k_{\mu } k_{\nu } \eta_{\mu\mu} \eta_{\mu\nu}^2 \eta_{\nu\nu} \left(\frac{i }{7 \pi } \left(140  k^4 m^3-448  k^2 m^5+1280  m^7\right)\right)+
\0\\ & \quad 
 +  \left(k_{\nu }^2 \eta_{\mu\mu} \eta_{\mu\nu}^3+k_{\mu }^2 \eta_{\mu\nu}^3 \eta_{\nu\nu}\right) \left(\frac{i }{3 \pi } \left(20  k^4 m^3-64  k^2 m^5+\frac{1280 }{7} m^7\right)\right)+
\0\\ & \quad 
 + k_{\mu } k_{\nu } \eta_{\mu\nu}^4 \left(\frac{i }{\pi } \left(-\frac{5 }{2} k^6 m-\frac{10 }{3} k^4 m^3-\frac{64 }{3} k^2 m^5+\frac{1280 }{21} m^7\right)\right)+
\0\\ & \quad 
 + \eta_{\mu\mu}^2 \eta_{\mu\nu} \eta_{\nu\nu}^2 \left(\frac{i }{7 \pi } \left(56  k^4 m^5-160  k^2 m^7+\frac{640 }{3} m^9\right)\right)+
\0\\ & \quad 
 + \eta_{\mu\mu} \eta_{\mu\nu}^3 \eta_{\nu\nu} \left(\frac{i }{3 \pi } \left(-20  k^6 m^3+64  k^4 m^5-\frac{1280 }{7} k^2 m^7+\frac{5120 }{21} m^9\right)\right)+
\0\\ & \quad 
 + \eta_{\mu\nu}^5 \left(\frac{i }{\pi } \left(\frac{1}{2} k^8 m+\frac{2 }{3} k^6 m^3+\frac{64 }{15} k^4 m^5-\frac{256 }{21} k^2 m^7+\frac{1024 }{63} m^9\right)\right) }
Scalars, spin 5 x 5, dimension 4:
\al{\label{eq:sc:5:5:4:}\tilde{T}_{5,5;4\text{D}}^{\text{s,t}} & = k^{10} \pi _{\mu\nu}^5 \left(\frac{i }{7 \pi ^2} \left( \left(-\frac{6508}{343035}+\frac{L_0}{198}\right)+ \left(-\frac{116687}{249480}+\frac{55 L_0}{72}\right)\frac{ m^2}{k^2}+
\right.\right.\0\\ & \quad \quad \quad \left.\left.
 +  \left(-\frac{270101}{55440}+\frac{15 L_0}{8}\right)\frac{ m^4}{k^4}+\frac{11072 }{1485}\frac{ m^6}{k^6}-\frac{4096 }{297}\frac{ m^8}{k^8}+\frac{1024 }{99}\frac{ m^{10}}{k^{10}}\right)+
\right.\0\\ & \quad \quad \left.
 + \frac{i  S}{693 \pi ^2} \left(\frac{1}{k}-20 \frac{ m^2}{k^3}+160 \frac{ m^4}{k^5}-640 \frac{ m^6}{k^7}+1280 \frac{ m^8}{k^9}-1024 \frac{ m^{10}}{k^{11}}\right)\right)+
\0\\ & \quad 
 + k^{10} \pi _{\mu\mu} \pi _{\mu\nu}^3 \pi _{\nu\nu} \left(\frac{i }{7 \pi ^2} \left( \left(-\frac{6508}{68607}+\frac{5 L_0}{198}\right)+ \left(\frac{12701}{6237}-\frac{5 L_0}{9}\right)\frac{ m^2}{k^2}-
\right.\right.\0\\ & \quad \quad \quad \left.\left.
 -  \left(\frac{25903}{5544}+\frac{15 L_0}{4}\right)\frac{ m^4}{k^4}+\frac{11072 }{297}\frac{ m^6}{k^6}-\frac{20480 }{297}\frac{ m^8}{k^8}+\frac{5120 }{99}\frac{ m^{10}}{k^{10}}\right)+
\right.\0\\ & \quad \quad \left.
 + \frac{i  S}{693 \pi ^2} \left(5 \frac{1}{k}-100 \frac{ m^2}{k^3}+800 \frac{ m^4}{k^5}-3200 \frac{ m^6}{k^7}+6400 \frac{ m^8}{k^9}-5120 \frac{ m^{10}}{k^{11}}\right)\right)+
\0\\ & \quad 
 + k^{10} \pi _{\mu\mu}^2 \pi _{\mu\nu} \pi _{\nu\nu}^2 \left(\frac{i }{7 \pi ^2} \left( \left(-\frac{1627}{45738}+\frac{5 L_0}{528}\right)+ \left(\frac{12701}{16632}-\frac{5 L_0}{24}\right)\frac{ m^2}{k^2}+
\right.\right.\0\\ & \quad \quad \quad \left.\left.
 +  \left(-\frac{24667}{3696}+\frac{15 L_0}{8}\right)\frac{ m^4}{k^4}+\frac{1384 }{99}\frac{ m^6}{k^6}-\frac{2560 }{99}\frac{ m^8}{k^8}+\frac{640 }{33}\frac{ m^{10}}{k^{10}}\right)+
\right.\0\\ & \quad \quad \left.
 + \frac{i  S}{231 \pi ^2} \left(\frac{5 }{8}\frac{1}{k}-\frac{25 }{2}\frac{ m^2}{k^3}+100 \frac{ m^4}{k^5}-400 \frac{ m^6}{k^7}+800 \frac{ m^8}{k^9}-640 \frac{ m^{10}}{k^{11}}\right)\right) }
\al{\label{eq:snc:5:5:4:}\tilde{T}_{5,5;4\text{D}}^{\text{s,nt}} & = k_{\mu }^4 k_{\nu }^4 \eta_{\mu\nu} \left(-\frac{5 i  L_1}{8 \pi ^2} m^2\right)+k_{\mu }^3 k_{\nu }^3 \eta_{\mu\mu} \eta_{\nu\nu} \left(-\frac{5 i  L_2}{4 \pi ^2} m^4\right)+
\0\\ & \quad 
 +  \left(k_{\mu }^2 k_{\nu }^4 \eta_{\mu\mu} \eta_{\mu\nu}+k_{\mu }^4 k_{\nu }^2 \eta_{\mu\nu} \eta_{\nu\nu}\right) \left(-\frac{15 i  L_2}{4 \pi ^2} m^4\right)+
\0\\ & \quad 
 + k_{\mu }^3 k_{\nu }^3 \eta_{\mu\nu}^2 \left(\frac{i }{2 \pi ^2} \left(\frac{5  L_1}{2} k^2 m^2-5  L_2 m^4\right)\right)+
\0\\ & \quad 
 +  \left(k_{\mu } k_{\nu }^3 \eta_{\mu\mu}^2 \eta_{\nu\nu}+k_{\mu }^3 k_{\nu } \eta_{\mu\mu} \eta_{\nu\nu}^2\right) \left(-\frac{5 i  L_3}{2 \pi ^2} m^6\right)+
\0\\ & \quad 
 +  \left(k_{\nu }^4 \eta_{\mu\mu}^2 \eta_{\mu\nu}+k_{\mu }^4 \eta_{\mu\nu} \eta_{\nu\nu}^2\right) \left(-\frac{5 i  L_3}{4 \pi ^2} m^6\right)+
\0\\ & \quad 
 + k_{\mu }^2 k_{\nu }^2 \eta_{\mu\mu} \eta_{\mu\nu} \eta_{\nu\nu} \left(\frac{i }{4 \pi ^2} \left(15  L_2 k^2 m^4-60  L_3 m^6\right)\right)+
\0\\ & \quad 
 +  \left(k_{\mu } k_{\nu }^3 \eta_{\mu\mu} \eta_{\mu\nu}^2+k_{\mu }^3 k_{\nu } \eta_{\mu\nu}^2 \eta_{\nu\nu}\right) \left(\frac{i }{4 \pi ^2} \left(15  L_2 k^2 m^4-40  L_3 m^6\right)\right)+
\0\\ & \quad 
 + k_{\mu }^2 k_{\nu }^2 \eta_{\mu\nu}^3 \left(\frac{i }{4 \pi ^2} \left(-5  L_1 k^4 m^2-40  L_3 m^6\right)\right)+
\0\\ & \quad 
 + k_{\mu } k_{\nu } \eta_{\mu\mu}^2 \eta_{\nu\nu}^2 \left(\frac{i }{4 \pi ^2} \left(5  L_3 k^2 m^6-\frac{25  L_4}{2} m^8\right)\right)+
\0\\ & \quad 
 +  \left(k_{\nu }^2 \eta_{\mu\mu}^2 \eta_{\mu\nu} \eta_{\nu\nu}+k_{\mu }^2 \eta_{\mu\mu} \eta_{\mu\nu} \eta_{\nu\nu}^2\right) \left(\frac{i }{2 \pi ^2} \left(5  L_3 k^2 m^6-\frac{25  L_4}{2} m^8\right)\right)+
\0\\ & \quad 
 + k_{\mu } k_{\nu } \eta_{\mu\mu} \eta_{\mu\nu}^2 \eta_{\nu\nu} \left(\frac{i }{4 \pi ^2} \left(-15  L_2 k^4 m^4+40  L_3 k^2 m^6-100  L_4 m^8\right)\right)+
\0\\ & \quad 
 +  \left(k_{\nu }^2 \eta_{\mu\mu} \eta_{\mu\nu}^3+k_{\mu }^2 \eta_{\mu\nu}^3 \eta_{\nu\nu}\right) \left(\frac{i }{\pi ^2} \left(-\frac{5  L_2}{4} k^4 m^4+\frac{10  L_3}{3} k^2 m^6-\frac{25  L_4}{3} m^8\right)\right)+
\0\\ & \quad 
 + k_{\mu } k_{\nu } \eta_{\mu\nu}^4 \left(\frac{i }{\pi ^2} \left(\frac{5  L_1}{8} k^6 m^2+\frac{5  L_2}{8} k^4 m^4+\frac{10  L_3}{3} k^2 m^6-\frac{25  L_4}{3} m^8\right)\right)+
\0\\ & \quad 
 + \eta_{\mu\mu}^2 \eta_{\mu\nu} \eta_{\nu\nu}^2 \left(\frac{i }{4 \pi ^2} \left(-5  L_3 k^4 m^6+\frac{25  L_4}{2} k^2 m^8-15  L_5 m^{10}\right)\right)+
\0\\ & \quad 
 + \eta_{\mu\mu} \eta_{\mu\nu}^3 \eta_{\nu\nu} \left(\frac{i }{\pi ^2} \left(\frac{5  L_2}{4} k^6 m^4-\frac{10  L_3}{3} k^4 m^6+\frac{25  L_4}{3} k^2 m^8-10  L_5 m^{10}\right)\right)+
\0\\ & \quad 
 + \eta_{\mu\nu}^5 \left(\frac{i }{\pi ^2} \left(-\frac{ L_1}{8} k^8 m^2-\frac{ L_2}{8} k^6 m^4-\frac{2  L_3}{3} k^4 m^6+\frac{5  L_4}{3} k^2 m^8-2  L_5 m^{10}\right)\right) }
Scalars, spin 5 x 5, dimension 5:
\al{\label{eq:sc:5:5:5:}\tilde{T}_{5,5;5\text{D}}^{\text{s,t}} & = k^{10} \pi _{\mu\nu}^5 \left(\frac{i }{2 \pi ^2} \left(\frac{1}{3072} m+\frac{367 }{2304}\frac{ m^3}{k^2}+\frac{97 }{480}\frac{ m^5}{k^4}+\frac{11 }{40}\frac{ m^7}{k^6}-\frac{17 }{36}\frac{ m^9}{k^8}+\frac{1}{3}\frac{ m^{11}}{k^{10}}\right)+
\right.\0\\ & \quad \quad \left.
 + \frac{ T}{2 \pi } \left(\frac{1}{12288} k-\frac{1}{512}\frac{ m^2}{k}+\frac{5 }{256}\frac{ m^4}{k^3}-\frac{5 }{48}\frac{ m^6}{k^5}+\frac{5 }{16}\frac{ m^8}{k^7}-\frac{1}{2}\frac{ m^{10}}{k^9}+
\right.\right.\0\\ & \quad \quad \quad \left.\left.
 + \frac{1}{3}\frac{ m^{12}}{k^{11}}\right)\right)+
\0\\ & \quad 
 + k^{10} \pi _{\mu\mu} \pi _{\mu\nu}^3 \pi _{\nu\nu} \left(\frac{i }{2 \pi ^2} \left(\frac{5 }{3072} m-\frac{85 }{2304}\frac{ m^3}{k^2}-\frac{95 }{96}\frac{ m^5}{k^4}+\frac{11 }{8}\frac{ m^7}{k^6}-\frac{85 }{36}\frac{ m^9}{k^8}+
\right.\right.\0\\ & \quad \quad \quad \left.\left.
 + \frac{5 }{3}\frac{ m^{11}}{k^{10}}\right)+
\right.\0\\ & \quad \quad \left.
 + \frac{ T}{2 \pi } \left(\frac{5 }{12288} k-\frac{5 }{512}\frac{ m^2}{k}+\frac{25 }{256}\frac{ m^4}{k^3}-\frac{25 }{48}\frac{ m^6}{k^5}+\frac{25 }{16}\frac{ m^8}{k^7}-\frac{5 }{2}\frac{ m^{10}}{k^9}+
\right.\right.\0\\ & \quad \quad \quad \left.\left.
 + \frac{5 }{3}\frac{ m^{12}}{k^{11}}\right)\right)+
\0\\ & \quad 
 + k^{10} \pi _{\mu\mu}^2 \pi _{\mu\nu} \pi _{\nu\nu}^2 \left(\frac{i }{16 \pi ^2} \left(\frac{5 }{1024} m-\frac{85 }{768}\frac{ m^3}{k^2}+\frac{33 }{32}\frac{ m^5}{k^4}+\frac{33 }{8}\frac{ m^7}{k^6}-\frac{85 }{12}\frac{ m^9}{k^8}+
\right.\right.\0\\ & \quad \quad \quad \left.\left.
 + 5 \frac{ m^{11}}{k^{10}}\right)+
\right.\0\\ & \quad \quad \left.
 + \frac{ T}{16 \pi } \left(\frac{5 }{4096} k-\frac{15 }{512}\frac{ m^2}{k}+\frac{75 }{256}\frac{ m^4}{k^3}-\frac{25 }{16}\frac{ m^6}{k^5}+\frac{75 }{16}\frac{ m^8}{k^7}-\frac{15 }{2}\frac{ m^{10}}{k^9}+
\right.\right.\0\\ & \quad \quad \quad \left.\left.
 + 5 \frac{ m^{12}}{k^{11}}\right)\right) }
\al{\label{eq:snc:5:5:5:}\tilde{T}_{5,5;5\text{D}}^{\text{s,nt}} & = k_{\mu }^4 k_{\nu }^4 \eta_{\mu\nu} \left(-\frac{5 i }{12 \pi ^2} m^3\right)+k_{\mu }^3 k_{\nu }^3 \eta_{\mu\mu} \eta_{\nu\nu} \left(-\frac{2 i }{3 \pi ^2} m^5\right)+
\0\\ & \quad 
 +  \left(k_{\mu }^2 k_{\nu }^4 \eta_{\mu\mu} \eta_{\mu\nu}+k_{\mu }^4 k_{\nu }^2 \eta_{\mu\nu} \eta_{\nu\nu}\right) \left(-\frac{2 i }{\pi ^2} m^5\right)+
\0\\ & \quad 
 + k_{\mu }^3 k_{\nu }^3 \eta_{\mu\nu}^2 \left(\frac{i }{3 \pi ^2} \left(\frac{5 }{2} k^2 m^3-4  m^5\right)\right)+
\0\\ & \quad 
 +  \left(k_{\mu } k_{\nu }^3 \eta_{\mu\mu}^2 \eta_{\nu\nu}+k_{\mu }^3 k_{\nu } \eta_{\mu\mu} \eta_{\nu\nu}^2\right) \left(-\frac{8 i }{7 \pi ^2} m^7\right)+
\0\\ & \quad 
 +  \left(k_{\nu }^4 \eta_{\mu\mu}^2 \eta_{\mu\nu}+k_{\mu }^4 \eta_{\mu\nu} \eta_{\nu\nu}^2\right) \left(-\frac{4 i }{7 \pi ^2} m^7\right)+
\0\\ & \quad 
 + k_{\mu }^2 k_{\nu }^2 \eta_{\mu\mu} \eta_{\mu\nu} \eta_{\nu\nu} \left(\frac{i }{7 \pi ^2} \left(14  k^2 m^5-48  m^7\right)\right)+
\0\\ & \quad 
 +  \left(k_{\mu } k_{\nu }^3 \eta_{\mu\mu} \eta_{\mu\nu}^2+k_{\mu }^3 k_{\nu } \eta_{\mu\nu}^2 \eta_{\nu\nu}\right) \left(\frac{i }{7 \pi ^2} \left(14  k^2 m^5-32  m^7\right)\right)+
\0\\ & \quad 
 + k_{\mu }^2 k_{\nu }^2 \eta_{\mu\nu}^3 \left(\frac{i }{\pi ^2} \left(-\frac{5 }{6} k^4 m^3-\frac{32 }{7} m^7\right)\right)+
\0\\ & \quad 
 + k_{\mu } k_{\nu } \eta_{\mu\mu}^2 \eta_{\nu\nu}^2 \left(\frac{i }{7 \pi ^2} \left(4  k^2 m^7-\frac{80 }{9} m^9\right)\right)+
\0\\ & \quad 
 +  \left(k_{\nu }^2 \eta_{\mu\mu}^2 \eta_{\mu\nu} \eta_{\nu\nu}+k_{\mu }^2 \eta_{\mu\mu} \eta_{\mu\nu} \eta_{\nu\nu}^2\right) \left(\frac{i }{7 \pi ^2} \left(8  k^2 m^7-\frac{160 }{9} m^9\right)\right)+
\0\\ & \quad 
 + k_{\mu } k_{\nu } \eta_{\mu\mu} \eta_{\mu\nu}^2 \eta_{\nu\nu} \left(\frac{i }{7 \pi ^2} \left(-14  k^4 m^5+32  k^2 m^7-\frac{640 }{9} m^9\right)\right)+
\0\\ & \quad 
 +  \left(k_{\nu }^2 \eta_{\mu\mu} \eta_{\mu\nu}^3+k_{\mu }^2 \eta_{\mu\nu}^3 \eta_{\nu\nu}\right) \left(\frac{i }{3 \pi ^2} \left(-2  k^4 m^5+\frac{32 }{7} k^2 m^7-\frac{640 }{63} m^9\right)\right)+
\0\\ & \quad 
 + k_{\mu } k_{\nu } \eta_{\mu\nu}^4 \left(\frac{i }{3 \pi ^2} \left(\frac{5 }{4} k^6 m^3+k^4 m^5+\frac{32 }{7} k^2 m^7-\frac{640 }{63} m^9\right)\right)+
\0\\ & \quad 
 + \eta_{\mu\mu}^2 \eta_{\mu\nu} \eta_{\nu\nu}^2 \left(\frac{i }{7 \pi ^2} \left(-4  k^4 m^7+\frac{80 }{9} k^2 m^9-\frac{320 }{33} m^{11}\right)\right)+
\0\\ & \quad 
 + \eta_{\mu\mu} \eta_{\mu\nu}^3 \eta_{\nu\nu} \left(\frac{i }{3 \pi ^2} \left(2  k^6 m^5-\frac{32 }{7} k^4 m^7+\frac{640 }{63} k^2 m^9-\frac{2560 }{231} m^{11}\right)\right)+
\0\\ & \quad 
 + \eta_{\mu\nu}^5 \left(\frac{i }{3 \pi ^2} \left(-\frac{1}{4} k^8 m^3-\frac{1}{5} k^6 m^5-\frac{32 }{35} k^4 m^7+\frac{128 }{63} k^2 m^9-\frac{512 }{231} m^{11}\right)\right) }
Scalars, spin 5 x 5, dimension 6:
\al{\label{eq:sc:5:5:6:}\tilde{T}_{5,5;6\text{D}}^{\text{s,t}} & = k^{10} \pi _{\mu\nu}^5 \left(\frac{i }{7 \pi ^3} \left( \left(-\frac{88069}{463783320}+\frac{L_0}{20592}\right) k^2+ \left(\frac{172673}{35675640}-\frac{L_0}{792}\right) m^2+
\right.\right.\0\\ & \quad \quad \quad \left.\left.
 +  \left(\frac{5813737}{51891840}-\frac{55 L_0}{576}\right)\frac{ m^4}{k^2}+ \left(\frac{3782563}{8648640}-\frac{5 L_0}{32}\right)\frac{ m^6}{k^4}-\frac{736 }{1755}\frac{ m^8}{k^6}+
\right.\right.\0\\ & \quad \quad \quad \left.\left.
 + \frac{2432 }{3861}\frac{ m^{10}}{k^8}-\frac{512 }{1287}\frac{ m^{12}}{k^{10}}\right)+
\right.\0\\ & \quad \quad \left.
 + \frac{i  S}{3003 \pi ^3} \left(\frac{1}{24} k-\frac{ m^2}{k}+10 \frac{ m^4}{k^3}-\frac{160 }{3}\frac{ m^6}{k^5}+160 \frac{ m^8}{k^7}-256 \frac{ m^{10}}{k^9}+
\right.\right.\0\\ & \quad \quad \quad \left.\left.
 + \frac{512 }{3}\frac{ m^{12}}{k^{11}}\right)\right)+
\0\\ & \quad 
 + k^{10} \pi _{\mu\mu} \pi _{\mu\nu}^3 \pi _{\nu\nu} \left(\frac{i }{7 \pi ^3} \left( \left(-\frac{88069}{92756664}+\frac{5 L_0}{20592}\right) k^2+
\right.\right.\0\\ & \quad \quad \quad \left.\left.
 +  \left(\frac{172673}{7135128}-\frac{5 L_0}{792}\right) m^2+ \left(-\frac{337471}{1297296}+\frac{5 L_0}{72}\right)\frac{ m^4}{k^2}+
\right.\right.\0\\ & \quad \quad \quad \left.\left.
 +  \left(\frac{157049}{864864}+\frac{5 L_0}{16}\right)\frac{ m^6}{k^4}-\frac{736 }{351}\frac{ m^8}{k^6}+\frac{12160 }{3861}\frac{ m^{10}}{k^8}-\frac{2560 }{1287}\frac{ m^{12}}{k^{10}}\right)+
\right.\0\\ & \quad \quad \left.
 + \frac{i  S}{3003 \pi ^3} \left(\frac{5 }{24} k-5 \frac{ m^2}{k}+50 \frac{ m^4}{k^3}-\frac{800 }{3}\frac{ m^6}{k^5}+800 \frac{ m^8}{k^7}-1280 \frac{ m^{10}}{k^9}+
\right.\right.\0\\ & \quad \quad \quad \left.\left.
 + \frac{2560 }{3}\frac{ m^{12}}{k^{11}}\right)\right)+
\0\\ & \quad 
 + k^{10} \pi _{\mu\mu}^2 \pi _{\mu\nu} \pi _{\nu\nu}^2 \left(\frac{i }{7 \pi ^3} \left( \left(-\frac{88069}{247351104}+\frac{5 L_0}{54912}\right) k^2+
\right.\right.\0\\ & \quad \quad \quad \left.\left.
 +  \left(\frac{172673}{19027008}-\frac{5 L_0}{2112}\right) m^2+ \left(-\frac{337471}{3459456}+\frac{5 L_0}{192}\right)\frac{ m^4}{k^2}+
\right.\right.\0\\ & \quad \quad \quad \left.\left.
 +  \left(\frac{328301}{576576}-\frac{5 L_0}{32}\right)\frac{ m^6}{k^4}-\frac{92 }{117}\frac{ m^8}{k^6}+\frac{1520 }{1287}\frac{ m^{10}}{k^8}-\frac{320 }{429}\frac{ m^{12}}{k^{10}}\right)+
\right.\0\\ & \quad \quad \left.
 + \frac{i  S}{1001 \pi ^3} \left(\frac{5 }{192} k-\frac{5 }{8}\frac{ m^2}{k}+\frac{25 }{4}\frac{ m^4}{k^3}-\frac{100 }{3}\frac{ m^6}{k^5}+100 \frac{ m^8}{k^7}-160 \frac{ m^{10}}{k^9}+
\right.\right.\0\\ & \quad \quad \quad \left.\left.
 + \frac{320 }{3}\frac{ m^{12}}{k^{11}}\right)\right) }
\al{\label{eq:snc:5:5:6:}\tilde{T}_{5,5;6\text{D}}^{\text{s,nt}} & = k_{\mu }^4 k_{\nu }^4 \eta_{\mu\nu} \left(\frac{5 i  L_2}{64 \pi ^3} m^4\right)+k_{\mu }^3 k_{\nu }^3 \eta_{\mu\mu} \eta_{\nu\nu} \left(\frac{5 i  L_3}{48 \pi ^3} m^6\right)+
\0\\ & \quad 
 +  \left(k_{\mu }^2 k_{\nu }^4 \eta_{\mu\mu} \eta_{\mu\nu}+k_{\mu }^4 k_{\nu }^2 \eta_{\mu\nu} \eta_{\nu\nu}\right) \left(\frac{5 i  L_3}{16 \pi ^3} m^6\right)+
\0\\ & \quad 
 + k_{\mu }^3 k_{\nu }^3 \eta_{\mu\nu}^2 \left(\frac{i }{8 \pi ^3} \left(-\frac{5  L_2}{4} k^2 m^4+\frac{5  L_3}{3} m^6\right)\right)+
\0\\ & \quad 
 +  \left(k_{\mu } k_{\nu }^3 \eta_{\mu\mu}^2 \eta_{\nu\nu}+k_{\mu }^3 k_{\nu } \eta_{\mu\mu} \eta_{\nu\nu}^2\right) \left(\frac{5 i  L_4}{32 \pi ^3} m^8\right)+
\0\\ & \quad 
 +  \left(k_{\nu }^4 \eta_{\mu\mu}^2 \eta_{\mu\nu}+k_{\mu }^4 \eta_{\mu\nu} \eta_{\nu\nu}^2\right) \left(\frac{5 i  L_4}{64 \pi ^3} m^8\right)+
\0\\ & \quad 
 + k_{\mu }^2 k_{\nu }^2 \eta_{\mu\mu} \eta_{\mu\nu} \eta_{\nu\nu} \left(\frac{i }{16 \pi ^3} \left(-5  L_3 k^2 m^6+15  L_4 m^8\right)\right)+
\0\\ & \quad 
 +  \left(k_{\mu } k_{\nu }^3 \eta_{\mu\mu} \eta_{\mu\nu}^2+k_{\mu }^3 k_{\nu } \eta_{\mu\nu}^2 \eta_{\nu\nu}\right) \left(\frac{i }{8 \pi ^3} \left(-\frac{5  L_3}{2} k^2 m^6+5  L_4 m^8\right)\right)+
\0\\ & \quad 
 + k_{\mu }^2 k_{\nu }^2 \eta_{\mu\nu}^3 \left(\frac{i }{8 \pi ^3} \left(\frac{5  L_2}{4} k^4 m^4+5  L_4 m^8\right)\right)+
\0\\ & \quad 
 + k_{\mu } k_{\nu } \eta_{\mu\mu}^2 \eta_{\nu\nu}^2 \left(\frac{i }{32 \pi ^3} \left(-\frac{5  L_4}{2} k^2 m^8+5  L_5 m^{10}\right)\right)+
\0\\ & \quad 
 +  \left(k_{\nu }^2 \eta_{\mu\mu}^2 \eta_{\mu\nu} \eta_{\nu\nu}+k_{\mu }^2 \eta_{\mu\mu} \eta_{\mu\nu} \eta_{\nu\nu}^2\right) \left(\frac{i }{16 \pi ^3} \left(-\frac{5  L_4}{2} k^2 m^8+5  L_5 m^{10}\right)\right)+
\0\\ & \quad 
 + k_{\mu } k_{\nu } \eta_{\mu\mu} \eta_{\mu\nu}^2 \eta_{\nu\nu} \left(\frac{i }{4 \pi ^3} \left(\frac{5  L_3}{4} k^4 m^6-\frac{5  L_4}{2} k^2 m^8+5  L_5 m^{10}\right)\right)+
\0\\ & \quad 
 +  \left(k_{\nu }^2 \eta_{\mu\mu} \eta_{\mu\nu}^3+k_{\mu }^2 \eta_{\mu\nu}^3 \eta_{\nu\nu}\right) \left(\frac{i }{12 \pi ^3} \left(\frac{5  L_3}{4} k^4 m^6-\frac{5  L_4}{2} k^2 m^8+5  L_5 m^{10}\right)\right)+
\0\\ & \quad 
 + k_{\mu } k_{\nu } \eta_{\mu\nu}^4 \left(\frac{i }{4 \pi ^3} \left(-\frac{5  L_2}{16} k^6 m^4-\frac{5  L_3}{24} k^4 m^6-\frac{5  L_4}{6} k^2 m^8+\frac{5  L_5}{3} m^{10}\right)\right)+
\0\\ & \quad 
 + \eta_{\mu\mu}^2 \eta_{\mu\nu} \eta_{\nu\nu}^2 \left(\frac{i }{32 \pi ^3} \left(\frac{5  L_4}{2} k^4 m^8-5  L_5 k^2 m^{10}+5  L_6 m^{12}\right)\right)+
\0\\ & \quad 
 + \eta_{\mu\mu} \eta_{\mu\nu}^3 \eta_{\nu\nu} \left(\frac{i }{12 \pi ^3} \left(-\frac{5  L_3}{4} k^6 m^6+\frac{5  L_4}{2} k^4 m^8-5  L_5 k^2 m^{10}+5  L_6 m^{12}\right)\right)+
\0\\ & \quad 
 + \eta_{\mu\nu}^5 \left(\frac{i }{4 \pi ^3} \left(\frac{ L_2}{16} k^8 m^4+\frac{ L_3}{24} k^6 m^6+\frac{ L_4}{6} k^4 m^8-\frac{ L_5}{3} k^2 m^{10}+\frac{ L_6}{3} m^{12}\right)\right) }
\subsection{Expansions in UV and IR for scalars}
Scalars, spin 0 x 0, dimension 3:
\al{\label{eq:sc:0:0:3:uv}\tilde{T}_{0,0;3\text{D}}^{\text{s,UV}} & = -\frac{1}{8}\frac{1}{k}+\frac{i }{2 \pi }\frac{ m}{k^2}+\frac{2 i }{3 \pi }\frac{ m^3}{k^4}+\frac{8 i }{5 \pi }\frac{ m^5}{k^6}+\frac{32 i }{7 \pi }\frac{ m^7}{k^8}+\frac{128 i }{9 \pi }\frac{ m^9}{k^{10}}+\frac{512 i }{11 \pi }\frac{ m^{11}}{k^{12}}+\ldots }
\al{\label{eq:sc:0:0:3:ir}\tilde{T}_{0,0;3\text{D}}^{\text{s,IR}} & = \frac{i }{8 \pi } \left(\frac{1}{m}+\frac{1}{12}\frac{ k^2}{m^3}+\frac{1}{80}\frac{ k^4}{m^5}+\frac{1}{448}\frac{ k^6}{m^7}+\frac{1}{2304}\frac{ k^8}{m^9}+\frac{1}{11264}\frac{ k^{10}}{m^{11}}+\ldots\right) }
\al{\label{eq:sc:0:0:3:uvir}\tilde{T}_{0,0;3\text{D}}^{\text{s,UV-IR}} & =  \ldots \textrm{(i.e.\ no overlap)} }
Scalars, spin 0 x 0, dimension 4:
\al{\label{eq:sc:0:0:4:uv}\tilde{T}_{0,0;4\text{D}}^{\text{s,UV}} & = \frac{i }{4 \pi ^2} \left( \left(\frac{1}{2}-\frac{P}{4}\right)+ \left(\frac{1}{2}+\frac{K}{2}\right)\frac{ m^2}{k^2}+ \left(-\frac{1}{4}+\frac{K}{2}\right)\frac{ m^4}{k^4}+ \left(-\frac{5}{6}+K\right)\frac{ m^6}{k^6}+
\right.\0\\ & \quad \quad \left.
 +  \left(-\frac{59}{24}+\frac{5 K}{2}\right)\frac{ m^8}{k^8}+ \left(-\frac{449}{60}+7 K\right)\frac{ m^{10}}{k^{10}}+ \left(-\frac{1417}{60}+21 K\right)\frac{ m^{12}}{k^{12}}+
\right.\0\\ & \quad \quad \left.
 + \ldots\right) }
\al{\label{eq:sc:0:0:4:ir}\tilde{T}_{0,0;4\text{D}}^{\text{s,IR}} & = \frac{i }{16 \pi ^2} \left(- L_0+\frac{1}{6}\frac{ k^2}{m^2}+\frac{1}{60}\frac{ k^4}{m^4}+\frac{1}{420}\frac{ k^6}{m^6}+\frac{1}{2520}\frac{ k^8}{m^8}+\frac{1}{13860}\frac{ k^{10}}{m^{10}}+
\right.\0\\ & \quad \quad \left.
 + \frac{1}{72072}\frac{ k^{12}}{m^{12}}+\ldots\right) }
\al{\label{eq:sc:0:0:4:uvir}\tilde{T}_{0,0;4\text{D}}^{\text{s,UV-IR}} & = \frac{i  }{8 \pi ^2}\left(1-\frac{K}{2}\right)+\ldots }
Scalars, spin 0 x 0, dimension 5:
\al{\label{eq:sc:0:0:5:uv}\tilde{T}_{0,0;5\text{D}}^{\text{s,UV}} & = \frac{1}{\pi ^2} \left(-\frac{ \pi }{128} k+\frac{ \pi }{32}\frac{ m^2}{k}-\frac{i }{12}\frac{ m^3}{k^2}-\frac{i }{15}\frac{ m^5}{k^4}-\frac{4 i }{35}\frac{ m^7}{k^6}-\frac{16 i }{63}\frac{ m^9}{k^8}-\frac{64 i }{99}\frac{ m^{11}}{k^{10}}+\ldots\right) }
\al{\label{eq:sc:0:0:5:ir}\tilde{T}_{0,0;5\text{D}}^{\text{s,IR}} & = \frac{i }{16 \pi ^2} \left(- m+\frac{1}{12}\frac{ k^2}{m}+\frac{1}{240}\frac{ k^4}{m^3}+\frac{1}{2240}\frac{ k^6}{m^5}+\frac{1}{16128}\frac{ k^8}{m^7}+\frac{1}{101376}\frac{ k^{10}}{m^9}+
\right.\0\\ & \quad \quad \left.
 + \frac{1}{585728}\frac{ k^{12}}{m^{11}}+\ldots\right) }
\al{\label{eq:sc:0:0:5:uvir}\tilde{T}_{0,0;5\text{D}}^{\text{s,UV-IR}} & =  \ldots \textrm{(i.e.\ no overlap)} }
Scalars, spin 0 x 0, dimension 6:
\al{\label{eq:sc:0:0:6:uv}\tilde{T}_{0,0;6\text{D}}^{\text{s,UV}} & = \frac{i }{16 \pi ^3} \left( \left(\frac{1}{9}-\frac{P}{24}\right) k^2+ \left(-\frac{1}{2}+\frac{P}{4}\right) m^2- \left(\frac{3}{8}+\frac{K}{4}\right)\frac{ m^4}{k^2}+ \left(\frac{1}{36}-\frac{K}{6}\right)\frac{ m^6}{k^4}+
\right.\0\\ & \quad \quad \left.
 +  \left(\frac{7}{48}-\frac{K}{4}\right)\frac{ m^8}{k^6}+ \left(\frac{47}{120}-\frac{K}{2}\right)\frac{ m^{10}}{k^8}+ \left(\frac{379}{360}-\frac{7 K}{6}\right)\frac{ m^{12}}{k^{10}}+\ldots\right) }
\al{\label{eq:sc:0:0:6:ir}\tilde{T}_{0,0;6\text{D}}^{\text{s,IR}} & = \frac{i }{64 \pi ^3} \left( \left(-1+L_0\right) m^2-\frac{ L_0}{6} k^2+\frac{1}{60}\frac{ k^4}{m^2}+\frac{1}{840}\frac{ k^6}{m^4}+\frac{1}{7560}\frac{ k^8}{m^6}+\frac{1}{55440}\frac{ k^{10}}{m^8}+
\right.\0\\ & \quad \quad \left.
 + \frac{1}{360360}\frac{ k^{12}}{m^{10}}+\frac{1}{2162160}\frac{ k^{14}}{m^{12}}+\ldots\right) }
\al{\label{eq:sc:0:0:6:uvir}\tilde{T}_{0,0;6\text{D}}^{\text{s,UV-IR}} & = \frac{i }{16 \pi ^3} \left( \left(\frac{1}{9}-\frac{K}{24}\right) k^2+ \left(-\frac{1}{4}+\frac{K}{4}\right) m^2\right)+\ldots }
Scalars, spin 0 x 2, dimension 3:
\al{\label{eq:sc:0:2:3:uv}\tilde{T}_{0,2;3\text{D}}^{\text{s,t,UV}} & = k^2 \pi _{\nu\nu} \left(\frac{1}{16}\frac{1}{k}-\frac{i }{2 \pi }\frac{ m}{k^2}-\frac{1}{4}\frac{ m^2}{k^3}+\frac{2 i }{3 \pi }\frac{ m^3}{k^4}+\frac{8 i }{15 \pi }\frac{ m^5}{k^6}+\frac{32 i }{35 \pi }\frac{ m^7}{k^8}+\frac{128 i }{63 \pi }\frac{ m^9}{k^{10}}+
\right.\0\\ & \quad \quad \left.
 + \frac{512 i }{99 \pi }\frac{ m^{11}}{k^{12}}+\ldots\right) }
\al{\label{eq:sc:0:2:3:ir}\tilde{T}_{0,2;3\text{D}}^{\text{s,t,IR}} & = k^2 \pi _{\nu\nu} \left(\frac{i }{8 \pi } \left(-\frac{1}{3}\frac{1}{m}-\frac{1}{60}\frac{ k^2}{m^3}-\frac{1}{560}\frac{ k^4}{m^5}-\frac{1}{4032}\frac{ k^6}{m^7}-\frac{1}{25344}\frac{ k^8}{m^9}-\frac{1}{146432}\frac{ k^{10}}{m^{11}}+
\right.\right.\0\\ & \quad \quad \quad \left.\left.
 + \ldots\right)\right) }
\al{\label{eq:sc:0:2:3:uvir}\tilde{T}_{0,2;3\text{D}}^{\text{s,UV-IR}} & =  \ldots \textrm{(i.e.\ no overlap)} }
Scalars, spin 0 x 2, dimension 4:
\al{\label{eq:sc:0:2:4:uv}\tilde{T}_{0,2;4\text{D}}^{\text{s,t,UV}} & = k^2 \pi _{\nu\nu} \left(\frac{i }{2 \pi ^2} \left( \left(-\frac{1}{9}+\frac{P}{24}\right)+ \left(\frac{1}{4}-\frac{K}{4}\right)\frac{ m^2}{k^2}+ \left(\frac{3}{8}+\frac{K}{4}\right)\frac{ m^4}{k^4}+
\right.\right.\0\\ & \quad \quad \quad \left.\left.
 +  \left(-\frac{1}{36}+\frac{K}{6}\right)\frac{ m^6}{k^6}+ \left(-\frac{7}{48}+\frac{K}{4}\right)\frac{ m^8}{k^8}+ \left(-\frac{47}{120}+\frac{K}{2}\right)\frac{ m^{10}}{k^{10}}+
\right.\right.\0\\ & \quad \quad \quad \left.\left.
 +  \left(-\frac{379}{360}+\frac{7 K}{6}\right)\frac{ m^{12}}{k^{12}}+\ldots\right)\right) }
\al{\label{eq:sc:0:2:4:ir}\tilde{T}_{0,2;4\text{D}}^{\text{s,t,IR}} & = k^2 \pi _{\nu\nu} \left(\frac{i }{48 \pi ^2} \left( L_0-\frac{1}{10}\frac{ k^2}{m^2}-\frac{1}{140}\frac{ k^4}{m^4}-\frac{1}{1260}\frac{ k^6}{m^6}-\frac{1}{9240}\frac{ k^8}{m^8}-\frac{1}{60060}\frac{ k^{10}}{m^{10}}-
\right.\right.\0\\ & \quad \quad \quad \left.\left.
 - \frac{1}{360360}\frac{ k^{12}}{m^{12}}+\ldots\right)\right) }
\al{\label{eq:sc:0:2:4:uvir}\tilde{T}_{0,2;4\text{D}}^{\text{s,UV-IR}} & = k^2 \pi _{\nu\nu} \frac{i  }{6 \pi ^2}\left(-\frac{1}{3}+\frac{K}{8}\right)+\ldots }
Scalars, spin 0 x 2, dimension 5:
\al{\label{eq:sc:0:2:5:uv}\tilde{T}_{0,2;5\text{D}}^{\text{s,t,UV}} & = k^2 \pi _{\nu\nu} \left(\frac{1}{\pi ^2} \left(\frac{ \pi }{512} k-\frac{ \pi }{64}\frac{ m^2}{k}+\frac{i }{12}\frac{ m^3}{k^2}+\frac{ \pi }{32}\frac{ m^4}{k^3}-\frac{i }{15}\frac{ m^5}{k^4}-\frac{4 i }{105}\frac{ m^7}{k^6}-\frac{16 i }{315}\frac{ m^9}{k^8}-
\right.\right.\0\\ & \quad \quad \quad \left.\left.
 - \frac{64 i }{693}\frac{ m^{11}}{k^{10}}+\ldots\right)\right) }
\al{\label{eq:sc:0:2:5:ir}\tilde{T}_{0,2;5\text{D}}^{\text{s,t,IR}} & = k^2 \pi _{\nu\nu} \left(\frac{i }{48 \pi ^2} \left(m-\frac{1}{20}\frac{ k^2}{m}-\frac{1}{560}\frac{ k^4}{m^3}-\frac{1}{6720}\frac{ k^6}{m^5}-\frac{1}{59136}\frac{ k^8}{m^7}-\frac{1}{439296}\frac{ k^{10}}{m^9}-
\right.\right.\0\\ & \quad \quad \quad \left.\left.
 - \frac{1}{2928640}\frac{ k^{12}}{m^{11}}+\ldots\right)\right) }
\al{\label{eq:sc:0:2:5:uvir}\tilde{T}_{0,2;5\text{D}}^{\text{s,UV-IR}} & =  \ldots \textrm{(i.e.\ no overlap)} }
Scalars, spin 0 x 2, dimension 6:
\al{\label{eq:sc:0:2:6:uv}\tilde{T}_{0,2;6\text{D}}^{\text{s,t,UV}} & = k^2 \pi _{\nu\nu} \left(\frac{i }{8 \pi ^3} \left( \left(-\frac{23}{1800}+\frac{P}{240}\right) k^2+ \left(\frac{1}{9}-\frac{P}{24}\right) m^2+ \left(-\frac{1}{16}+\frac{K}{8}\right)\frac{ m^4}{k^2}-
\right.\right.\0\\ & \quad \quad \quad \left.\left.
 -  \left(\frac{11}{72}+\frac{K}{12}\right)\frac{ m^6}{k^4}- \left(\frac{1}{288}+\frac{K}{24}\right)\frac{ m^8}{k^6}+ \left(\frac{23}{1200}-\frac{K}{20}\right)\frac{ m^{10}}{k^8}+
\right.\right.\0\\ & \quad \quad \quad \left.\left.
 +  \left(\frac{37}{720}-\frac{K}{12}\right)\frac{ m^{12}}{k^{10}}+\ldots\right)\right) }
\al{\label{eq:sc:0:2:6:ir}\tilde{T}_{0,2;6\text{D}}^{\text{s,t,IR}} & = k^2 \pi _{\nu\nu} \left(\frac{i }{192 \pi ^3} \left( \left(1-L_0\right) m^2+\frac{ L_0}{10} k^2-\frac{1}{140}\frac{ k^4}{m^2}-\frac{1}{2520}\frac{ k^6}{m^4}-\frac{1}{27720}\frac{ k^8}{m^6}-
\right.\right.\0\\ & \quad \quad \quad \left.\left.
 - \frac{1}{240240}\frac{ k^{10}}{m^8}-\frac{1}{1801800}\frac{ k^{12}}{m^{10}}-\frac{1}{12252240}\frac{ k^{14}}{m^{12}}+\ldots\right)\right) }
\al{\label{eq:sc:0:2:6:uvir}\tilde{T}_{0,2;6\text{D}}^{\text{s,UV-IR}} & = k^2 \pi _{\nu\nu} \left(\frac{i }{192 \pi ^3} \left( \left(-\frac{23}{75}+\frac{K}{10}\right) k^2+ \left(\frac{5}{3}-K\right) m^2\right)\right)+\ldots }
Scalars, spin 0 x 4, dimension 3:
\al{\label{eq:sc:0:4:3:uv}\tilde{T}_{0,4;3\text{D}}^{\text{s,t,UV}} & = k^4 \pi _{\nu\nu}^2 \left(-\frac{3 }{64}\frac{1}{k}+\frac{i }{2 \pi }\frac{ m}{k^2}+\frac{3 }{8}\frac{ m^2}{k^3}-\frac{2 i }{\pi }\frac{ m^3}{k^4}-\frac{3 }{4}\frac{ m^4}{k^5}+\frac{8 i }{5 \pi }\frac{ m^5}{k^6}+\frac{32 i }{35 \pi }\frac{ m^7}{k^8}+
\right.\0\\ & \quad \quad \left.
 + \frac{128 i }{105 \pi }\frac{ m^9}{k^{10}}+\frac{512 i }{231 \pi }\frac{ m^{11}}{k^{12}}+\ldots\right) }
\al{\label{eq:sc:0:4:3:ir}\tilde{T}_{0,4;3\text{D}}^{\text{s,t,IR}} & = k^4 \pi _{\nu\nu}^2 \left(\frac{i }{8 \pi } \left(\frac{1}{5}\frac{1}{m}+\frac{1}{140}\frac{ k^2}{m^3}+\frac{1}{1680}\frac{ k^4}{m^5}+\frac{1}{14784}\frac{ k^6}{m^7}+\frac{1}{109824}\frac{ k^8}{m^9}+
\right.\right.\0\\ & \quad \quad \quad \left.\left.
 + \frac{1}{732160}\frac{ k^{10}}{m^{11}}+\ldots\right)\right) }
\al{\label{eq:sc:0:4:3:uvir}\tilde{T}_{0,4;3\text{D}}^{\text{s,UV-IR}} & =  \ldots \textrm{(i.e.\ no overlap)} }
Scalars, spin 0 x 4, dimension 4:
\al{\label{eq:sc:0:4:4:uv}\tilde{T}_{0,4;4\text{D}}^{\text{s,t,UV}} & = k^4 \pi _{\nu\nu}^2 \left(\frac{i }{4 \pi ^2} \left( \left(\frac{23}{150}-\frac{P}{20}\right)+ \left(-\frac{5}{6}+\frac{K}{2}\right)\frac{ m^2}{k^2}+ \left(\frac{3}{4}-\frac{3 K}{2}\right)\frac{ m^4}{k^4}+
\right.\right.\0\\ & \quad \quad \quad \left.\left.
 +  \left(\frac{11}{6}+K\right)\frac{ m^6}{k^6}+ \left(\frac{1}{24}+\frac{K}{2}\right)\frac{ m^8}{k^8}+ \left(-\frac{23}{100}+\frac{3 K}{5}\right)\frac{ m^{10}}{k^{10}}+
\right.\right.\0\\ & \quad \quad \quad \left.\left.
 +  \left(-\frac{37}{60}+K\right)\frac{ m^{12}}{k^{12}}+\ldots\right)\right) }
\al{\label{eq:sc:0:4:4:ir}\tilde{T}_{0,4;4\text{D}}^{\text{s,t,IR}} & = k^4 \pi _{\nu\nu}^2 \left(\frac{i }{80 \pi ^2} \left(- L_0+\frac{1}{14}\frac{ k^2}{m^2}+\frac{1}{252}\frac{ k^4}{m^4}+\frac{1}{2772}\frac{ k^6}{m^6}+\frac{1}{24024}\frac{ k^8}{m^8}+
\right.\right.\0\\ & \quad \quad \quad \left.\left.
 + \frac{1}{180180}\frac{ k^{10}}{m^{10}}+\frac{1}{1225224}\frac{ k^{12}}{m^{12}}+\ldots\right)\right) }
\al{\label{eq:sc:0:4:4:uvir}\tilde{T}_{0,4;4\text{D}}^{\text{s,UV-IR}} & = k^4 \pi _{\nu\nu}^2 \frac{i  }{40 \pi ^2}\left(\frac{23}{15}-\frac{K}{2}\right)+\ldots }
Scalars, spin 0 x 4, dimension 5:
\al{\label{eq:sc:0:4:5:uv}\tilde{T}_{0,4;5\text{D}}^{\text{s,t,UV}} & = k^4 \pi _{\nu\nu}^2 \left(\frac{1}{\pi ^2} \left(-\frac{ \pi }{1024} k+\frac{3  \pi }{256}\frac{ m^2}{k}-\frac{i }{12}\frac{ m^3}{k^2}-\frac{3  \pi }{64}\frac{ m^4}{k^3}+\frac{i }{5}\frac{ m^5}{k^4}+\frac{ \pi }{16}\frac{ m^6}{k^5}-\frac{4 i }{35}\frac{ m^7}{k^6}-
\right.\right.\0\\ & \quad \quad \quad \left.\left.
 - \frac{16 i }{315}\frac{ m^9}{k^8}-\frac{64 i }{1155}\frac{ m^{11}}{k^{10}}+\ldots\right)\right) }
\al{\label{eq:sc:0:4:5:ir}\tilde{T}_{0,4;5\text{D}}^{\text{s,t,IR}} & = k^4 \pi _{\nu\nu}^2 \left(\frac{i }{16 \pi ^2} \left(-\frac{1}{5} m+\frac{1}{140}\frac{ k^2}{m}+\frac{1}{5040}\frac{ k^4}{m^3}+\frac{1}{73920}\frac{ k^6}{m^5}+\frac{1}{768768}\frac{ k^8}{m^7}+
\right.\right.\0\\ & \quad \quad \quad \left.\left.
 + \frac{1}{6589440}\frac{ k^{10}}{m^9}+\frac{1}{49786880}\frac{ k^{12}}{m^{11}}+\ldots\right)\right) }
\al{\label{eq:sc:0:4:5:uvir}\tilde{T}_{0,4;5\text{D}}^{\text{s,UV-IR}} & =  \ldots \textrm{(i.e.\ no overlap)} }
Scalars, spin 0 x 4, dimension 6:
\al{\label{eq:sc:0:4:6:uv}\tilde{T}_{0,4;6\text{D}}^{\text{s,t,UV}} & = k^4 \pi _{\nu\nu}^2 \left(\frac{i }{4 \pi ^3} \left( \left(\frac{11}{3675}-\frac{P}{1120}\right) k^2+ \left(-\frac{23}{600}+\frac{P}{80}\right) m^2+ \left(\frac{7}{96}-\frac{K}{16}\right)\frac{ m^4}{k^2}+
\right.\right.\0\\ & \quad \quad \quad \left.\left.
 +  \left(-\frac{1}{48}+\frac{K}{8}\right)\frac{ m^6}{k^4}- \left(\frac{25}{192}+\frac{K}{16}\right)\frac{ m^8}{k^6}- \left(\frac{17}{2400}+\frac{K}{40}\right)\frac{ m^{10}}{k^8}+
\right.\right.\0\\ & \quad \quad \quad \left.\left.
 +  \left(\frac{13}{2400}-\frac{K}{40}\right)\frac{ m^{12}}{k^{10}}+\ldots\right)\right) }
\al{\label{eq:sc:0:4:6:ir}\tilde{T}_{0,4;6\text{D}}^{\text{s,t,IR}} & = k^4 \pi _{\nu\nu}^2 \left(\frac{i }{320 \pi ^3} \left( \left(-1+L_0\right) m^2-\frac{ L_0}{14} k^2+\frac{1}{252}\frac{ k^4}{m^2}+\frac{1}{5544}\frac{ k^6}{m^4}+\frac{1}{72072}\frac{ k^8}{m^6}+
\right.\right.\0\\ & \quad \quad \quad \left.\left.
 + \frac{1}{720720}\frac{ k^{10}}{m^8}+\frac{1}{6126120}\frac{ k^{12}}{m^{10}}+\frac{1}{46558512}\frac{ k^{14}}{m^{12}}+\ldots\right)\right) }
\al{\label{eq:sc:0:4:6:uvir}\tilde{T}_{0,4;6\text{D}}^{\text{s,UV-IR}} & = k^4 \pi _{\nu\nu}^2 \left(\frac{i }{20 \pi ^3} \left( \left(\frac{11}{735}-\frac{K}{224}\right) k^2+ \left(-\frac{31}{240}+\frac{K}{16}\right) m^2\right)\right)+\ldots }
Scalars, spin 1 x 1, dimension 3:
\al{\label{eq:sc:1:1:3:uv}\tilde{T}_{1,1;3\text{D}}^{\text{s,t,UV}} & = k^2 \pi _{\mu\nu} \left(\frac{1}{16}\frac{1}{k}-\frac{i }{2 \pi }\frac{ m}{k^2}-\frac{1}{4}\frac{ m^2}{k^3}+\frac{2 i }{3 \pi }\frac{ m^3}{k^4}+\frac{8 i }{15 \pi }\frac{ m^5}{k^6}+\frac{32 i }{35 \pi }\frac{ m^7}{k^8}+\frac{128 i }{63 \pi }\frac{ m^9}{k^{10}}+
\right.\0\\ & \quad \quad \left.
 + \frac{512 i }{99 \pi }\frac{ m^{11}}{k^{12}}+\ldots\right) }
\al{\label{eq:sc:1:1:3:ir}\tilde{T}_{1,1;3\text{D}}^{\text{s,t,IR}} & = k^2 \pi _{\mu\nu} \left(\frac{i }{8 \pi } \left(-\frac{1}{3}\frac{1}{m}-\frac{1}{60}\frac{ k^2}{m^3}-\frac{1}{560}\frac{ k^4}{m^5}-\frac{1}{4032}\frac{ k^6}{m^7}-\frac{1}{25344}\frac{ k^8}{m^9}-\frac{1}{146432}\frac{ k^{10}}{m^{11}}+
\right.\right.\0\\ & \quad \quad \quad \left.\left.
 + \ldots\right)\right) }
\al{\label{eq:sc:1:1:3:uvir}\tilde{T}_{1,1;3\text{D}}^{\text{s,UV-IR}} & =  \ldots \textrm{(i.e.\ no overlap)} }
Scalars, spin 1 x 1, dimension 4:
\al{\label{eq:sc:1:1:4:uv}\tilde{T}_{1,1;4\text{D}}^{\text{s,t,UV}} & = k^2 \pi _{\mu\nu} \left(\frac{i }{2 \pi ^2} \left( \left(-\frac{1}{9}+\frac{P}{24}\right)+ \left(\frac{1}{4}-\frac{K}{4}\right)\frac{ m^2}{k^2}+ \left(\frac{3}{8}+\frac{K}{4}\right)\frac{ m^4}{k^4}+
\right.\right.\0\\ & \quad \quad \quad \left.\left.
 +  \left(-\frac{1}{36}+\frac{K}{6}\right)\frac{ m^6}{k^6}+ \left(-\frac{7}{48}+\frac{K}{4}\right)\frac{ m^8}{k^8}+ \left(-\frac{47}{120}+\frac{K}{2}\right)\frac{ m^{10}}{k^{10}}+
\right.\right.\0\\ & \quad \quad \quad \left.\left.
 +  \left(-\frac{379}{360}+\frac{7 K}{6}\right)\frac{ m^{12}}{k^{12}}+\ldots\right)\right) }
\al{\label{eq:sc:1:1:4:ir}\tilde{T}_{1,1;4\text{D}}^{\text{s,t,IR}} & = k^2 \pi _{\mu\nu} \left(\frac{i }{48 \pi ^2} \left( L_0-\frac{1}{10}\frac{ k^2}{m^2}-\frac{1}{140}\frac{ k^4}{m^4}-\frac{1}{1260}\frac{ k^6}{m^6}-\frac{1}{9240}\frac{ k^8}{m^8}-\frac{1}{60060}\frac{ k^{10}}{m^{10}}-
\right.\right.\0\\ & \quad \quad \quad \left.\left.
 - \frac{1}{360360}\frac{ k^{12}}{m^{12}}+\ldots\right)\right) }
\al{\label{eq:sc:1:1:4:uvir}\tilde{T}_{1,1;4\text{D}}^{\text{s,UV-IR}} & = k^2 \pi _{\mu\nu} \frac{i  }{6 \pi ^2}\left(-\frac{1}{3}+\frac{K}{8}\right)+\ldots }
Scalars, spin 1 x 1, dimension 5:
\al{\label{eq:sc:1:1:5:uv}\tilde{T}_{1,1;5\text{D}}^{\text{s,t,UV}} & = k^2 \pi _{\mu\nu} \left(\frac{1}{\pi ^2} \left(\frac{ \pi }{512} k-\frac{ \pi }{64}\frac{ m^2}{k}+\frac{i }{12}\frac{ m^3}{k^2}+\frac{ \pi }{32}\frac{ m^4}{k^3}-\frac{i }{15}\frac{ m^5}{k^4}-\frac{4 i }{105}\frac{ m^7}{k^6}-\frac{16 i }{315}\frac{ m^9}{k^8}-
\right.\right.\0\\ & \quad \quad \quad \left.\left.
 - \frac{64 i }{693}\frac{ m^{11}}{k^{10}}+\ldots\right)\right) }
\al{\label{eq:sc:1:1:5:ir}\tilde{T}_{1,1;5\text{D}}^{\text{s,t,IR}} & = k^2 \pi _{\mu\nu} \left(\frac{i }{48 \pi ^2} \left(m-\frac{1}{20}\frac{ k^2}{m}-\frac{1}{560}\frac{ k^4}{m^3}-\frac{1}{6720}\frac{ k^6}{m^5}-\frac{1}{59136}\frac{ k^8}{m^7}-\frac{1}{439296}\frac{ k^{10}}{m^9}-
\right.\right.\0\\ & \quad \quad \quad \left.\left.
 - \frac{1}{2928640}\frac{ k^{12}}{m^{11}}+\ldots\right)\right) }
\al{\label{eq:sc:1:1:5:uvir}\tilde{T}_{1,1;5\text{D}}^{\text{s,UV-IR}} & =  \ldots \textrm{(i.e.\ no overlap)} }
Scalars, spin 1 x 1, dimension 6:
\al{\label{eq:sc:1:1:6:uv}\tilde{T}_{1,1;6\text{D}}^{\text{s,t,UV}} & = k^2 \pi _{\mu\nu} \left(\frac{i }{8 \pi ^3} \left( \left(-\frac{23}{1800}+\frac{P}{240}\right) k^2+ \left(\frac{1}{9}-\frac{P}{24}\right) m^2+ \left(-\frac{1}{16}+\frac{K}{8}\right)\frac{ m^4}{k^2}-
\right.\right.\0\\ & \quad \quad \quad \left.\left.
 -  \left(\frac{11}{72}+\frac{K}{12}\right)\frac{ m^6}{k^4}- \left(\frac{1}{288}+\frac{K}{24}\right)\frac{ m^8}{k^6}+ \left(\frac{23}{1200}-\frac{K}{20}\right)\frac{ m^{10}}{k^8}+
\right.\right.\0\\ & \quad \quad \quad \left.\left.
 +  \left(\frac{37}{720}-\frac{K}{12}\right)\frac{ m^{12}}{k^{10}}+\ldots\right)\right) }
\al{\label{eq:sc:1:1:6:ir}\tilde{T}_{1,1;6\text{D}}^{\text{s,t,IR}} & = k^2 \pi _{\mu\nu} \left(\frac{i }{192 \pi ^3} \left( \left(1-L_0\right) m^2+\frac{ L_0}{10} k^2-\frac{1}{140}\frac{ k^4}{m^2}-\frac{1}{2520}\frac{ k^6}{m^4}-\frac{1}{27720}\frac{ k^8}{m^6}-
\right.\right.\0\\ & \quad \quad \quad \left.\left.
 - \frac{1}{240240}\frac{ k^{10}}{m^8}-\frac{1}{1801800}\frac{ k^{12}}{m^{10}}-\frac{1}{12252240}\frac{ k^{14}}{m^{12}}+\ldots\right)\right) }
\al{\label{eq:sc:1:1:6:uvir}\tilde{T}_{1,1;6\text{D}}^{\text{s,UV-IR}} & = k^2 \pi _{\mu\nu} \left(\frac{i }{192 \pi ^3} \left( \left(-\frac{23}{75}+\frac{K}{10}\right) k^2+ \left(\frac{5}{3}-K\right) m^2\right)\right)+\ldots }
Scalars, spin 1 x 3, dimension 3:
\al{\label{eq:sc:1:3:3:uv}\tilde{T}_{1,3;3\text{D}}^{\text{s,t,UV}} & = k^4 \pi _{\nu\nu} \pi _{\mu\nu} \left(-\frac{3 }{64}\frac{1}{k}+\frac{i }{2 \pi }\frac{ m}{k^2}+\frac{3 }{8}\frac{ m^2}{k^3}-\frac{2 i }{\pi }\frac{ m^3}{k^4}-\frac{3 }{4}\frac{ m^4}{k^5}+\frac{8 i }{5 \pi }\frac{ m^5}{k^6}+\frac{32 i }{35 \pi }\frac{ m^7}{k^8}+
\right.\0\\ & \quad \quad \left.
 + \frac{128 i }{105 \pi }\frac{ m^9}{k^{10}}+\frac{512 i }{231 \pi }\frac{ m^{11}}{k^{12}}+\ldots\right) }
\al{\label{eq:sc:1:3:3:ir}\tilde{T}_{1,3;3\text{D}}^{\text{s,t,IR}} & = k^4 \pi _{\nu\nu} \pi _{\mu\nu} \left(\frac{i }{8 \pi } \left(\frac{1}{5}\frac{1}{m}+\frac{1}{140}\frac{ k^2}{m^3}+\frac{1}{1680}\frac{ k^4}{m^5}+\frac{1}{14784}\frac{ k^6}{m^7}+\frac{1}{109824}\frac{ k^8}{m^9}+
\right.\right.\0\\ & \quad \quad \quad \left.\left.
 + \frac{1}{732160}\frac{ k^{10}}{m^{11}}+\ldots\right)\right) }
\al{\label{eq:sc:1:3:3:uvir}\tilde{T}_{1,3;3\text{D}}^{\text{s,UV-IR}} & =  \ldots \textrm{(i.e.\ no overlap)} }
Scalars, spin 1 x 3, dimension 4:
\al{\label{eq:sc:1:3:4:uv}\tilde{T}_{1,3;4\text{D}}^{\text{s,t,UV}} & = k^4 \pi _{\nu\nu} \pi _{\mu\nu} \left(\frac{i }{4 \pi ^2} \left( \left(\frac{23}{150}-\frac{P}{20}\right)+ \left(-\frac{5}{6}+\frac{K}{2}\right)\frac{ m^2}{k^2}+ \left(\frac{3}{4}-\frac{3 K}{2}\right)\frac{ m^4}{k^4}+
\right.\right.\0\\ & \quad \quad \quad \left.\left.
 +  \left(\frac{11}{6}+K\right)\frac{ m^6}{k^6}+ \left(\frac{1}{24}+\frac{K}{2}\right)\frac{ m^8}{k^8}+ \left(-\frac{23}{100}+\frac{3 K}{5}\right)\frac{ m^{10}}{k^{10}}+
\right.\right.\0\\ & \quad \quad \quad \left.\left.
 +  \left(-\frac{37}{60}+K\right)\frac{ m^{12}}{k^{12}}+\ldots\right)\right) }
\al{\label{eq:sc:1:3:4:ir}\tilde{T}_{1,3;4\text{D}}^{\text{s,t,IR}} & = k^4 \pi _{\nu\nu} \pi _{\mu\nu} \left(\frac{i }{80 \pi ^2} \left(- L_0+\frac{1}{14}\frac{ k^2}{m^2}+\frac{1}{252}\frac{ k^4}{m^4}+\frac{1}{2772}\frac{ k^6}{m^6}+\frac{1}{24024}\frac{ k^8}{m^8}+
\right.\right.\0\\ & \quad \quad \quad \left.\left.
 + \frac{1}{180180}\frac{ k^{10}}{m^{10}}+\frac{1}{1225224}\frac{ k^{12}}{m^{12}}+\ldots\right)\right) }
\al{\label{eq:sc:1:3:4:uvir}\tilde{T}_{1,3;4\text{D}}^{\text{s,UV-IR}} & = k^4 \pi _{\mu\nu} \pi _{\nu\nu} \frac{i  }{40 \pi ^2}\left(\frac{23}{15}-\frac{K}{2}\right)+\ldots }
Scalars, spin 1 x 3, dimension 5:
\al{\label{eq:sc:1:3:5:uv}\tilde{T}_{1,3;5\text{D}}^{\text{s,t,UV}} & = k^4 \pi _{\nu\nu} \pi _{\mu\nu} \left(\frac{1}{\pi ^2} \left(-\frac{ \pi }{1024} k+\frac{3  \pi }{256}\frac{ m^2}{k}-\frac{i }{12}\frac{ m^3}{k^2}-\frac{3  \pi }{64}\frac{ m^4}{k^3}+\frac{i }{5}\frac{ m^5}{k^4}+\frac{ \pi }{16}\frac{ m^6}{k^5}-
\right.\right.\0\\ & \quad \quad \quad \left.\left.
 - \frac{4 i }{35}\frac{ m^7}{k^6}-\frac{16 i }{315}\frac{ m^9}{k^8}-\frac{64 i }{1155}\frac{ m^{11}}{k^{10}}+\ldots\right)\right) }
\al{\label{eq:sc:1:3:5:ir}\tilde{T}_{1,3;5\text{D}}^{\text{s,t,IR}} & = k^4 \pi _{\nu\nu} \pi _{\mu\nu} \left(\frac{i }{16 \pi ^2} \left(-\frac{1}{5} m+\frac{1}{140}\frac{ k^2}{m}+\frac{1}{5040}\frac{ k^4}{m^3}+\frac{1}{73920}\frac{ k^6}{m^5}+\frac{1}{768768}\frac{ k^8}{m^7}+
\right.\right.\0\\ & \quad \quad \quad \left.\left.
 + \frac{1}{6589440}\frac{ k^{10}}{m^9}+\frac{1}{49786880}\frac{ k^{12}}{m^{11}}+\ldots\right)\right) }
\al{\label{eq:sc:1:3:5:uvir}\tilde{T}_{1,3;5\text{D}}^{\text{s,UV-IR}} & =  \ldots \textrm{(i.e.\ no overlap)} }
Scalars, spin 1 x 3, dimension 6:
\al{\label{eq:sc:1:3:6:uv}\tilde{T}_{1,3;6\text{D}}^{\text{s,t,UV}} & = k^4 \pi _{\nu\nu} \pi _{\mu\nu} \left(\frac{i }{4 \pi ^3} \left( \left(\frac{11}{3675}-\frac{P}{1120}\right) k^2+ \left(-\frac{23}{600}+\frac{P}{80}\right) m^2+ \left(\frac{7}{96}-\frac{K}{16}\right)\frac{ m^4}{k^2}+
\right.\right.\0\\ & \quad \quad \quad \left.\left.
 +  \left(-\frac{1}{48}+\frac{K}{8}\right)\frac{ m^6}{k^4}- \left(\frac{25}{192}+\frac{K}{16}\right)\frac{ m^8}{k^6}- \left(\frac{17}{2400}+\frac{K}{40}\right)\frac{ m^{10}}{k^8}+
\right.\right.\0\\ & \quad \quad \quad \left.\left.
 +  \left(\frac{13}{2400}-\frac{K}{40}\right)\frac{ m^{12}}{k^{10}}+\ldots\right)\right) }
\al{\label{eq:sc:1:3:6:ir}\tilde{T}_{1,3;6\text{D}}^{\text{s,t,IR}} & = k^4 \pi _{\nu\nu} \pi _{\mu\nu} \left(\frac{i }{320 \pi ^3} \left( \left(-1+L_0\right) m^2-\frac{ L_0}{14} k^2+\frac{1}{252}\frac{ k^4}{m^2}+\frac{1}{5544}\frac{ k^6}{m^4}+\frac{1}{72072}\frac{ k^8}{m^6}+
\right.\right.\0\\ & \quad \quad \quad \left.\left.
 + \frac{1}{720720}\frac{ k^{10}}{m^8}+\frac{1}{6126120}\frac{ k^{12}}{m^{10}}+\frac{1}{46558512}\frac{ k^{14}}{m^{12}}+\ldots\right)\right) }
\al{\label{eq:sc:1:3:6:uvir}\tilde{T}_{1,3;6\text{D}}^{\text{s,UV-IR}} & = k^4 \pi _{\mu\nu} \pi _{\nu\nu} \left(\frac{i }{20 \pi ^3} \left( \left(\frac{11}{735}-\frac{K}{224}\right) k^2+ \left(-\frac{31}{240}+\frac{K}{16}\right) m^2\right)\right)+\ldots }
Scalars, spin 1 x 5, dimension 3:
\al{\label{eq:sc:1:5:3:uv}\tilde{T}_{1,5;3\text{D}}^{\text{s,t,UV}} & = k^6 \pi _{\nu\nu}^2 \pi _{\mu\nu} \left(\frac{5 }{128}\frac{1}{k}-\frac{i }{2 \pi }\frac{ m}{k^2}-\frac{15 }{32}\frac{ m^2}{k^3}+\frac{10 i }{3 \pi }\frac{ m^3}{k^4}+\frac{15 }{8}\frac{ m^4}{k^5}-\frac{8 i }{\pi }\frac{ m^5}{k^6}-\frac{5 }{2}\frac{ m^6}{k^7}+
\right.\0\\ & \quad \quad \left.
 + \frac{32 i }{7 \pi }\frac{ m^7}{k^8}+\frac{128 i }{63 \pi }\frac{ m^9}{k^{10}}+\frac{512 i }{231 \pi }\frac{ m^{11}}{k^{12}}+\ldots\right) }
\al{\label{eq:sc:1:5:3:ir}\tilde{T}_{1,5;3\text{D}}^{\text{s,t,IR}} & = k^6 \pi _{\nu\nu}^2 \pi _{\mu\nu} \left(\frac{i }{8 \pi } \left(-\frac{1}{7}\frac{1}{m}-\frac{1}{252}\frac{ k^2}{m^3}-\frac{1}{3696}\frac{ k^4}{m^5}-\frac{5 }{192192}\frac{ k^6}{m^7}-\frac{1}{329472}\frac{ k^8}{m^9}-
\right.\right.\0\\ & \quad \quad \quad \left.\left.
 - \frac{1}{2489344}\frac{ k^{10}}{m^{11}}+\ldots\right)\right) }
\al{\label{eq:sc:1:5:3:uvir}\tilde{T}_{1,5;3\text{D}}^{\text{s,UV-IR}} & =  \ldots \textrm{(i.e.\ no overlap)} }
Scalars, spin 1 x 5, dimension 4:
\al{\label{eq:sc:1:5:4:uv}\tilde{T}_{1,5;4\text{D}}^{\text{s,t,UV}} & = k^6 \pi _{\nu\nu}^2 \pi _{\mu\nu} \left(\frac{i }{\pi ^2} \left( \left(-\frac{22}{735}+\frac{P}{112}\right)+ \left(\frac{31}{120}-\frac{K}{8}\right)\frac{ m^2}{k^2}+ \left(-\frac{35}{48}+\frac{5 K}{8}\right)\frac{ m^4}{k^4}+
\right.\right.\0\\ & \quad \quad \quad \left.\left.
 +  \left(\frac{5}{24}-\frac{5 K}{4}\right)\frac{ m^6}{k^6}+ \left(\frac{125}{96}+\frac{5 K}{8}\right)\frac{ m^8}{k^8}+ \left(\frac{17}{240}+\frac{K}{4}\right)\frac{ m^{10}}{k^{10}}+
\right.\right.\0\\ & \quad \quad \quad \left.\left.
 +  \left(-\frac{13}{240}+\frac{K}{4}\right)\frac{ m^{12}}{k^{12}}+\ldots\right)\right) }
\al{\label{eq:sc:1:5:4:ir}\tilde{T}_{1,5;4\text{D}}^{\text{s,t,IR}} & = k^6 \pi _{\nu\nu}^2 \pi _{\mu\nu} \left(\frac{i }{112 \pi ^2} \left( L_0-\frac{1}{18}\frac{ k^2}{m^2}-\frac{1}{396}\frac{ k^4}{m^4}-\frac{1}{5148}\frac{ k^6}{m^6}-\frac{1}{51480}\frac{ k^8}{m^8}-
\right.\right.\0\\ & \quad \quad \quad \left.\left.
 - \frac{1}{437580}\frac{ k^{10}}{m^{10}}-\frac{1}{3325608}\frac{ k^{12}}{m^{12}}+\ldots\right)\right) }
\al{\label{eq:sc:1:5:4:uvir}\tilde{T}_{1,5;4\text{D}}^{\text{s,UV-IR}} & = k^6 \pi _{\mu\nu} \pi _{\nu\nu}^2 \frac{i  }{7 \pi ^2}\left(-\frac{22}{105}+\frac{K}{16}\right)+\ldots }
Scalars, spin 1 x 5, dimension 5:
\al{\label{eq:sc:1:5:5:uv}\tilde{T}_{1,5;5\text{D}}^{\text{s,t,UV}} & = k^6 \pi _{\nu\nu}^2 \pi _{\mu\nu} \left(\frac{1}{\pi ^2} \left(\frac{5  \pi }{8192} k-\frac{5  \pi }{512}\frac{ m^2}{k}+\frac{i }{12}\frac{ m^3}{k^2}+\frac{15  \pi }{256}\frac{ m^4}{k^3}-\frac{i }{3}\frac{ m^5}{k^4}-\frac{5  \pi }{32}\frac{ m^6}{k^5}+
\right.\right.\0\\ & \quad \quad \quad \left.\left.
 + \frac{4 i }{7}\frac{ m^7}{k^6}+\frac{5  \pi }{32}\frac{ m^8}{k^7}-\frac{16 i }{63}\frac{ m^9}{k^8}-\frac{64 i }{693}\frac{ m^{11}}{k^{10}}+\ldots\right)\right) }
\al{\label{eq:sc:1:5:5:ir}\tilde{T}_{1,5;5\text{D}}^{\text{s,t,IR}} & = k^6 \pi _{\nu\nu}^2 \pi _{\mu\nu} \left(\frac{i }{16 \pi ^2} \left(\frac{1}{7} m-\frac{1}{252}\frac{ k^2}{m}-\frac{1}{11088}\frac{ k^4}{m^3}-\frac{1}{192192}\frac{ k^6}{m^5}-\frac{1}{2306304}\frac{ k^8}{m^7}-
\right.\right.\0\\ & \quad \quad \quad \left.\left.
 - \frac{1}{22404096}\frac{ k^{10}}{m^9}-\frac{1}{189190144}\frac{ k^{12}}{m^{11}}+\ldots\right)\right) }
\al{\label{eq:sc:1:5:5:uvir}\tilde{T}_{1,5;5\text{D}}^{\text{s,UV-IR}} & =  \ldots \textrm{(i.e.\ no overlap)} }
Scalars, spin 1 x 5, dimension 6:
\al{\label{eq:sc:1:5:6:uv}\tilde{T}_{1,5;6\text{D}}^{\text{s,t,UV}} & = k^6 \pi _{\nu\nu}^2 \pi _{\mu\nu} \left(\frac{i }{2 \pi ^3} \left( \left(-\frac{563}{635040}+\frac{P}{4032}\right) k^2+ \left(\frac{11}{735}-\frac{P}{224}\right) m^2+
\right.\right.\0\\ & \quad \quad \quad \left.\left.
 +  \left(-\frac{47}{960}+\frac{K}{32}\right)\frac{ m^4}{k^2}+ \left(\frac{25}{288}-\frac{5 K}{48}\right)\frac{ m^6}{k^4}+ \left(\frac{5}{384}+\frac{5 K}{32}\right)\frac{ m^8}{k^6}-
\right.\right.\0\\ & \quad \quad \quad \left.\left.
 -  \left(\frac{137}{960}+\frac{K}{16}\right)\frac{ m^{10}}{k^8}- \left(\frac{3}{320}+\frac{K}{48}\right)\frac{ m^{12}}{k^{10}}+\ldots\right)\right) }
\al{\label{eq:sc:1:5:6:ir}\tilde{T}_{1,5;6\text{D}}^{\text{s,t,IR}} & = k^6 \pi _{\nu\nu}^2 \pi _{\mu\nu} \left(\frac{i }{448 \pi ^3} \left( \left(1-L_0\right) m^2+\frac{ L_0}{18} k^2-\frac{1}{396}\frac{ k^4}{m^2}-\frac{1}{10296}\frac{ k^6}{m^4}-\frac{1}{154440}\frac{ k^8}{m^6}-
\right.\right.\0\\ & \quad \quad \quad \left.\left.
 - \frac{1}{1750320}\frac{ k^{10}}{m^8}-\frac{1}{16628040}\frac{ k^{12}}{m^{10}}-\frac{1}{139675536}\frac{ k^{14}}{m^{12}}+\ldots\right)\right) }
\al{\label{eq:sc:1:5:6:uvir}\tilde{T}_{1,5;6\text{D}}^{\text{s,UV-IR}} & = k^6 \pi _{\mu\nu} \pi _{\nu\nu}^2 \left(\frac{i }{448 \pi ^3} \left( \left(-\frac{563}{2835}+\frac{K}{18}\right) k^2+ \left(\frac{247}{105}-K\right) m^2\right)\right)+\ldots }
Scalars, spin 2 x 2, dimension 3:
\al{\label{eq:sc:2:2:3:uv}\tilde{T}_{2,2;3\text{D}}^{\text{s,t,UV}} & = k^4 \pi _{\mu\nu}^2 \left(-\frac{1}{32}\frac{1}{k}+\frac{i }{2 \pi }\frac{ m}{k^2}+\frac{1}{4}\frac{ m^2}{k^3}-\frac{4 i }{3 \pi }\frac{ m^3}{k^4}-\frac{1}{2}\frac{ m^4}{k^5}+\frac{16 i }{15 \pi }\frac{ m^5}{k^6}+\frac{64 i }{105 \pi }\frac{ m^7}{k^8}+
\right.\0\\ & \quad \quad \left.
 + \frac{256 i }{315 \pi }\frac{ m^9}{k^{10}}+\frac{1024 i }{693 \pi }\frac{ m^{11}}{k^{12}}+\ldots\right)+
\0\\ & \quad 
 + k^4 \pi _{\mu\mu} \pi _{\nu\nu} \left(-\frac{1}{64}\frac{1}{k}+\frac{1}{8}\frac{ m^2}{k^3}-\frac{2 i }{3 \pi }\frac{ m^3}{k^4}-\frac{1}{4}\frac{ m^4}{k^5}+\frac{8 i }{15 \pi }\frac{ m^5}{k^6}+\frac{32 i }{105 \pi }\frac{ m^7}{k^8}+
\right.\0\\ & \quad \quad \left.
 + \frac{128 i }{315 \pi }\frac{ m^9}{k^{10}}+\frac{512 i }{693 \pi }\frac{ m^{11}}{k^{12}}+\ldots\right) }
\al{\label{eq:sc:2:2:3:ir}\tilde{T}_{2,2;3\text{D}}^{\text{s,t,IR}} & = k^4 \pi _{\mu\nu}^2 \left(\frac{i }{6 \pi } \left(\frac{m}{k^2}+\frac{1}{10}\frac{1}{m}+\frac{1}{280}\frac{ k^2}{m^3}+\frac{1}{3360}\frac{ k^4}{m^5}+\frac{1}{29568}\frac{ k^6}{m^7}+\frac{1}{219648}\frac{ k^8}{m^9}+
\right.\right.\0\\ & \quad \quad \quad \left.\left.
 + \frac{1}{1464320}\frac{ k^{10}}{m^{11}}+\ldots\right)\right)+
\0\\ & \quad 
 + k^4 \pi _{\mu\mu} \pi _{\nu\nu} \left(\frac{i }{6 \pi } \left(-\frac{ m}{k^2}+\frac{1}{20}\frac{1}{m}+\frac{1}{560}\frac{ k^2}{m^3}+\frac{1}{6720}\frac{ k^4}{m^5}+\frac{1}{59136}\frac{ k^6}{m^7}+
\right.\right.\0\\ & \quad \quad \quad \left.\left.
 + \frac{1}{439296}\frac{ k^8}{m^9}+\frac{1}{2928640}\frac{ k^{10}}{m^{11}}+\ldots\right)\right) }
\al{\label{eq:sc:2:2:3:uvir}\tilde{T}_{2,2;3\text{D}}^{\text{s,UV-IR}} & = k^4 \pi _{\mu\nu}^2 \left(\frac{i }{3 \pi }\frac{ m}{k^2}\right)+\ldots }
Scalars, spin 2 x 2, dimension 4:
\al{\label{eq:sc:2:2:4:uv}\tilde{T}_{2,2;4\text{D}}^{\text{s,t,UV}} & = k^4 \pi _{\mu\nu}^2 \left(\frac{i }{2 \pi ^2} \left( \left(\frac{23}{450}-\frac{P}{60}\right)- \left(\frac{7}{36}-\frac{P}{6}+\frac{L_0}{4}\right)\frac{ m^2}{k^2}+ \left(\frac{1}{4}-\frac{K}{2}\right)\frac{ m^4}{k^4}+
\right.\right.\0\\ & \quad \quad \quad \left.\left.
 +  \left(\frac{11}{18}+\frac{K}{3}\right)\frac{ m^6}{k^6}+ \left(\frac{1}{72}+\frac{K}{6}\right)\frac{ m^8}{k^8}+ \left(-\frac{23}{300}+\frac{K}{5}\right)\frac{ m^{10}}{k^{10}}+
\right.\right.\0\\ & \quad \quad \quad \left.\left.
 +  \left(-\frac{37}{180}+\frac{K}{3}\right)\frac{ m^{12}}{k^{12}}+\ldots\right)\right)+
\0\\ & \quad 
 + k^4 \pi _{\mu\mu} \pi _{\nu\nu} \left(\frac{i }{\pi ^2} \left( \left(\frac{23}{1800}-\frac{P}{240}\right)+ \left(-\frac{1}{9}+\frac{P}{24}\right)\frac{ m^2}{k^2}+ \left(\frac{1}{16}-\frac{K}{8}\right)\frac{ m^4}{k^4}+
\right.\right.\0\\ & \quad \quad \quad \left.\left.
 +  \left(\frac{11}{72}+\frac{K}{12}\right)\frac{ m^6}{k^6}+ \left(\frac{1}{288}+\frac{K}{24}\right)\frac{ m^8}{k^8}+ \left(-\frac{23}{1200}+\frac{K}{20}\right)\frac{ m^{10}}{k^{10}}+
\right.\right.\0\\ & \quad \quad \quad \left.\left.
 +  \left(-\frac{37}{720}+\frac{K}{12}\right)\frac{ m^{12}}{k^{12}}+\ldots\right)\right) }
\al{\label{eq:sc:2:2:4:ir}\tilde{T}_{2,2;4\text{D}}^{\text{s,t,IR}} & = k^4 \pi _{\mu\nu}^2 \left(\frac{i }{24 \pi ^2} \left( \left(1-L_0\right)\frac{ m^2}{k^2}-\frac{ L_0}{5}+\frac{1}{70}\frac{ k^2}{m^2}+\frac{1}{1260}\frac{ k^4}{m^4}+\frac{1}{13860}\frac{ k^6}{m^6}+
\right.\right.\0\\ & \quad \quad \quad \left.\left.
 + \frac{1}{120120}\frac{ k^8}{m^8}+\frac{1}{900900}\frac{ k^{10}}{m^{10}}+\frac{1}{6126120}\frac{ k^{12}}{m^{12}}+\ldots\right)\right)+
\0\\ & \quad 
 + k^4 \pi _{\mu\mu} \pi _{\nu\nu} \left(\frac{i }{24 \pi ^2} \left( \left(-1+L_0\right)\frac{ m^2}{k^2}-\frac{ L_0}{10}+\frac{1}{140}\frac{ k^2}{m^2}+\frac{1}{2520}\frac{ k^4}{m^4}+\frac{1}{27720}\frac{ k^6}{m^6}+
\right.\right.\0\\ & \quad \quad \quad \left.\left.
 + \frac{1}{240240}\frac{ k^8}{m^8}+\frac{1}{1801800}\frac{ k^{10}}{m^{10}}+\frac{1}{12252240}\frac{ k^{12}}{m^{12}}+\ldots\right)\right) }
\al{\label{eq:sc:2:2:4:uvir}\tilde{T}_{2,2;4\text{D}}^{\text{s,UV-IR}} & = k^4 \pi _{\mu\nu}^2 \left(\frac{i }{12 \pi ^2} \left( \left(\frac{23}{75}-\frac{K}{10}\right)+ \left(-\frac{5}{3}+K\right)\frac{ m^2}{k^2}\right)\right)+
\0\\ & \quad 
 + k^4 \pi _{\mu\mu} \pi _{\nu\nu} \left(\frac{i }{24 \pi ^2} \left( \left(\frac{23}{75}-\frac{K}{10}\right)+ \left(-\frac{5}{3}+K\right)\frac{ m^2}{k^2}\right)\right)+\ldots }
Scalars, spin 2 x 2, dimension 5:
\al{\label{eq:sc:2:2:5:uv}\tilde{T}_{2,2;5\text{D}}^{\text{s,t,UV}} & = k^4 \pi _{\mu\nu}^2 \left(\frac{1}{\pi ^2} \left(-\frac{ \pi }{1536} k+\frac{ \pi }{128}\frac{ m^2}{k}-\frac{i }{12}\frac{ m^3}{k^2}-\frac{ \pi }{32}\frac{ m^4}{k^3}+\frac{2 i }{15}\frac{ m^5}{k^4}+\frac{ \pi }{24}\frac{ m^6}{k^5}-
\right.\right.\0\\ & \quad \quad \quad \left.\left.
 - \frac{8 i }{105}\frac{ m^7}{k^6}-\frac{32 i }{945}\frac{ m^9}{k^8}-\frac{128 i }{3465}\frac{ m^{11}}{k^{10}}+\ldots\right)\right)+
\0\\ & \quad 
 + k^4 \pi _{\mu\mu} \pi _{\nu\nu} \left(\frac{1}{\pi ^2} \left(-\frac{ \pi }{3072} k+\frac{ \pi }{256}\frac{ m^2}{k}-\frac{ \pi }{64}\frac{ m^4}{k^3}+\frac{i }{15}\frac{ m^5}{k^4}+\frac{ \pi }{48}\frac{ m^6}{k^5}-\frac{4 i }{105}\frac{ m^7}{k^6}-
\right.\right.\0\\ & \quad \quad \quad \left.\left.
 - \frac{16 i }{945}\frac{ m^9}{k^8}-\frac{64 i }{3465}\frac{ m^{11}}{k^{10}}+\ldots\right)\right) }
\al{\label{eq:sc:2:2:5:ir}\tilde{T}_{2,2;5\text{D}}^{\text{s,t,IR}} & = k^4 \pi _{\mu\nu}^2 \left(\frac{i }{12 \pi ^2} \left(-\frac{1}{3}\frac{ m^3}{k^2}-\frac{1}{10} m+\frac{1}{280}\frac{ k^2}{m}+\frac{1}{10080}\frac{ k^4}{m^3}+\frac{1}{147840}\frac{ k^6}{m^5}+
\right.\right.\0\\ & \quad \quad \quad \left.\left.
 + \frac{1}{1537536}\frac{ k^8}{m^7}+\frac{1}{13178880}\frac{ k^{10}}{m^9}+\frac{1}{99573760}\frac{ k^{12}}{m^{11}}+\ldots\right)\right)+
\0\\ & \quad 
 + k^4 \pi _{\mu\mu} \pi _{\nu\nu} \left(\frac{i }{12 \pi ^2} \left(\frac{1}{3}\frac{ m^3}{k^2}-\frac{1}{20} m+\frac{1}{560}\frac{ k^2}{m}+\frac{1}{20160}\frac{ k^4}{m^3}+\frac{1}{295680}\frac{ k^6}{m^5}+
\right.\right.\0\\ & \quad \quad \quad \left.\left.
 + \frac{1}{3075072}\frac{ k^8}{m^7}+\frac{1}{26357760}\frac{ k^{10}}{m^9}+\frac{1}{199147520}\frac{ k^{12}}{m^{11}}+\ldots\right)\right) }
\al{\label{eq:sc:2:2:5:uvir}\tilde{T}_{2,2;5\text{D}}^{\text{s,UV-IR}} & = k^4 \pi _{\mu\nu}^2 \left(-\frac{i }{18 \pi ^2}\frac{ m^3}{k^2}\right)+\ldots }
Scalars, spin 2 x 2, dimension 6:
\al{\label{eq:sc:2:2:6:uv}\tilde{T}_{2,2;6\text{D}}^{\text{s,t,UV}} & = k^4 \pi _{\mu\nu}^2 \left(\frac{i }{2 \pi ^3} \left( \left(\frac{11}{11025}-\frac{P}{3360}\right) k^2+ \left(-\frac{23}{1800}+\frac{P}{240}\right) m^2+
\right.\right.\0\\ & \quad \quad \quad \left.\left.
 +  \left(\frac{5}{576}-\frac{P}{48}+\frac{L_0}{32}\right)\frac{ m^4}{k^2}+ \left(-\frac{1}{144}+\frac{K}{24}\right)\frac{ m^6}{k^4}- \left(\frac{25}{576}+\frac{K}{48}\right)\frac{ m^8}{k^6}-
\right.\right.\0\\ & \quad \quad \quad \left.\left.
 -  \left(\frac{17}{7200}+\frac{K}{120}\right)\frac{ m^{10}}{k^8}+ \left(\frac{13}{7200}-\frac{K}{120}\right)\frac{ m^{12}}{k^{10}}+\ldots\right)\right)+
\0\\ & \quad 
 + k^4 \pi _{\mu\mu} \pi _{\nu\nu} \left(\frac{i }{12 \pi ^3} \left( \left(\frac{11}{3675}-\frac{P}{1120}\right) k^2+ \left(-\frac{23}{600}+\frac{P}{80}\right) m^2+
\right.\right.\0\\ & \quad \quad \quad \left.\left.
 +  \left(\frac{1}{6}-\frac{P}{16}\right)\frac{ m^4}{k^2}+ \left(-\frac{1}{48}+\frac{K}{8}\right)\frac{ m^6}{k^4}- \left(\frac{25}{192}+\frac{K}{16}\right)\frac{ m^8}{k^6}-
\right.\right.\0\\ & \quad \quad \quad \left.\left.
 -  \left(\frac{17}{2400}+\frac{K}{40}\right)\frac{ m^{10}}{k^8}+ \left(\frac{13}{2400}-\frac{K}{40}\right)\frac{ m^{12}}{k^{10}}+\ldots\right)\right) }
\al{\label{eq:sc:2:2:6:ir}\tilde{T}_{2,2;6\text{D}}^{\text{s,t,IR}} & = k^4 \pi _{\mu\nu}^2 \left(\frac{i }{32 \pi ^3} \left( \left(-\frac{1}{4}+\frac{L_0}{6}\right)\frac{ m^4}{k^2}+ \left(-\frac{1}{15}+\frac{L_0}{15}\right) m^2-\frac{ L_0}{210} k^2+\frac{1}{3780}\frac{ k^4}{m^2}+
\right.\right.\0\\ & \quad \quad \quad \left.\left.
 + \frac{1}{83160}\frac{ k^6}{m^4}+\frac{1}{1081080}\frac{ k^8}{m^6}+\frac{1}{10810800}\frac{ k^{10}}{m^8}+\frac{1}{91891800}\frac{ k^{12}}{m^{10}}+
\right.\right.\0\\ & \quad \quad \quad \left.\left.
 + \frac{1}{698377680}\frac{ k^{14}}{m^{12}}+\ldots\right)\right)+
\0\\ & \quad 
 + k^4 \pi _{\mu\mu} \pi _{\nu\nu} \left(\frac{i }{64 \pi ^3} \left( \left(\frac{1}{2}-\frac{L_0}{3}\right)\frac{ m^4}{k^2}+ \left(-\frac{1}{15}+\frac{L_0}{15}\right) m^2-\frac{ L_0}{210} k^2+
\right.\right.\0\\ & \quad \quad \quad \left.\left.
 + \frac{1}{3780}\frac{ k^4}{m^2}+\frac{1}{83160}\frac{ k^6}{m^4}+\frac{1}{1081080}\frac{ k^8}{m^6}+\frac{1}{10810800}\frac{ k^{10}}{m^8}+
\right.\right.\0\\ & \quad \quad \quad \left.\left.
 + \frac{1}{91891800}\frac{ k^{12}}{m^{10}}+\frac{1}{698377680}\frac{ k^{14}}{m^{12}}+\ldots\right)\right) }
\al{\label{eq:sc:2:2:6:uvir}\tilde{T}_{2,2;6\text{D}}^{\text{s,UV-IR}} & = k^4 \pi _{\mu\nu}^2 \left(\frac{i }{6 \pi ^3} \left( \left(\frac{11}{3675}-\frac{K}{1120}\right) k^2+ \left(-\frac{31}{1200}+\frac{K}{80}\right) m^2+
\right.\right.\0\\ & \quad \quad \quad \left.\left.
 +  \left(\frac{7}{96}-\frac{K}{16}\right)\frac{ m^4}{k^2}\right)\right)+
\0\\ & \quad 
 + k^4 \pi _{\mu\mu} \pi _{\nu\nu} \left(\frac{i }{12 \pi ^3} \left( \left(\frac{11}{3675}-\frac{K}{1120}\right) k^2+ \left(-\frac{31}{1200}+\frac{K}{80}\right) m^2+
\right.\right.\0\\ & \quad \quad \quad \left.\left.
 +  \left(\frac{7}{96}-\frac{K}{16}\right)\frac{ m^4}{k^2}\right)\right)+\ldots }
Scalars, spin 2 x 4, dimension 3:
\al{\label{eq:sc:2:4:3:uv}\tilde{T}_{2,4;3\text{D}}^{\text{s,t,UV}} & = k^6 \pi _{\mu\nu}^2 \pi _{\nu\nu} \left(\frac{1}{32}\frac{1}{k}-\frac{i }{2 \pi }\frac{ m}{k^2}-\frac{3 }{8}\frac{ m^2}{k^3}+\frac{8 i }{3 \pi }\frac{ m^3}{k^4}+\frac{3 }{2}\frac{ m^4}{k^5}-\frac{32 i }{5 \pi }\frac{ m^5}{k^6}-2 \frac{ m^6}{k^7}+
\right.\0\\ & \quad \quad \left.
 + \frac{128 i }{35 \pi }\frac{ m^7}{k^8}+\frac{512 i }{315 \pi }\frac{ m^9}{k^{10}}+\frac{2048 i }{1155 \pi }\frac{ m^{11}}{k^{12}}+\ldots\right)+
\0\\ & \quad 
 + k^6 \pi _{\mu\mu} \pi _{\nu\nu}^2 \left(\frac{1}{128}\frac{1}{k}-\frac{3 }{32}\frac{ m^2}{k^3}+\frac{2 i }{3 \pi }\frac{ m^3}{k^4}+\frac{3 }{8}\frac{ m^4}{k^5}-\frac{8 i }{5 \pi }\frac{ m^5}{k^6}-\frac{1}{2}\frac{ m^6}{k^7}+\frac{32 i }{35 \pi }\frac{ m^7}{k^8}+
\right.\0\\ & \quad \quad \left.
 + \frac{128 i }{315 \pi }\frac{ m^9}{k^{10}}+\frac{512 i }{1155 \pi }\frac{ m^{11}}{k^{12}}+\ldots\right) }
\al{\label{eq:sc:2:4:3:ir}\tilde{T}_{2,4;3\text{D}}^{\text{s,t,IR}} & = k^6 \pi _{\mu\nu}^2 \pi _{\nu\nu} \left(\frac{i }{2 \pi } \left(-\frac{1}{5}\frac{ m}{k^2}-\frac{1}{35}\frac{1}{m}-\frac{1}{1260}\frac{ k^2}{m^3}-\frac{1}{18480}\frac{ k^4}{m^5}-\frac{1}{192192}\frac{ k^6}{m^7}-
\right.\right.\0\\ & \quad \quad \quad \left.\left.
 - \frac{1}{1647360}\frac{ k^8}{m^9}-\frac{1}{12446720}\frac{ k^{10}}{m^{11}}+\ldots\right)\right)+
\0\\ & \quad 
 + k^6 \pi _{\mu\mu} \pi _{\nu\nu}^2 \left(\frac{i }{2 \pi } \left(\frac{1}{5}\frac{ m}{k^2}-\frac{1}{140}\frac{1}{m}-\frac{1}{5040}\frac{ k^2}{m^3}-\frac{1}{73920}\frac{ k^4}{m^5}-\frac{1}{768768}\frac{ k^6}{m^7}-
\right.\right.\0\\ & \quad \quad \quad \left.\left.
 - \frac{1}{6589440}\frac{ k^8}{m^9}-\frac{1}{49786880}\frac{ k^{10}}{m^{11}}+\ldots\right)\right) }
\al{\label{eq:sc:2:4:3:uvir}\tilde{T}_{2,4;3\text{D}}^{\text{s,UV-IR}} & = k^6 \pi _{\mu\nu}^2 \pi _{\nu\nu} \left(-\frac{2 i }{5 \pi }\frac{ m}{k^2}\right)+\ldots }
Scalars, spin 2 x 4, dimension 4:
\al{\label{eq:sc:2:4:4:uv}\tilde{T}_{2,4;4\text{D}}^{\text{s,t,UV}} & = k^6 \pi _{\mu\nu}^2 \pi _{\nu\nu} \left(\frac{i }{\pi ^2} \left( \left(-\frac{88}{3675}+\frac{P}{140}\right)+ \left(\frac{109}{600}-\frac{P}{10}+\frac{L_0}{8}\right)\frac{ m^2}{k^2}+
\right.\right.\0\\ & \quad \quad \quad \left.\left.
 +  \left(-\frac{7}{12}+\frac{K}{2}\right)\frac{ m^4}{k^4}+ \left(\frac{1}{6}-K\right)\frac{ m^6}{k^6}+ \left(\frac{25}{24}+\frac{K}{2}\right)\frac{ m^8}{k^8}+
\right.\right.\0\\ & \quad \quad \quad \left.\left.
 +  \left(\frac{17}{300}+\frac{K}{5}\right)\frac{ m^{10}}{k^{10}}+ \left(-\frac{13}{300}+\frac{K}{5}\right)\frac{ m^{12}}{k^{12}}+\ldots\right)\right)+
\0\\ & \quad 
 + k^6 \pi _{\mu\mu} \pi _{\nu\nu}^2 \left(\frac{i }{\pi ^2} \left( \left(-\frac{22}{3675}+\frac{P}{560}\right)+ \left(\frac{23}{300}-\frac{P}{40}\right)\frac{ m^2}{k^2}+
\right.\right.\0\\ & \quad \quad \quad \left.\left.
 +  \left(-\frac{7}{48}+\frac{K}{8}\right)\frac{ m^4}{k^4}+ \left(\frac{1}{24}-\frac{K}{4}\right)\frac{ m^6}{k^6}+ \left(\frac{25}{96}+\frac{K}{8}\right)\frac{ m^8}{k^8}+
\right.\right.\0\\ & \quad \quad \quad \left.\left.
 +  \left(\frac{17}{1200}+\frac{K}{20}\right)\frac{ m^{10}}{k^{10}}+ \left(-\frac{13}{1200}+\frac{K}{20}\right)\frac{ m^{12}}{k^{12}}+\ldots\right)\right) }
\al{\label{eq:sc:2:4:4:ir}\tilde{T}_{2,4;4\text{D}}^{\text{s,t,IR}} & = k^6 \pi _{\mu\nu}^2 \pi _{\nu\nu} \left(\frac{i }{20 \pi ^2} \left( \left(-\frac{1}{2}+\frac{L_0}{2}\right)\frac{ m^2}{k^2}+\frac{ L_0}{7}-\frac{1}{126}\frac{ k^2}{m^2}-\frac{1}{2772}\frac{ k^4}{m^4}-\frac{1}{36036}\frac{ k^6}{m^6}-
\right.\right.\0\\ & \quad \quad \quad \left.\left.
 - \frac{1}{360360}\frac{ k^8}{m^8}-\frac{1}{3063060}\frac{ k^{10}}{m^{10}}-\frac{1}{23279256}\frac{ k^{12}}{m^{12}}+\ldots\right)\right)+
\0\\ & \quad 
 + k^6 \pi _{\mu\mu} \pi _{\nu\nu}^2 \left(\frac{i }{40 \pi ^2} \left( \left(1-L_0\right)\frac{ m^2}{k^2}+\frac{ L_0}{14}-\frac{1}{252}\frac{ k^2}{m^2}-\frac{1}{5544}\frac{ k^4}{m^4}-\frac{1}{72072}\frac{ k^6}{m^6}-
\right.\right.\0\\ & \quad \quad \quad \left.\left.
 - \frac{1}{720720}\frac{ k^8}{m^8}-\frac{1}{6126120}\frac{ k^{10}}{m^{10}}-\frac{1}{46558512}\frac{ k^{12}}{m^{12}}+\ldots\right)\right) }
\al{\label{eq:sc:2:4:4:uvir}\tilde{T}_{2,4;4\text{D}}^{\text{s,UV-IR}} & = k^6 \pi _{\mu\nu}^2 \pi _{\nu\nu} \left(\frac{i }{5 \pi ^2} \left( \left(-\frac{88}{735}+\frac{K}{28}\right)+ \left(\frac{31}{30}-\frac{K}{2}\right)\frac{ m^2}{k^2}\right)\right)+
\0\\ & \quad 
 + k^6 \pi _{\mu\mu} \pi _{\nu\nu}^2 \left(\frac{i }{5 \pi ^2} \left( \left(-\frac{22}{735}+\frac{K}{112}\right)+ \left(\frac{31}{120}-\frac{K}{8}\right)\frac{ m^2}{k^2}\right)\right)+\ldots }
Scalars, spin 2 x 4, dimension 5:
\al{\label{eq:sc:2:4:5:uv}\tilde{T}_{2,4;5\text{D}}^{\text{s,t,UV}} & = k^6 \pi _{\mu\nu}^2 \pi _{\nu\nu} \left(\frac{1}{\pi ^2} \left(\frac{ \pi }{2048} k-\frac{ \pi }{128}\frac{ m^2}{k}+\frac{i }{12}\frac{ m^3}{k^2}+\frac{3  \pi }{64}\frac{ m^4}{k^3}-\frac{4 i }{15}\frac{ m^5}{k^4}-\frac{ \pi }{8}\frac{ m^6}{k^5}+
\right.\right.\0\\ & \quad \quad \quad \left.\left.
 + \frac{16 i }{35}\frac{ m^7}{k^6}+\frac{ \pi }{8}\frac{ m^8}{k^7}-\frac{64 i }{315}\frac{ m^9}{k^8}-\frac{256 i }{3465}\frac{ m^{11}}{k^{10}}+\ldots\right)\right)+
\0\\ & \quad 
 + k^6 \pi _{\mu\mu} \pi _{\nu\nu}^2 \left(\frac{1}{\pi ^2} \left(\frac{ \pi }{8192} k-\frac{ \pi }{512}\frac{ m^2}{k}+\frac{3  \pi }{256}\frac{ m^4}{k^3}-\frac{i }{15}\frac{ m^5}{k^4}-\frac{ \pi }{32}\frac{ m^6}{k^5}+\frac{4 i }{35}\frac{ m^7}{k^6}+
\right.\right.\0\\ & \quad \quad \quad \left.\left.
 + \frac{ \pi }{32}\frac{ m^8}{k^7}-\frac{16 i }{315}\frac{ m^9}{k^8}-\frac{64 i }{3465}\frac{ m^{11}}{k^{10}}+\ldots\right)\right) }
\al{\label{eq:sc:2:4:5:ir}\tilde{T}_{2,4;5\text{D}}^{\text{s,t,IR}} & = k^6 \pi _{\mu\nu}^2 \pi _{\nu\nu} \left(\frac{i }{20 \pi ^2} \left(\frac{1}{3}\frac{ m^3}{k^2}+\frac{1}{7} m-\frac{1}{252}\frac{ k^2}{m}-\frac{1}{11088}\frac{ k^4}{m^3}-\frac{1}{192192}\frac{ k^6}{m^5}-
\right.\right.\0\\ & \quad \quad \quad \left.\left.
 - \frac{1}{2306304}\frac{ k^8}{m^7}-\frac{1}{22404096}\frac{ k^{10}}{m^9}-\frac{1}{189190144}\frac{ k^{12}}{m^{11}}+\ldots\right)\right)+
\0\\ & \quad 
 + k^6 \pi _{\mu\mu} \pi _{\nu\nu}^2 \left(\frac{i }{20 \pi ^2} \left(-\frac{1}{3}\frac{ m^3}{k^2}+\frac{1}{28} m-\frac{1}{1008}\frac{ k^2}{m}-\frac{1}{44352}\frac{ k^4}{m^3}-\frac{1}{768768}\frac{ k^6}{m^5}-
\right.\right.\0\\ & \quad \quad \quad \left.\left.
 - \frac{1}{9225216}\frac{ k^8}{m^7}-\frac{1}{89616384}\frac{ k^{10}}{m^9}-\frac{1}{756760576}\frac{ k^{12}}{m^{11}}+\ldots\right)\right) }
\al{\label{eq:sc:2:4:5:uvir}\tilde{T}_{2,4;5\text{D}}^{\text{s,UV-IR}} & = k^6 \pi _{\mu\nu}^2 \pi _{\nu\nu} \left(\frac{i }{15 \pi ^2}\frac{ m^3}{k^2}\right)+\ldots }
Scalars, spin 2 x 4, dimension 6:
\al{\label{eq:sc:2:4:6:uv}\tilde{T}_{2,4;6\text{D}}^{\text{s,t,UV}} & = k^6 \pi _{\mu\nu}^2 \pi _{\nu\nu} \left(\frac{i }{\pi ^3} \left( \left(-\frac{563}{1587600}+\frac{P}{10080}\right) k^2+ \left(\frac{22}{3675}-\frac{P}{560}\right) m^2-
\right.\right.\0\\ & \quad \quad \quad \left.\left.
 -  \left(\frac{143}{9600}-\frac{P}{80}+\frac{L_0}{64}\right)\frac{ m^4}{k^2}+ \left(\frac{5}{144}-\frac{K}{24}\right)\frac{ m^6}{k^4}+ \left(\frac{1}{192}+\frac{K}{16}\right)\frac{ m^8}{k^6}-
\right.\right.\0\\ & \quad \quad \quad \left.\left.
 -  \left(\frac{137}{2400}+\frac{K}{40}\right)\frac{ m^{10}}{k^8}- \left(\frac{3}{800}+\frac{K}{120}\right)\frac{ m^{12}}{k^{10}}+\ldots\right)\right)+
\0\\ & \quad 
 + k^6 \pi _{\mu\mu} \pi _{\nu\nu}^2 \left(\frac{i }{2 \pi ^3} \left( \left(-\frac{563}{3175200}+\frac{P}{20160}\right) k^2+ \left(\frac{11}{3675}-\frac{P}{1120}\right) m^2+
\right.\right.\0\\ & \quad \quad \quad \left.\left.
 +  \left(-\frac{23}{1200}+\frac{P}{160}\right)\frac{ m^4}{k^2}+ \left(\frac{5}{288}-\frac{K}{48}\right)\frac{ m^6}{k^4}+ \left(\frac{1}{384}+\frac{K}{32}\right)\frac{ m^8}{k^6}-
\right.\right.\0\\ & \quad \quad \quad \left.\left.
 -  \left(\frac{137}{4800}+\frac{K}{80}\right)\frac{ m^{10}}{k^8}- \left(\frac{3}{1600}+\frac{K}{240}\right)\frac{ m^{12}}{k^{10}}+\ldots\right)\right) }
\al{\label{eq:sc:2:4:6:ir}\tilde{T}_{2,4;6\text{D}}^{\text{s,t,IR}} & = k^6 \pi _{\mu\nu}^2 \pi _{\nu\nu} \left(\frac{i }{80 \pi ^3} \left( \left(\frac{3}{8}-\frac{L_0}{4}\right)\frac{ m^4}{k^2}+ \left(\frac{1}{7}-\frac{L_0}{7}\right) m^2+\frac{ L_0}{126} k^2-\frac{1}{2772}\frac{ k^4}{m^2}-
\right.\right.\0\\ & \quad \quad \quad \left.\left.
 - \frac{1}{72072}\frac{ k^6}{m^4}-\frac{1}{1081080}\frac{ k^8}{m^6}-\frac{1}{12252240}\frac{ k^{10}}{m^8}-\frac{1}{116396280}\frac{ k^{12}}{m^{10}}-
\right.\right.\0\\ & \quad \quad \quad \left.\left.
 - \frac{1}{977728752}\frac{ k^{14}}{m^{12}}+\ldots\right)\right)+
\0\\ & \quad 
 + k^6 \pi _{\mu\mu} \pi _{\nu\nu}^2 \left(\frac{i }{320 \pi ^3} \left( \left(-\frac{3}{2}+L_0\right)\frac{ m^4}{k^2}+ \left(\frac{1}{7}-\frac{L_0}{7}\right) m^2+\frac{ L_0}{126} k^2-
\right.\right.\0\\ & \quad \quad \quad \left.\left.
 - \frac{1}{2772}\frac{ k^4}{m^2}-\frac{1}{72072}\frac{ k^6}{m^4}-\frac{1}{1081080}\frac{ k^8}{m^6}-\frac{1}{12252240}\frac{ k^{10}}{m^8}-
\right.\right.\0\\ & \quad \quad \quad \left.\left.
 - \frac{1}{116396280}\frac{ k^{12}}{m^{10}}-\frac{1}{977728752}\frac{ k^{14}}{m^{12}}+\ldots\right)\right) }
\al{\label{eq:sc:2:4:6:uvir}\tilde{T}_{2,4;6\text{D}}^{\text{s,UV-IR}} & = k^6 \pi _{\mu\nu}^2 \pi _{\nu\nu} \left(\frac{i }{80 \pi ^3} \left( \left(-\frac{563}{19845}+\frac{K}{126}\right) k^2+ \left(\frac{247}{735}-\frac{K}{7}\right) m^2+
\right.\right.\0\\ & \quad \quad \quad \left.\left.
 +  \left(-\frac{47}{30}+K\right)\frac{ m^4}{k^2}\right)\right)+
\0\\ & \quad 
 + k^6 \pi _{\mu\mu} \pi _{\nu\nu}^2 \left(\frac{i }{320 \pi ^3} \left( \left(-\frac{563}{19845}+\frac{K}{126}\right) k^2+ \left(\frac{247}{735}-\frac{K}{7}\right) m^2+
\right.\right.\0\\ & \quad \quad \quad \left.\left.
 +  \left(-\frac{47}{30}+K\right)\frac{ m^4}{k^2}\right)\right)+\ldots }
Scalars, spin 3 x 3, dimension 3:
\al{\label{eq:sc:3:3:3:uv}\tilde{T}_{3,3;3\text{D}}^{\text{s,t,UV}} & = k^6 \pi _{\mu\nu}^3 \left(\frac{1}{64}\frac{1}{k}-\frac{i }{2 \pi }\frac{ m}{k^2}-\frac{3 }{16}\frac{ m^2}{k^3}+\frac{4 i }{3 \pi }\frac{ m^3}{k^4}+\frac{3 }{4}\frac{ m^4}{k^5}-\frac{16 i }{5 \pi }\frac{ m^5}{k^6}-\frac{ m^6}{k^7}+\frac{64 i }{35 \pi }\frac{ m^7}{k^8}+
\right.\0\\ & \quad \quad \left.
 + \frac{256 i }{315 \pi }\frac{ m^9}{k^{10}}+\frac{1024 i }{1155 \pi }\frac{ m^{11}}{k^{12}}+\ldots\right)+
\0\\ & \quad 
 + k^6 \pi _{\mu\mu} \pi _{\mu\nu} \pi _{\nu\nu} \left(\frac{3 }{128}\frac{1}{k}-\frac{9 }{32}\frac{ m^2}{k^3}+\frac{2 i }{\pi }\frac{ m^3}{k^4}+\frac{9 }{8}\frac{ m^4}{k^5}-\frac{24 i }{5 \pi }\frac{ m^5}{k^6}-\frac{3 }{2}\frac{ m^6}{k^7}+
\right.\0\\ & \quad \quad \left.
 + \frac{96 i }{35 \pi }\frac{ m^7}{k^8}+\frac{128 i }{105 \pi }\frac{ m^9}{k^{10}}+\frac{512 i }{385 \pi }\frac{ m^{11}}{k^{12}}+\ldots\right) }
\al{\label{eq:sc:3:3:3:ir}\tilde{T}_{3,3;3\text{D}}^{\text{s,t,IR}} & = k^6 \pi _{\mu\nu}^3 \left(\frac{i }{2 \pi } \left(-\frac{3 }{5}\frac{ m}{k^2}-\frac{1}{70}\frac{1}{m}-\frac{1}{2520}\frac{ k^2}{m^3}-\frac{1}{36960}\frac{ k^4}{m^5}-\frac{1}{384384}\frac{ k^6}{m^7}-
\right.\right.\0\\ & \quad \quad \quad \left.\left.
 - \frac{1}{3294720}\frac{ k^8}{m^9}-\frac{1}{24893440}\frac{ k^{10}}{m^{11}}+\ldots\right)\right)+
\0\\ & \quad 
 + k^6 \pi _{\mu\mu} \pi _{\mu\nu} \pi _{\nu\nu} \left(\frac{i }{2 \pi } \left(\frac{3 }{5}\frac{ m}{k^2}-\frac{3 }{140}\frac{1}{m}-\frac{1}{1680}\frac{ k^2}{m^3}-\frac{1}{24640}\frac{ k^4}{m^5}-\frac{1}{256256}\frac{ k^6}{m^7}-
\right.\right.\0\\ & \quad \quad \quad \left.\left.
 - \frac{1}{2196480}\frac{ k^8}{m^9}-\frac{3 }{49786880}\frac{ k^{10}}{m^{11}}+\ldots\right)\right) }
\al{\label{eq:sc:3:3:3:uvir}\tilde{T}_{3,3;3\text{D}}^{\text{s,UV-IR}} & = k^6 \pi _{\mu\nu}^3 \left(-\frac{i }{5 \pi }\frac{ m}{k^2}\right)+\ldots }
Scalars, spin 3 x 3, dimension 4:
\al{\label{eq:sc:3:3:4:uv}\tilde{T}_{3,3;4\text{D}}^{\text{s,t,UV}} & = k^6 \pi _{\mu\nu}^3 \left(\frac{i }{\pi ^2} \left( \left(-\frac{44}{3675}+\frac{P}{280}\right)+ \left(\frac{17}{600}-\frac{P}{20}+\frac{L_0}{8}\right)\frac{ m^2}{k^2}+ \left(-\frac{7}{24}+\frac{K}{4}\right)\frac{ m^4}{k^4}+
\right.\right.\0\\ & \quad \quad \quad \left.\left.
 +  \left(\frac{1}{12}-\frac{K}{2}\right)\frac{ m^6}{k^6}+ \left(\frac{25}{48}+\frac{K}{4}\right)\frac{ m^8}{k^8}+ \left(\frac{17}{600}+\frac{K}{10}\right)\frac{ m^{10}}{k^{10}}+
\right.\right.\0\\ & \quad \quad \quad \left.\left.
 +  \left(-\frac{13}{600}+\frac{K}{10}\right)\frac{ m^{12}}{k^{12}}+\ldots\right)\right)+
\0\\ & \quad 
 + k^6 \pi _{\mu\mu} \pi _{\mu\nu} \pi _{\nu\nu} \left(\frac{i }{\pi ^2} \left( \left(-\frac{22}{1225}+\frac{3 P}{560}\right)+ \left(\frac{23}{100}-\frac{3 P}{40}\right)\frac{ m^2}{k^2}+
\right.\right.\0\\ & \quad \quad \quad \left.\left.
 +  \left(-\frac{7}{16}+\frac{3 K}{8}\right)\frac{ m^4}{k^4}+ \left(\frac{1}{8}-\frac{3 K}{4}\right)\frac{ m^6}{k^6}+ \left(\frac{25}{32}+\frac{3 K}{8}\right)\frac{ m^8}{k^8}+
\right.\right.\0\\ & \quad \quad \quad \left.\left.
 +  \left(\frac{17}{400}+\frac{3 K}{20}\right)\frac{ m^{10}}{k^{10}}+ \left(-\frac{13}{400}+\frac{3 K}{20}\right)\frac{ m^{12}}{k^{12}}+\ldots\right)\right) }
\al{\label{eq:sc:3:3:4:ir}\tilde{T}_{3,3;4\text{D}}^{\text{s,t,IR}} & = k^6 \pi _{\mu\nu}^3 \left(\frac{i }{40 \pi ^2} \left( \left(-3+3 L_0\right)\frac{ m^2}{k^2}+\frac{ L_0}{7}-\frac{1}{126}\frac{ k^2}{m^2}-\frac{1}{2772}\frac{ k^4}{m^4}-\frac{1}{36036}\frac{ k^6}{m^6}-
\right.\right.\0\\ & \quad \quad \quad \left.\left.
 - \frac{1}{360360}\frac{ k^8}{m^8}-\frac{1}{3063060}\frac{ k^{10}}{m^{10}}-\frac{1}{23279256}\frac{ k^{12}}{m^{12}}+\ldots\right)\right)+
\0\\ & \quad 
 + k^6 \pi _{\mu\mu} \pi _{\mu\nu} \pi _{\nu\nu} \left(\frac{i }{40 \pi ^2} \left( \left(3-3 L_0\right)\frac{ m^2}{k^2}+\frac{3  L_0}{14}-\frac{1}{84}\frac{ k^2}{m^2}-\frac{1}{1848}\frac{ k^4}{m^4}-
\right.\right.\0\\ & \quad \quad \quad \left.\left.
 - \frac{1}{24024}\frac{ k^6}{m^6}-\frac{1}{240240}\frac{ k^8}{m^8}-\frac{1}{2042040}\frac{ k^{10}}{m^{10}}-\frac{1}{15519504}\frac{ k^{12}}{m^{12}}+\ldots\right)\right) }
\al{\label{eq:sc:3:3:4:uvir}\tilde{T}_{3,3;4\text{D}}^{\text{s,UV-IR}} & = k^6 \pi _{\mu\nu}^3 \left(\frac{i }{5 \pi ^2} \left( \left(-\frac{44}{735}+\frac{K}{56}\right)+ \left(\frac{31}{60}-\frac{K}{4}\right)\frac{ m^2}{k^2}\right)\right)+
\0\\ & \quad 
 + k^6 \pi _{\mu\mu} \pi _{\mu\nu} \pi _{\nu\nu} \left(\frac{i }{5 \pi ^2} \left( \left(-\frac{22}{245}+\frac{3 K}{112}\right)+ \left(\frac{31}{40}-\frac{3 K}{8}\right)\frac{ m^2}{k^2}\right)\right)+\ldots }
Scalars, spin 3 x 3, dimension 5:
\al{\label{eq:sc:3:3:5:uv}\tilde{T}_{3,3;5\text{D}}^{\text{s,t,UV}} & = k^6 \pi _{\mu\nu}^3 \left(\frac{1}{\pi ^2} \left(\frac{ \pi }{4096} k-\frac{ \pi }{256}\frac{ m^2}{k}+\frac{i }{12}\frac{ m^3}{k^2}+\frac{3  \pi }{128}\frac{ m^4}{k^3}-\frac{2 i }{15}\frac{ m^5}{k^4}-\frac{ \pi }{16}\frac{ m^6}{k^5}+\frac{8 i }{35}\frac{ m^7}{k^6}+
\right.\right.\0\\ & \quad \quad \quad \left.\left.
 + \frac{ \pi }{16}\frac{ m^8}{k^7}-\frac{32 i }{315}\frac{ m^9}{k^8}-\frac{128 i }{3465}\frac{ m^{11}}{k^{10}}+\ldots\right)\right)+
\0\\ & \quad 
 + k^6 \pi _{\mu\mu} \pi _{\mu\nu} \pi _{\nu\nu} \left(\frac{1}{\pi ^2} \left(\frac{3  \pi }{8192} k-\frac{3  \pi }{512}\frac{ m^2}{k}+\frac{9  \pi }{256}\frac{ m^4}{k^3}-\frac{i }{5}\frac{ m^5}{k^4}-\frac{3  \pi }{32}\frac{ m^6}{k^5}+
\right.\right.\0\\ & \quad \quad \quad \left.\left.
 + \frac{12 i }{35}\frac{ m^7}{k^6}+\frac{3  \pi }{32}\frac{ m^8}{k^7}-\frac{16 i }{105}\frac{ m^9}{k^8}-\frac{64 i }{1155}\frac{ m^{11}}{k^{10}}+\ldots\right)\right) }
\al{\label{eq:sc:3:3:5:ir}\tilde{T}_{3,3;5\text{D}}^{\text{s,t,IR}} & = k^6 \pi _{\mu\nu}^3 \left(\frac{i }{20 \pi ^2} \left(\frac{m^3}{k^2}+\frac{1}{14} m-\frac{1}{504}\frac{ k^2}{m}-\frac{1}{22176}\frac{ k^4}{m^3}-\frac{1}{384384}\frac{ k^6}{m^5}-\frac{1}{4612608}\frac{ k^8}{m^7}-
\right.\right.\0\\ & \quad \quad \quad \left.\left.
 - \frac{1}{44808192}\frac{ k^{10}}{m^9}-\frac{1}{378380288}\frac{ k^{12}}{m^{11}}+\ldots\right)\right)+
\0\\ & \quad 
 + k^6 \pi _{\mu\mu} \pi _{\mu\nu} \pi _{\nu\nu} \left(\frac{i }{20 \pi ^2} \left(-\frac{ m^3}{k^2}+\frac{3 }{28} m-\frac{1}{336}\frac{ k^2}{m}-\frac{1}{14784}\frac{ k^4}{m^3}-\frac{1}{256256}\frac{ k^6}{m^5}-
\right.\right.\0\\ & \quad \quad \quad \left.\left.
 - \frac{1}{3075072}\frac{ k^8}{m^7}-\frac{1}{29872128}\frac{ k^{10}}{m^9}-\frac{3 }{756760576}\frac{ k^{12}}{m^{11}}+\ldots\right)\right) }
\al{\label{eq:sc:3:3:5:uvir}\tilde{T}_{3,3;5\text{D}}^{\text{s,UV-IR}} & = k^6 \pi _{\mu\nu}^3 \left(\frac{i }{30 \pi ^2}\frac{ m^3}{k^2}\right)+\ldots }
Scalars, spin 3 x 3, dimension 6:
\al{\label{eq:sc:3:3:6:uv}\tilde{T}_{3,3;6\text{D}}^{\text{s,t,UV}} & = k^6 \pi _{\mu\nu}^3 \left(\frac{i }{\pi ^3} \left( \left(-\frac{563}{3175200}+\frac{P}{20160}\right) k^2+ \left(\frac{11}{3675}-\frac{P}{1120}\right) m^2-
\right.\right.\0\\ & \quad \quad \quad \left.\left.
 -  \left(-\frac{41}{9600}-\frac{P}{160}+\frac{L_0}{64}\right)\frac{ m^4}{k^2}+ \left(\frac{5}{288}-\frac{K}{48}\right)\frac{ m^6}{k^4}+ \left(\frac{1}{384}+\frac{K}{32}\right)\frac{ m^8}{k^6}-
\right.\right.\0\\ & \quad \quad \quad \left.\left.
 -  \left(\frac{137}{4800}+\frac{K}{80}\right)\frac{ m^{10}}{k^8}- \left(\frac{3}{1600}+\frac{K}{240}\right)\frac{ m^{12}}{k^{10}}+\ldots\right)\right)+
\0\\ & \quad 
 + k^6 \pi _{\mu\mu} \pi _{\mu\nu} \pi _{\nu\nu} \left(\frac{i }{2 \pi ^3} \left( \left(-\frac{563}{1058400}+\frac{P}{6720}\right) k^2+ \left(\frac{11}{1225}-\frac{3 P}{1120}\right) m^2+
\right.\right.\0\\ & \quad \quad \quad \left.\left.
 +  \left(-\frac{23}{400}+\frac{3 P}{160}\right)\frac{ m^4}{k^2}+ \left(\frac{5}{96}-\frac{K}{16}\right)\frac{ m^6}{k^4}+ \left(\frac{1}{128}+\frac{3 K}{32}\right)\frac{ m^8}{k^6}-
\right.\right.\0\\ & \quad \quad \quad \left.\left.
 -  \left(\frac{137}{1600}+\frac{3 K}{80}\right)\frac{ m^{10}}{k^8}- \left(\frac{9}{1600}+\frac{K}{80}\right)\frac{ m^{12}}{k^{10}}+\ldots\right)\right) }
\al{\label{eq:sc:3:3:6:ir}\tilde{T}_{3,3;6\text{D}}^{\text{s,t,IR}} & = k^6 \pi _{\mu\nu}^3 \left(\frac{i }{160 \pi ^3} \left( \left(\frac{9}{4}-\frac{3 L_0}{2}\right)\frac{ m^4}{k^2}+ \left(\frac{1}{7}-\frac{L_0}{7}\right) m^2+\frac{ L_0}{126} k^2-\frac{1}{2772}\frac{ k^4}{m^2}-
\right.\right.\0\\ & \quad \quad \quad \left.\left.
 - \frac{1}{72072}\frac{ k^6}{m^4}-\frac{1}{1081080}\frac{ k^8}{m^6}-\frac{1}{12252240}\frac{ k^{10}}{m^8}-\frac{1}{116396280}\frac{ k^{12}}{m^{10}}-
\right.\right.\0\\ & \quad \quad \quad \left.\left.
 - \frac{1}{977728752}\frac{ k^{14}}{m^{12}}+\ldots\right)\right)+
\0\\ & \quad 
 + k^6 \pi _{\mu\mu} \pi _{\mu\nu} \pi _{\nu\nu} \left(\frac{i }{320 \pi ^3} \left( \left(-\frac{9}{2}+3 L_0\right)\frac{ m^4}{k^2}+ \left(\frac{3}{7}-\frac{3 L_0}{7}\right) m^2+\frac{ L_0}{42} k^2-
\right.\right.\0\\ & \quad \quad \quad \left.\left.
 - \frac{1}{924}\frac{ k^4}{m^2}-\frac{1}{24024}\frac{ k^6}{m^4}-\frac{1}{360360}\frac{ k^8}{m^6}-\frac{1}{4084080}\frac{ k^{10}}{m^8}-\frac{1}{38798760}\frac{ k^{12}}{m^{10}}-
\right.\right.\0\\ & \quad \quad \quad \left.\left.
 - \frac{1}{325909584}\frac{ k^{14}}{m^{12}}+\ldots\right)\right) }
\al{\label{eq:sc:3:3:6:uvir}\tilde{T}_{3,3;6\text{D}}^{\text{s,UV-IR}} & = k^6 \pi _{\mu\nu}^3 \left(\frac{i }{160 \pi ^3} \left( \left(-\frac{563}{19845}+\frac{K}{126}\right) k^2+ \left(\frac{247}{735}-\frac{K}{7}\right) m^2+
\right.\right.\0\\ & \quad \quad \quad \left.\left.
 +  \left(-\frac{47}{30}+K\right)\frac{ m^4}{k^2}\right)\right)+
\0\\ & \quad 
 + k^6 \pi _{\mu\mu} \pi _{\mu\nu} \pi _{\nu\nu} \left(\frac{i }{320 \pi ^3} \left( \left(-\frac{563}{6615}+\frac{K}{42}\right) k^2+ \left(\frac{247}{245}-\frac{3 K}{7}\right) m^2+
\right.\right.\0\\ & \quad \quad \quad \left.\left.
 +  \left(-\frac{47}{10}+3 K\right)\frac{ m^4}{k^2}\right)\right)+\ldots }
Scalars, spin 3 x 5, dimension 3:
\al{\label{eq:sc:3:5:3:uv}\tilde{T}_{3,5;3\text{D}}^{\text{s,t,UV}} & = k^8 \pi _{\mu\nu}^3 \pi _{\nu\nu} \left(-\frac{5 }{256}\frac{1}{k}+\frac{i }{2 \pi }\frac{ m}{k^2}+\frac{5 }{16}\frac{ m^2}{k^3}-\frac{8 i }{3 \pi }\frac{ m^3}{k^4}-\frac{15 }{8}\frac{ m^4}{k^5}+\frac{32 i }{3 \pi }\frac{ m^5}{k^6}+5 \frac{ m^6}{k^7}-
\right.\0\\ & \quad \quad \left.
 - \frac{128 i }{7 \pi }\frac{ m^7}{k^8}-5 \frac{ m^8}{k^9}+\frac{512 i }{63 \pi }\frac{ m^9}{k^{10}}+\frac{2048 i }{693 \pi }\frac{ m^{11}}{k^{12}}+\ldots\right)+
\0\\ & \quad 
 + k^8 \pi _{\mu\mu} \pi _{\mu\nu} \pi _{\nu\nu}^2 \left(-\frac{15 }{1024}\frac{1}{k}+\frac{15 }{64}\frac{ m^2}{k^3}-\frac{2 i }{\pi }\frac{ m^3}{k^4}-\frac{45 }{32}\frac{ m^4}{k^5}+\frac{8 i }{\pi }\frac{ m^5}{k^6}+\frac{15 }{4}\frac{ m^6}{k^7}-
\right.\0\\ & \quad \quad \left.
 - \frac{96 i }{7 \pi }\frac{ m^7}{k^8}-\frac{15 }{4}\frac{ m^8}{k^9}+\frac{128 i }{21 \pi }\frac{ m^9}{k^{10}}+\frac{512 i }{231 \pi }\frac{ m^{11}}{k^{12}}+\ldots\right) }
\al{\label{eq:sc:3:5:3:ir}\tilde{T}_{3,5;3\text{D}}^{\text{s,t,IR}} & = k^8 \pi _{\mu\nu}^3 \pi _{\nu\nu} \left(\frac{i }{2 \pi } \left(\frac{3 }{7}\frac{ m}{k^2}+\frac{1}{63}\frac{1}{m}+\frac{1}{2772}\frac{ k^2}{m^3}+\frac{1}{48048}\frac{ k^4}{m^5}+\frac{1}{576576}\frac{ k^6}{m^7}+
\right.\right.\0\\ & \quad \quad \quad \left.\left.
 + \frac{1}{5601024}\frac{ k^8}{m^9}+\frac{1}{47297536}\frac{ k^{10}}{m^{11}}+\ldots\right)\right)+
\0\\ & \quad 
 + k^8 \pi _{\mu\mu} \pi _{\mu\nu} \pi _{\nu\nu}^2 \left(\frac{i }{2 \pi } \left(-\frac{3 }{7}\frac{ m}{k^2}+\frac{1}{84}\frac{1}{m}+\frac{1}{3696}\frac{ k^2}{m^3}+\frac{1}{64064}\frac{ k^4}{m^5}+\frac{1}{768768}\frac{ k^6}{m^7}+
\right.\right.\0\\ & \quad \quad \quad \left.\left.
 + \frac{1}{7468032}\frac{ k^8}{m^9}+\frac{3 }{189190144}\frac{ k^{10}}{m^{11}}+\ldots\right)\right) }
\al{\label{eq:sc:3:5:3:uvir}\tilde{T}_{3,5;3\text{D}}^{\text{s,UV-IR}} & = k^8 \pi _{\mu\nu}^3 \pi _{\nu\nu} \left(\frac{2 i }{7 \pi }\frac{ m}{k^2}\right)+\ldots }
Scalars, spin 3 x 5, dimension 4:
\al{\label{eq:sc:3:5:4:uv}\tilde{T}_{3,5;4\text{D}}^{\text{s,t,UV}} & = k^8 \pi _{\mu\nu}^3 \pi _{\nu\nu} \left(\frac{i }{\pi ^2} \left( \left(\frac{563}{39690}-\frac{P}{252}\right)- \left(\frac{673}{5880}-\frac{P}{14}+\frac{L_0}{8}\right)\frac{ m^2}{k^2}+
\right.\right.\0\\ & \quad \quad \quad \left.\left.
 +  \left(\frac{47}{60}-\frac{K}{2}\right)\frac{ m^4}{k^4}+ \left(-\frac{25}{18}+\frac{5 K}{3}\right)\frac{ m^6}{k^6}- \left(\frac{5}{24}+\frac{5 K}{2}\right)\frac{ m^8}{k^8}+
\right.\right.\0\\ & \quad \quad \quad \left.\left.
 +  \left(\frac{137}{60}+K\right)\frac{ m^{10}}{k^{10}}+ \left(\frac{3}{20}+\frac{K}{3}\right)\frac{ m^{12}}{k^{12}}+\ldots\right)\right)+
\0\\ & \quad 
 + k^8 \pi _{\mu\mu} \pi _{\mu\nu} \pi _{\nu\nu}^2 \left(\frac{i }{\pi ^2} \left( \left(\frac{563}{52920}-\frac{P}{336}\right)+ \left(-\frac{44}{245}+\frac{3 P}{56}\right)\frac{ m^2}{k^2}+
\right.\right.\0\\ & \quad \quad \quad \left.\left.
 +  \left(\frac{47}{80}-\frac{3 K}{8}\right)\frac{ m^4}{k^4}+ \left(-\frac{25}{24}+\frac{5 K}{4}\right)\frac{ m^6}{k^6}- \left(\frac{5}{32}+\frac{15 K}{8}\right)\frac{ m^8}{k^8}+
\right.\right.\0\\ & \quad \quad \quad \left.\left.
 +  \left(\frac{137}{80}+\frac{3 K}{4}\right)\frac{ m^{10}}{k^{10}}+ \left(\frac{9}{80}+\frac{K}{4}\right)\frac{ m^{12}}{k^{12}}+\ldots\right)\right) }
\al{\label{eq:sc:3:5:4:ir}\tilde{T}_{3,5;4\text{D}}^{\text{s,t,IR}} & = k^8 \pi _{\mu\nu}^3 \pi _{\nu\nu} \left(\frac{i }{28 \pi ^2} \left( \left(\frac{3}{2}-\frac{3 L_0}{2}\right)\frac{ m^2}{k^2}-\frac{ L_0}{9}+\frac{1}{198}\frac{ k^2}{m^2}+\frac{1}{5148}\frac{ k^4}{m^4}+\frac{1}{77220}\frac{ k^6}{m^6}+
\right.\right.\0\\ & \quad \quad \quad \left.\left.
 + \frac{1}{875160}\frac{ k^8}{m^8}+\frac{1}{8314020}\frac{ k^{10}}{m^{10}}+\frac{1}{69837768}\frac{ k^{12}}{m^{12}}+\ldots\right)\right)+
\0\\ & \quad 
 + k^8 \pi _{\mu\mu} \pi _{\mu\nu} \pi _{\nu\nu}^2 \left(\frac{i }{56 \pi ^2} \left( \left(-3+3 L_0\right)\frac{ m^2}{k^2}-\frac{ L_0}{6}+\frac{1}{132}\frac{ k^2}{m^2}+\frac{1}{3432}\frac{ k^4}{m^4}+
\right.\right.\0\\ & \quad \quad \quad \left.\left.
 + \frac{1}{51480}\frac{ k^6}{m^6}+\frac{1}{583440}\frac{ k^8}{m^8}+\frac{1}{5542680}\frac{ k^{10}}{m^{10}}+\frac{1}{46558512}\frac{ k^{12}}{m^{12}}+\ldots\right)\right) }
\al{\label{eq:sc:3:5:4:uvir}\tilde{T}_{3,5;4\text{D}}^{\text{s,UV-IR}} & = k^8 \pi _{\mu\nu}^3 \pi _{\nu\nu} \left(\frac{i }{14 \pi ^2} \left( \left(\frac{563}{2835}-\frac{K}{18}\right)+ \left(-\frac{247}{105}+K\right)\frac{ m^2}{k^2}\right)\right)+
\0\\ & \quad 
 + k^8 \pi _{\mu\mu} \pi _{\mu\nu} \pi _{\nu\nu}^2 \left(\frac{i }{56 \pi ^2} \left( \left(\frac{563}{945}-\frac{K}{6}\right)+ \left(-\frac{247}{35}+3 K\right)\frac{ m^2}{k^2}\right)\right)+\ldots }
Scalars, spin 3 x 5, dimension 5:
\al{\label{eq:sc:3:5:5:uv}\tilde{T}_{3,5;5\text{D}}^{\text{s,t,UV}} & = k^8 \pi _{\mu\nu}^3 \pi _{\nu\nu} \left(\frac{1}{\pi ^2} \left(-\frac{ \pi }{4096} k+\frac{5  \pi }{1024}\frac{ m^2}{k}-\frac{i }{12}\frac{ m^3}{k^2}-\frac{5  \pi }{128}\frac{ m^4}{k^3}+\frac{4 i }{15}\frac{ m^5}{k^4}+\frac{5  \pi }{32}\frac{ m^6}{k^5}-
\right.\right.\0\\ & \quad \quad \quad \left.\left.
 - \frac{16 i }{21}\frac{ m^7}{k^6}-\frac{5  \pi }{16}\frac{ m^8}{k^7}+\frac{64 i }{63}\frac{ m^9}{k^8}+\frac{ \pi }{4}\frac{ m^{10}}{k^9}-\frac{256 i }{693}\frac{ m^{11}}{k^{10}}+\ldots\right)\right)+
\0\\ & \quad 
 + k^8 \pi _{\mu\mu} \pi _{\mu\nu} \pi _{\nu\nu}^2 \left(\frac{1}{\pi ^2} \left(-\frac{3  \pi }{16384} k+\frac{15  \pi }{4096}\frac{ m^2}{k}-\frac{15  \pi }{512}\frac{ m^4}{k^3}+\frac{i }{5}\frac{ m^5}{k^4}+\frac{15  \pi }{128}\frac{ m^6}{k^5}-
\right.\right.\0\\ & \quad \quad \quad \left.\left.
 - \frac{4 i }{7}\frac{ m^7}{k^6}-\frac{15  \pi }{64}\frac{ m^8}{k^7}+\frac{16 i }{21}\frac{ m^9}{k^8}+\frac{3  \pi }{16}\frac{ m^{10}}{k^9}-\frac{64 i }{231}\frac{ m^{11}}{k^{10}}+\ldots\right)\right) }
\al{\label{eq:sc:3:5:5:ir}\tilde{T}_{3,5;5\text{D}}^{\text{s,t,IR}} & = k^8 \pi _{\mu\nu}^3 \pi _{\nu\nu} \left(\frac{i }{4 \pi ^2} \left(-\frac{1}{7}\frac{ m^3}{k^2}-\frac{1}{63} m+\frac{1}{2772}\frac{ k^2}{m}+\frac{1}{144144}\frac{ k^4}{m^3}+\frac{1}{2882880}\frac{ k^6}{m^5}+
\right.\right.\0\\ & \quad \quad \quad \left.\left.
 + \frac{1}{39207168}\frac{ k^8}{m^7}+\frac{1}{425677824}\frac{ k^{10}}{m^9}+\frac{1}{3972993024}\frac{ k^{12}}{m^{11}}+\ldots\right)\right)+
\0\\ & \quad 
 + k^8 \pi _{\mu\mu} \pi _{\mu\nu} \pi _{\nu\nu}^2 \left(\frac{i }{4 \pi ^2} \left(\frac{1}{7}\frac{ m^3}{k^2}-\frac{1}{84} m+\frac{1}{3696}\frac{ k^2}{m}+\frac{1}{192192}\frac{ k^4}{m^3}+
\right.\right.\0\\ & \quad \quad \quad \left.\left.
 + \frac{1}{3843840}\frac{ k^6}{m^5}+\frac{1}{52276224}\frac{ k^8}{m^7}+\frac{1}{567570432}\frac{ k^{10}}{m^9}+\frac{1}{5297324032}\frac{ k^{12}}{m^{11}}+
\right.\right.\0\\ & \quad \quad \quad \left.\left.
 + \ldots\right)\right) }
\al{\label{eq:sc:3:5:5:uvir}\tilde{T}_{3,5;5\text{D}}^{\text{s,UV-IR}} & = k^8 \pi _{\mu\nu}^3 \pi _{\nu\nu} \left(-\frac{i }{21 \pi ^2}\frac{ m^3}{k^2}\right)+\ldots }
Scalars, spin 3 x 5, dimension 6:
\al{\label{eq:sc:3:5:6:uv}\tilde{T}_{3,5;6\text{D}}^{\text{s,t,UV}} & = k^8 \pi _{\mu\nu}^3 \pi _{\nu\nu} \left(\frac{i }{4 \pi ^3} \left( \left(\frac{1627}{2401245}-\frac{P}{5544}\right) k^2+ \left(-\frac{563}{39690}+\frac{P}{252}\right) m^2+
\right.\right.\0\\ & \quad \quad \quad \left.\left.
 +  \left(\frac{611}{23520}-\frac{P}{28}+\frac{L_0}{16}\right)\frac{ m^4}{k^2}+ \left(-\frac{37}{180}+\frac{K}{6}\right)\frac{ m^6}{k^4}+ \left(\frac{35}{144}-\frac{5 K}{12}\right)\frac{ m^8}{k^6}+
\right.\right.\0\\ & \quad \quad \quad \left.\left.
 +  \left(\frac{17}{120}+\frac{K}{2}\right)\frac{ m^{10}}{k^8}- \left(\frac{49}{120}+\frac{K}{6}\right)\frac{ m^{12}}{k^{10}}+\ldots\right)\right)+
\0\\ & \quad 
 + k^8 \pi _{\mu\mu} \pi _{\mu\nu} \pi _{\nu\nu}^2 \left(\frac{i }{2 \pi ^3} \left( \left(\frac{1627}{6403320}-\frac{P}{14784}\right) k^2+ \left(-\frac{563}{105840}+\frac{P}{672}\right) m^2+
\right.\right.\0\\ & \quad \quad \quad \left.\left.
 +  \left(\frac{11}{245}-\frac{3 P}{224}\right)\frac{ m^4}{k^2}+ \left(-\frac{37}{480}+\frac{K}{16}\right)\frac{ m^6}{k^4}+ \left(\frac{35}{384}-\frac{5 K}{32}\right)\frac{ m^8}{k^6}+
\right.\right.\0\\ & \quad \quad \quad \left.\left.
 +  \left(\frac{17}{320}+\frac{3 K}{16}\right)\frac{ m^{10}}{k^8}- \left(\frac{49}{320}+\frac{K}{16}\right)\frac{ m^{12}}{k^{10}}+\ldots\right)\right) }
\al{\label{eq:sc:3:5:6:ir}\tilde{T}_{3,5;6\text{D}}^{\text{s,t,IR}} & = k^8 \pi _{\mu\nu}^3 \pi _{\nu\nu} \left(\frac{i }{112 \pi ^3} \left( \left(-\frac{9}{8}+\frac{3 L_0}{4}\right)\frac{ m^4}{k^2}+ \left(-\frac{1}{9}+\frac{L_0}{9}\right) m^2-\frac{ L_0}{198} k^2+\frac{1}{5148}\frac{ k^4}{m^2}+
\right.\right.\0\\ & \quad \quad \quad \left.\left.
 + \frac{1}{154440}\frac{ k^6}{m^4}+\frac{1}{2625480}\frac{ k^8}{m^6}+\frac{1}{33256080}\frac{ k^{10}}{m^8}+\frac{1}{349188840}\frac{ k^{12}}{m^{10}}+
\right.\right.\0\\ & \quad \quad \quad \left.\left.
 + \frac{1}{3212537328}\frac{ k^{14}}{m^{12}}+\ldots\right)\right)+
\0\\ & \quad 
 + k^8 \pi _{\mu\mu} \pi _{\mu\nu} \pi _{\nu\nu}^2 \left(\frac{i }{448 \pi ^3} \left( \left(\frac{9}{2}-3 L_0\right)\frac{ m^4}{k^2}+ \left(-\frac{1}{3}+\frac{L_0}{3}\right) m^2-\frac{ L_0}{66} k^2+
\right.\right.\0\\ & \quad \quad \quad \left.\left.
 + \frac{1}{1716}\frac{ k^4}{m^2}+\frac{1}{51480}\frac{ k^6}{m^4}+\frac{1}{875160}\frac{ k^8}{m^6}+\frac{1}{11085360}\frac{ k^{10}}{m^8}+
\right.\right.\0\\ & \quad \quad \quad \left.\left.
 + \frac{1}{116396280}\frac{ k^{12}}{m^{10}}+\frac{1}{1070845776}\frac{ k^{14}}{m^{12}}+\ldots\right)\right) }
\al{\label{eq:sc:3:5:6:uvir}\tilde{T}_{3,5;6\text{D}}^{\text{s,UV-IR}} & = k^8 \pi _{\mu\nu}^3 \pi _{\nu\nu} \left(\frac{i }{28 \pi ^3} \left( \left(\frac{1627}{343035}-\frac{K}{792}\right) k^2+ \left(-\frac{811}{11340}+\frac{K}{36}\right) m^2+
\right.\right.\0\\ & \quad \quad \quad \left.\left.
 +  \left(\frac{389}{840}-\frac{K}{4}\right)\frac{ m^4}{k^2}\right)\right)+
\0\\ & \quad 
 + k^8 \pi _{\mu\mu} \pi _{\mu\nu} \pi _{\nu\nu}^2 \left(\frac{i }{112 \pi ^3} \left( \left(\frac{1627}{114345}-\frac{K}{264}\right) k^2+ \left(-\frac{811}{3780}+\frac{K}{12}\right) m^2+
\right.\right.\0\\ & \quad \quad \quad \left.\left.
 +  \left(\frac{389}{280}-\frac{3 K}{4}\right)\frac{ m^4}{k^2}\right)\right)+\ldots }
Scalars, spin 4 x 4, dimension 3:
\al{\label{eq:sc:4:4:3:uv}\tilde{T}_{4,4;3\text{D}}^{\text{s,t,UV}} & = k^8 \pi _{\mu\nu}^4 \left(-\frac{1}{128}\frac{1}{k}+\frac{i }{2 \pi }\frac{ m}{k^2}+\frac{1}{8}\frac{ m^2}{k^3}-\frac{2 i }{3 \pi }\frac{ m^3}{k^4}-\frac{3 }{4}\frac{ m^4}{k^5}+\frac{64 i }{15 \pi }\frac{ m^5}{k^6}+2 \frac{ m^6}{k^7}-
\right.\0\\ & \quad \quad \left.
 - \frac{256 i }{35 \pi }\frac{ m^7}{k^8}-2 \frac{ m^8}{k^9}+\frac{1024 i }{315 \pi }\frac{ m^9}{k^{10}}+\frac{4096 i }{3465 \pi }\frac{ m^{11}}{k^{12}}+\ldots\right)+
\0\\ & \quad 
 + k^8 \pi _{\mu\mu} \pi _{\mu\nu}^2 \pi _{\nu\nu} \left(-\frac{3 }{128}\frac{1}{k}+\frac{3 }{8}\frac{ m^2}{k^3}-\frac{4 i }{\pi }\frac{ m^3}{k^4}-\frac{9 }{4}\frac{ m^4}{k^5}+\frac{64 i }{5 \pi }\frac{ m^5}{k^6}+6 \frac{ m^6}{k^7}-
\right.\0\\ & \quad \quad \left.
 - \frac{768 i }{35 \pi }\frac{ m^7}{k^8}-6 \frac{ m^8}{k^9}+\frac{1024 i }{105 \pi }\frac{ m^9}{k^{10}}+\frac{4096 i }{1155 \pi }\frac{ m^{11}}{k^{12}}+\ldots\right)+
\0\\ & \quad 
 + k^8 \pi _{\mu\mu}^2 \pi _{\nu\nu}^2 \left(-\frac{3 }{1024}\frac{1}{k}+\frac{3 }{64}\frac{ m^2}{k^3}-\frac{9 }{32}\frac{ m^4}{k^5}+\frac{8 i }{5 \pi }\frac{ m^5}{k^6}+\frac{3 }{4}\frac{ m^6}{k^7}-\frac{96 i }{35 \pi }\frac{ m^7}{k^8}-
\right.\0\\ & \quad \quad \left.
 - \frac{3 }{4}\frac{ m^8}{k^9}+\frac{128 i }{105 \pi }\frac{ m^9}{k^{10}}+\frac{512 i }{1155 \pi }\frac{ m^{11}}{k^{12}}+\ldots\right) }
\al{\label{eq:sc:4:4:3:ir}\tilde{T}_{4,4;3\text{D}}^{\text{s,t,IR}} & = k^8 \pi _{\mu\nu}^4 \left(\frac{i }{5 \pi } \left(2 \frac{ m^3}{k^4}+\frac{27 }{14}\frac{ m}{k^2}+\frac{1}{63}\frac{1}{m}+\frac{1}{2772}\frac{ k^2}{m^3}+\frac{1}{48048}\frac{ k^4}{m^5}+\frac{1}{576576}\frac{ k^6}{m^7}+
\right.\right.\0\\ & \quad \quad \quad \left.\left.
 + \frac{1}{5601024}\frac{ k^8}{m^9}+\frac{1}{47297536}\frac{ k^{10}}{m^{11}}+\ldots\right)\right)+
\0\\ & \quad 
 + k^8 \pi _{\mu\mu} \pi _{\mu\nu}^2 \pi _{\nu\nu} \left(\frac{i }{5 \pi } \left(-4 \frac{ m^3}{k^4}-\frac{12 }{7}\frac{ m}{k^2}+\frac{1}{21}\frac{1}{m}+\frac{1}{924}\frac{ k^2}{m^3}+\frac{1}{16016}\frac{ k^4}{m^5}+
\right.\right.\0\\ & \quad \quad \quad \left.\left.
 + \frac{1}{192192}\frac{ k^6}{m^7}+\frac{1}{1867008}\frac{ k^8}{m^9}+\frac{3 }{47297536}\frac{ k^{10}}{m^{11}}+\ldots\right)\right)+
\0\\ & \quad 
 + k^8 \pi _{\mu\mu}^2 \pi _{\nu\nu}^2 \left(\frac{i }{5 \pi } \left(2 \frac{ m^3}{k^4}-\frac{3 }{14}\frac{ m}{k^2}+\frac{1}{168}\frac{1}{m}+\frac{1}{7392}\frac{ k^2}{m^3}+\frac{1}{128128}\frac{ k^4}{m^5}+
\right.\right.\0\\ & \quad \quad \quad \left.\left.
 + \frac{1}{1537536}\frac{ k^6}{m^7}+\frac{1}{14936064}\frac{ k^8}{m^9}+\frac{3 }{378380288}\frac{ k^{10}}{m^{11}}+\ldots\right)\right) }
\al{\label{eq:sc:4:4:3:uvir}\tilde{T}_{4,4;3\text{D}}^{\text{s,UV-IR}} & = k^8 \pi _{\mu\nu}^4 \left(\frac{i }{5 \pi } \left(\frac{4 }{7}\frac{ m}{k^2}-\frac{16 }{3}\frac{ m^3}{k^4}\right)\right)+k^8 \pi _{\mu\mu} \pi _{\mu\nu}^2 \pi _{\nu\nu} \left(-\frac{16 i }{5 \pi }\frac{ m^3}{k^4}\right)+\ldots }
Scalars, spin 4 x 4, dimension 4:
\al{\label{eq:sc:4:4:4:uv}\tilde{T}_{4,4;4\text{D}}^{\text{s,t,UV}} & = k^8 \pi _{\mu\nu}^4 \left(\frac{i }{\pi ^2} \left( \left(\frac{563}{99225}-\frac{P}{630}\right)- \left(-\frac{859}{29400}-\frac{P}{35}+\frac{L_0}{8}\right)\frac{ m^2}{k^2}+
\right.\right.\0\\ & \quad \quad \quad \left.\left.
 +  \left(\frac{511}{1200}-\frac{P}{5}+\frac{L_0}{8}\right)\frac{ m^4}{k^4}+ \left(-\frac{5}{9}+\frac{2 K}{3}\right)\frac{ m^6}{k^6}- \left(\frac{1}{12}+K\right)\frac{ m^8}{k^8}+
\right.\right.\0\\ & \quad \quad \quad \left.\left.
 +  \left(\frac{137}{150}+\frac{2 K}{5}\right)\frac{ m^{10}}{k^{10}}+ \left(\frac{3}{50}+\frac{2 K}{15}\right)\frac{ m^{12}}{k^{12}}+\ldots\right)\right)+
\0\\ & \quad 
 + k^8 \pi _{\mu\mu} \pi _{\mu\nu}^2 \pi _{\nu\nu} \left(\frac{i }{\pi ^2} \left( \left(\frac{563}{33075}-\frac{P}{210}\right)+ \left(-\frac{352}{1225}+\frac{3 P}{35}\right)\frac{ m^2}{k^2}+
\right.\right.\0\\ & \quad \quad \quad \left.\left.
 +  \left(\frac{143}{200}-\frac{3 P}{5}+\frac{3 L_0}{4}\right)\frac{ m^4}{k^4}+ \left(-\frac{5}{3}+2 K\right)\frac{ m^6}{k^6}- \left(\frac{1}{4}+3 K\right)\frac{ m^8}{k^8}+
\right.\right.\0\\ & \quad \quad \quad \left.\left.
 +  \left(\frac{137}{50}+\frac{6 K}{5}\right)\frac{ m^{10}}{k^{10}}+ \left(\frac{9}{50}+\frac{2 K}{5}\right)\frac{ m^{12}}{k^{12}}+\ldots\right)\right)+
\0\\ & \quad 
 + k^8 \pi _{\mu\mu}^2 \pi _{\nu\nu}^2 \left(\frac{i }{\pi ^2} \left( \left(\frac{563}{264600}-\frac{P}{1680}\right)+ \left(-\frac{44}{1225}+\frac{3 P}{280}\right)\frac{ m^2}{k^2}+
\right.\right.\0\\ & \quad \quad \quad \left.\left.
 +  \left(\frac{23}{100}-\frac{3 P}{40}\right)\frac{ m^4}{k^4}+ \left(-\frac{5}{24}+\frac{K}{4}\right)\frac{ m^6}{k^6}- \left(\frac{1}{32}+\frac{3 K}{8}\right)\frac{ m^8}{k^8}+
\right.\right.\0\\ & \quad \quad \quad \left.\left.
 +  \left(\frac{137}{400}+\frac{3 K}{20}\right)\frac{ m^{10}}{k^{10}}+ \left(\frac{9}{400}+\frac{K}{20}\right)\frac{ m^{12}}{k^{12}}+\ldots\right)\right) }
\al{\label{eq:sc:4:4:4:ir}\tilde{T}_{4,4;4\text{D}}^{\text{s,t,IR}} & = k^8 \pi _{\mu\nu}^4 \left(\frac{i }{10 \pi ^2} \left( \left(\frac{9}{8}-\frac{3 L_0}{4}\right)\frac{ m^4}{k^4}+ \left(\frac{27}{28}-\frac{27 L_0}{28}\right)\frac{ m^2}{k^2}-\frac{ L_0}{63}+\frac{1}{1386}\frac{ k^2}{m^2}+
\right.\right.\0\\ & \quad \quad \quad \left.\left.
 + \frac{1}{36036}\frac{ k^4}{m^4}+\frac{1}{540540}\frac{ k^6}{m^6}+\frac{1}{6126120}\frac{ k^8}{m^8}+\frac{1}{58198140}\frac{ k^{10}}{m^{10}}+
\right.\right.\0\\ & \quad \quad \quad \left.\left.
 + \frac{1}{488864376}\frac{ k^{12}}{m^{12}}+\ldots\right)\right)+
\0\\ & \quad 
 + k^8 \pi _{\mu\mu} \pi _{\mu\nu}^2 \pi _{\nu\nu} \left(\frac{i }{5 \pi ^2} \left( \left(-\frac{9}{8}+\frac{3 L_0}{4}\right)\frac{ m^4}{k^4}+ \left(-\frac{3}{7}+\frac{3 L_0}{7}\right)\frac{ m^2}{k^2}-\frac{ L_0}{42}+
\right.\right.\0\\ & \quad \quad \quad \left.\left.
 + \frac{1}{924}\frac{ k^2}{m^2}+\frac{1}{24024}\frac{ k^4}{m^4}+\frac{1}{360360}\frac{ k^6}{m^6}+\frac{1}{4084080}\frac{ k^8}{m^8}+\frac{1}{38798760}\frac{ k^{10}}{m^{10}}+
\right.\right.\0\\ & \quad \quad \quad \left.\left.
 + \frac{1}{325909584}\frac{ k^{12}}{m^{12}}+\ldots\right)\right)+
\0\\ & \quad 
 + k^8 \pi _{\mu\mu}^2 \pi _{\nu\nu}^2 \left(\frac{i }{40 \pi ^2} \left( \left(\frac{9}{2}-3 L_0\right)\frac{ m^4}{k^4}+ \left(-\frac{3}{7}+\frac{3 L_0}{7}\right)\frac{ m^2}{k^2}-\frac{ L_0}{42}+\frac{1}{924}\frac{ k^2}{m^2}+
\right.\right.\0\\ & \quad \quad \quad \left.\left.
 + \frac{1}{24024}\frac{ k^4}{m^4}+\frac{1}{360360}\frac{ k^6}{m^6}+\frac{1}{4084080}\frac{ k^8}{m^8}+\frac{1}{38798760}\frac{ k^{10}}{m^{10}}+
\right.\right.\0\\ & \quad \quad \quad \left.\left.
 + \frac{1}{325909584}\frac{ k^{12}}{m^{12}}+\ldots\right)\right) }
\al{\label{eq:sc:4:4:4:uvir}\tilde{T}_{4,4;4\text{D}}^{\text{s,UV-IR}} & = k^8 \pi _{\mu\nu}^4 \left(\frac{i }{5 \pi ^2} \left( \left(\frac{563}{19845}-\frac{K}{126}\right)+ \left(-\frac{247}{735}+\frac{K}{7}\right)\frac{ m^2}{k^2}+ \left(\frac{47}{30}-K\right)\frac{ m^4}{k^4}\right)\right)+
\0\\ & \quad 
 + k^8 \pi _{\mu\mu} \pi _{\mu\nu}^2 \pi _{\nu\nu} \left(\frac{i }{5 \pi ^2} \left( \left(\frac{563}{6615}-\frac{K}{42}\right)+ \left(-\frac{247}{245}+\frac{3 K}{7}\right)\frac{ m^2}{k^2}+
\right.\right.\0\\ & \quad \quad \quad \left.\left.
 +  \left(\frac{47}{10}-3 K\right)\frac{ m^4}{k^4}\right)\right)+
\0\\ & \quad 
 + k^8 \pi _{\mu\mu}^2 \pi _{\nu\nu}^2 \left(\frac{i }{40 \pi ^2} \left( \left(\frac{563}{6615}-\frac{K}{42}\right)+ \left(-\frac{247}{245}+\frac{3 K}{7}\right)\frac{ m^2}{k^2}+
\right.\right.\0\\ & \quad \quad \quad \left.\left.
 +  \left(\frac{47}{10}-3 K\right)\frac{ m^4}{k^4}\right)\right)+\ldots }
Scalars, spin 4 x 4, dimension 5:
\al{\label{eq:sc:4:4:5:uv}\tilde{T}_{4,4;5\text{D}}^{\text{s,t,UV}} & = k^8 \pi _{\mu\nu}^4 \left(\frac{1}{\pi ^2} \left(-\frac{ \pi }{10240} k+\frac{ \pi }{512}\frac{ m^2}{k}-\frac{i }{12}\frac{ m^3}{k^2}-\frac{ \pi }{64}\frac{ m^4}{k^3}+\frac{i }{15}\frac{ m^5}{k^4}+\frac{ \pi }{16}\frac{ m^6}{k^5}-
\right.\right.\0\\ & \quad \quad \quad \left.\left.
 - \frac{32 i }{105}\frac{ m^7}{k^6}-\frac{ \pi }{8}\frac{ m^8}{k^7}+\frac{128 i }{315}\frac{ m^9}{k^8}+\frac{ \pi }{10}\frac{ m^{10}}{k^9}-\frac{512 i }{3465}\frac{ m^{11}}{k^{10}}+\ldots\right)\right)+
\0\\ & \quad 
 + k^8 \pi _{\mu\mu} \pi _{\mu\nu}^2 \pi _{\nu\nu} \left(\frac{1}{\pi ^2} \left(-\frac{3  \pi }{10240} k+\frac{3  \pi }{512}\frac{ m^2}{k}-\frac{3  \pi }{64}\frac{ m^4}{k^3}+\frac{2 i }{5}\frac{ m^5}{k^4}+\frac{3  \pi }{16}\frac{ m^6}{k^5}-
\right.\right.\0\\ & \quad \quad \quad \left.\left.
 - \frac{32 i }{35}\frac{ m^7}{k^6}-\frac{3  \pi }{8}\frac{ m^8}{k^7}+\frac{128 i }{105}\frac{ m^9}{k^8}+\frac{3  \pi }{10}\frac{ m^{10}}{k^9}-\frac{512 i }{1155}\frac{ m^{11}}{k^{10}}+\ldots\right)\right)+
\0\\ & \quad 
 + k^8 \pi _{\mu\mu}^2 \pi _{\nu\nu}^2 \left(\frac{1}{\pi ^2} \left(-\frac{3  \pi }{81920} k+\frac{3  \pi }{4096}\frac{ m^2}{k}-\frac{3  \pi }{512}\frac{ m^4}{k^3}+\frac{3  \pi }{128}\frac{ m^6}{k^5}-\frac{4 i }{35}\frac{ m^7}{k^6}-
\right.\right.\0\\ & \quad \quad \quad \left.\left.
 - \frac{3  \pi }{64}\frac{ m^8}{k^7}+\frac{16 i }{105}\frac{ m^9}{k^8}+\frac{3  \pi }{80}\frac{ m^{10}}{k^9}-\frac{64 i }{1155}\frac{ m^{11}}{k^{10}}+\ldots\right)\right) }
\al{\label{eq:sc:4:4:5:ir}\tilde{T}_{4,4;5\text{D}}^{\text{s,t,IR}} & = k^8 \pi _{\mu\nu}^4 \left(\frac{i }{5 \pi ^2} \left(-\frac{1}{5}\frac{ m^5}{k^4}-\frac{9 }{28}\frac{ m^3}{k^2}-\frac{1}{126} m+\frac{1}{5544}\frac{ k^2}{m}+\frac{1}{288288}\frac{ k^4}{m^3}+
\right.\right.\0\\ & \quad \quad \quad \left.\left.
 + \frac{1}{5765760}\frac{ k^6}{m^5}+\frac{1}{78414336}\frac{ k^8}{m^7}+\frac{1}{851355648}\frac{ k^{10}}{m^9}+\frac{1}{7945986048}\frac{ k^{12}}{m^{11}}+
\right.\right.\0\\ & \quad \quad \quad \left.\left.
 + \ldots\right)\right)+
\0\\ & \quad 
 + k^8 \pi _{\mu\mu} \pi _{\mu\nu}^2 \pi _{\nu\nu} \left(\frac{i }{5 \pi ^2} \left(\frac{2 }{5}\frac{ m^5}{k^4}+\frac{2 }{7}\frac{ m^3}{k^2}-\frac{1}{42} m+\frac{1}{1848}\frac{ k^2}{m}+\frac{1}{96096}\frac{ k^4}{m^3}+
\right.\right.\0\\ & \quad \quad \quad \left.\left.
 + \frac{1}{1921920}\frac{ k^6}{m^5}+\frac{1}{26138112}\frac{ k^8}{m^7}+\frac{1}{283785216}\frac{ k^{10}}{m^9}+\frac{1}{2648662016}\frac{ k^{12}}{m^{11}}+
\right.\right.\0\\ & \quad \quad \quad \left.\left.
 + \ldots\right)\right)+
\0\\ & \quad 
 + k^8 \pi _{\mu\mu}^2 \pi _{\nu\nu}^2 \left(\frac{i }{5 \pi ^2} \left(-\frac{1}{5}\frac{ m^5}{k^4}+\frac{1}{28}\frac{ m^3}{k^2}-\frac{1}{336} m+\frac{1}{14784}\frac{ k^2}{m}+\frac{1}{768768}\frac{ k^4}{m^3}+
\right.\right.\0\\ & \quad \quad \quad \left.\left.
 + \frac{1}{15375360}\frac{ k^6}{m^5}+\frac{1}{209104896}\frac{ k^8}{m^7}+\frac{1}{2270281728}\frac{ k^{10}}{m^9}+
\right.\right.\0\\ & \quad \quad \quad \left.\left.
 + \frac{1}{21189296128}\frac{ k^{12}}{m^{11}}+\ldots\right)\right) }
\al{\label{eq:sc:4:4:5:uvir}\tilde{T}_{4,4;5\text{D}}^{\text{s,UV-IR}} & = k^8 \pi _{\mu\nu}^4 \left(\frac{i }{15 \pi ^2} \left(-\frac{2 }{7}\frac{ m^3}{k^2}+\frac{8 }{5}\frac{ m^5}{k^4}\right)\right)+k^8 \pi _{\mu\mu} \pi _{\mu\nu}^2 \pi _{\nu\nu} \left(\frac{8 i }{25 \pi ^2}\frac{ m^5}{k^4}\right)+\ldots }
Scalars, spin 4 x 4, dimension 6:
\al{\label{eq:sc:4:4:6:uv}\tilde{T}_{4,4;6\text{D}}^{\text{s,t,UV}} & = k^8 \pi _{\mu\nu}^4 \left(\frac{i }{2 \pi ^3} \left( \left(\frac{1627}{12006225}-\frac{P}{27720}\right) k^2+ \left(-\frac{563}{198450}+\frac{P}{1260}\right) m^2+
\right.\right.\0\\ & \quad \quad \quad \left.\left.
 +  \left(-\frac{5393}{235200}-\frac{P}{140}+\frac{L_0}{32}\right)\frac{ m^4}{k^2}- \left(\frac{461}{7200}-\frac{P}{30}+\frac{L_0}{48}\right)\frac{ m^6}{k^4}+
\right.\right.\0\\ & \quad \quad \quad \left.\left.
 +  \left(\frac{7}{144}-\frac{K}{12}\right)\frac{ m^8}{k^6}+ \left(\frac{17}{600}+\frac{K}{10}\right)\frac{ m^{10}}{k^8}- \left(\frac{49}{600}+\frac{K}{30}\right)\frac{ m^{12}}{k^{10}}+\ldots\right)\right)+
\0\\ & \quad 
 + k^8 \pi _{\mu\mu} \pi _{\mu\nu}^2 \pi _{\nu\nu} \left(\frac{i }{\pi ^3} \left( \left(\frac{1627}{8004150}-\frac{P}{18480}\right) k^2+ \left(-\frac{563}{132300}+\frac{P}{840}\right) m^2+
\right.\right.\0\\ & \quad \quad \quad \left.\left.
 +  \left(\frac{44}{1225}-\frac{3 P}{280}\right)\frac{ m^4}{k^2}- \left(\frac{31}{800}-\frac{P}{20}+\frac{L_0}{16}\right)\frac{ m^6}{k^4}+ \left(\frac{7}{96}-\frac{K}{8}\right)\frac{ m^8}{k^6}+
\right.\right.\0\\ & \quad \quad \quad \left.\left.
 +  \left(\frac{17}{400}+\frac{3 K}{20}\right)\frac{ m^{10}}{k^8}- \left(\frac{49}{400}+\frac{K}{20}\right)\frac{ m^{12}}{k^{10}}+\ldots\right)\right)+
\0\\ & \quad 
 + k^8 \pi _{\mu\mu}^2 \pi _{\nu\nu}^2 \left(\frac{i }{2 \pi ^3} \left( \left(\frac{1627}{32016600}-\frac{P}{73920}\right) k^2+ \left(-\frac{563}{529200}+\frac{P}{3360}\right) m^2+
\right.\right.\0\\ & \quad \quad \quad \left.\left.
 +  \left(\frac{11}{1225}-\frac{3 P}{1120}\right)\frac{ m^4}{k^2}+ \left(-\frac{23}{600}+\frac{P}{80}\right)\frac{ m^6}{k^4}+ \left(\frac{7}{384}-\frac{K}{32}\right)\frac{ m^8}{k^6}+
\right.\right.\0\\ & \quad \quad \quad \left.\left.
 +  \left(\frac{17}{1600}+\frac{3 K}{80}\right)\frac{ m^{10}}{k^8}- \left(\frac{49}{1600}+\frac{K}{80}\right)\frac{ m^{12}}{k^{10}}+\ldots\right)\right) }
\al{\label{eq:sc:4:4:6:ir}\tilde{T}_{4,4;6\text{D}}^{\text{s,t,IR}} & = k^8 \pi _{\mu\nu}^4 \left(\frac{i }{40 \pi ^3} \left( \left(-\frac{11}{24}+\frac{L_0}{4}\right)\frac{ m^6}{k^4}+ \left(-\frac{81}{112}+\frac{27 L_0}{56}\right)\frac{ m^4}{k^2}+ \left(-\frac{1}{63}+\frac{L_0}{63}\right) m^2-
\right.\right.\0\\ & \quad \quad \quad \left.\left.
 - \frac{ L_0}{1386} k^2+\frac{1}{36036}\frac{ k^4}{m^2}+\frac{1}{1081080}\frac{ k^6}{m^4}+\frac{1}{18378360}\frac{ k^8}{m^6}+\frac{1}{232792560}\frac{ k^{10}}{m^8}+
\right.\right.\0\\ & \quad \quad \quad \left.\left.
 + \frac{1}{2444321880}\frac{ k^{12}}{m^{10}}+\frac{1}{22487761296}\frac{ k^{14}}{m^{12}}+\ldots\right)\right)+
\0\\ & \quad 
 + k^8 \pi _{\mu\mu} \pi _{\mu\nu}^2 \pi _{\nu\nu} \left(\frac{i }{40 \pi ^3} \left( \left(\frac{11}{12}-\frac{L_0}{2}\right)\frac{ m^6}{k^4}+ \left(\frac{9}{14}-\frac{3 L_0}{7}\right)\frac{ m^4}{k^2}+
\right.\right.\0\\ & \quad \quad \quad \left.\left.
 +  \left(-\frac{1}{21}+\frac{L_0}{21}\right) m^2-\frac{ L_0}{462} k^2+\frac{1}{12012}\frac{ k^4}{m^2}+\frac{1}{360360}\frac{ k^6}{m^4}+\frac{1}{6126120}\frac{ k^8}{m^6}+
\right.\right.\0\\ & \quad \quad \quad \left.\left.
 + \frac{1}{77597520}\frac{ k^{10}}{m^8}+\frac{1}{814773960}\frac{ k^{12}}{m^{10}}+\frac{1}{7495920432}\frac{ k^{14}}{m^{12}}+\ldots\right)\right)+
\0\\ & \quad 
 + k^8 \pi _{\mu\mu}^2 \pi _{\nu\nu}^2 \left(\frac{i }{160 \pi ^3} \left( \left(-\frac{11}{6}+L_0\right)\frac{ m^6}{k^4}+ \left(\frac{9}{28}-\frac{3 L_0}{14}\right)\frac{ m^4}{k^2}+
\right.\right.\0\\ & \quad \quad \quad \left.\left.
 +  \left(-\frac{1}{42}+\frac{L_0}{42}\right) m^2-\frac{ L_0}{924} k^2+\frac{1}{24024}\frac{ k^4}{m^2}+\frac{1}{720720}\frac{ k^6}{m^4}+
\right.\right.\0\\ & \quad \quad \quad \left.\left.
 + \frac{1}{12252240}\frac{ k^8}{m^6}+\frac{1}{155195040}\frac{ k^{10}}{m^8}+\frac{1}{1629547920}\frac{ k^{12}}{m^{10}}+
\right.\right.\0\\ & \quad \quad \quad \left.\left.
 + \frac{1}{14991840864}\frac{ k^{14}}{m^{12}}+\ldots\right)\right) }
\al{\label{eq:sc:4:4:6:uvir}\tilde{T}_{4,4;6\text{D}}^{\text{s,UV-IR}} & = k^8 \pi _{\mu\nu}^4 \left(\frac{i }{10 \pi ^3} \left( \left(\frac{1627}{2401245}-\frac{K}{5544}\right) k^2+ \left(-\frac{811}{79380}+\frac{K}{252}\right) m^2+
\right.\right.\0\\ & \quad \quad \quad \left.\left.
 +  \left(\frac{389}{5880}-\frac{K}{28}\right)\frac{ m^4}{k^2}+ \left(-\frac{37}{180}+\frac{K}{6}\right)\frac{ m^6}{k^4}\right)\right)+
\0\\ & \quad 
 + k^8 \pi _{\mu\mu} \pi _{\mu\nu}^2 \pi _{\nu\nu} \left(\frac{i }{10 \pi ^3} \left( \left(\frac{1627}{800415}-\frac{K}{1848}\right) k^2+ \left(-\frac{811}{26460}+\frac{K}{84}\right) m^2+
\right.\right.\0\\ & \quad \quad \quad \left.\left.
 +  \left(\frac{389}{1960}-\frac{3 K}{28}\right)\frac{ m^4}{k^2}+ \left(-\frac{37}{60}+\frac{K}{2}\right)\frac{ m^6}{k^4}\right)\right)+
\0\\ & \quad 
 + k^8 \pi _{\mu\mu}^2 \pi _{\nu\nu}^2 \left(\frac{i }{80 \pi ^3} \left( \left(\frac{1627}{800415}-\frac{K}{1848}\right) k^2+ \left(-\frac{811}{26460}+\frac{K}{84}\right) m^2+
\right.\right.\0\\ & \quad \quad \quad \left.\left.
 +  \left(\frac{389}{1960}-\frac{3 K}{28}\right)\frac{ m^4}{k^2}+ \left(-\frac{37}{60}+\frac{K}{2}\right)\frac{ m^6}{k^4}\right)\right)+\ldots }
Scalars, spin 5 x 5, dimension 3:
\al{\label{eq:sc:5:5:3:uv}\tilde{T}_{5,5;3\text{D}}^{\text{s,t,UV}} & = k^{10} \pi _{\mu\nu}^5 \left(\frac{1}{256}\frac{1}{k}-\frac{i }{2 \pi }\frac{ m}{k^2}-\frac{5 }{64}\frac{ m^2}{k^3}-\frac{2 i }{3 \pi }\frac{ m^3}{k^4}+\frac{5 }{8}\frac{ m^4}{k^5}-\frac{64 i }{15 \pi }\frac{ m^5}{k^6}-\frac{5 }{2}\frac{ m^6}{k^7}+
\right.\0\\ & \quad \quad \left.
 + \frac{256 i }{21 \pi }\frac{ m^7}{k^8}+5 \frac{ m^8}{k^9}-\frac{1024 i }{63 \pi }\frac{ m^9}{k^{10}}-4 \frac{ m^{10}}{k^{11}}+\frac{4096 i }{693 \pi }\frac{ m^{11}}{k^{12}}+\ldots\right)+
\0\\ & \quad 
 + k^{10} \pi _{\mu\mu} \pi _{\mu\nu}^3 \pi _{\nu\nu} \left(\frac{5 }{256}\frac{1}{k}-\frac{25 }{64}\frac{ m^2}{k^3}+\frac{20 i }{3 \pi }\frac{ m^3}{k^4}+\frac{25 }{8}\frac{ m^4}{k^5}-\frac{64 i }{3 \pi }\frac{ m^5}{k^6}-\frac{25 }{2}\frac{ m^6}{k^7}+
\right.\0\\ & \quad \quad \left.
 + \frac{1280 i }{21 \pi }\frac{ m^7}{k^8}+25 \frac{ m^8}{k^9}-\frac{5120 i }{63 \pi }\frac{ m^9}{k^{10}}-20 \frac{ m^{10}}{k^{11}}+\frac{20480 i }{693 \pi }\frac{ m^{11}}{k^{12}}+\ldots\right)+
\0\\ & \quad 
 + k^{10} \pi _{\mu\mu}^2 \pi _{\mu\nu} \pi _{\nu\nu}^2 \left(\frac{15 }{2048}\frac{1}{k}-\frac{75 }{512}\frac{ m^2}{k^3}+\frac{75 }{64}\frac{ m^4}{k^5}-\frac{8 i }{\pi }\frac{ m^5}{k^6}-\frac{75 }{16}\frac{ m^6}{k^7}+\frac{160 i }{7 \pi }\frac{ m^7}{k^8}+
\right.\0\\ & \quad \quad \left.
 + \frac{75 }{8}\frac{ m^8}{k^9}-\frac{640 i }{21 \pi }\frac{ m^9}{k^{10}}-\frac{15 }{2}\frac{ m^{10}}{k^{11}}+\frac{2560 i }{231 \pi }\frac{ m^{11}}{k^{12}}+\ldots\right) }
\al{\label{eq:sc:5:5:3:ir}\tilde{T}_{5,5;3\text{D}}^{\text{s,t,IR}} & = k^{10} \pi _{\mu\nu}^5 \left(\frac{i }{\pi } \left(-\frac{10 }{7}\frac{ m^3}{k^4}-\frac{55 }{126}\frac{ m}{k^2}-\frac{1}{693}\frac{1}{m}-\frac{1}{36036}\frac{ k^2}{m^3}-\frac{1}{720720}\frac{ k^4}{m^5}-
\right.\right.\0\\ & \quad \quad \quad \left.\left.
 - \frac{1}{9801792}\frac{ k^6}{m^7}-\frac{1}{106419456}\frac{ k^8}{m^9}-\frac{1}{993248256}\frac{ k^{10}}{m^{11}}+\ldots\right)\right)+
\0\\ & \quad 
 + k^{10} \pi _{\mu\mu} \pi _{\mu\nu}^3 \pi _{\nu\nu} \left(\frac{i }{\pi } \left(\frac{20 }{7}\frac{ m^3}{k^4}+\frac{20 }{63}\frac{ m}{k^2}-\frac{5 }{693}\frac{1}{m}-\frac{5 }{36036}\frac{ k^2}{m^3}-\frac{1}{144144}\frac{ k^4}{m^5}-
\right.\right.\0\\ & \quad \quad \quad \left.\left.
 - \frac{5 }{9801792}\frac{ k^6}{m^7}-\frac{5 }{106419456}\frac{ k^8}{m^9}-\frac{5 }{993248256}\frac{ k^{10}}{m^{11}}+\ldots\right)\right)+
\0\\ & \quad 
 + k^{10} \pi _{\mu\mu}^2 \pi _{\mu\nu} \pi _{\nu\nu}^2 \left(\frac{i }{\pi } \left(-\frac{10 }{7}\frac{ m^3}{k^4}+\frac{5 }{42}\frac{ m}{k^2}-\frac{5 }{1848}\frac{1}{m}-\frac{5 }{96096}\frac{ k^2}{m^3}-\frac{1}{384384}\frac{ k^4}{m^5}-
\right.\right.\0\\ & \quad \quad \quad \left.\left.
 - \frac{5 }{26138112}\frac{ k^6}{m^7}-\frac{5 }{283785216}\frac{ k^8}{m^9}-\frac{5 }{2648662016}\frac{ k^{10}}{m^{11}}+\ldots\right)\right) }
\al{\label{eq:sc:5:5:3:uvir}\tilde{T}_{5,5;3\text{D}}^{\text{s,UV-IR}} & = k^{10} \pi _{\mu\nu}^5 \left(\frac{i }{21 \pi } \left(-\frac{4 }{3}\frac{ m}{k^2}+16 \frac{ m^3}{k^4}\right)\right)+k^{10} \pi _{\mu\mu} \pi _{\mu\nu}^3 \pi _{\nu\nu} \left(\frac{80 i }{21 \pi }\frac{ m^3}{k^4}\right)+\ldots }
Scalars, spin 5 x 5, dimension 4:
\al{\label{eq:sc:5:5:4:uv}\tilde{T}_{5,5;4\text{D}}^{\text{s,t,UV}} & = k^{10} \pi _{\mu\nu}^5 \left(\frac{i }{\pi ^2} \left( \left(-\frac{6508}{2401245}+\frac{P}{1386}\right)+ \left(-\frac{10837}{158760}-\frac{P}{63}+\frac{L_0}{8}\right)\frac{ m^2}{k^2}+
\right.\right.\0\\ & \quad \quad \quad \left.\left.
 +  \left(-\frac{7837}{11760}+\frac{P}{7}+\frac{L_0}{8}\right)\frac{ m^4}{k^4}+ \left(\frac{37}{45}-\frac{2 K}{3}\right)\frac{ m^6}{k^6}+ \left(-\frac{35}{36}+\frac{5 K}{3}\right)\frac{ m^8}{k^8}-
\right.\right.\0\\ & \quad \quad \quad \left.\left.
 -  \left(\frac{17}{30}+2 K\right)\frac{ m^{10}}{k^{10}}+ \left(\frac{49}{30}+\frac{2 K}{3}\right)\frac{ m^{12}}{k^{12}}+\ldots\right)\right)+
\0\\ & \quad 
 + k^{10} \pi _{\mu\mu} \pi _{\mu\nu}^3 \pi _{\nu\nu} \left(\frac{i }{\pi ^2} \left( \left(-\frac{6508}{480249}+\frac{5 P}{1386}\right)+ \left(\frac{1126}{3969}-\frac{5 P}{63}\right)\frac{ m^2}{k^2}-
\right.\right.\0\\ & \quad \quad \quad \left.\left.
 -  \left(\frac{611}{1176}-\frac{5 P}{7}+\frac{5 L_0}{4}\right)\frac{ m^4}{k^4}+ \left(\frac{37}{9}-\frac{10 K}{3}\right)\frac{ m^6}{k^6}+
\right.\right.\0\\ & \quad \quad \quad \left.\left.
 +  \left(-\frac{175}{36}+\frac{25 K}{3}\right)\frac{ m^8}{k^8}- \left(\frac{17}{6}+10 K\right)\frac{ m^{10}}{k^{10}}+ \left(\frac{49}{6}+\frac{10 K}{3}\right)\frac{ m^{12}}{k^{12}}+
\right.\right.\0\\ & \quad \quad \quad \left.\left.
 + \ldots\right)\right)+
\0\\ & \quad 
 + k^{10} \pi _{\mu\mu}^2 \pi _{\mu\nu} \pi _{\nu\nu}^2 \left(\frac{i }{\pi ^2} \left( \left(-\frac{1627}{320166}+\frac{5 P}{3696}\right)+ \left(\frac{563}{5292}-\frac{5 P}{168}\right)\frac{ m^2}{k^2}+
\right.\right.\0\\ & \quad \quad \quad \left.\left.
 +  \left(-\frac{44}{49}+\frac{15 P}{56}\right)\frac{ m^4}{k^4}+ \left(\frac{37}{24}-\frac{5 K}{4}\right)\frac{ m^6}{k^6}+ \left(-\frac{175}{96}+\frac{25 K}{8}\right)\frac{ m^8}{k^8}-
\right.\right.\0\\ & \quad \quad \quad \left.\left.
 -  \left(\frac{17}{16}+\frac{15 K}{4}\right)\frac{ m^{10}}{k^{10}}+ \left(\frac{49}{16}+\frac{5 K}{4}\right)\frac{ m^{12}}{k^{12}}+\ldots\right)\right) }
\al{\label{eq:sc:5:5:4:ir}\tilde{T}_{5,5;4\text{D}}^{\text{s,t,IR}} & = k^{10} \pi _{\mu\nu}^5 \left(\frac{i }{14 \pi ^2} \left( \left(-\frac{45}{8}+\frac{15 L_0}{4}\right)\frac{ m^4}{k^4}+ \left(-\frac{55}{36}+\frac{55 L_0}{36}\right)\frac{ m^2}{k^2}+\frac{ L_0}{99}-\frac{1}{2574}\frac{ k^2}{m^2}-
\right.\right.\0\\ & \quad \quad \quad \left.\left.
 - \frac{1}{77220}\frac{ k^4}{m^4}-\frac{1}{1312740}\frac{ k^6}{m^6}-\frac{1}{16628040}\frac{ k^8}{m^8}-\frac{1}{174594420}\frac{ k^{10}}{m^{10}}-
\right.\right.\0\\ & \quad \quad \quad \left.\left.
 - \frac{1}{1606268664}\frac{ k^{12}}{m^{12}}+\ldots\right)\right)+
\0\\ & \quad 
 + k^{10} \pi _{\mu\mu} \pi _{\mu\nu}^3 \pi _{\nu\nu} \left(\frac{i }{7 \pi ^2} \left( \left(\frac{45}{8}-\frac{15 L_0}{4}\right)\frac{ m^4}{k^4}+ \left(\frac{5}{9}-\frac{5 L_0}{9}\right)\frac{ m^2}{k^2}+\frac{5  L_0}{198}-
\right.\right.\0\\ & \quad \quad \quad \left.\left.
 - \frac{5 }{5148}\frac{ k^2}{m^2}-\frac{1}{30888}\frac{ k^4}{m^4}-\frac{1}{525096}\frac{ k^6}{m^6}-\frac{1}{6651216}\frac{ k^8}{m^8}-\frac{1}{69837768}\frac{ k^{10}}{m^{10}}-
\right.\right.\0\\ & \quad \quad \quad \left.\left.
 - \frac{5 }{3212537328}\frac{ k^{12}}{m^{12}}+\ldots\right)\right)+
\0\\ & \quad 
 + k^{10} \pi _{\mu\mu}^2 \pi _{\mu\nu} \pi _{\nu\nu}^2 \left(\frac{i }{56 \pi ^2} \left( \left(-\frac{45}{2}+15 L_0\right)\frac{ m^4}{k^4}+ \left(\frac{5}{3}-\frac{5 L_0}{3}\right)\frac{ m^2}{k^2}+\frac{5  L_0}{66}-
\right.\right.\0\\ & \quad \quad \quad \left.\left.
 - \frac{5 }{1716}\frac{ k^2}{m^2}-\frac{1}{10296}\frac{ k^4}{m^4}-\frac{1}{175032}\frac{ k^6}{m^6}-\frac{1}{2217072}\frac{ k^8}{m^8}-\frac{1}{23279256}\frac{ k^{10}}{m^{10}}-
\right.\right.\0\\ & \quad \quad \quad \left.\left.
 - \frac{5 }{1070845776}\frac{ k^{12}}{m^{12}}+\ldots\right)\right) }
\al{\label{eq:sc:5:5:4:uvir}\tilde{T}_{5,5;4\text{D}}^{\text{s,UV-IR}} & = k^{10} \pi _{\mu\nu}^5 \left(\frac{i }{7 \pi ^2} \left( \left(-\frac{6508}{343035}+\frac{K}{198}\right)+ \left(\frac{811}{2835}-\frac{K}{9}\right)\frac{ m^2}{k^2}+
\right.\right.\0\\ & \quad \quad \quad \left.\left.
 +  \left(-\frac{389}{210}+K\right)\frac{ m^4}{k^4}\right)\right)+
\0\\ & \quad 
 + k^{10} \pi _{\mu\mu} \pi _{\mu\nu}^3 \pi _{\nu\nu} \left(\frac{i }{7 \pi ^2} \left( \left(-\frac{6508}{68607}+\frac{5 K}{198}\right)+ \left(\frac{811}{567}-\frac{5 K}{9}\right)\frac{ m^2}{k^2}+
\right.\right.\0\\ & \quad \quad \quad \left.\left.
 +  \left(-\frac{389}{42}+5 K\right)\frac{ m^4}{k^4}\right)\right)+
\0\\ & \quad 
 + k^{10} \pi _{\mu\mu}^2 \pi _{\mu\nu} \pi _{\nu\nu}^2 \left(\frac{i }{14 \pi ^2} \left( \left(-\frac{1627}{22869}+\frac{5 K}{264}\right)+ \left(\frac{811}{756}-\frac{5 K}{12}\right)\frac{ m^2}{k^2}+
\right.\right.\0\\ & \quad \quad \quad \left.\left.
 +  \left(-\frac{389}{56}+\frac{15 K}{4}\right)\frac{ m^4}{k^4}\right)\right)+\ldots }
Scalars, spin 5 x 5, dimension 5:
\al{\label{eq:sc:5:5:5:uv}\tilde{T}_{5,5;5\text{D}}^{\text{s,t,UV}} & = k^{10} \pi _{\mu\nu}^5 \left(\frac{1}{\pi ^2} \left(\frac{ \pi }{24576} k-\frac{ \pi }{1024}\frac{ m^2}{k}+\frac{i }{12}\frac{ m^3}{k^2}+\frac{5  \pi }{512}\frac{ m^4}{k^3}+\frac{i }{15}\frac{ m^5}{k^4}-\frac{5  \pi }{96}\frac{ m^6}{k^5}+
\right.\right.\0\\ & \quad \quad \quad \left.\left.
 + \frac{32 i }{105}\frac{ m^7}{k^6}+\frac{5  \pi }{32}\frac{ m^8}{k^7}-\frac{128 i }{189}\frac{ m^9}{k^8}-\frac{ \pi }{4}\frac{ m^{10}}{k^9}+\frac{512 i }{693}\frac{ m^{11}}{k^{10}}+\frac{ \pi }{6}\frac{ m^{12}}{k^{11}}+\ldots\right)\right)+
\0\\ & \quad 
 + k^{10} \pi _{\mu\mu} \pi _{\mu\nu}^3 \pi _{\nu\nu} \left(\frac{1}{\pi ^2} \left(\frac{5  \pi }{24576} k-\frac{5  \pi }{1024}\frac{ m^2}{k}+\frac{25  \pi }{512}\frac{ m^4}{k^3}-\frac{2 i }{3}\frac{ m^5}{k^4}-\frac{25  \pi }{96}\frac{ m^6}{k^5}+
\right.\right.\0\\ & \quad \quad \quad \left.\left.
 + \frac{32 i }{21}\frac{ m^7}{k^6}+\frac{25  \pi }{32}\frac{ m^8}{k^7}-\frac{640 i }{189}\frac{ m^9}{k^8}-\frac{5  \pi }{4}\frac{ m^{10}}{k^9}+\frac{2560 i }{693}\frac{ m^{11}}{k^{10}}+\frac{5  \pi }{6}\frac{ m^{12}}{k^{11}}+
\right.\right.\0\\ & \quad \quad \quad \left.\left.
 + \ldots\right)\right)+
\0\\ & \quad 
 + k^{10} \pi _{\mu\mu}^2 \pi _{\mu\nu} \pi _{\nu\nu}^2 \left(\frac{1}{\pi ^2} \left(\frac{5  \pi }{65536} k-\frac{15  \pi }{8192}\frac{ m^2}{k}+\frac{75  \pi }{4096}\frac{ m^4}{k^3}-\frac{25  \pi }{256}\frac{ m^6}{k^5}+\frac{4 i }{7}\frac{ m^7}{k^6}+
\right.\right.\0\\ & \quad \quad \quad \left.\left.
 + \frac{75  \pi }{256}\frac{ m^8}{k^7}-\frac{80 i }{63}\frac{ m^9}{k^8}-\frac{15  \pi }{32}\frac{ m^{10}}{k^9}+\frac{320 i }{231}\frac{ m^{11}}{k^{10}}+\frac{5  \pi }{16}\frac{ m^{12}}{k^{11}}+\ldots\right)\right) }
\al{\label{eq:sc:5:5:5:ir}\tilde{T}_{5,5;5\text{D}}^{\text{s,t,IR}} & = k^{10} \pi _{\mu\nu}^5 \left(\frac{i }{7 \pi ^2} \left(\frac{m^5}{k^4}+\frac{55 }{108}\frac{ m^3}{k^2}+\frac{1}{198} m-\frac{1}{10296}\frac{ k^2}{m}-\frac{1}{617760}\frac{ k^4}{m^3}-
\right.\right.\0\\ & \quad \quad \quad \left.\left.
 - \frac{1}{14002560}\frac{ k^6}{m^5}-\frac{1}{212838912}\frac{ k^8}{m^7}-\frac{1}{2554066944}\frac{ k^{10}}{m^9}-
\right.\right.\0\\ & \quad \quad \quad \left.\left.
 - \frac{1}{26108239872}\frac{ k^{12}}{m^{11}}+\ldots\right)\right)+
\0\\ & \quad 
 + k^{10} \pi _{\mu\mu} \pi _{\mu\nu}^3 \pi _{\nu\nu} \left(\frac{i }{7 \pi ^2} \left(-2 \frac{ m^5}{k^4}-\frac{10 }{27}\frac{ m^3}{k^2}+\frac{5 }{198} m-\frac{5 }{10296}\frac{ k^2}{m}-\frac{1}{123552}\frac{ k^4}{m^3}-
\right.\right.\0\\ & \quad \quad \quad \left.\left.
 - \frac{1}{2800512}\frac{ k^6}{m^5}-\frac{5 }{212838912}\frac{ k^8}{m^7}-\frac{5 }{2554066944}\frac{ k^{10}}{m^9}-\frac{5 }{26108239872}\frac{ k^{12}}{m^{11}}+
\right.\right.\0\\ & \quad \quad \quad \left.\left.
 + \ldots\right)\right)+
\0\\ & \quad 
 + k^{10} \pi _{\mu\mu}^2 \pi _{\mu\nu} \pi _{\nu\nu}^2 \left(\frac{i }{7 \pi ^2} \left(\frac{m^5}{k^4}-\frac{5 }{36}\frac{ m^3}{k^2}+\frac{5 }{528} m-\frac{5 }{27456}\frac{ k^2}{m}-\frac{1}{329472}\frac{ k^4}{m^3}-
\right.\right.\0\\ & \quad \quad \quad \left.\left.
 - \frac{1}{7468032}\frac{ k^6}{m^5}-\frac{5 }{567570432}\frac{ k^8}{m^7}-\frac{5 }{6810845184}\frac{ k^{10}}{m^9}-\frac{5 }{69621972992}\frac{ k^{12}}{m^{11}}+
\right.\right.\0\\ & \quad \quad \quad \left.\left.
 + \ldots\right)\right) }
\al{\label{eq:sc:5:5:5:uvir}\tilde{T}_{5,5;5\text{D}}^{\text{s,UV-IR}} & = k^{10} \pi _{\mu\nu}^5 \left(\frac{i }{21 \pi ^2} \left(\frac{2 }{9}\frac{ m^3}{k^2}-\frac{8 }{5}\frac{ m^5}{k^4}\right)\right)+k^{10} \pi _{\mu\mu} \pi _{\mu\nu}^3 \pi _{\nu\nu} \left(-\frac{8 i }{21 \pi ^2}\frac{ m^5}{k^4}\right)+\ldots }
Scalars, spin 5 x 5, dimension 6:
\al{\label{eq:sc:5:5:6:uv}\tilde{T}_{5,5;6\text{D}}^{\text{s,t,UV}} & = k^{10} \pi _{\mu\nu}^5 \left(\frac{i }{\pi ^3} \left( \left(-\frac{88069}{3246483240}+\frac{P}{144144}\right) k^2+ \left(\frac{1627}{2401245}-\frac{P}{5544}\right) m^2-
\right.\right.\0\\ & \quad \quad \quad \left.\left.
 -  \left(-\frac{41519}{2540160}-\frac{P}{504}+\frac{L_0}{64}\right)\frac{ m^4}{k^2}- \left(-\frac{8327}{141120}+\frac{P}{84}+\frac{L_0}{96}\right)\frac{ m^6}{k^4}+
\right.\right.\0\\ & \quad \quad \quad \left.\left.
 +  \left(-\frac{59}{1440}+\frac{K}{24}\right)\frac{ m^8}{k^6}+ \left(\frac{23}{720}-\frac{K}{12}\right)\frac{ m^{10}}{k^8}+ \left(\frac{3}{80}+\frac{K}{12}\right)\frac{ m^{12}}{k^{10}}+
\right.\right.\0\\ & \quad \quad \quad \left.\left.
 + \ldots\right)\right)+
\0\\ & \quad 
 + k^{10} \pi _{\mu\mu} \pi _{\mu\nu}^3 \pi _{\nu\nu} \left(\frac{i }{\pi ^3} \left( \left(-\frac{88069}{649296648}+\frac{5 P}{144144}\right) k^2+
\right.\right.\0\\ & \quad \quad \quad \left.\left.
 +  \left(\frac{1627}{480249}-\frac{5 P}{5544}\right) m^2+ \left(-\frac{563}{15876}+\frac{5 P}{504}\right)\frac{ m^4}{k^2}+
\right.\right.\0\\ & \quad \quad \quad \left.\left.
 +  \left(\frac{121}{14112}-\frac{5 P}{84}+\frac{5 L_0}{48}\right)\frac{ m^6}{k^4}+ \left(-\frac{59}{288}+\frac{5 K}{24}\right)\frac{ m^8}{k^6}+
\right.\right.\0\\ & \quad \quad \quad \left.\left.
 +  \left(\frac{23}{144}-\frac{5 K}{12}\right)\frac{ m^{10}}{k^8}+ \left(\frac{3}{16}+\frac{5 K}{12}\right)\frac{ m^{12}}{k^{10}}+\ldots\right)\right)+
\0\\ & \quad 
 + k^{10} \pi _{\mu\mu}^2 \pi _{\mu\nu} \pi _{\nu\nu}^2 \left(\frac{i }{\pi ^3} \left( \left(-\frac{88069}{1731457728}+\frac{5 P}{384384}\right) k^2+
\right.\right.\0\\ & \quad \quad \quad \left.\left.
 +  \left(\frac{1627}{1280664}-\frac{5 P}{14784}\right) m^2+ \left(-\frac{563}{42336}+\frac{5 P}{1344}\right)\frac{ m^4}{k^2}+
\right.\right.\0\\ & \quad \quad \quad \left.\left.
 +  \left(\frac{11}{147}-\frac{5 P}{224}\right)\frac{ m^6}{k^4}+ \left(-\frac{59}{768}+\frac{5 K}{64}\right)\frac{ m^8}{k^6}+ \left(\frac{23}{384}-\frac{5 K}{32}\right)\frac{ m^{10}}{k^8}+
\right.\right.\0\\ & \quad \quad \quad \left.\left.
 +  \left(\frac{9}{128}+\frac{5 K}{32}\right)\frac{ m^{12}}{k^{10}}+\ldots\right)\right) }
\al{\label{eq:sc:5:5:6:ir}\tilde{T}_{5,5;6\text{D}}^{\text{s,t,IR}} & = k^{10} \pi _{\mu\nu}^5 \left(\frac{i }{56 \pi ^3} \left( \left(\frac{55}{24}-\frac{5 L_0}{4}\right)\frac{ m^6}{k^4}+ \left(\frac{55}{48}-\frac{55 L_0}{72}\right)\frac{ m^4}{k^2}+ \left(\frac{1}{99}-\frac{L_0}{99}\right) m^2+
\right.\right.\0\\ & \quad \quad \quad \left.\left.
 + \frac{ L_0}{2574} k^2-\frac{1}{77220}\frac{ k^4}{m^2}-\frac{1}{2625480}\frac{ k^6}{m^4}-\frac{1}{49884120}\frac{ k^8}{m^6}-\frac{1}{698377680}\frac{ k^{10}}{m^8}-
\right.\right.\0\\ & \quad \quad \quad \left.\left.
 - \frac{1}{8031343320}\frac{ k^{12}}{m^{10}}-\frac{1}{80313433200}\frac{ k^{14}}{m^{12}}+\ldots\right)\right)+
\0\\ & \quad 
 + k^{10} \pi _{\mu\mu} \pi _{\mu\nu}^3 \pi _{\nu\nu} \left(\frac{i }{56 \pi ^3} \left( \left(-\frac{55}{12}+\frac{5 L_0}{2}\right)\frac{ m^6}{k^4}+ \left(-\frac{5}{6}+\frac{5 L_0}{9}\right)\frac{ m^4}{k^2}+
\right.\right.\0\\ & \quad \quad \quad \left.\left.
 +  \left(\frac{5}{99}-\frac{5 L_0}{99}\right) m^2+\frac{5  L_0}{2574} k^2-\frac{1}{15444}\frac{ k^4}{m^2}-\frac{1}{525096}\frac{ k^6}{m^4}-\frac{1}{9976824}\frac{ k^8}{m^6}-
\right.\right.\0\\ & \quad \quad \quad \left.\left.
 - \frac{1}{139675536}\frac{ k^{10}}{m^8}-\frac{1}{1606268664}\frac{ k^{12}}{m^{10}}-\frac{1}{16062686640}\frac{ k^{14}}{m^{12}}+\ldots\right)\right)+
\0\\ & \quad 
 + k^{10} \pi _{\mu\mu}^2 \pi _{\mu\nu} \pi _{\nu\nu}^2 \left(\frac{i }{224 \pi ^3} \left( \left(\frac{55}{6}-5 L_0\right)\frac{ m^6}{k^4}+ \left(-\frac{5}{4}+\frac{5 L_0}{6}\right)\frac{ m^4}{k^2}+
\right.\right.\0\\ & \quad \quad \quad \left.\left.
 +  \left(\frac{5}{66}-\frac{5 L_0}{66}\right) m^2+\frac{5  L_0}{1716} k^2-\frac{1}{10296}\frac{ k^4}{m^2}-\frac{1}{350064}\frac{ k^6}{m^4}-\frac{1}{6651216}\frac{ k^8}{m^6}-
\right.\right.\0\\ & \quad \quad \quad \left.\left.
 - \frac{1}{93117024}\frac{ k^{10}}{m^8}-\frac{1}{1070845776}\frac{ k^{12}}{m^{10}}-\frac{1}{10708457760}\frac{ k^{14}}{m^{12}}+\ldots\right)\right) }
\al{\label{eq:sc:5:5:6:uvir}\tilde{T}_{5,5;6\text{D}}^{\text{s,UV-IR}} & = k^{10} \pi _{\mu\nu}^5 \left(\frac{i }{84 \pi ^3} \left( \left(-\frac{88069}{38648610}+\frac{K}{1716}\right) k^2+ \left(\frac{9551}{228690}-\frac{K}{66}\right) m^2+
\right.\right.\0\\ & \quad \quad \quad \left.\left.
 +  \left(-\frac{1307}{3780}+\frac{K}{6}\right)\frac{ m^4}{k^2}+ \left(\frac{319}{210}-K\right)\frac{ m^6}{k^4}\right)\right)+
\0\\ & \quad 
 + k^{10} \pi _{\mu\mu} \pi _{\mu\nu}^3 \pi _{\nu\nu} \left(\frac{i }{84 \pi ^3} \left( \left(-\frac{88069}{7729722}+\frac{5 K}{1716}\right) k^2+ \left(\frac{9551}{45738}-\frac{5 K}{66}\right) m^2+
\right.\right.\0\\ & \quad \quad \quad \left.\left.
 +  \left(-\frac{1307}{756}+\frac{5 K}{6}\right)\frac{ m^4}{k^2}+ \left(\frac{319}{42}-5 K\right)\frac{ m^6}{k^4}\right)\right)+
\0\\ & \quad 
 + k^{10} \pi _{\mu\mu}^2 \pi _{\mu\nu} \pi _{\nu\nu}^2 \left(\frac{i }{224 \pi ^3} \left( \left(-\frac{88069}{7729722}+\frac{5 K}{1716}\right) k^2+ \left(\frac{9551}{45738}-\frac{5 K}{66}\right) m^2+
\right.\right.\0\\ & \quad \quad \quad \left.\left.
 +  \left(-\frac{1307}{756}+\frac{5 K}{6}\right)\frac{ m^4}{k^2}+ \left(\frac{319}{42}-5 K\right)\frac{ m^6}{k^4}\right)\right)+\ldots }
\subsection{Divergences of the scalar amplitudes}
Scalars, spin 0 x 2:
\al{\label{eq:snc:0:2:3:divr}\tilde{T}_{0,2;3\text{D}}^{\text{s,nt}}\cdot k & = k_{\nu } \left(\frac{i }{2 \pi } m\right) }
\al{\label{eq:snc:0:2:4:divr}\tilde{T}_{0,2;4\text{D}}^{\text{s,nt}}\cdot k & = k_{\nu } \left(-\frac{i  L_1}{8 \pi ^2} m^2\right) }
\al{\label{eq:snc:0:2:5:divr}\tilde{T}_{0,2;5\text{D}}^{\text{s,nt}}\cdot k & = k_{\nu } \left(-\frac{i }{12 \pi ^2} m^3\right) }
\al{\label{eq:snc:0:2:6:divr}\tilde{T}_{0,2;6\text{D}}^{\text{s,nt}}\cdot k & = k_{\nu } \left(\frac{i  L_2}{64 \pi ^3} m^4\right) }
Scalars, spin 0 x 4:
\al{\label{eq:snc:0:4:3:divr}\tilde{T}_{0,4;3\text{D}}^{\text{s,nt}}\cdot k & = k_{\nu }^3 \left(\frac{i }{2 \pi } m\right)+k_{\nu } \eta_{\nu\nu} \left(\frac{2 i }{\pi } m^3\right) }
\al{\label{eq:snc:0:4:4:divr}\tilde{T}_{0,4;4\text{D}}^{\text{s,nt}}\cdot k & = k_{\nu }^3 \left(-\frac{i  L_1}{8 \pi ^2} m^2\right)+k_{\nu } \eta_{\nu\nu} \left(-\frac{3 i  L_2}{8 \pi ^2} m^4\right) }
\al{\label{eq:snc:0:4:5:divr}\tilde{T}_{0,4;5\text{D}}^{\text{s,nt}}\cdot k & = k_{\nu }^3 \left(-\frac{i }{12 \pi ^2} m^3\right)+k_{\nu } \eta_{\nu\nu} \left(-\frac{i }{5 \pi ^2} m^5\right) }
\al{\label{eq:snc:0:4:6:divr}\tilde{T}_{0,4;6\text{D}}^{\text{s,nt}}\cdot k & = k_{\nu }^3 \left(\frac{i  L_2}{64 \pi ^3} m^4\right)+k_{\nu } \eta_{\nu\nu} \left(\frac{i  L_3}{32 \pi ^3} m^6\right) }
Scalars, spin 1 x 1:
\al{\label{eq:snc:1:1:3:div}k\cdot \tilde{T}_{1,1;3\text{D}}^{\text{s,nt}} & = k_{\nu } \left(\frac{i }{2 \pi } m\right) }
\al{\label{eq:snc:1:1:4:div}k\cdot \tilde{T}_{1,1;4\text{D}}^{\text{s,nt}} & = k_{\nu } \left(-\frac{i  L_1}{8 \pi ^2} m^2\right) }
\al{\label{eq:snc:1:1:5:div}k\cdot \tilde{T}_{1,1;5\text{D}}^{\text{s,nt}} & = k_{\nu } \left(-\frac{i }{12 \pi ^2} m^3\right) }
\al{\label{eq:snc:1:1:6:div}k\cdot \tilde{T}_{1,1;6\text{D}}^{\text{s,nt}} & = k_{\nu } \left(\frac{i  L_2}{64 \pi ^3} m^4\right) }
Scalars, spin 1 x 3:
\al{\label{eq:snc:1:3:3:divl}k\cdot \tilde{T}_{1,3;3\text{D}}^{\text{s,nt}} & = k_{\nu }^3 \left(\frac{i }{2 \pi } m\right)+k_{\nu } \eta_{\nu\nu} \left(\frac{2 i }{\pi } m^3\right) }
\al{\label{eq:snc:1:3:3:divr}\tilde{T}_{1,3;3\text{D}}^{\text{s,nt}}\cdot k & = k_{\mu } k_{\nu }^2 \left(\frac{i }{2 \pi } m\right)+k_{\mu } \eta_{\nu\nu} \left(\frac{2 i }{3 \pi } m^3\right)+k_{\nu } \eta_{\mu\nu} \left(\frac{4 i }{3 \pi } m^3\right) }
\al{\label{eq:snc:1:3:4:divl}k\cdot \tilde{T}_{1,3;4\text{D}}^{\text{s,nt}} & = k_{\nu }^3 \left(-\frac{i  L_1}{8 \pi ^2} m^2\right)+k_{\nu } \eta_{\nu\nu} \left(-\frac{3 i  L_2}{8 \pi ^2} m^4\right) }
\al{\label{eq:snc:1:3:4:divr}\tilde{T}_{1,3;4\text{D}}^{\text{s,nt}}\cdot k & = k_{\mu } k_{\nu }^2 \left(-\frac{i  L_1}{8 \pi ^2} m^2\right)+k_{\mu } \eta_{\nu\nu} \left(-\frac{i  L_2}{8 \pi ^2} m^4\right)+k_{\nu } \eta_{\mu\nu} \left(-\frac{i  L_2}{4 \pi ^2} m^4\right) }
\al{\label{eq:snc:1:3:5:divl}k\cdot \tilde{T}_{1,3;5\text{D}}^{\text{s,nt}} & = k_{\nu }^3 \left(-\frac{i }{12 \pi ^2} m^3\right)+k_{\nu } \eta_{\nu\nu} \left(-\frac{i }{5 \pi ^2} m^5\right) }
\al{\label{eq:snc:1:3:5:divr}\tilde{T}_{1,3;5\text{D}}^{\text{s,nt}}\cdot k & = k_{\mu } k_{\nu }^2 \left(-\frac{i }{12 \pi ^2} m^3\right)+k_{\mu } \eta_{\nu\nu} \left(-\frac{i }{15 \pi ^2} m^5\right)+k_{\nu } \eta_{\mu\nu} \left(-\frac{2 i }{15 \pi ^2} m^5\right) }
\al{\label{eq:snc:1:3:6:divl}k\cdot \tilde{T}_{1,3;6\text{D}}^{\text{s,nt}} & = k_{\nu }^3 \left(\frac{i  L_2}{64 \pi ^3} m^4\right)+k_{\nu } \eta_{\nu\nu} \left(\frac{i  L_3}{32 \pi ^3} m^6\right) }
\al{\label{eq:snc:1:3:6:divr}\tilde{T}_{1,3;6\text{D}}^{\text{s,nt}}\cdot k & = k_{\mu } k_{\nu }^2 \left(\frac{i  L_2}{64 \pi ^3} m^4\right)+k_{\mu } \eta_{\nu\nu} \left(\frac{i  L_3}{96 \pi ^3} m^6\right)+k_{\nu } \eta_{\mu\nu} \left(\frac{i  L_3}{48 \pi ^3} m^6\right) }
Scalars, spin 1 x 5:
\al{\label{eq:snc:1:5:3:divl}k\cdot \tilde{T}_{1,5;3\text{D}}^{\text{s,nt}} & = k_{\nu }^5 \left(\frac{i }{2 \pi } m\right)+k_{\nu }^3 \eta_{\nu\nu} \left(\frac{20 i }{3 \pi } m^3\right)+k_{\nu } \eta_{\nu\nu}^2 \left(\frac{8 i }{\pi } m^5\right) }
\al{\label{eq:snc:1:5:3:divr}\tilde{T}_{1,5;3\text{D}}^{\text{s,nt}}\cdot k & = k_{\mu } k_{\nu }^4 \left(\frac{i }{2 \pi } m\right)+k_{\mu } k_{\nu }^2 \eta_{\nu\nu} \left(\frac{4 i }{\pi } m^3\right)+k_{\nu }^3 \eta_{\mu\nu} \left(\frac{8 i }{3 \pi } m^3\right)+k_{\mu } \eta_{\nu\nu}^2 \left(\frac{8 i }{5 \pi } m^5\right)+
\0\\ & \quad 
 + k_{\nu } \eta_{\mu\nu} \eta_{\nu\nu} \left(\frac{32 i }{5 \pi } m^5\right) }
\al{\label{eq:snc:1:5:4:divl}k\cdot \tilde{T}_{1,5;4\text{D}}^{\text{s,nt}} & = k_{\nu }^5 \left(-\frac{i  L_1}{8 \pi ^2} m^2\right)+k_{\nu }^3 \eta_{\nu\nu} \left(-\frac{5 i  L_2}{4 \pi ^2} m^4\right)+k_{\nu } \eta_{\nu\nu}^2 \left(-\frac{5 i  L_3}{4 \pi ^2} m^6\right) }
\al{\label{eq:snc:1:5:4:divr}\tilde{T}_{1,5;4\text{D}}^{\text{s,nt}}\cdot k & = k_{\mu } k_{\nu }^4 \left(-\frac{i  L_1}{8 \pi ^2} m^2\right)+k_{\mu } k_{\nu }^2 \eta_{\nu\nu} \left(-\frac{3 i  L_2}{4 \pi ^2} m^4\right)+k_{\nu }^3 \eta_{\mu\nu} \left(-\frac{i  L_2}{2 \pi ^2} m^4\right)+
\0\\ & \quad 
 + k_{\mu } \eta_{\nu\nu}^2 \left(-\frac{i  L_3}{4 \pi ^2} m^6\right)+k_{\nu } \eta_{\mu\nu} \eta_{\nu\nu} \left(-\frac{i  L_3}{\pi ^2} m^6\right) }
\al{\label{eq:snc:1:5:5:divl}k\cdot \tilde{T}_{1,5;5\text{D}}^{\text{s,nt}} & = k_{\nu }^5 \left(-\frac{i }{12 \pi ^2} m^3\right)+k_{\nu }^3 \eta_{\nu\nu} \left(-\frac{2 i }{3 \pi ^2} m^5\right)+k_{\nu } \eta_{\nu\nu}^2 \left(-\frac{4 i }{7 \pi ^2} m^7\right) }
\al{\label{eq:snc:1:5:5:divr}\tilde{T}_{1,5;5\text{D}}^{\text{s,nt}}\cdot k & = k_{\mu } k_{\nu }^4 \left(-\frac{i }{12 \pi ^2} m^3\right)+k_{\mu } k_{\nu }^2 \eta_{\nu\nu} \left(-\frac{2 i }{5 \pi ^2} m^5\right)+k_{\nu }^3 \eta_{\mu\nu} \left(-\frac{4 i }{15 \pi ^2} m^5\right)+
\0\\ & \quad 
 + k_{\mu } \eta_{\nu\nu}^2 \left(-\frac{4 i }{35 \pi ^2} m^7\right)+k_{\nu } \eta_{\mu\nu} \eta_{\nu\nu} \left(-\frac{16 i }{35 \pi ^2} m^7\right) }
\al{\label{eq:snc:1:5:6:divl}k\cdot \tilde{T}_{1,5;6\text{D}}^{\text{s,nt}} & = k_{\nu }^5 \left(\frac{i  L_2}{64 \pi ^3} m^4\right)+k_{\nu }^3 \eta_{\nu\nu} \left(\frac{5 i  L_3}{48 \pi ^3} m^6\right)+k_{\nu } \eta_{\nu\nu}^2 \left(\frac{5 i  L_4}{64 \pi ^3} m^8\right) }
\al{\label{eq:snc:1:5:6:divr}\tilde{T}_{1,5;6\text{D}}^{\text{s,nt}}\cdot k & = k_{\mu } k_{\nu }^4 \left(\frac{i  L_2}{64 \pi ^3} m^4\right)+k_{\mu } k_{\nu }^2 \eta_{\nu\nu} \left(\frac{i  L_3}{16 \pi ^3} m^6\right)+k_{\nu }^3 \eta_{\mu\nu} \left(\frac{i  L_3}{24 \pi ^3} m^6\right)+
\0\\ & \quad 
 + k_{\mu } \eta_{\nu\nu}^2 \left(\frac{i  L_4}{64 \pi ^3} m^8\right)+k_{\nu } \eta_{\mu\nu} \eta_{\nu\nu} \left(\frac{i  L_4}{16 \pi ^3} m^8\right) }
Scalars, spin 2 x 2:
\al{\label{eq:snc:2:2:3:div}k\cdot \tilde{T}_{2,2;3\text{D}}^{\text{s,nt}} & = k_{\mu } k_{\nu }^2 \left(\frac{i }{2 \pi } m\right)+k_{\mu } \eta_{\nu\nu} \left(\frac{2 i }{3 \pi } m^3\right)+k_{\nu } \eta_{\mu\nu} \left(\frac{4 i }{3 \pi } m^3\right) }
\al{\label{eq:snc:2:2:4:div}k\cdot \tilde{T}_{2,2;4\text{D}}^{\text{s,nt}} & = k_{\mu } k_{\nu }^2 \left(-\frac{i  L_1}{8 \pi ^2} m^2\right)+k_{\mu } \eta_{\nu\nu} \left(-\frac{i  L_2}{8 \pi ^2} m^4\right)+k_{\nu } \eta_{\mu\nu} \left(-\frac{i  L_2}{4 \pi ^2} m^4\right) }
\al{\label{eq:snc:2:2:5:div}k\cdot \tilde{T}_{2,2;5\text{D}}^{\text{s,nt}} & = k_{\mu } k_{\nu }^2 \left(-\frac{i }{12 \pi ^2} m^3\right)+k_{\mu } \eta_{\nu\nu} \left(-\frac{i }{15 \pi ^2} m^5\right)+k_{\nu } \eta_{\mu\nu} \left(-\frac{2 i }{15 \pi ^2} m^5\right) }
\al{\label{eq:snc:2:2:6:div}k\cdot \tilde{T}_{2,2;6\text{D}}^{\text{s,nt}} & = k_{\mu } k_{\nu }^2 \left(\frac{i  L_2}{64 \pi ^3} m^4\right)+k_{\mu } \eta_{\nu\nu} \left(\frac{i  L_3}{96 \pi ^3} m^6\right)+k_{\nu } \eta_{\mu\nu} \left(\frac{i  L_3}{48 \pi ^3} m^6\right) }
Scalars, spin 2 x 4:
\al{\label{eq:snc:2:4:3:divl}k\cdot \tilde{T}_{2,4;3\text{D}}^{\text{s,nt}} & = k_{\mu } k_{\nu }^4 \left(\frac{i }{2 \pi } m\right)+k_{\mu } k_{\nu }^2 \eta_{\nu\nu} \left(\frac{4 i }{\pi } m^3\right)+k_{\nu }^3 \eta_{\mu\nu} \left(\frac{8 i }{3 \pi } m^3\right)+k_{\mu } \eta_{\nu\nu}^2 \left(\frac{8 i }{5 \pi } m^5\right)+
\0\\ & \quad 
 + k_{\nu } \eta_{\mu\nu} \eta_{\nu\nu} \left(\frac{32 i }{5 \pi } m^5\right) }
\al{\label{eq:snc:2:4:3:divr}\tilde{T}_{2,4;3\text{D}}^{\text{s,nt}}\cdot k & = k_{\mu }^2 k_{\nu }^3 \left(\frac{i }{2 \pi } m\right)+k_{\nu }^3 \eta_{\mu\mu} \left(\frac{2 i }{3 \pi } m^3\right)+k_{\mu }^2 k_{\nu } \eta_{\nu\nu} \left(\frac{2 i }{\pi } m^3\right)+k_{\mu } k_{\nu }^2 \eta_{\mu\nu} \left(\frac{4 i }{\pi } m^3\right)+
\0\\ & \quad 
 + k_{\nu } \eta_{\mu\mu} \eta_{\nu\nu} \left(\frac{8 i }{5 \pi } m^5\right)+ \left(k_{\nu } \eta_{\mu\nu}^2+k_{\mu } \eta_{\mu\nu} \eta_{\nu\nu}\right) \left(\frac{16 i }{5 \pi } m^5\right) }
\al{\label{eq:snc:2:4:4:divl}k\cdot \tilde{T}_{2,4;4\text{D}}^{\text{s,nt}} & = k_{\mu } k_{\nu }^4 \left(-\frac{i  L_1}{8 \pi ^2} m^2\right)+k_{\mu } k_{\nu }^2 \eta_{\nu\nu} \left(-\frac{3 i  L_2}{4 \pi ^2} m^4\right)+k_{\nu }^3 \eta_{\mu\nu} \left(-\frac{i  L_2}{2 \pi ^2} m^4\right)+
\0\\ & \quad 
 + k_{\mu } \eta_{\nu\nu}^2 \left(-\frac{i  L_3}{4 \pi ^2} m^6\right)+k_{\nu } \eta_{\mu\nu} \eta_{\nu\nu} \left(-\frac{i  L_3}{\pi ^2} m^6\right) }
\al{\label{eq:snc:2:4:4:divr}\tilde{T}_{2,4;4\text{D}}^{\text{s,nt}}\cdot k & = k_{\mu }^2 k_{\nu }^3 \left(-\frac{i  L_1}{8 \pi ^2} m^2\right)+k_{\nu }^3 \eta_{\mu\mu} \left(-\frac{i  L_2}{8 \pi ^2} m^4\right)+k_{\mu }^2 k_{\nu } \eta_{\nu\nu} \left(-\frac{3 i  L_2}{8 \pi ^2} m^4\right)+
\0\\ & \quad 
 + k_{\mu } k_{\nu }^2 \eta_{\mu\nu} \left(-\frac{3 i  L_2}{4 \pi ^2} m^4\right)+k_{\nu } \eta_{\mu\mu} \eta_{\nu\nu} \left(-\frac{i  L_3}{4 \pi ^2} m^6\right)+
\0\\ & \quad 
 +  \left(k_{\nu } \eta_{\mu\nu}^2+k_{\mu } \eta_{\mu\nu} \eta_{\nu\nu}\right) \left(-\frac{i  L_3}{2 \pi ^2} m^6\right) }
\al{\label{eq:snc:2:4:5:divl}k\cdot \tilde{T}_{2,4;5\text{D}}^{\text{s,nt}} & = k_{\mu } k_{\nu }^4 \left(-\frac{i }{12 \pi ^2} m^3\right)+k_{\mu } k_{\nu }^2 \eta_{\nu\nu} \left(-\frac{2 i }{5 \pi ^2} m^5\right)+k_{\nu }^3 \eta_{\mu\nu} \left(-\frac{4 i }{15 \pi ^2} m^5\right)+
\0\\ & \quad 
 + k_{\mu } \eta_{\nu\nu}^2 \left(-\frac{4 i }{35 \pi ^2} m^7\right)+k_{\nu } \eta_{\mu\nu} \eta_{\nu\nu} \left(-\frac{16 i }{35 \pi ^2} m^7\right) }
\al{\label{eq:snc:2:4:5:divr}\tilde{T}_{2,4;5\text{D}}^{\text{s,nt}}\cdot k & = k_{\mu }^2 k_{\nu }^3 \left(-\frac{i }{12 \pi ^2} m^3\right)+k_{\nu }^3 \eta_{\mu\mu} \left(-\frac{i }{15 \pi ^2} m^5\right)+k_{\mu }^2 k_{\nu } \eta_{\nu\nu} \left(-\frac{i }{5 \pi ^2} m^5\right)+
\0\\ & \quad 
 + k_{\mu } k_{\nu }^2 \eta_{\mu\nu} \left(-\frac{2 i }{5 \pi ^2} m^5\right)+k_{\nu } \eta_{\mu\mu} \eta_{\nu\nu} \left(-\frac{4 i }{35 \pi ^2} m^7\right)+
\0\\ & \quad 
 +  \left(k_{\nu } \eta_{\mu\nu}^2+k_{\mu } \eta_{\mu\nu} \eta_{\nu\nu}\right) \left(-\frac{8 i }{35 \pi ^2} m^7\right) }
\al{\label{eq:snc:2:4:6:divl}k\cdot \tilde{T}_{2,4;6\text{D}}^{\text{s,nt}} & = k_{\mu } k_{\nu }^4 \left(\frac{i  L_2}{64 \pi ^3} m^4\right)+k_{\mu } k_{\nu }^2 \eta_{\nu\nu} \left(\frac{i  L_3}{16 \pi ^3} m^6\right)+k_{\nu }^3 \eta_{\mu\nu} \left(\frac{i  L_3}{24 \pi ^3} m^6\right)+
\0\\ & \quad 
 + k_{\mu } \eta_{\nu\nu}^2 \left(\frac{i  L_4}{64 \pi ^3} m^8\right)+k_{\nu } \eta_{\mu\nu} \eta_{\nu\nu} \left(\frac{i  L_4}{16 \pi ^3} m^8\right) }
\al{\label{eq:snc:2:4:6:divr}\tilde{T}_{2,4;6\text{D}}^{\text{s,nt}}\cdot k & = k_{\mu }^2 k_{\nu }^3 \left(\frac{i  L_2}{64 \pi ^3} m^4\right)+k_{\nu }^3 \eta_{\mu\mu} \left(\frac{i  L_3}{96 \pi ^3} m^6\right)+k_{\mu }^2 k_{\nu } \eta_{\nu\nu} \left(\frac{i  L_3}{32 \pi ^3} m^6\right)+
\0\\ & \quad 
 + k_{\mu } k_{\nu }^2 \eta_{\mu\nu} \left(\frac{i  L_3}{16 \pi ^3} m^6\right)+k_{\nu } \eta_{\mu\mu} \eta_{\nu\nu} \left(\frac{i  L_4}{64 \pi ^3} m^8\right)+
\0\\ & \quad 
 +  \left(k_{\nu } \eta_{\mu\nu}^2+k_{\mu } \eta_{\mu\nu} \eta_{\nu\nu}\right) \left(\frac{i  L_4}{32 \pi ^3} m^8\right) }
Scalars, spin 3 x 3:
\al{\label{eq:snc:3:3:3:div}k\cdot \tilde{T}_{3,3;3\text{D}}^{\text{s,nt}} & = k_{\mu }^2 k_{\nu }^3 \left(\frac{i }{2 \pi } m\right)+k_{\nu }^3 \eta_{\mu\mu} \left(\frac{2 i }{3 \pi } m^3\right)+k_{\mu }^2 k_{\nu } \eta_{\nu\nu} \left(\frac{2 i }{\pi } m^3\right)+k_{\mu } k_{\nu }^2 \eta_{\mu\nu} \left(\frac{4 i }{\pi } m^3\right)+
\0\\ & \quad 
 + k_{\nu } \eta_{\mu\mu} \eta_{\nu\nu} \left(\frac{8 i }{5 \pi } m^5\right)+ \left(k_{\nu } \eta_{\mu\nu}^2+k_{\mu } \eta_{\mu\nu} \eta_{\nu\nu}\right) \left(\frac{16 i }{5 \pi } m^5\right) }
\al{\label{eq:snc:3:3:4:div}k\cdot \tilde{T}_{3,3;4\text{D}}^{\text{s,nt}} & = k_{\mu }^2 k_{\nu }^3 \left(-\frac{i  L_1}{8 \pi ^2} m^2\right)+k_{\nu }^3 \eta_{\mu\mu} \left(-\frac{i  L_2}{8 \pi ^2} m^4\right)+k_{\mu }^2 k_{\nu } \eta_{\nu\nu} \left(-\frac{3 i  L_2}{8 \pi ^2} m^4\right)+
\0\\ & \quad 
 + k_{\mu } k_{\nu }^2 \eta_{\mu\nu} \left(-\frac{3 i  L_2}{4 \pi ^2} m^4\right)+k_{\nu } \eta_{\mu\mu} \eta_{\nu\nu} \left(-\frac{i  L_3}{4 \pi ^2} m^6\right)+
\0\\ & \quad 
 +  \left(k_{\nu } \eta_{\mu\nu}^2+k_{\mu } \eta_{\mu\nu} \eta_{\nu\nu}\right) \left(-\frac{i  L_3}{2 \pi ^2} m^6\right) }
\al{\label{eq:snc:3:3:5:div}k\cdot \tilde{T}_{3,3;5\text{D}}^{\text{s,nt}} & = k_{\mu }^2 k_{\nu }^3 \left(-\frac{i }{12 \pi ^2} m^3\right)+k_{\nu }^3 \eta_{\mu\mu} \left(-\frac{i }{15 \pi ^2} m^5\right)+k_{\mu }^2 k_{\nu } \eta_{\nu\nu} \left(-\frac{i }{5 \pi ^2} m^5\right)+
\0\\ & \quad 
 + k_{\mu } k_{\nu }^2 \eta_{\mu\nu} \left(-\frac{2 i }{5 \pi ^2} m^5\right)+k_{\nu } \eta_{\mu\mu} \eta_{\nu\nu} \left(-\frac{4 i }{35 \pi ^2} m^7\right)+
\0\\ & \quad 
 +  \left(k_{\nu } \eta_{\mu\nu}^2+k_{\mu } \eta_{\mu\nu} \eta_{\nu\nu}\right) \left(-\frac{8 i }{35 \pi ^2} m^7\right) }
\al{\label{eq:snc:3:3:6:div}k\cdot \tilde{T}_{3,3;6\text{D}}^{\text{s,nt}} & = k_{\mu }^2 k_{\nu }^3 \left(\frac{i  L_2}{64 \pi ^3} m^4\right)+k_{\nu }^3 \eta_{\mu\mu} \left(\frac{i  L_3}{96 \pi ^3} m^6\right)+k_{\mu }^2 k_{\nu } \eta_{\nu\nu} \left(\frac{i  L_3}{32 \pi ^3} m^6\right)+
\0\\ & \quad 
 + k_{\mu } k_{\nu }^2 \eta_{\mu\nu} \left(\frac{i  L_3}{16 \pi ^3} m^6\right)+k_{\nu } \eta_{\mu\mu} \eta_{\nu\nu} \left(\frac{i  L_4}{64 \pi ^3} m^8\right)+
\0\\ & \quad 
 +  \left(k_{\nu } \eta_{\mu\nu}^2+k_{\mu } \eta_{\mu\nu} \eta_{\nu\nu}\right) \left(\frac{i  L_4}{32 \pi ^3} m^8\right) }
Scalars, spin 3 x 5:
\al{\label{eq:snc:3:5:3:divl}k\cdot \tilde{T}_{3,5;3\text{D}}^{\text{s,nt}} & = k_{\mu }^2 k_{\nu }^5 \left(\frac{i }{2 \pi } m\right)+k_{\nu }^5 \eta_{\mu\mu} \left(\frac{2 i }{3 \pi } m^3\right)+ \left(k_{\mu } k_{\nu }^4 \eta_{\mu\nu}+k_{\mu }^2 k_{\nu }^3 \eta_{\nu\nu}\right) \left(\frac{20 i }{3 \pi } m^3\right)+
\0\\ & \quad 
 + k_{\nu }^3 \eta_{\mu\mu} \eta_{\nu\nu} \left(\frac{16 i }{3 \pi } m^5\right)+k_{\mu }^2 k_{\nu } \eta_{\nu\nu}^2 \left(\frac{8 i }{\pi } m^5\right)+k_{\mu } k_{\nu }^2 \eta_{\mu\nu} \eta_{\nu\nu} \left(\frac{32 i }{\pi } m^5\right)+
\0\\ & \quad 
 + k_{\nu }^3 \eta_{\mu\nu}^2 \left(\frac{32 i }{3 \pi } m^5\right)+k_{\nu } \eta_{\mu\mu} \eta_{\nu\nu}^2 \left(\frac{32 i }{7 \pi } m^7\right)+k_{\mu } \eta_{\mu\nu} \eta_{\nu\nu}^2 \left(\frac{64 i }{7 \pi } m^7\right)+
\0\\ & \quad 
 + k_{\nu } \eta_{\mu\nu}^2 \eta_{\nu\nu} \left(\frac{128 i }{7 \pi } m^7\right) }
\al{\label{eq:snc:3:5:3:divr}\tilde{T}_{3,5;3\text{D}}^{\text{s,nt}}\cdot k & = k_{\mu }^3 k_{\nu }^4 \left(\frac{i }{2 \pi } m\right)+k_{\mu } k_{\nu }^4 \eta_{\mu\mu} \left(\frac{2 i }{\pi } m^3\right)+k_{\mu }^3 k_{\nu }^2 \eta_{\nu\nu} \left(\frac{4 i }{\pi } m^3\right)+k_{\mu }^2 k_{\nu }^3 \eta_{\mu\nu} \left(\frac{8 i }{\pi } m^3\right)+
\0\\ & \quad 
 + k_{\mu } k_{\nu }^2 \eta_{\mu\mu} \eta_{\nu\nu} \left(\frac{48 i }{5 \pi } m^5\right)+k_{\mu }^3 \eta_{\nu\nu}^2 \left(\frac{8 i }{5 \pi } m^5\right)+k_{\nu }^3 \eta_{\mu\mu} \eta_{\mu\nu} \left(\frac{32 i }{5 \pi } m^5\right)+
\0\\ & \quad 
 +  \left(k_{\mu } k_{\nu }^2 \eta_{\mu\nu}^2+k_{\mu }^2 k_{\nu } \eta_{\mu\nu} \eta_{\nu\nu}\right) \left(\frac{96 i }{5 \pi } m^5\right)+k_{\mu } \eta_{\mu\mu} \eta_{\nu\nu}^2 \left(\frac{96 i }{35 \pi } m^7\right)+
\0\\ & \quad 
 +  \left(k_{\nu } \eta_{\mu\mu} \eta_{\mu\nu} \eta_{\nu\nu}+k_{\mu } \eta_{\mu\nu}^2 \eta_{\nu\nu}\right) \left(\frac{384 i }{35 \pi } m^7\right)+k_{\nu } \eta_{\mu\nu}^3 \left(\frac{256 i }{35 \pi } m^7\right) }
\al{\label{eq:snc:3:5:4:divl}k\cdot \tilde{T}_{3,5;4\text{D}}^{\text{s,nt}} & = k_{\mu }^2 k_{\nu }^5 \left(-\frac{i  L_1}{8 \pi ^2} m^2\right)+k_{\nu }^5 \eta_{\mu\mu} \left(-\frac{i  L_2}{8 \pi ^2} m^4\right)+
\0\\ & \quad 
 +  \left(k_{\mu } k_{\nu }^4 \eta_{\mu\nu}+k_{\mu }^2 k_{\nu }^3 \eta_{\nu\nu}\right) \left(-\frac{5 i  L_2}{4 \pi ^2} m^4\right)+k_{\nu }^3 \eta_{\mu\mu} \eta_{\nu\nu} \left(-\frac{5 i  L_3}{6 \pi ^2} m^6\right)+
\0\\ & \quad 
 + k_{\mu }^2 k_{\nu } \eta_{\nu\nu}^2 \left(-\frac{5 i  L_3}{4 \pi ^2} m^6\right)+k_{\mu } k_{\nu }^2 \eta_{\mu\nu} \eta_{\nu\nu} \left(-\frac{5 i  L_3}{\pi ^2} m^6\right)+
\0\\ & \quad 
 + k_{\nu }^3 \eta_{\mu\nu}^2 \left(-\frac{5 i  L_3}{3 \pi ^2} m^6\right)+k_{\nu } \eta_{\mu\mu} \eta_{\nu\nu}^2 \left(-\frac{5 i  L_4}{8 \pi ^2} m^8\right)+k_{\mu } \eta_{\mu\nu} \eta_{\nu\nu}^2 \left(-\frac{5 i  L_4}{4 \pi ^2} m^8\right)+
\0\\ & \quad 
 + k_{\nu } \eta_{\mu\nu}^2 \eta_{\nu\nu} \left(-\frac{5 i  L_4}{2 \pi ^2} m^8\right) }
\al{\label{eq:snc:3:5:4:divr}\tilde{T}_{3,5;4\text{D}}^{\text{s,nt}}\cdot k & = k_{\mu }^3 k_{\nu }^4 \left(-\frac{i  L_1}{8 \pi ^2} m^2\right)+k_{\mu } k_{\nu }^4 \eta_{\mu\mu} \left(-\frac{3 i  L_2}{8 \pi ^2} m^4\right)+k_{\mu }^3 k_{\nu }^2 \eta_{\nu\nu} \left(-\frac{3 i  L_2}{4 \pi ^2} m^4\right)+
\0\\ & \quad 
 + k_{\mu }^2 k_{\nu }^3 \eta_{\mu\nu} \left(-\frac{3 i  L_2}{2 \pi ^2} m^4\right)+k_{\mu } k_{\nu }^2 \eta_{\mu\mu} \eta_{\nu\nu} \left(-\frac{3 i  L_3}{2 \pi ^2} m^6\right)+
\0\\ & \quad 
 + k_{\mu }^3 \eta_{\nu\nu}^2 \left(-\frac{i  L_3}{4 \pi ^2} m^6\right)+k_{\nu }^3 \eta_{\mu\mu} \eta_{\mu\nu} \left(-\frac{i  L_3}{\pi ^2} m^6\right)+
\0\\ & \quad 
 +  \left(k_{\mu } k_{\nu }^2 \eta_{\mu\nu}^2+k_{\mu }^2 k_{\nu } \eta_{\mu\nu} \eta_{\nu\nu}\right) \left(-\frac{3 i  L_3}{\pi ^2} m^6\right)+k_{\mu } \eta_{\mu\mu} \eta_{\nu\nu}^2 \left(-\frac{3 i  L_4}{8 \pi ^2} m^8\right)+
\0\\ & \quad 
 +  \left(k_{\nu } \eta_{\mu\mu} \eta_{\mu\nu} \eta_{\nu\nu}+k_{\mu } \eta_{\mu\nu}^2 \eta_{\nu\nu}\right) \left(-\frac{3 i  L_4}{2 \pi ^2} m^8\right)+k_{\nu } \eta_{\mu\nu}^3 \left(-\frac{i  L_4}{\pi ^2} m^8\right) }
\al{\label{eq:snc:3:5:5:divl}k\cdot \tilde{T}_{3,5;5\text{D}}^{\text{s,nt}} & = k_{\mu }^2 k_{\nu }^5 \left(-\frac{i }{12 \pi ^2} m^3\right)+k_{\nu }^5 \eta_{\mu\mu} \left(-\frac{i }{15 \pi ^2} m^5\right)+
\0\\ & \quad 
 +  \left(k_{\mu } k_{\nu }^4 \eta_{\mu\nu}+k_{\mu }^2 k_{\nu }^3 \eta_{\nu\nu}\right) \left(-\frac{2 i }{3 \pi ^2} m^5\right)+k_{\nu }^3 \eta_{\mu\mu} \eta_{\nu\nu} \left(-\frac{8 i }{21 \pi ^2} m^7\right)+
\0\\ & \quad 
 + k_{\mu }^2 k_{\nu } \eta_{\nu\nu}^2 \left(-\frac{4 i }{7 \pi ^2} m^7\right)+k_{\mu } k_{\nu }^2 \eta_{\mu\nu} \eta_{\nu\nu} \left(-\frac{16 i }{7 \pi ^2} m^7\right)+k_{\nu }^3 \eta_{\mu\nu}^2 \left(-\frac{16 i }{21 \pi ^2} m^7\right)+
\0\\ & \quad 
 + k_{\nu } \eta_{\mu\mu} \eta_{\nu\nu}^2 \left(-\frac{16 i }{63 \pi ^2} m^9\right)+k_{\mu } \eta_{\mu\nu} \eta_{\nu\nu}^2 \left(-\frac{32 i }{63 \pi ^2} m^9\right)+
\0\\ & \quad 
 + k_{\nu } \eta_{\mu\nu}^2 \eta_{\nu\nu} \left(-\frac{64 i }{63 \pi ^2} m^9\right) }
\al{\label{eq:snc:3:5:5:divr}\tilde{T}_{3,5;5\text{D}}^{\text{s,nt}}\cdot k & = k_{\mu }^3 k_{\nu }^4 \left(-\frac{i }{12 \pi ^2} m^3\right)+k_{\mu } k_{\nu }^4 \eta_{\mu\mu} \left(-\frac{i }{5 \pi ^2} m^5\right)+k_{\mu }^3 k_{\nu }^2 \eta_{\nu\nu} \left(-\frac{2 i }{5 \pi ^2} m^5\right)+
\0\\ & \quad 
 + k_{\mu }^2 k_{\nu }^3 \eta_{\mu\nu} \left(-\frac{4 i }{5 \pi ^2} m^5\right)+k_{\mu } k_{\nu }^2 \eta_{\mu\mu} \eta_{\nu\nu} \left(-\frac{24 i }{35 \pi ^2} m^7\right)+
\0\\ & \quad 
 + k_{\mu }^3 \eta_{\nu\nu}^2 \left(-\frac{4 i }{35 \pi ^2} m^7\right)+k_{\nu }^3 \eta_{\mu\mu} \eta_{\mu\nu} \left(-\frac{16 i }{35 \pi ^2} m^7\right)+
\0\\ & \quad 
 +  \left(k_{\mu } k_{\nu }^2 \eta_{\mu\nu}^2+k_{\mu }^2 k_{\nu } \eta_{\mu\nu} \eta_{\nu\nu}\right) \left(-\frac{48 i }{35 \pi ^2} m^7\right)+k_{\mu } \eta_{\mu\mu} \eta_{\nu\nu}^2 \left(-\frac{16 i }{105 \pi ^2} m^9\right)+
\0\\ & \quad 
 +  \left(k_{\nu } \eta_{\mu\mu} \eta_{\mu\nu} \eta_{\nu\nu}+k_{\mu } \eta_{\mu\nu}^2 \eta_{\nu\nu}\right) \left(-\frac{64 i }{105 \pi ^2} m^9\right)+k_{\nu } \eta_{\mu\nu}^3 \left(-\frac{128 i }{315 \pi ^2} m^9\right) }
\al{\label{eq:snc:3:5:6:divl}k\cdot \tilde{T}_{3,5;6\text{D}}^{\text{s,nt}} & = k_{\mu }^2 k_{\nu }^5 \left(\frac{i  L_2}{64 \pi ^3} m^4\right)+k_{\nu }^5 \eta_{\mu\mu} \left(\frac{i  L_3}{96 \pi ^3} m^6\right)+ \left(k_{\mu } k_{\nu }^4 \eta_{\mu\nu}+k_{\mu }^2 k_{\nu }^3 \eta_{\nu\nu}\right) \left(\frac{5 i  L_3}{48 \pi ^3} m^6\right)+
\0\\ & \quad 
 + k_{\nu }^3 \eta_{\mu\mu} \eta_{\nu\nu} \left(\frac{5 i  L_4}{96 \pi ^3} m^8\right)+k_{\mu }^2 k_{\nu } \eta_{\nu\nu}^2 \left(\frac{5 i  L_4}{64 \pi ^3} m^8\right)+k_{\mu } k_{\nu }^2 \eta_{\mu\nu} \eta_{\nu\nu} \left(\frac{5 i  L_4}{16 \pi ^3} m^8\right)+
\0\\ & \quad 
 + k_{\nu }^3 \eta_{\mu\nu}^2 \left(\frac{5 i  L_4}{48 \pi ^3} m^8\right)+k_{\nu } \eta_{\mu\mu} \eta_{\nu\nu}^2 \left(\frac{i  L_5}{32 \pi ^3} m^{10}\right)+k_{\mu } \eta_{\mu\nu} \eta_{\nu\nu}^2 \left(\frac{i  L_5}{16 \pi ^3} m^{10}\right)+
\0\\ & \quad 
 + k_{\nu } \eta_{\mu\nu}^2 \eta_{\nu\nu} \left(\frac{i  L_5}{8 \pi ^3} m^{10}\right) }
\al{\label{eq:snc:3:5:6:divr}\tilde{T}_{3,5;6\text{D}}^{\text{s,nt}}\cdot k & = k_{\mu }^3 k_{\nu }^4 \left(\frac{i  L_2}{64 \pi ^3} m^4\right)+k_{\mu } k_{\nu }^4 \eta_{\mu\mu} \left(\frac{i  L_3}{32 \pi ^3} m^6\right)+k_{\mu }^3 k_{\nu }^2 \eta_{\nu\nu} \left(\frac{i  L_3}{16 \pi ^3} m^6\right)+
\0\\ & \quad 
 + k_{\mu }^2 k_{\nu }^3 \eta_{\mu\nu} \left(\frac{i  L_3}{8 \pi ^3} m^6\right)+k_{\mu } k_{\nu }^2 \eta_{\mu\mu} \eta_{\nu\nu} \left(\frac{3 i  L_4}{32 \pi ^3} m^8\right)+k_{\mu }^3 \eta_{\nu\nu}^2 \left(\frac{i  L_4}{64 \pi ^3} m^8\right)+
\0\\ & \quad 
 + k_{\nu }^3 \eta_{\mu\mu} \eta_{\mu\nu} \left(\frac{i  L_4}{16 \pi ^3} m^8\right)+ \left(k_{\mu } k_{\nu }^2 \eta_{\mu\nu}^2+k_{\mu }^2 k_{\nu } \eta_{\mu\nu} \eta_{\nu\nu}\right) \left(\frac{3 i  L_4}{16 \pi ^3} m^8\right)+
\0\\ & \quad 
 + k_{\mu } \eta_{\mu\mu} \eta_{\nu\nu}^2 \left(\frac{3 i  L_5}{160 \pi ^3} m^{10}\right)+ \left(k_{\nu } \eta_{\mu\mu} \eta_{\mu\nu} \eta_{\nu\nu}+k_{\mu } \eta_{\mu\nu}^2 \eta_{\nu\nu}\right) \left(\frac{3 i  L_5}{40 \pi ^3} m^{10}\right)+
\0\\ & \quad 
 + k_{\nu } \eta_{\mu\nu}^3 \left(\frac{i  L_5}{20 \pi ^3} m^{10}\right) }
Scalars, spin 4 x 4:
\al{\label{eq:snc:4:4:3:div}k\cdot \tilde{T}_{4,4;3\text{D}}^{\text{s,nt}} & = k_{\mu }^3 k_{\nu }^4 \left(\frac{i }{2 \pi } m\right)+k_{\mu } k_{\nu }^4 \eta_{\mu\mu} \left(\frac{2 i }{\pi } m^3\right)+k_{\mu }^3 k_{\nu }^2 \eta_{\nu\nu} \left(\frac{4 i }{\pi } m^3\right)+k_{\mu }^2 k_{\nu }^3 \eta_{\mu\nu} \left(\frac{8 i }{\pi } m^3\right)+
\0\\ & \quad 
 + k_{\mu } k_{\nu }^2 \eta_{\mu\mu} \eta_{\nu\nu} \left(\frac{48 i }{5 \pi } m^5\right)+k_{\mu }^3 \eta_{\nu\nu}^2 \left(\frac{8 i }{5 \pi } m^5\right)+k_{\nu }^3 \eta_{\mu\mu} \eta_{\mu\nu} \left(\frac{32 i }{5 \pi } m^5\right)+
\0\\ & \quad 
 +  \left(k_{\mu } k_{\nu }^2 \eta_{\mu\nu}^2+k_{\mu }^2 k_{\nu } \eta_{\mu\nu} \eta_{\nu\nu}\right) \left(\frac{96 i }{5 \pi } m^5\right)+k_{\mu } \eta_{\mu\mu} \eta_{\nu\nu}^2 \left(\frac{96 i }{35 \pi } m^7\right)+
\0\\ & \quad 
 +  \left(k_{\nu } \eta_{\mu\mu} \eta_{\mu\nu} \eta_{\nu\nu}+k_{\mu } \eta_{\mu\nu}^2 \eta_{\nu\nu}\right) \left(\frac{384 i }{35 \pi } m^7\right)+k_{\nu } \eta_{\mu\nu}^3 \left(\frac{256 i }{35 \pi } m^7\right) }
\al{\label{eq:snc:4:4:4:div}k\cdot \tilde{T}_{4,4;4\text{D}}^{\text{s,nt}} & = k_{\mu }^3 k_{\nu }^4 \left(-\frac{i  L_1}{8 \pi ^2} m^2\right)+k_{\mu } k_{\nu }^4 \eta_{\mu\mu} \left(-\frac{3 i  L_2}{8 \pi ^2} m^4\right)+k_{\mu }^3 k_{\nu }^2 \eta_{\nu\nu} \left(-\frac{3 i  L_2}{4 \pi ^2} m^4\right)+
\0\\ & \quad 
 + k_{\mu }^2 k_{\nu }^3 \eta_{\mu\nu} \left(-\frac{3 i  L_2}{2 \pi ^2} m^4\right)+k_{\mu } k_{\nu }^2 \eta_{\mu\mu} \eta_{\nu\nu} \left(-\frac{3 i  L_3}{2 \pi ^2} m^6\right)+
\0\\ & \quad 
 + k_{\mu }^3 \eta_{\nu\nu}^2 \left(-\frac{i  L_3}{4 \pi ^2} m^6\right)+k_{\nu }^3 \eta_{\mu\mu} \eta_{\mu\nu} \left(-\frac{i  L_3}{\pi ^2} m^6\right)+
\0\\ & \quad 
 +  \left(k_{\mu } k_{\nu }^2 \eta_{\mu\nu}^2+k_{\mu }^2 k_{\nu } \eta_{\mu\nu} \eta_{\nu\nu}\right) \left(-\frac{3 i  L_3}{\pi ^2} m^6\right)+k_{\mu } \eta_{\mu\mu} \eta_{\nu\nu}^2 \left(-\frac{3 i  L_4}{8 \pi ^2} m^8\right)+
\0\\ & \quad 
 +  \left(k_{\nu } \eta_{\mu\mu} \eta_{\mu\nu} \eta_{\nu\nu}+k_{\mu } \eta_{\mu\nu}^2 \eta_{\nu\nu}\right) \left(-\frac{3 i  L_4}{2 \pi ^2} m^8\right)+k_{\nu } \eta_{\mu\nu}^3 \left(-\frac{i  L_4}{\pi ^2} m^8\right) }
\al{\label{eq:snc:4:4:5:div}k\cdot \tilde{T}_{4,4;5\text{D}}^{\text{s,nt}} & = k_{\mu }^3 k_{\nu }^4 \left(-\frac{i }{12 \pi ^2} m^3\right)+k_{\mu } k_{\nu }^4 \eta_{\mu\mu} \left(-\frac{i }{5 \pi ^2} m^5\right)+k_{\mu }^3 k_{\nu }^2 \eta_{\nu\nu} \left(-\frac{2 i }{5 \pi ^2} m^5\right)+
\0\\ & \quad 
 + k_{\mu }^2 k_{\nu }^3 \eta_{\mu\nu} \left(-\frac{4 i }{5 \pi ^2} m^5\right)+k_{\mu } k_{\nu }^2 \eta_{\mu\mu} \eta_{\nu\nu} \left(-\frac{24 i }{35 \pi ^2} m^7\right)+
\0\\ & \quad 
 + k_{\mu }^3 \eta_{\nu\nu}^2 \left(-\frac{4 i }{35 \pi ^2} m^7\right)+k_{\nu }^3 \eta_{\mu\mu} \eta_{\mu\nu} \left(-\frac{16 i }{35 \pi ^2} m^7\right)+
\0\\ & \quad 
 +  \left(k_{\mu } k_{\nu }^2 \eta_{\mu\nu}^2+k_{\mu }^2 k_{\nu } \eta_{\mu\nu} \eta_{\nu\nu}\right) \left(-\frac{48 i }{35 \pi ^2} m^7\right)+k_{\mu } \eta_{\mu\mu} \eta_{\nu\nu}^2 \left(-\frac{16 i }{105 \pi ^2} m^9\right)+
\0\\ & \quad 
 +  \left(k_{\nu } \eta_{\mu\mu} \eta_{\mu\nu} \eta_{\nu\nu}+k_{\mu } \eta_{\mu\nu}^2 \eta_{\nu\nu}\right) \left(-\frac{64 i }{105 \pi ^2} m^9\right)+k_{\nu } \eta_{\mu\nu}^3 \left(-\frac{128 i }{315 \pi ^2} m^9\right) }
\al{\label{eq:snc:4:4:6:div}k\cdot \tilde{T}_{4,4;6\text{D}}^{\text{s,nt}} & = k_{\mu }^3 k_{\nu }^4 \left(\frac{i  L_2}{64 \pi ^3} m^4\right)+k_{\mu } k_{\nu }^4 \eta_{\mu\mu} \left(\frac{i  L_3}{32 \pi ^3} m^6\right)+k_{\mu }^3 k_{\nu }^2 \eta_{\nu\nu} \left(\frac{i  L_3}{16 \pi ^3} m^6\right)+
\0\\ & \quad 
 + k_{\mu }^2 k_{\nu }^3 \eta_{\mu\nu} \left(\frac{i  L_3}{8 \pi ^3} m^6\right)+k_{\mu } k_{\nu }^2 \eta_{\mu\mu} \eta_{\nu\nu} \left(\frac{3 i  L_4}{32 \pi ^3} m^8\right)+k_{\mu }^3 \eta_{\nu\nu}^2 \left(\frac{i  L_4}{64 \pi ^3} m^8\right)+
\0\\ & \quad 
 + k_{\nu }^3 \eta_{\mu\mu} \eta_{\mu\nu} \left(\frac{i  L_4}{16 \pi ^3} m^8\right)+ \left(k_{\mu } k_{\nu }^2 \eta_{\mu\nu}^2+k_{\mu }^2 k_{\nu } \eta_{\mu\nu} \eta_{\nu\nu}\right) \left(\frac{3 i  L_4}{16 \pi ^3} m^8\right)+
\0\\ & \quad 
 + k_{\mu } \eta_{\mu\mu} \eta_{\nu\nu}^2 \left(\frac{3 i  L_5}{160 \pi ^3} m^{10}\right)+ \left(k_{\nu } \eta_{\mu\mu} \eta_{\mu\nu} \eta_{\nu\nu}+k_{\mu } \eta_{\mu\nu}^2 \eta_{\nu\nu}\right) \left(\frac{3 i  L_5}{40 \pi ^3} m^{10}\right)+
\0\\ & \quad 
 + k_{\nu } \eta_{\mu\nu}^3 \left(\frac{i  L_5}{20 \pi ^3} m^{10}\right) }
Scalars, spin 5 x 5:
\al{\label{eq:snc:5:5:3:div}k\cdot \tilde{T}_{5,5;3\text{D}}^{\text{s,nt}} & = k_{\mu }^4 k_{\nu }^5 \left(\frac{i }{2 \pi } m\right)+k_{\mu }^2 k_{\nu }^5 \eta_{\mu\mu} \left(\frac{4 i }{\pi } m^3\right)+k_{\mu }^4 k_{\nu }^3 \eta_{\nu\nu} \left(\frac{20 i }{3 \pi } m^3\right)+
\0\\ & \quad 
 + k_{\mu }^3 k_{\nu }^4 \eta_{\mu\nu} \left(\frac{40 i }{3 \pi } m^3\right)+k_{\nu }^5 \eta_{\mu\mu}^2 \left(\frac{8 i }{5 \pi } m^5\right)+k_{\mu }^4 k_{\nu } \eta_{\nu\nu}^2 \left(\frac{8 i }{\pi } m^5\right)+
\0\\ & \quad 
 +  \left(k_{\mu } k_{\nu }^4 \eta_{\mu\mu} \eta_{\mu\nu}+k_{\mu }^2 k_{\nu }^3 \eta_{\mu\mu} \eta_{\nu\nu}\right) \left(\frac{32 i }{\pi } m^5\right)+
\0\\ & \quad 
 +  \left(k_{\mu }^2 k_{\nu }^3 \eta_{\mu\nu}^2+k_{\mu }^3 k_{\nu }^2 \eta_{\mu\nu} \eta_{\nu\nu}\right) \left(\frac{64 i }{\pi } m^5\right)+k_{\nu }^3 \eta_{\mu\mu}^2 \eta_{\nu\nu} \left(\frac{64 i }{7 \pi } m^7\right)+
\0\\ & \quad 
 + k_{\mu }^2 k_{\nu } \eta_{\mu\mu} \eta_{\nu\nu}^2 \left(\frac{192 i }{7 \pi } m^7\right)+ \left(k_{\mu } k_{\nu }^2 \eta_{\mu\mu} \eta_{\mu\nu} \eta_{\nu\nu}+k_{\mu }^2 k_{\nu } \eta_{\mu\nu}^2 \eta_{\nu\nu}\right) \left(\frac{768 i }{7 \pi } m^7\right)+
\0\\ & \quad 
 + k_{\mu }^3 \eta_{\mu\nu} \eta_{\nu\nu}^2 \left(\frac{128 i }{7 \pi } m^7\right)+k_{\nu }^3 \eta_{\mu\mu} \eta_{\mu\nu}^2 \left(\frac{256 i }{7 \pi } m^7\right)+k_{\mu } k_{\nu }^2 \eta_{\mu\nu}^3 \left(\frac{512 i }{7 \pi } m^7\right)+
\0\\ & \quad 
 + k_{\nu } \eta_{\mu\mu}^2 \eta_{\nu\nu}^2 \left(\frac{128 i }{21 \pi } m^9\right)+k_{\mu } \eta_{\mu\mu} \eta_{\mu\nu} \eta_{\nu\nu}^2 \left(\frac{512 i }{21 \pi } m^9\right)+
\0\\ & \quad 
 + k_{\nu } \eta_{\mu\mu} \eta_{\mu\nu}^2 \eta_{\nu\nu} \left(\frac{1024 i }{21 \pi } m^9\right)+k_{\mu } \eta_{\mu\nu}^3 \eta_{\nu\nu} \left(\frac{2048 i }{63 \pi } m^9\right)+k_{\nu } \eta_{\mu\nu}^4 \left(\frac{1024 i }{63 \pi } m^9\right) }
\al{\label{eq:snc:5:5:4:div}k\cdot \tilde{T}_{5,5;4\text{D}}^{\text{s,nt}} & = k_{\mu }^4 k_{\nu }^5 \left(-\frac{i  L_1}{8 \pi ^2} m^2\right)+k_{\mu }^2 k_{\nu }^5 \eta_{\mu\mu} \left(-\frac{3 i  L_2}{4 \pi ^2} m^4\right)+k_{\mu }^4 k_{\nu }^3 \eta_{\nu\nu} \left(-\frac{5 i  L_2}{4 \pi ^2} m^4\right)+
\0\\ & \quad 
 + k_{\mu }^3 k_{\nu }^4 \eta_{\mu\nu} \left(-\frac{5 i  L_2}{2 \pi ^2} m^4\right)+k_{\nu }^5 \eta_{\mu\mu}^2 \left(-\frac{i  L_3}{4 \pi ^2} m^6\right)+k_{\mu }^4 k_{\nu } \eta_{\nu\nu}^2 \left(-\frac{5 i  L_3}{4 \pi ^2} m^6\right)+
\0\\ & \quad 
 +  \left(k_{\mu } k_{\nu }^4 \eta_{\mu\mu} \eta_{\mu\nu}+k_{\mu }^2 k_{\nu }^3 \eta_{\mu\mu} \eta_{\nu\nu}\right) \left(-\frac{5 i  L_3}{\pi ^2} m^6\right)+
\0\\ & \quad 
 +  \left(k_{\mu }^2 k_{\nu }^3 \eta_{\mu\nu}^2+k_{\mu }^3 k_{\nu }^2 \eta_{\mu\nu} \eta_{\nu\nu}\right) \left(-\frac{10 i  L_3}{\pi ^2} m^6\right)+k_{\nu }^3 \eta_{\mu\mu}^2 \eta_{\nu\nu} \left(-\frac{5 i  L_4}{4 \pi ^2} m^8\right)+
\0\\ & \quad 
 + k_{\mu }^2 k_{\nu } \eta_{\mu\mu} \eta_{\nu\nu}^2 \left(-\frac{15 i  L_4}{4 \pi ^2} m^8\right)+
\0\\ & \quad 
 +  \left(k_{\mu } k_{\nu }^2 \eta_{\mu\mu} \eta_{\mu\nu} \eta_{\nu\nu}+k_{\mu }^2 k_{\nu } \eta_{\mu\nu}^2 \eta_{\nu\nu}\right) \left(-\frac{15 i  L_4}{\pi ^2} m^8\right)+
\0\\ & \quad 
 + k_{\mu }^3 \eta_{\mu\nu} \eta_{\nu\nu}^2 \left(-\frac{5 i  L_4}{2 \pi ^2} m^8\right)+k_{\nu }^3 \eta_{\mu\mu} \eta_{\mu\nu}^2 \left(-\frac{5 i  L_4}{\pi ^2} m^8\right)+
\0\\ & \quad 
 + k_{\mu } k_{\nu }^2 \eta_{\mu\nu}^3 \left(-\frac{10 i  L_4}{\pi ^2} m^8\right)+k_{\nu } \eta_{\mu\mu}^2 \eta_{\nu\nu}^2 \left(-\frac{3 i  L_5}{4 \pi ^2} m^{10}\right)+
\0\\ & \quad 
 + k_{\mu } \eta_{\mu\mu} \eta_{\mu\nu} \eta_{\nu\nu}^2 \left(-\frac{3 i  L_5}{\pi ^2} m^{10}\right)+k_{\nu } \eta_{\mu\mu} \eta_{\mu\nu}^2 \eta_{\nu\nu} \left(-\frac{6 i  L_5}{\pi ^2} m^{10}\right)+
\0\\ & \quad 
 + k_{\mu } \eta_{\mu\nu}^3 \eta_{\nu\nu} \left(-\frac{4 i  L_5}{\pi ^2} m^{10}\right)+k_{\nu } \eta_{\mu\nu}^4 \left(-\frac{2 i  L_5}{\pi ^2} m^{10}\right) }
\al{\label{eq:snc:5:5:5:div}k\cdot \tilde{T}_{5,5;5\text{D}}^{\text{s,nt}} & = k_{\mu }^4 k_{\nu }^5 \left(-\frac{i }{12 \pi ^2} m^3\right)+k_{\mu }^2 k_{\nu }^5 \eta_{\mu\mu} \left(-\frac{2 i }{5 \pi ^2} m^5\right)+k_{\mu }^4 k_{\nu }^3 \eta_{\nu\nu} \left(-\frac{2 i }{3 \pi ^2} m^5\right)+
\0\\ & \quad 
 + k_{\mu }^3 k_{\nu }^4 \eta_{\mu\nu} \left(-\frac{4 i }{3 \pi ^2} m^5\right)+k_{\nu }^5 \eta_{\mu\mu}^2 \left(-\frac{4 i }{35 \pi ^2} m^7\right)+k_{\mu }^4 k_{\nu } \eta_{\nu\nu}^2 \left(-\frac{4 i }{7 \pi ^2} m^7\right)+
\0\\ & \quad 
 +  \left(k_{\mu } k_{\nu }^4 \eta_{\mu\mu} \eta_{\mu\nu}+k_{\mu }^2 k_{\nu }^3 \eta_{\mu\mu} \eta_{\nu\nu}\right) \left(-\frac{16 i }{7 \pi ^2} m^7\right)+
\0\\ & \quad 
 +  \left(k_{\mu }^2 k_{\nu }^3 \eta_{\mu\nu}^2+k_{\mu }^3 k_{\nu }^2 \eta_{\mu\nu} \eta_{\nu\nu}\right) \left(-\frac{32 i }{7 \pi ^2} m^7\right)+k_{\nu }^3 \eta_{\mu\mu}^2 \eta_{\nu\nu} \left(-\frac{32 i }{63 \pi ^2} m^9\right)+
\0\\ & \quad 
 + k_{\mu }^2 k_{\nu } \eta_{\mu\mu} \eta_{\nu\nu}^2 \left(-\frac{32 i }{21 \pi ^2} m^9\right)+
\0\\ & \quad 
 +  \left(k_{\mu } k_{\nu }^2 \eta_{\mu\mu} \eta_{\mu\nu} \eta_{\nu\nu}+k_{\mu }^2 k_{\nu } \eta_{\mu\nu}^2 \eta_{\nu\nu}\right) \left(-\frac{128 i }{21 \pi ^2} m^9\right)+
\0\\ & \quad 
 + k_{\mu }^3 \eta_{\mu\nu} \eta_{\nu\nu}^2 \left(-\frac{64 i }{63 \pi ^2} m^9\right)+k_{\nu }^3 \eta_{\mu\mu} \eta_{\mu\nu}^2 \left(-\frac{128 i }{63 \pi ^2} m^9\right)+
\0\\ & \quad 
 + k_{\mu } k_{\nu }^2 \eta_{\mu\nu}^3 \left(-\frac{256 i }{63 \pi ^2} m^9\right)+k_{\nu } \eta_{\mu\mu}^2 \eta_{\nu\nu}^2 \left(-\frac{64 i }{231 \pi ^2} m^{11}\right)+
\0\\ & \quad 
 + k_{\mu } \eta_{\mu\mu} \eta_{\mu\nu} \eta_{\nu\nu}^2 \left(-\frac{256 i }{231 \pi ^2} m^{11}\right)+k_{\nu } \eta_{\mu\mu} \eta_{\mu\nu}^2 \eta_{\nu\nu} \left(-\frac{512 i }{231 \pi ^2} m^{11}\right)+
\0\\ & \quad 
 + k_{\mu } \eta_{\mu\nu}^3 \eta_{\nu\nu} \left(-\frac{1024 i }{693 \pi ^2} m^{11}\right)+k_{\nu } \eta_{\mu\nu}^4 \left(-\frac{512 i }{693 \pi ^2} m^{11}\right) }
\al{\label{eq:snc:5:5:6:div}k\cdot \tilde{T}_{5,5;6\text{D}}^{\text{s,nt}} & = k_{\mu }^4 k_{\nu }^5 \left(\frac{i  L_2}{64 \pi ^3} m^4\right)+k_{\mu }^2 k_{\nu }^5 \eta_{\mu\mu} \left(\frac{i  L_3}{16 \pi ^3} m^6\right)+k_{\mu }^4 k_{\nu }^3 \eta_{\nu\nu} \left(\frac{5 i  L_3}{48 \pi ^3} m^6\right)+
\0\\ & \quad 
 + k_{\mu }^3 k_{\nu }^4 \eta_{\mu\nu} \left(\frac{5 i  L_3}{24 \pi ^3} m^6\right)+k_{\nu }^5 \eta_{\mu\mu}^2 \left(\frac{i  L_4}{64 \pi ^3} m^8\right)+k_{\mu }^4 k_{\nu } \eta_{\nu\nu}^2 \left(\frac{5 i  L_4}{64 \pi ^3} m^8\right)+
\0\\ & \quad 
 +  \left(k_{\mu } k_{\nu }^4 \eta_{\mu\mu} \eta_{\mu\nu}+k_{\mu }^2 k_{\nu }^3 \eta_{\mu\mu} \eta_{\nu\nu}\right) \left(\frac{5 i  L_4}{16 \pi ^3} m^8\right)+
\0\\ & \quad 
 +  \left(k_{\mu }^2 k_{\nu }^3 \eta_{\mu\nu}^2+k_{\mu }^3 k_{\nu }^2 \eta_{\mu\nu} \eta_{\nu\nu}\right) \left(\frac{5 i  L_4}{8 \pi ^3} m^8\right)+k_{\nu }^3 \eta_{\mu\mu}^2 \eta_{\nu\nu} \left(\frac{i  L_5}{16 \pi ^3} m^{10}\right)+
\0\\ & \quad 
 + k_{\mu }^2 k_{\nu } \eta_{\mu\mu} \eta_{\nu\nu}^2 \left(\frac{3 i  L_5}{16 \pi ^3} m^{10}\right)+
\0\\ & \quad 
 +  \left(k_{\mu } k_{\nu }^2 \eta_{\mu\mu} \eta_{\mu\nu} \eta_{\nu\nu}+k_{\mu }^2 k_{\nu } \eta_{\mu\nu}^2 \eta_{\nu\nu}\right) \left(\frac{3 i  L_5}{4 \pi ^3} m^{10}\right)+k_{\mu }^3 \eta_{\mu\nu} \eta_{\nu\nu}^2 \left(\frac{i  L_5}{8 \pi ^3} m^{10}\right)+
\0\\ & \quad 
 + k_{\nu }^3 \eta_{\mu\mu} \eta_{\mu\nu}^2 \left(\frac{i  L_5}{4 \pi ^3} m^{10}\right)+k_{\mu } k_{\nu }^2 \eta_{\mu\nu}^3 \left(\frac{i  L_5}{2 \pi ^3} m^{10}\right)+k_{\nu } \eta_{\mu\mu}^2 \eta_{\nu\nu}^2 \left(\frac{i  L_6}{32 \pi ^3} m^{12}\right)+
\0\\ & \quad 
 + k_{\mu } \eta_{\mu\mu} \eta_{\mu\nu} \eta_{\nu\nu}^2 \left(\frac{i  L_6}{8 \pi ^3} m^{12}\right)+k_{\nu } \eta_{\mu\mu} \eta_{\mu\nu}^2 \eta_{\nu\nu} \left(\frac{i  L_6}{4 \pi ^3} m^{12}\right)+
\0\\ & \quad 
 + k_{\mu } \eta_{\mu\nu}^3 \eta_{\nu\nu} \left(\frac{i  L_6}{6 \pi ^3} m^{12}\right)+k_{\nu } \eta_{\mu\nu}^4 \left(\frac{i  L_6}{12 \pi ^3} m^{12}\right) }

%% file: fs12t.tex
\subsection{Fermion amplitudes}
\label{FA}
Fermions, spin 0 x 0, dimension 3:
\al{\label{eq:fc:0:0:3:}\tilde{T}_{0,0;3\text{D}}^{\text{f}} & = -\frac{i }{2 \pi } m+ T \left(-\frac{1}{8} k+\frac{1}{2}\frac{ m^2}{k}\right) }
Fermions, spin 0 x 0, dimension 4:
\al{\label{eq:fc:0:0:4:}\tilde{T}_{0,0;4\text{D}}^{\text{f}} & = \frac{i }{4 \pi ^2} \left( \left(1-\frac{L_0}{2}\right) k^2+ \left(-5+3 L_0\right) m^2\right)+\frac{i  S}{4 \pi ^2} \left(- k+4 \frac{ m^2}{k}\right) }
Fermions, spin 0 x 0, dimension 5:
\al{\label{eq:fc:0:0:5:}\tilde{T}_{0,0;5\text{D}}^{\text{f}} & = \frac{i }{4 \pi ^2} \left(-\frac{1}{4} k^2 m+\frac{5 }{3} m^3\right)+\frac{ T}{4 \pi } \left(-\frac{1}{16} k^3+\frac{1}{2} k m^2-\frac{ m^4}{k}\right) }
Fermions, spin 0 x 0, dimension 6:
\al{\label{eq:fc:0:0:6:}\tilde{T}_{0,0;6\text{D}}^{\text{f}} & = \frac{i }{4 \pi ^3} \left( \left(\frac{1}{9}-\frac{L_0}{24}\right) k^4+ \left(-\frac{37}{36}+\frac{5 L_0}{12}\right) k^2 m^2+ \left(\frac{65}{24}-\frac{5 L_0}{4}\right) m^4\right)+
\0\\ & \quad 
 + \frac{i  S}{3 \pi ^3} \left(-\frac{1}{16} k^3+\frac{1}{2} k m^2-\frac{ m^4}{k}\right) }
Fermions, spin 0 x 2, dimension 3:
\al{\label{eq:fc:0:2:3:}\tilde{T}_{0,2;3\text{D}}^{\text{f,t}} & = k^2 \pi _{\nu\nu} \left(\frac{i }{2 \pi }\frac{ m^2}{k^2}+ T \left(-\frac{1}{8}\frac{ m}{k}+\frac{1}{2}\frac{ m^3}{k^3}\right)\right) }
\al{\label{eq:fnc:0:2:3:}\tilde{T}_{0,2;3\text{D}}^{\text{f,nt}} & = \eta_{\nu\nu} \left(-\frac{i }{\pi } m^2\right) }
Fermions, spin 0 x 2, dimension 4:
\al{\label{eq:fc:0:2:4:}\tilde{T}_{0,2;4\text{D}}^{\text{f,t}} & = k^2 \pi _{\nu\nu} \left(\frac{i }{3 \pi ^2} \left( \left(\frac{2}{3}-\frac{L_0}{4}\right) m-2 \frac{ m^3}{k^2}\right)+\frac{i  S}{3 \pi ^2} \left(-\frac{1}{2}\frac{ m}{k}+2 \frac{ m^3}{k^3}\right)\right) }
\al{\label{eq:fnc:0:2:4:}\tilde{T}_{0,2;4\text{D}}^{\text{f,nt}} & = \eta_{\nu\nu} \left(\frac{i  L_1}{2 \pi ^2} m^3\right) }
Fermions, spin 0 x 2, dimension 5:
\al{\label{eq:fc:0:2:5:}\tilde{T}_{0,2;5\text{D}}^{\text{f,t}} & = k^2 \pi _{\nu\nu} \left(\frac{i }{8 \pi ^2} \left(-\frac{1}{4} m^2-\frac{ m^4}{k^2}\right)+\frac{ T}{8 \pi } \left(-\frac{1}{16} k m+\frac{1}{2}\frac{ m^3}{k}-\frac{ m^5}{k^3}\right)\right) }
\al{\label{eq:fnc:0:2:5:}\tilde{T}_{0,2;5\text{D}}^{\text{f,nt}} & = \eta_{\nu\nu} \left(\frac{i }{3 \pi ^2} m^4\right) }
Fermions, spin 0 x 2, dimension 6:
\al{\label{eq:fc:0:2:6:}\tilde{T}_{0,2;6\text{D}}^{\text{f,t}} & = k^2 \pi _{\nu\nu} \left(\frac{i }{3 \pi ^3} \left( \left(\frac{23}{600}-\frac{L_0}{80}\right) k^2 m+ \left(-\frac{43}{120}+\frac{L_0}{8}\right) m^3+\frac{2 }{5}\frac{ m^5}{k^2}\right)+
\right.\0\\ & \quad \quad \left.
 + \frac{i  S}{15 \pi ^3} \left(-\frac{1}{8} k m+\frac{m^3}{k}-2 \frac{ m^5}{k^3}\right)\right) }
\al{\label{eq:fnc:0:2:6:}\tilde{T}_{0,2;6\text{D}}^{\text{f,nt}} & = \eta_{\nu\nu} \left(-\frac{i  L_2}{8 \pi ^3} m^5\right) }
Fermions, spin 0 x 4, dimension 3:
\al{\label{eq:fc:0:4:3:}\tilde{T}_{0,4;3\text{D}}^{\text{f,t}} & = k^4 \pi _{\nu\nu}^2 \left(\frac{i }{2 \pi } \left(-\frac{5 }{4}\frac{ m^2}{k^2}+3 \frac{ m^4}{k^4}\right)+ T \left(\frac{3 }{32}\frac{ m}{k}-\frac{3 }{4}\frac{ m^3}{k^3}+\frac{3 }{2}\frac{ m^5}{k^5}\right)\right) }
\al{\label{eq:fnc:0:4:3:}\tilde{T}_{0,4;3\text{D}}^{\text{f,nt}} & = k_{\nu }^2 \eta_{\nu\nu} \left(-\frac{2 i }{\pi } m^2\right)+\eta_{\nu\nu}^2 \left(\frac{i }{\pi } \left(k^2 m^2-4  m^4\right)\right) }
Fermions, spin 0 x 4, dimension 4:
\al{\label{eq:fc:0:4:4:}\tilde{T}_{0,4;4\text{D}}^{\text{f,t}} & = k^4 \pi _{\nu\nu}^2 \left(\frac{i }{5 \pi ^2} \left( \left(-\frac{23}{30}+\frac{L_0}{4}\right) m+\frac{14 }{3}\frac{ m^3}{k^2}-8 \frac{ m^5}{k^4}\right)+
\right.\0\\ & \quad \quad \left.
 + \frac{i  S}{5 \pi ^2} \left(\frac{1}{2}\frac{ m}{k}-4 \frac{ m^3}{k^3}+8 \frac{ m^5}{k^5}\right)\right) }
\al{\label{eq:fnc:0:4:4:}\tilde{T}_{0,4;4\text{D}}^{\text{f,nt}} & = k_{\nu }^2 \eta_{\nu\nu} \left(\frac{i  L_1}{\pi ^2} m^3\right)+\eta_{\nu\nu}^2 \left(\frac{i }{2 \pi ^2} \left(- L_1 k^2 m^3+3  L_2 m^5\right)\right) }
Fermions, spin 0 x 4, dimension 5:
\al{\label{eq:fc:0:4:5:}\tilde{T}_{0,4;5\text{D}}^{\text{f,t}} & = k^4 \pi _{\nu\nu}^2 \left(\frac{i }{2 \pi ^2} \left(\frac{1}{32} m^2+\frac{1}{3}\frac{ m^4}{k^2}-\frac{1}{2}\frac{ m^6}{k^4}\right)+\frac{ T}{4 \pi } \left(\frac{1}{64} k m-\frac{3 }{16}\frac{ m^3}{k}+\frac{3 }{4}\frac{ m^5}{k^3}-\frac{ m^7}{k^5}\right)\right) }
\al{\label{eq:fnc:0:4:5:}\tilde{T}_{0,4;5\text{D}}^{\text{f,nt}} & = k_{\nu }^2 \eta_{\nu\nu} \left(\frac{2 i }{3 \pi ^2} m^4\right)+\eta_{\nu\nu}^2 \left(\frac{i }{\pi ^2} \left(-\frac{1}{3} k^2 m^4+\frac{4 }{5} m^6\right)\right) }
Fermions, spin 0 x 4, dimension 6:
\al{\label{eq:fc:0:4:6:}\tilde{T}_{0,4;6\text{D}}^{\text{f,t}} & = k^4 \pi _{\nu\nu}^2 \left(\frac{i }{\pi ^3} \left( \left(-\frac{22}{3675}+\frac{L_0}{560}\right) k^2 m+ \left(\frac{337}{4200}-\frac{L_0}{40}\right) m^3-\frac{4 }{21}\frac{ m^5}{k^2}+\frac{8 }{35}\frac{ m^7}{k^4}\right)+
\right.\0\\ & \quad \quad \left.
 + \frac{i  S}{35 \pi ^3} \left(\frac{1}{8} k m-\frac{3 }{2}\frac{ m^3}{k}+6 \frac{ m^5}{k^3}-8 \frac{ m^7}{k^5}\right)\right) }
\al{\label{eq:fnc:0:4:6:}\tilde{T}_{0,4;6\text{D}}^{\text{f,nt}} & = k_{\nu }^2 \eta_{\nu\nu} \left(-\frac{i  L_2}{4 \pi ^3} m^5\right)+\eta_{\nu\nu}^2 \left(\frac{i }{4 \pi ^3} \left(\frac{ L_2}{2} k^2 m^5- L_3 m^7\right)\right) }
Fermions, spin 1 x 1, dimension 3:
\al{\label{eq:fc:1:1:3:}\tilde{T}_{1,1;3\text{D}}^{\text{f,t}} & = k^2 \pi _{\mu\nu} \left(\frac{i }{4 \pi }\frac{ m}{k^2}+ T \left(\frac{1}{16}\frac{1}{k}+\frac{1}{4}\frac{ m^2}{k^3}\right)\right)+(k\cdot\epsilon)_{\mu\nu} \left(-\frac{i  T}{4}\frac{ m}{k}\right) }
\al{\label{eq:fnc:1:1:3:}\tilde{T}_{1,1;3\text{D}}^{\text{f,nt}} & = 0 }
Fermions, spin 1 x 1, dimension 4:
\al{\label{eq:fc:1:1:4:}\tilde{T}_{1,1;4\text{D}}^{\text{f,t}} & = k^2 \pi _{\mu\nu} \left(\frac{i }{3 \pi ^2} \left( \left(-\frac{5}{12}+\frac{L_0}{4}\right)-\frac{ m^2}{k^2}\right)+\frac{i  S}{3 \pi ^2} \left(\frac{1}{2}\frac{1}{k}+\frac{m^2}{k^3}\right)\right) }
\al{\label{eq:fnc:1:1:4:}\tilde{T}_{1,1;4\text{D}}^{\text{f,nt}} & = 0 }
Fermions, spin 1 x 1, dimension 5:
\al{\label{eq:fc:1:1:5:}\tilde{T}_{1,1;5\text{D}}^{\text{f,t}} & = k^2 \pi _{\mu\nu} \left(\frac{i }{16 \pi ^2} \left(\frac{3 }{4} m-\frac{ m^3}{k^2}\right)+\frac{ T}{16 \pi } \left(\frac{3 }{16} k-\frac{1}{2}\frac{ m^2}{k}-\frac{ m^4}{k^3}\right)\right) }
\al{\label{eq:fnc:1:1:5:}\tilde{T}_{1,1;5\text{D}}^{\text{f,nt}} & = 0 }
Fermions, spin 1 x 1, dimension 6:
\al{\label{eq:fc:1:1:6:}\tilde{T}_{1,1;6\text{D}}^{\text{f,t}} & = k^2 \pi _{\mu\nu} \left(\frac{i }{3 \pi ^3} \left( \left(-\frac{77}{1200}+\frac{L_0}{40}\right) k^2+ \left(\frac{31}{120}-\frac{L_0}{8}\right) m^2+\frac{1}{5}\frac{ m^4}{k^2}\right)+
\right.\0\\ & \quad \quad \left.
 + \frac{i  S}{5 \pi ^3} \left(\frac{1}{12} k-\frac{1}{4}\frac{ m^2}{k}-\frac{1}{3}\frac{ m^4}{k^3}\right)\right) }
\al{\label{eq:fnc:1:1:6:}\tilde{T}_{1,1;6\text{D}}^{\text{f,nt}} & = 0 }
Fermions, spin 1 x 3, dimension 3:
\al{\label{eq:fc:1:3:3:}\tilde{T}_{1,3;3\text{D}}^{\text{f,t}} & = k^4 \pi _{\mu\nu} \pi _{\nu\nu} \left(\frac{i }{4 \pi } \left(-\frac{1}{4}\frac{ m}{k^2}+3 \frac{ m^3}{k^4}\right)+ T \left(-\frac{1}{64}\frac{1}{k}-\frac{1}{8}\frac{ m^2}{k^3}+\frac{3 }{4}\frac{ m^4}{k^5}\right)\right)+
\0\\ & \quad 
 + k^2 (k\cdot\epsilon)_{\mu\nu} \pi _{\nu\nu} \left(\frac{1}{2 \pi }\frac{ m^2}{k^2}+i  T \left(\frac{1}{8}\frac{ m}{k}-\frac{1}{2}\frac{ m^3}{k^3}\right)\right) }
\al{\label{eq:fnc:1:3:3:}\tilde{T}_{1,3;3\text{D}}^{\text{f,nt}} & = \eta_{\mu\nu} \eta_{\nu\nu} \left(-\frac{4 i }{3 \pi } m^3\right)+(k\cdot\epsilon)_{\mu\nu} \eta_{\nu\nu} \left(-\frac{1}{\pi } m^2\right) }
Fermions, spin 1 x 3, dimension 4:
\al{\label{eq:fc:1:3:4:}\tilde{T}_{1,3;4\text{D}}^{\text{f,t}} & = k^4 \pi _{\nu\nu} \pi _{\mu\nu} \left(\frac{i }{5 \pi ^2} \left( \left(\frac{31}{180}-\frac{L_0}{12}\right)+\frac{2 }{3}\frac{ m^2}{k^2}-4 \frac{ m^4}{k^4}\right)+
\right.\0\\ & \quad \quad \left.
 + \frac{i  S}{5 \pi ^2} \left(-\frac{1}{6}\frac{1}{k}-\frac{1}{3}\frac{ m^2}{k^3}+4 \frac{ m^4}{k^5}\right)\right) }
\al{\label{eq:fnc:1:3:4:}\tilde{T}_{1,3;4\text{D}}^{\text{f,nt}} & = \eta_{\nu\nu} \eta_{\mu\nu} \left(\frac{i  L_2}{2 \pi ^2} m^4\right) }
Fermions, spin 1 x 3, dimension 5:
\al{\label{eq:fc:1:3:5:}\tilde{T}_{1,3;5\text{D}}^{\text{f,t}} & = k^4 \pi _{\nu\nu} \pi _{\mu\nu} \left(\frac{i }{8 \pi ^2} \left(-\frac{1}{16} m+\frac{1}{6}\frac{ m^3}{k^2}-\frac{ m^5}{k^4}\right)+
\right.\0\\ & \quad \quad \left.
 + \frac{ T}{8 \pi } \left(-\frac{1}{64} k+\frac{1}{16}\frac{ m^2}{k}+\frac{1}{4}\frac{ m^4}{k^3}-\frac{ m^6}{k^5}\right)\right) }
\al{\label{eq:fnc:1:3:5:}\tilde{T}_{1,3;5\text{D}}^{\text{f,nt}} & = \eta_{\nu\nu} \eta_{\mu\nu} \left(\frac{4 i }{15 \pi ^2} m^5\right) }
Fermions, spin 1 x 3, dimension 6:
\al{\label{eq:fc:1:3:6:}\tilde{T}_{1,3;6\text{D}}^{\text{f,t}} & = k^4 \pi _{\nu\nu} \pi _{\mu\nu} \left(\frac{i }{5 \pi ^3} \left( \left(\frac{599}{35280}-\frac{L_0}{168}\right) k^2+ \left(-\frac{247}{2520}+\frac{L_0}{24}\right) m^2-\frac{1}{7}\frac{ m^4}{k^2}+\frac{4 }{7}\frac{ m^6}{k^4}\right)+
\right.\0\\ & \quad \quad \left.
 + \frac{i  S}{7 \pi ^3} \left(-\frac{1}{60} k+\frac{1}{12}\frac{ m^2}{k}+\frac{2 }{15}\frac{ m^4}{k^3}-\frac{4 }{5}\frac{ m^6}{k^5}\right)\right) }
\al{\label{eq:fnc:1:3:6:}\tilde{T}_{1,3;6\text{D}}^{\text{f,nt}} & = \eta_{\nu\nu} \eta_{\mu\nu} \left(-\frac{i  L_3}{12 \pi ^3} m^6\right) }
Fermions, spin 1 x 5, dimension 3:
\al{\label{eq:fc:1:5:3:}\tilde{T}_{1,5;3\text{D}}^{\text{f,t}} & = k^6 \pi _{\mu\nu} \pi _{\nu\nu}^2 \left(\frac{i }{2 \pi } \left(\frac{1}{16}\frac{ m}{k^2}-\frac{11 }{6}\frac{ m^3}{k^4}+5 \frac{ m^5}{k^6}\right)+
\right.\0\\ & \quad \quad \left.
 +  T \left(\frac{1}{128}\frac{1}{k}+\frac{3 }{32}\frac{ m^2}{k^3}-\frac{9 }{8}\frac{ m^4}{k^5}+\frac{5 }{2}\frac{ m^6}{k^7}\right)\right)+
\0\\ & \quad 
 + k^4 (k\cdot\epsilon)_{\mu\nu} \pi _{\nu\nu}^2 \left(\frac{1}{2 \pi } \left(-\frac{5 }{4}\frac{ m^2}{k^2}+3 \frac{ m^4}{k^4}\right)+i  T \left(-\frac{3 }{32}\frac{ m}{k}+\frac{3 }{4}\frac{ m^3}{k^3}-\frac{3 }{2}\frac{ m^5}{k^5}\right)\right) }
\al{\label{eq:fnc:1:5:3:}\tilde{T}_{1,5;3\text{D}}^{\text{f,nt}} & = k_{\mu } k_{\nu } \eta_{\nu\nu}^2 \left(-\frac{4 i }{3 \pi } m^3\right)+k_{\nu }^2 \eta_{\mu\nu} \eta_{\nu\nu} \left(-\frac{8 i }{3 \pi } m^3\right)+\eta_{\mu\nu} \eta_{\nu\nu}^2 \left(\frac{i }{\pi } \left(\frac{4 }{3} k^2 m^3-\frac{32 }{5} m^5\right)\right)+
\0\\ & \quad 
 + k_{\nu }^2 (k\cdot\epsilon)_{\mu\nu} \eta_{\nu\nu} \left(-\frac{2 }{\pi } m^2\right)+(k\cdot\epsilon)_{\mu\nu} \eta_{\nu\nu}^2 \left(\frac{1}{\pi } \left(k^2 m^2-4  m^4\right)\right) }
Fermions, spin 1 x 5, dimension 4:
\al{\label{eq:fc:1:5:4:}\tilde{T}_{1,5;4\text{D}}^{\text{f,t}} & = k^6 \pi _{\nu\nu}^2 \pi _{\mu\nu} \left(\frac{i }{7 \pi ^2} \left( \left(-\frac{247}{2100}+\frac{L_0}{20}\right)-\frac{3 }{5}\frac{ m^2}{k^2}+\frac{116 }{15}\frac{ m^4}{k^4}-16 \frac{ m^6}{k^6}\right)+
\right.\0\\ & \quad \quad \left.
 + \frac{i  S}{7 \pi ^2} \left(\frac{1}{10}\frac{1}{k}+\frac{1}{5}\frac{ m^2}{k^3}-\frac{32 }{5}\frac{ m^4}{k^5}+16 \frac{ m^6}{k^7}\right)\right) }
\al{\label{eq:fnc:1:5:4:}\tilde{T}_{1,5;4\text{D}}^{\text{f,nt}} & = k_{\mu } k_{\nu } \eta_{\nu\nu}^2 \left(\frac{i  L_2}{2 \pi ^2} m^4\right)+k_{\nu }^2 \eta_{\mu\nu} \eta_{\nu\nu} \left(\frac{i  L_2}{\pi ^2} m^4\right)+
\0\\ & \quad 
 + \eta_{\mu\nu} \eta_{\nu\nu}^2 \left(\frac{i }{2 \pi ^2} \left(- L_2 k^2 m^4+4  L_3 m^6\right)\right) }
Fermions, spin 1 x 5, dimension 5:
\al{\label{eq:fc:1:5:5:}\tilde{T}_{1,5;5\text{D}}^{\text{f,t}} & = k^6 \pi _{\nu\nu}^2 \pi _{\mu\nu} \left(\frac{i }{16 \pi ^2} \left(\frac{3 }{64} m-\frac{3 }{16}\frac{ m^3}{k^2}+\frac{31 }{12}\frac{ m^5}{k^4}-5 \frac{ m^7}{k^6}\right)+
\right.\0\\ & \quad \quad \left.
 + \frac{ T}{16 \pi } \left(\frac{3 }{256} k-\frac{1}{16}\frac{ m^2}{k}-\frac{3 }{8}\frac{ m^4}{k^3}+3 \frac{ m^6}{k^5}-5 \frac{ m^8}{k^7}\right)\right) }
\al{\label{eq:fnc:1:5:5:}\tilde{T}_{1,5;5\text{D}}^{\text{f,nt}} & = k_{\mu } k_{\nu } \eta_{\nu\nu}^2 \left(\frac{4 i }{15 \pi ^2} m^5\right)+k_{\nu }^2 \eta_{\mu\nu} \eta_{\nu\nu} \left(\frac{8 i }{15 \pi ^2} m^5\right)+
\0\\ & \quad 
 + \eta_{\mu\nu} \eta_{\nu\nu}^2 \left(\frac{i }{5 \pi ^2} \left(-\frac{4 }{3} k^2 m^5+\frac{32 }{7} m^7\right)\right) }
Fermions, spin 1 x 5, dimension 6:
\al{\label{eq:fc:1:5:6:}\tilde{T}_{1,5;6\text{D}}^{\text{f,t}} & = k^6 \pi _{\nu\nu}^2 \pi _{\mu\nu} \left(\frac{i }{7 \pi ^3} \left( \left(-\frac{1937}{226800}+\frac{L_0}{360}\right) k^2+ \left(\frac{811}{12600}-\frac{L_0}{40}\right) m^2+\frac{2 }{15}\frac{ m^4}{k^2}-
\right.\right.\0\\ & \quad \quad \quad \left.\left.
 - \frac{152 }{135}\frac{ m^6}{k^4}+\frac{16 }{9}\frac{ m^8}{k^6}\right)+
\right.\0\\ & \quad \quad \left.
 + \frac{i  S}{3 \pi ^3} \left(\frac{1}{420} k-\frac{1}{60}\frac{ m^2}{k}-\frac{1}{35}\frac{ m^4}{k^3}+\frac{44 }{105}\frac{ m^6}{k^5}-\frac{16 }{21}\frac{ m^8}{k^7}\right)\right) }
\al{\label{eq:fnc:1:5:6:}\tilde{T}_{1,5;6\text{D}}^{\text{f,nt}} & = k_{\mu } k_{\nu } \eta_{\nu\nu}^2 \left(-\frac{i  L_3}{12 \pi ^3} m^6\right)+k_{\nu }^2 \eta_{\mu\nu} \eta_{\nu\nu} \left(-\frac{i  L_3}{6 \pi ^3} m^6\right)+
\0\\ & \quad 
 + \eta_{\mu\nu} \eta_{\nu\nu}^2 \left(\frac{i }{4 \pi ^3} \left(\frac{ L_3}{3} k^2 m^6- L_4 m^8\right)\right) }
Fermions, spin 2 x 2, dimension 3:
\al{\label{eq:fc:2:2:3:}\tilde{T}_{2,2;3\text{D}}^{\text{f,t}} & = k^4 \pi _{\mu\nu}^2 \left(\frac{i }{2 \pi } \left(-\frac{1}{4}\frac{ m}{k^2}+\frac{m^3}{k^4}\right)+ T \left(-\frac{1}{32}\frac{1}{k}+\frac{1}{2}\frac{ m^4}{k^5}\right)\right)+
\0\\ & \quad 
 + k^4 \pi _{\mu\mu} \pi _{\nu\nu} \left(\frac{i }{4 \pi } \left(\frac{1}{4}\frac{ m}{k^2}+\frac{m^3}{k^4}\right)+ T \left(\frac{1}{64}\frac{1}{k}-\frac{1}{8}\frac{ m^2}{k^3}+\frac{1}{4}\frac{ m^4}{k^5}\right)\right)+
\0\\ & \quad 
 + k^2 (k\cdot\epsilon)_{\mu\nu} \pi _{\mu\nu} \left(\frac{1}{2 \pi }\frac{ m^2}{k^2}+i  T \left(\frac{1}{8}\frac{ m}{k}-\frac{1}{2}\frac{ m^3}{k^3}\right)\right) }
\al{\label{eq:fnc:2:2:3:}\tilde{T}_{2,2;3\text{D}}^{\text{f,nt}} & =  \left(\eta_{\mu\nu}^2+\eta_{\mu\mu} \eta_{\nu\nu}\right) \left(-\frac{2 i }{3 \pi } m^3\right)+(k\cdot\epsilon)_{\mu\nu} \eta_{\mu\nu} \left(-\frac{1}{\pi } m^2\right) }
Fermions, spin 2 x 2, dimension 4:
\al{\label{eq:fc:2:2:4:}\tilde{T}_{2,2;4\text{D}}^{\text{f,t}} & = k^4 \pi _{\mu\nu}^2 \left(\frac{i }{\pi ^2} \left( \left(\frac{3}{50}-\frac{L_0}{40}\right)+ \left(-\frac{19}{180}+\frac{L_0}{12}\right)\frac{ m^2}{k^2}-\frac{8 }{15}\frac{ m^4}{k^4}\right)+
\right.\0\\ & \quad \quad \left.
 + \frac{i  S}{5 \pi ^2} \left(-\frac{1}{4}\frac{1}{k}+\frac{1}{3}\frac{ m^2}{k^3}+\frac{8 }{3}\frac{ m^4}{k^5}\right)\right)+
\0\\ & \quad 
 + k^4 \pi _{\mu\mu} \pi _{\nu\nu} \left(\frac{i }{3 \pi ^2} \left( \left(-\frac{23}{300}+\frac{L_0}{40}\right)+ \left(\frac{43}{60}-\frac{L_0}{4}\right)\frac{ m^2}{k^2}-\frac{4 }{5}\frac{ m^4}{k^4}\right)+
\right.\0\\ & \quad \quad \left.
 + \frac{i  S}{15 \pi ^2} \left(\frac{1}{4}\frac{1}{k}-2 \frac{ m^2}{k^3}+4 \frac{ m^4}{k^5}\right)\right) }
\al{\label{eq:fnc:2:2:4:}\tilde{T}_{2,2;4\text{D}}^{\text{f,nt}} & =  \left(\eta_{\mu\nu}^2+\eta_{\mu\mu} \eta_{\nu\nu}\right) \left(\frac{i  L_2}{4 \pi ^2} m^4\right) }
Fermions, spin 2 x 2, dimension 5:
\al{\label{eq:fc:2:2:5:}\tilde{T}_{2,2;5\text{D}}^{\text{f,t}} & = k^4 \pi _{\mu\nu}^2 \left(\frac{i }{12 \pi ^2} \left(-\frac{1}{8} m+\frac{7 }{12}\frac{ m^3}{k^2}-\frac{ m^5}{k^4}\right)+\frac{ T}{4 \pi } \left(-\frac{1}{96} k+\frac{1}{16}\frac{ m^2}{k}-\frac{1}{3}\frac{ m^6}{k^5}\right)\right)+
\0\\ & \quad 
 + k^4 \pi _{\mu\mu} \pi _{\nu\nu} \left(\frac{i }{12 \pi ^2} \left(\frac{1}{32} m-\frac{1}{3}\frac{ m^3}{k^2}-\frac{1}{2}\frac{ m^5}{k^4}\right)+
\right.\0\\ & \quad \quad \left.
 + \frac{ T}{8 \pi } \left(\frac{1}{192} k-\frac{1}{16}\frac{ m^2}{k}+\frac{1}{4}\frac{ m^4}{k^3}-\frac{1}{3}\frac{ m^6}{k^5}\right)\right) }
\al{\label{eq:fnc:2:2:5:}\tilde{T}_{2,2;5\text{D}}^{\text{f,nt}} & =  \left(\eta_{\mu\nu}^2+\eta_{\mu\mu} \eta_{\nu\nu}\right) \left(\frac{2 i }{15 \pi ^2} m^5\right) }
Fermions, spin 2 x 2, dimension 6:
\al{\label{eq:fc:2:2:6:}\tilde{T}_{2,2;6\text{D}}^{\text{f,t}} & = k^4 \pi _{\mu\nu}^2 \left(\frac{i }{\pi ^3} \left( \left(\frac{31}{7056}-\frac{L_0}{672}\right) k^2+ \left(-\frac{277}{8400}+\frac{L_0}{80}\right) m^2+ \left(\frac{347}{10080}-\frac{L_0}{48}\right)\frac{ m^4}{k^2}+
\right.\right.\0\\ & \quad \quad \quad \left.\left.
 + \frac{8 }{105}\frac{ m^6}{k^4}\right)+\frac{i  S}{21 \pi ^3} \left(-\frac{1}{16} k+\frac{2 }{5}\frac{ m^2}{k}-\frac{1}{5}\frac{ m^4}{k^3}-\frac{8 }{5}\frac{ m^6}{k^5}\right)\right)+
\0\\ & \quad 
 + k^4 \pi _{\mu\mu} \pi _{\nu\nu} \left(\frac{i }{3 \pi ^3} \left( \left(-\frac{11}{3675}+\frac{L_0}{1120}\right) k^2+ \left(\frac{337}{8400}-\frac{L_0}{80}\right) m^2+
\right.\right.\0\\ & \quad \quad \quad \left.\left.
 +  \left(-\frac{127}{672}+\frac{L_0}{16}\right)\frac{ m^4}{k^2}+\frac{4 }{35}\frac{ m^6}{k^4}\right)+\frac{i  S}{35 \pi ^3} \left(\frac{1}{48} k-\frac{1}{4}\frac{ m^2}{k}+\frac{m^4}{k^3}-\frac{4 }{3}\frac{ m^6}{k^5}\right)\right) }
\al{\label{eq:fnc:2:2:6:}\tilde{T}_{2,2;6\text{D}}^{\text{f,nt}} & =  \left(\eta_{\mu\nu}^2+\eta_{\mu\mu} \eta_{\nu\nu}\right) \left(-\frac{i  L_3}{24 \pi ^3} m^6\right) }
Fermions, spin 2 x 4, dimension 3:
\al{\label{eq:fc:2:4:3:}\tilde{T}_{2,4;3\text{D}}^{\text{f,t}} & = k^6 \pi _{\mu\nu}^2 \pi _{\nu\nu} \left(\frac{i }{4 \pi } \left(\frac{1}{4}\frac{ m}{k^2}-\frac{7 }{3}\frac{ m^3}{k^4}+8 \frac{ m^5}{k^6}\right)+ T \left(\frac{1}{64}\frac{1}{k}-\frac{3 }{4}\frac{ m^4}{k^5}+2 \frac{ m^6}{k^7}\right)\right)+
\0\\ & \quad 
 + k^6 \pi _{\mu\mu} \pi _{\nu\nu}^2 \left(\frac{i }{\pi } \left(-\frac{1}{32}\frac{ m}{k^2}-\frac{1}{3}\frac{ m^3}{k^4}+\frac{1}{2}\frac{ m^5}{k^6}\right)+
\right.\0\\ & \quad \quad \left.
 +  T \left(-\frac{1}{128}\frac{1}{k}+\frac{3 }{32}\frac{ m^2}{k^3}-\frac{3 }{8}\frac{ m^4}{k^5}+\frac{1}{2}\frac{ m^6}{k^7}\right)\right)+
\0\\ & \quad 
 + k^4 (k\cdot\epsilon)_{\mu\nu} \pi _{\mu\nu} \pi _{\nu\nu} \left(\frac{1}{2 \pi } \left(-\frac{5 }{4}\frac{ m^2}{k^2}+3 \frac{ m^4}{k^4}\right)+i  T \left(-\frac{3 }{32}\frac{ m}{k}+\frac{3 }{4}\frac{ m^3}{k^3}-\frac{3 }{2}\frac{ m^5}{k^5}\right)\right) }
\al{\label{eq:fnc:2:4:3:}\tilde{T}_{2,4;3\text{D}}^{\text{f,nt}} & =  \left(k_{\nu }^2 \eta_{\mu\mu} \eta_{\nu\nu}+k_{\mu } k_{\nu } \eta_{\mu\nu} \eta_{\nu\nu}\right) \left(-\frac{4 i }{3 \pi } m^3\right)+ \left(k_{\nu }^2 \eta_{\mu\nu}^2+k_{\mu }^2 \eta_{\nu\nu}^2\right) \left(-\frac{2 i }{3 \pi } m^3\right)+
\0\\ & \quad 
 + \eta_{\mu\mu} \eta_{\nu\nu}^2 \left(\frac{i }{\pi } \left(\frac{2 }{3} k^2 m^3-\frac{8 }{5} m^5\right)\right)+\eta_{\mu\nu}^2 \eta_{\nu\nu} \left(\frac{i }{\pi } \left(\frac{2 }{3} k^2 m^3-\frac{24 }{5} m^5\right)\right)+
\0\\ & \quad 
 +  \left(k_{\nu }^2 (k\cdot\epsilon)_{\mu\nu} \eta_{\mu\nu}+k_{\mu } k_{\nu } (k\cdot\epsilon)_{\mu\nu} \eta_{\nu\nu}\right) \left(-\frac{1}{\pi } m^2\right)+
\0\\ & \quad 
 + (k\cdot\epsilon)_{\mu\nu} \eta_{\mu\nu} \eta_{\nu\nu} \left(\frac{1}{\pi } \left(k^2 m^2-4  m^4\right)\right) }
Fermions, spin 2 x 4, dimension 4:
\al{\label{eq:fc:2:4:4:}\tilde{T}_{2,4;4\text{D}}^{\text{f,t}} & = k^6 \pi _{\mu\nu}^2 \pi _{\nu\nu} \left(\frac{i }{5 \pi ^2} \left( \left(-\frac{141}{980}+\frac{3 L_0}{56}\right)+ \left(\frac{157}{420}-\frac{L_0}{4}\right)\frac{ m^2}{k^2}+\frac{76 }{21}\frac{ m^4}{k^4}-\frac{64 }{7}\frac{ m^6}{k^6}\right)+
\right.\0\\ & \quad \quad \left.
 + \frac{i  S}{7 \pi ^2} \left(\frac{3 }{20}\frac{1}{k}-\frac{2 }{5}\frac{ m^2}{k^3}-4 \frac{ m^4}{k^5}+\frac{64 }{5}\frac{ m^6}{k^7}\right)\right)+
\0\\ & \quad 
 + k^6 \pi _{\mu\mu} \pi _{\nu\nu}^2 \left(\frac{i }{\pi ^2} \left( \left(\frac{44}{3675}-\frac{L_0}{280}\right)+ \left(-\frac{337}{2100}+\frac{L_0}{20}\right)\frac{ m^2}{k^2}+\frac{8 }{21}\frac{ m^4}{k^4}-
\right.\right.\0\\ & \quad \quad \quad \left.\left.
 - \frac{16 }{35}\frac{ m^6}{k^6}\right)+\frac{i  S}{35 \pi ^2} \left(-\frac{1}{4}\frac{1}{k}+3 \frac{ m^2}{k^3}-12 \frac{ m^4}{k^5}+16 \frac{ m^6}{k^7}\right)\right) }
\al{\label{eq:fnc:2:4:4:}\tilde{T}_{2,4;4\text{D}}^{\text{f,nt}} & =  \left(k_{\nu }^2 \eta_{\mu\mu} \eta_{\nu\nu}+k_{\mu } k_{\nu } \eta_{\mu\nu} \eta_{\nu\nu}\right) \left(\frac{i  L_2}{2 \pi ^2} m^4\right)+ \left(k_{\nu }^2 \eta_{\mu\nu}^2+k_{\mu }^2 \eta_{\nu\nu}^2\right) \left(\frac{i  L_2}{4 \pi ^2} m^4\right)+
\0\\ & \quad 
 + \eta_{\mu\mu} \eta_{\nu\nu}^2 \left(\frac{i }{2 \pi ^2} \left(-\frac{ L_2}{2} k^2 m^4+ L_3 m^6\right)\right)+
\0\\ & \quad 
 + \eta_{\mu\nu}^2 \eta_{\nu\nu} \left(\frac{i }{2 \pi ^2} \left(-\frac{ L_2}{2} k^2 m^4+3  L_3 m^6\right)\right) }
Fermions, spin 2 x 4, dimension 5:
\al{\label{eq:fc:2:4:5:}\tilde{T}_{2,4;5\text{D}}^{\text{f,t}} & = k^6 \pi _{\mu\nu}^2 \pi _{\nu\nu} \left(\frac{i }{4 \pi ^2} \left(\frac{1}{64} m-\frac{5 }{48}\frac{ m^3}{k^2}+\frac{5 }{12}\frac{ m^5}{k^4}-\frac{ m^7}{k^6}\right)+
\right.\0\\ & \quad \quad \left.
 + \frac{ T}{4 \pi } \left(\frac{1}{256} k-\frac{1}{32}\frac{ m^2}{k}+\frac{1}{2}\frac{ m^6}{k^5}-\frac{ m^8}{k^7}\right)\right)+
\0\\ & \quad 
 + k^6 \pi _{\mu\mu} \pi _{\nu\nu}^2 \left(\frac{i }{16 \pi ^2} \left(-\frac{1}{64} m+\frac{11 }{48}\frac{ m^3}{k^2}+\frac{11 }{12}\frac{ m^5}{k^4}-\frac{ m^7}{k^6}\right)+
\right.\0\\ & \quad \quad \left.
 + \frac{ T}{16 \pi } \left(-\frac{1}{256} k+\frac{1}{16}\frac{ m^2}{k}-\frac{3 }{8}\frac{ m^4}{k^3}+\frac{m^6}{k^5}-\frac{ m^8}{k^7}\right)\right) }
\al{\label{eq:fnc:2:4:5:}\tilde{T}_{2,4;5\text{D}}^{\text{f,nt}} & =  \left(k_{\nu }^2 \eta_{\mu\mu} \eta_{\nu\nu}+k_{\mu } k_{\nu } \eta_{\mu\nu} \eta_{\nu\nu}\right) \left(\frac{4 i }{15 \pi ^2} m^5\right)+ \left(k_{\nu }^2 \eta_{\mu\nu}^2+k_{\mu }^2 \eta_{\nu\nu}^2\right) \left(\frac{2 i }{15 \pi ^2} m^5\right)+
\0\\ & \quad 
 + \eta_{\mu\mu} \eta_{\nu\nu}^2 \left(\frac{i }{5 \pi ^2} \left(-\frac{2 }{3} k^2 m^5+\frac{8 }{7} m^7\right)\right)+\eta_{\mu\nu}^2 \eta_{\nu\nu} \left(\frac{i }{5 \pi ^2} \left(-\frac{2 }{3} k^2 m^5+\frac{24 }{7} m^7\right)\right) }
Fermions, spin 2 x 4, dimension 6:
\al{\label{eq:fc:2:4:6:}\tilde{T}_{2,4;6\text{D}}^{\text{f,t}} & = k^6 \pi _{\mu\nu}^2 \pi _{\nu\nu} \left(\frac{i }{\pi ^3} \left( \left(-\frac{25}{15876}+\frac{L_0}{2016}\right) k^2+ \left(\frac{2713}{176400}-\frac{3 L_0}{560}\right) m^2+
\right.\right.\0\\ & \quad \quad \quad \left.\left.
 +  \left(-\frac{379}{16800}+\frac{L_0}{80}\right)\frac{ m^4}{k^2}-\frac{20 }{189}\frac{ m^6}{k^4}+\frac{64 }{315}\frac{ m^8}{k^6}\right)+
\right.\0\\ & \quad \quad \left.
 + \frac{i  S}{3 \pi ^3} \left(\frac{1}{336} k-\frac{11 }{420}\frac{ m^2}{k}+\frac{1}{35}\frac{ m^4}{k^3}+\frac{4 }{15}\frac{ m^6}{k^5}-\frac{64 }{105}\frac{ m^8}{k^7}\right)\right)+
\0\\ & \quad 
 + k^6 \pi _{\mu\mu} \pi _{\nu\nu}^2 \left(\frac{i }{5 \pi ^3} \left( \left(\frac{563}{317520}-\frac{L_0}{2016}\right) k^2+ \left(-\frac{1091}{35280}+\frac{L_0}{112}\right) m^2+
\right.\right.\0\\ & \quad \quad \quad \left.\left.
 +  \left(\frac{233}{1120}-\frac{L_0}{16}\right)\frac{ m^4}{k^2}-\frac{52 }{189}\frac{ m^6}{k^4}+\frac{16 }{63}\frac{ m^8}{k^6}\right)+
\right.\0\\ & \quad \quad \left.
 + \frac{i  S}{105 \pi ^3} \left(-\frac{1}{48} k+\frac{1}{3}\frac{ m^2}{k}-2 \frac{ m^4}{k^3}+\frac{16 }{3}\frac{ m^6}{k^5}-\frac{16 }{3}\frac{ m^8}{k^7}\right)\right) }
\al{\label{eq:fnc:2:4:6:}\tilde{T}_{2,4;6\text{D}}^{\text{f,nt}} & =  \left(k_{\nu }^2 \eta_{\mu\mu} \eta_{\nu\nu}+k_{\mu } k_{\nu } \eta_{\mu\nu} \eta_{\nu\nu}\right) \left(-\frac{i  L_3}{12 \pi ^3} m^6\right)+ \left(k_{\nu }^2 \eta_{\mu\nu}^2+k_{\mu }^2 \eta_{\nu\nu}^2\right) \left(-\frac{i  L_3}{24 \pi ^3} m^6\right)+
\0\\ & \quad 
 + \eta_{\mu\mu} \eta_{\nu\nu}^2 \left(\frac{i }{8 \pi ^3} \left(\frac{ L_3}{3} k^2 m^6-\frac{ L_4}{2} m^8\right)\right)+\eta_{\mu\nu}^2 \eta_{\nu\nu} \left(\frac{i }{8 \pi ^3} \left(\frac{ L_3}{3} k^2 m^6-\frac{3  L_4}{2} m^8\right)\right) }
Fermions, spin 3 x 3, dimension 3:
\al{\label{eq:fc:3:3:3:}\tilde{T}_{3,3;3\text{D}}^{\text{f,t}} & = k^6 \pi _{\mu\nu}^3 \left(\frac{i }{2 \pi } \left(\frac{1}{8}\frac{ m}{k^2}-\frac{1}{3}\frac{ m^3}{k^4}+2 \frac{ m^5}{k^6}\right)+ T \left(\frac{1}{64}\frac{1}{k}-\frac{1}{16}\frac{ m^2}{k^3}-\frac{1}{4}\frac{ m^4}{k^5}+\frac{m^6}{k^7}\right)\right)+
\0\\ & \quad 
 + k^6 \pi _{\mu\mu} \pi _{\mu\nu} \pi _{\nu\nu} \left(\frac{i }{2 \pi } \left(-\frac{1}{16}\frac{ m}{k^2}-\frac{3 }{2}\frac{ m^3}{k^4}+3 \frac{ m^5}{k^6}\right)+
\right.\0\\ & \quad \quad \left.
 +  T \left(-\frac{1}{128}\frac{1}{k}+\frac{5 }{32}\frac{ m^2}{k^3}-\frac{7 }{8}\frac{ m^4}{k^5}+\frac{3 }{2}\frac{ m^6}{k^7}\right)\right)+
\0\\ & \quad 
 + k^4 (k\cdot\epsilon)_{\mu\nu} \pi _{\mu\nu}^2 \left(\frac{1}{4 \pi } \left(-3 \frac{ m^2}{k^2}+4 \frac{ m^4}{k^4}\right)+i  T \left(-\frac{1}{16}\frac{ m}{k}+\frac{1}{2}\frac{ m^3}{k^3}-\frac{ m^5}{k^5}\right)\right)+
\0\\ & \quad 
 + k^4 (k\cdot\epsilon)_{\mu\nu} \pi _{\mu\mu} \pi _{\nu\nu} \left(\frac{1}{2 \pi } \left(\frac{1}{4}\frac{ m^2}{k^2}+\frac{m^4}{k^4}\right)+i  T \left(-\frac{1}{32}\frac{ m}{k}+\frac{1}{4}\frac{ m^3}{k^3}-\frac{1}{2}\frac{ m^5}{k^5}\right)\right) }
\al{\label{eq:fnc:3:3:3:}\tilde{T}_{3,3;3\text{D}}^{\text{f,nt}} & =  \left(k_{\nu }^2 \eta_{\mu\mu} \eta_{\mu\nu}+k_{\mu } k_{\nu } \eta_{\mu\mu} \eta_{\nu\nu}+k_{\mu }^2 \eta_{\mu\nu} \eta_{\nu\nu}\right) \left(-\frac{4 i }{3 \pi } m^3\right)+
\0\\ & \quad 
 + \eta_{\mu\mu} \eta_{\mu\nu} \eta_{\nu\nu} \left(\frac{i }{3 \pi } \left(4  k^2 m^3-\frac{64 }{5} m^5\right)\right)+\eta_{\mu\nu}^3 \left(-\frac{32 i }{15 \pi } m^5\right)+
\0\\ & \quad 
 + k_{\mu } k_{\nu } (k\cdot\epsilon)_{\mu\nu} \eta_{\mu\nu} \left(-\frac{2 }{\pi } m^2\right)+(k\cdot\epsilon)_{\mu\nu} \eta_{\mu\mu} \eta_{\nu\nu} \left(-\frac{4 }{3 \pi } m^4\right)+
\0\\ & \quad 
 + (k\cdot\epsilon)_{\mu\nu} \eta_{\mu\nu}^2 \left(\frac{1}{3 \pi } \left(3  k^2 m^2-8  m^4\right)\right) }
Fermions, spin 3 x 3, dimension 4:
\al{\label{eq:fc:3:3:4:}\tilde{T}_{3,3;4\text{D}}^{\text{f,t}} & = k^6 \pi _{\mu\nu}^3 \left(\frac{i }{5 \pi ^2} \left( \left(-\frac{599}{4410}+\frac{L_0}{21}\right)+ \left(\frac{247}{315}-\frac{L_0}{3}\right)\frac{ m^2}{k^2}+\frac{8 }{7}\frac{ m^4}{k^4}-\frac{32 }{7}\frac{ m^6}{k^6}\right)+
\right.\0\\ & \quad \quad \left.
 + \frac{i  S}{7 \pi ^2} \left(\frac{2 }{15}\frac{1}{k}-\frac{2 }{3}\frac{ m^2}{k^3}-\frac{16 }{15}\frac{ m^4}{k^5}+\frac{32 }{5}\frac{ m^6}{k^7}\right)\right)+
\0\\ & \quad 
 + k^6 \pi _{\mu\mu} \pi _{\mu\nu} \pi _{\nu\nu} \left(\frac{i }{5 \pi ^2} \left( \left(\frac{457}{8820}-\frac{L_0}{84}\right)+ \left(-\frac{382}{315}+\frac{L_0}{3}\right)\frac{ m^2}{k^2}+\frac{92 }{21}\frac{ m^4}{k^4}-
\right.\right.\0\\ & \quad \quad \quad \left.\left.
 - \frac{48 }{7}\frac{ m^6}{k^6}\right)+\frac{i  S}{7 \pi ^2} \left(-\frac{1}{30}\frac{1}{k}+\frac{13 }{15}\frac{ m^2}{k^3}-\frac{16 }{3}\frac{ m^4}{k^5}+\frac{48 }{5}\frac{ m^6}{k^7}\right)\right) }
\al{\label{eq:fnc:3:3:4:}\tilde{T}_{3,3;4\text{D}}^{\text{f,nt}} & =  \left(k_{\nu }^2 \eta_{\mu\mu} \eta_{\mu\nu}+k_{\mu } k_{\nu } \eta_{\mu\mu} \eta_{\nu\nu}+k_{\mu }^2 \eta_{\mu\nu} \eta_{\nu\nu}\right) \left(\frac{i  L_2}{2 \pi ^2} m^4\right)+
\0\\ & \quad 
 + \eta_{\mu\mu} \eta_{\mu\nu} \eta_{\nu\nu} \left(\frac{i }{\pi ^2} \left(-\frac{ L_2}{2} k^2 m^4+\frac{4  L_3}{3} m^6\right)\right)+\eta_{\mu\nu}^3 \left(\frac{2 i  L_3}{3 \pi ^2} m^6\right) }
Fermions, spin 3 x 3, dimension 5:
\al{\label{eq:fc:3:3:5:}\tilde{T}_{3,3;5\text{D}}^{\text{f,t}} & = k^6 \pi _{\mu\nu}^3 \left(\frac{i }{8 \pi ^2} \left(\frac{5 }{192} m-\frac{31 }{144}\frac{ m^3}{k^2}+\frac{1}{4}\frac{ m^5}{k^4}-\frac{ m^7}{k^6}\right)+
\right.\0\\ & \quad \quad \left.
 + \frac{ T}{8 \pi } \left(\frac{5 }{768} k-\frac{1}{16}\frac{ m^2}{k}+\frac{1}{8}\frac{ m^4}{k^3}+\frac{1}{3}\frac{ m^6}{k^5}-\frac{ m^8}{k^7}\right)\right)+
\0\\ & \quad 
 + k^6 \pi _{\mu\mu} \pi _{\mu\nu} \pi _{\nu\nu} \left(\frac{i }{16 \pi ^2} \left(-\frac{1}{192} m+\frac{35 }{144}\frac{ m^3}{k^2}+\frac{25 }{12}\frac{ m^5}{k^4}-3 \frac{ m^7}{k^6}\right)+
\right.\0\\ & \quad \quad \left.
 + \frac{ T}{16 \pi } \left(-\frac{1}{768} k+\frac{1}{16}\frac{ m^2}{k}-\frac{5 }{8}\frac{ m^4}{k^3}+\frac{7 }{3}\frac{ m^6}{k^5}-3 \frac{ m^8}{k^7}\right)\right) }
\al{\label{eq:fnc:3:3:5:}\tilde{T}_{3,3;5\text{D}}^{\text{f,nt}} & =  \left(k_{\nu }^2 \eta_{\mu\mu} \eta_{\mu\nu}+k_{\mu } k_{\nu } \eta_{\mu\mu} \eta_{\nu\nu}+k_{\mu }^2 \eta_{\mu\nu} \eta_{\nu\nu}\right) \left(\frac{4 i }{15 \pi ^2} m^5\right)+
\0\\ & \quad 
 + \eta_{\mu\mu} \eta_{\mu\nu} \eta_{\nu\nu} \left(\frac{i }{15 \pi ^2} \left(-4  k^2 m^5+\frac{64 }{7} m^7\right)\right)+\eta_{\mu\nu}^3 \left(\frac{32 i }{105 \pi ^2} m^7\right) }
Fermions, spin 3 x 3, dimension 6:
\al{\label{eq:fc:3:3:6:}\tilde{T}_{3,3;6\text{D}}^{\text{f,t}} & = k^6 \pi _{\mu\nu}^3 \left(\frac{i }{15 \pi ^3} \left( \left(-\frac{1021}{52920}+\frac{L_0}{168}\right) k^2+ \left(\frac{317}{1470}-\frac{L_0}{14}\right) m^2+
\right.\right.\0\\ & \quad \quad \quad \left.\left.
 +  \left(-\frac{77}{120}+\frac{L_0}{4}\right)\frac{ m^4}{k^2}-\frac{32 }{63}\frac{ m^6}{k^4}+\frac{32 }{21}\frac{ m^8}{k^6}\right)+
\right.\0\\ & \quad \quad \left.
 + \frac{i  S}{21 \pi ^3} \left(\frac{1}{60} k-\frac{1}{6}\frac{ m^2}{k}+\frac{2 }{5}\frac{ m^4}{k^3}+\frac{8 }{15}\frac{ m^6}{k^5}-\frac{32 }{15}\frac{ m^8}{k^7}\right)\right)+
\0\\ & \quad 
 + k^6 \pi _{\mu\mu} \pi _{\mu\nu} \pi _{\nu\nu} \left(\frac{i }{3 \pi ^3} \left(\frac{1}{5040} k^2+ \left(-\frac{457}{29400}+\frac{L_0}{280}\right) m^2+
\right.\right.\0\\ & \quad \quad \quad \left.\left.
 +  \left(\frac{779}{4200}-\frac{L_0}{20}\right)\frac{ m^4}{k^2}-\frac{8 }{21}\frac{ m^6}{k^4}+\frac{16 }{35}\frac{ m^8}{k^6}\right)+
\right.\0\\ & \quad \quad \left.
 + \frac{i  S}{35 \pi ^3} \left(\frac{1}{12}\frac{ m^2}{k}-\frac{ m^4}{k^3}+4 \frac{ m^6}{k^5}-\frac{16 }{3}\frac{ m^8}{k^7}\right)\right) }
\al{\label{eq:fnc:3:3:6:}\tilde{T}_{3,3;6\text{D}}^{\text{f,nt}} & =  \left(k_{\nu }^2 \eta_{\mu\mu} \eta_{\mu\nu}+k_{\mu } k_{\nu } \eta_{\mu\mu} \eta_{\nu\nu}+k_{\mu }^2 \eta_{\mu\nu} \eta_{\nu\nu}\right) \left(-\frac{i  L_3}{12 \pi ^3} m^6\right)+
\0\\ & \quad 
 + \eta_{\mu\mu} \eta_{\mu\nu} \eta_{\nu\nu} \left(\frac{i }{6 \pi ^3} \left(\frac{ L_3}{2} k^2 m^6- L_4 m^8\right)\right)+\eta_{\mu\nu}^3 \left(-\frac{i  L_4}{12 \pi ^3} m^8\right) }
Fermions, spin 3 x 5, dimension 3:
\al{\label{eq:fc:3:5:3:}\tilde{T}_{3,5;3\text{D}}^{\text{f,t}} & = k^8 \pi _{\mu\nu}^3 \pi _{\nu\nu} \left(\frac{i }{4 \pi } \left(-\frac{3 }{16}\frac{ m}{k^2}+\frac{3 }{4}\frac{ m^3}{k^4}-\frac{31 }{3}\frac{ m^5}{k^6}+20 \frac{ m^7}{k^8}\right)+
\right.\0\\ & \quad \quad \left.
 +  T \left(-\frac{3 }{256}\frac{1}{k}+\frac{1}{16}\frac{ m^2}{k^3}+\frac{3 }{8}\frac{ m^4}{k^5}-3 \frac{ m^6}{k^7}+5 \frac{ m^8}{k^9}\right)\right)+
\0\\ & \quad 
 + k^8 \pi _{\mu\mu} \pi _{\mu\nu} \pi _{\nu\nu}^2 \left(\frac{i }{4 \pi } \left(\frac{7 }{64}\frac{ m}{k^2}+\frac{155 }{48}\frac{ m^3}{k^4}-\frac{47 }{4}\frac{ m^5}{k^6}+15 \frac{ m^7}{k^8}\right)+
\right.\0\\ & \quad \quad \left.
 +  T \left(\frac{7 }{1024}\frac{1}{k}-\frac{9 }{64}\frac{ m^2}{k^3}+\frac{33 }{32}\frac{ m^4}{k^5}-\frac{13 }{4}\frac{ m^6}{k^7}+\frac{15 }{4}\frac{ m^8}{k^9}\right)\right)+
\0\\ & \quad 
 + k^6 (k\cdot\epsilon)_{\mu\nu} \pi _{\mu\nu}^2 \pi _{\nu\nu} \left(\frac{1}{\pi } \left(\frac{3 }{4}\frac{ m^2}{k^2}-\frac{8 }{3}\frac{ m^4}{k^4}+4 \frac{ m^6}{k^6}\right)+
\right.\0\\ & \quad \quad \left.
 + i  T \left(\frac{1}{16}\frac{ m}{k}-\frac{3 }{4}\frac{ m^3}{k^3}+3 \frac{ m^5}{k^5}-4 \frac{ m^7}{k^7}\right)\right)+
\0\\ & \quad 
 + k^6 (k\cdot\epsilon)_{\mu\nu} \pi _{\mu\mu} \pi _{\nu\nu}^2 \left(\frac{1}{\pi } \left(-\frac{1}{16}\frac{ m^2}{k^2}-\frac{2 }{3}\frac{ m^4}{k^4}+\frac{m^6}{k^6}\right)+
\right.\0\\ & \quad \quad \left.
 + i  T \left(\frac{1}{64}\frac{ m}{k}-\frac{3 }{16}\frac{ m^3}{k^3}+\frac{3 }{4}\frac{ m^5}{k^5}-\frac{ m^7}{k^7}\right)\right) }
\al{\label{eq:fnc:3:5:3:}\tilde{T}_{3,5;3\text{D}}^{\text{f,nt}} & =  \left(k_{\mu } k_{\nu }^3 \eta_{\mu\mu} \eta_{\nu\nu}+k_{\mu }^2 k_{\nu }^2 \eta_{\mu\nu} \eta_{\nu\nu}\right) \left(-\frac{8 i }{3 \pi } m^3\right)+ \left(k_{\nu }^4 \eta_{\mu\mu} \eta_{\mu\nu}+k_{\mu }^3 k_{\nu } \eta_{\nu\nu}^2\right) \left(-\frac{4 i }{3 \pi } m^3\right)+
\0\\ & \quad 
 +  \left(k_{\mu } k_{\nu } \eta_{\mu\mu} \eta_{\nu\nu}^2+k_{\mu }^2 \eta_{\mu\nu} \eta_{\nu\nu}^2\right) \left(\frac{i }{\pi } \left(\frac{4 }{3} k^2 m^3-\frac{32 }{5} m^5\right)\right)+
\0\\ & \quad 
 + k_{\nu }^2 \eta_{\mu\mu} \eta_{\mu\nu} \eta_{\nu\nu} \left(\frac{i }{\pi } \left(\frac{8 }{3} k^2 m^3-\frac{64 }{5} m^5\right)\right)+k_{\mu } k_{\nu } \eta_{\mu\nu}^2 \eta_{\nu\nu} \left(-\frac{64 i }{5 \pi } m^5\right)+
\0\\ & \quad 
 + k_{\nu }^2 \eta_{\mu\nu}^3 \left(-\frac{64 i }{15 \pi } m^5\right)+\eta_{\mu\mu} \eta_{\mu\nu} \eta_{\nu\nu}^2 \left(\frac{i }{\pi } \left(-\frac{4 }{3} k^4 m^3+\frac{32 }{5} k^2 m^5-\frac{64 }{5} m^7\right)\right)+
\0\\ & \quad 
 + \eta_{\mu\nu}^3 \eta_{\nu\nu} \left(\frac{i }{5 \pi } \left(\frac{64 }{3} k^2 m^5-\frac{512 }{7} m^7\right)\right)+k_{\mu }^2 k_{\nu }^2 (k\cdot\epsilon)_{\mu\nu} \eta_{\nu\nu} \left(-\frac{1}{\pi } m^2\right)+
\0\\ & \quad 
 + k_{\mu } k_{\nu }^3 (k\cdot\epsilon)_{\mu\nu} \eta_{\mu\nu} \left(-\frac{2 }{\pi } m^2\right)+k_{\nu }^2 (k\cdot\epsilon)_{\mu\nu} \eta_{\mu\mu} \eta_{\nu\nu} \left(-\frac{8 }{3 \pi } m^4\right)+
\0\\ & \quad 
 + k_{\mu }^2 (k\cdot\epsilon)_{\mu\nu} \eta_{\nu\nu}^2 \left(-\frac{4 }{3 \pi } m^4\right)+k_{\mu } k_{\nu } (k\cdot\epsilon)_{\mu\nu} \eta_{\mu\nu} \eta_{\nu\nu} \left(\frac{1}{3 \pi } \left(6  k^2 m^2-32  m^4\right)\right)+
\0\\ & \quad 
 + k_{\nu }^2 (k\cdot\epsilon)_{\mu\nu} \eta_{\mu\nu}^2 \left(\frac{1}{3 \pi } \left(3  k^2 m^2-16  m^4\right)\right)+
\0\\ & \quad 
 + (k\cdot\epsilon)_{\mu\nu} \eta_{\mu\mu} \eta_{\nu\nu}^2 \left(\frac{1}{\pi } \left(\frac{4 }{3} k^2 m^4-\frac{16 }{5} m^6\right)\right)+
\0\\ & \quad 
 + (k\cdot\epsilon)_{\mu\nu} \eta_{\mu\nu}^2 \eta_{\nu\nu} \left(\frac{1}{\pi } \left(- k^4 m^2+\frac{16 }{3} k^2 m^4-\frac{64 }{5} m^6\right)\right) }
Fermions, spin 3 x 5, dimension 4:
\al{\label{eq:fc:3:5:4:}\tilde{T}_{3,5;4\text{D}}^{\text{f,t}} & = k^8 \pi _{\mu\nu}^3 \pi _{\nu\nu} \left(\frac{i }{7 \pi ^2} \left( \left(\frac{1937}{14175}-\frac{2 L_0}{45}\right)+ \left(-\frac{1622}{1575}+\frac{2 L_0}{5}\right)\frac{ m^2}{k^2}-\frac{32 }{15}\frac{ m^4}{k^4}+
\right.\right.\0\\ & \quad \quad \quad \left.\left.
 + \frac{2432 }{135}\frac{ m^6}{k^6}-\frac{256 }{9}\frac{ m^8}{k^8}\right)+
\right.\0\\ & \quad \quad \left.
 + \frac{i  S}{3 \pi ^2} \left(-\frac{4 }{105}\frac{1}{k}+\frac{4 }{15}\frac{ m^2}{k^3}+\frac{16 }{35}\frac{ m^4}{k^5}-\frac{704 }{105}\frac{ m^6}{k^7}+\frac{256 }{21}\frac{ m^8}{k^9}\right)\right)+
\0\\ & \quad 
 + k^8 \pi _{\mu\mu} \pi _{\mu\nu} \pi _{\nu\nu}^2 \left(\frac{i }{\pi ^2} \left( \left(-\frac{1231}{132300}+\frac{L_0}{420}\right)+ \left(\frac{258}{1225}-\frac{2 L_0}{35}\right)\frac{ m^2}{k^2}-\frac{104 }{105}\frac{ m^4}{k^4}+
\right.\right.\0\\ & \quad \quad \quad \left.\left.
 + \frac{128 }{45}\frac{ m^6}{k^6}-\frac{64 }{21}\frac{ m^8}{k^8}\right)+
\right.\0\\ & \quad \quad \left.
 + \frac{i  S}{\pi ^2} \left(\frac{1}{210}\frac{1}{k}-\frac{11 }{105}\frac{ m^2}{k^3}+\frac{4 }{5}\frac{ m^4}{k^5}-\frac{272 }{105}\frac{ m^6}{k^7}+\frac{64 }{21}\frac{ m^8}{k^9}\right)\right) }
\al{\label{eq:fnc:3:5:4:}\tilde{T}_{3,5;4\text{D}}^{\text{f,nt}} & =  \left(k_{\mu } k_{\nu }^3 \eta_{\mu\mu} \eta_{\nu\nu}+k_{\mu }^2 k_{\nu }^2 \eta_{\mu\nu} \eta_{\nu\nu}\right) \left(\frac{i  L_2}{\pi ^2} m^4\right)+ \left(k_{\nu }^4 \eta_{\mu\mu} \eta_{\mu\nu}+k_{\mu }^3 k_{\nu } \eta_{\nu\nu}^2\right) \left(\frac{i  L_2}{2 \pi ^2} m^4\right)+
\0\\ & \quad 
 +  \left(k_{\mu } k_{\nu } \eta_{\mu\mu} \eta_{\nu\nu}^2+k_{\mu }^2 \eta_{\mu\nu} \eta_{\nu\nu}^2\right) \left(\frac{i }{2 \pi ^2} \left(- L_2 k^2 m^4+4  L_3 m^6\right)\right)+
\0\\ & \quad 
 + k_{\nu }^2 \eta_{\mu\mu} \eta_{\mu\nu} \eta_{\nu\nu} \left(\frac{i }{\pi ^2} \left(- L_2 k^2 m^4+4  L_3 m^6\right)\right)+k_{\mu } k_{\nu } \eta_{\mu\nu}^2 \eta_{\nu\nu} \left(\frac{4 i  L_3}{\pi ^2} m^6\right)+
\0\\ & \quad 
 + k_{\nu }^2 \eta_{\mu\nu}^3 \left(\frac{4 i  L_3}{3 \pi ^2} m^6\right)+\eta_{\mu\mu} \eta_{\mu\nu} \eta_{\nu\nu}^2 \left(\frac{i }{2 \pi ^2} \left( L_2 k^4 m^4-4  L_3 k^2 m^6+7  L_4 m^8\right)\right)+
\0\\ & \quad 
 + \eta_{\mu\nu}^3 \eta_{\nu\nu} \left(\frac{i }{3 \pi ^2} \left(-4  L_3 k^2 m^6+12  L_4 m^8\right)\right) }
Fermions, spin 3 x 5, dimension 5:
\al{\label{eq:fc:3:5:5:}\tilde{T}_{3,5;5\text{D}}^{\text{f,t}} & = k^8 \pi _{\mu\nu}^3 \pi _{\nu\nu} \left(\frac{i }{\pi ^2} \left(-\frac{1}{512} m+\frac{1}{48}\frac{ m^3}{k^2}-\frac{3 }{80}\frac{ m^5}{k^4}+\frac{1}{3}\frac{ m^7}{k^6}-\frac{1}{2}\frac{ m^9}{k^8}\right)+
\right.\0\\ & \quad \quad \left.
 + \frac{ T}{2 \pi } \left(-\frac{1}{1024} k+\frac{3 }{256}\frac{ m^2}{k}-\frac{1}{32}\frac{ m^4}{k^3}-\frac{1}{8}\frac{ m^6}{k^5}+\frac{3 }{4}\frac{ m^8}{k^7}-\frac{ m^{10}}{k^9}\right)\right)+
\0\\ & \quad 
 + k^8 \pi _{\mu\mu} \pi _{\mu\nu} \pi _{\nu\nu}^2 \left(\frac{i }{8 \pi ^2} \left(\frac{1}{256} m-\frac{5 }{48}\frac{ m^3}{k^2}-\frac{137 }{120}\frac{ m^5}{k^4}+3 \frac{ m^7}{k^6}-3 \frac{ m^9}{k^8}\right)+
\right.\0\\ & \quad \quad \left.
 + \frac{ T}{8 \pi } \left(\frac{1}{1024} k-\frac{7 }{256}\frac{ m^2}{k}+\frac{9 }{32}\frac{ m^4}{k^3}-\frac{11 }{8}\frac{ m^6}{k^5}+\frac{13 }{4}\frac{ m^8}{k^7}-3 \frac{ m^{10}}{k^9}\right)\right) }
\al{\label{eq:fnc:3:5:5:}\tilde{T}_{3,5;5\text{D}}^{\text{f,nt}} & =  \left(k_{\mu } k_{\nu }^3 \eta_{\mu\mu} \eta_{\nu\nu}+k_{\mu }^2 k_{\nu }^2 \eta_{\mu\nu} \eta_{\nu\nu}\right) \left(\frac{8 i }{15 \pi ^2} m^5\right)+ \left(k_{\nu }^4 \eta_{\mu\mu} \eta_{\mu\nu}+k_{\mu }^3 k_{\nu } \eta_{\nu\nu}^2\right) \left(\frac{4 i }{15 \pi ^2} m^5\right)+
\0\\ & \quad 
 +  \left(k_{\mu } k_{\nu } \eta_{\mu\mu} \eta_{\nu\nu}^2+k_{\mu }^2 \eta_{\mu\nu} \eta_{\nu\nu}^2\right) \left(\frac{i }{5 \pi ^2} \left(-\frac{4 }{3} k^2 m^5+\frac{32 }{7} m^7\right)\right)+
\0\\ & \quad 
 + k_{\nu }^2 \eta_{\mu\mu} \eta_{\mu\nu} \eta_{\nu\nu} \left(\frac{i }{5 \pi ^2} \left(-\frac{8 }{3} k^2 m^5+\frac{64 }{7} m^7\right)\right)+k_{\mu } k_{\nu } \eta_{\mu\nu}^2 \eta_{\nu\nu} \left(\frac{64 i }{35 \pi ^2} m^7\right)+
\0\\ & \quad 
 + k_{\nu }^2 \eta_{\mu\nu}^3 \left(\frac{64 i }{105 \pi ^2} m^7\right)+\eta_{\mu\mu} \eta_{\mu\nu} \eta_{\nu\nu}^2 \left(\frac{i }{5 \pi ^2} \left(\frac{4 }{3} k^4 m^5-\frac{32 }{7} k^2 m^7+\frac{64 }{9} m^9\right)\right)+
\0\\ & \quad 
 + \eta_{\mu\nu}^3 \eta_{\nu\nu} \left(\frac{i }{105 \pi ^2} \left(-64  k^2 m^7+\frac{512 }{3} m^9\right)\right) }
Fermions, spin 3 x 5, dimension 6:
\al{\label{eq:fc:3:5:6:}\tilde{T}_{3,5;6\text{D}}^{\text{f,t}} & = k^8 \pi _{\mu\nu}^3 \pi _{\nu\nu} \left(\frac{i }{7 \pi ^3} \left( \left(\frac{11861}{2286900}-\frac{L_0}{660}\right) k^2+ \left(-\frac{11126}{155925}+\frac{L_0}{45}\right) m^2+
\right.\right.\0\\ & \quad \quad \quad \left.\left.
 +  \left(\frac{19067}{69300}-\frac{L_0}{10}\right)\frac{ m^4}{k^2}+\frac{32 }{99}\frac{ m^6}{k^4}-\frac{3008 }{1485}\frac{ m^8}{k^6}+\frac{256 }{99}\frac{ m^{10}}{k^8}\right)+
\right.\0\\ & \quad \quad \left.
 + \frac{i  S}{33 \pi ^3} \left(-\frac{1}{70} k+\frac{19 }{105}\frac{ m^2}{k}-\frac{64 }{105}\frac{ m^4}{k^3}-\frac{32 }{35}\frac{ m^6}{k^5}+\frac{128 }{15}\frac{ m^8}{k^7}-\frac{256 }{21}\frac{ m^{10}}{k^9}\right)\right)+
\0\\ & \quad 
 + k^8 \pi _{\mu\mu} \pi _{\mu\nu} \pi _{\nu\nu}^2 \left(\frac{i }{7 \pi ^3} \left( \left(-\frac{29497}{27442800}+\frac{L_0}{3960}\right) k^2+ \left(\frac{13751}{415800}-\frac{L_0}{120}\right) m^2+
\right.\right.\0\\ & \quad \quad \quad \left.\left.
 +  \left(-\frac{8689}{23100}+\frac{L_0}{10}\right)\frac{ m^4}{k^2}+\frac{1504 }{1485}\frac{ m^6}{k^4}-\frac{368 }{165}\frac{ m^8}{k^6}+\frac{64 }{33}\frac{ m^{10}}{k^8}\right)+
\right.\0\\ & \quad \quad \left.
 + \frac{i  S}{33 \pi ^3} \left(\frac{1}{420} k-\frac{31 }{420}\frac{ m^2}{k}+\frac{4 }{5}\frac{ m^4}{k^3}-\frac{424 }{105}\frac{ m^6}{k^5}+\frac{1024 }{105}\frac{ m^8}{k^7}-\frac{64 }{7}\frac{ m^{10}}{k^9}\right)\right) }
\al{\label{eq:fnc:3:5:6:}\tilde{T}_{3,5;6\text{D}}^{\text{f,nt}} & =  \left(k_{\mu } k_{\nu }^3 \eta_{\mu\mu} \eta_{\nu\nu}+k_{\mu }^2 k_{\nu }^2 \eta_{\mu\nu} \eta_{\nu\nu}\right) \left(-\frac{i  L_3}{6 \pi ^3} m^6\right)+
\0\\ & \quad 
 +  \left(k_{\nu }^4 \eta_{\mu\mu} \eta_{\mu\nu}+k_{\mu }^3 k_{\nu } \eta_{\nu\nu}^2\right) \left(-\frac{i  L_3}{12 \pi ^3} m^6\right)+
\0\\ & \quad 
 +  \left(k_{\mu } k_{\nu } \eta_{\mu\mu} \eta_{\nu\nu}^2+k_{\mu }^2 \eta_{\mu\nu} \eta_{\nu\nu}^2\right) \left(\frac{i }{4 \pi ^3} \left(\frac{ L_3}{3} k^2 m^6- L_4 m^8\right)\right)+
\0\\ & \quad 
 + k_{\nu }^2 \eta_{\mu\mu} \eta_{\mu\nu} \eta_{\nu\nu} \left(\frac{i }{2 \pi ^3} \left(\frac{ L_3}{3} k^2 m^6- L_4 m^8\right)\right)+k_{\mu } k_{\nu } \eta_{\mu\nu}^2 \eta_{\nu\nu} \left(-\frac{i  L_4}{2 \pi ^3} m^8\right)+
\0\\ & \quad 
 + k_{\nu }^2 \eta_{\mu\nu}^3 \left(-\frac{i  L_4}{6 \pi ^3} m^8\right)+
\0\\ & \quad 
 + \eta_{\mu\mu} \eta_{\mu\nu} \eta_{\nu\nu}^2 \left(\frac{i }{4 \pi ^3} \left(-\frac{ L_3}{3} k^4 m^6+ L_4 k^2 m^8-\frac{7  L_5}{5} m^{10}\right)\right)+
\0\\ & \quad 
 + \eta_{\mu\nu}^3 \eta_{\nu\nu} \left(\frac{i }{\pi ^3} \left(\frac{ L_4}{6} k^2 m^8-\frac{2  L_5}{5} m^{10}\right)\right) }
Fermions, spin 4 x 4, dimension 3:
\al{\label{eq:fc:4:4:3:}\tilde{T}_{4,4;3\text{D}}^{\text{f,t}} & = k^8 \pi _{\mu\nu}^4 \left(\frac{i }{2 \pi } \left(-\frac{1}{16}\frac{ m}{k^2}-\frac{11 }{12}\frac{ m^3}{k^4}-\frac{5 }{3}\frac{ m^5}{k^6}+4 \frac{ m^7}{k^8}\right)+
\right.\0\\ & \quad \quad \left.
 +  T \left(-\frac{1}{128}\frac{1}{k}+\frac{1}{16}\frac{ m^2}{k^3}-\frac{ m^6}{k^7}+2 \frac{ m^8}{k^9}\right)\right)+
\0\\ & \quad 
 + k^8 \pi _{\mu\mu} \pi _{\mu\nu}^2 \pi _{\nu\nu} \left(\frac{i }{8 \pi } \left(13 \frac{ m^3}{k^4}-32 \frac{ m^5}{k^6}+48 \frac{ m^7}{k^8}\right)+
\right.\0\\ & \quad \quad \left.
 +  T \left(-\frac{3 }{32}\frac{ m^2}{k^3}+\frac{9 }{8}\frac{ m^4}{k^5}-\frac{9 }{2}\frac{ m^6}{k^7}+6 \frac{ m^8}{k^9}\right)\right)+
\0\\ & \quad 
 + k^8 \pi _{\mu\mu}^2 \pi _{\nu\nu}^2 \left(\frac{i }{4 \pi } \left(\frac{3 }{64}\frac{ m}{k^2}-\frac{11 }{16}\frac{ m^3}{k^4}-\frac{11 }{4}\frac{ m^5}{k^6}+3 \frac{ m^7}{k^8}\right)+
\right.\0\\ & \quad \quad \left.
 +  T \left(\frac{3 }{1024}\frac{1}{k}-\frac{3 }{64}\frac{ m^2}{k^3}+\frac{9 }{32}\frac{ m^4}{k^5}-\frac{3 }{4}\frac{ m^6}{k^7}+\frac{3 }{4}\frac{ m^8}{k^9}\right)\right)+
\0\\ & \quad 
 + k^6 (k\cdot\epsilon)_{\mu\nu} \pi _{\mu\nu}^3 \left(\frac{1}{\pi } \left(\frac{7 }{8}\frac{ m^2}{k^2}-\frac{4 }{3}\frac{ m^4}{k^4}+2 \frac{ m^6}{k^6}\right)+
\right.\0\\ & \quad \quad \left.
 + i  T \left(\frac{1}{32}\frac{ m}{k}-\frac{3 }{8}\frac{ m^3}{k^3}+\frac{3 }{2}\frac{ m^5}{k^5}-2 \frac{ m^7}{k^7}\right)\right)+
\0\\ & \quad 
 + k^6 (k\cdot\epsilon)_{\mu\nu} \pi _{\mu\mu} \pi _{\mu\nu} \pi _{\nu\nu} \left(\frac{1}{16 \pi } \left(-3 \frac{ m^2}{k^2}-32 \frac{ m^4}{k^4}+48 \frac{ m^6}{k^6}\right)+
\right.\0\\ & \quad \quad \left.
 + i  T \left(\frac{3 }{64}\frac{ m}{k}-\frac{9 }{16}\frac{ m^3}{k^3}+\frac{9 }{4}\frac{ m^5}{k^5}-3 \frac{ m^7}{k^7}\right)\right) }
\al{\label{eq:fnc:4:4:3:}\tilde{T}_{4,4;3\text{D}}^{\text{f,nt}} & = k_{\mu }^2 k_{\nu }^2 \eta_{\mu\mu} \eta_{\nu\nu} \left(-\frac{2 i }{\pi } m^3\right)+ \left(k_{\mu } k_{\nu }^3 \eta_{\mu\mu} \eta_{\mu\nu}+k_{\mu }^3 k_{\nu } \eta_{\mu\nu} \eta_{\nu\nu}\right) \left(-\frac{4 i }{\pi } m^3\right)+
\0\\ & \quad 
 + k_{\mu }^2 k_{\nu }^2 \eta_{\mu\nu}^2 \left(\frac{2 i }{\pi } m^3\right)+ \left(k_{\nu }^2 \eta_{\mu\mu}^2 \eta_{\nu\nu}+k_{\mu }^2 \eta_{\mu\mu} \eta_{\nu\nu}^2\right) \left(-\frac{16 i }{5 \pi } m^5\right)+
\0\\ & \quad 
 + k_{\mu } k_{\nu } \eta_{\mu\mu} \eta_{\mu\nu} \eta_{\nu\nu} \left(\frac{i }{\pi } \left(4  k^2 m^3-16  m^5\right)\right)+
\0\\ & \quad 
 +  \left(k_{\nu }^2 \eta_{\mu\mu} \eta_{\mu\nu}^2+k_{\mu }^2 \eta_{\mu\nu}^2 \eta_{\nu\nu}\right) \left(\frac{i }{\pi } \left(2  k^2 m^3-8  m^5\right)\right)+
\0\\ & \quad 
 + k_{\mu } k_{\nu } \eta_{\mu\nu}^3 \left(\frac{i }{3 \pi } \left(-8  k^2 m^3-\frac{64 }{5} m^5\right)\right)+\eta_{\mu\mu}^2 \eta_{\nu\nu}^2 \left(\frac{i }{5 \pi } \left(8  k^2 m^5-\frac{96 }{7} m^7\right)\right)+
\0\\ & \quad 
 + \eta_{\mu\mu} \eta_{\mu\nu}^2 \eta_{\nu\nu} \left(\frac{i }{5 \pi } \left(-10  k^4 m^3+40  k^2 m^5-96  m^7\right)\right)+
\0\\ & \quad 
 + \eta_{\mu\nu}^4 \left(\frac{i }{\pi } \left(\frac{2 }{3} k^4 m^3+\frac{16 }{15} k^2 m^5-\frac{192 }{35} m^7\right)\right)+k_{\mu }^2 k_{\nu }^2 (k\cdot\epsilon)_{\mu\nu} \eta_{\mu\nu} \left(-\frac{3 }{\pi } m^2\right)+
\0\\ & \quad 
 +  \left(k_{\nu }^2 (k\cdot\epsilon)_{\mu\nu} \eta_{\mu\mu} \eta_{\mu\nu}+k_{\mu } k_{\nu } (k\cdot\epsilon)_{\mu\nu} \eta_{\mu\mu} \eta_{\nu\nu}+k_{\mu }^2 (k\cdot\epsilon)_{\mu\nu} \eta_{\mu\nu} \eta_{\nu\nu}\right) \left(-\frac{4 }{\pi } m^4\right)+
\0\\ & \quad 
 + k_{\mu } k_{\nu } (k\cdot\epsilon)_{\mu\nu} \eta_{\mu\nu}^2 \left(\frac{1}{\pi } \left(3  k^2 m^2-8  m^4\right)\right)+
\0\\ & \quad 
 + (k\cdot\epsilon)_{\mu\nu} \eta_{\mu\mu} \eta_{\mu\nu} \eta_{\nu\nu} \left(\frac{1}{5 \pi } \left(20  k^2 m^4-48  m^6\right)\right)+
\0\\ & \quad 
 + (k\cdot\epsilon)_{\mu\nu} \eta_{\mu\nu}^3 \left(\frac{1}{\pi } \left(- k^4 m^2+\frac{8 }{3} k^2 m^4-\frac{32 }{5} m^6\right)\right) }
Fermions, spin 4 x 4, dimension 4:
\al{\label{eq:fc:4:4:4:}\tilde{T}_{4,4;4\text{D}}^{\text{f,t}} & = k^8 \pi _{\mu\nu}^4 \left(\frac{i }{\pi ^2} \left( \left(\frac{50}{3969}-\frac{L_0}{252}\right)+ \left(-\frac{2713}{22050}+\frac{3 L_0}{70}\right)\frac{ m^2}{k^2}+
\right.\right.\0\\ & \quad \quad \quad \left.\left.
 +  \left(-\frac{817}{4200}+\frac{3 L_0}{20}\right)\frac{ m^4}{k^4}+\frac{160 }{189}\frac{ m^6}{k^6}-\frac{512 }{315}\frac{ m^8}{k^8}\right)+
\right.\0\\ & \quad \quad \left.
 + \frac{i  S}{3 \pi ^2} \left(-\frac{1}{42}\frac{1}{k}+\frac{22 }{105}\frac{ m^2}{k^3}-\frac{8 }{35}\frac{ m^4}{k^5}-\frac{32 }{15}\frac{ m^6}{k^7}+\frac{512 }{105}\frac{ m^8}{k^9}\right)\right)+
\0\\ & \quad 
 + k^8 \pi _{\mu\mu} \pi _{\mu\nu}^2 \pi _{\nu\nu} \left(\frac{i }{5 \pi ^2} \left( \left(\frac{62}{6615}-\frac{L_0}{168}\right)+ \left(\frac{1651}{2940}-\frac{3 L_0}{28}\right)\frac{ m^2}{k^2}-
\right.\right.\0\\ & \quad \quad \quad \left.\left.
 -  \left(\frac{421}{140}+\frac{3 L_0}{2}\right)\frac{ m^4}{k^4}+\frac{176 }{9}\frac{ m^6}{k^6}-\frac{512 }{21}\frac{ m^8}{k^8}\right)+
\right.\0\\ & \quad \quad \left.
 + \frac{i  S}{\pi ^2} \left(-\frac{1}{420}\frac{1}{k}-\frac{1}{21}\frac{ m^2}{k^3}+\frac{4 }{5}\frac{ m^4}{k^5}-\frac{368 }{105}\frac{ m^6}{k^7}+\frac{512 }{105}\frac{ m^8}{k^9}\right)\right)+
\0\\ & \quad 
 + k^8 \pi _{\mu\mu}^2 \pi _{\nu\nu}^2 \left(\frac{i }{5 \pi ^2} \left( \left(-\frac{563}{26460}+\frac{L_0}{168}\right)+ \left(\frac{1091}{2940}-\frac{3 L_0}{28}\right)\frac{ m^2}{k^2}+
\right.\right.\0\\ & \quad \quad \quad \left.\left.
 +  \left(-\frac{699}{280}+\frac{3 L_0}{4}\right)\frac{ m^4}{k^4}+\frac{208 }{63}\frac{ m^6}{k^6}-\frac{64 }{21}\frac{ m^8}{k^8}\right)+
\right.\0\\ & \quad \quad \left.
 + \frac{i  S}{35 \pi ^2} \left(\frac{1}{12}\frac{1}{k}-\frac{4 }{3}\frac{ m^2}{k^3}+8 \frac{ m^4}{k^5}-\frac{64 }{3}\frac{ m^6}{k^7}+\frac{64 }{3}\frac{ m^8}{k^9}\right)\right) }
\al{\label{eq:fnc:4:4:4:}\tilde{T}_{4,4;4\text{D}}^{\text{f,nt}} & = k_{\mu }^2 k_{\nu }^2 \eta_{\mu\mu} \eta_{\nu\nu} \left(\frac{3 i  L_2}{4 \pi ^2} m^4\right)+ \left(k_{\mu } k_{\nu }^3 \eta_{\mu\mu} \eta_{\mu\nu}+k_{\mu }^3 k_{\nu } \eta_{\mu\nu} \eta_{\nu\nu}\right) \left(\frac{3 i  L_2}{2 \pi ^2} m^4\right)+
\0\\ & \quad 
 + k_{\mu }^2 k_{\nu }^2 \eta_{\mu\nu}^2 \left(-\frac{3 i  L_2}{4 \pi ^2} m^4\right)+ \left(k_{\nu }^2 \eta_{\mu\mu}^2 \eta_{\nu\nu}+k_{\mu }^2 \eta_{\mu\mu} \eta_{\nu\nu}^2\right) \left(\frac{i  L_3}{\pi ^2} m^6\right)+
\0\\ & \quad 
 + k_{\mu } k_{\nu } \eta_{\mu\mu} \eta_{\mu\nu} \eta_{\nu\nu} \left(\frac{i }{2 \pi ^2} \left(-3  L_2 k^2 m^4+10  L_3 m^6\right)\right)+
\0\\ & \quad 
 +  \left(k_{\nu }^2 \eta_{\mu\mu} \eta_{\mu\nu}^2+k_{\mu }^2 \eta_{\mu\nu}^2 \eta_{\nu\nu}\right) \left(\frac{i }{2 \pi ^2} \left(-\frac{3  L_2}{2} k^2 m^4+5  L_3 m^6\right)\right)+
\0\\ & \quad 
 + k_{\mu } k_{\nu } \eta_{\mu\nu}^3 \left(\frac{i }{3 \pi ^2} \left(3  L_2 k^2 m^4+4  L_3 m^6\right)\right)+
\0\\ & \quad 
 + \eta_{\mu\mu}^2 \eta_{\nu\nu}^2 \left(\frac{i }{2 \pi ^2} \left(- L_3 k^2 m^6+\frac{3  L_4}{2} m^8\right)\right)+
\0\\ & \quad 
 + \eta_{\mu\mu} \eta_{\mu\nu}^2 \eta_{\nu\nu} \left(\frac{i }{2 \pi ^2} \left(\frac{3  L_2}{2} k^4 m^4-5  L_3 k^2 m^6+\frac{21  L_4}{2} m^8\right)\right)+
\0\\ & \quad 
 + \eta_{\mu\nu}^4 \left(\frac{i }{\pi ^2} \left(-\frac{ L_2}{4} k^4 m^4-\frac{ L_3}{3} k^2 m^6+\frac{3  L_4}{2} m^8\right)\right) }
Fermions, spin 4 x 4, dimension 5:
\al{\label{eq:fc:4:4:5:}\tilde{T}_{4,4;5\text{D}}^{\text{f,t}} & = k^8 \pi _{\mu\nu}^4 \left(\frac{i }{5 \pi ^2} \left(-\frac{3 }{512} m+\frac{9 }{128}\frac{ m^3}{k^2}+\frac{211 }{480}\frac{ m^5}{k^4}+\frac{13 }{24}\frac{ m^7}{k^6}-\frac{ m^9}{k^8}\right)+
\right.\0\\ & \quad \quad \left.
 + \frac{ T}{\pi } \left(-\frac{3 }{10240} k+\frac{1}{256}\frac{ m^2}{k}-\frac{1}{64}\frac{ m^4}{k^3}+\frac{1}{8}\frac{ m^8}{k^7}-\frac{1}{5}\frac{ m^{10}}{k^9}\right)\right)+
\0\\ & \quad 
 + k^8 \pi _{\mu\mu} \pi _{\mu\nu}^2 \pi _{\nu\nu} \left(\frac{i }{5 \pi ^2} \left(-\frac{3 }{1024} m-\frac{1}{256}\frac{ m^3}{k^2}-\frac{493 }{320}\frac{ m^5}{k^4}+\frac{41 }{16}\frac{ m^7}{k^6}-3 \frac{ m^9}{k^8}\right)+
\right.\0\\ & \quad \quad \left.
 + \frac{ T}{\pi } \left(-\frac{3 }{20480} k+\frac{3 }{128}\frac{ m^4}{k^3}-\frac{3 }{16}\frac{ m^6}{k^5}+\frac{9 }{16}\frac{ m^8}{k^7}-\frac{3 }{5}\frac{ m^{10}}{k^9}\right)\right)+
\0\\ & \quad 
 + k^8 \pi _{\mu\mu}^2 \pi _{\nu\nu}^2 \left(\frac{i }{5 \pi ^2} \left(\frac{3 }{2048} m-\frac{7 }{256}\frac{ m^3}{k^2}+\frac{1}{5}\frac{ m^5}{k^4}+\frac{7 }{16}\frac{ m^7}{k^6}-\frac{3 }{8}\frac{ m^9}{k^8}\right)+
\right.\0\\ & \quad \quad \left.
 + \frac{ T}{8 \pi } \left(\frac{3 }{5120} k-\frac{3 }{256}\frac{ m^2}{k}+\frac{3 }{32}\frac{ m^4}{k^3}-\frac{3 }{8}\frac{ m^6}{k^5}+\frac{3 }{4}\frac{ m^8}{k^7}-\frac{3 }{5}\frac{ m^{10}}{k^9}\right)\right) }
\al{\label{eq:fnc:4:4:5:}\tilde{T}_{4,4;5\text{D}}^{\text{f,nt}} & = k_{\mu }^2 k_{\nu }^2 \eta_{\mu\mu} \eta_{\nu\nu} \left(\frac{2 i }{5 \pi ^2} m^5\right)+ \left(k_{\mu } k_{\nu }^3 \eta_{\mu\mu} \eta_{\mu\nu}+k_{\mu }^3 k_{\nu } \eta_{\mu\nu} \eta_{\nu\nu}\right) \left(\frac{4 i }{5 \pi ^2} m^5\right)+
\0\\ & \quad 
 + k_{\mu }^2 k_{\nu }^2 \eta_{\mu\nu}^2 \left(-\frac{2 i }{5 \pi ^2} m^5\right)+ \left(k_{\nu }^2 \eta_{\mu\mu}^2 \eta_{\nu\nu}+k_{\mu }^2 \eta_{\mu\mu} \eta_{\nu\nu}^2\right) \left(\frac{16 i }{35 \pi ^2} m^7\right)+
\0\\ & \quad 
 + k_{\mu } k_{\nu } \eta_{\mu\mu} \eta_{\mu\nu} \eta_{\nu\nu} \left(\frac{i }{\pi ^2} \left(-\frac{4 }{5} k^2 m^5+\frac{16 }{7} m^7\right)\right)+
\0\\ & \quad 
 +  \left(k_{\nu }^2 \eta_{\mu\mu} \eta_{\mu\nu}^2+k_{\mu }^2 \eta_{\mu\nu}^2 \eta_{\nu\nu}\right) \left(\frac{i }{\pi ^2} \left(-\frac{2 }{5} k^2 m^5+\frac{8 }{7} m^7\right)\right)+
\0\\ & \quad 
 + k_{\mu } k_{\nu } \eta_{\mu\nu}^3 \left(\frac{i }{15 \pi ^2} \left(8  k^2 m^5+\frac{64 }{7} m^7\right)\right)+
\0\\ & \quad 
 + \eta_{\mu\mu}^2 \eta_{\nu\nu}^2 \left(\frac{i }{35 \pi ^2} \left(-8  k^2 m^7+\frac{32 }{3} m^9\right)\right)+
\0\\ & \quad 
 + \eta_{\mu\mu} \eta_{\mu\nu}^2 \eta_{\nu\nu} \left(\frac{i }{\pi ^2} \left(\frac{2 }{5} k^4 m^5-\frac{8 }{7} k^2 m^7+\frac{32 }{15} m^9\right)\right)+
\0\\ & \quad 
 + \eta_{\mu\nu}^4 \left(\frac{i }{15 \pi ^2} \left(-2  k^4 m^5-\frac{16 }{7} k^2 m^7+\frac{64 }{7} m^9\right)\right) }
Fermions, spin 4 x 4, dimension 6:
\al{\label{eq:fc:4:4:6:}\tilde{T}_{4,4;6\text{D}}^{\text{f,t}} & = k^8 \pi _{\mu\nu}^4 \left(\frac{i }{\pi ^3} \left( \left(\frac{859}{1960200}-\frac{L_0}{7920}\right) k^2+ \left(-\frac{11441}{1746360}+\frac{L_0}{504}\right) m^2+
\right.\right.\0\\ & \quad \quad \quad \left.\left.
 +  \left(\frac{62801}{1940400}-\frac{3 L_0}{280}\right)\frac{ m^4}{k^2}+ \left(\frac{35171}{831600}-\frac{L_0}{40}\right)\frac{ m^6}{k^4}-\frac{992 }{10395}\frac{ m^8}{k^6}+
\right.\right.\0\\ & \quad \quad \quad \left.\left.
 + \frac{512 }{3465}\frac{ m^{10}}{k^8}\right)+
\right.\0\\ & \quad \quad \left.
 + \frac{i  S}{55 \pi ^3} \left(-\frac{1}{72} k+\frac{4 }{21}\frac{ m^2}{k}-\frac{52 }{63}\frac{ m^4}{k^3}+\frac{32 }{63}\frac{ m^6}{k^5}+\frac{32 }{7}\frac{ m^8}{k^7}-\frac{512 }{63}\frac{ m^{10}}{k^9}\right)\right)+
\0\\ & \quad 
 + k^8 \pi _{\mu\mu} \pi _{\mu\nu}^2 \pi _{\nu\nu} \left(\frac{i }{\pi ^3} \left( \left(\frac{5353}{21344400}-\frac{L_0}{12320}\right) k^2+ \left(-\frac{6401}{5821200}+\frac{L_0}{1680}\right) m^2+
\right.\right.\0\\ & \quad \quad \quad \left.\left.
 +  \left(-\frac{35867}{1293600}+\frac{3 L_0}{560}\right)\frac{ m^4}{k^2}+ \left(\frac{967}{15400}+\frac{L_0}{20}\right)\frac{ m^6}{k^4}-\frac{304 }{693}\frac{ m^8}{k^6}+
\right.\right.\0\\ & \quad \quad \quad \left.\left.
 + \frac{512 }{1155}\frac{ m^{10}}{k^8}\right)+
\right.\0\\ & \quad \quad \left.
 + \frac{i  S}{55 \pi ^3} \left(-\frac{1}{112} k+\frac{1}{21}\frac{ m^2}{k}+\frac{2 }{3}\frac{ m^4}{k^3}-\frac{48 }{7}\frac{ m^6}{k^5}+\frac{464 }{21}\frac{ m^8}{k^7}-\frac{512 }{21}\frac{ m^{10}}{k^9}\right)\right)+
\0\\ & \quad 
 + k^8 \pi _{\mu\mu}^2 \pi _{\nu\nu}^2 \left(\frac{i }{5 \pi ^3} \left( \left(-\frac{1627}{3201660}+\frac{L_0}{7392}\right) k^2+ \left(\frac{12701}{1164240}-\frac{L_0}{336}\right) m^2+
\right.\right.\0\\ & \quad \quad \quad \left.\left.
 +  \left(-\frac{24667}{258720}+\frac{3 L_0}{112}\right)\frac{ m^4}{k^2}+ \left(\frac{23777}{55440}-\frac{L_0}{8}\right)\frac{ m^6}{k^4}-\frac{256 }{693}\frac{ m^8}{k^6}+
\right.\right.\0\\ & \quad \quad \quad \left.\left.
 + \frac{64 }{231}\frac{ m^{10}}{k^8}\right)+
\right.\0\\ & \quad \quad \left.
 + \frac{i  S}{231 \pi ^3} \left(\frac{1}{80} k-\frac{1}{4}\frac{ m^2}{k}+2 \frac{ m^4}{k^3}-8 \frac{ m^6}{k^5}+16 \frac{ m^8}{k^7}-\frac{64 }{5}\frac{ m^{10}}{k^9}\right)\right) }
\al{\label{eq:fnc:4:4:6:}\tilde{T}_{4,4;6\text{D}}^{\text{f,nt}} & = k_{\mu }^2 k_{\nu }^2 \eta_{\mu\mu} \eta_{\nu\nu} \left(-\frac{i  L_3}{8 \pi ^3} m^6\right)+ \left(k_{\mu } k_{\nu }^3 \eta_{\mu\mu} \eta_{\mu\nu}+k_{\mu }^3 k_{\nu } \eta_{\mu\nu} \eta_{\nu\nu}\right) \left(-\frac{i  L_3}{4 \pi ^3} m^6\right)+
\0\\ & \quad 
 + k_{\mu }^2 k_{\nu }^2 \eta_{\mu\nu}^2 \left(\frac{i  L_3}{8 \pi ^3} m^6\right)+ \left(k_{\nu }^2 \eta_{\mu\mu}^2 \eta_{\nu\nu}+k_{\mu }^2 \eta_{\mu\mu} \eta_{\nu\nu}^2\right) \left(-\frac{i  L_4}{8 \pi ^3} m^8\right)+
\0\\ & \quad 
 + k_{\mu } k_{\nu } \eta_{\mu\mu} \eta_{\mu\nu} \eta_{\nu\nu} \left(\frac{i }{4 \pi ^3} \left( L_3 k^2 m^6-\frac{5  L_4}{2} m^8\right)\right)+
\0\\ & \quad 
 +  \left(k_{\nu }^2 \eta_{\mu\mu} \eta_{\mu\nu}^2+k_{\mu }^2 \eta_{\mu\nu}^2 \eta_{\nu\nu}\right) \left(\frac{i }{8 \pi ^3} \left( L_3 k^2 m^6-\frac{5  L_4}{2} m^8\right)\right)+
\0\\ & \quad 
 + k_{\mu } k_{\nu } \eta_{\mu\nu}^3 \left(\frac{i }{6 \pi ^3} \left(- L_3 k^2 m^6- L_4 m^8\right)\right)+
\0\\ & \quad 
 + \eta_{\mu\mu}^2 \eta_{\nu\nu}^2 \left(\frac{i }{8 \pi ^3} \left(\frac{ L_4}{2} k^2 m^8-\frac{3  L_5}{5} m^{10}\right)\right)+
\0\\ & \quad 
 + \eta_{\mu\mu} \eta_{\mu\nu}^2 \eta_{\nu\nu} \left(\frac{i }{8 \pi ^3} \left(- L_3 k^4 m^6+\frac{5  L_4}{2} k^2 m^8-\frac{21  L_5}{5} m^{10}\right)\right)+
\0\\ & \quad 
 + \eta_{\mu\nu}^4 \left(\frac{i }{4 \pi ^3} \left(\frac{ L_3}{6} k^4 m^6+\frac{ L_4}{6} k^2 m^8-\frac{3  L_5}{5} m^{10}\right)\right) }
Fermions, spin 5 x 5, dimension 3:
\al{\label{eq:fc:5:5:3:}\tilde{T}_{5,5;3\text{D}}^{\text{f,t}} & = k^{10} \pi _{\mu\nu}^5 \left(\frac{i }{\pi } \left(\frac{1}{64}\frac{ m}{k^2}+\frac{7 }{6}\frac{ m^3}{k^4}+\frac{3 }{10}\frac{ m^5}{k^6}-\frac{8 }{3}\frac{ m^7}{k^8}+4 \frac{ m^9}{k^{10}}\right)+
\right.\0\\ & \quad \quad \left.
 +  T \left(\frac{1}{256}\frac{1}{k}-\frac{3 }{64}\frac{ m^2}{k^3}+\frac{1}{8}\frac{ m^4}{k^5}+\frac{1}{2}\frac{ m^6}{k^7}-3 \frac{ m^8}{k^9}+4 \frac{ m^{10}}{k^{11}}\right)\right)+
\0\\ & \quad 
 + k^{10} \pi _{\mu\mu} \pi _{\mu\nu}^3 \pi _{\nu\nu} \left(\frac{i }{\pi } \left(\frac{1}{64}\frac{ m}{k^2}-\frac{31 }{12}\frac{ m^3}{k^4}+\frac{31 }{6}\frac{ m^5}{k^6}-\frac{52 }{3}\frac{ m^7}{k^8}+20 \frac{ m^9}{k^{10}}\right)+
\right.\0\\ & \quad \quad \left.
 +  T \left(\frac{1}{256}\frac{1}{k}+\frac{1}{64}\frac{ m^2}{k^3}-\frac{7 }{8}\frac{ m^4}{k^5}+\frac{13 }{2}\frac{ m^6}{k^7}-19 \frac{ m^8}{k^9}+20 \frac{ m^{10}}{k^{11}}\right)\right)+
\0\\ & \quad 
 + k^{10} \pi _{\mu\mu}^2 \pi _{\mu\nu} \pi _{\nu\nu}^2 \left(\frac{i }{2 \pi } \left(-\frac{9 }{256}\frac{ m}{k^2}+\frac{3 }{4}\frac{ m^3}{k^4}+\frac{53 }{8}\frac{ m^5}{k^6}-16 \frac{ m^7}{k^8}+15 \frac{ m^9}{k^{10}}\right)+
\right.\0\\ & \quad \quad \left.
 +  T \left(-\frac{9 }{2048}\frac{1}{k}+\frac{51 }{512}\frac{ m^2}{k^3}-\frac{57 }{64}\frac{ m^4}{k^5}+\frac{63 }{16}\frac{ m^6}{k^7}-\frac{69 }{8}\frac{ m^8}{k^9}+\frac{15 }{2}\frac{ m^{10}}{k^{11}}\right)\right)+
\0\\ & \quad 
 + k^8 (k\cdot\epsilon)_{\mu\nu} \pi _{\mu\nu}^4 \left(\frac{1}{\pi } \left(-\frac{15 }{16}\frac{ m^2}{k^2}+\frac{5 }{12}\frac{ m^4}{k^4}-\frac{11 }{3}\frac{ m^6}{k^6}+4 \frac{ m^8}{k^8}\right)+
\right.\0\\ & \quad \quad \left.
 + i  T \left(-\frac{1}{64}\frac{ m}{k}+\frac{1}{4}\frac{ m^3}{k^3}-\frac{3 }{2}\frac{ m^5}{k^5}+4 \frac{ m^7}{k^7}-4 \frac{ m^9}{k^9}\right)\right)+
\0\\ & \quad 
 + k^8 (k\cdot\epsilon)_{\mu\nu} \pi _{\mu\mu} \pi _{\mu\nu}^2 \pi _{\nu\nu} \left(\frac{1}{4 \pi } \left(\frac{3 }{4}\frac{ m^2}{k^2}+21 \frac{ m^4}{k^4}-44 \frac{ m^6}{k^6}+48 \frac{ m^8}{k^8}\right)+
\right.\0\\ & \quad \quad \left.
 + i  T \left(-\frac{3 }{64}\frac{ m}{k}+\frac{3 }{4}\frac{ m^3}{k^3}-\frac{9 }{2}\frac{ m^5}{k^5}+12 \frac{ m^7}{k^7}-12 \frac{ m^9}{k^9}\right)\right)+
\0\\ & \quad 
 + k^8 (k\cdot\epsilon)_{\mu\nu} \pi _{\mu\mu}^2 \pi _{\nu\nu}^2 \left(\frac{1}{2 \pi } \left(\frac{3 }{64}\frac{ m^2}{k^2}-\frac{11 }{16}\frac{ m^4}{k^4}-\frac{11 }{4}\frac{ m^6}{k^6}+3 \frac{ m^8}{k^8}\right)+
\right.\0\\ & \quad \quad \left.
 + i  T \left(-\frac{3 }{512}\frac{ m}{k}+\frac{3 }{32}\frac{ m^3}{k^3}-\frac{9 }{16}\frac{ m^5}{k^5}+\frac{3 }{2}\frac{ m^7}{k^7}-\frac{3 }{2}\frac{ m^9}{k^9}\right)\right) }
\al{\label{eq:fnc:5:5:3:}\tilde{T}_{5,5;3\text{D}}^{\text{f,nt}} & = k_{\mu }^3 k_{\nu }^3 \eta_{\mu\mu} \eta_{\nu\nu} \left(-\frac{8 i }{3 \pi } m^3\right)+ \left(k_{\mu }^2 k_{\nu }^4 \eta_{\mu\mu} \eta_{\mu\nu}+k_{\mu }^4 k_{\nu }^2 \eta_{\mu\nu} \eta_{\nu\nu}\right) \left(-\frac{8 i }{\pi } m^3\right)+
\0\\ & \quad 
 + k_{\mu }^3 k_{\nu }^3 \eta_{\mu\nu}^2 \left(\frac{16 i }{3 \pi } m^3\right)+ \left(k_{\mu } k_{\nu }^3 \eta_{\mu\mu}^2 \eta_{\nu\nu}+k_{\mu }^3 k_{\nu } \eta_{\mu\mu} \eta_{\nu\nu}^2\right) \left(-\frac{64 i }{5 \pi } m^5\right)+
\0\\ & \quad 
 +  \left(k_{\nu }^4 \eta_{\mu\mu}^2 \eta_{\mu\nu}+k_{\mu }^4 \eta_{\mu\nu} \eta_{\nu\nu}^2\right) \left(-\frac{32 i }{5 \pi } m^5\right)+
\0\\ & \quad 
 + k_{\mu }^2 k_{\nu }^2 \eta_{\mu\mu} \eta_{\mu\nu} \eta_{\nu\nu} \left(\frac{i }{5 \pi } \left(40  k^2 m^3-256  m^5\right)\right)+
\0\\ & \quad 
 +  \left(k_{\mu } k_{\nu }^3 \eta_{\mu\mu} \eta_{\mu\nu}^2+k_{\mu }^3 k_{\nu } \eta_{\mu\nu}^2 \eta_{\nu\nu}\right) \left(\frac{i }{5 \pi } \left(40  k^2 m^3-128  m^5\right)\right)+
\0\\ & \quad 
 + k_{\mu }^2 k_{\nu }^2 \eta_{\mu\nu}^3 \left(\frac{i }{3 \pi } \left(-32  k^2 m^3-\frac{128 }{5} m^5\right)\right)+
\0\\ & \quad 
 + k_{\mu } k_{\nu } \eta_{\mu\mu}^2 \eta_{\nu\nu}^2 \left(\frac{i }{5 \pi } \left(32  k^2 m^5-\frac{704 }{7} m^7\right)\right)+
\0\\ & \quad 
 +  \left(k_{\nu }^2 \eta_{\mu\mu}^2 \eta_{\mu\nu} \eta_{\nu\nu}+k_{\mu }^2 \eta_{\mu\mu} \eta_{\mu\nu} \eta_{\nu\nu}^2\right) \left(\frac{i }{5 \pi } \left(64  k^2 m^5-\frac{1408 }{7} m^7\right)\right)+
\0\\ & \quad 
 + k_{\mu } k_{\nu } \eta_{\mu\mu} \eta_{\mu\nu}^2 \eta_{\nu\nu} \left(\frac{i }{5 \pi } \left(-40  k^4 m^3+128  k^2 m^5-\frac{4096 }{7} m^7\right)\right)+
\0\\ & \quad 
 +  \left(k_{\nu }^2 \eta_{\mu\mu} \eta_{\mu\nu}^3+k_{\mu }^2 \eta_{\mu\nu}^3 \eta_{\nu\nu}\right) \left(\frac{i }{3 \pi } \left(-8  k^4 m^3+\frac{128 }{5} k^2 m^5-\frac{4096 }{35} m^7\right)\right)+
\0\\ & \quad 
 + k_{\mu } k_{\nu } \eta_{\mu\nu}^4 \left(\frac{i }{3 \pi } \left(20  k^4 m^3-\frac{512 }{7} m^7\right)\right)+
\0\\ & \quad 
 + \eta_{\mu\mu}^2 \eta_{\mu\nu} \eta_{\nu\nu}^2 \left(\frac{i }{5 \pi } \left(-32  k^4 m^5+\frac{704 }{7} k^2 m^7-\frac{1024 }{7} m^9\right)\right)+
\0\\ & \quad 
 + \eta_{\mu\mu} \eta_{\mu\nu}^3 \eta_{\nu\nu} \left(\frac{i }{3 \pi } \left(8  k^6 m^3-\frac{128 }{5} k^4 m^5+\frac{4096 }{35} k^2 m^7-\frac{22528 }{105} m^9\right)\right)+
\0\\ & \quad 
 + \eta_{\mu\nu}^5 \left(\frac{i }{3 \pi } \left(-4  k^6 m^3+\frac{512 }{35} k^2 m^7-\frac{4096 }{105} m^9\right)\right)+
\0\\ & \quad 
 + k_{\mu }^3 k_{\nu }^3 (k\cdot\epsilon)_{\mu\nu} \eta_{\mu\nu} \left(-\frac{4 }{\pi } m^2\right)+k_{\mu }^2 k_{\nu }^2 (k\cdot\epsilon)_{\mu\nu} \eta_{\mu\mu} \eta_{\nu\nu} \left(-\frac{8 }{\pi } m^4\right)+
\0\\ & \quad 
 +  \left(k_{\mu } k_{\nu }^3 (k\cdot\epsilon)_{\mu\nu} \eta_{\mu\mu} \eta_{\mu\nu}+k_{\mu }^3 k_{\nu } (k\cdot\epsilon)_{\mu\nu} \eta_{\mu\nu} \eta_{\nu\nu}\right) \left(-\frac{16 }{\pi } m^4\right)+
\0\\ & \quad 
 + k_{\mu }^2 k_{\nu }^2 (k\cdot\epsilon)_{\mu\nu} \eta_{\mu\nu}^2 \left(\frac{1}{\pi } \left(6  k^2 m^2-16  m^4\right)\right)+
\0\\ & \quad 
 +  \left(k_{\nu }^2 (k\cdot\epsilon)_{\mu\nu} \eta_{\mu\mu}^2 \eta_{\nu\nu}+k_{\mu }^2 (k\cdot\epsilon)_{\mu\nu} \eta_{\mu\mu} \eta_{\nu\nu}^2\right) \left(-\frac{32 }{5 \pi } m^6\right)+
\0\\ & \quad 
 + k_{\mu } k_{\nu } (k\cdot\epsilon)_{\mu\nu} \eta_{\mu\mu} \eta_{\mu\nu} \eta_{\nu\nu} \left(\frac{1}{5 \pi } \left(80  k^2 m^4-256  m^6\right)\right)+
\0\\ & \quad 
 +  \left(k_{\nu }^2 (k\cdot\epsilon)_{\mu\nu} \eta_{\mu\mu} \eta_{\mu\nu}^2+k_{\mu }^2 (k\cdot\epsilon)_{\mu\nu} \eta_{\mu\nu}^2 \eta_{\nu\nu}\right) \left(\frac{1}{5 \pi } \left(40  k^2 m^4-128  m^6\right)\right)+
\0\\ & \quad 
 + k_{\mu } k_{\nu } (k\cdot\epsilon)_{\mu\nu} \eta_{\mu\nu}^3 \left(\frac{1}{3 \pi } \left(-12  k^4 m^2+16  k^2 m^4-\frac{512 }{5} m^6\right)\right)+
\0\\ & \quad 
 + (k\cdot\epsilon)_{\mu\nu} \eta_{\mu\mu}^2 \eta_{\nu\nu}^2 \left(\frac{1}{5 \pi } \left(16  k^2 m^6-\frac{192 }{7} m^8\right)\right)+
\0\\ & \quad 
 + (k\cdot\epsilon)_{\mu\nu} \eta_{\mu\mu} \eta_{\mu\nu}^2 \eta_{\nu\nu} \left(\frac{1}{5 \pi } \left(-40  k^4 m^4+128  k^2 m^6-\frac{1536 }{7} m^8\right)\right)+
\0\\ & \quad 
 + (k\cdot\epsilon)_{\mu\nu} \eta_{\mu\nu}^4 \left(\frac{1}{\pi } \left(k^6 m^2-\frac{4 }{3} k^4 m^4+\frac{128 }{15} k^2 m^6-\frac{512 }{35} m^8\right)\right) }
Fermions, spin 5 x 5, dimension 4:
\al{\label{eq:fc:5:5:4:}\tilde{T}_{5,5;4\text{D}}^{\text{f,t}} & = k^{10} \pi _{\mu\nu}^5 \left(\frac{i }{7 \pi ^2} \left( \left(-\frac{23722}{571725}+\frac{2 L_0}{165}\right)+ \left(\frac{89008}{155925}-\frac{8 L_0}{45}\right)\frac{ m^2}{k^2}+
\right.\right.\0\\ & \quad \quad \quad \left.\left.
 +  \left(\frac{211289}{69300}-\frac{27 L_0}{10}\right)\frac{ m^4}{k^4}-\frac{256 }{99}\frac{ m^6}{k^6}+\frac{24064 }{1485}\frac{ m^8}{k^8}-\frac{2048 }{99}\frac{ m^{10}}{k^{10}}\right)+
\right.\0\\ & \quad \quad \left.
 + \frac{i  S}{33 \pi ^2} \left(\frac{4 }{35}\frac{1}{k}-\frac{152 }{105}\frac{ m^2}{k^3}+\frac{512 }{105}\frac{ m^4}{k^5}+\frac{256 }{35}\frac{ m^6}{k^7}-\frac{1024 }{15}\frac{ m^8}{k^9}+\frac{2048 }{21}\frac{ m^{10}}{k^{11}}\right)\right)+
\0\\ & \quad 
 + k^{10} \pi _{\mu\mu} \pi _{\mu\nu}^3 \pi _{\nu\nu} \left(\frac{i }{\pi ^2} \left( \left(-\frac{83338}{12006225}+\frac{8 L_0}{3465}\right)+ \left(\frac{13004}{1091475}-\frac{4 L_0}{315}\right)\frac{ m^2}{k^2}+
\right.\right.\0\\ & \quad \quad \quad \left.\left.
 +  \left(-\frac{99289}{242550}+\frac{27 L_0}{35}\right)\frac{ m^4}{k^4}-\frac{55808 }{10395}\frac{ m^6}{k^6}+\frac{22016 }{1485}\frac{ m^8}{k^8}-\frac{10240 }{693}\frac{ m^{10}}{k^{10}}\right)+
\right.\0\\ & \quad \quad \left.
 + \frac{i  S}{99 \pi ^2} \left(\frac{16 }{35}\frac{1}{k}-\frac{8 }{5}\frac{ m^2}{k^3}-\frac{1664 }{35}\frac{ m^4}{k^5}+\frac{15104 }{35}\frac{ m^6}{k^7}-\frac{47104 }{35}\frac{ m^8}{k^9}+
\right.\right.\0\\ & \quad \quad \quad \left.\left.
 + \frac{10240 }{7}\frac{ m^{10}}{k^{11}}\right)\right)+
\0\\ & \quad 
 + k^{10} \pi _{\mu\mu}^2 \pi _{\mu\nu} \pi _{\nu\nu}^2 \left(\frac{i }{\pi ^2} \left( \left(\frac{13511}{2286900}-\frac{L_0}{660}\right)+ \left(-\frac{52379}{363825}+\frac{4 L_0}{105}\right)\frac{ m^2}{k^2}+
\right.\right.\0\\ & \quad \quad \quad \left.\left.
 +  \left(\frac{227603}{161700}-\frac{27 L_0}{70}\right)\frac{ m^4}{k^4}-\frac{11552 }{3465}\frac{ m^6}{k^6}+\frac{23488 }{3465}\frac{ m^8}{k^8}-\frac{1280 }{231}\frac{ m^{10}}{k^{10}}\right)+
\right.\0\\ & \quad \quad \left.
 + \frac{i  S}{11 \pi ^2} \left(-\frac{1}{30}\frac{1}{k}+\frac{27 }{35}\frac{ m^2}{k^3}-\frac{736 }{105}\frac{ m^4}{k^5}+\frac{3296 }{105}\frac{ m^6}{k^7}-\frac{2432 }{35}\frac{ m^8}{k^9}+\frac{1280 }{21}\frac{ m^{10}}{k^{11}}\right)\right) }
\al{\label{eq:fnc:5:5:4:}\tilde{T}_{5,5;4\text{D}}^{\text{f,nt}} & = k_{\mu }^3 k_{\nu }^3 \eta_{\mu\mu} \eta_{\nu\nu} \left(\frac{i  L_2}{\pi ^2} m^4\right)+ \left(k_{\mu }^2 k_{\nu }^4 \eta_{\mu\mu} \eta_{\mu\nu}+k_{\mu }^4 k_{\nu }^2 \eta_{\mu\nu} \eta_{\nu\nu}\right) \left(\frac{3 i  L_2}{\pi ^2} m^4\right)+
\0\\ & \quad 
 + k_{\mu }^3 k_{\nu }^3 \eta_{\mu\nu}^2 \left(-\frac{2 i  L_2}{\pi ^2} m^4\right)+ \left(k_{\mu } k_{\nu }^3 \eta_{\mu\mu}^2 \eta_{\nu\nu}+k_{\mu }^3 k_{\nu } \eta_{\mu\mu} \eta_{\nu\nu}^2\right) \left(\frac{4 i  L_3}{\pi ^2} m^6\right)+
\0\\ & \quad 
 +  \left(k_{\nu }^4 \eta_{\mu\mu}^2 \eta_{\mu\nu}+k_{\mu }^4 \eta_{\mu\nu} \eta_{\nu\nu}^2\right) \left(\frac{2 i  L_3}{\pi ^2} m^6\right)+
\0\\ & \quad 
 + k_{\mu }^2 k_{\nu }^2 \eta_{\mu\mu} \eta_{\mu\nu} \eta_{\nu\nu} \left(\frac{i }{\pi ^2} \left(-3  L_2 k^2 m^4+16  L_3 m^6\right)\right)+
\0\\ & \quad 
 +  \left(k_{\mu } k_{\nu }^3 \eta_{\mu\mu} \eta_{\mu\nu}^2+k_{\mu }^3 k_{\nu } \eta_{\mu\nu}^2 \eta_{\nu\nu}\right) \left(\frac{i }{\pi ^2} \left(-3  L_2 k^2 m^4+8  L_3 m^6\right)\right)+
\0\\ & \quad 
 + k_{\mu }^2 k_{\nu }^2 \eta_{\mu\nu}^3 \left(\frac{i }{3 \pi ^2} \left(12  L_2 k^2 m^4+8  L_3 m^6\right)\right)+
\0\\ & \quad 
 + k_{\mu } k_{\nu } \eta_{\mu\mu}^2 \eta_{\nu\nu}^2 \left(\frac{i }{2 \pi ^2} \left(-4  L_3 k^2 m^6+11  L_4 m^8\right)\right)+
\0\\ & \quad 
 +  \left(k_{\nu }^2 \eta_{\mu\mu}^2 \eta_{\mu\nu} \eta_{\nu\nu}+k_{\mu }^2 \eta_{\mu\mu} \eta_{\mu\nu} \eta_{\nu\nu}^2\right) \left(\frac{i }{\pi ^2} \left(-4  L_3 k^2 m^6+11  L_4 m^8\right)\right)+
\0\\ & \quad 
 + k_{\mu } k_{\nu } \eta_{\mu\mu} \eta_{\mu\nu}^2 \eta_{\nu\nu} \left(\frac{i }{\pi ^2} \left(3  L_2 k^4 m^4-8  L_3 k^2 m^6+32  L_4 m^8\right)\right)+
\0\\ & \quad 
 +  \left(k_{\nu }^2 \eta_{\mu\mu} \eta_{\mu\nu}^3+k_{\mu }^2 \eta_{\mu\nu}^3 \eta_{\nu\nu}\right) \left(\frac{i }{3 \pi ^2} \left(3  L_2 k^4 m^4-8  L_3 k^2 m^6+32  L_4 m^8\right)\right)+
\0\\ & \quad 
 + k_{\mu } k_{\nu } \eta_{\mu\nu}^4 \left(\frac{i }{\pi ^2} \left(-\frac{5  L_2}{2} k^4 m^4+\frac{20  L_4}{3} m^8\right)\right)+
\0\\ & \quad 
 + \eta_{\mu\mu}^2 \eta_{\mu\nu} \eta_{\nu\nu}^2 \left(\frac{i }{\pi ^2} \left(2  L_3 k^4 m^6-\frac{11  L_4}{2} k^2 m^8+\frac{36  L_5}{5} m^{10}\right)\right)+
\0\\ & \quad 
 + \eta_{\mu\mu} \eta_{\mu\nu}^3 \eta_{\nu\nu} \left(\frac{i }{\pi ^2} \left(- L_2 k^6 m^4+\frac{8  L_3}{3} k^4 m^6-\frac{32  L_4}{3} k^2 m^8+\frac{88  L_5}{5} m^{10}\right)\right)+
\0\\ & \quad 
 + \eta_{\mu\nu}^5 \left(\frac{i }{\pi ^2} \left(\frac{ L_2}{2} k^6 m^4-\frac{4  L_4}{3} k^2 m^8+\frac{16  L_5}{5} m^{10}\right)\right) }
Fermions, spin 5 x 5, dimension 5:
\al{\label{eq:fc:5:5:5:}\tilde{T}_{5,5;5\text{D}}^{\text{f,t}} & = k^{10} \pi _{\mu\nu}^5 \left(\frac{i }{\pi ^2} \left(\frac{7 }{15360} m-\frac{83 }{11520}\frac{ m^3}{k^2}-\frac{183 }{800}\frac{ m^5}{k^4}-\frac{1}{24}\frac{ m^7}{k^6}+\frac{49 }{180}\frac{ m^9}{k^8}-\frac{1}{3}\frac{ m^{11}}{k^{10}}\right)+
\right.\0\\ & \quad \quad \left.
 + \frac{ T}{\pi } \left(\frac{7 }{61440} k-\frac{1}{512}\frac{ m^2}{k}+\frac{3 }{256}\frac{ m^4}{k^3}-\frac{1}{48}\frac{ m^6}{k^5}-\frac{1}{16}\frac{ m^8}{k^7}+\frac{3 }{10}\frac{ m^{10}}{k^9}-
\right.\right.\0\\ & \quad \quad \quad \left.\left.
 - \frac{1}{3}\frac{ m^{12}}{k^{11}}\right)\right)+
\0\\ & \quad 
 + k^{10} \pi _{\mu\mu} \pi _{\mu\nu}^3 \pi _{\nu\nu} \left(\frac{i }{\pi ^2} \left(\frac{11 }{15360} m-\frac{79 }{11520}\frac{ m^3}{k^2}+\frac{1223 }{2400}\frac{ m^5}{k^4}-\frac{27 }{40}\frac{ m^7}{k^6}+\frac{317 }{180}\frac{ m^9}{k^8}-
\right.\right.\0\\ & \quad \quad \quad \left.\left.
 - \frac{5 }{3}\frac{ m^{11}}{k^{10}}\right)+
\right.\0\\ & \quad \quad \left.
 + \frac{ T}{\pi } \left(\frac{11 }{61440} k-\frac{1}{512}\frac{ m^2}{k}-\frac{1}{256}\frac{ m^4}{k^3}+\frac{7 }{48}\frac{ m^6}{k^5}-\frac{13 }{16}\frac{ m^8}{k^7}+\frac{19 }{10}\frac{ m^{10}}{k^9}-
\right.\right.\0\\ & \quad \quad \quad \left.\left.
 - \frac{5 }{3}\frac{ m^{12}}{k^{11}}\right)\right)+
\0\\ & \quad 
 + k^{10} \pi _{\mu\mu}^2 \pi _{\mu\nu} \pi _{\nu\nu}^2 \left(\frac{i }{8 \pi ^2} \left(-\frac{13 }{5120} m+\frac{257 }{3840}\frac{ m^3}{k^2}-\frac{569 }{800}\frac{ m^5}{k^4}-\frac{137 }{40}\frac{ m^7}{k^6}+
\right.\right.\0\\ & \quad \quad \quad \left.\left.
 + \frac{389 }{60}\frac{ m^9}{k^8}-5 \frac{ m^{11}}{k^{10}}\right)+
\right.\0\\ & \quad \quad \left.
 + \frac{ T}{8 \pi } \left(-\frac{13 }{20480} k+\frac{9 }{512}\frac{ m^2}{k}-\frac{51 }{256}\frac{ m^4}{k^3}+\frac{19 }{16}\frac{ m^6}{k^5}-\frac{63 }{16}\frac{ m^8}{k^7}+\frac{69 }{10}\frac{ m^{10}}{k^9}-
\right.\right.\0\\ & \quad \quad \quad \left.\left.
 - 5 \frac{ m^{12}}{k^{11}}\right)\right) }
\al{\label{eq:fnc:5:5:5:}\tilde{T}_{5,5;5\text{D}}^{\text{f,nt}} & = k_{\mu }^3 k_{\nu }^3 \eta_{\mu\mu} \eta_{\nu\nu} \left(\frac{8 i }{15 \pi ^2} m^5\right)+ \left(k_{\mu }^2 k_{\nu }^4 \eta_{\mu\mu} \eta_{\mu\nu}+k_{\mu }^4 k_{\nu }^2 \eta_{\mu\nu} \eta_{\nu\nu}\right) \left(\frac{8 i }{5 \pi ^2} m^5\right)+
\0\\ & \quad 
 + k_{\mu }^3 k_{\nu }^3 \eta_{\mu\nu}^2 \left(-\frac{16 i }{15 \pi ^2} m^5\right)+ \left(k_{\mu } k_{\nu }^3 \eta_{\mu\mu}^2 \eta_{\nu\nu}+k_{\mu }^3 k_{\nu } \eta_{\mu\mu} \eta_{\nu\nu}^2\right) \left(\frac{64 i }{35 \pi ^2} m^7\right)+
\0\\ & \quad 
 +  \left(k_{\nu }^4 \eta_{\mu\mu}^2 \eta_{\mu\nu}+k_{\mu }^4 \eta_{\mu\nu} \eta_{\nu\nu}^2\right) \left(\frac{32 i }{35 \pi ^2} m^7\right)+
\0\\ & \quad 
 + k_{\mu }^2 k_{\nu }^2 \eta_{\mu\mu} \eta_{\mu\nu} \eta_{\nu\nu} \left(\frac{i }{5 \pi ^2} \left(-8  k^2 m^5+\frac{256 }{7} m^7\right)\right)+
\0\\ & \quad 
 +  \left(k_{\mu } k_{\nu }^3 \eta_{\mu\mu} \eta_{\mu\nu}^2+k_{\mu }^3 k_{\nu } \eta_{\mu\nu}^2 \eta_{\nu\nu}\right) \left(\frac{i }{5 \pi ^2} \left(-8  k^2 m^5+\frac{128 }{7} m^7\right)\right)+
\0\\ & \quad 
 + k_{\mu }^2 k_{\nu }^2 \eta_{\mu\nu}^3 \left(\frac{i }{15 \pi ^2} \left(32  k^2 m^5+\frac{128 }{7} m^7\right)\right)+
\0\\ & \quad 
 + k_{\mu } k_{\nu } \eta_{\mu\mu}^2 \eta_{\nu\nu}^2 \left(\frac{i }{35 \pi ^2} \left(-32  k^2 m^7+\frac{704 }{9} m^9\right)\right)+
\0\\ & \quad 
 +  \left(k_{\nu }^2 \eta_{\mu\mu}^2 \eta_{\mu\nu} \eta_{\nu\nu}+k_{\mu }^2 \eta_{\mu\mu} \eta_{\mu\nu} \eta_{\nu\nu}^2\right) \left(\frac{i }{35 \pi ^2} \left(-64  k^2 m^7+\frac{1408 }{9} m^9\right)\right)+
\0\\ & \quad 
 + k_{\mu } k_{\nu } \eta_{\mu\mu} \eta_{\mu\nu}^2 \eta_{\nu\nu} \left(\frac{i }{5 \pi ^2} \left(8  k^4 m^5-\frac{128 }{7} k^2 m^7+\frac{4096 }{63} m^9\right)\right)+
\0\\ & \quad 
 +  \left(k_{\nu }^2 \eta_{\mu\mu} \eta_{\mu\nu}^3+k_{\mu }^2 \eta_{\mu\nu}^3 \eta_{\nu\nu}\right) \left(\frac{i }{15 \pi ^2} \left(8  k^4 m^5-\frac{128 }{7} k^2 m^7+\frac{4096 }{63} m^9\right)\right)+
\0\\ & \quad 
 + k_{\mu } k_{\nu } \eta_{\mu\nu}^4 \left(\frac{i }{3 \pi ^2} \left(-4  k^4 m^5+\frac{512 }{63} m^9\right)\right)+
\0\\ & \quad 
 + \eta_{\mu\mu}^2 \eta_{\mu\nu} \eta_{\nu\nu}^2 \left(\frac{i }{35 \pi ^2} \left(32  k^4 m^7-\frac{704 }{9} k^2 m^9+\frac{1024 }{11} m^{11}\right)\right)+
\0\\ & \quad 
 + \eta_{\mu\mu} \eta_{\mu\nu}^3 \eta_{\nu\nu} \left(\frac{i }{15 \pi ^2} \left(-8  k^6 m^5+\frac{128 }{7} k^4 m^7-\frac{4096 }{63} k^2 m^9+\frac{2048 }{21} m^{11}\right)\right)+
\0\\ & \quad 
 + \eta_{\mu\nu}^5 \left(\frac{i }{15 \pi ^2} \left(4  k^6 m^5-\frac{512 }{63} k^2 m^9+\frac{4096 }{231} m^{11}\right)\right) }
Fermions, spin 5 x 5, dimension 6:
\al{\label{eq:fc:5:5:6:}\tilde{T}_{5,5;6\text{D}}^{\text{f,t}} & = k^{10} \pi _{\mu\nu}^5 \left(\frac{i }{\pi ^3} \left( \left(-\frac{659507}{4058104050}+\frac{2 L_0}{45045}\right) k^2+ \left(\frac{158813}{52026975}-\frac{L_0}{1155}\right) m^2+
\right.\right.\0\\ & \quad \quad \quad \left.\left.
 +  \left(-\frac{299671}{14189175}+\frac{2 L_0}{315}\right)\frac{ m^4}{k^2}+ \left(-\frac{3691967}{37837800}+\frac{9 L_0}{140}\right)\frac{ m^6}{k^4}+\frac{128 }{3003}\frac{ m^8}{k^6}-
\right.\right.\0\\ & \quad \quad \quad \left.\left.
 - \frac{4096 }{19305}\frac{ m^{10}}{k^8}+\frac{2048 }{9009}\frac{ m^{12}}{k^{10}}\right)+
\right.\0\\ & \quad \quad \left.
 + \frac{i  S}{39 \pi ^3} \left(\frac{4 }{1155} k-\frac{2 }{33}\frac{ m^2}{k}+\frac{8 }{21}\frac{ m^4}{k^3}-\frac{64 }{77}\frac{ m^6}{k^5}-\frac{256 }{231}\frac{ m^8}{k^7}+\frac{8704 }{1155}\frac{ m^{10}}{k^9}-
\right.\right.\0\\ & \quad \quad \quad \left.\left.
 - \frac{2048 }{231}\frac{ m^{12}}{k^{11}}\right)\right)+
\0\\ & \quad 
 + k^{10} \pi _{\mu\mu} \pi _{\mu\nu}^3 \pi _{\nu\nu} \left(\frac{i }{\pi ^3} \left( \left(-\frac{156833}{579729150}+\frac{L_0}{12870}\right) k^2+
\right.\right.\0\\ & \quad \quad \quad \left.\left.
 +  \left(\frac{565952}{156080925}-\frac{4 L_0}{3465}\right) m^2+ \left(-\frac{10061}{2579850}+\frac{L_0}{315}\right)\frac{ m^4}{k^2}+
\right.\right.\0\\ & \quad \quad \quad \left.\left.
 +  \left(\frac{2347967}{18918900}-\frac{9 L_0}{70}\right)\frac{ m^6}{k^4}+\frac{82048 }{135135}\frac{ m^8}{k^6}-\frac{183296 }{135135}\frac{ m^{10}}{k^8}+\frac{10240 }{9009}\frac{ m^{12}}{k^{10}}\right)+
\right.\0\\ & \quad \quad \left.
 + \frac{i  S}{13 \pi ^3} \left(\frac{1}{495} k-\frac{2 }{77}\frac{ m^2}{k}+\frac{8 }{231}\frac{ m^4}{k^3}+\frac{64 }{63}\frac{ m^6}{k^5}-\frac{512 }{77}\frac{ m^8}{k^7}+\frac{18944 }{1155}\frac{ m^{10}}{k^9}-
\right.\right.\0\\ & \quad \quad \quad \left.\left.
 - \frac{10240 }{693}\frac{ m^{12}}{k^{11}}\right)\right)+
\0\\ & \quad 
 + k^{10} \pi _{\mu\mu}^2 \pi _{\mu\nu} \pi _{\nu\nu}^2 \left(\frac{i }{\pi ^3} \left( \left(\frac{367291}{3607203600}-\frac{L_0}{40040}\right) k^2+
\right.\right.\0\\ & \quad \quad \quad \left.\left.
 +  \left(-\frac{178613}{59459400}+\frac{L_0}{1320}\right) m^2+ \left(\frac{346921}{9459450}-\frac{L_0}{105}\right)\frac{ m^4}{k^2}+
\right.\right.\0\\ & \quad \quad \quad \left.\left.
 +  \left(-\frac{91573}{382200}+\frac{9 L_0}{140}\right)\frac{ m^6}{k^4}+\frac{2416 }{6435}\frac{ m^8}{k^6}-\frac{27904 }{45045}\frac{ m^{10}}{k^8}+\frac{1280 }{3003}\frac{ m^{12}}{k^{10}}\right)+
\right.\0\\ & \quad \quad \left.
 + \frac{i  S}{143 \pi ^3} \left(-\frac{1}{140} k+\frac{17 }{84}\frac{ m^2}{k}-\frac{7 }{3}\frac{ m^4}{k^3}+\frac{296 }{21}\frac{ m^6}{k^5}-\frac{992 }{21}\frac{ m^8}{k^7}+\frac{8768 }{105}\frac{ m^{10}}{k^9}-
\right.\right.\0\\ & \quad \quad \quad \left.\left.
 - \frac{1280 }{21}\frac{ m^{12}}{k^{11}}\right)\right) }
\al{\label{eq:fnc:5:5:6:}\tilde{T}_{5,5;6\text{D}}^{\text{f,nt}} & = k_{\mu }^3 k_{\nu }^3 \eta_{\mu\mu} \eta_{\nu\nu} \left(-\frac{i  L_3}{6 \pi ^3} m^6\right)+ \left(k_{\mu }^2 k_{\nu }^4 \eta_{\mu\mu} \eta_{\mu\nu}+k_{\mu }^4 k_{\nu }^2 \eta_{\mu\nu} \eta_{\nu\nu}\right) \left(-\frac{i  L_3}{2 \pi ^3} m^6\right)+
\0\\ & \quad 
 + k_{\mu }^3 k_{\nu }^3 \eta_{\mu\nu}^2 \left(\frac{i  L_3}{3 \pi ^3} m^6\right)+ \left(k_{\mu } k_{\nu }^3 \eta_{\mu\mu}^2 \eta_{\nu\nu}+k_{\mu }^3 k_{\nu } \eta_{\mu\mu} \eta_{\nu\nu}^2\right) \left(-\frac{i  L_4}{2 \pi ^3} m^8\right)+
\0\\ & \quad 
 +  \left(k_{\nu }^4 \eta_{\mu\mu}^2 \eta_{\mu\nu}+k_{\mu }^4 \eta_{\mu\nu} \eta_{\nu\nu}^2\right) \left(-\frac{i  L_4}{4 \pi ^3} m^8\right)+
\0\\ & \quad 
 + k_{\mu }^2 k_{\nu }^2 \eta_{\mu\mu} \eta_{\mu\nu} \eta_{\nu\nu} \left(\frac{i }{2 \pi ^3} \left( L_3 k^2 m^6-4  L_4 m^8\right)\right)+
\0\\ & \quad 
 +  \left(k_{\mu } k_{\nu }^3 \eta_{\mu\mu} \eta_{\mu\nu}^2+k_{\mu }^3 k_{\nu } \eta_{\mu\nu}^2 \eta_{\nu\nu}\right) \left(\frac{i }{2 \pi ^3} \left( L_3 k^2 m^6-2  L_4 m^8\right)\right)+
\0\\ & \quad 
 + k_{\mu }^2 k_{\nu }^2 \eta_{\mu\nu}^3 \left(\frac{i }{3 \pi ^3} \left(-2  L_3 k^2 m^6- L_4 m^8\right)\right)+
\0\\ & \quad 
 + k_{\mu } k_{\nu } \eta_{\mu\mu}^2 \eta_{\nu\nu}^2 \left(\frac{i }{4 \pi ^3} \left( L_4 k^2 m^8-\frac{11  L_5}{5} m^{10}\right)\right)+
\0\\ & \quad 
 +  \left(k_{\nu }^2 \eta_{\mu\mu}^2 \eta_{\mu\nu} \eta_{\nu\nu}+k_{\mu }^2 \eta_{\mu\mu} \eta_{\mu\nu} \eta_{\nu\nu}^2\right) \left(\frac{i }{2 \pi ^3} \left( L_4 k^2 m^8-\frac{11  L_5}{5} m^{10}\right)\right)+
\0\\ & \quad 
 + k_{\mu } k_{\nu } \eta_{\mu\mu} \eta_{\mu\nu}^2 \eta_{\nu\nu} \left(\frac{i }{\pi ^3} \left(-\frac{ L_3}{2} k^4 m^6+ L_4 k^2 m^8-\frac{16  L_5}{5} m^{10}\right)\right)+
\0\\ & \quad 
 +  \left(k_{\nu }^2 \eta_{\mu\mu} \eta_{\mu\nu}^3+k_{\mu }^2 \eta_{\mu\nu}^3 \eta_{\nu\nu}\right) \left(\frac{i }{3 \pi ^3} \left(-\frac{ L_3}{2} k^4 m^6+ L_4 k^2 m^8-\frac{16  L_5}{5} m^{10}\right)\right)+
\0\\ & \quad 
 + k_{\mu } k_{\nu } \eta_{\mu\nu}^4 \left(\frac{i }{3 \pi ^3} \left(\frac{5  L_3}{4} k^4 m^6-2  L_5 m^{10}\right)\right)+
\0\\ & \quad 
 + \eta_{\mu\mu}^2 \eta_{\mu\nu} \eta_{\nu\nu}^2 \left(\frac{i }{\pi ^3} \left(-\frac{ L_4}{4} k^4 m^8+\frac{11  L_5}{20} k^2 m^{10}-\frac{3  L_6}{5} m^{12}\right)\right)+
\0\\ & \quad 
 + \eta_{\mu\mu} \eta_{\mu\nu}^3 \eta_{\nu\nu} \left(\frac{i }{3 \pi ^3} \left(\frac{ L_3}{2} k^6 m^6- L_4 k^4 m^8+\frac{16  L_5}{5} k^2 m^{10}-\frac{22  L_6}{5} m^{12}\right)\right)+
\0\\ & \quad 
 + \eta_{\mu\nu}^5 \left(\frac{i }{3 \pi ^3} \left(-\frac{ L_3}{4} k^6 m^6+\frac{2  L_5}{5} k^2 m^{10}-\frac{4  L_6}{5} m^{12}\right)\right) }
\subsection{Expansions in UV and IR for fermions}
Fermions, spin 0 x 0, dimension 3:
\al{\label{eq:fc:0:0:3:uv}\tilde{T}_{0,0;3\text{D}}^{\text{f,UV}} & = -\frac{1}{8} k+\frac{1}{2}\frac{ m^2}{k}-\frac{4 i }{3 \pi }\frac{ m^3}{k^2}-\frac{16 i }{15 \pi }\frac{ m^5}{k^4}-\frac{64 i }{35 \pi }\frac{ m^7}{k^6}-\frac{256 i }{63 \pi }\frac{ m^9}{k^8}-\frac{1024 i }{99 \pi }\frac{ m^{11}}{k^{10}}+\ldots }
\al{\label{eq:fc:0:0:3:ir}\tilde{T}_{0,0;3\text{D}}^{\text{f,IR}} & = \frac{i }{4 \pi } \left(-4  m+\frac{1}{3}\frac{ k^2}{m}+\frac{1}{60}\frac{ k^4}{m^3}+\frac{1}{560}\frac{ k^6}{m^5}+\frac{1}{4032}\frac{ k^8}{m^7}+\frac{1}{25344}\frac{ k^{10}}{m^9}+\frac{1}{146432}\frac{ k^{12}}{m^{11}}+
\right.\0\\ & \quad \quad \left.
 + \ldots\right) }
\al{\label{eq:fc:0:0:3:uvir}\tilde{T}_{0,0;3\text{D}}^{\text{f,UV-IR}} & =  \ldots \textrm{(i.e.\ no overlap)} }
Fermions, spin 0 x 0, dimension 4:
\al{\label{eq:fc:0:0:4:uv}\tilde{T}_{0,0;4\text{D}}^{\text{f,UV}} & = \frac{i }{2 \pi ^2} \left( \left(\frac{1}{2}-\frac{P}{4}\right) k^2+ \left(-2+\frac{3 P}{2}\right) m^2- \left(\frac{9}{4}+\frac{3 K}{2}\right)\frac{ m^4}{k^2}+ \left(\frac{1}{6}-K\right)\frac{ m^6}{k^4}+
\right.\0\\ & \quad \quad \left.
 +  \left(\frac{7}{8}-\frac{3 K}{2}\right)\frac{ m^8}{k^6}+ \left(\frac{47}{20}-3 K\right)\frac{ m^{10}}{k^8}+ \left(\frac{379}{60}-7 K\right)\frac{ m^{12}}{k^{10}}+\ldots\right) }
\al{\label{eq:fc:0:0:4:ir}\tilde{T}_{0,0;4\text{D}}^{\text{f,IR}} & = \frac{i }{4 \pi ^2} \left( \left(-1+3 L_0\right) m^2- \left(\frac{1}{3}+\frac{L_0}{2}\right) k^2+\frac{1}{20}\frac{ k^4}{m^2}+\frac{1}{280}\frac{ k^6}{m^4}+\frac{1}{2520}\frac{ k^8}{m^6}+
\right.\0\\ & \quad \quad \left.
 + \frac{1}{18480}\frac{ k^{10}}{m^8}+\frac{1}{120120}\frac{ k^{12}}{m^{10}}+\frac{1}{720720}\frac{ k^{14}}{m^{12}}+\ldots\right) }
\al{\label{eq:fc:0:0:4:uvir}\tilde{T}_{0,0;4\text{D}}^{\text{f,UV-IR}} & = \frac{i }{\pi ^2} \left( \left(\frac{1}{3}-\frac{K}{8}\right) k^2+ \left(-\frac{3}{4}+\frac{3 K}{4}\right) m^2\right)+\ldots }
Fermions, spin 0 x 0, dimension 5:
\al{\label{eq:fc:0:0:5:uv}\tilde{T}_{0,0;5\text{D}}^{\text{f,UV}} & = \frac{1}{\pi ^2} \left(-\frac{ \pi }{64} k^3+\frac{ \pi }{8} k m^2-\frac{ \pi }{4}\frac{ m^4}{k}+\frac{8 i }{15}\frac{ m^5}{k^2}+\frac{32 i }{105}\frac{ m^7}{k^4}+\frac{128 i }{315}\frac{ m^9}{k^6}+\frac{512 i }{693}\frac{ m^{11}}{k^8}+\ldots\right) }
\al{\label{eq:fc:0:0:5:ir}\tilde{T}_{0,0;5\text{D}}^{\text{f,IR}} & = \frac{i }{3 \pi ^2} \left(2  m^3-\frac{1}{2} k^2 m+\frac{1}{40}\frac{ k^4}{m}+\frac{1}{1120}\frac{ k^6}{m^3}+\frac{1}{13440}\frac{ k^8}{m^5}+\frac{1}{118272}\frac{ k^{10}}{m^7}+
\right.\0\\ & \quad \quad \left.
 + \frac{1}{878592}\frac{ k^{12}}{m^9}+\frac{1}{5857280}\frac{ k^{14}}{m^{11}}+\ldots\right) }
\al{\label{eq:fc:0:0:5:uvir}\tilde{T}_{0,0;5\text{D}}^{\text{f,UV-IR}} & =  \ldots \textrm{(i.e.\ no overlap)} }
Fermions, spin 0 x 0, dimension 6:
\al{\label{eq:fc:0:0:6:uv}\tilde{T}_{0,0;6\text{D}}^{\text{f,UV}} & = \frac{i }{2 \pi ^3} \left( \left(\frac{1}{18}-\frac{P}{48}\right) k^4+ \left(-\frac{17}{36}+\frac{5 P}{24}\right) k^2 m^2+ \left(1-\frac{5 P}{8}\right) m^4+
\right.\0\\ & \quad \quad \left.
 +  \left(\frac{55}{72}+\frac{5 K}{12}\right)\frac{ m^6}{k^2}+ \left(\frac{5}{288}+\frac{5 K}{24}\right)\frac{ m^8}{k^4}+ \left(-\frac{23}{240}+\frac{K}{4}\right)\frac{ m^{10}}{k^6}+
\right.\0\\ & \quad \quad \left.
 +  \left(-\frac{37}{144}+\frac{5 K}{12}\right)\frac{ m^{12}}{k^8}+\ldots\right) }
\al{\label{eq:fc:0:0:6:ir}\tilde{T}_{0,0;6\text{D}}^{\text{f,IR}} & = \frac{i }{16 \pi ^3} \left( \left(\frac{11}{2}-5 L_0\right) m^4+ \left(-1+\frac{5 L_0}{3}\right) k^2 m^2- \left(\frac{1}{15}+\frac{L_0}{6}\right) k^4+\frac{1}{84}\frac{ k^6}{m^2}+
\right.\0\\ & \quad \quad \left.
 + \frac{1}{1512}\frac{ k^8}{m^4}+\frac{1}{16632}\frac{ k^{10}}{m^6}+\frac{1}{144144}\frac{ k^{12}}{m^8}+\frac{1}{1081080}\frac{ k^{14}}{m^{10}}+\frac{1}{7351344}\frac{ k^{16}}{m^{12}}+
\right.\0\\ & \quad \quad \left.
 + \ldots\right) }
\al{\label{eq:fc:0:0:6:uvir}\tilde{T}_{0,0;6\text{D}}^{\text{f,UV-IR}} & = \frac{i }{16 \pi ^3} \left( \left(\frac{23}{45}-\frac{K}{6}\right) k^4+ \left(-\frac{25}{9}+\frac{5 K}{3}\right) k^2 m^2+ \left(\frac{5}{2}-5 K\right) m^4\right)+\ldots }
Fermions, spin 0 x 2, dimension 3:
\al{\label{eq:fc:0:2:3:uv}\tilde{T}_{0,2;3\text{D}}^{\text{f,t,UV}} & = k^2 \pi _{\nu\nu} \left(-\frac{1}{8}\frac{ m}{k}+\frac{i }{\pi }\frac{ m^2}{k^2}+\frac{1}{2}\frac{ m^3}{k^3}-\frac{4 i }{3 \pi }\frac{ m^4}{k^4}-\frac{16 i }{15 \pi }\frac{ m^6}{k^6}-\frac{64 i }{35 \pi }\frac{ m^8}{k^8}-\frac{256 i }{63 \pi }\frac{ m^{10}}{k^{10}}-
\right.\0\\ & \quad \quad \left.
 - \frac{1024 i }{99 \pi }\frac{ m^{12}}{k^{12}}+\ldots\right) }
\al{\label{eq:fc:0:2:3:ir}\tilde{T}_{0,2;3\text{D}}^{\text{f,t,IR}} & = k^2 \pi _{\nu\nu} \left(\frac{i }{4 \pi } \left(\frac{1}{3}+\frac{1}{60}\frac{ k^2}{m^2}+\frac{1}{560}\frac{ k^4}{m^4}+\frac{1}{4032}\frac{ k^6}{m^6}+\frac{1}{25344}\frac{ k^8}{m^8}+\frac{1}{146432}\frac{ k^{10}}{m^{10}}+
\right.\right.\0\\ & \quad \quad \quad \left.\left.
 + \frac{1}{798720}\frac{ k^{12}}{m^{12}}+\ldots\right)\right) }
\al{\label{eq:fc:0:2:3:uvir}\tilde{T}_{0,2;3\text{D}}^{\text{f,UV-IR}} & =  \ldots \textrm{(i.e.\ no overlap)} }
Fermions, spin 0 x 2, dimension 4:
\al{\label{eq:fc:0:2:4:uv}\tilde{T}_{0,2;4\text{D}}^{\text{f,t,UV}} & = k^2 \pi _{\nu\nu} \left(\frac{i }{\pi ^2} \left( \left(\frac{2}{9}-\frac{P}{12}\right) m+ \left(-\frac{1}{2}+\frac{K}{2}\right)\frac{ m^3}{k^2}- \left(\frac{3}{4}+\frac{K}{2}\right)\frac{ m^5}{k^4}+
\right.\right.\0\\ & \quad \quad \quad \left.\left.
 +  \left(\frac{1}{18}-\frac{K}{3}\right)\frac{ m^7}{k^6}+ \left(\frac{7}{24}-\frac{K}{2}\right)\frac{ m^9}{k^8}+ \left(\frac{47}{60}-K\right)\frac{ m^{11}}{k^{10}}+\ldots\right)\right) }
\al{\label{eq:fc:0:2:4:ir}\tilde{T}_{0,2;4\text{D}}^{\text{f,t,IR}} & = k^2 \pi _{\nu\nu} \left(\frac{i }{12 \pi ^2} \left(- L_0 m+\frac{1}{10}\frac{ k^2}{m}+\frac{1}{140}\frac{ k^4}{m^3}+\frac{1}{1260}\frac{ k^6}{m^5}+\frac{1}{9240}\frac{ k^8}{m^7}+\frac{1}{60060}\frac{ k^{10}}{m^9}+
\right.\right.\0\\ & \quad \quad \quad \left.\left.
 + \frac{1}{360360}\frac{ k^{12}}{m^{11}}+\ldots\right)\right) }
\al{\label{eq:fc:0:2:4:uvir}\tilde{T}_{0,2;4\text{D}}^{\text{f,UV-IR}} & = k^2 \pi _{\nu\nu} \left(\frac{i  }{3 \pi ^2}\left(\frac{2}{3}-\frac{K}{4}\right) m\right)+\ldots }
Fermions, spin 0 x 2, dimension 5:
\al{\label{eq:fc:0:2:5:uv}\tilde{T}_{0,2;5\text{D}}^{\text{f,t,UV}} & = k^2 \pi _{\nu\nu} \left(\frac{1}{\pi ^2} \left(-\frac{ \pi }{128} k m+\frac{ \pi }{16}\frac{ m^3}{k}-\frac{i }{3}\frac{ m^4}{k^2}-\frac{ \pi }{8}\frac{ m^5}{k^3}+\frac{4 i }{15}\frac{ m^6}{k^4}+\frac{16 i }{105}\frac{ m^8}{k^6}+
\right.\right.\0\\ & \quad \quad \quad \left.\left.
 + \frac{64 i }{315}\frac{ m^{10}}{k^8}+\frac{256 i }{693}\frac{ m^{12}}{k^{10}}+\ldots\right)\right) }
\al{\label{eq:fc:0:2:5:ir}\tilde{T}_{0,2;5\text{D}}^{\text{f,t,IR}} & = k^2 \pi _{\nu\nu} \left(\frac{i }{12 \pi ^2} \left(- m^2+\frac{1}{20} k^2+\frac{1}{560}\frac{ k^4}{m^2}+\frac{1}{6720}\frac{ k^6}{m^4}+\frac{1}{59136}\frac{ k^8}{m^6}+\frac{1}{439296}\frac{ k^{10}}{m^8}+
\right.\right.\0\\ & \quad \quad \quad \left.\left.
 + \frac{1}{2928640}\frac{ k^{12}}{m^{10}}+\frac{1}{18104320}\frac{ k^{14}}{m^{12}}+\ldots\right)\right) }
\al{\label{eq:fc:0:2:5:uvir}\tilde{T}_{0,2;5\text{D}}^{\text{f,UV-IR}} & =  \ldots \textrm{(i.e.\ no overlap)} }
Fermions, spin 0 x 2, dimension 6:
\al{\label{eq:fc:0:2:6:uv}\tilde{T}_{0,2;6\text{D}}^{\text{f,t,UV}} & = k^2 \pi _{\nu\nu} \left(\frac{i }{\pi ^3} \left( \left(\frac{23}{1800}-\frac{P}{240}\right) k^2 m+ \left(-\frac{1}{9}+\frac{P}{24}\right) m^3+ \left(\frac{1}{16}-\frac{K}{8}\right)\frac{ m^5}{k^2}+
\right.\right.\0\\ & \quad \quad \quad \left.\left.
 +  \left(\frac{11}{72}+\frac{K}{12}\right)\frac{ m^7}{k^4}+ \left(\frac{1}{288}+\frac{K}{24}\right)\frac{ m^9}{k^6}+ \left(-\frac{23}{1200}+\frac{K}{20}\right)\frac{ m^{11}}{k^8}+\ldots\right)\right) }
\al{\label{eq:fc:0:2:6:ir}\tilde{T}_{0,2;6\text{D}}^{\text{f,t,IR}} & = k^2 \pi _{\nu\nu} \left(\frac{i }{24 \pi ^3} \left( \left(-1+L_0\right) m^3-\frac{ L_0}{10} k^2 m+\frac{1}{140}\frac{ k^4}{m}+\frac{1}{2520}\frac{ k^6}{m^3}+\frac{1}{27720}\frac{ k^8}{m^5}+
\right.\right.\0\\ & \quad \quad \quad \left.\left.
 + \frac{1}{240240}\frac{ k^{10}}{m^7}+\frac{1}{1801800}\frac{ k^{12}}{m^9}+\frac{1}{12252240}\frac{ k^{14}}{m^{11}}+\ldots\right)\right) }
\al{\label{eq:fc:0:2:6:uvir}\tilde{T}_{0,2;6\text{D}}^{\text{f,UV-IR}} & = k^2 \pi _{\nu\nu} \left(\frac{i }{24 \pi ^3} \left( \left(\frac{23}{75}-\frac{K}{10}\right) k^2 m+ \left(-\frac{5}{3}+K\right) m^3\right)\right)+\ldots }
Fermions, spin 0 x 4, dimension 3:
\al{\label{eq:fc:0:4:3:uv}\tilde{T}_{0,4;3\text{D}}^{\text{f,t,UV}} & = k^4 \pi _{\nu\nu}^2 \left(\frac{3 }{32}\frac{ m}{k}-\frac{i }{\pi }\frac{ m^2}{k^2}-\frac{3 }{4}\frac{ m^3}{k^3}+\frac{4 i }{\pi }\frac{ m^4}{k^4}+\frac{3 }{2}\frac{ m^5}{k^5}-\frac{16 i }{5 \pi }\frac{ m^6}{k^6}-\frac{64 i }{35 \pi }\frac{ m^8}{k^8}-
\right.\0\\ & \quad \quad \left.
 - \frac{256 i }{105 \pi }\frac{ m^{10}}{k^{10}}-\frac{1024 i }{231 \pi }\frac{ m^{12}}{k^{12}}+\ldots\right) }
\al{\label{eq:fc:0:4:3:ir}\tilde{T}_{0,4;3\text{D}}^{\text{f,t,IR}} & = k^4 \pi _{\nu\nu}^2 \left(\frac{i }{4 \pi } \left(-\frac{1}{5}-\frac{1}{140}\frac{ k^2}{m^2}-\frac{1}{1680}\frac{ k^4}{m^4}-\frac{1}{14784}\frac{ k^6}{m^6}-\frac{1}{109824}\frac{ k^8}{m^8}-
\right.\right.\0\\ & \quad \quad \quad \left.\left.
 - \frac{1}{732160}\frac{ k^{10}}{m^{10}}-\frac{1}{4526080}\frac{ k^{12}}{m^{12}}+\ldots\right)\right) }
\al{\label{eq:fc:0:4:3:uvir}\tilde{T}_{0,4;3\text{D}}^{\text{f,UV-IR}} & =  \ldots \textrm{(i.e.\ no overlap)} }
Fermions, spin 0 x 4, dimension 4:
\al{\label{eq:fc:0:4:4:uv}\tilde{T}_{0,4;4\text{D}}^{\text{f,t,UV}} & = k^4 \pi _{\nu\nu}^2 \left(\frac{i }{\pi ^2} \left( \left(-\frac{23}{150}+\frac{P}{20}\right) m+ \left(\frac{5}{6}-\frac{K}{2}\right)\frac{ m^3}{k^2}+ \left(-\frac{3}{4}+\frac{3 K}{2}\right)\frac{ m^5}{k^4}-
\right.\right.\0\\ & \quad \quad \quad \left.\left.
 -  \left(\frac{11}{6}+K\right)\frac{ m^7}{k^6}- \left(\frac{1}{24}+\frac{K}{2}\right)\frac{ m^9}{k^8}+ \left(\frac{23}{100}-\frac{3 K}{5}\right)\frac{ m^{11}}{k^{10}}+\ldots\right)\right) }
\al{\label{eq:fc:0:4:4:ir}\tilde{T}_{0,4;4\text{D}}^{\text{f,t,IR}} & = k^4 \pi _{\nu\nu}^2 \left(\frac{i }{20 \pi ^2} \left( L_0 m-\frac{1}{14}\frac{ k^2}{m}-\frac{1}{252}\frac{ k^4}{m^3}-\frac{1}{2772}\frac{ k^6}{m^5}-\frac{1}{24024}\frac{ k^8}{m^7}-\frac{1}{180180}\frac{ k^{10}}{m^9}-
\right.\right.\0\\ & \quad \quad \quad \left.\left.
 - \frac{1}{1225224}\frac{ k^{12}}{m^{11}}+\ldots\right)\right) }
\al{\label{eq:fc:0:4:4:uvir}\tilde{T}_{0,4;4\text{D}}^{\text{f,UV-IR}} & = k^4 \pi _{\nu\nu}^2 \left(\frac{i  }{10 \pi ^2}\left(-\frac{23}{15}+\frac{K}{2}\right) m\right)+\ldots }
Fermions, spin 0 x 4, dimension 5:
\al{\label{eq:fc:0:4:5:uv}\tilde{T}_{0,4;5\text{D}}^{\text{f,t,UV}} & = k^4 \pi _{\nu\nu}^2 \left(\frac{1}{\pi ^2} \left(\frac{ \pi }{256} k m-\frac{3  \pi }{64}\frac{ m^3}{k}+\frac{i }{3}\frac{ m^4}{k^2}+\frac{3  \pi }{16}\frac{ m^5}{k^3}-\frac{4 i }{5}\frac{ m^6}{k^4}-\frac{ \pi }{4}\frac{ m^7}{k^5}+\frac{16 i }{35}\frac{ m^8}{k^6}+
\right.\right.\0\\ & \quad \quad \quad \left.\left.
 + \frac{64 i }{315}\frac{ m^{10}}{k^8}+\frac{256 i }{1155}\frac{ m^{12}}{k^{10}}+\ldots\right)\right) }
\al{\label{eq:fc:0:4:5:ir}\tilde{T}_{0,4;5\text{D}}^{\text{f,t,IR}} & = k^4 \pi _{\nu\nu}^2 \left(\frac{i }{4 \pi ^2} \left(\frac{1}{5} m^2-\frac{1}{140} k^2-\frac{1}{5040}\frac{ k^4}{m^2}-\frac{1}{73920}\frac{ k^6}{m^4}-\frac{1}{768768}\frac{ k^8}{m^6}-
\right.\right.\0\\ & \quad \quad \quad \left.\left.
 - \frac{1}{6589440}\frac{ k^{10}}{m^8}-\frac{1}{49786880}\frac{ k^{12}}{m^{10}}-\frac{1}{343982080}\frac{ k^{14}}{m^{12}}+\ldots\right)\right) }
\al{\label{eq:fc:0:4:5:uvir}\tilde{T}_{0,4;5\text{D}}^{\text{f,UV-IR}} & =  \ldots \textrm{(i.e.\ no overlap)} }
Fermions, spin 0 x 4, dimension 6:
\al{\label{eq:fc:0:4:6:uv}\tilde{T}_{0,4;6\text{D}}^{\text{f,t,UV}} & = k^4 \pi _{\nu\nu}^2 \left(\frac{i }{\pi ^3} \left( \left(-\frac{22}{3675}+\frac{P}{560}\right) k^2 m+ \left(\frac{23}{300}-\frac{P}{40}\right) m^3+ \left(-\frac{7}{48}+\frac{K}{8}\right)\frac{ m^5}{k^2}+
\right.\right.\0\\ & \quad \quad \quad \left.\left.
 +  \left(\frac{1}{24}-\frac{K}{4}\right)\frac{ m^7}{k^4}+ \left(\frac{25}{96}+\frac{K}{8}\right)\frac{ m^9}{k^6}+ \left(\frac{17}{1200}+\frac{K}{20}\right)\frac{ m^{11}}{k^8}+\ldots\right)\right) }
\al{\label{eq:fc:0:4:6:ir}\tilde{T}_{0,4;6\text{D}}^{\text{f,t,IR}} & = k^4 \pi _{\nu\nu}^2 \left(\frac{i }{40 \pi ^3} \left( \left(1-L_0\right) m^3+\frac{ L_0}{14} k^2 m-\frac{1}{252}\frac{ k^4}{m}-\frac{1}{5544}\frac{ k^6}{m^3}-\frac{1}{72072}\frac{ k^8}{m^5}-
\right.\right.\0\\ & \quad \quad \quad \left.\left.
 - \frac{1}{720720}\frac{ k^{10}}{m^7}-\frac{1}{6126120}\frac{ k^{12}}{m^9}-\frac{1}{46558512}\frac{ k^{14}}{m^{11}}+\ldots\right)\right) }
\al{\label{eq:fc:0:4:6:uvir}\tilde{T}_{0,4;6\text{D}}^{\text{f,UV-IR}} & = k^4 \pi _{\nu\nu}^2 \left(\frac{i }{5 \pi ^3} \left( \left(-\frac{22}{735}+\frac{K}{112}\right) k^2 m+ \left(\frac{31}{120}-\frac{K}{8}\right) m^3\right)\right)+\ldots }
Fermions, spin 1 x 1, dimension 3:
\al{\label{eq:fc:1:1:3:uv}\tilde{T}_{1,1;3\text{D}}^{\text{f,t,UV}} & = k^2 \pi _{\mu\nu} \left(\frac{1}{16}\frac{1}{k}+\frac{1}{4}\frac{ m^2}{k^3}-\frac{4 i }{3 \pi }\frac{ m^3}{k^4}-\frac{32 i }{15 \pi }\frac{ m^5}{k^6}-\frac{192 i }{35 \pi }\frac{ m^7}{k^8}-\frac{1024 i }{63 \pi }\frac{ m^9}{k^{10}}-\frac{5120 i }{99 \pi }\frac{ m^{11}}{k^{12}}+
\right.\0\\ & \quad \quad \left.
 + \ldots\right)+
\0\\ & \quad 
 + (k\cdot\epsilon)_{\mu\nu} \left(-\frac{i }{4}\frac{ m}{k}-\frac{1}{\pi }\frac{ m^2}{k^2}-\frac{4 }{3 \pi }\frac{ m^4}{k^4}-\frac{16 }{5 \pi }\frac{ m^6}{k^6}-\frac{64 }{7 \pi }\frac{ m^8}{k^8}-\frac{256 }{9 \pi }\frac{ m^{10}}{k^{10}}-
\right.\0\\ & \quad \quad \left.
 - \frac{1024 }{11 \pi }\frac{ m^{12}}{k^{12}}+\ldots\right) }
\al{\label{eq:fc:1:1:3:ir}\tilde{T}_{1,1;3\text{D}}^{\text{f,t,IR}} & = k^2 \pi _{\mu\nu} \left(\frac{i }{4 \pi } \left(-\frac{1}{3}\frac{1}{m}-\frac{1}{30}\frac{ k^2}{m^3}-\frac{3 }{560}\frac{ k^4}{m^5}-\frac{1}{1008}\frac{ k^6}{m^7}-\frac{5 }{25344}\frac{ k^8}{m^9}-\frac{3 }{73216}\frac{ k^{10}}{m^{11}}+
\right.\right.\0\\ & \quad \quad \quad \left.\left.
 + \ldots\right)\right)+
\0\\ & \quad 
 + (k\cdot\epsilon)_{\mu\nu} \left(\frac{1}{4 \pi } \left(- 1-\frac{1}{12}\frac{ k^2}{m^2}-\frac{1}{80}\frac{ k^4}{m^4}-\frac{1}{448}\frac{ k^6}{m^6}-\frac{1}{2304}\frac{ k^8}{m^8}-\frac{1}{11264}\frac{ k^{10}}{m^{10}}-
\right.\right.\0\\ & \quad \quad \quad \left.\left.
 - \frac{1}{53248}\frac{ k^{12}}{m^{12}}+\ldots\right)\right) }
\al{\label{eq:fc:1:1:3:uvir}\tilde{T}_{1,1;3\text{D}}^{\text{f,UV-IR}} & =  \ldots \textrm{(i.e.\ no overlap)} }
Fermions, spin 1 x 1, dimension 4:
\al{\label{eq:fc:1:1:4:uv}\tilde{T}_{1,1;4\text{D}}^{\text{f,t,UV}} & = k^2 \pi _{\mu\nu} \left(\frac{i }{\pi ^2} \left( \left(-\frac{5}{36}+\frac{P}{12}\right)-\frac{1}{2}\frac{ m^2}{k^2}- \left(\frac{1}{4}+\frac{K}{2}\right)\frac{ m^4}{k^4}+ \left(\frac{4}{9}-\frac{2 K}{3}\right)\frac{ m^6}{k^6}+
\right.\right.\0\\ & \quad \quad \quad \left.\left.
 +  \left(\frac{11}{8}-\frac{3 K}{2}\right)\frac{ m^8}{k^8}+ \left(\frac{62}{15}-4 K\right)\frac{ m^{10}}{k^{10}}+ \left(\frac{463}{36}-\frac{35 K}{3}\right)\frac{ m^{12}}{k^{12}}+\ldots\right)\right) }
\al{\label{eq:fc:1:1:4:ir}\tilde{T}_{1,1;4\text{D}}^{\text{f,t,IR}} & = k^2 \pi _{\mu\nu} \left(\frac{i }{4 \pi ^2} \left(\frac{ L_0}{3}-\frac{1}{15}\frac{ k^2}{m^2}-\frac{1}{140}\frac{ k^4}{m^4}-\frac{1}{945}\frac{ k^6}{m^6}-\frac{1}{5544}\frac{ k^8}{m^8}-\frac{1}{30030}\frac{ k^{10}}{m^{10}}-
\right.\right.\0\\ & \quad \quad \quad \left.\left.
 - \frac{1}{154440}\frac{ k^{12}}{m^{12}}+\ldots\right)\right) }
\al{\label{eq:fc:1:1:4:uvir}\tilde{T}_{1,1;4\text{D}}^{\text{f,UV-IR}} & = k^2 \pi _{\mu\nu} \frac{i  }{12 \pi ^2}\left(-\frac{5}{3}+K\right)+\ldots }
Fermions, spin 1 x 1, dimension 5:
\al{\label{eq:fc:1:1:5:uv}\tilde{T}_{1,1;5\text{D}}^{\text{f,t,UV}} & = k^2 \pi _{\mu\nu} \left(\frac{1}{\pi ^2} \left(\frac{3  \pi }{256} k-\frac{ \pi }{32}\frac{ m^2}{k}-\frac{ \pi }{16}\frac{ m^4}{k^3}+\frac{4 i }{15}\frac{ m^5}{k^4}+\frac{32 i }{105}\frac{ m^7}{k^6}+\frac{64 i }{105}\frac{ m^9}{k^8}+
\right.\right.\0\\ & \quad \quad \quad \left.\left.
 + \frac{1024 i }{693}\frac{ m^{11}}{k^{10}}+\ldots\right)\right) }
\al{\label{eq:fc:1:1:5:ir}\tilde{T}_{1,1;5\text{D}}^{\text{f,t,IR}} & = k^2 \pi _{\mu\nu} \left(\frac{i }{4 \pi ^2} \left(\frac{1}{3} m-\frac{1}{30}\frac{ k^2}{m}-\frac{1}{560}\frac{ k^4}{m^3}-\frac{1}{5040}\frac{ k^6}{m^5}-\frac{5 }{177408}\frac{ k^8}{m^7}-\frac{1}{219648}\frac{ k^{10}}{m^9}-
\right.\right.\0\\ & \quad \quad \quad \left.\left.
 - \frac{7 }{8785920}\frac{ k^{12}}{m^{11}}+\ldots\right)\right) }
\al{\label{eq:fc:1:1:5:uvir}\tilde{T}_{1,1;5\text{D}}^{\text{f,UV-IR}} & =  \ldots \textrm{(i.e.\ no overlap)} }
Fermions, spin 1 x 1, dimension 6:
\al{\label{eq:fc:1:1:6:uv}\tilde{T}_{1,1;6\text{D}}^{\text{f,t,UV}} & = k^2 \pi _{\mu\nu} \left(\frac{i }{\pi ^3} \left( \left(-\frac{77}{3600}+\frac{P}{120}\right) k^2+ \left(\frac{5}{72}-\frac{P}{24}\right) m^2+\frac{1}{8}\frac{ m^4}{k^2}+
\right.\right.\0\\ & \quad \quad \quad \left.\left.
 +  \left(\frac{5}{72}+\frac{K}{12}\right)\frac{ m^6}{k^4}+ \left(-\frac{5}{144}+\frac{K}{12}\right)\frac{ m^8}{k^6}+ \left(-\frac{43}{400}+\frac{3 K}{20}\right)\frac{ m^{10}}{k^8}+
\right.\right.\0\\ & \quad \quad \quad \left.\left.
 +  \left(-\frac{13}{45}+\frac{K}{3}\right)\frac{ m^{12}}{k^{10}}+\ldots\right)\right) }
\al{\label{eq:fc:1:1:6:ir}\tilde{T}_{1,1;6\text{D}}^{\text{f,t,IR}} & = k^2 \pi _{\mu\nu} \left(\frac{i }{8 \pi ^3} \left( \left(\frac{1}{3}-\frac{L_0}{3}\right) m^2+\frac{ L_0}{15} k^2-\frac{1}{140}\frac{ k^4}{m^2}-\frac{1}{1890}\frac{ k^6}{m^4}-\frac{1}{16632}\frac{ k^8}{m^6}-
\right.\right.\0\\ & \quad \quad \quad \left.\left.
 - \frac{1}{120120}\frac{ k^{10}}{m^8}-\frac{1}{772200}\frac{ k^{12}}{m^{10}}-\frac{1}{4594590}\frac{ k^{14}}{m^{12}}+\ldots\right)\right) }
\al{\label{eq:fc:1:1:6:uvir}\tilde{T}_{1,1;6\text{D}}^{\text{f,UV-IR}} & = k^2 \pi _{\mu\nu} \left(\frac{i }{12 \pi ^3} \left( \left(-\frac{77}{300}+\frac{K}{10}\right) k^2+ \left(\frac{1}{3}-\frac{K}{2}\right) m^2\right)\right)+\ldots }
Fermions, spin 1 x 3, dimension 3:
\al{\label{eq:fc:1:3:3:uv}\tilde{T}_{1,3;3\text{D}}^{\text{f,t,UV}} & = k^4 \pi _{\mu\nu} \pi _{\nu\nu} \left(-\frac{1}{64}\frac{1}{k}-\frac{1}{8}\frac{ m^2}{k^3}+\frac{4 i }{3 \pi }\frac{ m^3}{k^4}+\frac{3 }{4}\frac{ m^4}{k^5}-\frac{32 i }{15 \pi }\frac{ m^5}{k^6}-\frac{64 i }{35 \pi }\frac{ m^7}{k^8}-\frac{1024 i }{315 \pi }\frac{ m^9}{k^{10}}-
\right.\0\\ & \quad \quad \left.
 - \frac{5120 i }{693 \pi }\frac{ m^{11}}{k^{12}}+\ldots\right)+
\0\\ & \quad 
 + k^2 (k\cdot\epsilon)_{\mu\nu} \pi _{\nu\nu} \left(\frac{i }{8}\frac{ m}{k}+\frac{1}{\pi }\frac{ m^2}{k^2}-\frac{i }{2}\frac{ m^3}{k^3}-\frac{4 }{3 \pi }\frac{ m^4}{k^4}-\frac{16 }{15 \pi }\frac{ m^6}{k^6}-\frac{64 }{35 \pi }\frac{ m^8}{k^8}-
\right.\0\\ & \quad \quad \left.
 - \frac{256 }{63 \pi }\frac{ m^{10}}{k^{10}}-\frac{1024 }{99 \pi }\frac{ m^{12}}{k^{12}}+\ldots\right) }
\al{\label{eq:fc:1:3:3:ir}\tilde{T}_{1,3;3\text{D}}^{\text{f,t,IR}} & = k^4 \pi _{\mu\nu} \pi _{\nu\nu} \left(\frac{i }{4 \pi } \left(\frac{1}{15}\frac{1}{m}+\frac{1}{210}\frac{ k^2}{m^3}+\frac{1}{1680}\frac{ k^4}{m^5}+\frac{1}{11088}\frac{ k^6}{m^7}+\frac{5 }{329472}\frac{ k^8}{m^9}+
\right.\right.\0\\ & \quad \quad \quad \left.\left.
 + \frac{1}{366080}\frac{ k^{10}}{m^{11}}+\ldots\right)\right)+
\0\\ & \quad 
 + k^2 (k\cdot\epsilon)_{\mu\nu} \pi _{\nu\nu} \left(\frac{1}{4 \pi } \left(\frac{1}{3}+\frac{1}{60}\frac{ k^2}{m^2}+\frac{1}{560}\frac{ k^4}{m^4}+\frac{1}{4032}\frac{ k^6}{m^6}+\frac{1}{25344}\frac{ k^8}{m^8}+
\right.\right.\0\\ & \quad \quad \quad \left.\left.
 + \frac{1}{146432}\frac{ k^{10}}{m^{10}}+\frac{1}{798720}\frac{ k^{12}}{m^{12}}+\ldots\right)\right) }
\al{\label{eq:fc:1:3:3:uvir}\tilde{T}_{1,3;3\text{D}}^{\text{f,UV-IR}} & =  \ldots \textrm{(i.e.\ no overlap)} }
Fermions, spin 1 x 3, dimension 4:
\al{\label{eq:fc:1:3:4:uv}\tilde{T}_{1,3;4\text{D}}^{\text{f,t,UV}} & = k^4 \pi _{\nu\nu} \pi _{\mu\nu} \left(\frac{i }{\pi ^2} \left( \left(\frac{31}{900}-\frac{P}{60}\right)+\frac{1}{6}\frac{ m^2}{k^2}+ \left(-\frac{3}{4}+\frac{K}{2}\right)\frac{ m^4}{k^4}- \left(\frac{8}{9}+\frac{2 K}{3}\right)\frac{ m^6}{k^6}+
\right.\right.\0\\ & \quad \quad \quad \left.\left.
 +  \left(\frac{1}{8}-\frac{K}{2}\right)\frac{ m^8}{k^8}+ \left(\frac{38}{75}-\frac{4 K}{5}\right)\frac{ m^{10}}{k^{10}}+ \left(\frac{49}{36}-\frac{5 K}{3}\right)\frac{ m^{12}}{k^{12}}+\ldots\right)\right) }
\al{\label{eq:fc:1:3:4:ir}\tilde{T}_{1,3;4\text{D}}^{\text{f,t,IR}} & = k^4 \pi _{\nu\nu} \pi _{\mu\nu} \left(\frac{i }{12 \pi ^2} \left(-\frac{ L_0}{5}+\frac{1}{35}\frac{ k^2}{m^2}+\frac{1}{420}\frac{ k^4}{m^4}+\frac{1}{3465}\frac{ k^6}{m^6}+\frac{1}{24024}\frac{ k^8}{m^8}+
\right.\right.\0\\ & \quad \quad \quad \left.\left.
 + \frac{1}{150150}\frac{ k^{10}}{m^{10}}+\frac{1}{875160}\frac{ k^{12}}{m^{12}}+\ldots\right)\right) }
\al{\label{eq:fc:1:3:4:uvir}\tilde{T}_{1,3;4\text{D}}^{\text{f,UV-IR}} & = k^4 \pi _{\mu\nu} \pi _{\nu\nu} \frac{i  }{60 \pi ^2}\left(\frac{31}{15}-K\right)+\ldots }
Fermions, spin 1 x 3, dimension 5:
\al{\label{eq:fc:1:3:5:uv}\tilde{T}_{1,3;5\text{D}}^{\text{f,t,UV}} & = k^4 \pi _{\nu\nu} \pi _{\mu\nu} \left(\frac{1}{\pi ^2} \left(-\frac{ \pi }{512} k+\frac{ \pi }{128}\frac{ m^2}{k}+\frac{ \pi }{32}\frac{ m^4}{k^3}-\frac{4 i }{15}\frac{ m^5}{k^4}-\frac{ \pi }{8}\frac{ m^6}{k^5}+\frac{32 i }{105}\frac{ m^7}{k^6}+
\right.\right.\0\\ & \quad \quad \quad \left.\left.
 + \frac{64 i }{315}\frac{ m^9}{k^8}+\frac{1024 i }{3465}\frac{ m^{11}}{k^{10}}+\ldots\right)\right) }
\al{\label{eq:fc:1:3:5:ir}\tilde{T}_{1,3;5\text{D}}^{\text{f,t,IR}} & = k^4 \pi _{\nu\nu} \pi _{\mu\nu} \left(\frac{i }{12 \pi ^2} \left(-\frac{1}{5} m+\frac{1}{70}\frac{ k^2}{m}+\frac{1}{1680}\frac{ k^4}{m^3}+\frac{1}{18480}\frac{ k^6}{m^5}+\frac{5 }{768768}\frac{ k^8}{m^7}+
\right.\right.\0\\ & \quad \quad \quad \left.\left.
 + \frac{1}{1098240}\frac{ k^{10}}{m^9}+\frac{7 }{49786880}\frac{ k^{12}}{m^{11}}+\ldots\right)\right) }
\al{\label{eq:fc:1:3:5:uvir}\tilde{T}_{1,3;5\text{D}}^{\text{f,UV-IR}} & =  \ldots \textrm{(i.e.\ no overlap)} }
Fermions, spin 1 x 3, dimension 6:
\al{\label{eq:fc:1:3:6:uv}\tilde{T}_{1,3;6\text{D}}^{\text{f,t,UV}} & = k^4 \pi _{\nu\nu} \pi _{\mu\nu} \left(\frac{i }{\pi ^3} \left( \left(\frac{599}{176400}-\frac{P}{840}\right) k^2+ \left(-\frac{31}{1800}+\frac{P}{120}\right) m^2-\frac{1}{24}\frac{ m^4}{k^2}+
\right.\right.\0\\ & \quad \quad \quad \left.\left.
 +  \left(\frac{7}{72}-\frac{K}{12}\right)\frac{ m^6}{k^4}+ \left(\frac{19}{144}+\frac{K}{12}\right)\frac{ m^8}{k^6}+ \left(-\frac{1}{400}+\frac{K}{20}\right)\frac{ m^{10}}{k^8}+
\right.\right.\0\\ & \quad \quad \quad \left.\left.
 +  \left(-\frac{7}{225}+\frac{K}{15}\right)\frac{ m^{12}}{k^{10}}+\ldots\right)\right) }
\al{\label{eq:fc:1:3:6:ir}\tilde{T}_{1,3;6\text{D}}^{\text{f,t,IR}} & = k^4 \pi _{\nu\nu} \pi _{\mu\nu} \left(\frac{i }{24 \pi ^3} \left( \left(-\frac{1}{5}+\frac{L_0}{5}\right) m^2-\frac{ L_0}{35} k^2+\frac{1}{420}\frac{ k^4}{m^2}+\frac{1}{6930}\frac{ k^6}{m^4}+\frac{1}{72072}\frac{ k^8}{m^6}+
\right.\right.\0\\ & \quad \quad \quad \left.\left.
 + \frac{1}{600600}\frac{ k^{10}}{m^8}+\frac{1}{4375800}\frac{ k^{12}}{m^{10}}+\frac{1}{29099070}\frac{ k^{14}}{m^{12}}+\ldots\right)\right) }
\al{\label{eq:fc:1:3:6:uvir}\tilde{T}_{1,3;6\text{D}}^{\text{f,UV-IR}} & = k^4 \pi _{\mu\nu} \pi _{\nu\nu} \left(\frac{i }{15 \pi ^3} \left( \left(\frac{599}{11760}-\frac{K}{56}\right) k^2+ \left(-\frac{2}{15}+\frac{K}{8}\right) m^2\right)\right)+\ldots }
Fermions, spin 1 x 5, dimension 3:
\al{\label{eq:fc:1:5:3:uv}\tilde{T}_{1,5;3\text{D}}^{\text{f,t,UV}} & = k^6 \pi _{\mu\nu} \pi _{\nu\nu}^2 \left(\frac{1}{128}\frac{1}{k}+\frac{3 }{32}\frac{ m^2}{k^3}-\frac{4 i }{3 \pi }\frac{ m^3}{k^4}-\frac{9 }{8}\frac{ m^4}{k^5}+\frac{32 i }{5 \pi }\frac{ m^5}{k^6}+\frac{5 }{2}\frac{ m^6}{k^7}-\frac{192 i }{35 \pi }\frac{ m^7}{k^8}-
\right.\0\\ & \quad \quad \left.
 - \frac{1024 i }{315 \pi }\frac{ m^9}{k^{10}}-\frac{1024 i }{231 \pi }\frac{ m^{11}}{k^{12}}+\ldots\right)+
\0\\ & \quad 
 + k^4 (k\cdot\epsilon)_{\mu\nu} \pi _{\nu\nu}^2 \left(-\frac{3 i }{32}\frac{ m}{k}-\frac{1}{\pi }\frac{ m^2}{k^2}+\frac{3 i }{4}\frac{ m^3}{k^3}+\frac{4 }{\pi }\frac{ m^4}{k^4}-\frac{3 i }{2}\frac{ m^5}{k^5}-\frac{16 }{5 \pi }\frac{ m^6}{k^6}-
\right.\0\\ & \quad \quad \left.
 - \frac{64 }{35 \pi }\frac{ m^8}{k^8}-\frac{256 }{105 \pi }\frac{ m^{10}}{k^{10}}-\frac{1024 }{231 \pi }\frac{ m^{12}}{k^{12}}+\ldots\right) }
\al{\label{eq:fc:1:5:3:ir}\tilde{T}_{1,5;3\text{D}}^{\text{f,t,IR}} & = k^6 \pi _{\mu\nu} \pi _{\nu\nu}^2 \left(\frac{i }{4 \pi } \left(-\frac{1}{35}\frac{1}{m}-\frac{1}{630}\frac{ k^2}{m^3}-\frac{1}{6160}\frac{ k^4}{m^5}-\frac{1}{48048}\frac{ k^6}{m^7}-\frac{1}{329472}\frac{ k^8}{m^9}-
\right.\right.\0\\ & \quad \quad \quad \left.\left.
 - \frac{3 }{6223360}\frac{ k^{10}}{m^{11}}+\ldots\right)\right)+
\0\\ & \quad 
 + k^4 (k\cdot\epsilon)_{\mu\nu} \pi _{\nu\nu}^2 \left(\frac{1}{4 \pi } \left(-\frac{1}{5}-\frac{1}{140}\frac{ k^2}{m^2}-\frac{1}{1680}\frac{ k^4}{m^4}-\frac{1}{14784}\frac{ k^6}{m^6}-\frac{1}{109824}\frac{ k^8}{m^8}-
\right.\right.\0\\ & \quad \quad \quad \left.\left.
 - \frac{1}{732160}\frac{ k^{10}}{m^{10}}-\frac{1}{4526080}\frac{ k^{12}}{m^{12}}+\ldots\right)\right) }
\al{\label{eq:fc:1:5:3:uvir}\tilde{T}_{1,5;3\text{D}}^{\text{f,UV-IR}} & =  \ldots \textrm{(i.e.\ no overlap)} }
Fermions, spin 1 x 5, dimension 4:
\al{\label{eq:fc:1:5:4:uv}\tilde{T}_{1,5;4\text{D}}^{\text{f,t,UV}} & = k^6 \pi _{\nu\nu}^2 \pi _{\mu\nu} \left(\frac{i }{\pi ^2} \left( \left(-\frac{247}{14700}+\frac{P}{140}\right)-\frac{1}{10}\frac{ m^2}{k^2}+ \left(\frac{13}{12}-\frac{K}{2}\right)\frac{ m^4}{k^4}+
\right.\right.\0\\ & \quad \quad \quad \left.\left.
 +  \left(-\frac{4}{3}+2 K\right)\frac{ m^6}{k^6}- \left(\frac{21}{8}+\frac{3 K}{2}\right)\frac{ m^8}{k^8}- \left(\frac{2}{75}+\frac{4 K}{5}\right)\frac{ m^{10}}{k^{10}}+
\right.\right.\0\\ & \quad \quad \quad \left.\left.
 +  \left(\frac{5}{12}-K\right)\frac{ m^{12}}{k^{12}}+\ldots\right)\right) }
\al{\label{eq:fc:1:5:4:ir}\tilde{T}_{1,5;4\text{D}}^{\text{f,t,IR}} & = k^6 \pi _{\nu\nu}^2 \pi _{\mu\nu} \left(\frac{i }{20 \pi ^2} \left(\frac{ L_0}{7}-\frac{1}{63}\frac{ k^2}{m^2}-\frac{1}{924}\frac{ k^4}{m^4}-\frac{1}{9009}\frac{ k^6}{m^6}-\frac{1}{72072}\frac{ k^8}{m^8}-
\right.\right.\0\\ & \quad \quad \quad \left.\left.
 - \frac{1}{510510}\frac{ k^{10}}{m^{10}}-\frac{1}{3325608}\frac{ k^{12}}{m^{12}}+\ldots\right)\right) }
\al{\label{eq:fc:1:5:4:uvir}\tilde{T}_{1,5;4\text{D}}^{\text{f,UV-IR}} & = k^6 \pi _{\mu\nu} \pi _{\nu\nu}^2 \frac{i  }{140 \pi ^2}\left(-\frac{247}{105}+K\right)+\ldots }
Fermions, spin 1 x 5, dimension 5:
\al{\label{eq:fc:1:5:5:uv}\tilde{T}_{1,5;5\text{D}}^{\text{f,t,UV}} & = k^6 \pi _{\nu\nu}^2 \pi _{\mu\nu} \left(\frac{1}{\pi ^2} \left(\frac{3  \pi }{4096} k-\frac{ \pi }{256}\frac{ m^2}{k}-\frac{3  \pi }{128}\frac{ m^4}{k^3}+\frac{4 i }{15}\frac{ m^5}{k^4}+\frac{3  \pi }{16}\frac{ m^6}{k^5}-\frac{32 i }{35}\frac{ m^7}{k^6}-
\right.\right.\0\\ & \quad \quad \quad \left.\left.
 - \frac{5  \pi }{16}\frac{ m^8}{k^7}+\frac{64 i }{105}\frac{ m^9}{k^8}+\frac{1024 i }{3465}\frac{ m^{11}}{k^{10}}+\ldots\right)\right) }
\al{\label{eq:fc:1:5:5:ir}\tilde{T}_{1,5;5\text{D}}^{\text{f,t,IR}} & = k^6 \pi _{\nu\nu}^2 \pi _{\mu\nu} \left(\frac{i }{4 \pi ^2} \left(\frac{1}{35} m-\frac{1}{630}\frac{ k^2}{m}-\frac{1}{18480}\frac{ k^4}{m^3}-\frac{1}{240240}\frac{ k^6}{m^5}-\frac{1}{2306304}\frac{ k^8}{m^7}-
\right.\right.\0\\ & \quad \quad \quad \left.\left.
 - \frac{1}{18670080}\frac{ k^{10}}{m^9}-\frac{7 }{945950720}\frac{ k^{12}}{m^{11}}+\ldots\right)\right) }
\al{\label{eq:fc:1:5:5:uvir}\tilde{T}_{1,5;5\text{D}}^{\text{f,UV-IR}} & =  \ldots \textrm{(i.e.\ no overlap)} }
Fermions, spin 1 x 5, dimension 6:
\al{\label{eq:fc:1:5:6:uv}\tilde{T}_{1,5;6\text{D}}^{\text{f,t,UV}} & = k^6 \pi _{\nu\nu}^2 \pi _{\mu\nu} \left(\frac{i }{\pi ^3} \left( \left(-\frac{1937}{1587600}+\frac{P}{2520}\right) k^2+ \left(\frac{247}{29400}-\frac{P}{280}\right) m^2+\frac{1}{40}\frac{ m^4}{k^2}+
\right.\right.\0\\ & \quad \quad \quad \left.\left.
 +  \left(-\frac{11}{72}+\frac{K}{12}\right)\frac{ m^6}{k^4}+ \left(\frac{5}{48}-\frac{K}{4}\right)\frac{ m^8}{k^6}+ \left(\frac{117}{400}+\frac{3 K}{20}\right)\frac{ m^{10}}{k^8}+
\right.\right.\0\\ & \quad \quad \quad \left.\left.
 +  \left(\frac{1}{75}+\frac{K}{15}\right)\frac{ m^{12}}{k^{10}}+\ldots\right)\right) }
\al{\label{eq:fc:1:5:6:ir}\tilde{T}_{1,5;6\text{D}}^{\text{f,t,IR}} & = k^6 \pi _{\nu\nu}^2 \pi _{\mu\nu} \left(\frac{i }{40 \pi ^3} \left( \left(\frac{1}{7}-\frac{L_0}{7}\right) m^2+\frac{ L_0}{63} k^2-\frac{1}{924}\frac{ k^4}{m^2}-\frac{1}{18018}\frac{ k^6}{m^4}-
\right.\right.\0\\ & \quad \quad \quad \left.\left.
 - \frac{1}{216216}\frac{ k^8}{m^6}-\frac{1}{2042040}\frac{ k^{10}}{m^8}-\frac{1}{16628040}\frac{ k^{12}}{m^{10}}-\frac{1}{122216094}\frac{ k^{14}}{m^{12}}+
\right.\right.\0\\ & \quad \quad \quad \left.\left.
 + \ldots\right)\right) }
\al{\label{eq:fc:1:5:6:uvir}\tilde{T}_{1,5;6\text{D}}^{\text{f,UV-IR}} & = k^6 \pi _{\mu\nu} \pi _{\nu\nu}^2 \left(\frac{i }{140 \pi ^3} \left( \left(-\frac{1937}{11340}+\frac{K}{18}\right) k^2+ \left(\frac{71}{105}-\frac{K}{2}\right) m^2\right)\right)+\ldots }
Fermions, spin 2 x 2, dimension 3:
\al{\label{eq:fc:2:2:3:uv}\tilde{T}_{2,2;3\text{D}}^{\text{f,t,UV}} & = k^4 \pi _{\mu\nu}^2 \left(-\frac{1}{32}\frac{1}{k}+\frac{2 i }{3 \pi }\frac{ m^3}{k^4}+\frac{1}{2}\frac{ m^4}{k^5}-\frac{8 i }{5 \pi }\frac{ m^5}{k^6}-\frac{32 i }{21 \pi }\frac{ m^7}{k^8}-\frac{128 i }{45 \pi }\frac{ m^9}{k^{10}}-\frac{512 i }{77 \pi }\frac{ m^{11}}{k^{12}}+
\right.\0\\ & \quad \quad \left.
 + \ldots\right)+
\0\\ & \quad 
 + k^4 \pi _{\mu\mu} \pi _{\nu\nu} \left(\frac{1}{64}\frac{1}{k}-\frac{1}{8}\frac{ m^2}{k^3}+\frac{2 i }{3 \pi }\frac{ m^3}{k^4}+\frac{1}{4}\frac{ m^4}{k^5}-\frac{8 i }{15 \pi }\frac{ m^5}{k^6}-\frac{32 i }{105 \pi }\frac{ m^7}{k^8}-
\right.\0\\ & \quad \quad \left.
 - \frac{128 i }{315 \pi }\frac{ m^9}{k^{10}}-\frac{512 i }{693 \pi }\frac{ m^{11}}{k^{12}}+\ldots\right)+
\0\\ & \quad 
 + k^2 (k\cdot\epsilon)_{\mu\nu} \pi _{\mu\nu} \left(\frac{i }{8}\frac{ m}{k}+\frac{1}{\pi }\frac{ m^2}{k^2}-\frac{i }{2}\frac{ m^3}{k^3}-\frac{4 }{3 \pi }\frac{ m^4}{k^4}-\frac{16 }{15 \pi }\frac{ m^6}{k^6}-\frac{64 }{35 \pi }\frac{ m^8}{k^8}-
\right.\0\\ & \quad \quad \left.
 - \frac{256 }{63 \pi }\frac{ m^{10}}{k^{10}}-\frac{1024 }{99 \pi }\frac{ m^{12}}{k^{12}}+\ldots\right) }
\al{\label{eq:fc:2:2:3:ir}\tilde{T}_{2,2;3\text{D}}^{\text{f,t,IR}} & = k^4 \pi _{\mu\nu}^2 \left(\frac{i }{2 \pi } \left(-\frac{1}{3}\frac{ m}{k^2}+\frac{1}{20}\frac{1}{m}+\frac{1}{336}\frac{ k^2}{m^3}+\frac{1}{2880}\frac{ k^4}{m^5}+\frac{1}{19712}\frac{ k^6}{m^7}+\frac{1}{119808}\frac{ k^8}{m^9}+
\right.\right.\0\\ & \quad \quad \quad \left.\left.
 + \frac{1}{675840}\frac{ k^{10}}{m^{11}}+\ldots\right)\right)+
\0\\ & \quad 
 + k^4 \pi _{\mu\mu} \pi _{\nu\nu} \left(\frac{i }{6 \pi } \left(\frac{m}{k^2}-\frac{1}{20}\frac{1}{m}-\frac{1}{560}\frac{ k^2}{m^3}-\frac{1}{6720}\frac{ k^4}{m^5}-\frac{1}{59136}\frac{ k^6}{m^7}-
\right.\right.\0\\ & \quad \quad \quad \left.\left.
 - \frac{1}{439296}\frac{ k^8}{m^9}-\frac{1}{2928640}\frac{ k^{10}}{m^{11}}+\ldots\right)\right)+
\0\\ & \quad 
 + k^2 (k\cdot\epsilon)_{\mu\nu} \pi _{\mu\nu} \left(\frac{1}{4 \pi } \left(\frac{1}{3}+\frac{1}{60}\frac{ k^2}{m^2}+\frac{1}{560}\frac{ k^4}{m^4}+\frac{1}{4032}\frac{ k^6}{m^6}+\frac{1}{25344}\frac{ k^8}{m^8}+
\right.\right.\0\\ & \quad \quad \quad \left.\left.
 + \frac{1}{146432}\frac{ k^{10}}{m^{10}}+\frac{1}{798720}\frac{ k^{12}}{m^{12}}+\ldots\right)\right) }
\al{\label{eq:fc:2:2:3:uvir}\tilde{T}_{2,2;3\text{D}}^{\text{f,UV-IR}} & =  \ldots \textrm{(i.e.\ no overlap)} }
Fermions, spin 2 x 2, dimension 4:
\al{\label{eq:fc:2:2:4:uv}\tilde{T}_{2,2;4\text{D}}^{\text{f,t,UV}} & = k^4 \pi _{\mu\nu}^2 \left(\frac{i }{2 \pi ^2} \left( \left(\frac{3}{25}-\frac{P}{20}\right)+ \left(-\frac{1}{9}+\frac{P}{6}\right)\frac{ m^2}{k^2}+ \left(-\frac{5}{4}+\frac{K}{2}\right)\frac{ m^4}{k^4}-
\right.\right.\0\\ & \quad \quad \quad \left.\left.
 -  \left(\frac{7}{6}+K\right)\frac{ m^6}{k^6}+ \left(\frac{19}{72}-\frac{5 K}{6}\right)\frac{ m^8}{k^8}+ \left(\frac{281}{300}-\frac{7 K}{5}\right)\frac{ m^{10}}{k^{10}}+
\right.\right.\0\\ & \quad \quad \quad \left.\left.
 +  \left(\frac{151}{60}-3 K\right)\frac{ m^{12}}{k^{12}}+\ldots\right)\right)+
\0\\ & \quad 
 + k^4 \pi _{\mu\mu} \pi _{\nu\nu} \left(\frac{i }{\pi ^2} \left( \left(-\frac{23}{900}+\frac{P}{120}\right)+ \left(\frac{2}{9}-\frac{P}{12}\right)\frac{ m^2}{k^2}+ \left(-\frac{1}{8}+\frac{K}{4}\right)\frac{ m^4}{k^4}-
\right.\right.\0\\ & \quad \quad \quad \left.\left.
 -  \left(\frac{11}{36}+\frac{K}{6}\right)\frac{ m^6}{k^6}- \left(\frac{1}{144}+\frac{K}{12}\right)\frac{ m^8}{k^8}+ \left(\frac{23}{600}-\frac{K}{10}\right)\frac{ m^{10}}{k^{10}}+
\right.\right.\0\\ & \quad \quad \quad \left.\left.
 +  \left(\frac{37}{360}-\frac{K}{6}\right)\frac{ m^{12}}{k^{12}}+\ldots\right)\right) }
\al{\label{eq:fc:2:2:4:ir}\tilde{T}_{2,2;4\text{D}}^{\text{f,t,IR}} & = k^4 \pi _{\mu\nu}^2 \left(\frac{i }{4 \pi ^2} \left( \left(-\frac{1}{3}+\frac{L_0}{3}\right)\frac{ m^2}{k^2}-\frac{ L_0}{10}+\frac{1}{84}\frac{ k^2}{m^2}+\frac{1}{1080}\frac{ k^4}{m^4}+\frac{1}{9240}\frac{ k^6}{m^6}+
\right.\right.\0\\ & \quad \quad \quad \left.\left.
 + \frac{1}{65520}\frac{ k^8}{m^8}+\frac{1}{415800}\frac{ k^{10}}{m^{10}}+\frac{1}{2450448}\frac{ k^{12}}{m^{12}}+\ldots\right)\right)+
\0\\ & \quad 
 + k^4 \pi _{\mu\mu} \pi _{\nu\nu} \left(\frac{i }{12 \pi ^2} \left( \left(1-L_0\right)\frac{ m^2}{k^2}+\frac{ L_0}{10}-\frac{1}{140}\frac{ k^2}{m^2}-\frac{1}{2520}\frac{ k^4}{m^4}-\frac{1}{27720}\frac{ k^6}{m^6}-
\right.\right.\0\\ & \quad \quad \quad \left.\left.
 - \frac{1}{240240}\frac{ k^8}{m^8}-\frac{1}{1801800}\frac{ k^{10}}{m^{10}}-\frac{1}{12252240}\frac{ k^{12}}{m^{12}}+\ldots\right)\right) }
\al{\label{eq:fc:2:2:4:uvir}\tilde{T}_{2,2;4\text{D}}^{\text{f,UV-IR}} & = k^4 \pi _{\mu\nu}^2 \left(\frac{i }{2 \pi ^2} \left( \left(\frac{3}{25}-\frac{K}{20}\right)+ \left(\frac{1}{18}+\frac{K}{6}\right)\frac{ m^2}{k^2}\right)\right)+
\0\\ & \quad 
 + k^4 \pi _{\mu\mu} \pi _{\nu\nu} \left(\frac{i }{12 \pi ^2} \left( \left(-\frac{23}{75}+\frac{K}{10}\right)+ \left(\frac{5}{3}-K\right)\frac{ m^2}{k^2}\right)\right)+\ldots }
Fermions, spin 2 x 2, dimension 5:
\al{\label{eq:fc:2:2:5:uv}\tilde{T}_{2,2;5\text{D}}^{\text{f,t,UV}} & = k^4 \pi _{\mu\nu}^2 \left(\frac{1}{\pi ^2} \left(-\frac{ \pi }{384} k+\frac{ \pi }{64}\frac{ m^2}{k}-\frac{2 i }{15}\frac{ m^5}{k^4}-\frac{ \pi }{12}\frac{ m^6}{k^5}+\frac{8 i }{35}\frac{ m^7}{k^6}+\frac{32 i }{189}\frac{ m^9}{k^8}+
\right.\right.\0\\ & \quad \quad \quad \left.\left.
 + \frac{128 i }{495}\frac{ m^{11}}{k^{10}}+\ldots\right)\right)+
\0\\ & \quad 
 + k^4 \pi _{\mu\mu} \pi _{\nu\nu} \left(\frac{1}{\pi ^2} \left(\frac{ \pi }{1536} k-\frac{ \pi }{128}\frac{ m^2}{k}+\frac{ \pi }{32}\frac{ m^4}{k^3}-\frac{2 i }{15}\frac{ m^5}{k^4}-\frac{ \pi }{24}\frac{ m^6}{k^5}+\frac{8 i }{105}\frac{ m^7}{k^6}+
\right.\right.\0\\ & \quad \quad \quad \left.\left.
 + \frac{32 i }{945}\frac{ m^9}{k^8}+\frac{128 i }{3465}\frac{ m^{11}}{k^{10}}+\ldots\right)\right) }
\al{\label{eq:fc:2:2:5:ir}\tilde{T}_{2,2;5\text{D}}^{\text{f,t,IR}} & = k^4 \pi _{\mu\nu}^2 \left(\frac{i }{2 \pi ^2} \left(\frac{1}{9}\frac{ m^3}{k^2}-\frac{1}{20} m+\frac{1}{336}\frac{ k^2}{m}+\frac{1}{8640}\frac{ k^4}{m^3}+\frac{1}{98560}\frac{ k^6}{m^5}+\frac{1}{838656}\frac{ k^8}{m^7}+
\right.\right.\0\\ & \quad \quad \quad \left.\left.
 + \frac{1}{6082560}\frac{ k^{10}}{m^9}+\frac{1}{39829504}\frac{ k^{12}}{m^{11}}+\ldots\right)\right)+
\0\\ & \quad 
 + k^4 \pi _{\mu\mu} \pi _{\nu\nu} \left(\frac{i }{6 \pi ^2} \left(-\frac{1}{3}\frac{ m^3}{k^2}+\frac{1}{20} m-\frac{1}{560}\frac{ k^2}{m}-\frac{1}{20160}\frac{ k^4}{m^3}-\frac{1}{295680}\frac{ k^6}{m^5}-
\right.\right.\0\\ & \quad \quad \quad \left.\left.
 - \frac{1}{3075072}\frac{ k^8}{m^7}-\frac{1}{26357760}\frac{ k^{10}}{m^9}-\frac{1}{199147520}\frac{ k^{12}}{m^{11}}+\ldots\right)\right) }
\al{\label{eq:fc:2:2:5:uvir}\tilde{T}_{2,2;5\text{D}}^{\text{f,UV-IR}} & =  \ldots \textrm{(i.e.\ no overlap)} }
Fermions, spin 2 x 2, dimension 6:
\al{\label{eq:fc:2:2:6:uv}\tilde{T}_{2,2;6\text{D}}^{\text{f,t,UV}} & = k^4 \pi _{\mu\nu}^2 \left(\frac{i }{4 \pi ^3} \left( \left(\frac{31}{1764}-\frac{P}{168}\right) k^2+ \left(-\frac{3}{25}+\frac{P}{20}\right) m^2+ \left(\frac{1}{18}-\frac{P}{12}\right)\frac{ m^4}{k^2}+
\right.\right.\0\\ & \quad \quad \quad \left.\left.
 +  \left(\frac{13}{36}-\frac{K}{6}\right)\frac{ m^6}{k^4}+ \left(\frac{17}{48}+\frac{K}{4}\right)\frac{ m^8}{k^6}+ \left(-\frac{7}{360}+\frac{K}{6}\right)\frac{ m^{10}}{k^8}+
\right.\right.\0\\ & \quad \quad \quad \left.\left.
 +  \left(-\frac{211}{1800}+\frac{7 K}{30}\right)\frac{ m^{12}}{k^{10}}+\ldots\right)\right)+
\0\\ & \quad 
 + k^4 \pi _{\mu\mu} \pi _{\nu\nu} \left(\frac{i }{3 \pi ^3} \left( \left(-\frac{11}{3675}+\frac{P}{1120}\right) k^2+ \left(\frac{23}{600}-\frac{P}{80}\right) m^2+
\right.\right.\0\\ & \quad \quad \quad \left.\left.
 +  \left(-\frac{1}{6}+\frac{P}{16}\right)\frac{ m^4}{k^2}+ \left(\frac{1}{48}-\frac{K}{8}\right)\frac{ m^6}{k^4}+ \left(\frac{25}{192}+\frac{K}{16}\right)\frac{ m^8}{k^6}+
\right.\right.\0\\ & \quad \quad \quad \left.\left.
 +  \left(\frac{17}{2400}+\frac{K}{40}\right)\frac{ m^{10}}{k^8}+ \left(-\frac{13}{2400}+\frac{K}{40}\right)\frac{ m^{12}}{k^{10}}+\ldots\right)\right) }
\al{\label{eq:fc:2:2:6:ir}\tilde{T}_{2,2;6\text{D}}^{\text{f,t,IR}} & = k^4 \pi _{\mu\nu}^2 \left(\frac{i }{16 \pi ^3} \left( \left(\frac{1}{2}-\frac{L_0}{3}\right)\frac{ m^4}{k^2}+ \left(-\frac{1}{5}+\frac{L_0}{5}\right) m^2-\frac{ L_0}{42} k^2+\frac{1}{540}\frac{ k^4}{m^2}+
\right.\right.\0\\ & \quad \quad \quad \left.\left.
 + \frac{1}{9240}\frac{ k^6}{m^4}+\frac{1}{98280}\frac{ k^8}{m^6}+\frac{1}{831600}\frac{ k^{10}}{m^8}+\frac{1}{6126120}\frac{ k^{12}}{m^{10}}+\frac{1}{41081040}\frac{ k^{14}}{m^{12}}+
\right.\right.\0\\ & \quad \quad \quad \left.\left.
 + \ldots\right)\right)+
\0\\ & \quad 
 + k^4 \pi _{\mu\mu} \pi _{\nu\nu} \left(\frac{i }{16 \pi ^3} \left( \left(-\frac{1}{2}+\frac{L_0}{3}\right)\frac{ m^4}{k^2}+ \left(\frac{1}{15}-\frac{L_0}{15}\right) m^2+\frac{ L_0}{210} k^2-
\right.\right.\0\\ & \quad \quad \quad \left.\left.
 - \frac{1}{3780}\frac{ k^4}{m^2}-\frac{1}{83160}\frac{ k^6}{m^4}-\frac{1}{1081080}\frac{ k^8}{m^6}-\frac{1}{10810800}\frac{ k^{10}}{m^8}-
\right.\right.\0\\ & \quad \quad \quad \left.\left.
 - \frac{1}{91891800}\frac{ k^{12}}{m^{10}}-\frac{1}{698377680}\frac{ k^{14}}{m^{12}}+\ldots\right)\right) }
\al{\label{eq:fc:2:2:6:uvir}\tilde{T}_{2,2;6\text{D}}^{\text{f,UV-IR}} & = k^4 \pi _{\mu\nu}^2 \left(\frac{i }{16 \pi ^3} \left( \left(\frac{31}{441}-\frac{K}{42}\right) k^2+ \left(-\frac{7}{25}+\frac{K}{5}\right) m^2- \left(\frac{5}{18}+\frac{K}{3}\right)\frac{ m^4}{k^2}\right)\right)+
\0\\ & \quad 
 + k^4 \pi _{\mu\mu} \pi _{\nu\nu} \left(\frac{i }{3 \pi ^3} \left( \left(-\frac{11}{3675}+\frac{K}{1120}\right) k^2+ \left(\frac{31}{1200}-\frac{K}{80}\right) m^2+
\right.\right.\0\\ & \quad \quad \quad \left.\left.
 +  \left(-\frac{7}{96}+\frac{K}{16}\right)\frac{ m^4}{k^2}\right)\right)+\ldots }
Fermions, spin 2 x 4, dimension 3:
\al{\label{eq:fc:2:4:3:uv}\tilde{T}_{2,4;3\text{D}}^{\text{f,t,UV}} & = k^6 \pi _{\mu\nu}^2 \pi _{\nu\nu} \left(\frac{1}{64}\frac{1}{k}-\frac{2 i }{3 \pi }\frac{ m^3}{k^4}-\frac{3 }{4}\frac{ m^4}{k^5}+\frac{24 i }{5 \pi }\frac{ m^5}{k^6}+2 \frac{ m^6}{k^7}-\frac{32 i }{7 \pi }\frac{ m^7}{k^8}-\frac{128 i }{45 \pi }\frac{ m^9}{k^{10}}-
\right.\0\\ & \quad \quad \left.
 - \frac{1536 i }{385 \pi }\frac{ m^{11}}{k^{12}}+\ldots\right)+
\0\\ & \quad 
 + k^6 \pi _{\mu\mu} \pi _{\nu\nu}^2 \left(-\frac{1}{128}\frac{1}{k}+\frac{3 }{32}\frac{ m^2}{k^3}-\frac{2 i }{3 \pi }\frac{ m^3}{k^4}-\frac{3 }{8}\frac{ m^4}{k^5}+\frac{8 i }{5 \pi }\frac{ m^5}{k^6}+\frac{1}{2}\frac{ m^6}{k^7}-
\right.\0\\ & \quad \quad \left.
 - \frac{32 i }{35 \pi }\frac{ m^7}{k^8}-\frac{128 i }{315 \pi }\frac{ m^9}{k^{10}}-\frac{512 i }{1155 \pi }\frac{ m^{11}}{k^{12}}+\ldots\right)+
\0\\ & \quad 
 + k^4 (k\cdot\epsilon)_{\mu\nu} \pi _{\mu\nu} \pi _{\nu\nu} \left(-\frac{3 i }{32}\frac{ m}{k}-\frac{1}{\pi }\frac{ m^2}{k^2}+\frac{3 i }{4}\frac{ m^3}{k^3}+\frac{4 }{\pi }\frac{ m^4}{k^4}-\frac{3 i }{2}\frac{ m^5}{k^5}-\frac{16 }{5 \pi }\frac{ m^6}{k^6}-
\right.\0\\ & \quad \quad \left.
 - \frac{64 }{35 \pi }\frac{ m^8}{k^8}-\frac{256 }{105 \pi }\frac{ m^{10}}{k^{10}}-\frac{1024 }{231 \pi }\frac{ m^{12}}{k^{12}}+\ldots\right) }
\al{\label{eq:fc:2:4:3:ir}\tilde{T}_{2,4;3\text{D}}^{\text{f,t,IR}} & = k^6 \pi _{\mu\nu}^2 \pi _{\nu\nu} \left(\frac{i }{2 \pi } \left(\frac{1}{5}\frac{ m}{k^2}-\frac{3 }{140}\frac{1}{m}-\frac{1}{1008}\frac{ k^2}{m^3}-\frac{1}{10560}\frac{ k^4}{m^5}-\frac{3 }{256256}\frac{ k^6}{m^7}-
\right.\right.\0\\ & \quad \quad \quad \left.\left.
 - \frac{1}{599040}\frac{ k^8}{m^9}-\frac{1}{3829760}\frac{ k^{10}}{m^{11}}+\ldots\right)\right)+
\0\\ & \quad 
 + k^6 \pi _{\mu\mu} \pi _{\nu\nu}^2 \left(\frac{i }{2 \pi } \left(-\frac{1}{5}\frac{ m}{k^2}+\frac{1}{140}\frac{1}{m}+\frac{1}{5040}\frac{ k^2}{m^3}+\frac{1}{73920}\frac{ k^4}{m^5}+\frac{1}{768768}\frac{ k^6}{m^7}+
\right.\right.\0\\ & \quad \quad \quad \left.\left.
 + \frac{1}{6589440}\frac{ k^8}{m^9}+\frac{1}{49786880}\frac{ k^{10}}{m^{11}}+\ldots\right)\right)+
\0\\ & \quad 
 + k^4 (k\cdot\epsilon)_{\mu\nu} \pi _{\mu\nu} \pi _{\nu\nu} \left(\frac{1}{4 \pi } \left(-\frac{1}{5}-\frac{1}{140}\frac{ k^2}{m^2}-\frac{1}{1680}\frac{ k^4}{m^4}-\frac{1}{14784}\frac{ k^6}{m^6}-
\right.\right.\0\\ & \quad \quad \quad \left.\left.
 - \frac{1}{109824}\frac{ k^8}{m^8}-\frac{1}{732160}\frac{ k^{10}}{m^{10}}-\frac{1}{4526080}\frac{ k^{12}}{m^{12}}+\ldots\right)\right) }
\al{\label{eq:fc:2:4:3:uvir}\tilde{T}_{2,4;3\text{D}}^{\text{f,UV-IR}} & =  \ldots \textrm{(i.e.\ no overlap)} }
Fermions, spin 2 x 4, dimension 4:
\al{\label{eq:fc:2:4:4:uv}\tilde{T}_{2,4;4\text{D}}^{\text{f,t,UV}} & = k^6 \pi _{\mu\nu}^2 \pi _{\nu\nu} \left(\frac{i }{\pi ^2} \left( \left(-\frac{141}{4900}+\frac{3 P}{280}\right)+ \left(\frac{4}{75}-\frac{P}{20}\right)\frac{ m^2}{k^2}+ \left(\frac{19}{24}-\frac{K}{4}\right)\frac{ m^4}{k^4}+
\right.\right.\0\\ & \quad \quad \quad \left.\left.
 +  \left(-\frac{5}{4}+\frac{3 K}{2}\right)\frac{ m^6}{k^6}- \left(\frac{101}{48}+\frac{5 K}{4}\right)\frac{ m^8}{k^8}+ \left(\frac{1}{600}-\frac{7 K}{10}\right)\frac{ m^{10}}{k^{10}}+
\right.\right.\0\\ & \quad \quad \quad \left.\left.
 +  \left(\frac{79}{200}-\frac{9 K}{10}\right)\frac{ m^{12}}{k^{12}}+\ldots\right)\right)+
\0\\ & \quad 
 + k^6 \pi _{\mu\mu} \pi _{\nu\nu}^2 \left(\frac{i }{\pi ^2} \left( \left(\frac{44}{3675}-\frac{P}{280}\right)+ \left(-\frac{23}{150}+\frac{P}{20}\right)\frac{ m^2}{k^2}+ \left(\frac{7}{24}-\frac{K}{4}\right)\frac{ m^4}{k^4}+
\right.\right.\0\\ & \quad \quad \quad \left.\left.
 +  \left(-\frac{1}{12}+\frac{K}{2}\right)\frac{ m^6}{k^6}- \left(\frac{25}{48}+\frac{K}{4}\right)\frac{ m^8}{k^8}- \left(\frac{17}{600}+\frac{K}{10}\right)\frac{ m^{10}}{k^{10}}+
\right.\right.\0\\ & \quad \quad \quad \left.\left.
 +  \left(\frac{13}{600}-\frac{K}{10}\right)\frac{ m^{12}}{k^{12}}+\ldots\right)\right) }
\al{\label{eq:fc:2:4:4:ir}\tilde{T}_{2,4;4\text{D}}^{\text{f,t,IR}} & = k^6 \pi _{\mu\nu}^2 \pi _{\nu\nu} \left(\frac{i }{4 \pi ^2} \left( \left(\frac{1}{5}-\frac{L_0}{5}\right)\frac{ m^2}{k^2}+\frac{3  L_0}{70}-\frac{1}{252}\frac{ k^2}{m^2}-\frac{1}{3960}\frac{ k^4}{m^4}-\frac{1}{40040}\frac{ k^6}{m^6}-
\right.\right.\0\\ & \quad \quad \quad \left.\left.
 - \frac{1}{327600}\frac{ k^8}{m^8}-\frac{1}{2356200}\frac{ k^{10}}{m^{10}}-\frac{1}{15519504}\frac{ k^{12}}{m^{12}}+\ldots\right)\right)+
\0\\ & \quad 
 + k^6 \pi _{\mu\mu} \pi _{\nu\nu}^2 \left(\frac{i }{20 \pi ^2} \left( \left(-1+L_0\right)\frac{ m^2}{k^2}-\frac{ L_0}{14}+\frac{1}{252}\frac{ k^2}{m^2}+\frac{1}{5544}\frac{ k^4}{m^4}+\frac{1}{72072}\frac{ k^6}{m^6}+
\right.\right.\0\\ & \quad \quad \quad \left.\left.
 + \frac{1}{720720}\frac{ k^8}{m^8}+\frac{1}{6126120}\frac{ k^{10}}{m^{10}}+\frac{1}{46558512}\frac{ k^{12}}{m^{12}}+\ldots\right)\right) }
\al{\label{eq:fc:2:4:4:uvir}\tilde{T}_{2,4;4\text{D}}^{\text{f,UV-IR}} & = k^6 \pi _{\mu\nu}^2 \pi _{\nu\nu} \left(\frac{i }{20 \pi ^2} \left( \left(-\frac{141}{245}+\frac{3 K}{14}\right)+ \left(\frac{1}{15}-K\right)\frac{ m^2}{k^2}\right)\right)+
\0\\ & \quad 
 + k^6 \pi _{\mu\mu} \pi _{\nu\nu}^2 \left(\frac{i }{5 \pi ^2} \left( \left(\frac{44}{735}-\frac{K}{56}\right)+ \left(-\frac{31}{60}+\frac{K}{4}\right)\frac{ m^2}{k^2}\right)\right)+\ldots }
Fermions, spin 2 x 4, dimension 5:
\al{\label{eq:fc:2:4:5:uv}\tilde{T}_{2,4;5\text{D}}^{\text{f,t,UV}} & = k^6 \pi _{\mu\nu}^2 \pi _{\nu\nu} \left(\frac{1}{\pi ^2} \left(\frac{ \pi }{1024} k-\frac{ \pi }{128}\frac{ m^2}{k}+\frac{2 i }{15}\frac{ m^5}{k^4}+\frac{ \pi }{8}\frac{ m^6}{k^5}-\frac{24 i }{35}\frac{ m^7}{k^6}-\frac{ \pi }{4}\frac{ m^8}{k^7}+
\right.\right.\0\\ & \quad \quad \quad \left.\left.
 + \frac{32 i }{63}\frac{ m^9}{k^8}+\frac{128 i }{495}\frac{ m^{11}}{k^{10}}+\ldots\right)\right)+
\0\\ & \quad 
 + k^6 \pi _{\mu\mu} \pi _{\nu\nu}^2 \left(\frac{1}{\pi ^2} \left(-\frac{ \pi }{4096} k+\frac{ \pi }{256}\frac{ m^2}{k}-\frac{3  \pi }{128}\frac{ m^4}{k^3}+\frac{2 i }{15}\frac{ m^5}{k^4}+\frac{ \pi }{16}\frac{ m^6}{k^5}-\frac{8 i }{35}\frac{ m^7}{k^6}-
\right.\right.\0\\ & \quad \quad \quad \left.\left.
 - \frac{ \pi }{16}\frac{ m^8}{k^7}+\frac{32 i }{315}\frac{ m^9}{k^8}+\frac{128 i }{3465}\frac{ m^{11}}{k^{10}}+\ldots\right)\right) }
\al{\label{eq:fc:2:4:5:ir}\tilde{T}_{2,4;5\text{D}}^{\text{f,t,IR}} & = k^6 \pi _{\mu\nu}^2 \pi _{\nu\nu} \left(\frac{i }{2 \pi ^2} \left(-\frac{1}{15}\frac{ m^3}{k^2}+\frac{3 }{140} m-\frac{1}{1008}\frac{ k^2}{m}-\frac{1}{31680}\frac{ k^4}{m^3}-\frac{3 }{1281280}\frac{ k^6}{m^5}-
\right.\right.\0\\ & \quad \quad \quad \left.\left.
 - \frac{1}{4193280}\frac{ k^8}{m^7}-\frac{1}{34467840}\frac{ k^{10}}{m^9}-\frac{3 }{756760576}\frac{ k^{12}}{m^{11}}+\ldots\right)\right)+
\0\\ & \quad 
 + k^6 \pi _{\mu\mu} \pi _{\nu\nu}^2 \left(\frac{i }{10 \pi ^2} \left(\frac{1}{3}\frac{ m^3}{k^2}-\frac{1}{28} m+\frac{1}{1008}\frac{ k^2}{m}+\frac{1}{44352}\frac{ k^4}{m^3}+\frac{1}{768768}\frac{ k^6}{m^5}+
\right.\right.\0\\ & \quad \quad \quad \left.\left.
 + \frac{1}{9225216}\frac{ k^8}{m^7}+\frac{1}{89616384}\frac{ k^{10}}{m^9}+\frac{1}{756760576}\frac{ k^{12}}{m^{11}}+\ldots\right)\right) }
\al{\label{eq:fc:2:4:5:uvir}\tilde{T}_{2,4;5\text{D}}^{\text{f,UV-IR}} & =  \ldots \textrm{(i.e.\ no overlap)} }
Fermions, spin 2 x 4, dimension 6:
\al{\label{eq:fc:2:4:6:uv}\tilde{T}_{2,4;6\text{D}}^{\text{f,t,UV}} & = k^6 \pi _{\mu\nu}^2 \pi _{\nu\nu} \left(\frac{i }{\pi ^3} \left( \left(-\frac{25}{15876}+\frac{P}{2016}\right) k^2+ \left(\frac{141}{9800}-\frac{3 P}{560}\right) m^2+
\right.\right.\0\\ & \quad \quad \quad \left.\left.
 +  \left(-\frac{1}{75}+\frac{P}{80}\right)\frac{ m^4}{k^2}+ \left(-\frac{17}{144}+\frac{K}{24}\right)\frac{ m^6}{k^4}+ \left(\frac{7}{64}-\frac{3 K}{16}\right)\frac{ m^8}{k^6}+
\right.\right.\0\\ & \quad \quad \quad \left.\left.
 +  \left(\frac{113}{480}+\frac{K}{8}\right)\frac{ m^{10}}{k^8}+ \left(\frac{23}{2400}+\frac{7 K}{120}\right)\frac{ m^{12}}{k^{10}}+\ldots\right)\right)+
\0\\ & \quad 
 + k^6 \pi _{\mu\mu} \pi _{\nu\nu}^2 \left(\frac{i }{\pi ^3} \left( \left(\frac{563}{1587600}-\frac{P}{10080}\right) k^2+ \left(-\frac{22}{3675}+\frac{P}{560}\right) m^2+
\right.\right.\0\\ & \quad \quad \quad \left.\left.
 +  \left(\frac{23}{600}-\frac{P}{80}\right)\frac{ m^4}{k^2}+ \left(-\frac{5}{144}+\frac{K}{24}\right)\frac{ m^6}{k^4}- \left(\frac{1}{192}+\frac{K}{16}\right)\frac{ m^8}{k^6}+
\right.\right.\0\\ & \quad \quad \quad \left.\left.
 +  \left(\frac{137}{2400}+\frac{K}{40}\right)\frac{ m^{10}}{k^8}+ \left(\frac{3}{800}+\frac{K}{120}\right)\frac{ m^{12}}{k^{10}}+\ldots\right)\right) }
\al{\label{eq:fc:2:4:6:ir}\tilde{T}_{2,4;6\text{D}}^{\text{f,t,IR}} & = k^6 \pi _{\mu\nu}^2 \pi _{\nu\nu} \left(\frac{i }{16 \pi ^3} \left( \left(-\frac{3}{10}+\frac{L_0}{5}\right)\frac{ m^4}{k^2}+ \left(\frac{3}{35}-\frac{3 L_0}{35}\right) m^2+\frac{ L_0}{126} k^2-\frac{1}{1980}\frac{ k^4}{m^2}-
\right.\right.\0\\ & \quad \quad \quad \left.\left.
 - \frac{1}{40040}\frac{ k^6}{m^4}-\frac{1}{491400}\frac{ k^8}{m^6}-\frac{1}{4712400}\frac{ k^{10}}{m^8}-\frac{1}{38798760}\frac{ k^{12}}{m^{10}}-
\right.\right.\0\\ & \quad \quad \quad \left.\left.
 - \frac{1}{287567280}\frac{ k^{14}}{m^{12}}+\ldots\right)\right)+
\0\\ & \quad 
 + k^6 \pi _{\mu\mu} \pi _{\nu\nu}^2 \left(\frac{i }{80 \pi ^3} \left( \left(\frac{3}{2}-L_0\right)\frac{ m^4}{k^2}+ \left(-\frac{1}{7}+\frac{L_0}{7}\right) m^2-\frac{ L_0}{126} k^2+\frac{1}{2772}\frac{ k^4}{m^2}+
\right.\right.\0\\ & \quad \quad \quad \left.\left.
 + \frac{1}{72072}\frac{ k^6}{m^4}+\frac{1}{1081080}\frac{ k^8}{m^6}+\frac{1}{12252240}\frac{ k^{10}}{m^8}+\frac{1}{116396280}\frac{ k^{12}}{m^{10}}+
\right.\right.\0\\ & \quad \quad \quad \left.\left.
 + \frac{1}{977728752}\frac{ k^{14}}{m^{12}}+\ldots\right)\right) }
\al{\label{eq:fc:2:4:6:uvir}\tilde{T}_{2,4;6\text{D}}^{\text{f,UV-IR}} & = k^6 \pi _{\mu\nu}^2 \pi _{\nu\nu} \left(\frac{i }{4 \pi ^3} \left( \left(-\frac{25}{3969}+\frac{K}{504}\right) k^2+ \left(\frac{177}{4900}-\frac{3 K}{140}\right) m^2+
\right.\right.\0\\ & \quad \quad \quad \left.\left.
 +  \left(\frac{13}{600}+\frac{K}{20}\right)\frac{ m^4}{k^2}\right)\right)+
\0\\ & \quad 
 + k^6 \pi _{\mu\mu} \pi _{\nu\nu}^2 \left(\frac{i }{80 \pi ^3} \left( \left(\frac{563}{19845}-\frac{K}{126}\right) k^2+ \left(-\frac{247}{735}+\frac{K}{7}\right) m^2+
\right.\right.\0\\ & \quad \quad \quad \left.\left.
 +  \left(\frac{47}{30}-K\right)\frac{ m^4}{k^2}\right)\right)+\ldots }
Fermions, spin 3 x 3, dimension 3:
\al{\label{eq:fc:3:3:3:uv}\tilde{T}_{3,3;3\text{D}}^{\text{f,t,UV}} & = k^6 \pi _{\mu\nu}^3 \left(\frac{1}{64}\frac{1}{k}-\frac{1}{16}\frac{ m^2}{k^3}-\frac{1}{4}\frac{ m^4}{k^5}+\frac{32 i }{15 \pi }\frac{ m^5}{k^6}+\frac{m^6}{k^7}-\frac{256 i }{105 \pi }\frac{ m^7}{k^8}-\frac{512 i }{315 \pi }\frac{ m^9}{k^{10}}-
\right.\0\\ & \quad \quad \left.
 - \frac{8192 i }{3465 \pi }\frac{ m^{11}}{k^{12}}+\ldots\right)+
\0\\ & \quad 
 + k^6 \pi _{\mu\mu} \pi _{\mu\nu} \pi _{\nu\nu} \left(-\frac{1}{128}\frac{1}{k}+\frac{5 }{32}\frac{ m^2}{k^3}-\frac{4 i }{3 \pi }\frac{ m^3}{k^4}-\frac{7 }{8}\frac{ m^4}{k^5}+\frac{64 i }{15 \pi }\frac{ m^5}{k^6}+\frac{3 }{2}\frac{ m^6}{k^7}-
\right.\0\\ & \quad \quad \left.
 - \frac{64 i }{21 \pi }\frac{ m^7}{k^8}-\frac{512 i }{315 \pi }\frac{ m^9}{k^{10}}-\frac{1024 i }{495 \pi }\frac{ m^{11}}{k^{12}}+\ldots\right)+
\0\\ & \quad 
 + k^4 (k\cdot\epsilon)_{\mu\nu} \pi _{\mu\nu}^2 \left(-\frac{i }{16}\frac{ m}{k}-\frac{1}{\pi }\frac{ m^2}{k^2}+\frac{i }{2}\frac{ m^3}{k^3}+\frac{8 }{3 \pi }\frac{ m^4}{k^4}-i \frac{ m^5}{k^5}-\frac{32 }{15 \pi }\frac{ m^6}{k^6}-
\right.\0\\ & \quad \quad \left.
 - \frac{128 }{105 \pi }\frac{ m^8}{k^8}-\frac{512 }{315 \pi }\frac{ m^{10}}{k^{10}}-\frac{2048 }{693 \pi }\frac{ m^{12}}{k^{12}}+\ldots\right)+
\0\\ & \quad 
 + k^4 (k\cdot\epsilon)_{\mu\nu} \pi _{\mu\mu} \pi _{\nu\nu} \left(-\frac{i }{32}\frac{ m}{k}+\frac{i }{4}\frac{ m^3}{k^3}+\frac{4 }{3 \pi }\frac{ m^4}{k^4}-\frac{i }{2}\frac{ m^5}{k^5}-\frac{16 }{15 \pi }\frac{ m^6}{k^6}-
\right.\0\\ & \quad \quad \left.
 - \frac{64 }{105 \pi }\frac{ m^8}{k^8}-\frac{256 }{315 \pi }\frac{ m^{10}}{k^{10}}-\frac{1024 }{693 \pi }\frac{ m^{12}}{k^{12}}+\ldots\right) }
\al{\label{eq:fc:3:3:3:ir}\tilde{T}_{3,3;3\text{D}}^{\text{f,t,IR}} & = k^6 \pi _{\mu\nu}^3 \left(\frac{i }{3 \pi } \left(\frac{2 }{5}\frac{ m}{k^2}-\frac{1}{35}\frac{1}{m}-\frac{1}{840}\frac{ k^2}{m^3}-\frac{1}{9240}\frac{ k^4}{m^5}-\frac{5 }{384384}\frac{ k^6}{m^7}-\frac{1}{549120}\frac{ k^8}{m^9}-
\right.\right.\0\\ & \quad \quad \quad \left.\left.
 - \frac{7 }{24893440}\frac{ k^{10}}{m^{11}}+\ldots\right)\right)+
\0\\ & \quad 
 + k^6 \pi _{\mu\mu} \pi _{\mu\nu} \pi _{\nu\nu} \left(\frac{i }{3 \pi } \left(-\frac{2 }{5}\frac{ m}{k^2}+\frac{1}{140}\frac{1}{m}-\frac{1}{73920}\frac{ k^4}{m^5}-\frac{1}{384384}\frac{ k^6}{m^7}-
\right.\right.\0\\ & \quad \quad \quad \left.\left.
 - \frac{1}{2196480}\frac{ k^8}{m^9}-\frac{1}{12446720}\frac{ k^{10}}{m^{11}}+\ldots\right)\right)+
\0\\ & \quad 
 + k^4 (k\cdot\epsilon)_{\mu\nu} \pi _{\mu\nu}^2 \left(\frac{1}{3 \pi } \left(-\frac{ m^2}{k^2}-\frac{1}{10}-\frac{1}{280}\frac{ k^2}{m^2}-\frac{1}{3360}\frac{ k^4}{m^4}-\frac{1}{29568}\frac{ k^6}{m^6}-
\right.\right.\0\\ & \quad \quad \quad \left.\left.
 - \frac{1}{219648}\frac{ k^8}{m^8}-\frac{1}{1464320}\frac{ k^{10}}{m^{10}}-\frac{1}{9052160}\frac{ k^{12}}{m^{12}}+\ldots\right)\right)+
\0\\ & \quad 
 + k^4 (k\cdot\epsilon)_{\mu\nu} \pi _{\mu\mu} \pi _{\nu\nu} \left(\frac{1}{3 \pi } \left(\frac{m^2}{k^2}-\frac{1}{20}-\frac{1}{560}\frac{ k^2}{m^2}-\frac{1}{6720}\frac{ k^4}{m^4}-\frac{1}{59136}\frac{ k^6}{m^6}-
\right.\right.\0\\ & \quad \quad \quad \left.\left.
 - \frac{1}{439296}\frac{ k^8}{m^8}-\frac{1}{2928640}\frac{ k^{10}}{m^{10}}-\frac{1}{18104320}\frac{ k^{12}}{m^{12}}+\ldots\right)\right) }
\al{\label{eq:fc:3:3:3:uvir}\tilde{T}_{3,3;3\text{D}}^{\text{f,UV-IR}} & = k^4 (k\cdot\epsilon)_{\mu\nu} \pi _{\mu\nu}^2 \left(-\frac{2 }{3 \pi }\frac{ m^2}{k^2}\right)+\ldots }
Fermions, spin 3 x 3, dimension 4:
\al{\label{eq:fc:3:3:4:uv}\tilde{T}_{3,3;4\text{D}}^{\text{f,t,UV}} & = k^6 \pi _{\mu\nu}^3 \left(\frac{i }{\pi ^2} \left( \left(-\frac{599}{22050}+\frac{P}{105}\right)+ \left(\frac{31}{225}-\frac{P}{15}\right)\frac{ m^2}{k^2}+\frac{1}{3}\frac{ m^4}{k^4}+
\right.\right.\0\\ & \quad \quad \quad \left.\left.
 +  \left(-\frac{7}{9}+\frac{2 K}{3}\right)\frac{ m^6}{k^6}- \left(\frac{19}{18}+\frac{2 K}{3}\right)\frac{ m^8}{k^8}+ \left(\frac{1}{50}-\frac{2 K}{5}\right)\frac{ m^{10}}{k^{10}}+
\right.\right.\0\\ & \quad \quad \quad \left.\left.
 +  \left(\frac{56}{225}-\frac{8 K}{15}\right)\frac{ m^{12}}{k^{12}}+\ldots\right)\right)+
\0\\ & \quad 
 + k^6 \pi _{\mu\mu} \pi _{\mu\nu} \pi _{\nu\nu} \left(\frac{i }{\pi ^2} \left( \left(\frac{457}{44100}-\frac{P}{420}\right)+ \left(-\frac{107}{450}+\frac{P}{15}\right)\frac{ m^2}{k^2}+
\right.\right.\0\\ & \quad \quad \quad \left.\left.
 +  \left(\frac{3}{4}-\frac{K}{2}\right)\frac{ m^4}{k^4}+ \left(-\frac{5}{9}+\frac{4 K}{3}\right)\frac{ m^6}{k^6}- \left(\frac{113}{72}+\frac{5 K}{6}\right)\frac{ m^8}{k^8}-
\right.\right.\0\\ & \quad \quad \quad \left.\left.
 -  \left(\frac{7}{150}+\frac{2 K}{5}\right)\frac{ m^{10}}{k^{10}}+ \left(\frac{151}{900}-\frac{7 K}{15}\right)\frac{ m^{12}}{k^{12}}+\ldots\right)\right) }
\al{\label{eq:fc:3:3:4:ir}\tilde{T}_{3,3;4\text{D}}^{\text{f,t,IR}} & = k^6 \pi _{\mu\nu}^3 \left(\frac{i }{3 \pi ^2} \left( \left(\frac{1}{5}-\frac{L_0}{5}\right)\frac{ m^2}{k^2}+\frac{ L_0}{35}-\frac{1}{420}\frac{ k^2}{m^2}-\frac{1}{6930}\frac{ k^4}{m^4}-\frac{1}{72072}\frac{ k^6}{m^6}-
\right.\right.\0\\ & \quad \quad \quad \left.\left.
 - \frac{1}{600600}\frac{ k^8}{m^8}-\frac{1}{4375800}\frac{ k^{10}}{m^{10}}-\frac{1}{29099070}\frac{ k^{12}}{m^{12}}+\ldots\right)\right)+
\0\\ & \quad 
 + k^6 \pi _{\mu\mu} \pi _{\mu\nu} \pi _{\nu\nu} \left(\frac{i }{3 \pi ^2} \left( \left(-\frac{1}{5}+\frac{L_0}{5}\right)\frac{ m^2}{k^2}-\frac{ L_0}{140}-\frac{1}{55440}\frac{ k^4}{m^4}-\frac{1}{360360}\frac{ k^6}{m^6}-
\right.\right.\0\\ & \quad \quad \quad \left.\left.
 - \frac{1}{2402400}\frac{ k^8}{m^8}-\frac{1}{15315300}\frac{ k^{10}}{m^{10}}-\frac{1}{93117024}\frac{ k^{12}}{m^{12}}+\ldots\right)\right) }
\al{\label{eq:fc:3:3:4:uvir}\tilde{T}_{3,3;4\text{D}}^{\text{f,UV-IR}} & = k^6 \pi _{\mu\nu}^3 \left(\frac{i }{15 \pi ^2} \left( \left(-\frac{599}{1470}+\frac{K}{7}\right)+ \left(\frac{16}{15}-K\right)\frac{ m^2}{k^2}\right)\right)+
\0\\ & \quad 
 + k^6 \pi _{\mu\mu} \pi _{\mu\nu} \pi _{\nu\nu} \left(\frac{i }{15 \pi ^2} \left( \left(\frac{457}{2940}-\frac{K}{28}\right)+ \left(-\frac{77}{30}+K\right)\frac{ m^2}{k^2}\right)\right)+\ldots }
Fermions, spin 3 x 3, dimension 5:
\al{\label{eq:fc:3:3:5:uv}\tilde{T}_{3,3;5\text{D}}^{\text{f,t,UV}} & = k^6 \pi _{\mu\nu}^3 \left(\frac{1}{\pi ^2} \left(\frac{5  \pi }{6144} k-\frac{ \pi }{128}\frac{ m^2}{k}+\frac{ \pi }{64}\frac{ m^4}{k^3}+\frac{ \pi }{24}\frac{ m^6}{k^5}-\frac{32 i }{105}\frac{ m^7}{k^6}-\frac{ \pi }{8}\frac{ m^8}{k^7}+
\right.\right.\0\\ & \quad \quad \quad \left.\left.
 + \frac{256 i }{945}\frac{ m^9}{k^8}+\frac{512 i }{3465}\frac{ m^{11}}{k^{10}}+\ldots\right)\right)+
\0\\ & \quad 
 + k^6 \pi _{\mu\mu} \pi _{\mu\nu} \pi _{\nu\nu} \left(\frac{1}{\pi ^2} \left(-\frac{ \pi }{12288} k+\frac{ \pi }{256}\frac{ m^2}{k}-\frac{5  \pi }{128}\frac{ m^4}{k^3}+\frac{4 i }{15}\frac{ m^5}{k^4}+\frac{7  \pi }{48}\frac{ m^6}{k^5}-
\right.\right.\0\\ & \quad \quad \quad \left.\left.
 - \frac{64 i }{105}\frac{ m^7}{k^6}-\frac{3  \pi }{16}\frac{ m^8}{k^7}+\frac{64 i }{189}\frac{ m^9}{k^8}+\frac{512 i }{3465}\frac{ m^{11}}{k^{10}}+\ldots\right)\right) }
\al{\label{eq:fc:3:3:5:ir}\tilde{T}_{3,3;5\text{D}}^{\text{f,t,IR}} & = k^6 \pi _{\mu\nu}^3 \left(\frac{i }{3 \pi ^2} \left(-\frac{2 }{15}\frac{ m^3}{k^2}+\frac{1}{35} m-\frac{1}{840}\frac{ k^2}{m}-\frac{1}{27720}\frac{ k^4}{m^3}-\frac{1}{384384}\frac{ k^6}{m^5}-
\right.\right.\0\\ & \quad \quad \quad \left.\left.
 - \frac{1}{3843840}\frac{ k^8}{m^7}-\frac{7 }{224040960}\frac{ k^{10}}{m^9}-\frac{1}{236487680}\frac{ k^{12}}{m^{11}}+\ldots\right)\right)+
\0\\ & \quad 
 + k^6 \pi _{\mu\mu} \pi _{\mu\nu} \pi _{\nu\nu} \left(\frac{i }{3 \pi ^2} \left(\frac{2 }{15}\frac{ m^3}{k^2}-\frac{1}{140} m-\frac{1}{221760}\frac{ k^4}{m^3}-\frac{1}{1921920}\frac{ k^6}{m^5}-
\right.\right.\0\\ & \quad \quad \quad \left.\left.
 - \frac{1}{15375360}\frac{ k^8}{m^7}-\frac{1}{112020480}\frac{ k^{10}}{m^9}-\frac{1}{756760576}\frac{ k^{12}}{m^{11}}+\ldots\right)\right) }
\al{\label{eq:fc:3:3:5:uvir}\tilde{T}_{3,3;5\text{D}}^{\text{f,UV-IR}} & =  \ldots \textrm{(i.e.\ no overlap)} }
Fermions, spin 3 x 3, dimension 6:
\al{\label{eq:fc:3:3:6:uv}\tilde{T}_{3,3;6\text{D}}^{\text{f,t,UV}} & = k^6 \pi _{\mu\nu}^3 \left(\frac{i }{3 \pi ^3} \left( \left(-\frac{1021}{264600}+\frac{P}{840}\right) k^2+ \left(\frac{599}{14700}-\frac{P}{70}\right) m^2+
\right.\right.\0\\ & \quad \quad \quad \left.\left.
 +  \left(-\frac{31}{300}+\frac{P}{20}\right)\frac{ m^4}{k^2}-\frac{1}{6}\frac{ m^6}{k^4}+ \left(\frac{11}{48}-\frac{K}{4}\right)\frac{ m^8}{k^6}+ \left(\frac{107}{300}+\frac{K}{5}\right)\frac{ m^{10}}{k^8}+
\right.\right.\0\\ & \quad \quad \quad \left.\left.
 +  \left(\frac{7}{600}+\frac{K}{10}\right)\frac{ m^{12}}{k^{10}}+\ldots\right)\right)+
\0\\ & \quad 
 + k^6 \pi _{\mu\mu} \pi _{\mu\nu} \pi _{\nu\nu} \left(\frac{i }{6 \pi ^3} \left(\frac{1}{2520} k^2+ \left(-\frac{457}{14700}+\frac{P}{140}\right) m^2+ \left(\frac{107}{300}-\frac{P}{10}\right)\frac{ m^4}{k^2}+
\right.\right.\0\\ & \quad \quad \quad \left.\left.
 +  \left(-\frac{7}{12}+\frac{K}{2}\right)\frac{ m^6}{k^4}+ \left(\frac{1}{6}-K\right)\frac{ m^8}{k^6}+ \left(\frac{25}{24}+\frac{K}{2}\right)\frac{ m^{10}}{k^8}+
\right.\right.\0\\ & \quad \quad \quad \left.\left.
 +  \left(\frac{17}{300}+\frac{K}{5}\right)\frac{ m^{12}}{k^{10}}+\ldots\right)\right) }
\al{\label{eq:fc:3:3:6:ir}\tilde{T}_{3,3;6\text{D}}^{\text{f,t,IR}} & = k^6 \pi _{\mu\nu}^3 \left(\frac{i }{2 \pi ^3} \left( \left(-\frac{1}{20}+\frac{L_0}{30}\right)\frac{ m^4}{k^2}+ \left(\frac{1}{105}-\frac{L_0}{105}\right) m^2+\frac{ L_0}{1260} k^2-\frac{1}{20790}\frac{ k^4}{m^2}-
\right.\right.\0\\ & \quad \quad \quad \left.\left.
 - \frac{1}{432432}\frac{ k^6}{m^4}-\frac{1}{5405400}\frac{ k^8}{m^6}-\frac{1}{52509600}\frac{ k^{10}}{m^8}-\frac{1}{436486050}\frac{ k^{12}}{m^{10}}-
\right.\right.\0\\ & \quad \quad \quad \left.\left.
 - \frac{1}{3259095840}\frac{ k^{14}}{m^{12}}+\ldots\right)\right)+
\0\\ & \quad 
 + k^6 \pi _{\mu\mu} \pi _{\mu\nu} \pi _{\nu\nu} \left(\frac{i }{20 \pi ^3} \left( \left(\frac{1}{2}-\frac{L_0}{3}\right)\frac{ m^4}{k^2}+ \left(-\frac{1}{42}+\frac{L_0}{42}\right) m^2-\frac{1}{16632}\frac{ k^4}{m^2}-
\right.\right.\0\\ & \quad \quad \quad \left.\left.
 - \frac{1}{216216}\frac{ k^6}{m^4}-\frac{1}{2162160}\frac{ k^8}{m^6}-\frac{1}{18378360}\frac{ k^{10}}{m^8}-\frac{1}{139675536}\frac{ k^{12}}{m^{10}}-
\right.\right.\0\\ & \quad \quad \quad \left.\left.
 - \frac{1}{977728752}\frac{ k^{14}}{m^{12}}+\ldots\right)\right) }
\al{\label{eq:fc:3:3:6:uvir}\tilde{T}_{3,3;6\text{D}}^{\text{f,UV-IR}} & = k^6 \pi _{\mu\nu}^3 \left(\frac{i }{30 \pi ^3} \left( \left(-\frac{1021}{26460}+\frac{K}{84}\right) k^2+ \left(\frac{389}{1470}-\frac{K}{7}\right) m^2+
\right.\right.\0\\ & \quad \quad \quad \left.\left.
 +  \left(-\frac{17}{60}+\frac{K}{2}\right)\frac{ m^4}{k^2}\right)\right)+
\0\\ & \quad 
 + k^6 \pi _{\mu\mu} \pi _{\mu\nu} \pi _{\nu\nu} \left(\frac{i }{15 \pi ^3} \left( \left(-\frac{44}{735}+\frac{K}{56}\right) m^2+ \left(\frac{31}{60}-\frac{K}{4}\right)\frac{ m^4}{k^2}\right)\right)+\ldots }
Fermions, spin 3 x 5, dimension 3:
\al{\label{eq:fc:3:5:3:uv}\tilde{T}_{3,5;3\text{D}}^{\text{f,t,UV}} & = k^8 \pi _{\mu\nu}^3 \pi _{\nu\nu} \left(-\frac{3 }{256}\frac{1}{k}+\frac{1}{16}\frac{ m^2}{k^3}+\frac{3 }{8}\frac{ m^4}{k^5}-\frac{64 i }{15 \pi }\frac{ m^5}{k^6}-3 \frac{ m^6}{k^7}+\frac{512 i }{35 \pi }\frac{ m^7}{k^8}+5 \frac{ m^8}{k^9}-
\right.\0\\ & \quad \quad \left.
 - \frac{1024 i }{105 \pi }\frac{ m^9}{k^{10}}-\frac{16384 i }{3465 \pi }\frac{ m^{11}}{k^{12}}+\ldots\right)+
\0\\ & \quad 
 + k^8 \pi _{\mu\mu} \pi _{\mu\nu} \pi _{\nu\nu}^2 \left(\frac{7 }{1024}\frac{1}{k}-\frac{9 }{64}\frac{ m^2}{k^3}+\frac{4 i }{3 \pi }\frac{ m^3}{k^4}+\frac{33 }{32}\frac{ m^4}{k^5}-\frac{32 i }{5 \pi }\frac{ m^5}{k^6}-\frac{13 }{4}\frac{ m^6}{k^7}+
\right.\0\\ & \quad \quad \left.
 + \frac{64 i }{5 \pi }\frac{ m^7}{k^8}+\frac{15 }{4}\frac{ m^8}{k^9}-\frac{2048 i }{315 \pi }\frac{ m^9}{k^{10}}-\frac{1024 i }{385 \pi }\frac{ m^{11}}{k^{12}}+\ldots\right)+
\0\\ & \quad 
 + k^6 (k\cdot\epsilon)_{\mu\nu} \pi _{\mu\nu}^2 \pi _{\nu\nu} \left(\frac{i }{16}\frac{ m}{k}+\frac{1}{\pi }\frac{ m^2}{k^2}-\frac{3 i }{4}\frac{ m^3}{k^3}-\frac{16 }{3 \pi }\frac{ m^4}{k^4}+3 i \frac{ m^5}{k^5}+\frac{64 }{5 \pi }\frac{ m^6}{k^6}-
\right.\0\\ & \quad \quad \left.
 - 4 i \frac{ m^7}{k^7}-\frac{256 }{35 \pi }\frac{ m^8}{k^8}-\frac{1024 }{315 \pi }\frac{ m^{10}}{k^{10}}-\frac{4096 }{1155 \pi }\frac{ m^{12}}{k^{12}}+\ldots\right)+
\0\\ & \quad 
 + k^6 (k\cdot\epsilon)_{\mu\nu} \pi _{\mu\mu} \pi _{\nu\nu}^2 \left(\frac{i }{64}\frac{ m}{k}-\frac{3 i }{16}\frac{ m^3}{k^3}-\frac{4 }{3 \pi }\frac{ m^4}{k^4}+\frac{3 i }{4}\frac{ m^5}{k^5}+\frac{16 }{5 \pi }\frac{ m^6}{k^6}-i \frac{ m^7}{k^7}-
\right.\0\\ & \quad \quad \left.
 - \frac{64 }{35 \pi }\frac{ m^8}{k^8}-\frac{256 }{315 \pi }\frac{ m^{10}}{k^{10}}-\frac{1024 }{1155 \pi }\frac{ m^{12}}{k^{12}}+\ldots\right) }
\al{\label{eq:fc:3:5:3:ir}\tilde{T}_{3,5;3\text{D}}^{\text{f,t,IR}} & = k^8 \pi _{\mu\nu}^3 \pi _{\nu\nu} \left(\frac{i }{\pi } \left(-\frac{4 }{35}\frac{ m}{k^2}+\frac{2 }{315}\frac{1}{m}+\frac{1}{4620}\frac{ k^2}{m^3}+\frac{1}{60060}\frac{ k^4}{m^5}+\frac{1}{576576}\frac{ k^6}{m^7}+
\right.\right.\0\\ & \quad \quad \quad \left.\left.
 + \frac{1}{4667520}\frac{ k^8}{m^9}+\frac{7 }{236487680}\frac{ k^{10}}{m^{11}}+\ldots\right)\right)+
\0\\ & \quad 
 + k^8 \pi _{\mu\mu} \pi _{\mu\nu} \pi _{\nu\nu}^2 \left(\frac{i }{5 \pi } \left(\frac{4 }{7}\frac{ m}{k^2}-\frac{1}{84}\frac{1}{m}-\frac{1}{5544}\frac{ k^2}{m^3}-\frac{1}{192192}\frac{ k^4}{m^5}+
\right.\right.\0\\ & \quad \quad \quad \left.\left.
 + \frac{1}{22404096}\frac{ k^8}{m^9}+\frac{1}{94595072}\frac{ k^{10}}{m^{11}}+\ldots\right)\right)+
\0\\ & \quad 
 + k^6 (k\cdot\epsilon)_{\mu\nu} \pi _{\mu\nu}^2 \pi _{\nu\nu} \left(\frac{1}{\pi } \left(\frac{1}{5}\frac{ m^2}{k^2}+\frac{1}{35}+\frac{1}{1260}\frac{ k^2}{m^2}+\frac{1}{18480}\frac{ k^4}{m^4}+\frac{1}{192192}\frac{ k^6}{m^6}+
\right.\right.\0\\ & \quad \quad \quad \left.\left.
 + \frac{1}{1647360}\frac{ k^8}{m^8}+\frac{1}{12446720}\frac{ k^{10}}{m^{10}}+\frac{1}{85995520}\frac{ k^{12}}{m^{12}}+\ldots\right)\right)+
\0\\ & \quad 
 + k^6 (k\cdot\epsilon)_{\mu\nu} \pi _{\mu\mu} \pi _{\nu\nu}^2 \left(\frac{1}{\pi } \left(-\frac{1}{5}\frac{ m^2}{k^2}+\frac{1}{140}+\frac{1}{5040}\frac{ k^2}{m^2}+\frac{1}{73920}\frac{ k^4}{m^4}+
\right.\right.\0\\ & \quad \quad \quad \left.\left.
 + \frac{1}{768768}\frac{ k^6}{m^6}+\frac{1}{6589440}\frac{ k^8}{m^8}+\frac{1}{49786880}\frac{ k^{10}}{m^{10}}+\frac{1}{343982080}\frac{ k^{12}}{m^{12}}+\ldots\right)\right) }
\al{\label{eq:fc:3:5:3:uvir}\tilde{T}_{3,5;3\text{D}}^{\text{f,UV-IR}} & = k^6 (k\cdot\epsilon)_{\mu\nu} \pi _{\mu\nu}^2 \pi _{\nu\nu} \left(\frac{4 }{5 \pi }\frac{ m^2}{k^2}\right)+\ldots }
Fermions, spin 3 x 5, dimension 4:
\al{\label{eq:fc:3:5:4:uv}\tilde{T}_{3,5;4\text{D}}^{\text{f,t,UV}} & = k^8 \pi _{\mu\nu}^3 \pi _{\nu\nu} \left(\frac{i }{\pi ^2} \left( \left(\frac{1937}{99225}-\frac{2 P}{315}\right)+ \left(-\frac{494}{3675}+\frac{2 P}{35}\right)\frac{ m^2}{k^2}-\frac{2 }{5}\frac{ m^4}{k^4}+
\right.\right.\0\\ & \quad \quad \quad \left.\left.
 +  \left(\frac{22}{9}-\frac{4 K}{3}\right)\frac{ m^6}{k^6}+ \left(-\frac{5}{3}+4 K\right)\frac{ m^8}{k^8}- \left(\frac{117}{25}+\frac{12 K}{5}\right)\frac{ m^{10}}{k^{10}}-
\right.\right.\0\\ & \quad \quad \quad \left.\left.
 -  \left(\frac{16}{75}+\frac{16 K}{15}\right)\frac{ m^{12}}{k^{12}}+\ldots\right)\right)+
\0\\ & \quad 
 + k^8 \pi _{\mu\mu} \pi _{\mu\nu} \pi _{\nu\nu}^2 \left(\frac{i }{\pi ^2} \left( \left(-\frac{1231}{132300}+\frac{P}{420}\right)+ \left(\frac{1513}{7350}-\frac{2 P}{35}\right)\frac{ m^2}{k^2}+
\right.\right.\0\\ & \quad \quad \quad \left.\left.
 +  \left(-\frac{53}{60}+\frac{K}{2}\right)\frac{ m^4}{k^4}+ (2-2 K)\frac{ m^6}{k^6}+ \left(-\frac{5}{24}+\frac{7 K}{2}\right)\frac{ m^8}{k^8}-
\right.\right.\0\\ & \quad \quad \quad \left.\left.
 -  \left(\frac{259}{75}+\frac{8 K}{5}\right)\frac{ m^{10}}{k^{10}}- \left(\frac{61}{300}+\frac{3 K}{5}\right)\frac{ m^{12}}{k^{12}}+\ldots\right)\right) }
\al{\label{eq:fc:3:5:4:ir}\tilde{T}_{3,5;4\text{D}}^{\text{f,t,IR}} & = k^8 \pi _{\mu\nu}^3 \pi _{\nu\nu} \left(\frac{i }{5 \pi ^2} \left( \left(-\frac{2}{7}+\frac{2 L_0}{7}\right)\frac{ m^2}{k^2}-\frac{2  L_0}{63}+\frac{1}{462}\frac{ k^2}{m^2}+\frac{1}{9009}\frac{ k^4}{m^4}+
\right.\right.\0\\ & \quad \quad \quad \left.\left.
 + \frac{1}{108108}\frac{ k^6}{m^6}+\frac{1}{1021020}\frac{ k^8}{m^8}+\frac{1}{8314020}\frac{ k^{10}}{m^{10}}+\frac{1}{61108047}\frac{ k^{12}}{m^{12}}+\ldots\right)\right)+
\0\\ & \quad 
 + k^8 \pi _{\mu\mu} \pi _{\mu\nu} \pi _{\nu\nu}^2 \left(\frac{i }{35 \pi ^2} \left( \left(2-2 L_0\right)\frac{ m^2}{k^2}+\frac{ L_0}{12}-\frac{1}{396}\frac{ k^2}{m^2}-\frac{1}{20592}\frac{ k^4}{m^4}+
\right.\right.\0\\ & \quad \quad \quad \left.\left.
 + \frac{1}{3500640}\frac{ k^8}{m^8}+\frac{1}{16628040}\frac{ k^{10}}{m^{10}}+\frac{1}{93117024}\frac{ k^{12}}{m^{12}}+\ldots\right)\right) }
\al{\label{eq:fc:3:5:4:uvir}\tilde{T}_{3,5;4\text{D}}^{\text{f,UV-IR}} & = k^8 \pi _{\mu\nu}^3 \pi _{\nu\nu} \left(\frac{i }{35 \pi ^2} \left( \left(\frac{1937}{2835}-\frac{2 K}{9}\right)+ \left(-\frac{284}{105}+2 K\right)\frac{ m^2}{k^2}\right)\right)+
\0\\ & \quad 
 + k^8 \pi _{\mu\mu} \pi _{\mu\nu} \pi _{\nu\nu}^2 \left(\frac{i }{35 \pi ^2} \left( \left(-\frac{1231}{3780}+\frac{K}{12}\right)+ \left(\frac{1093}{210}-2 K\right)\frac{ m^2}{k^2}\right)\right)+\ldots }
Fermions, spin 3 x 5, dimension 5:
\al{\label{eq:fc:3:5:5:uv}\tilde{T}_{3,5;5\text{D}}^{\text{f,t,UV}} & = k^8 \pi _{\mu\nu}^3 \pi _{\nu\nu} \left(\frac{1}{\pi ^2} \left(-\frac{ \pi }{2048} k+\frac{3  \pi }{512}\frac{ m^2}{k}-\frac{ \pi }{64}\frac{ m^4}{k^3}-\frac{ \pi }{16}\frac{ m^6}{k^5}+\frac{64 i }{105}\frac{ m^7}{k^6}+\frac{3  \pi }{8}\frac{ m^8}{k^7}-
\right.\right.\0\\ & \quad \quad \quad \left.\left.
 - \frac{512 i }{315}\frac{ m^9}{k^8}-\frac{ \pi }{2}\frac{ m^{10}}{k^9}+\frac{1024 i }{1155}\frac{ m^{11}}{k^{10}}+\ldots\right)\right)+
\0\\ & \quad 
 + k^8 \pi _{\mu\mu} \pi _{\mu\nu} \pi _{\nu\nu}^2 \left(\frac{1}{\pi ^2} \left(\frac{ \pi }{8192} k-\frac{7  \pi }{2048}\frac{ m^2}{k}+\frac{9  \pi }{256}\frac{ m^4}{k^3}-\frac{4 i }{15}\frac{ m^5}{k^4}-\frac{11  \pi }{64}\frac{ m^6}{k^5}+
\right.\right.\0\\ & \quad \quad \quad \left.\left.
 + \frac{32 i }{35}\frac{ m^7}{k^6}+\frac{13  \pi }{32}\frac{ m^8}{k^7}-\frac{64 i }{45}\frac{ m^9}{k^8}-\frac{3  \pi }{8}\frac{ m^{10}}{k^9}+\frac{2048 i }{3465}\frac{ m^{11}}{k^{10}}+\ldots\right)\right) }
\al{\label{eq:fc:3:5:5:ir}\tilde{T}_{3,5;5\text{D}}^{\text{f,t,IR}} & = k^8 \pi _{\mu\nu}^3 \pi _{\nu\nu} \left(\frac{i }{15 \pi ^2} \left(\frac{4 }{7}\frac{ m^3}{k^2}-\frac{2 }{21} m+\frac{1}{308}\frac{ k^2}{m}+\frac{1}{12012}\frac{ k^4}{m^3}+\frac{1}{192192}\frac{ k^6}{m^5}+
\right.\right.\0\\ & \quad \quad \quad \left.\left.
 + \frac{1}{2178176}\frac{ k^8}{m^7}+\frac{7 }{141892608}\frac{ k^{10}}{m^9}+\frac{1}{165541376}\frac{ k^{12}}{m^{11}}+\ldots\right)\right)+
\0\\ & \quad 
 + k^8 \pi _{\mu\mu} \pi _{\mu\nu} \pi _{\nu\nu}^2 \left(\frac{i }{5 \pi ^2} \left(-\frac{4 }{21}\frac{ m^3}{k^2}+\frac{1}{84} m-\frac{1}{5544}\frac{ k^2}{m}-\frac{1}{576576}\frac{ k^4}{m^3}+
\right.\right.\0\\ & \quad \quad \quad \left.\left.
 + \frac{1}{156828672}\frac{ k^8}{m^7}+\frac{1}{851355648}\frac{ k^{10}}{m^9}+\frac{1}{5297324032}\frac{ k^{12}}{m^{11}}+\ldots\right)\right) }
\al{\label{eq:fc:3:5:5:uvir}\tilde{T}_{3,5;5\text{D}}^{\text{f,UV-IR}} & =  \ldots \textrm{(i.e.\ no overlap)} }
Fermions, spin 3 x 5, dimension 6:
\al{\label{eq:fc:3:5:6:uv}\tilde{T}_{3,5;6\text{D}}^{\text{f,t,UV}} & = k^8 \pi _{\mu\nu}^3 \pi _{\nu\nu} \left(\frac{i }{\pi ^3} \left( \left(\frac{11861}{16008300}-\frac{P}{4620}\right) k^2+ \left(-\frac{1937}{198450}+\frac{P}{315}\right) m^2+
\right.\right.\0\\ & \quad \quad \quad \left.\left.
 +  \left(\frac{247}{7350}-\frac{P}{70}\right)\frac{ m^4}{k^2}+\frac{1}{15}\frac{ m^6}{k^4}+ \left(-\frac{19}{72}+\frac{K}{6}\right)\frac{ m^8}{k^6}+
\right.\right.\0\\ & \quad \quad \quad \left.\left.
 +  \left(\frac{13}{150}-\frac{2 K}{5}\right)\frac{ m^{10}}{k^8}+ \left(\frac{127}{300}+\frac{K}{5}\right)\frac{ m^{12}}{k^{10}}+\ldots\right)\right)+
\0\\ & \quad 
 + k^8 \pi _{\mu\mu} \pi _{\mu\nu} \pi _{\nu\nu}^2 \left(\frac{i }{\pi ^3} \left( \left(-\frac{29497}{192099600}+\frac{P}{27720}\right) k^2+ \left(\frac{1231}{264600}-\frac{P}{840}\right) m^2+
\right.\right.\0\\ & \quad \quad \quad \left.\left.
 +  \left(-\frac{1513}{29400}+\frac{P}{70}\right)\frac{ m^4}{k^2}+ \left(\frac{43}{360}-\frac{K}{12}\right)\frac{ m^6}{k^4}+ \left(-\frac{3}{16}+\frac{K}{4}\right)\frac{ m^8}{k^6}-
\right.\right.\0\\ & \quad \quad \quad \left.\left.
 -  \left(\frac{59}{1200}+\frac{7 K}{20}\right)\frac{ m^{10}}{k^8}+ \left(\frac{31}{100}+\frac{2 K}{15}\right)\frac{ m^{12}}{k^{10}}+\ldots\right)\right) }
\al{\label{eq:fc:3:5:6:ir}\tilde{T}_{3,5;6\text{D}}^{\text{f,t,IR}} & = k^8 \pi _{\mu\nu}^3 \pi _{\nu\nu} \left(\frac{i }{5 \pi ^3} \left( \left(\frac{3}{28}-\frac{L_0}{14}\right)\frac{ m^4}{k^2}+ \left(-\frac{1}{63}+\frac{L_0}{63}\right) m^2-\frac{ L_0}{924} k^2+\frac{1}{18018}\frac{ k^4}{m^2}+
\right.\right.\0\\ & \quad \quad \quad \left.\left.
 + \frac{1}{432432}\frac{ k^6}{m^4}+\frac{1}{6126120}\frac{ k^8}{m^6}+\frac{1}{66512160}\frac{ k^{10}}{m^8}+\frac{1}{611080470}\frac{ k^{12}}{m^{10}}+
\right.\right.\0\\ & \quad \quad \quad \left.\left.
 + \frac{1}{4997280288}\frac{ k^{14}}{m^{12}}+\ldots\right)\right)+
\0\\ & \quad 
 + k^8 \pi _{\mu\mu} \pi _{\mu\nu} \pi _{\nu\nu}^2 \left(\frac{i }{70 \pi ^3} \left( \left(-\frac{3}{2}+L_0\right)\frac{ m^4}{k^2}+ \left(\frac{1}{12}-\frac{L_0}{12}\right) m^2+\frac{ L_0}{396} k^2-
\right.\right.\0\\ & \quad \quad \quad \left.\left.
 - \frac{1}{20592}\frac{ k^4}{m^2}+\frac{1}{10501920}\frac{ k^8}{m^6}+\frac{1}{66512160}\frac{ k^{10}}{m^8}+\frac{1}{465585120}\frac{ k^{12}}{m^{10}}+
\right.\right.\0\\ & \quad \quad \quad \left.\left.
 + \frac{1}{3212537328}\frac{ k^{14}}{m^{12}}+\ldots\right)\right) }
\al{\label{eq:fc:3:5:6:uvir}\tilde{T}_{3,5;6\text{D}}^{\text{f,UV-IR}} & = k^8 \pi _{\mu\nu}^3 \pi _{\nu\nu} \left(\frac{i }{35 \pi ^3} \left( \left(\frac{11861}{457380}-\frac{K}{132}\right) k^2+ \left(-\frac{1307}{5670}+\frac{K}{9}\right) m^2+
\right.\right.\0\\ & \quad \quad \quad \left.\left.
 +  \left(\frac{179}{420}-\frac{K}{2}\right)\frac{ m^4}{k^2}\right)\right)+
\0\\ & \quad 
 + k^8 \pi _{\mu\mu} \pi _{\mu\nu} \pi _{\nu\nu}^2 \left(\frac{i }{70 \pi ^3} \left( \left(-\frac{29497}{2744280}+\frac{K}{396}-\frac{\log (2)}{198}-\frac{\log (\pi )}{396}+
\right.\right.\right.\0\\ & \quad \quad \quad \quad \left.\left.\left.
 + \frac{1}{396} \log (4 \pi )\right) k^2+ \left(\frac{229}{945}-\frac{K}{12}\right) m^2+ \left(-\frac{883}{420}+K\right)\frac{ m^4}{k^2}\right)\right)+\ldots }
Fermions, spin 4 x 4, dimension 3:
\al{\label{eq:fc:4:4:3:uv}\tilde{T}_{4,4;3\text{D}}^{\text{f,t,UV}} & = k^8 \pi _{\mu\nu}^4 \left(-\frac{1}{128}\frac{1}{k}+\frac{1}{16}\frac{ m^2}{k^3}-\frac{2 i }{3 \pi }\frac{ m^3}{k^4}-\frac{16 i }{15 \pi }\frac{ m^5}{k^6}-\frac{ m^6}{k^7}+\frac{192 i }{35 \pi }\frac{ m^7}{k^8}+2 \frac{ m^8}{k^9}-
\right.\0\\ & \quad \quad \left.
 - \frac{256 i }{63 \pi }\frac{ m^9}{k^{10}}-\frac{1024 i }{495 \pi }\frac{ m^{11}}{k^{12}}+\ldots\right)+
\0\\ & \quad 
 + k^8 \pi _{\mu\mu} \pi _{\mu\nu}^2 \pi _{\nu\nu} \left(-\frac{3 }{32}\frac{ m^2}{k^3}+\frac{2 i }{\pi }\frac{ m^3}{k^4}+\frac{9 }{8}\frac{ m^4}{k^5}-\frac{8 i }{\pi }\frac{ m^5}{k^6}-\frac{9 }{2}\frac{ m^6}{k^7}+\frac{96 i }{5 \pi }\frac{ m^7}{k^8}+
\right.\0\\ & \quad \quad \left.
 + 6 \frac{ m^8}{k^9}-\frac{384 i }{35 \pi }\frac{ m^9}{k^{10}}-\frac{512 i }{105 \pi }\frac{ m^{11}}{k^{12}}+\ldots\right)+
\0\\ & \quad 
 + k^8 \pi _{\mu\mu}^2 \pi _{\nu\nu}^2 \left(\frac{3 }{1024}\frac{1}{k}-\frac{3 }{64}\frac{ m^2}{k^3}+\frac{9 }{32}\frac{ m^4}{k^5}-\frac{8 i }{5 \pi }\frac{ m^5}{k^6}-\frac{3 }{4}\frac{ m^6}{k^7}+\frac{96 i }{35 \pi }\frac{ m^7}{k^8}+\frac{3 }{4}\frac{ m^8}{k^9}-
\right.\0\\ & \quad \quad \left.
 - \frac{128 i }{105 \pi }\frac{ m^9}{k^{10}}-\frac{512 i }{1155 \pi }\frac{ m^{11}}{k^{12}}+\ldots\right)+
\0\\ & \quad 
 + k^6 (k\cdot\epsilon)_{\mu\nu} \pi _{\mu\nu}^3 \left(\frac{i }{32}\frac{ m}{k}+\frac{1}{\pi }\frac{ m^2}{k^2}-\frac{3 i }{8}\frac{ m^3}{k^3}-\frac{8 }{3 \pi }\frac{ m^4}{k^4}+\frac{3 i }{2}\frac{ m^5}{k^5}+\frac{32 }{5 \pi }\frac{ m^6}{k^6}-
\right.\0\\ & \quad \quad \left.
 - 2 i \frac{ m^7}{k^7}-\frac{128 }{35 \pi }\frac{ m^8}{k^8}-\frac{512 }{315 \pi }\frac{ m^{10}}{k^{10}}-\frac{2048 }{1155 \pi }\frac{ m^{12}}{k^{12}}+\ldots\right)+
\0\\ & \quad 
 + k^6 (k\cdot\epsilon)_{\mu\nu} \pi _{\mu\mu} \pi _{\mu\nu} \pi _{\nu\nu} \left(\frac{3 i }{64}\frac{ m}{k}-\frac{9 i }{16}\frac{ m^3}{k^3}-\frac{4 }{\pi }\frac{ m^4}{k^4}+\frac{9 i }{4}\frac{ m^5}{k^5}+\frac{48 }{5 \pi }\frac{ m^6}{k^6}-3 i \frac{ m^7}{k^7}-
\right.\0\\ & \quad \quad \left.
 - \frac{192 }{35 \pi }\frac{ m^8}{k^8}-\frac{256 }{105 \pi }\frac{ m^{10}}{k^{10}}-\frac{1024 }{385 \pi }\frac{ m^{12}}{k^{12}}+\ldots\right) }
\al{\label{eq:fc:4:4:3:ir}\tilde{T}_{4,4;3\text{D}}^{\text{f,t,IR}} & = k^8 \pi _{\mu\nu}^4 \left(\frac{i }{\pi } \left(-\frac{2 }{5}\frac{ m^3}{k^4}-\frac{3 }{35}\frac{ m}{k^2}+\frac{1}{252}\frac{1}{m}+\frac{1}{7920}\frac{ k^2}{m^3}+\frac{3 }{320320}\frac{ k^4}{m^5}+\frac{1}{1048320}\frac{ k^6}{m^7}+
\right.\right.\0\\ & \quad \quad \quad \left.\left.
 + \frac{1}{8616960}\frac{ k^8}{m^9}+\frac{3 }{189190144}\frac{ k^{10}}{m^{11}}+\ldots\right)\right)+
\0\\ & \quad 
 + k^8 \pi _{\mu\mu} \pi _{\mu\nu}^2 \pi _{\nu\nu} \left(\frac{i }{\pi } \left(\frac{4 }{5}\frac{ m^3}{k^4}+\frac{3 }{70}\frac{ m}{k^2}+\frac{1}{840}\frac{1}{m}+\frac{1}{12320}\frac{ k^2}{m^3}+\frac{1}{128128}\frac{ k^4}{m^5}+
\right.\right.\0\\ & \quad \quad \quad \left.\left.
 + \frac{1}{1098240}\frac{ k^6}{m^7}+\frac{3 }{24893440}\frac{ k^8}{m^9}+\frac{3 }{171991040}\frac{ k^{10}}{m^{11}}+\ldots\right)\right)+
\0\\ & \quad 
 + k^8 \pi _{\mu\mu}^2 \pi _{\nu\nu}^2 \left(\frac{i }{5 \pi } \left(-2 \frac{ m^3}{k^4}+\frac{3 }{14}\frac{ m}{k^2}-\frac{1}{168}\frac{1}{m}-\frac{1}{7392}\frac{ k^2}{m^3}-\frac{1}{128128}\frac{ k^4}{m^5}-
\right.\right.\0\\ & \quad \quad \quad \left.\left.
 - \frac{1}{1537536}\frac{ k^6}{m^7}-\frac{1}{14936064}\frac{ k^8}{m^9}-\frac{3 }{378380288}\frac{ k^{10}}{m^{11}}+\ldots\right)\right)+
\0\\ & \quad 
 + k^6 (k\cdot\epsilon)_{\mu\nu} \pi _{\mu\nu}^3 \left(\frac{1}{\pi } \left(\frac{3 }{5}\frac{ m^2}{k^2}+\frac{1}{70}+\frac{1}{2520}\frac{ k^2}{m^2}+\frac{1}{36960}\frac{ k^4}{m^4}+\frac{1}{384384}\frac{ k^6}{m^6}+
\right.\right.\0\\ & \quad \quad \quad \left.\left.
 + \frac{1}{3294720}\frac{ k^8}{m^8}+\frac{1}{24893440}\frac{ k^{10}}{m^{10}}+\frac{1}{171991040}\frac{ k^{12}}{m^{12}}+\ldots\right)\right)+
\0\\ & \quad 
 + k^6 (k\cdot\epsilon)_{\mu\nu} \pi _{\mu\mu} \pi _{\mu\nu} \pi _{\nu\nu} \left(\frac{1}{\pi } \left(-\frac{3 }{5}\frac{ m^2}{k^2}+\frac{3 }{140}+\frac{1}{1680}\frac{ k^2}{m^2}+\frac{1}{24640}\frac{ k^4}{m^4}+
\right.\right.\0\\ & \quad \quad \quad \left.\left.
 + \frac{1}{256256}\frac{ k^6}{m^6}+\frac{1}{2196480}\frac{ k^8}{m^8}+\frac{3 }{49786880}\frac{ k^{10}}{m^{10}}+\frac{3 }{343982080}\frac{ k^{12}}{m^{12}}+\ldots\right)\right) }
\al{\label{eq:fc:4:4:3:uvir}\tilde{T}_{4,4;3\text{D}}^{\text{f,UV-IR}} & = k^8 \pi _{\mu\nu}^4 \left(-\frac{4 i }{15 \pi }\frac{ m^3}{k^4}\right)+k^8 \pi _{\mu\mu} \pi _{\mu\nu}^2 \pi _{\nu\nu} \left(\frac{6 i }{5 \pi }\frac{ m^3}{k^4}\right)+k^6 (k\cdot\epsilon)_{\mu\nu} \pi _{\mu\nu}^3 \left(\frac{2 }{5 \pi }\frac{ m^2}{k^2}\right)+\ldots }
Fermions, spin 4 x 4, dimension 4:
\al{\label{eq:fc:4:4:4:uv}\tilde{T}_{4,4;4\text{D}}^{\text{f,t,UV}} & = k^8 \pi _{\mu\nu}^4 \left(\frac{i }{\pi ^2} \left( \left(\frac{50}{3969}-\frac{P}{252}\right)+ \left(-\frac{141}{1225}+\frac{3 P}{70}\right)\frac{ m^2}{k^2}+
\right.\right.\0\\ & \quad \quad \quad \left.\left.
 +  \left(-\frac{161}{600}-\frac{P}{10}+\frac{L_0}{4}\right)\frac{ m^4}{k^4}+ \left(\frac{17}{18}-\frac{K}{3}\right)\frac{ m^6}{k^6}+ \left(-\frac{7}{8}+\frac{3 K}{2}\right)\frac{ m^8}{k^8}-
\right.\right.\0\\ & \quad \quad \quad \left.\left.
 -  \left(\frac{113}{60}+K\right)\frac{ m^{10}}{k^{10}}- \left(\frac{23}{300}+\frac{7 K}{15}\right)\frac{ m^{12}}{k^{12}}+\ldots\right)\right)+
\0\\ & \quad 
 + k^8 \pi _{\mu\mu} \pi _{\mu\nu}^2 \pi _{\nu\nu} \left(\frac{i }{\pi ^2} \left( \left(\frac{62}{33075}-\frac{P}{840}\right)+ \left(\frac{281}{2450}-\frac{3 P}{140}\right)\frac{ m^2}{k^2}-
\right.\right.\0\\ & \quad \quad \quad \left.\left.
 -  \left(\frac{111}{200}-\frac{9 P}{20}+\frac{3 L_0}{4}\right)\frac{ m^4}{k^4}+ \left(\frac{37}{12}-\frac{5 K}{2}\right)\frac{ m^6}{k^6}+ \left(-\frac{17}{16}+\frac{21 K}{4}\right)\frac{ m^8}{k^8}-
\right.\right.\0\\ & \quad \quad \quad \left.\left.
 -  \left(\frac{1113}{200}+\frac{27 K}{10}\right)\frac{ m^{10}}{k^{10}}- \left(\frac{59}{200}+\frac{11 K}{10}\right)\frac{ m^{12}}{k^{12}}+\ldots\right)\right)+
\0\\ & \quad 
 + k^8 \pi _{\mu\mu}^2 \pi _{\nu\nu}^2 \left(\frac{i }{\pi ^2} \left( \left(-\frac{563}{132300}+\frac{P}{840}\right)+ \left(\frac{88}{1225}-\frac{3 P}{140}\right)\frac{ m^2}{k^2}+
\right.\right.\0\\ & \quad \quad \quad \left.\left.
 +  \left(-\frac{23}{50}+\frac{3 P}{20}\right)\frac{ m^4}{k^4}+ \left(\frac{5}{12}-\frac{K}{2}\right)\frac{ m^6}{k^6}+ \left(\frac{1}{16}+\frac{3 K}{4}\right)\frac{ m^8}{k^8}-
\right.\right.\0\\ & \quad \quad \quad \left.\left.
 -  \left(\frac{137}{200}+\frac{3 K}{10}\right)\frac{ m^{10}}{k^{10}}- \left(\frac{9}{200}+\frac{K}{10}\right)\frac{ m^{12}}{k^{12}}+\ldots\right)\right) }
\al{\label{eq:fc:4:4:4:ir}\tilde{T}_{4,4;4\text{D}}^{\text{f,t,IR}} & = k^8 \pi _{\mu\nu}^4 \left(\frac{i }{2 \pi ^2} \left( \left(-\frac{9}{20}+\frac{3 L_0}{10}\right)\frac{ m^4}{k^4}+ \left(-\frac{3}{35}+\frac{3 L_0}{35}\right)\frac{ m^2}{k^2}-\frac{ L_0}{126}+\frac{1}{1980}\frac{ k^2}{m^2}+
\right.\right.\0\\ & \quad \quad \quad \left.\left.
 + \frac{1}{40040}\frac{ k^4}{m^4}+\frac{1}{491400}\frac{ k^6}{m^6}+\frac{1}{4712400}\frac{ k^8}{m^8}+\frac{1}{38798760}\frac{ k^{10}}{m^{10}}+
\right.\right.\0\\ & \quad \quad \quad \left.\left.
 + \frac{1}{287567280}\frac{ k^{12}}{m^{12}}+\ldots\right)\right)+
\0\\ & \quad 
 + k^8 \pi _{\mu\mu} \pi _{\mu\nu}^2 \pi _{\nu\nu} \left(\frac{i }{2 \pi ^2} \left( \left(\frac{9}{10}-\frac{3 L_0}{5}\right)\frac{ m^4}{k^4}+ \left(\frac{3}{70}-\frac{3 L_0}{70}\right)\frac{ m^2}{k^2}-\frac{ L_0}{420}+
\right.\right.\0\\ & \quad \quad \quad \left.\left.
 + \frac{1}{3080}\frac{ k^2}{m^2}+\frac{1}{48048}\frac{ k^4}{m^4}+\frac{1}{514800}\frac{ k^6}{m^6}+\frac{3 }{13613600}\frac{ k^8}{m^8}+\frac{1}{35271600}\frac{ k^{10}}{m^{10}}+
\right.\right.\0\\ & \quad \quad \quad \left.\left.
 + \frac{1}{250699680}\frac{ k^{12}}{m^{12}}+\ldots\right)\right)+
\0\\ & \quad 
 + k^8 \pi _{\mu\mu}^2 \pi _{\nu\nu}^2 \left(\frac{i }{20 \pi ^2} \left( \left(-\frac{9}{2}+3 L_0\right)\frac{ m^4}{k^4}+ \left(\frac{3}{7}-\frac{3 L_0}{7}\right)\frac{ m^2}{k^2}+\frac{ L_0}{42}-\frac{1}{924}\frac{ k^2}{m^2}-
\right.\right.\0\\ & \quad \quad \quad \left.\left.
 - \frac{1}{24024}\frac{ k^4}{m^4}-\frac{1}{360360}\frac{ k^6}{m^6}-\frac{1}{4084080}\frac{ k^8}{m^8}-\frac{1}{38798760}\frac{ k^{10}}{m^{10}}-
\right.\right.\0\\ & \quad \quad \quad \left.\left.
 - \frac{1}{325909584}\frac{ k^{12}}{m^{12}}+\ldots\right)\right) }
\al{\label{eq:fc:4:4:4:uvir}\tilde{T}_{4,4;4\text{D}}^{\text{f,UV-IR}} & = k^8 \pi _{\mu\nu}^4 \left(\frac{i }{\pi ^2} \left( \left(\frac{50}{3969}-\frac{K}{252}\right)+ \left(-\frac{177}{2450}+\frac{3 K}{70}\right)\frac{ m^2}{k^2}- \left(\frac{13}{300}+\frac{K}{10}\right)\frac{ m^4}{k^4}\right)\right)+
\0\\ & \quad 
 + k^8 \pi _{\mu\mu} \pi _{\mu\nu}^2 \pi _{\nu\nu} \left(\frac{i }{5 \pi ^2} \left( \left(\frac{62}{6615}-\frac{K}{168}\right)+ \left(\frac{457}{980}-\frac{3 K}{28}\right)\frac{ m^2}{k^2}+
\right.\right.\0\\ & \quad \quad \quad \left.\left.
 +  \left(-\frac{201}{40}+\frac{9 K}{4}\right)\frac{ m^4}{k^4}\right)\right)+
\0\\ & \quad 
 + k^8 \pi _{\mu\mu}^2 \pi _{\nu\nu}^2 \left(\frac{i }{20 \pi ^2} \left( \left(-\frac{563}{6615}+\frac{K}{42}\right)+ \left(\frac{247}{245}-\frac{3 K}{7}\right)\frac{ m^2}{k^2}+
\right.\right.\0\\ & \quad \quad \quad \left.\left.
 +  \left(-\frac{47}{10}+3 K\right)\frac{ m^4}{k^4}\right)\right)+\ldots }
Fermions, spin 4 x 4, dimension 5:
\al{\label{eq:fc:4:4:5:uv}\tilde{T}_{4,4;5\text{D}}^{\text{f,t,UV}} & = k^8 \pi _{\mu\nu}^4 \left(\frac{1}{\pi ^2} \left(-\frac{3  \pi }{10240} k+\frac{ \pi }{256}\frac{ m^2}{k}-\frac{ \pi }{64}\frac{ m^4}{k^3}+\frac{2 i }{15}\frac{ m^5}{k^4}+\frac{16 i }{105}\frac{ m^7}{k^6}+\frac{ \pi }{8}\frac{ m^8}{k^7}-
\right.\right.\0\\ & \quad \quad \quad \left.\left.
 - \frac{64 i }{105}\frac{ m^9}{k^8}-\frac{ \pi }{5}\frac{ m^{10}}{k^9}+\frac{256 i }{693}\frac{ m^{11}}{k^{10}}+\ldots\right)\right)+
\0\\ & \quad 
 + k^8 \pi _{\mu\mu} \pi _{\mu\nu}^2 \pi _{\nu\nu} \left(\frac{1}{\pi ^2} \left(-\frac{3  \pi }{20480} k+\frac{3  \pi }{128}\frac{ m^4}{k^3}-\frac{2 i }{5}\frac{ m^5}{k^4}-\frac{3  \pi }{16}\frac{ m^6}{k^5}+\frac{8 i }{7}\frac{ m^7}{k^6}+
\right.\right.\0\\ & \quad \quad \quad \left.\left.
 + \frac{9  \pi }{16}\frac{ m^8}{k^7}-\frac{32 i }{15}\frac{ m^9}{k^8}-\frac{3  \pi }{5}\frac{ m^{10}}{k^9}+\frac{384 i }{385}\frac{ m^{11}}{k^{10}}+\ldots\right)\right)+
\0\\ & \quad 
 + k^8 \pi _{\mu\mu}^2 \pi _{\nu\nu}^2 \left(\frac{1}{\pi ^2} \left(\frac{3  \pi }{40960} k-\frac{3  \pi }{2048}\frac{ m^2}{k}+\frac{3  \pi }{256}\frac{ m^4}{k^3}-\frac{3  \pi }{64}\frac{ m^6}{k^5}+\frac{8 i }{35}\frac{ m^7}{k^6}+
\right.\right.\0\\ & \quad \quad \quad \left.\left.
 + \frac{3  \pi }{32}\frac{ m^8}{k^7}-\frac{32 i }{105}\frac{ m^9}{k^8}-\frac{3  \pi }{40}\frac{ m^{10}}{k^9}+\frac{128 i }{1155}\frac{ m^{11}}{k^{10}}+\ldots\right)\right) }
\al{\label{eq:fc:4:4:5:ir}\tilde{T}_{4,4;5\text{D}}^{\text{f,t,IR}} & = k^8 \pi _{\mu\nu}^4 \left(\frac{i }{\pi ^2} \left(\frac{2 }{25}\frac{ m^5}{k^4}+\frac{1}{35}\frac{ m^3}{k^2}-\frac{1}{252} m+\frac{1}{7920}\frac{ k^2}{m}+\frac{1}{320320}\frac{ k^4}{m^3}+\frac{1}{5241600}\frac{ k^6}{m^5}+
\right.\right.\0\\ & \quad \quad \quad \left.\left.
 + \frac{1}{60318720}\frac{ k^8}{m^7}+\frac{1}{567570432}\frac{ k^{10}}{m^9}+\frac{1}{4674109440}\frac{ k^{12}}{m^{11}}+\ldots\right)\right)+
\0\\ & \quad 
 + k^8 \pi _{\mu\mu} \pi _{\mu\nu}^2 \pi _{\nu\nu} \left(\frac{i }{\pi ^2} \left(-\frac{4 }{25}\frac{ m^5}{k^4}-\frac{1}{70}\frac{ m^3}{k^2}-\frac{1}{840} m+\frac{1}{12320}\frac{ k^2}{m}+\frac{1}{384384}\frac{ k^4}{m^3}+
\right.\right.\0\\ & \quad \quad \quad \left.\left.
 + \frac{1}{5491200}\frac{ k^6}{m^5}+\frac{3 }{174254080}\frac{ k^8}{m^7}+\frac{1}{515973120}\frac{ k^{10}}{m^9}+\frac{1}{4074864640}\frac{ k^{12}}{m^{11}}+
\right.\right.\0\\ & \quad \quad \quad \left.\left.
 + \ldots\right)\right)+
\0\\ & \quad 
 + k^8 \pi _{\mu\mu}^2 \pi _{\nu\nu}^2 \left(\frac{i }{5 \pi ^2} \left(\frac{2 }{5}\frac{ m^5}{k^4}-\frac{1}{14}\frac{ m^3}{k^2}+\frac{1}{168} m-\frac{1}{7392}\frac{ k^2}{m}-\frac{1}{384384}\frac{ k^4}{m^3}-
\right.\right.\0\\ & \quad \quad \quad \left.\left.
 - \frac{1}{7687680}\frac{ k^6}{m^5}-\frac{1}{104552448}\frac{ k^8}{m^7}-\frac{1}{1135140864}\frac{ k^{10}}{m^9}-\frac{1}{10594648064}\frac{ k^{12}}{m^{11}}+
\right.\right.\0\\ & \quad \quad \quad \left.\left.
 + \ldots\right)\right) }
\al{\label{eq:fc:4:4:5:uvir}\tilde{T}_{4,4;5\text{D}}^{\text{f,UV-IR}} & = k^8 \pi _{\mu\nu}^4 \left(\frac{4 i }{75 \pi ^2}\frac{ m^5}{k^4}\right)+k^8 \pi _{\mu\mu} \pi _{\mu\nu}^2 \pi _{\nu\nu} \left(-\frac{6 i }{25 \pi ^2}\frac{ m^5}{k^4}\right)+\ldots }
Fermions, spin 4 x 4, dimension 6:
\al{\label{eq:fc:4:4:6:uv}\tilde{T}_{4,4;6\text{D}}^{\text{f,t,UV}} & = k^8 \pi _{\mu\nu}^4 \left(\frac{i }{\pi ^3} \left( \left(\frac{859}{1960200}-\frac{P}{7920}\right) k^2+ \left(-\frac{25}{3969}+\frac{P}{504}\right) m^2+
\right.\right.\0\\ & \quad \quad \quad \left.\left.
 +  \left(\frac{141}{4900}-\frac{3 P}{280}\right)\frac{ m^4}{k^2}- \left(-\frac{211}{3600}-\frac{P}{60}+\frac{L_0}{24}\right)\frac{ m^6}{k^4}+
\right.\right.\0\\ & \quad \quad \quad \left.\left.
 +  \left(-\frac{31}{288}+\frac{K}{24}\right)\frac{ m^8}{k^6}+ \left(\frac{23}{400}-\frac{3 K}{20}\right)\frac{ m^{10}}{k^8}+ \left(\frac{41}{240}+\frac{K}{12}\right)\frac{ m^{12}}{k^{10}}+
\right.\right.\0\\ & \quad \quad \quad \left.\left.
 + \ldots\right)\right)+
\0\\ & \quad 
 + k^8 \pi _{\mu\mu} \pi _{\mu\nu}^2 \pi _{\nu\nu} \left(\frac{i }{\pi ^3} \left( \left(\frac{5353}{21344400}-\frac{P}{12320}\right) k^2+ \left(-\frac{31}{33075}+\frac{P}{1680}\right) m^2+
\right.\right.\0\\ & \quad \quad \quad \left.\left.
 +  \left(-\frac{281}{9800}+\frac{3 P}{560}\right)\frac{ m^4}{k^2}+ \left(\frac{61}{1200}-\frac{3 P}{40}+\frac{L_0}{8}\right)\frac{ m^6}{k^4}+
\right.\right.\0\\ & \quad \quad \quad \left.\left.
 +  \left(-\frac{59}{192}+\frac{5 K}{16}\right)\frac{ m^8}{k^6}+ \left(\frac{1}{800}-\frac{21 K}{40}\right)\frac{ m^{10}}{k^8}+ \left(\frac{401}{800}+\frac{9 K}{40}\right)\frac{ m^{12}}{k^{10}}+
\right.\right.\0\\ & \quad \quad \quad \left.\left.
 + \ldots\right)\right)+
\0\\ & \quad 
 + k^8 \pi _{\mu\mu}^2 \pi _{\nu\nu}^2 \left(\frac{i }{\pi ^3} \left( \left(-\frac{1627}{16008300}+\frac{P}{36960}\right) k^2+ \left(\frac{563}{264600}-\frac{P}{1680}\right) m^2+
\right.\right.\0\\ & \quad \quad \quad \left.\left.
 +  \left(-\frac{22}{1225}+\frac{3 P}{560}\right)\frac{ m^4}{k^2}+ \left(\frac{23}{300}-\frac{P}{40}\right)\frac{ m^6}{k^4}+ \left(-\frac{7}{192}+\frac{K}{16}\right)\frac{ m^8}{k^6}-
\right.\right.\0\\ & \quad \quad \quad \left.\left.
 -  \left(\frac{17}{800}+\frac{3 K}{40}\right)\frac{ m^{10}}{k^8}+ \left(\frac{49}{800}+\frac{K}{40}\right)\frac{ m^{12}}{k^{10}}+\ldots\right)\right) }
\al{\label{eq:fc:4:4:6:ir}\tilde{T}_{4,4;6\text{D}}^{\text{f,t,IR}} & = k^8 \pi _{\mu\nu}^4 \left(\frac{i }{8 \pi ^3} \left( \left(\frac{11}{30}-\frac{L_0}{5}\right)\frac{ m^6}{k^4}+ \left(\frac{9}{70}-\frac{3 L_0}{35}\right)\frac{ m^4}{k^2}+ \left(-\frac{1}{63}+\frac{L_0}{63}\right) m^2-
\right.\right.\0\\ & \quad \quad \quad \left.\left.
 - \frac{ L_0}{990} k^2+\frac{1}{20020}\frac{ k^4}{m^2}+\frac{1}{491400}\frac{ k^6}{m^4}+\frac{1}{7068600}\frac{ k^8}{m^6}+\frac{1}{77597520}\frac{ k^{10}}{m^8}+
\right.\right.\0\\ & \quad \quad \quad \left.\left.
 + \frac{1}{718918200}\frac{ k^{12}}{m^{10}}+\frac{1}{5917831920}\frac{ k^{14}}{m^{12}}+\ldots\right)\right)+
\0\\ & \quad 
 + k^8 \pi _{\mu\mu} \pi _{\mu\nu}^2 \pi _{\nu\nu} \left(\frac{i }{4 \pi ^3} \left( \left(-\frac{11}{30}+\frac{L_0}{5}\right)\frac{ m^6}{k^4}+ \left(-\frac{9}{280}+\frac{3 L_0}{140}\right)\frac{ m^4}{k^2}+
\right.\right.\0\\ & \quad \quad \quad \left.\left.
 +  \left(-\frac{1}{420}+\frac{L_0}{420}\right) m^2-\frac{ L_0}{3080} k^2+\frac{1}{48048}\frac{ k^4}{m^2}+\frac{1}{1029600}\frac{ k^6}{m^4}+
\right.\right.\0\\ & \quad \quad \quad \left.\left.
 + \frac{1}{13613600}\frac{ k^8}{m^6}+\frac{1}{141086400}\frac{ k^{10}}{m^8}+\frac{1}{1253498400}\frac{ k^{12}}{m^{10}}+
\right.\right.\0\\ & \quad \quad \quad \left.\left.
 + \frac{1}{9994560576}\frac{ k^{14}}{m^{12}}+\ldots\right)\right)+
\0\\ & \quad 
 + k^8 \pi _{\mu\mu}^2 \pi _{\nu\nu}^2 \left(\frac{i }{40 \pi ^3} \left( \left(\frac{11}{6}-L_0\right)\frac{ m^6}{k^4}+ \left(-\frac{9}{28}+\frac{3 L_0}{14}\right)\frac{ m^4}{k^2}+
\right.\right.\0\\ & \quad \quad \quad \left.\left.
 +  \left(\frac{1}{42}-\frac{L_0}{42}\right) m^2+\frac{ L_0}{924} k^2-\frac{1}{24024}\frac{ k^4}{m^2}-\frac{1}{720720}\frac{ k^6}{m^4}-\frac{1}{12252240}\frac{ k^8}{m^6}-
\right.\right.\0\\ & \quad \quad \quad \left.\left.
 - \frac{1}{155195040}\frac{ k^{10}}{m^8}-\frac{1}{1629547920}\frac{ k^{12}}{m^{10}}-\frac{1}{14991840864}\frac{ k^{14}}{m^{12}}+\ldots\right)\right) }
\al{\label{eq:fc:4:4:6:uvir}\tilde{T}_{4,4;6\text{D}}^{\text{f,UV-IR}} & = k^8 \pi _{\mu\nu}^4 \left(\frac{i }{4 \pi ^3} \left( \left(\frac{859}{490050}-\frac{K}{1980}\right) k^2+ \left(-\frac{137}{7938}+\frac{K}{126}\right) m^2+
\right.\right.\0\\ & \quad \quad \quad \left.\left.
 +  \left(\frac{249}{4900}-\frac{3 K}{70}\right)\frac{ m^4}{k^2}+ \left(\frac{23}{450}+\frac{K}{15}\right)\frac{ m^6}{k^4}\right)\right)+
\0\\ & \quad 
 + k^8 \pi _{\mu\mu} \pi _{\mu\nu}^2 \pi _{\nu\nu} \left(\frac{i }{40 \pi ^3} \left( \left(\frac{5353}{533610}-\frac{K}{308}\right) k^2+ \left(-\frac{181}{13230}+\frac{K}{42}\right) m^2+
\right.\right.\0\\ & \quad \quad \quad \left.\left.
 +  \left(-\frac{809}{980}+\frac{3 K}{14}\right)\frac{ m^4}{k^2}+ \left(\frac{57}{10}-3 K\right)\frac{ m^6}{k^4}\right)\right)+
\0\\ & \quad 
 + k^8 \pi _{\mu\mu}^2 \pi _{\nu\nu}^2 \left(\frac{i }{20 \pi ^3} \left( \left(-\frac{1627}{800415}+\frac{K}{1848}\right) k^2+ \left(\frac{811}{26460}-\frac{K}{84}\right) m^2+
\right.\right.\0\\ & \quad \quad \quad \left.\left.
 +  \left(-\frac{389}{1960}+\frac{3 K}{28}\right)\frac{ m^4}{k^2}+ \left(\frac{37}{60}-\frac{K}{2}\right)\frac{ m^6}{k^4}\right)\right)+\ldots }
Fermions, spin 5 x 5, dimension 3:
\al{\label{eq:fc:5:5:3:uv}\tilde{T}_{5,5;3\text{D}}^{\text{f,t,UV}} & = k^{10} \pi _{\mu\nu}^5 \left(\frac{1}{256}\frac{1}{k}-\frac{3 }{64}\frac{ m^2}{k^3}+\frac{4 i }{3 \pi }\frac{ m^3}{k^4}+\frac{1}{8}\frac{ m^4}{k^5}+\frac{1}{2}\frac{ m^6}{k^7}-\frac{512 i }{105 \pi }\frac{ m^7}{k^8}-3 \frac{ m^8}{k^9}+
\right.\0\\ & \quad \quad \left.
 + \frac{4096 i }{315 \pi }\frac{ m^9}{k^{10}}+4 \frac{ m^{10}}{k^{11}}-\frac{8192 i }{1155 \pi }\frac{ m^{11}}{k^{12}}+\ldots\right)+
\0\\ & \quad 
 + k^{10} \pi _{\mu\mu} \pi _{\mu\nu}^3 \pi _{\nu\nu} \left(\frac{1}{256}\frac{1}{k}+\frac{1}{64}\frac{ m^2}{k^3}-\frac{8 i }{3 \pi }\frac{ m^3}{k^4}-\frac{7 }{8}\frac{ m^4}{k^5}+\frac{128 i }{15 \pi }\frac{ m^5}{k^6}+\frac{13 }{2}\frac{ m^6}{k^7}-
\right.\0\\ & \quad \quad \left.
 - \frac{4096 i }{105 \pi }\frac{ m^7}{k^8}-19 \frac{ m^8}{k^9}+\frac{22528 i }{315 \pi }\frac{ m^9}{k^{10}}+20 \frac{ m^{10}}{k^{11}}-\frac{16384 i }{495 \pi }\frac{ m^{11}}{k^{12}}+\ldots\right)+
\0\\ & \quad 
 + k^{10} \pi _{\mu\mu}^2 \pi _{\mu\nu} \pi _{\nu\nu}^2 \left(-\frac{9 }{2048}\frac{1}{k}+\frac{51 }{512}\frac{ m^2}{k^3}-\frac{57 }{64}\frac{ m^4}{k^5}+\frac{32 i }{5 \pi }\frac{ m^5}{k^6}+\frac{63 }{16}\frac{ m^6}{k^7}-
\right.\0\\ & \quad \quad \left.
 - \frac{704 i }{35 \pi }\frac{ m^7}{k^8}-\frac{69 }{8}\frac{ m^8}{k^9}+\frac{1024 i }{35 \pi }\frac{ m^9}{k^{10}}+\frac{15 }{2}\frac{ m^{10}}{k^{11}}-\frac{13312 i }{1155 \pi }\frac{ m^{11}}{k^{12}}+\ldots\right)+
\0\\ & \quad 
 + k^8 (k\cdot\epsilon)_{\mu\nu} \pi _{\mu\nu}^4 \left(-\frac{i }{64}\frac{ m}{k}-\frac{1}{\pi }\frac{ m^2}{k^2}+\frac{i }{4}\frac{ m^3}{k^3}+\frac{4 }{3 \pi }\frac{ m^4}{k^4}-\frac{3 i }{2}\frac{ m^5}{k^5}-\frac{128 }{15 \pi }\frac{ m^6}{k^6}+
\right.\0\\ & \quad \quad \left.
 + 4 i \frac{ m^7}{k^7}+\frac{512 }{35 \pi }\frac{ m^8}{k^8}-4 i \frac{ m^9}{k^9}-\frac{2048 }{315 \pi }\frac{ m^{10}}{k^{10}}-\frac{8192 }{3465 \pi }\frac{ m^{12}}{k^{12}}+\ldots\right)+
\0\\ & \quad 
 + k^8 (k\cdot\epsilon)_{\mu\nu} \pi _{\mu\mu} \pi _{\mu\nu}^2 \pi _{\nu\nu} \left(-\frac{3 i }{64}\frac{ m}{k}+\frac{3 i }{4}\frac{ m^3}{k^3}+\frac{8 }{\pi }\frac{ m^4}{k^4}-\frac{9 i }{2}\frac{ m^5}{k^5}-\frac{128 }{5 \pi }\frac{ m^6}{k^6}+
\right.\0\\ & \quad \quad \left.
 + 12 i \frac{ m^7}{k^7}+\frac{1536 }{35 \pi }\frac{ m^8}{k^8}-12 i \frac{ m^9}{k^9}-\frac{2048 }{105 \pi }\frac{ m^{10}}{k^{10}}-\frac{8192 }{1155 \pi }\frac{ m^{12}}{k^{12}}+\ldots\right)+
\0\\ & \quad 
 + k^8 (k\cdot\epsilon)_{\mu\nu} \pi _{\mu\mu}^2 \pi _{\nu\nu}^2 \left(-\frac{3 i }{512}\frac{ m}{k}+\frac{3 i }{32}\frac{ m^3}{k^3}-\frac{9 i }{16}\frac{ m^5}{k^5}-\frac{16 }{5 \pi }\frac{ m^6}{k^6}+\frac{3 i }{2}\frac{ m^7}{k^7}+
\right.\0\\ & \quad \quad \left.
 + \frac{192 }{35 \pi }\frac{ m^8}{k^8}-\frac{3 i }{2}\frac{ m^9}{k^9}-\frac{256 }{105 \pi }\frac{ m^{10}}{k^{10}}-\frac{1024 }{1155 \pi }\frac{ m^{12}}{k^{12}}+\ldots\right) }
\al{\label{eq:fc:5:5:3:ir}\tilde{T}_{5,5;3\text{D}}^{\text{f,t,IR}} & = k^{10} \pi _{\mu\nu}^5 \left(\frac{i }{5 \pi } \left(\frac{36 }{7}\frac{ m^3}{k^4}+\frac{16 }{63}\frac{ m}{k^2}-\frac{2 }{231}\frac{1}{m}-\frac{2 }{9009}\frac{ k^2}{m^3}-\frac{1}{72072}\frac{ k^4}{m^5}-\frac{1}{816816}\frac{ k^6}{m^7}-
\right.\right.\0\\ & \quad \quad \quad \left.\left.
 - \frac{7 }{53209728}\frac{ k^8}{m^9}-\frac{1}{62078016}\frac{ k^{10}}{m^{11}}+\ldots\right)\right)+
\0\\ & \quad 
 + k^{10} \pi _{\mu\mu} \pi _{\mu\nu}^3 \pi _{\nu\nu} \left(\frac{i }{5 \pi } \left(-\frac{72 }{7}\frac{ m^3}{k^4}+\frac{8 }{63}\frac{ m}{k^2}-\frac{8 }{693}\frac{1}{m}-\frac{1}{2574}\frac{ k^2}{m^3}-\frac{1}{36036}\frac{ k^4}{m^5}-
\right.\right.\0\\ & \quad \quad \quad \left.\left.
 - \frac{1}{376992}\frac{ k^6}{m^7}-\frac{1}{3325608}\frac{ k^8}{m^9}-\frac{1}{26138112}\frac{ k^{10}}{m^{11}}+\ldots\right)\right)+
\0\\ & \quad 
 + k^{10} \pi _{\mu\mu}^2 \pi _{\mu\nu} \pi _{\nu\nu}^2 \left(\frac{i }{5 \pi } \left(\frac{36 }{7}\frac{ m^3}{k^4}-\frac{8 }{21}\frac{ m}{k^2}+\frac{1}{132}\frac{1}{m}+\frac{1}{8008}\frac{ k^2}{m^3}+\frac{1}{192192}\frac{ k^4}{m^5}+
\right.\right.\0\\ & \quad \quad \quad \left.\left.
 + \frac{1}{3267264}\frac{ k^6}{m^7}+\frac{1}{47297536}\frac{ k^8}{m^9}+\frac{1}{662165504}\frac{ k^{10}}{m^{11}}+\ldots\right)\right)+
\0\\ & \quad 
 + k^8 (k\cdot\epsilon)_{\mu\nu} \pi _{\mu\nu}^4 \left(\frac{1}{5 \pi } \left(-4 \frac{ m^4}{k^4}-\frac{27 }{7}\frac{ m^2}{k^2}-\frac{2 }{63}-\frac{1}{1386}\frac{ k^2}{m^2}-\frac{1}{24024}\frac{ k^4}{m^4}-
\right.\right.\0\\ & \quad \quad \quad \left.\left.
 - \frac{1}{288288}\frac{ k^6}{m^6}-\frac{1}{2800512}\frac{ k^8}{m^8}-\frac{1}{23648768}\frac{ k^{10}}{m^{10}}-\frac{1}{180590592}\frac{ k^{12}}{m^{12}}+
\right.\right.\0\\ & \quad \quad \quad \left.\left.
 + \ldots\right)\right)+
\0\\ & \quad 
 + k^8 (k\cdot\epsilon)_{\mu\nu} \pi _{\mu\mu} \pi _{\mu\nu}^2 \pi _{\nu\nu} \left(\frac{1}{5 \pi } \left(8 \frac{ m^4}{k^4}+\frac{24 }{7}\frac{ m^2}{k^2}-\frac{2 }{21}-\frac{1}{462}\frac{ k^2}{m^2}-\frac{1}{8008}\frac{ k^4}{m^4}-
\right.\right.\0\\ & \quad \quad \quad \left.\left.
 - \frac{1}{96096}\frac{ k^6}{m^6}-\frac{1}{933504}\frac{ k^8}{m^8}-\frac{3 }{23648768}\frac{ k^{10}}{m^{10}}-\frac{1}{60196864}\frac{ k^{12}}{m^{12}}+\ldots\right)\right)+
\0\\ & \quad 
 + k^8 (k\cdot\epsilon)_{\mu\nu} \pi _{\mu\mu}^2 \pi _{\nu\nu}^2 \left(\frac{1}{5 \pi } \left(-4 \frac{ m^4}{k^4}+\frac{3 }{7}\frac{ m^2}{k^2}-\frac{1}{84}-\frac{1}{3696}\frac{ k^2}{m^2}-\frac{1}{64064}\frac{ k^4}{m^4}-
\right.\right.\0\\ & \quad \quad \quad \left.\left.
 - \frac{1}{768768}\frac{ k^6}{m^6}-\frac{1}{7468032}\frac{ k^8}{m^8}-\frac{3 }{189190144}\frac{ k^{10}}{m^{10}}-\frac{1}{481574912}\frac{ k^{12}}{m^{12}}+
\right.\right.\0\\ & \quad \quad \quad \left.\left.
 + \ldots\right)\right) }
\al{\label{eq:fc:5:5:3:uvir}\tilde{T}_{5,5;3\text{D}}^{\text{f,UV-IR}} & = k^{10} \pi _{\mu\nu}^5 \left(\frac{32 i }{105 \pi }\frac{ m^3}{k^4}\right)+k^{10} \pi _{\mu\mu} \pi _{\mu\nu}^3 \pi _{\nu\nu} \left(-\frac{64 i }{105 \pi }\frac{ m^3}{k^4}\right)+
\0\\ & \quad 
 + k^8 (k\cdot\epsilon)_{\mu\nu} \pi _{\mu\nu}^4 \left(\frac{1}{5 \pi } \left(-\frac{8 }{7}\frac{ m^2}{k^2}+\frac{32 }{3}\frac{ m^4}{k^4}\right)\right)+
\0\\ & \quad 
 + k^8 (k\cdot\epsilon)_{\mu\nu} \pi _{\mu\mu} \pi _{\mu\nu}^2 \pi _{\nu\nu} \left(\frac{32 }{5 \pi }\frac{ m^4}{k^4}\right)+\ldots }
Fermions, spin 5 x 5, dimension 4:
\al{\label{eq:fc:5:5:4:uv}\tilde{T}_{5,5;4\text{D}}^{\text{f,t,UV}} & = k^{10} \pi _{\mu\nu}^5 \left(\frac{i }{\pi ^2} \left( \left(-\frac{23722}{4002075}+\frac{2 P}{1155}\right)+ \left(\frac{7748}{99225}-\frac{8 P}{315}\right)\frac{ m^2}{k^2}-
\right.\right.\0\\ & \quad \quad \quad \left.\left.
 -  \left(-\frac{7073}{14700}-\frac{4 P}{35}+\frac{L_0}{2}\right)\frac{ m^4}{k^4}-\frac{8 }{15}\frac{ m^6}{k^6}+ \left(\frac{19}{9}-\frac{4 K}{3}\right)\frac{ m^8}{k^8}+
\right.\right.\0\\ & \quad \quad \quad \left.\left.
 +  \left(-\frac{52}{75}+\frac{16 K}{5}\right)\frac{ m^{10}}{k^{10}}- \left(\frac{254}{75}+\frac{8 K}{5}\right)\frac{ m^{12}}{k^{12}}+\ldots\right)\right)+
\0\\ & \quad 
 + k^{10} \pi _{\mu\mu} \pi _{\mu\nu}^3 \pi _{\nu\nu} \left(\frac{i }{\pi ^2} \left( \left(-\frac{83338}{12006225}+\frac{8 P}{3465}\right)+ \left(\frac{724}{99225}-\frac{4 P}{315}\right)\frac{ m^2}{k^2}+
\right.\right.\0\\ & \quad \quad \quad \left.\left.
 +  \left(-\frac{2873}{7350}-\frac{8 P}{35}+L_0\right)\frac{ m^4}{k^4}+ \left(-\frac{44}{9}+\frac{8 K}{3}\right)\frac{ m^6}{k^6}+ \left(\frac{92}{9}-\frac{32 K}{3}\right)\frac{ m^8}{k^8}+
\right.\right.\0\\ & \quad \quad \quad \left.\left.
 +  \left(\frac{14}{75}+\frac{88 K}{5}\right)\frac{ m^{10}}{k^{10}}- \left(\frac{1252}{75}+\frac{112 K}{15}\right)\frac{ m^{12}}{k^{12}}+\ldots\right)\right)+
\0\\ & \quad 
 + k^{10} \pi _{\mu\mu}^2 \pi _{\mu\nu} \pi _{\nu\nu}^2 \left(\frac{i }{\pi ^2} \left( \left(\frac{13511}{2286900}-\frac{P}{660}\right)+ \left(-\frac{9323}{66150}+\frac{4 P}{105}\right)\frac{ m^2}{k^2}+
\right.\right.\0\\ & \quad \quad \quad \left.\left.
 +  \left(\frac{3273}{2450}-\frac{27 P}{70}\right)\frac{ m^4}{k^4}+ \left(-\frac{8}{3}+2 K\right)\frac{ m^6}{k^6}+ \left(\frac{89}{24}-\frac{11 K}{2}\right)\frac{ m^8}{k^8}+
\right.\right.\0\\ & \quad \quad \quad \left.\left.
 +  \left(\frac{36}{25}+\frac{36 K}{5}\right)\frac{ m^{10}}{k^{10}}- \left(\frac{617}{100}+\frac{13 K}{5}\right)\frac{ m^{12}}{k^{12}}+\ldots\right)\right) }
\al{\label{eq:fc:5:5:4:ir}\tilde{T}_{5,5;4\text{D}}^{\text{f,t,IR}} & = k^{10} \pi _{\mu\nu}^5 \left(\frac{i }{5 \pi ^2} \left( \left(\frac{81}{28}-\frac{27 L_0}{14}\right)\frac{ m^4}{k^4}+ \left(\frac{8}{63}-\frac{8 L_0}{63}\right)\frac{ m^2}{k^2}+\frac{2  L_0}{231}-\frac{4 }{9009}\frac{ k^2}{m^2}-
\right.\right.\0\\ & \quad \quad \quad \left.\left.
 - \frac{1}{54054}\frac{ k^4}{m^4}-\frac{1}{765765}\frac{ k^6}{m^6}-\frac{1}{8314020}\frac{ k^8}{m^8}-\frac{4 }{305540235}\frac{ k^{10}}{m^{10}}-
\right.\right.\0\\ & \quad \quad \quad \left.\left.
 - \frac{1}{624660036}\frac{ k^{12}}{m^{12}}+\ldots\right)\right)+
\0\\ & \quad 
 + k^{10} \pi _{\mu\mu} \pi _{\mu\nu}^3 \pi _{\nu\nu} \left(\frac{i }{5 \pi ^2} \left( \left(-\frac{81}{14}+\frac{27 L_0}{7}\right)\frac{ m^4}{k^4}+ \left(\frac{4}{63}-\frac{4 L_0}{63}\right)\frac{ m^2}{k^2}+\frac{8  L_0}{693}-
\right.\right.\0\\ & \quad \quad \quad \left.\left.
 - \frac{1}{1287}\frac{ k^2}{m^2}-\frac{1}{27027}\frac{ k^4}{m^4}-\frac{1}{353430}\frac{ k^6}{m^6}-\frac{4 }{14549535}\frac{ k^8}{m^8}-\frac{1}{32162130}\frac{ k^{10}}{m^{10}}-
\right.\right.\0\\ & \quad \quad \quad \left.\left.
 - \frac{1}{255542742}\frac{ k^{12}}{m^{12}}+\ldots\right)\right)+
\0\\ & \quad 
 + k^{10} \pi _{\mu\mu}^2 \pi _{\mu\nu} \pi _{\nu\nu}^2 \left(\frac{i }{5 \pi ^2} \left( \left(\frac{81}{28}-\frac{27 L_0}{14}\right)\frac{ m^4}{k^4}+ \left(-\frac{4}{21}+\frac{4 L_0}{21}\right)\frac{ m^2}{k^2}-\frac{ L_0}{132}+
\right.\right.\0\\ & \quad \quad \quad \left.\left.
 + \frac{1}{4004}\frac{ k^2}{m^2}+\frac{1}{144144}\frac{ k^4}{m^4}+\frac{1}{3063060}\frac{ k^6}{m^6}+\frac{1}{51731680}\frac{ k^8}{m^8}+
\right.\right.\0\\ & \quad \quad \quad \left.\left.
 + \frac{1}{814773960}\frac{ k^{10}}{m^{10}}+\frac{1}{14991840864}\frac{ k^{12}}{m^{12}}+\ldots\right)\right) }
\al{\label{eq:fc:5:5:4:uvir}\tilde{T}_{5,5;4\text{D}}^{\text{f,UV-IR}} & = k^{10} \pi _{\mu\nu}^5 \left(\frac{i }{35 \pi ^2} \left( \left(-\frac{23722}{114345}+\frac{2 K}{33}\right)+ \left(\frac{5228}{2835}-\frac{8 K}{9}\right)\frac{ m^2}{k^2}+
\right.\right.\0\\ & \quad \quad \quad \left.\left.
 +  \left(-\frac{358}{105}+4 K\right)\frac{ m^4}{k^4}\right)\right)+
\0\\ & \quad 
 + k^{10} \pi _{\mu\mu} \pi _{\mu\nu}^3 \pi _{\nu\nu} \left(\frac{i }{35 \pi ^2} \left( \left(-\frac{83338}{343035}+\frac{8 K}{99}\right)- \left(\frac{536}{2835}+\frac{4 K}{9}\right)\frac{ m^2}{k^2}+
\right.\right.\0\\ & \quad \quad \quad \left.\left.
 +  \left(\frac{2816}{105}-8 K\right)\frac{ m^4}{k^4}\right)\right)+
\0\\ & \quad 
 + k^{10} \pi _{\mu\mu}^2 \pi _{\mu\nu} \pi _{\nu\nu}^2 \left(\frac{i }{5 \pi ^2} \left( \left(\frac{13511}{457380}-\frac{K}{132}\right)+ \left(-\frac{6803}{13230}+\frac{4 K}{21}\right)\frac{ m^2}{k^2}+
\right.\right.\0\\ & \quad \quad \quad \left.\left.
 +  \left(\frac{3711}{980}-\frac{27 K}{14}\right)\frac{ m^4}{k^4}\right)\right)+\ldots }
Fermions, spin 5 x 5, dimension 5:
\al{\label{eq:fc:5:5:5:uv}\tilde{T}_{5,5;5\text{D}}^{\text{f,t,UV}} & = k^{10} \pi _{\mu\nu}^5 \left(\frac{1}{\pi ^2} \left(\frac{7  \pi }{61440} k-\frac{ \pi }{512}\frac{ m^2}{k}+\frac{3  \pi }{256}\frac{ m^4}{k^3}-\frac{4 i }{15}\frac{ m^5}{k^4}-\frac{ \pi }{48}\frac{ m^6}{k^5}-\frac{ \pi }{16}\frac{ m^8}{k^7}+
\right.\right.\0\\ & \quad \quad \quad \left.\left.
 + \frac{512 i }{945}\frac{ m^9}{k^8}+\frac{3  \pi }{10}\frac{ m^{10}}{k^9}-\frac{4096 i }{3465}\frac{ m^{11}}{k^{10}}-\frac{ \pi }{3}\frac{ m^{12}}{k^{11}}+\ldots\right)\right)+
\0\\ & \quad 
 + k^{10} \pi _{\mu\mu} \pi _{\mu\nu}^3 \pi _{\nu\nu} \left(\frac{1}{\pi ^2} \left(\frac{11  \pi }{61440} k-\frac{ \pi }{512}\frac{ m^2}{k}-\frac{ \pi }{256}\frac{ m^4}{k^3}+\frac{8 i }{15}\frac{ m^5}{k^4}+\frac{7  \pi }{48}\frac{ m^6}{k^5}-
\right.\right.\0\\ & \quad \quad \quad \left.\left.
 - \frac{128 i }{105}\frac{ m^7}{k^6}-\frac{13  \pi }{16}\frac{ m^8}{k^7}+\frac{4096 i }{945}\frac{ m^9}{k^8}+\frac{19  \pi }{10}\frac{ m^{10}}{k^9}-\frac{2048 i }{315}\frac{ m^{11}}{k^{10}}-\frac{5  \pi }{3}\frac{ m^{12}}{k^{11}}+
\right.\right.\0\\ & \quad \quad \quad \left.\left.
 + \ldots\right)\right)+
\0\\ & \quad 
 + k^{10} \pi _{\mu\mu}^2 \pi _{\mu\nu} \pi _{\nu\nu}^2 \left(\frac{1}{\pi ^2} \left(-\frac{13  \pi }{163840} k+\frac{9  \pi }{4096}\frac{ m^2}{k}-\frac{51  \pi }{2048}\frac{ m^4}{k^3}+\frac{19  \pi }{128}\frac{ m^6}{k^5}-
\right.\right.\0\\ & \quad \quad \quad \left.\left.
 - \frac{32 i }{35}\frac{ m^7}{k^6}-\frac{63  \pi }{128}\frac{ m^8}{k^7}+\frac{704 i }{315}\frac{ m^9}{k^8}+\frac{69  \pi }{80}\frac{ m^{10}}{k^9}-\frac{1024 i }{385}\frac{ m^{11}}{k^{10}}-\frac{5  \pi }{8}\frac{ m^{12}}{k^{11}}+
\right.\right.\0\\ & \quad \quad \quad \left.\left.
 + \ldots\right)\right) }
\al{\label{eq:fc:5:5:5:ir}\tilde{T}_{5,5;5\text{D}}^{\text{f,t,IR}} & = k^{10} \pi _{\mu\nu}^5 \left(\frac{i }{5 \pi ^2} \left(-\frac{36 }{35}\frac{ m^5}{k^4}-\frac{16 }{189}\frac{ m^3}{k^2}+\frac{2 }{231} m-\frac{2 }{9009}\frac{ k^2}{m}-\frac{1}{216216}\frac{ k^4}{m^3}-
\right.\right.\0\\ & \quad \quad \quad \left.\left.
 - \frac{1}{4084080}\frac{ k^6}{m^5}-\frac{1}{53209728}\frac{ k^8}{m^7}-\frac{1}{558702144}\frac{ k^{10}}{m^9}-\frac{3 }{15229806592}\frac{ k^{12}}{m^{11}}+
\right.\right.\0\\ & \quad \quad \quad \left.\left.
 + \ldots\right)\right)+
\0\\ & \quad 
 + k^{10} \pi _{\mu\mu} \pi _{\mu\nu}^3 \pi _{\nu\nu} \left(\frac{i }{5 \pi ^2} \left(\frac{72 }{35}\frac{ m^5}{k^4}-\frac{8 }{189}\frac{ m^3}{k^2}+\frac{8 }{693} m-\frac{1}{2574}\frac{ k^2}{m}-\frac{1}{108108}\frac{ k^4}{m^3}-
\right.\right.\0\\ & \quad \quad \quad \left.\left.
 - \frac{1}{1884960}\frac{ k^6}{m^5}-\frac{1}{23279256}\frac{ k^8}{m^7}-\frac{1}{235243008}\frac{ k^{10}}{m^9}-\frac{1}{2076791808}\frac{ k^{12}}{m^{11}}+
\right.\right.\0\\ & \quad \quad \quad \left.\left.
 + \ldots\right)\right)+
\0\\ & \quad 
 + k^{10} \pi _{\mu\mu}^2 \pi _{\mu\nu} \pi _{\nu\nu}^2 \left(\frac{i }{5 \pi ^2} \left(-\frac{36 }{35}\frac{ m^5}{k^4}+\frac{8 }{63}\frac{ m^3}{k^2}-\frac{1}{132} m+\frac{1}{8008}\frac{ k^2}{m}+\frac{1}{576576}\frac{ k^4}{m^3}+
\right.\right.\0\\ & \quad \quad \quad \left.\left.
 + \frac{1}{16336320}\frac{ k^6}{m^5}+\frac{1}{331082752}\frac{ k^8}{m^7}+\frac{1}{5959489536}\frac{ k^{10}}{m^9}+
\right.\right.\0\\ & \quad \quad \quad \left.\left.
 + \frac{1}{121838452736}\frac{ k^{12}}{m^{11}}+\ldots\right)\right) }
\al{\label{eq:fc:5:5:5:uvir}\tilde{T}_{5,5;5\text{D}}^{\text{f,UV-IR}} & = k^{10} \pi _{\mu\nu}^5 \left(-\frac{32 i }{525 \pi ^2}\frac{ m^5}{k^4}\right)+k^{10} \pi _{\mu\mu} \pi _{\mu\nu}^3 \pi _{\nu\nu} \left(\frac{64 i }{525 \pi ^2}\frac{ m^5}{k^4}\right)+\ldots }
Fermions, spin 5 x 5, dimension 6:
\al{\label{eq:fc:5:5:6:uv}\tilde{T}_{5,5;6\text{D}}^{\text{f,t,UV}} & = k^{10} \pi _{\mu\nu}^5 \left(\frac{i }{3 \pi ^3} \left( \left(-\frac{659507}{1352701350}+\frac{2 P}{15015}\right) k^2+ \left(\frac{11861}{1334025}-\frac{P}{385}\right) m^2+
\right.\right.\0\\ & \quad \quad \quad \left.\left.
 +  \left(-\frac{1937}{33075}+\frac{2 P}{105}\right)\frac{ m^4}{k^2}+ \left(-\frac{9523}{29400}-\frac{2 P}{35}+\frac{L_0}{4}\right)\frac{ m^6}{k^4}+\frac{1}{5}\frac{ m^8}{k^6}+
\right.\right.\0\\ & \quad \quad \quad \left.\left.
 +  \left(-\frac{83}{150}+\frac{2 K}{5}\right)\frac{ m^{10}}{k^8}+ \left(\frac{1}{25}-\frac{4 K}{5}\right)\frac{ m^{12}}{k^{10}}+\ldots\right)\right)+
\0\\ & \quad 
 + k^{10} \pi _{\mu\mu} \pi _{\mu\nu}^3 \pi _{\nu\nu} \left(\frac{i }{\pi ^3} \left( \left(-\frac{156833}{579729150}+\frac{P}{12870}\right) k^2+
\right.\right.\0\\ & \quad \quad \quad \left.\left.
 +  \left(\frac{41669}{12006225}-\frac{4 P}{3465}\right) m^2+ \left(-\frac{181}{99225}+\frac{P}{315}\right)\frac{ m^4}{k^2}-
\right.\right.\0\\ & \quad \quad \quad \left.\left.
 -  \left(-\frac{5323}{44100}-\frac{4 P}{105}+\frac{L_0}{6}\right)\frac{ m^6}{k^4}+ \left(\frac{19}{36}-\frac{K}{3}\right)\frac{ m^8}{k^6}+
\right.\right.\0\\ & \quad \quad \quad \left.\left.
 +  \left(-\frac{182}{225}+\frac{16 K}{15}\right)\frac{ m^{10}}{k^8}- \left(\frac{13}{50}+\frac{22 K}{15}\right)\frac{ m^{12}}{k^{10}}+\ldots\right)\right)+
\0\\ & \quad 
 + k^{10} \pi _{\mu\mu}^2 \pi _{\mu\nu} \pi _{\nu\nu}^2 \left(\frac{i }{\pi ^3} \left( \left(\frac{367291}{3607203600}-\frac{P}{40040}\right) k^2+
\right.\right.\0\\ & \quad \quad \quad \left.\left.
 +  \left(-\frac{13511}{4573800}+\frac{P}{1320}\right) m^2+ \left(\frac{9323}{264600}-\frac{P}{105}\right)\frac{ m^4}{k^2}+
\right.\right.\0\\ & \quad \quad \quad \left.\left.
 +  \left(-\frac{1091}{4900}+\frac{9 P}{140}\right)\frac{ m^6}{k^4}+ \left(\frac{13}{48}-\frac{K}{4}\right)\frac{ m^8}{k^6}+ \left(-\frac{313}{1200}+\frac{11 K}{20}\right)\frac{ m^{10}}{k^8}-
\right.\right.\0\\ & \quad \quad \quad \left.\left.
 -  \left(\frac{11}{50}+\frac{3 K}{5}\right)\frac{ m^{12}}{k^{10}}+\ldots\right)\right) }
\al{\label{eq:fc:5:5:6:ir}\tilde{T}_{5,5;6\text{D}}^{\text{f,t,IR}} & = k^{10} \pi _{\mu\nu}^5 \left(\frac{i }{5 \pi ^3} \left( \left(-\frac{33}{56}+\frac{9 L_0}{28}\right)\frac{ m^6}{k^4}+ \left(-\frac{1}{21}+\frac{2 L_0}{63}\right)\frac{ m^4}{k^2}+ \left(\frac{1}{231}-\frac{L_0}{231}\right) m^2+
\right.\right.\0\\ & \quad \quad \quad \left.\left.
 + \frac{2  L_0}{9009} k^2-\frac{1}{108108}\frac{ k^4}{m^2}-\frac{1}{3063060}\frac{ k^6}{m^4}-\frac{1}{49884120}\frac{ k^8}{m^6}-
\right.\right.\0\\ & \quad \quad \quad \left.\left.
 - \frac{1}{611080470}\frac{ k^{10}}{m^8}-\frac{1}{6246600360}\frac{ k^{12}}{m^{10}}-\frac{1}{56219403240}\frac{ k^{14}}{m^{12}}+\ldots\right)\right)+
\0\\ & \quad 
 + k^{10} \pi _{\mu\mu} \pi _{\mu\nu}^3 \pi _{\nu\nu} \left(\frac{i }{5 \pi ^3} \left( \left(\frac{33}{28}-\frac{9 L_0}{14}\right)\frac{ m^6}{k^4}+ \left(-\frac{1}{42}+\frac{L_0}{63}\right)\frac{ m^4}{k^2}+
\right.\right.\0\\ & \quad \quad \quad \left.\left.
 +  \left(\frac{4}{693}-\frac{4 L_0}{693}\right) m^2+\frac{ L_0}{2574} k^2-\frac{1}{54054}\frac{ k^4}{m^2}-\frac{1}{1413720}\frac{ k^6}{m^4}-
\right.\right.\0\\ & \quad \quad \quad \left.\left.
 - \frac{2 }{43648605}\frac{ k^8}{m^6}-\frac{1}{257297040}\frac{ k^{10}}{m^8}-\frac{1}{2555427420}\frac{ k^{12}}{m^{10}}-
\right.\right.\0\\ & \quad \quad \quad \left.\left.
 - \frac{1}{22487761296}\frac{ k^{14}}{m^{12}}+\ldots\right)\right)+
\0\\ & \quad 
 + k^{10} \pi _{\mu\mu}^2 \pi _{\mu\nu} \pi _{\nu\nu}^2 \left(\frac{i }{5 \pi ^3} \left( \left(-\frac{33}{56}+\frac{9 L_0}{28}\right)\frac{ m^6}{k^4}+ \left(\frac{1}{14}-\frac{L_0}{21}\right)\frac{ m^4}{k^2}+
\right.\right.\0\\ & \quad \quad \quad \left.\left.
 +  \left(-\frac{1}{264}+\frac{L_0}{264}\right) m^2-\frac{ L_0}{8008} k^2+\frac{1}{288288}\frac{ k^4}{m^2}+\frac{1}{12252240}\frac{ k^6}{m^4}+
\right.\right.\0\\ & \quad \quad \quad \left.\left.
 + \frac{1}{310390080}\frac{ k^8}{m^6}+\frac{1}{6518191680}\frac{ k^{10}}{m^8}+\frac{1}{149918408640}\frac{ k^{12}}{m^{10}}+\ldots\right)\right) }
\al{\label{eq:fc:5:5:6:uvir}\tilde{T}_{5,5;6\text{D}}^{\text{f,UV-IR}} & = k^{10} \pi _{\mu\nu}^5 \left(\frac{i }{105 \pi ^3} \left( \left(-\frac{659507}{38648610}+\frac{2 K}{429}\right) k^2+ \left(\frac{8396}{38115}-\frac{K}{11}\right) m^2+
\right.\right.\0\\ & \quad \quad \quad \left.\left.
 +  \left(-\frac{992}{945}+\frac{2 K}{3}\right)\frac{ m^4}{k^2}+ \left(\frac{109}{105}-2 K\right)\frac{ m^6}{k^4}\right)\right)+
\0\\ & \quad 
 + k^{10} \pi _{\mu\mu} \pi _{\mu\nu}^3 \pi _{\nu\nu} \left(\frac{i }{15 \pi ^3} \left( \left(-\frac{156833}{38648610}+\frac{K}{858}\right) k^2+ \left(\frac{27809}{800415}-\frac{4 K}{231}\right) m^2+
\right.\right.\0\\ & \quad \quad \quad \left.\left.
 +  \left(\frac{583}{13230}+\frac{K}{21}\right)\frac{ m^4}{k^2}+ \left(-\frac{1268}{735}+\frac{4 K}{7}\right)\frac{ m^6}{k^4}\right)\right)+
\0\\ & \quad 
 + k^{10} \pi _{\mu\mu}^2 \pi _{\mu\nu} \pi _{\nu\nu}^2 \left(\frac{i }{5 \pi ^3} \left( \left(\frac{367291}{721440720}-\frac{K}{8008}\right) k^2+
\right.\right.\0\\ & \quad \quad \quad \left.\left.
 +  \left(-\frac{5023}{457380}+\frac{K}{264}\right) m^2+ \left(\frac{5543}{52920}-\frac{K}{21}\right)\frac{ m^4}{k^2}+
\right.\right.\0\\ & \quad \quad \quad \left.\left.
 +  \left(-\frac{1027}{1960}+\frac{9 K}{28}\right)\frac{ m^6}{k^4}\right)\right)+\ldots }
\subsection{Divergences of the fermion amplitudes}
Fermions, spin 0 x 2:
\al{\label{eq:fnc:0:2:3:divr}\tilde{T}_{0,2;3\text{D}}^{\text{f,nt}}\cdot k & = k_{\nu } \left(-\frac{i }{\pi } m^2\right) }
\al{\label{eq:fnc:0:2:4:divr}\tilde{T}_{0,2;4\text{D}}^{\text{f,nt}}\cdot k & = k_{\nu } \left(\frac{i  L_1}{2 \pi ^2} m^3\right) }
\al{\label{eq:fnc:0:2:5:divr}\tilde{T}_{0,2;5\text{D}}^{\text{f,nt}}\cdot k & = k_{\nu } \left(\frac{i }{3 \pi ^2} m^4\right) }
\al{\label{eq:fnc:0:2:6:divr}\tilde{T}_{0,2;6\text{D}}^{\text{f,nt}}\cdot k & = k_{\nu } \left(-\frac{i  L_2}{8 \pi ^3} m^5\right) }
Fermions, spin 0 x 4:
\al{\label{eq:fnc:0:4:3:divr}\tilde{T}_{0,4;3\text{D}}^{\text{f,nt}}\cdot k & = k_{\nu }^3 \left(-\frac{i }{\pi } m^2\right)+k_{\nu } \eta_{\nu\nu} \left(-\frac{4 i }{\pi } m^4\right) }
\al{\label{eq:fnc:0:4:4:divr}\tilde{T}_{0,4;4\text{D}}^{\text{f,nt}}\cdot k & = k_{\nu }^3 \left(\frac{i  L_1}{2 \pi ^2} m^3\right)+k_{\nu } \eta_{\nu\nu} \left(\frac{3 i  L_2}{2 \pi ^2} m^5\right) }
\al{\label{eq:fnc:0:4:5:divr}\tilde{T}_{0,4;5\text{D}}^{\text{f,nt}}\cdot k & = k_{\nu }^3 \left(\frac{i }{3 \pi ^2} m^4\right)+k_{\nu } \eta_{\nu\nu} \left(\frac{4 i }{5 \pi ^2} m^6\right) }
\al{\label{eq:fnc:0:4:6:divr}\tilde{T}_{0,4;6\text{D}}^{\text{f,nt}}\cdot k & = k_{\nu }^3 \left(-\frac{i  L_2}{8 \pi ^3} m^5\right)+k_{\nu } \eta_{\nu\nu} \left(-\frac{i  L_3}{4 \pi ^3} m^7\right) }
Fermions, spin 1 x 1:
\al{\label{eq:fnc:1:1:3:div}k\cdot \tilde{T}_{1,1;3\text{D}}^{\text{f,nt}} & = 0 }
\al{\label{eq:fnc:1:1:4:div}k\cdot \tilde{T}_{1,1;4\text{D}}^{\text{f,nt}} & = 0 }
\al{\label{eq:fnc:1:1:5:div}k\cdot \tilde{T}_{1,1;5\text{D}}^{\text{f,nt}} & = 0 }
\al{\label{eq:fnc:1:1:6:div}k\cdot \tilde{T}_{1,1;6\text{D}}^{\text{f,nt}} & = 0 }
Fermions, spin 1 x 3:
\al{\label{eq:fnc:1:3:3:divl}k\cdot \tilde{T}_{1,3;3\text{D}}^{\text{f,nt}} & = k_{\nu } \eta_{\nu\nu} \left(-\frac{4 i }{3 \pi } m^3\right) }
\al{\label{eq:fnc:1:3:3:divr}\tilde{T}_{1,3;3\text{D}}^{\text{f,nt}}\cdot k & = k_{\mu } \eta_{\nu\nu} \left(-\frac{4 i }{9 \pi } m^3\right)+k_{\nu } \eta_{\mu\nu} \left(-\frac{8 i }{9 \pi } m^3\right)+k_{\nu } (k\cdot\epsilon)_{\mu\nu} \left(-\frac{2 }{3 \pi } m^2\right) }
\al{\label{eq:fnc:1:3:4:divl}k\cdot \tilde{T}_{1,3;4\text{D}}^{\text{f,nt}} & = k_{\nu } \eta_{\nu\nu} \left(\frac{i  L_2}{2 \pi ^2} m^4\right) }
\al{\label{eq:fnc:1:3:4:divr}\tilde{T}_{1,3;4\text{D}}^{\text{f,nt}}\cdot k & = k_{\mu } \eta_{\nu\nu} \left(\frac{i  L_2}{6 \pi ^2} m^4\right)+k_{\nu } \eta_{\mu\nu} \left(\frac{i  L_2}{3 \pi ^2} m^4\right) }
\al{\label{eq:fnc:1:3:5:divl}k\cdot \tilde{T}_{1,3;5\text{D}}^{\text{f,nt}} & = k_{\nu } \eta_{\nu\nu} \left(\frac{4 i }{15 \pi ^2} m^5\right) }
\al{\label{eq:fnc:1:3:5:divr}\tilde{T}_{1,3;5\text{D}}^{\text{f,nt}}\cdot k & = k_{\mu } \eta_{\nu\nu} \left(\frac{4 i }{45 \pi ^2} m^5\right)+k_{\nu } \eta_{\mu\nu} \left(\frac{8 i }{45 \pi ^2} m^5\right) }
\al{\label{eq:fnc:1:3:6:divl}k\cdot \tilde{T}_{1,3;6\text{D}}^{\text{f,nt}} & = k_{\nu } \eta_{\nu\nu} \left(-\frac{i  L_3}{12 \pi ^3} m^6\right) }
\al{\label{eq:fnc:1:3:6:divr}\tilde{T}_{1,3;6\text{D}}^{\text{f,nt}}\cdot k & = k_{\mu } \eta_{\nu\nu} \left(-\frac{i  L_3}{36 \pi ^3} m^6\right)+k_{\nu } \eta_{\mu\nu} \left(-\frac{i  L_3}{18 \pi ^3} m^6\right) }
Fermions, spin 1 x 5:
\al{\label{eq:fnc:1:5:3:divl}k\cdot \tilde{T}_{1,5;3\text{D}}^{\text{f,nt}} & = k_{\nu }^3 \eta_{\nu\nu} \left(-\frac{8 i }{3 \pi } m^3\right)+k_{\nu } \eta_{\nu\nu}^2 \left(-\frac{32 i }{5 \pi } m^5\right) }
\al{\label{eq:fnc:1:5:3:divr}\tilde{T}_{1,5;3\text{D}}^{\text{f,nt}}\cdot k & = k_{\mu } k_{\nu }^2 \eta_{\nu\nu} \left(-\frac{8 i }{5 \pi } m^3\right)+k_{\nu }^3 \eta_{\mu\nu} \left(-\frac{16 i }{15 \pi } m^3\right)+k_{\mu } \eta_{\nu\nu}^2 \left(-\frac{32 i }{25 \pi } m^5\right)+
\0\\ & \quad 
 + k_{\nu } \eta_{\mu\nu} \eta_{\nu\nu} \left(-\frac{128 i }{25 \pi } m^5\right)+k_{\nu }^3 (k\cdot\epsilon)_{\mu\nu} \left(-\frac{4 }{5 \pi } m^2\right)+
\0\\ & \quad 
 + k_{\nu } (k\cdot\epsilon)_{\mu\nu} \eta_{\nu\nu} \left(-\frac{16 }{5 \pi } m^4\right) }
\al{\label{eq:fnc:1:5:4:divl}k\cdot \tilde{T}_{1,5;4\text{D}}^{\text{f,nt}} & = k_{\nu }^3 \eta_{\nu\nu} \left(\frac{i  L_2}{\pi ^2} m^4\right)+k_{\nu } \eta_{\nu\nu}^2 \left(\frac{2 i  L_3}{\pi ^2} m^6\right) }
\al{\label{eq:fnc:1:5:4:divr}\tilde{T}_{1,5;4\text{D}}^{\text{f,nt}}\cdot k & = k_{\mu } k_{\nu }^2 \eta_{\nu\nu} \left(\frac{3 i  L_2}{5 \pi ^2} m^4\right)+k_{\nu }^3 \eta_{\mu\nu} \left(\frac{2 i  L_2}{5 \pi ^2} m^4\right)+k_{\mu } \eta_{\nu\nu}^2 \left(\frac{2 i  L_3}{5 \pi ^2} m^6\right)+
\0\\ & \quad 
 + k_{\nu } \eta_{\mu\nu} \eta_{\nu\nu} \left(\frac{8 i  L_3}{5 \pi ^2} m^6\right) }
\al{\label{eq:fnc:1:5:5:divl}k\cdot \tilde{T}_{1,5;5\text{D}}^{\text{f,nt}} & = k_{\nu }^3 \eta_{\nu\nu} \left(\frac{8 i }{15 \pi ^2} m^5\right)+k_{\nu } \eta_{\nu\nu}^2 \left(\frac{32 i }{35 \pi ^2} m^7\right) }
\al{\label{eq:fnc:1:5:5:divr}\tilde{T}_{1,5;5\text{D}}^{\text{f,nt}}\cdot k & = k_{\mu } k_{\nu }^2 \eta_{\nu\nu} \left(\frac{8 i }{25 \pi ^2} m^5\right)+k_{\nu }^3 \eta_{\mu\nu} \left(\frac{16 i }{75 \pi ^2} m^5\right)+k_{\mu } \eta_{\nu\nu}^2 \left(\frac{32 i }{175 \pi ^2} m^7\right)+
\0\\ & \quad 
 + k_{\nu } \eta_{\mu\nu} \eta_{\nu\nu} \left(\frac{128 i }{175 \pi ^2} m^7\right) }
\al{\label{eq:fnc:1:5:6:divl}k\cdot \tilde{T}_{1,5;6\text{D}}^{\text{f,nt}} & = k_{\nu }^3 \eta_{\nu\nu} \left(-\frac{i  L_3}{6 \pi ^3} m^6\right)+k_{\nu } \eta_{\nu\nu}^2 \left(-\frac{i  L_4}{4 \pi ^3} m^8\right) }
\al{\label{eq:fnc:1:5:6:divr}\tilde{T}_{1,5;6\text{D}}^{\text{f,nt}}\cdot k & = k_{\mu } k_{\nu }^2 \eta_{\nu\nu} \left(-\frac{i  L_3}{10 \pi ^3} m^6\right)+k_{\nu }^3 \eta_{\mu\nu} \left(-\frac{i  L_3}{15 \pi ^3} m^6\right)+k_{\mu } \eta_{\nu\nu}^2 \left(-\frac{i  L_4}{20 \pi ^3} m^8\right)+
\0\\ & \quad 
 + k_{\nu } \eta_{\mu\nu} \eta_{\nu\nu} \left(-\frac{i  L_4}{5 \pi ^3} m^8\right) }
Fermions, spin 2 x 2:
\al{\label{eq:fnc:2:2:3:div}k\cdot \tilde{T}_{2,2;3\text{D}}^{\text{f,nt}} & =  \left(k_{\nu } \eta_{\mu\nu}+k_{\mu } \eta_{\nu\nu}\right) \left(-\frac{2 i }{3 \pi } m^3\right)+k_{\nu } (k\cdot\epsilon)_{\mu\nu} \left(-\frac{1}{2 \pi } m^2\right) }
\al{\label{eq:fnc:2:2:4:div}k\cdot \tilde{T}_{2,2;4\text{D}}^{\text{f,nt}} & =  \left(k_{\nu } \eta_{\mu\nu}+k_{\mu } \eta_{\nu\nu}\right) \left(\frac{i  L_2}{4 \pi ^2} m^4\right) }
\al{\label{eq:fnc:2:2:5:div}k\cdot \tilde{T}_{2,2;5\text{D}}^{\text{f,nt}} & =  \left(k_{\nu } \eta_{\mu\nu}+k_{\mu } \eta_{\nu\nu}\right) \left(\frac{2 i }{15 \pi ^2} m^5\right) }
\al{\label{eq:fnc:2:2:6:div}k\cdot \tilde{T}_{2,2;6\text{D}}^{\text{f,nt}} & =  \left(k_{\nu } \eta_{\mu\nu}+k_{\mu } \eta_{\nu\nu}\right) \left(-\frac{i  L_3}{24 \pi ^3} m^6\right) }
Fermions, spin 2 x 4:
\al{\label{eq:fnc:2:4:3:divl}k\cdot \tilde{T}_{2,4;3\text{D}}^{\text{f,nt}} & = k_{\mu } k_{\nu }^2 \eta_{\nu\nu} \left(-\frac{2 i }{\pi } m^3\right)+k_{\nu }^3 \eta_{\mu\nu} \left(-\frac{2 i }{3 \pi } m^3\right)+k_{\mu } \eta_{\nu\nu}^2 \left(-\frac{8 i }{5 \pi } m^5\right)+
\0\\ & \quad 
 + k_{\nu } \eta_{\mu\nu} \eta_{\nu\nu} \left(-\frac{24 i }{5 \pi } m^5\right)+k_{\nu }^3 (k\cdot\epsilon)_{\mu\nu} \left(-\frac{1}{2 \pi } m^2\right)+k_{\nu } (k\cdot\epsilon)_{\mu\nu} \eta_{\nu\nu} \left(-\frac{2 }{\pi } m^4\right) }
\al{\label{eq:fnc:2:4:3:divr}\tilde{T}_{2,4;3\text{D}}^{\text{f,nt}}\cdot k & = k_{\nu }^3 \eta_{\mu\mu} \left(-\frac{2 i }{3 \pi } m^3\right)+ \left(k_{\mu } k_{\nu }^2 \eta_{\mu\nu}+k_{\mu }^2 k_{\nu } \eta_{\nu\nu}\right) \left(-\frac{i }{\pi } m^3\right)+
\0\\ & \quad 
 + k_{\nu } \eta_{\mu\mu} \eta_{\nu\nu} \left(-\frac{8 i }{5 \pi } m^5\right)+ \left(k_{\nu } \eta_{\mu\nu}^2+k_{\mu } \eta_{\mu\nu} \eta_{\nu\nu}\right) \left(-\frac{12 i }{5 \pi } m^5\right)+
\0\\ & \quad 
 + k_{\mu } k_{\nu }^2 (k\cdot\epsilon)_{\mu\nu} \left(-\frac{3 }{4 \pi } m^2\right)+k_{\mu } (k\cdot\epsilon)_{\mu\nu} \eta_{\nu\nu} \left(-\frac{1}{\pi } m^4\right)+
\0\\ & \quad 
 + k_{\nu } (k\cdot\epsilon)_{\mu\nu} \eta_{\mu\nu} \left(-\frac{2 }{\pi } m^4\right) }
\al{\label{eq:fnc:2:4:4:divl}k\cdot \tilde{T}_{2,4;4\text{D}}^{\text{f,nt}} & = k_{\mu } k_{\nu }^2 \eta_{\nu\nu} \left(\frac{3 i  L_2}{4 \pi ^2} m^4\right)+k_{\nu }^3 \eta_{\mu\nu} \left(\frac{i  L_2}{4 \pi ^2} m^4\right)+k_{\mu } \eta_{\nu\nu}^2 \left(\frac{i  L_3}{2 \pi ^2} m^6\right)+
\0\\ & \quad 
 + k_{\nu } \eta_{\mu\nu} \eta_{\nu\nu} \left(\frac{3 i  L_3}{2 \pi ^2} m^6\right) }
\al{\label{eq:fnc:2:4:4:divr}\tilde{T}_{2,4;4\text{D}}^{\text{f,nt}}\cdot k & = k_{\nu }^3 \eta_{\mu\mu} \left(\frac{i  L_2}{4 \pi ^2} m^4\right)+ \left(k_{\mu } k_{\nu }^2 \eta_{\mu\nu}+k_{\mu }^2 k_{\nu } \eta_{\nu\nu}\right) \left(\frac{3 i  L_2}{8 \pi ^2} m^4\right)+
\0\\ & \quad 
 + k_{\nu } \eta_{\mu\mu} \eta_{\nu\nu} \left(\frac{i  L_3}{2 \pi ^2} m^6\right)+ \left(k_{\nu } \eta_{\mu\nu}^2+k_{\mu } \eta_{\mu\nu} \eta_{\nu\nu}\right) \left(\frac{3 i  L_3}{4 \pi ^2} m^6\right) }
\al{\label{eq:fnc:2:4:5:divl}k\cdot \tilde{T}_{2,4;5\text{D}}^{\text{f,nt}} & = k_{\mu } k_{\nu }^2 \eta_{\nu\nu} \left(\frac{2 i }{5 \pi ^2} m^5\right)+k_{\nu }^3 \eta_{\mu\nu} \left(\frac{2 i }{15 \pi ^2} m^5\right)+k_{\mu } \eta_{\nu\nu}^2 \left(\frac{8 i }{35 \pi ^2} m^7\right)+
\0\\ & \quad 
 + k_{\nu } \eta_{\mu\nu} \eta_{\nu\nu} \left(\frac{24 i }{35 \pi ^2} m^7\right) }
\al{\label{eq:fnc:2:4:5:divr}\tilde{T}_{2,4;5\text{D}}^{\text{f,nt}}\cdot k & = k_{\nu }^3 \eta_{\mu\mu} \left(\frac{2 i }{15 \pi ^2} m^5\right)+ \left(k_{\mu } k_{\nu }^2 \eta_{\mu\nu}+k_{\mu }^2 k_{\nu } \eta_{\nu\nu}\right) \left(\frac{i }{5 \pi ^2} m^5\right)+
\0\\ & \quad 
 + k_{\nu } \eta_{\mu\mu} \eta_{\nu\nu} \left(\frac{8 i }{35 \pi ^2} m^7\right)+ \left(k_{\nu } \eta_{\mu\nu}^2+k_{\mu } \eta_{\mu\nu} \eta_{\nu\nu}\right) \left(\frac{12 i }{35 \pi ^2} m^7\right) }
\al{\label{eq:fnc:2:4:6:divl}k\cdot \tilde{T}_{2,4;6\text{D}}^{\text{f,nt}} & = k_{\mu } k_{\nu }^2 \eta_{\nu\nu} \left(-\frac{i  L_3}{8 \pi ^3} m^6\right)+k_{\nu }^3 \eta_{\mu\nu} \left(-\frac{i  L_3}{24 \pi ^3} m^6\right)+k_{\mu } \eta_{\nu\nu}^2 \left(-\frac{i  L_4}{16 \pi ^3} m^8\right)+
\0\\ & \quad 
 + k_{\nu } \eta_{\mu\nu} \eta_{\nu\nu} \left(-\frac{3 i  L_4}{16 \pi ^3} m^8\right) }
\al{\label{eq:fnc:2:4:6:divr}\tilde{T}_{2,4;6\text{D}}^{\text{f,nt}}\cdot k & = k_{\nu }^3 \eta_{\mu\mu} \left(-\frac{i  L_3}{24 \pi ^3} m^6\right)+ \left(k_{\mu } k_{\nu }^2 \eta_{\mu\nu}+k_{\mu }^2 k_{\nu } \eta_{\nu\nu}\right) \left(-\frac{i  L_3}{16 \pi ^3} m^6\right)+
\0\\ & \quad 
 + k_{\nu } \eta_{\mu\mu} \eta_{\nu\nu} \left(-\frac{i  L_4}{16 \pi ^3} m^8\right)+ \left(k_{\nu } \eta_{\mu\nu}^2+k_{\mu } \eta_{\mu\nu} \eta_{\nu\nu}\right) \left(-\frac{3 i  L_4}{32 \pi ^3} m^8\right) }
Fermions, spin 3 x 3:
\al{\label{eq:fnc:3:3:3:div}k\cdot \tilde{T}_{3,3;3\text{D}}^{\text{f,nt}} & = k_{\nu }^3 \eta_{\mu\mu} \left(-\frac{4 i }{9 \pi } m^3\right)+k_{\mu }^2 k_{\nu } \eta_{\nu\nu} \left(-\frac{4 i }{3 \pi } m^3\right)+k_{\mu } k_{\nu }^2 \eta_{\mu\nu} \left(-\frac{8 i }{9 \pi } m^3\right)+
\0\\ & \quad 
 + k_{\nu } \eta_{\mu\mu} \eta_{\nu\nu} \left(-\frac{64 i }{45 \pi } m^5\right)+k_{\mu } \eta_{\mu\nu} \eta_{\nu\nu} \left(-\frac{128 i }{45 \pi } m^5\right)+k_{\nu } \eta_{\mu\nu}^2 \left(-\frac{32 i }{15 \pi } m^5\right)+
\0\\ & \quad 
 + k_{\mu } k_{\nu }^2 (k\cdot\epsilon)_{\mu\nu} \left(-\frac{2 }{3 \pi } m^2\right)+k_{\mu } (k\cdot\epsilon)_{\mu\nu} \eta_{\nu\nu} \left(-\frac{8 }{9 \pi } m^4\right)+
\0\\ & \quad 
 + k_{\nu } (k\cdot\epsilon)_{\mu\nu} \eta_{\mu\nu} \left(-\frac{16 }{9 \pi } m^4\right) }
\al{\label{eq:fnc:3:3:4:div}k\cdot \tilde{T}_{3,3;4\text{D}}^{\text{f,nt}} & = k_{\nu }^3 \eta_{\mu\mu} \left(\frac{i  L_2}{6 \pi ^2} m^4\right)+k_{\mu }^2 k_{\nu } \eta_{\nu\nu} \left(\frac{i  L_2}{2 \pi ^2} m^4\right)+k_{\mu } k_{\nu }^2 \eta_{\mu\nu} \left(\frac{i  L_2}{3 \pi ^2} m^4\right)+
\0\\ & \quad 
 + k_{\nu } \eta_{\mu\mu} \eta_{\nu\nu} \left(\frac{4 i  L_3}{9 \pi ^2} m^6\right)+k_{\mu } \eta_{\mu\nu} \eta_{\nu\nu} \left(\frac{8 i  L_3}{9 \pi ^2} m^6\right)+k_{\nu } \eta_{\mu\nu}^2 \left(\frac{2 i  L_3}{3 \pi ^2} m^6\right) }
\al{\label{eq:fnc:3:3:5:div}k\cdot \tilde{T}_{3,3;5\text{D}}^{\text{f,nt}} & = k_{\nu }^3 \eta_{\mu\mu} \left(\frac{4 i }{45 \pi ^2} m^5\right)+k_{\mu }^2 k_{\nu } \eta_{\nu\nu} \left(\frac{4 i }{15 \pi ^2} m^5\right)+k_{\mu } k_{\nu }^2 \eta_{\mu\nu} \left(\frac{8 i }{45 \pi ^2} m^5\right)+
\0\\ & \quad 
 + k_{\nu } \eta_{\mu\mu} \eta_{\nu\nu} \left(\frac{64 i }{315 \pi ^2} m^7\right)+k_{\mu } \eta_{\mu\nu} \eta_{\nu\nu} \left(\frac{128 i }{315 \pi ^2} m^7\right)+k_{\nu } \eta_{\mu\nu}^2 \left(\frac{32 i }{105 \pi ^2} m^7\right) }
\al{\label{eq:fnc:3:3:6:div}k\cdot \tilde{T}_{3,3;6\text{D}}^{\text{f,nt}} & = k_{\nu }^3 \eta_{\mu\mu} \left(-\frac{i  L_3}{36 \pi ^3} m^6\right)+k_{\mu }^2 k_{\nu } \eta_{\nu\nu} \left(-\frac{i  L_3}{12 \pi ^3} m^6\right)+k_{\mu } k_{\nu }^2 \eta_{\mu\nu} \left(-\frac{i  L_3}{18 \pi ^3} m^6\right)+
\0\\ & \quad 
 + k_{\nu } \eta_{\mu\mu} \eta_{\nu\nu} \left(-\frac{i  L_4}{18 \pi ^3} m^8\right)+k_{\mu } \eta_{\mu\nu} \eta_{\nu\nu} \left(-\frac{i  L_4}{9 \pi ^3} m^8\right)+k_{\nu } \eta_{\mu\nu}^2 \left(-\frac{i  L_4}{12 \pi ^3} m^8\right) }
Fermions, spin 3 x 5:
\al{\label{eq:fnc:3:5:3:divl}k\cdot \tilde{T}_{3,5;3\text{D}}^{\text{f,nt}} & = k_{\nu }^5 \eta_{\mu\mu} \left(-\frac{4 i }{9 \pi } m^3\right)+k_{\mu }^2 k_{\nu }^3 \eta_{\nu\nu} \left(-\frac{8 i }{3 \pi } m^3\right)+k_{\mu } k_{\nu }^4 \eta_{\mu\nu} \left(-\frac{8 i }{9 \pi } m^3\right)+
\0\\ & \quad 
 + k_{\mu }^2 k_{\nu } \eta_{\nu\nu}^2 \left(-\frac{32 i }{5 \pi } m^5\right)+k_{\mu } k_{\nu }^2 \eta_{\mu\nu} \eta_{\nu\nu} \left(-\frac{256 i }{15 \pi } m^5\right)+
\0\\ & \quad 
 +  \left(k_{\nu }^3 \eta_{\mu\nu}^2+k_{\nu }^3 \eta_{\mu\mu} \eta_{\nu\nu}\right) \left(-\frac{64 i }{15 \pi } m^5\right)+k_{\nu } \eta_{\mu\mu} \eta_{\nu\nu}^2 \left(-\frac{64 i }{15 \pi } m^7\right)+
\0\\ & \quad 
 + k_{\mu } \eta_{\mu\nu} \eta_{\nu\nu}^2 \left(-\frac{128 i }{15 \pi } m^7\right)+k_{\nu } \eta_{\mu\nu}^2 \eta_{\nu\nu} \left(-\frac{512 i }{35 \pi } m^7\right)+
\0\\ & \quad 
 + k_{\mu } k_{\nu }^4 (k\cdot\epsilon)_{\mu\nu} \left(-\frac{2 }{3 \pi } m^2\right)+k_{\mu } k_{\nu }^2 (k\cdot\epsilon)_{\mu\nu} \eta_{\nu\nu} \left(-\frac{16 }{3 \pi } m^4\right)+
\0\\ & \quad 
 + k_{\nu }^3 (k\cdot\epsilon)_{\mu\nu} \eta_{\mu\nu} \left(-\frac{32 }{9 \pi } m^4\right)+k_{\mu } (k\cdot\epsilon)_{\mu\nu} \eta_{\nu\nu}^2 \left(-\frac{32 }{15 \pi } m^6\right)+
\0\\ & \quad 
 + k_{\nu } (k\cdot\epsilon)_{\mu\nu} \eta_{\mu\nu} \eta_{\nu\nu} \left(-\frac{128 }{15 \pi } m^6\right) }
\al{\label{eq:fnc:3:5:3:divr}\tilde{T}_{3,5;3\text{D}}^{\text{f,nt}}\cdot k & = k_{\mu } k_{\nu }^4 \eta_{\mu\mu} \left(-\frac{4 i }{3 \pi } m^3\right)+k_{\mu }^3 k_{\nu }^2 \eta_{\nu\nu} \left(-\frac{8 i }{5 \pi } m^3\right)+k_{\mu }^2 k_{\nu }^3 \eta_{\mu\nu} \left(-\frac{16 i }{15 \pi } m^3\right)+
\0\\ & \quad 
 + k_{\mu }^3 \eta_{\nu\nu}^2 \left(-\frac{32 i }{25 \pi } m^5\right)+k_{\nu }^3 \eta_{\mu\mu} \eta_{\mu\nu} \left(-\frac{128 i }{25 \pi } m^5\right)+
\0\\ & \quad 
 + k_{\mu }^2 k_{\nu } \eta_{\mu\nu} \eta_{\nu\nu} \left(-\frac{256 i }{25 \pi } m^5\right)+ \left(k_{\mu } k_{\nu }^2 \eta_{\mu\nu}^2+k_{\mu } k_{\nu }^2 \eta_{\mu\mu} \eta_{\nu\nu}\right) \left(-\frac{192 i }{25 \pi } m^5\right)+
\0\\ & \quad 
 + k_{\mu } \eta_{\mu\mu} \eta_{\nu\nu}^2 \left(-\frac{64 i }{25 \pi } m^7\right)+k_{\nu } \eta_{\mu\mu} \eta_{\mu\nu} \eta_{\nu\nu} \left(-\frac{256 i }{25 \pi } m^7\right)+
\0\\ & \quad 
 + k_{\mu } \eta_{\mu\nu}^2 \eta_{\nu\nu} \left(-\frac{1536 i }{175 \pi } m^7\right)+k_{\nu } \eta_{\mu\nu}^3 \left(-\frac{1024 i }{175 \pi } m^7\right)+
\0\\ & \quad 
 + k_{\mu }^2 k_{\nu }^3 (k\cdot\epsilon)_{\mu\nu} \left(-\frac{4 }{5 \pi } m^2\right)+k_{\nu }^3 (k\cdot\epsilon)_{\mu\nu} \eta_{\mu\mu} \left(-\frac{16 }{15 \pi } m^4\right)+
\0\\ & \quad 
 + k_{\mu }^2 k_{\nu } (k\cdot\epsilon)_{\mu\nu} \eta_{\nu\nu} \left(-\frac{16 }{5 \pi } m^4\right)+k_{\mu } k_{\nu }^2 (k\cdot\epsilon)_{\mu\nu} \eta_{\mu\nu} \left(-\frac{32 }{5 \pi } m^4\right)+
\0\\ & \quad 
 + k_{\nu } (k\cdot\epsilon)_{\mu\nu} \eta_{\mu\mu} \eta_{\nu\nu} \left(-\frac{64 }{25 \pi } m^6\right)+
\0\\ & \quad 
 +  \left(k_{\nu } (k\cdot\epsilon)_{\mu\nu} \eta_{\mu\nu}^2+k_{\mu } (k\cdot\epsilon)_{\mu\nu} \eta_{\mu\nu} \eta_{\nu\nu}\right) \left(-\frac{128 }{25 \pi } m^6\right) }
\al{\label{eq:fnc:3:5:4:divl}k\cdot \tilde{T}_{3,5;4\text{D}}^{\text{f,nt}} & = k_{\nu }^5 \eta_{\mu\mu} \left(\frac{i  L_2}{6 \pi ^2} m^4\right)+k_{\mu }^2 k_{\nu }^3 \eta_{\nu\nu} \left(\frac{i  L_2}{\pi ^2} m^4\right)+k_{\mu } k_{\nu }^4 \eta_{\mu\nu} \left(\frac{i  L_2}{3 \pi ^2} m^4\right)+
\0\\ & \quad 
 + k_{\mu }^2 k_{\nu } \eta_{\nu\nu}^2 \left(\frac{2 i  L_3}{\pi ^2} m^6\right)+k_{\mu } k_{\nu }^2 \eta_{\mu\nu} \eta_{\nu\nu} \left(\frac{16 i  L_3}{3 \pi ^2} m^6\right)+
\0\\ & \quad 
 +  \left(k_{\nu }^3 \eta_{\mu\nu}^2+k_{\nu }^3 \eta_{\mu\mu} \eta_{\nu\nu}\right) \left(\frac{4 i  L_3}{3 \pi ^2} m^6\right)+k_{\nu } \eta_{\mu\mu} \eta_{\nu\nu}^2 \left(\frac{7 i  L_4}{6 \pi ^2} m^8\right)+
\0\\ & \quad 
 + k_{\mu } \eta_{\mu\nu} \eta_{\nu\nu}^2 \left(\frac{7 i  L_4}{3 \pi ^2} m^8\right)+k_{\nu } \eta_{\mu\nu}^2 \eta_{\nu\nu} \left(\frac{4 i  L_4}{\pi ^2} m^8\right) }
\al{\label{eq:fnc:3:5:4:divr}\tilde{T}_{3,5;4\text{D}}^{\text{f,nt}}\cdot k & = k_{\mu } k_{\nu }^4 \eta_{\mu\mu} \left(\frac{i  L_2}{2 \pi ^2} m^4\right)+k_{\mu }^3 k_{\nu }^2 \eta_{\nu\nu} \left(\frac{3 i  L_2}{5 \pi ^2} m^4\right)+k_{\mu }^2 k_{\nu }^3 \eta_{\mu\nu} \left(\frac{2 i  L_2}{5 \pi ^2} m^4\right)+
\0\\ & \quad 
 + k_{\mu }^3 \eta_{\nu\nu}^2 \left(\frac{2 i  L_3}{5 \pi ^2} m^6\right)+k_{\nu }^3 \eta_{\mu\mu} \eta_{\mu\nu} \left(\frac{8 i  L_3}{5 \pi ^2} m^6\right)+k_{\mu }^2 k_{\nu } \eta_{\mu\nu} \eta_{\nu\nu} \left(\frac{16 i  L_3}{5 \pi ^2} m^6\right)+
\0\\ & \quad 
 +  \left(k_{\mu } k_{\nu }^2 \eta_{\mu\nu}^2+k_{\mu } k_{\nu }^2 \eta_{\mu\mu} \eta_{\nu\nu}\right) \left(\frac{12 i  L_3}{5 \pi ^2} m^6\right)+k_{\mu } \eta_{\mu\mu} \eta_{\nu\nu}^2 \left(\frac{7 i  L_4}{10 \pi ^2} m^8\right)+
\0\\ & \quad 
 + k_{\nu } \eta_{\mu\mu} \eta_{\mu\nu} \eta_{\nu\nu} \left(\frac{14 i  L_4}{5 \pi ^2} m^8\right)+k_{\mu } \eta_{\mu\nu}^2 \eta_{\nu\nu} \left(\frac{12 i  L_4}{5 \pi ^2} m^8\right)+k_{\nu } \eta_{\mu\nu}^3 \left(\frac{8 i  L_4}{5 \pi ^2} m^8\right) }
\al{\label{eq:fnc:3:5:5:divl}k\cdot \tilde{T}_{3,5;5\text{D}}^{\text{f,nt}} & = k_{\nu }^5 \eta_{\mu\mu} \left(\frac{4 i }{45 \pi ^2} m^5\right)+k_{\mu }^2 k_{\nu }^3 \eta_{\nu\nu} \left(\frac{8 i }{15 \pi ^2} m^5\right)+k_{\mu } k_{\nu }^4 \eta_{\mu\nu} \left(\frac{8 i }{45 \pi ^2} m^5\right)+
\0\\ & \quad 
 + k_{\mu }^2 k_{\nu } \eta_{\nu\nu}^2 \left(\frac{32 i }{35 \pi ^2} m^7\right)+k_{\mu } k_{\nu }^2 \eta_{\mu\nu} \eta_{\nu\nu} \left(\frac{256 i }{105 \pi ^2} m^7\right)+
\0\\ & \quad 
 +  \left(k_{\nu }^3 \eta_{\mu\nu}^2+k_{\nu }^3 \eta_{\mu\mu} \eta_{\nu\nu}\right) \left(\frac{64 i }{105 \pi ^2} m^7\right)+k_{\nu } \eta_{\mu\mu} \eta_{\nu\nu}^2 \left(\frac{64 i }{135 \pi ^2} m^9\right)+
\0\\ & \quad 
 + k_{\mu } \eta_{\mu\nu} \eta_{\nu\nu}^2 \left(\frac{128 i }{135 \pi ^2} m^9\right)+k_{\nu } \eta_{\mu\nu}^2 \eta_{\nu\nu} \left(\frac{512 i }{315 \pi ^2} m^9\right) }
\al{\label{eq:fnc:3:5:5:divr}\tilde{T}_{3,5;5\text{D}}^{\text{f,nt}}\cdot k & = k_{\mu } k_{\nu }^4 \eta_{\mu\mu} \left(\frac{4 i }{15 \pi ^2} m^5\right)+k_{\mu }^3 k_{\nu }^2 \eta_{\nu\nu} \left(\frac{8 i }{25 \pi ^2} m^5\right)+k_{\mu }^2 k_{\nu }^3 \eta_{\mu\nu} \left(\frac{16 i }{75 \pi ^2} m^5\right)+
\0\\ & \quad 
 + k_{\mu }^3 \eta_{\nu\nu}^2 \left(\frac{32 i }{175 \pi ^2} m^7\right)+k_{\nu }^3 \eta_{\mu\mu} \eta_{\mu\nu} \left(\frac{128 i }{175 \pi ^2} m^7\right)+
\0\\ & \quad 
 + k_{\mu }^2 k_{\nu } \eta_{\mu\nu} \eta_{\nu\nu} \left(\frac{256 i }{175 \pi ^2} m^7\right)+ \left(k_{\mu } k_{\nu }^2 \eta_{\mu\nu}^2+k_{\mu } k_{\nu }^2 \eta_{\mu\mu} \eta_{\nu\nu}\right) \left(\frac{192 i }{175 \pi ^2} m^7\right)+
\0\\ & \quad 
 + k_{\mu } \eta_{\mu\mu} \eta_{\nu\nu}^2 \left(\frac{64 i }{225 \pi ^2} m^9\right)+k_{\nu } \eta_{\mu\mu} \eta_{\mu\nu} \eta_{\nu\nu} \left(\frac{256 i }{225 \pi ^2} m^9\right)+
\0\\ & \quad 
 + k_{\mu } \eta_{\mu\nu}^2 \eta_{\nu\nu} \left(\frac{512 i }{525 \pi ^2} m^9\right)+k_{\nu } \eta_{\mu\nu}^3 \left(\frac{1024 i }{1575 \pi ^2} m^9\right) }
\al{\label{eq:fnc:3:5:6:divl}k\cdot \tilde{T}_{3,5;6\text{D}}^{\text{f,nt}} & = k_{\nu }^5 \eta_{\mu\mu} \left(-\frac{i  L_3}{36 \pi ^3} m^6\right)+k_{\mu }^2 k_{\nu }^3 \eta_{\nu\nu} \left(-\frac{i  L_3}{6 \pi ^3} m^6\right)+k_{\mu } k_{\nu }^4 \eta_{\mu\nu} \left(-\frac{i  L_3}{18 \pi ^3} m^6\right)+
\0\\ & \quad 
 + k_{\mu }^2 k_{\nu } \eta_{\nu\nu}^2 \left(-\frac{i  L_4}{4 \pi ^3} m^8\right)+k_{\mu } k_{\nu }^2 \eta_{\mu\nu} \eta_{\nu\nu} \left(-\frac{2 i  L_4}{3 \pi ^3} m^8\right)+
\0\\ & \quad 
 +  \left(k_{\nu }^3 \eta_{\mu\nu}^2+k_{\nu }^3 \eta_{\mu\mu} \eta_{\nu\nu}\right) \left(-\frac{i  L_4}{6 \pi ^3} m^8\right)+k_{\nu } \eta_{\mu\mu} \eta_{\nu\nu}^2 \left(-\frac{7 i  L_5}{60 \pi ^3} m^{10}\right)+
\0\\ & \quad 
 + k_{\mu } \eta_{\mu\nu} \eta_{\nu\nu}^2 \left(-\frac{7 i  L_5}{30 \pi ^3} m^{10}\right)+k_{\nu } \eta_{\mu\nu}^2 \eta_{\nu\nu} \left(-\frac{2 i  L_5}{5 \pi ^3} m^{10}\right) }
\al{\label{eq:fnc:3:5:6:divr}\tilde{T}_{3,5;6\text{D}}^{\text{f,nt}}\cdot k & = k_{\mu } k_{\nu }^4 \eta_{\mu\mu} \left(-\frac{i  L_3}{12 \pi ^3} m^6\right)+k_{\mu }^3 k_{\nu }^2 \eta_{\nu\nu} \left(-\frac{i  L_3}{10 \pi ^3} m^6\right)+k_{\mu }^2 k_{\nu }^3 \eta_{\mu\nu} \left(-\frac{i  L_3}{15 \pi ^3} m^6\right)+
\0\\ & \quad 
 + k_{\mu }^3 \eta_{\nu\nu}^2 \left(-\frac{i  L_4}{20 \pi ^3} m^8\right)+k_{\nu }^3 \eta_{\mu\mu} \eta_{\mu\nu} \left(-\frac{i  L_4}{5 \pi ^3} m^8\right)+
\0\\ & \quad 
 + k_{\mu }^2 k_{\nu } \eta_{\mu\nu} \eta_{\nu\nu} \left(-\frac{2 i  L_4}{5 \pi ^3} m^8\right)+ \left(k_{\mu } k_{\nu }^2 \eta_{\mu\nu}^2+k_{\mu } k_{\nu }^2 \eta_{\mu\mu} \eta_{\nu\nu}\right) \left(-\frac{3 i  L_4}{10 \pi ^3} m^8\right)+
\0\\ & \quad 
 + k_{\mu } \eta_{\mu\mu} \eta_{\nu\nu}^2 \left(-\frac{7 i  L_5}{100 \pi ^3} m^{10}\right)+k_{\nu } \eta_{\mu\mu} \eta_{\mu\nu} \eta_{\nu\nu} \left(-\frac{7 i  L_5}{25 \pi ^3} m^{10}\right)+
\0\\ & \quad 
 + k_{\mu } \eta_{\mu\nu}^2 \eta_{\nu\nu} \left(-\frac{6 i  L_5}{25 \pi ^3} m^{10}\right)+k_{\nu } \eta_{\mu\nu}^3 \left(-\frac{4 i  L_5}{25 \pi ^3} m^{10}\right) }
Fermions, spin 4 x 4:
\al{\label{eq:fnc:4:4:3:div}k\cdot \tilde{T}_{4,4;3\text{D}}^{\text{f,nt}} & =  \left(k_{\mu } k_{\nu }^4 \eta_{\mu\mu}+k_{\mu }^2 k_{\nu }^3 \eta_{\mu\nu}\right) \left(-\frac{i }{\pi } m^3\right)+k_{\mu }^3 k_{\nu }^2 \eta_{\nu\nu} \left(-\frac{2 i }{\pi } m^3\right)+k_{\mu }^3 \eta_{\nu\nu}^2 \left(-\frac{8 i }{5 \pi } m^5\right)+
\0\\ & \quad 
 + k_{\nu }^3 \eta_{\mu\mu} \eta_{\mu\nu} \left(-\frac{4 i }{\pi } m^5\right)+k_{\mu }^2 k_{\nu } \eta_{\mu\nu} \eta_{\nu\nu} \left(-\frac{12 i }{\pi } m^5\right)+
\0\\ & \quad 
 +  \left(k_{\mu } k_{\nu }^2 \eta_{\mu\nu}^2+k_{\mu } k_{\nu }^2 \eta_{\mu\mu} \eta_{\nu\nu}\right) \left(-\frac{36 i }{5 \pi } m^5\right)+k_{\mu } \eta_{\mu\mu} \eta_{\nu\nu}^2 \left(-\frac{96 i }{35 \pi } m^7\right)+
\0\\ & \quad 
 +  \left(k_{\nu } \eta_{\mu\mu} \eta_{\mu\nu} \eta_{\nu\nu}+k_{\mu } \eta_{\mu\nu}^2 \eta_{\nu\nu}\right) \left(-\frac{48 i }{5 \pi } m^7\right)+k_{\nu } \eta_{\mu\nu}^3 \left(-\frac{192 i }{35 \pi } m^7\right)+
\0\\ & \quad 
 + k_{\mu }^2 k_{\nu }^3 (k\cdot\epsilon)_{\mu\nu} \left(-\frac{3 }{4 \pi } m^2\right)+k_{\nu }^3 (k\cdot\epsilon)_{\mu\nu} \eta_{\mu\mu} \left(-\frac{1}{\pi } m^4\right)+
\0\\ & \quad 
 + k_{\mu }^2 k_{\nu } (k\cdot\epsilon)_{\mu\nu} \eta_{\nu\nu} \left(-\frac{3 }{\pi } m^4\right)+k_{\mu } k_{\nu }^2 (k\cdot\epsilon)_{\mu\nu} \eta_{\mu\nu} \left(-\frac{6 }{\pi } m^4\right)+
\0\\ & \quad 
 + k_{\nu } (k\cdot\epsilon)_{\mu\nu} \eta_{\mu\mu} \eta_{\nu\nu} \left(-\frac{12 }{5 \pi } m^6\right)+
\0\\ & \quad 
 +  \left(k_{\nu } (k\cdot\epsilon)_{\mu\nu} \eta_{\mu\nu}^2+k_{\mu } (k\cdot\epsilon)_{\mu\nu} \eta_{\mu\nu} \eta_{\nu\nu}\right) \left(-\frac{24 }{5 \pi } m^6\right) }
\al{\label{eq:fnc:4:4:4:div}k\cdot \tilde{T}_{4,4;4\text{D}}^{\text{f,nt}} & =  \left(k_{\mu } k_{\nu }^4 \eta_{\mu\mu}+k_{\mu }^2 k_{\nu }^3 \eta_{\mu\nu}\right) \left(\frac{3 i  L_2}{8 \pi ^2} m^4\right)+k_{\mu }^3 k_{\nu }^2 \eta_{\nu\nu} \left(\frac{3 i  L_2}{4 \pi ^2} m^4\right)+
\0\\ & \quad 
 + k_{\mu }^3 \eta_{\nu\nu}^2 \left(\frac{i  L_3}{2 \pi ^2} m^6\right)+k_{\nu }^3 \eta_{\mu\mu} \eta_{\mu\nu} \left(\frac{5 i  L_3}{4 \pi ^2} m^6\right)+k_{\mu }^2 k_{\nu } \eta_{\mu\nu} \eta_{\nu\nu} \left(\frac{15 i  L_3}{4 \pi ^2} m^6\right)+
\0\\ & \quad 
 +  \left(k_{\mu } k_{\nu }^2 \eta_{\mu\nu}^2+k_{\mu } k_{\nu }^2 \eta_{\mu\mu} \eta_{\nu\nu}\right) \left(\frac{9 i  L_3}{4 \pi ^2} m^6\right)+k_{\mu } \eta_{\mu\mu} \eta_{\nu\nu}^2 \left(\frac{3 i  L_4}{4 \pi ^2} m^8\right)+
\0\\ & \quad 
 +  \left(k_{\nu } \eta_{\mu\mu} \eta_{\mu\nu} \eta_{\nu\nu}+k_{\mu } \eta_{\mu\nu}^2 \eta_{\nu\nu}\right) \left(\frac{21 i  L_4}{8 \pi ^2} m^8\right)+k_{\nu } \eta_{\mu\nu}^3 \left(\frac{3 i  L_4}{2 \pi ^2} m^8\right) }
\al{\label{eq:fnc:4:4:5:div}k\cdot \tilde{T}_{4,4;5\text{D}}^{\text{f,nt}} & =  \left(k_{\mu } k_{\nu }^4 \eta_{\mu\mu}+k_{\mu }^2 k_{\nu }^3 \eta_{\mu\nu}\right) \left(\frac{i }{5 \pi ^2} m^5\right)+k_{\mu }^3 k_{\nu }^2 \eta_{\nu\nu} \left(\frac{2 i }{5 \pi ^2} m^5\right)+k_{\mu }^3 \eta_{\nu\nu}^2 \left(\frac{8 i }{35 \pi ^2} m^7\right)+
\0\\ & \quad 
 + k_{\nu }^3 \eta_{\mu\mu} \eta_{\mu\nu} \left(\frac{4 i }{7 \pi ^2} m^7\right)+k_{\mu }^2 k_{\nu } \eta_{\mu\nu} \eta_{\nu\nu} \left(\frac{12 i }{7 \pi ^2} m^7\right)+
\0\\ & \quad 
 +  \left(k_{\mu } k_{\nu }^2 \eta_{\mu\nu}^2+k_{\mu } k_{\nu }^2 \eta_{\mu\mu} \eta_{\nu\nu}\right) \left(\frac{36 i }{35 \pi ^2} m^7\right)+k_{\mu } \eta_{\mu\mu} \eta_{\nu\nu}^2 \left(\frac{32 i }{105 \pi ^2} m^9\right)+
\0\\ & \quad 
 +  \left(k_{\nu } \eta_{\mu\mu} \eta_{\mu\nu} \eta_{\nu\nu}+k_{\mu } \eta_{\mu\nu}^2 \eta_{\nu\nu}\right) \left(\frac{16 i }{15 \pi ^2} m^9\right)+k_{\nu } \eta_{\mu\nu}^3 \left(\frac{64 i }{105 \pi ^2} m^9\right) }
\al{\label{eq:fnc:4:4:6:div}k\cdot \tilde{T}_{4,4;6\text{D}}^{\text{f,nt}} & =  \left(k_{\mu } k_{\nu }^4 \eta_{\mu\mu}+k_{\mu }^2 k_{\nu }^3 \eta_{\mu\nu}\right) \left(-\frac{i  L_3}{16 \pi ^3} m^6\right)+k_{\mu }^3 k_{\nu }^2 \eta_{\nu\nu} \left(-\frac{i  L_3}{8 \pi ^3} m^6\right)+
\0\\ & \quad 
 + k_{\mu }^3 \eta_{\nu\nu}^2 \left(-\frac{i  L_4}{16 \pi ^3} m^8\right)+k_{\nu }^3 \eta_{\mu\mu} \eta_{\mu\nu} \left(-\frac{5 i  L_4}{32 \pi ^3} m^8\right)+
\0\\ & \quad 
 + k_{\mu }^2 k_{\nu } \eta_{\mu\nu} \eta_{\nu\nu} \left(-\frac{15 i  L_4}{32 \pi ^3} m^8\right)+ \left(k_{\mu } k_{\nu }^2 \eta_{\mu\nu}^2+k_{\mu } k_{\nu }^2 \eta_{\mu\mu} \eta_{\nu\nu}\right) \left(-\frac{9 i  L_4}{32 \pi ^3} m^8\right)+
\0\\ & \quad 
 + k_{\mu } \eta_{\mu\mu} \eta_{\nu\nu}^2 \left(-\frac{3 i  L_5}{40 \pi ^3} m^{10}\right)+ \left(k_{\nu } \eta_{\mu\mu} \eta_{\mu\nu} \eta_{\nu\nu}+k_{\mu } \eta_{\mu\nu}^2 \eta_{\nu\nu}\right) \left(-\frac{21 i  L_5}{80 \pi ^3} m^{10}\right)+
\0\\ & \quad 
 + k_{\nu } \eta_{\mu\nu}^3 \left(-\frac{3 i  L_5}{20 \pi ^3} m^{10}\right) }
Fermions, spin 5 x 5:
\al{\label{eq:fnc:5:5:3:div}k\cdot \tilde{T}_{5,5;3\text{D}}^{\text{f,nt}} & = k_{\mu }^2 k_{\nu }^5 \eta_{\mu\mu} \left(-\frac{8 i }{5 \pi } m^3\right)+k_{\mu }^4 k_{\nu }^3 \eta_{\nu\nu} \left(-\frac{8 i }{3 \pi } m^3\right)+k_{\mu }^3 k_{\nu }^4 \eta_{\mu\nu} \left(-\frac{16 i }{15 \pi } m^3\right)+
\0\\ & \quad 
 + k_{\nu }^5 \eta_{\mu\mu}^2 \left(-\frac{32 i }{25 \pi } m^5\right)+k_{\mu }^2 k_{\nu }^3 \eta_{\mu\mu} \eta_{\nu\nu} \left(-\frac{512 i }{25 \pi } m^5\right)+k_{\mu }^4 k_{\nu } \eta_{\nu\nu}^2 \left(-\frac{32 i }{5 \pi } m^5\right)+
\0\\ & \quad 
 +  \left(k_{\mu } k_{\nu }^4 \eta_{\mu\mu} \eta_{\mu\nu}+k_{\mu }^2 k_{\nu }^3 \eta_{\mu\nu}^2\right) \left(-\frac{384 i }{25 \pi } m^5\right)+k_{\mu }^3 k_{\nu }^2 \eta_{\mu\nu} \eta_{\nu\nu} \left(-\frac{768 i }{25 \pi } m^5\right)+
\0\\ & \quad 
 + k_{\nu }^3 \eta_{\mu\mu}^2 \eta_{\nu\nu} \left(-\frac{1408 i }{175 \pi } m^7\right)+k_{\mu }^2 k_{\nu } \eta_{\mu\mu} \eta_{\nu\nu}^2 \left(-\frac{4224 i }{175 \pi } m^7\right)+
\0\\ & \quad 
 + k_{\mu } k_{\nu }^2 \eta_{\mu\mu} \eta_{\mu\nu} \eta_{\nu\nu} \left(-\frac{13824 i }{175 \pi } m^7\right)+k_{\mu }^3 \eta_{\mu\nu} \eta_{\nu\nu}^2 \left(-\frac{2816 i }{175 \pi } m^7\right)+
\0\\ & \quad 
 + k_{\nu }^3 \eta_{\mu\mu} \eta_{\mu\nu}^2 \left(-\frac{4096 i }{175 \pi } m^7\right)+k_{\mu }^2 k_{\nu } \eta_{\mu\nu}^2 \eta_{\nu\nu} \left(-\frac{12288 i }{175 \pi } m^7\right)+
\0\\ & \quad 
 + k_{\mu } k_{\nu }^2 \eta_{\mu\nu}^3 \left(-\frac{6144 i }{175 \pi } m^7\right)+k_{\nu } \eta_{\mu\mu}^2 \eta_{\nu\nu}^2 \left(-\frac{1024 i }{175 \pi } m^9\right)+
\0\\ & \quad 
 + k_{\mu } \eta_{\mu\mu} \eta_{\mu\nu} \eta_{\nu\nu}^2 \left(-\frac{4096 i }{175 \pi } m^9\right)+k_{\nu } \eta_{\mu\mu} \eta_{\mu\nu}^2 \eta_{\nu\nu} \left(-\frac{22528 i }{525 \pi } m^9\right)+
\0\\ & \quad 
 + k_{\mu } \eta_{\mu\nu}^3 \eta_{\nu\nu} \left(-\frac{45056 i }{1575 \pi } m^9\right)+k_{\nu } \eta_{\mu\nu}^4 \left(-\frac{4096 i }{315 \pi } m^9\right)+
\0\\ & \quad 
 + k_{\mu }^3 k_{\nu }^4 (k\cdot\epsilon)_{\mu\nu} \left(-\frac{4 }{5 \pi } m^2\right)+k_{\mu } k_{\nu }^4 (k\cdot\epsilon)_{\mu\nu} \eta_{\mu\mu} \left(-\frac{16 }{5 \pi } m^4\right)+
\0\\ & \quad 
 + k_{\mu }^3 k_{\nu }^2 (k\cdot\epsilon)_{\mu\nu} \eta_{\nu\nu} \left(-\frac{32 }{5 \pi } m^4\right)+k_{\mu }^2 k_{\nu }^3 (k\cdot\epsilon)_{\mu\nu} \eta_{\mu\nu} \left(-\frac{64 }{5 \pi } m^4\right)+
\0\\ & \quad 
 + k_{\mu } k_{\nu }^2 (k\cdot\epsilon)_{\mu\nu} \eta_{\mu\mu} \eta_{\nu\nu} \left(-\frac{384 }{25 \pi } m^6\right)+k_{\mu }^3 (k\cdot\epsilon)_{\mu\nu} \eta_{\nu\nu}^2 \left(-\frac{64 }{25 \pi } m^6\right)+
\0\\ & \quad 
 + k_{\nu }^3 (k\cdot\epsilon)_{\mu\nu} \eta_{\mu\mu} \eta_{\mu\nu} \left(-\frac{256 }{25 \pi } m^6\right)+
\0\\ & \quad 
 +  \left(k_{\mu } k_{\nu }^2 (k\cdot\epsilon)_{\mu\nu} \eta_{\mu\nu}^2+k_{\mu }^2 k_{\nu } (k\cdot\epsilon)_{\mu\nu} \eta_{\mu\nu} \eta_{\nu\nu}\right) \left(-\frac{768 }{25 \pi } m^6\right)+
\0\\ & \quad 
 + k_{\mu } (k\cdot\epsilon)_{\mu\nu} \eta_{\mu\mu} \eta_{\nu\nu}^2 \left(-\frac{768 }{175 \pi } m^8\right)+
\0\\ & \quad 
 +  \left(k_{\nu } (k\cdot\epsilon)_{\mu\nu} \eta_{\mu\mu} \eta_{\mu\nu} \eta_{\nu\nu}+k_{\mu } (k\cdot\epsilon)_{\mu\nu} \eta_{\mu\nu}^2 \eta_{\nu\nu}\right) \left(-\frac{3072 }{175 \pi } m^8\right)+
\0\\ & \quad 
 + k_{\nu } (k\cdot\epsilon)_{\mu\nu} \eta_{\mu\nu}^3 \left(-\frac{2048 }{175 \pi } m^8\right) }
\al{\label{eq:fnc:5:5:4:div}k\cdot \tilde{T}_{5,5;4\text{D}}^{\text{f,nt}} & = k_{\mu }^2 k_{\nu }^5 \eta_{\mu\mu} \left(\frac{3 i  L_2}{5 \pi ^2} m^4\right)+k_{\mu }^4 k_{\nu }^3 \eta_{\nu\nu} \left(\frac{i  L_2}{\pi ^2} m^4\right)+k_{\mu }^3 k_{\nu }^4 \eta_{\mu\nu} \left(\frac{2 i  L_2}{5 \pi ^2} m^4\right)+
\0\\ & \quad 
 + k_{\nu }^5 \eta_{\mu\mu}^2 \left(\frac{2 i  L_3}{5 \pi ^2} m^6\right)+k_{\mu }^2 k_{\nu }^3 \eta_{\mu\mu} \eta_{\nu\nu} \left(\frac{32 i  L_3}{5 \pi ^2} m^6\right)+k_{\mu }^4 k_{\nu } \eta_{\nu\nu}^2 \left(\frac{2 i  L_3}{\pi ^2} m^6\right)+
\0\\ & \quad 
 +  \left(k_{\mu } k_{\nu }^4 \eta_{\mu\mu} \eta_{\mu\nu}+k_{\mu }^2 k_{\nu }^3 \eta_{\mu\nu}^2\right) \left(\frac{24 i  L_3}{5 \pi ^2} m^6\right)+k_{\mu }^3 k_{\nu }^2 \eta_{\mu\nu} \eta_{\nu\nu} \left(\frac{48 i  L_3}{5 \pi ^2} m^6\right)+
\0\\ & \quad 
 + k_{\nu }^3 \eta_{\mu\mu}^2 \eta_{\nu\nu} \left(\frac{11 i  L_4}{5 \pi ^2} m^8\right)+k_{\mu }^2 k_{\nu } \eta_{\mu\mu} \eta_{\nu\nu}^2 \left(\frac{33 i  L_4}{5 \pi ^2} m^8\right)+
\0\\ & \quad 
 + k_{\mu } k_{\nu }^2 \eta_{\mu\mu} \eta_{\mu\nu} \eta_{\nu\nu} \left(\frac{108 i  L_4}{5 \pi ^2} m^8\right)+k_{\mu }^3 \eta_{\mu\nu} \eta_{\nu\nu}^2 \left(\frac{22 i  L_4}{5 \pi ^2} m^8\right)+
\0\\ & \quad 
 + k_{\nu }^3 \eta_{\mu\mu} \eta_{\mu\nu}^2 \left(\frac{32 i  L_4}{5 \pi ^2} m^8\right)+k_{\mu }^2 k_{\nu } \eta_{\mu\nu}^2 \eta_{\nu\nu} \left(\frac{96 i  L_4}{5 \pi ^2} m^8\right)+
\0\\ & \quad 
 + k_{\mu } k_{\nu }^2 \eta_{\mu\nu}^3 \left(\frac{48 i  L_4}{5 \pi ^2} m^8\right)+k_{\nu } \eta_{\mu\mu}^2 \eta_{\nu\nu}^2 \left(\frac{36 i  L_5}{25 \pi ^2} m^{10}\right)+
\0\\ & \quad 
 + k_{\mu } \eta_{\mu\mu} \eta_{\mu\nu} \eta_{\nu\nu}^2 \left(\frac{144 i  L_5}{25 \pi ^2} m^{10}\right)+k_{\nu } \eta_{\mu\mu} \eta_{\mu\nu}^2 \eta_{\nu\nu} \left(\frac{264 i  L_5}{25 \pi ^2} m^{10}\right)+
\0\\ & \quad 
 + k_{\mu } \eta_{\mu\nu}^3 \eta_{\nu\nu} \left(\frac{176 i  L_5}{25 \pi ^2} m^{10}\right)+k_{\nu } \eta_{\mu\nu}^4 \left(\frac{16 i  L_5}{5 \pi ^2} m^{10}\right) }
\al{\label{eq:fnc:5:5:5:div}k\cdot \tilde{T}_{5,5;5\text{D}}^{\text{f,nt}} & = k_{\mu }^2 k_{\nu }^5 \eta_{\mu\mu} \left(\frac{8 i }{25 \pi ^2} m^5\right)+k_{\mu }^4 k_{\nu }^3 \eta_{\nu\nu} \left(\frac{8 i }{15 \pi ^2} m^5\right)+k_{\mu }^3 k_{\nu }^4 \eta_{\mu\nu} \left(\frac{16 i }{75 \pi ^2} m^5\right)+
\0\\ & \quad 
 + k_{\nu }^5 \eta_{\mu\mu}^2 \left(\frac{32 i }{175 \pi ^2} m^7\right)+k_{\mu }^2 k_{\nu }^3 \eta_{\mu\mu} \eta_{\nu\nu} \left(\frac{512 i }{175 \pi ^2} m^7\right)+k_{\mu }^4 k_{\nu } \eta_{\nu\nu}^2 \left(\frac{32 i }{35 \pi ^2} m^7\right)+
\0\\ & \quad 
 +  \left(k_{\mu } k_{\nu }^4 \eta_{\mu\mu} \eta_{\mu\nu}+k_{\mu }^2 k_{\nu }^3 \eta_{\mu\nu}^2\right) \left(\frac{384 i }{175 \pi ^2} m^7\right)+k_{\mu }^3 k_{\nu }^2 \eta_{\mu\nu} \eta_{\nu\nu} \left(\frac{768 i }{175 \pi ^2} m^7\right)+
\0\\ & \quad 
 + k_{\nu }^3 \eta_{\mu\mu}^2 \eta_{\nu\nu} \left(\frac{1408 i }{1575 \pi ^2} m^9\right)+k_{\mu }^2 k_{\nu } \eta_{\mu\mu} \eta_{\nu\nu}^2 \left(\frac{1408 i }{525 \pi ^2} m^9\right)+
\0\\ & \quad 
 + k_{\mu } k_{\nu }^2 \eta_{\mu\mu} \eta_{\mu\nu} \eta_{\nu\nu} \left(\frac{1536 i }{175 \pi ^2} m^9\right)+k_{\mu }^3 \eta_{\mu\nu} \eta_{\nu\nu}^2 \left(\frac{2816 i }{1575 \pi ^2} m^9\right)+
\0\\ & \quad 
 + k_{\nu }^3 \eta_{\mu\mu} \eta_{\mu\nu}^2 \left(\frac{4096 i }{1575 \pi ^2} m^9\right)+k_{\mu }^2 k_{\nu } \eta_{\mu\nu}^2 \eta_{\nu\nu} \left(\frac{4096 i }{525 \pi ^2} m^9\right)+
\0\\ & \quad 
 + k_{\mu } k_{\nu }^2 \eta_{\mu\nu}^3 \left(\frac{2048 i }{525 \pi ^2} m^9\right)+k_{\nu } \eta_{\mu\mu}^2 \eta_{\nu\nu}^2 \left(\frac{1024 i }{1925 \pi ^2} m^{11}\right)+
\0\\ & \quad 
 + k_{\mu } \eta_{\mu\mu} \eta_{\mu\nu} \eta_{\nu\nu}^2 \left(\frac{4096 i }{1925 \pi ^2} m^{11}\right)+k_{\nu } \eta_{\mu\mu} \eta_{\mu\nu}^2 \eta_{\nu\nu} \left(\frac{2048 i }{525 \pi ^2} m^{11}\right)+
\0\\ & \quad 
 + k_{\mu } \eta_{\mu\nu}^3 \eta_{\nu\nu} \left(\frac{4096 i }{1575 \pi ^2} m^{11}\right)+k_{\nu } \eta_{\mu\nu}^4 \left(\frac{4096 i }{3465 \pi ^2} m^{11}\right) }
\al{\label{eq:fnc:5:5:6:div}k\cdot \tilde{T}_{5,5;6\text{D}}^{\text{f,nt}} & = k_{\mu }^2 k_{\nu }^5 \eta_{\mu\mu} \left(-\frac{i  L_3}{10 \pi ^3} m^6\right)+k_{\mu }^4 k_{\nu }^3 \eta_{\nu\nu} \left(-\frac{i  L_3}{6 \pi ^3} m^6\right)+k_{\mu }^3 k_{\nu }^4 \eta_{\mu\nu} \left(-\frac{i  L_3}{15 \pi ^3} m^6\right)+
\0\\ & \quad 
 + k_{\nu }^5 \eta_{\mu\mu}^2 \left(-\frac{i  L_4}{20 \pi ^3} m^8\right)+k_{\mu }^2 k_{\nu }^3 \eta_{\mu\mu} \eta_{\nu\nu} \left(-\frac{4 i  L_4}{5 \pi ^3} m^8\right)+
\0\\ & \quad 
 + k_{\mu }^4 k_{\nu } \eta_{\nu\nu}^2 \left(-\frac{i  L_4}{4 \pi ^3} m^8\right)+ \left(k_{\mu } k_{\nu }^4 \eta_{\mu\mu} \eta_{\mu\nu}+k_{\mu }^2 k_{\nu }^3 \eta_{\mu\nu}^2\right) \left(-\frac{3 i  L_4}{5 \pi ^3} m^8\right)+
\0\\ & \quad 
 + k_{\mu }^3 k_{\nu }^2 \eta_{\mu\nu} \eta_{\nu\nu} \left(-\frac{6 i  L_4}{5 \pi ^3} m^8\right)+k_{\nu }^3 \eta_{\mu\mu}^2 \eta_{\nu\nu} \left(-\frac{11 i  L_5}{50 \pi ^3} m^{10}\right)+
\0\\ & \quad 
 + k_{\mu }^2 k_{\nu } \eta_{\mu\mu} \eta_{\nu\nu}^2 \left(-\frac{33 i  L_5}{50 \pi ^3} m^{10}\right)+k_{\mu } k_{\nu }^2 \eta_{\mu\mu} \eta_{\mu\nu} \eta_{\nu\nu} \left(-\frac{54 i  L_5}{25 \pi ^3} m^{10}\right)+
\0\\ & \quad 
 + k_{\mu }^3 \eta_{\mu\nu} \eta_{\nu\nu}^2 \left(-\frac{11 i  L_5}{25 \pi ^3} m^{10}\right)+k_{\nu }^3 \eta_{\mu\mu} \eta_{\mu\nu}^2 \left(-\frac{16 i  L_5}{25 \pi ^3} m^{10}\right)+
\0\\ & \quad 
 + k_{\mu }^2 k_{\nu } \eta_{\mu\nu}^2 \eta_{\nu\nu} \left(-\frac{48 i  L_5}{25 \pi ^3} m^{10}\right)+k_{\mu } k_{\nu }^2 \eta_{\mu\nu}^3 \left(-\frac{24 i  L_5}{25 \pi ^3} m^{10}\right)+
\0\\ & \quad 
 + k_{\nu } \eta_{\mu\mu}^2 \eta_{\nu\nu}^2 \left(-\frac{3 i  L_6}{25 \pi ^3} m^{12}\right)+k_{\mu } \eta_{\mu\mu} \eta_{\mu\nu} \eta_{\nu\nu}^2 \left(-\frac{12 i  L_6}{25 \pi ^3} m^{12}\right)+
\0\\ & \quad 
 + k_{\nu } \eta_{\mu\mu} \eta_{\mu\nu}^2 \eta_{\nu\nu} \left(-\frac{22 i  L_6}{25 \pi ^3} m^{12}\right)+k_{\mu } \eta_{\mu\nu}^3 \eta_{\nu\nu} \left(-\frac{44 i  L_6}{75 \pi ^3} m^{12}\right)+
\0\\ & \quad 
 + k_{\nu } \eta_{\mu\nu}^4 \left(-\frac{4 i  L_6}{15 \pi ^3} m^{12}\right) }